\documentclass[%
 aip,
 amsmath,amssymb,
 reprint,%
 cha,%
floatfix
]{revtex4-2}

\usepackage{graphicx}
\usepackage{xcolor}
\usepackage[utf8]{inputenc}
\usepackage[T1]{fontenc}
\usepackage[english]{babel}
\usepackage{lineno}
\usepackage{siunitx} 
\usepackage{amsmath}
\usepackage{physics}

\usepackage{enumitem}
\makeatletter
\def\@email#1#2{%
 \endgroup
 \patchcmd{\titleblock@produce}
  {\frontmatter@RRAPformat}
  {\frontmatter@RRAPformat{\produce@RRAP{*#1\href{mailto:#2}{#2}}}\frontmatter@RRAPformat}
  {}{}
}%
\makeatother
\begin{document}
\title[Perspectives on adaptive dynamical systems]{Perspectives on adaptive dynamical systems}

\author{Jakub Sawicki$^*$}
\affiliation{$^*$\textbf{Authors who contributed equally and to whom correspondence should be addressed. JS: zergon@gmx.net; RB: {rico.berner@physik.hu-berlin.de}; SL: {sl2127@cam.ac.uk}.}}
\affiliation{Potsdam Institute for Climate Impact Research, Telegrafenberg, 14473 Potsdam, Germany}
\affiliation{Fachhochschule Nordwestschweiz FHNW, Leonhardsstrasse 6, 4009 Basel, Switzerland}

\author{Rico Berner$^*$}
\affiliation{$^*$\textbf{Authors who contributed equally and to whom correspondence should be addressed. JS: zergon@gmx.net; RB: {rico.berner@physik.hu-berlin.de}; SL: {sl2127@cam.ac.uk}.}}
\affiliation{
Department of Physics, Humboldt-Universit\"at zu Berlin, Newtonstraße 15, 12489 Berlin, Germany
}

\author{Sarah A. M. Loos$^*$}
\affiliation{$^*$\textbf{Authors who contributed equally and to whom correspondence should be addressed. JS: zergon@gmx.net; RB: {rico.berner@physik.hu-berlin.de}; SL: {sl2127@cam.ac.uk}.}}
\affiliation{DAMTP, 
University of Cambridge, Wilberforce Road, Cambridge CB3 0WA, United Kingdom
}

\author{Mehrnaz Anvari}
\affiliation{Potsdam Institute for Climate Impact Research, Telegrafenberg, 14473 Potsdam, Germany}
\affiliation{Fraunhofer Institute for Algorithms and Scientific Computing, Schloss Birlinghoven, 53757 Sankt-Augustin, Germany}
\author{Rolf Bader}
\affiliation{Institute of Systematic Musicology, University of Hamburg, Germany}
\author{Wolfram Barfuss}
\affiliation{Transdisciplinary Research Area: Sustainable Futures, University of Bonn, 53113 Bonn, Germany}
\affiliation{Center for Development Research (ZEF), University of Bonn, 53113 Bonn, Germany}
\author{Nicola Botta}
\affiliation{Potsdam Institute for Climate Impact Research, Telegrafenberg, 14473 Potsdam, Germany}
\affiliation{Department of Computer Science and Engineering, Chalmers University of Technology, 412 96 G\"oteborg, Sweden}
\author{Nuria Brede}
\affiliation{Potsdam Institute for Climate Impact Research, Telegrafenberg, 14473 Potsdam, Germany}
\affiliation{Department of Computer Science, University of Potsdam, An der Bahn 2, 14476 Potsdam, Germany}
\author{Igor Franovi\'c}
\affiliation{Scientific Computing Laboratory, Center for the Study of Complex Systems,
Institute of Physics Belgrade, University of Belgrade, Pregrevica 118, 11080 Belgrade, Serbia}
\author{Daniel J. Gauthier}
\affiliation{The Ohio State University, Department of Physics and Electrical and Computer Engineering, 191 West Woodruff Ave. Columbus, OH 43210, USA}
\affiliation{ResCon Technologies, LLC, 1275 Kinnear Rd., Suite 239, Columbus, OH 43212, USA}
\author{Sebastian Goldt}
\affiliation{Department of Physics, International School of Advanced Studies (SISSA), Trieste, Italy}
\author{Aida Hajizadeh}
\affiliation{Research Group Comparative Neuroscience, Leibniz Institute for Neurobiology, Magdeburg, Germany} 
\author{Philipp H\"ovel}
\affiliation{Theoretical Physics, Saarland University, 66123 Saarbr\"ucken, Germany}
\affiliation{Center for Biophysics, Saarland University, 66123 Saarbr\"ucken, Germany}
\author{Omer Karin}
\affiliation{Department of Mathematics, Imperial College London, UK}
\author{Philipp Lorenz-Spreen}
\affiliation{Center for Adaptive Rationality, Max Planck Institute for Human Development, Lentzeallee 94, 14195 Berlin, Germany}
\author{Christoph Miehl}
\affiliation{Max Planck Institute for Brain Research, Max-von-Laue-Straße 4, 60438 Frankfurt am Main, Germany}
\affiliation{Technical University of Munich, School of Life Sciences, Alte Akademie 8, 85354 Freising, Germany}
\author{Jan M\"olter}
\affiliation{Department of Mathematics, School of Computation, Information and Technology, Technical University of Munich, Boltzmannstra{\ss}e 3, 85748 Garching bei M\"unchen, Germany}
\author{Simona Olmi}
\affiliation{CNR, Consiglio Nazionale delle Ricerche, Istituto dei Sistemi Complessi, via Madonna del Piano 10, 50019 Sesto Fiorentino, Italy}
\author{Eckehard Sch\"oll}
\affiliation{Potsdam Institute for Climate Impact Research, Telegrafenberg, 14473 Potsdam, Germany}
\affiliation{Institut f{\"u}r Theoretische Physik, Technische Universit{\"a}t Berlin, Hardenbergstra\ss{}e 36, 10623 Berlin, Germany}
\affiliation{Bernstein Center for Computational Neuroscience Berlin, Humboldt-Universit{\"a}t, 10115 Berlin, Germany}
\author{Alireza Seif}
\affiliation{Pritzker School of Molecular Engineering, The University of Chicago, Illinois 60637, USA}
\author{Peter A. Tass}
\affiliation{Department of Neurosurgery, Stanford University School of Medicine, Stanford, CA, USA}
\author{Giovanni Volpe}
\affiliation{Department of Physics, University of Gothenburg, Gothenburg, Sweden}
\author{Serhiy Yanchuk}
\affiliation{Potsdam Institute for Climate Impact Research, Telegrafenberg, 14473 Potsdam, Germany}
\affiliation{Department of Mathematics, Humboldt-Universit\"at zu Berlin, Rudower Chaussee 25, 12489 Berlin, Germany}
\author{J\"urgen Kurths}
\affiliation{Potsdam Institute for Climate Impact Research, Telegrafenberg, 14473 Potsdam, Germany}
\affiliation{
Department of Physics, Humboldt-Universit\"at zu Berlin, Newtonstraße 15, 12489 Berlin, Germany}

\date{\today}

\begin{abstract}
Adaptivity is a dynamical feature that is omnipresent in nature, socio-economics, and technology. For example, adaptive couplings appear in various real-world systems like the power grid, social, and neural networks, and they form the backbone of closed-loop control strategies and machine learning algorithms. 
In this article, we provide an interdisciplinary perspective on adaptive systems. We reflect on the notion and terminology of adaptivity in different disciplines and discuss which role adaptivity plays for various fields. We highlight common open challenges, and give perspectives on future research directions, looking to inspire interdisciplinary approaches.
\tableofcontents
\end{abstract}

\maketitle
\begin{quotation}
Charles Darwin taught us that it is not the strongest of a species that survive -- but the ones who are most \textit{adaptable} to change. Likewise, the process of learning can be considered to be ``any change in a system that produces a more or less permanent change in its capacity for adapting to its environment.''\cite{SIM69}. 
These two statements clearly underline the importance of adaptivity for life. Simply speaking, one could say:
``To live means to adapt.'' At the same time, adaptive mechanisms are also the essential features of (`intelligent') artificial systems, from state-of-the-art control techniques for complex systems, to machine learning approaches and robotic systems. 
Perhaps the most basic notion of adaptivity is the ability to adjust to condition or change over time. This ability is an essential component of various natural and artificial processes considered in different research fields. 
It is also the key property of the human mind to perceive and enjoy music and visual arts, and to create and invent, and thus is the driving force behind all cultural achievements.
Adaptive mechanisms take place on a wide range of spatial and temporal scales, from the adaptation of a single neuron, over the ability of a social system to adjust to a changing environment, 
up to the adaptation of the Earth system's climate.
Over the last decades, substantial know-how to describe and control complex systems has been developed in different scientific areas.
With the increasing potential of modern technology, on the one hand, and the enormous challenges facing humanity as a large social system, on the other hand, there is a renewed interest to take an interdisciplinary approach to adaptivity. 
This article gives an overview of the role of adaptive systems in different scientific fields and highlights prospects for future research directions on adaptivity.

\end{quotation}



\section{\label{sec:intro}Introduction}
A widespread feature of natural and artificial complex systems is their adaptivity. There is lively interest in modeling and understanding the various forms of adaptive mechanisms appearing in real-world systems and to develop new control strategies based on adaptive mechanisms.

Such control strategies play an essential role, especially in \emph{complex systems science}, as they reflect to some extent the understanding we have of a complex system.  Because of their interactions, relationships, dependencies, nonlinearities, and high-dimensionalities,
the behavior of complex systems is inherently difficult to model. Machine Learning tools are often used to solve predictions about complex systems. However, applying Machine Learning to complex systems is quite challenging because the training data set has to reflect the diverse dynamics. This usually results in the data set being very large, making such methods well suited for so-called Big Data. 

\begin{figure}
    \centering
    \includegraphics[width=.5\textwidth]{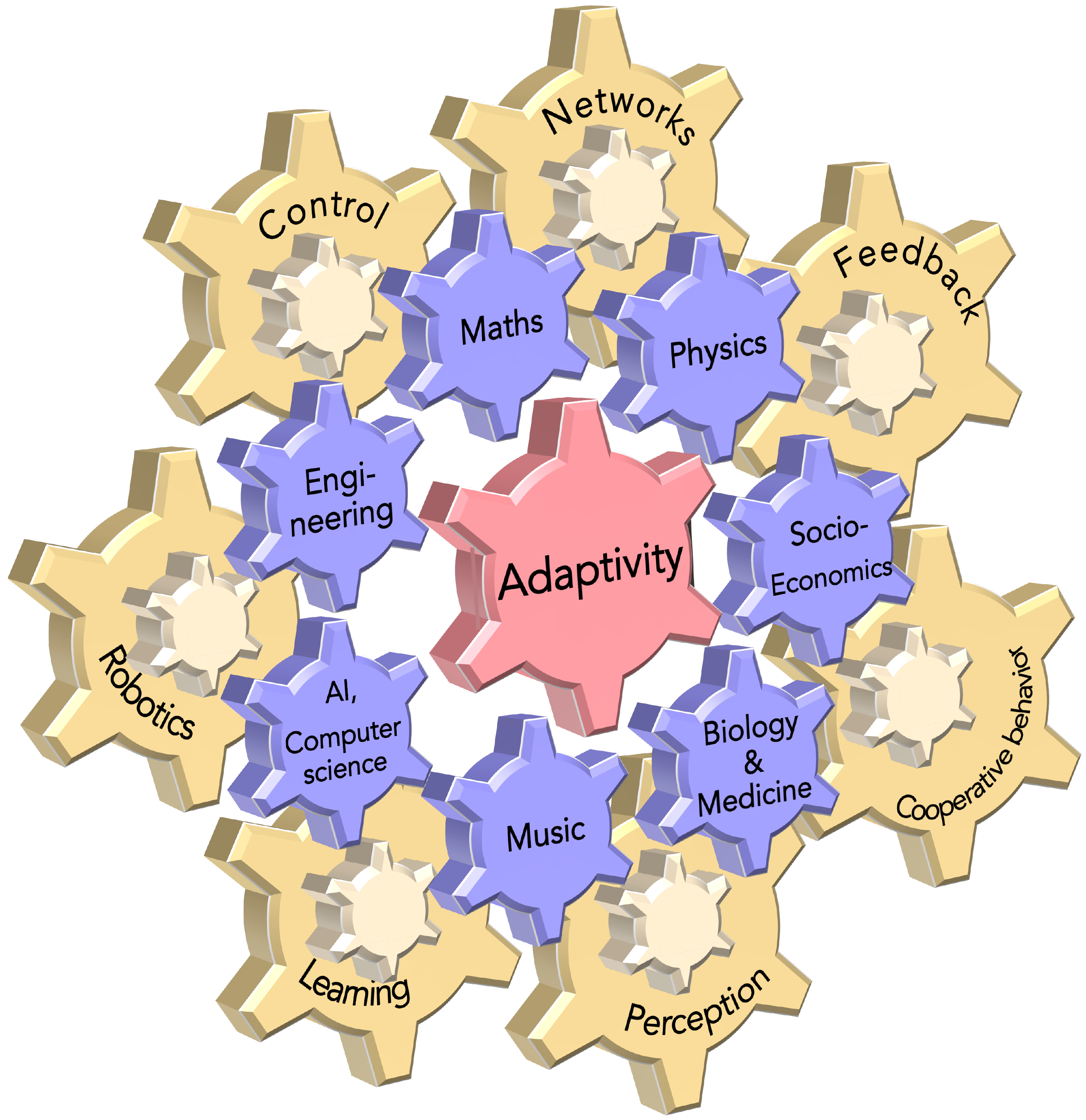}
    \caption{Adaptivity across different scientific disciplines (blue) and applications (yellow) as well as its strong interlinking and interlocking, similar to a system of gears.}
    \label{fig:gears}
\end{figure}

Moreover, the focus today is not only on complex systems consisting of many interacting components, but as an interdisciplinary field, complex systems actually attract contributions from many different fields. Despite the strong drive for innovation and application of adaptive complex systems in various scientific fields, as conceptualized in Fig.\,\ref{fig:gears}, cross-fertilization between different disciplines is hardly promoted. A partial answer towards a mathematical theory of adaptive systems has been developed since the 1960s for control and optimization problems~\cite{YAK68,YAK68a,fomin1981adaptive,ANN21,FRAD22}, including stochastic systems \cite{tsoepkin1971adaptation}, a systematic exposition of the interrelations and interplay between adaptation and learning~\cite{tsoepkin1971adaptation} as well as the use of the speed gradient method \cite{Fradkov2007} in adaptive control of network topology~\cite{LEH14}. In this review we discuss recent interdisciplinary applications of adaptive dynamical systems and focus on collecting ideas that would allow for including modern research fields such as complex network theory, power grid modeling, or climate systems where a full mathematical theory is still elusive.


This perspective article aims to make a first step in opening a dialogue between different scientific communities and the diverse formalism of their languages. It summarizes different perspectives on the concept of adaptivity and shows which open challenges are waiting to be taken up. To this end, it brings together the viewpoints  on the topic of adaptivity of researchers from a wide range of backgrounds including physics, biology, mathematics, computer and social science, and musicology.
%
This perspective article 
features a collection of contributions from experts representing various scientific disciplines.
The individual contributions are guided by the following questions:
\begin{enumerate}
    \item What role do adaptive mechanisms play in their respective field? How can one \emph{define} adaptivity? What \emph{methods} are related to adaptivity? What \emph{applications} are related to adaptivity?

    \item Which \emph{challenges} can be solved by using adaptive mechanisms? Are there \emph{open research questions} related to adaptivity? What are the \emph{future perspectives}?
\end{enumerate}

The article consists of four main topical parts: Network perspective and the modeling of adaptivity (Sec.\,\ref{SEC:NETWORKS-ADAPTIVTY}), Perception and neural adaptivity (Sec.\,\ref{SEC:NEURAL}), Adaptivity and artificial learning (Sec.\,\ref{SEC:ARTIFICIAL-LEARNING}), and Adaptivity in socio-economic systems (Sec.\,\ref{SEC:SOCIO-ECONOMIC}). Each part contains perspectives from several specialists active in the respective area of research.

\newblock
In the first part (Sec.\,\ref{SEC:NETWORKS-ADAPTIVTY}),  we discuss different ideas on the definition of adaptivity from the perspective of nonlinear dynamics, control theory and network science, and how adaptive systems can be used to understand real-world systems of interacting units (networks). In the beginning, a generic viewpoint on adaptivity with regards to the interplay of structure and function in dynamical network theory is introduced (Sec.\,\ref{SEC:YANCHUK}). Building upon this idea, adaptation is discussed as a slowly evolving feedback mechanism (Sec.\,\ref{SEC:FRANOVIC}). Further highlighted are the interplay of adaptivity and noise as well as the role of adaptive control mechanisms in inducing critical transitions. Complementing the discussion on the notion of adaptivity, the question is raised: Is adaptivity in nonlinear dynamics, neuroscience, artificial intelligence and socio-economic dynamics instances of the same abstract notion? To answer this question, the framework of dependent type theory is introduced and suggested to be utilized for comparing different notions of adaptivity (Sec.\,\ref{SEC:BOTTA}). The last section summarizes the first part from the complex networks perspective where the interplay between dynamics and network topology is in the center of interest (Sec.\,\ref{SEC:SCHOELL}). Here, various connections between models featuring adaptivity  are shown and adaptive network models are highlighted as a powerful modeling approach towards real-world dynamical systems.

\newblock
The second part (Sec.~\ref{SEC:NEURAL}) focuses on the important role of adaptation in physiology, especially in the form of perception mechanisms and neuronal plasticity. Evolution tends to come up with similar solutions to related problems. 
The physiological properties of biological systems can be seen as complex networks of interactions which are known as regulatory networks. Under similar contexts, such regulatory networks of distinct systems share similarities -- these are so called adaptation motifs, where specific adaption motifs have distinct functional significance (Sec.\,\ref{SEC:KARIN}).
Organisms, and hence their brains, have developed strategies to adapt to modifications in the environment across timescales, from adaptation to sudden changes in sensory stimuli to long timescales of evolutionary processes. Also, learning and memory formation can be viewed as adaptive processes, where learning in neuronal circuits relies on short- and specifically long-term synaptic plasticity (Sec.\,\ref{SEC:MIEHL}).
Neuronal systems often consist of millions of neurons whose individual dynamics are often not accessible with mathematical methods. However, for the macroscopic collective dynamics emerging in such systems, several methodologies have been developed. A powerful method is the next generation neural mass approach which allows for a low-dimensional reduction of neuronal populations equipped with frequency adaptation and short term plasticity (Sec.\,\ref{SEC:OLMI}).
Computational models have proven to be useful for understanding the mechanisms underlying adaptation mechanisms in the brain. In medicine, for example deep brain stimulation is the gold standard for treating medically refractory Parkinson’s patients who suffer from various motor and non-motor-symptoms and display an abnormal neuronal synchrony. Considering synaptic plasticity in computational modeling enables to design appropriate therapeutic stimulation (Sec.\,\ref{SEC:TASS}).
Music is a constant adaptation process, where adaptations are active processes, including changing strategies, emotional reactions, or the development of new abilities. A physical culture theory is assuming music as an adaptive system to be represented by spatio-temporal electric fields in the brain, consisting of impulses, physical energy bursts, sent out, returning with a certain damping, thereby causing new impulses (Sec.\,\ref{SEC:BADER}).
In experiments, the magnitude of the neural response in the auditory cortex is decreasing if the same stimulus is presented repetitively with a constant stimulus onset interval. The gradual reduction of the magnitude is termed adaptation and is is suggested to be due to modulations of synaptic coupling between neurons (Sec.\,\ref{SEC:HAJIZADEH}).

\newblock
Another wide field where adaptivity plays a key role is artificial intelligence and machine learning. We illuminate this field in the third part of the article, Section~\ref{SEC:ARTIFICIAL-LEARNING}. Indeed, at its very heart, ``learning'' means ``adapting'' to input data. The adapting system can thereby be for example a real or ``artificial brain'' such as a neural network, and the adaptation rules may depend on the learning task, network architecture, and learning algorithm.
Section \ref{SEC:ARTIFICIAL-LEARNING} provides a variety of perspectives on adaptivity in artificial learning, discussing current research, new applications, and open challenges. The methods span from deep neural networks (Sec.\,\ref{SEC:GOLDT}) and recurrent neural networks (Sec.\,\ref{SEC:SEIF}), reinforcement learning (Sec.\,\ref{SEC:BARFUSS}) to
reservoir computing (Sec.\,\ref{SEC:GAUTHIER}). 
A common focus throughout the section \ref{SEC:ARTIFICIAL-LEARNING} is the two-way relationship between natural sciences and machine learning.
On the one hand, 
tools from theoretical physics may provide insights into the functionality of machine learning algorithms, pushing our understanding beyond the `black box' paradigm. In particular, concepts from statistical physics are explored to address fundamental questions, like reconciling the success of artificial learning with the curse of dimensionality (see Sec.\,\ref{SEC:GOLDT}). Second, simple models inspired from physics are used to generate training data to probe specific features of machine learning algorithms, such as their ability to extract and utilize memory of a given input sequence (see Sec.\,\ref{SEC:SEIF}).
On the other hand, the usage of
machine learning tools to investigate (Secs.\,\ref{SEC:SEIF}, \ref{SEC:BARFUSS}) or to control (Secs.\,\ref{SEC:GAUTHIER}, \ref{SEC:VOLPE}) complex physical systems is a field of rapidly growing relevance. A sticking example is how reservoir computing techniques open up new strategies to control chaotic nonlinear dynamics (Sec.\,\ref{SEC:GAUTHIER}). In this context, another major challenge concerns the exploration of the rules of (and the control of) the collective or co-operative behavior of self-organizing multi-agent systems; from the design of new algorithms (Sec.\,\ref{SEC:BARFUSS}) to the control of real-world microscopic `biomimetic' intelligent particles and swarms of robots (Sec.\,\ref{SEC:VOLPE}).

\newblock

The last part of the article (Sec.~\ref{SEC:SOCIO-ECONOMIC}) is devoted to the large field of socio-economic system. Here, adaptive mechanisms appear naturally and play an important role for their modeling. Adaptive networks also play a central role not only for realistic investigations of spreading dynamics but can help to study and design interventions for disease containment, mitigation, and eradication. Elaborating on this, in the last section of this fourth part, an overview on adaptivity in epidemiology is provided (Sec.\,\ref{SEC:HOEVEL}). Another interesting topic is the interaction of social and epidemic systems where also the coevolutionary (adaptive) dynamics of the interaction structure and the dynamical units is in the focus of recent research (Sec.\,\ref{SEC:MOELTER}). Apart from the connection to epidemiology, social systems themselves are adaptive. Here, adaptivity can be regarded as the process of changing social systems through external influences. In this context, understanding these changes induced by an increasing connectivity through online platforms or an increasing availability of information are driving research questions (Sec.\,\ref{SEC:LORENZSPREEN}). The human factor is also considerably important for the (adaptive) control of power grids, e.g., considering a temporally changing energy consumption (Sec.\,\ref{SEC:ANVARI}). The challenges in order to be compatible with new circumstances are discussed from different viewpoints. In power grid systems, we find the adaptation of both the topology and dynamics of the grid. On the other side, there is an anthropogenic influences on Earth system (Sec.\,\ref{SEC:KURTHS}). Here, we can learn much from the past about adaptive mechanisms in this complex system and the perturbations to which it is subjected. With this the last section of this article provides challenging open research questions that could be solved by using adaptivity one or the other way. 
%
%

\section{Network perspective and models of adaptivity} 
In this section, different ideas are discussed on how adaptivity can be defined in the context of nonlinear dynamics, control theory and network science, and how adaptive systems could be used to understand real-world systems of interacting units (networks). Perspectives are provided on how different dynamical models featuring adaptive mechanism are related and how these models can be used to investigate the dynamics of natural or man-made systems.
\label{SEC:NETWORKS-ADAPTIVTY}
\subsection{Structural adaptivity in dynamical networks -- by Serhiy Yanchuk}\label{SEC:YANCHUK}

Adaptivity is a general concept commonly understood as a process or
ability of a system to adjust itself to changing (external) conditions.
Thus, when speaking of adaptivity, one implicitly distinguishes the
`conditions' $(X)$ and the adaptation property $(Y)$. In the
following, an attempt is made to define these two variables (components)
with special reference to the theory of adaptive dynamical networks.
\begin{itemize}
\item \textbf{The structure} $Y$ is the adaptation matter, the part of
the system responsible for the adaptation properties. In adaptive
dynamical networks, this is usually understood as a network structure
represented by connectivity and/or connection weights. By analogy
with dynamical networks and neuroscience in general, we refer to this
variable as \emph{structure}. 
\item \textbf{The function} $X$ represents the conditions that trigger
the adaptation. In adaptive dynamical networks, this is usually the
dynamic state of the network, i.e., the collective and individual
dynamics of the nodes. This factor may also include stochastic or
external perturbations. These variables usually change with time,
i.e., $X(t)$ in the case of temporal adaptation. Following the terminology
of the dynamical networks, we generally refer to this variable as \emph{function}. 
\end{itemize}
The non-adaptive systems correspond to a constant structure $Y=Y_{0}$
which is independent of the function $X(t)$. By assuming that $X$
is governed by a system of differential equations, a general representation
of a \emph{non-adaptive} system is 
\begin{align}
\dot{X}(t) & =f(X,Y),\label{eq:sy-1}\\
\dot{Y} & =0.\label{eq:sy-2}
\end{align}
We assume here the general case that the structure $Y$ influences
the function $X$. Systems (\ref{eq:sy-1})--(\ref{eq:sy-2}) are
often used for modeling neural networks with fixed connectivity
$Y$. An example of a \emph{non-adaptive} dynamical network is the
coupled system 
\[
\dot{x}_{i}=f_{i}(x_{i},t)+\sum_{j=1}^{N}\kappa_{ij}g_{ij}(x_{i},x_{j}),
\]
where $x_{i}(t)$ determines the state of node $i=1,\dots,N$ and
$\kappa_{ij}$ is the connection weight ($\kappa_{ij}=0$ if there is no
connection). The absence of network adaptivity is indicated by the
fixed structure $\kappa_{ij}$. The function variable in this example is
$X=\left(x_{1},\dots,x_{N}\right)$ while the structure variable is
$Y=\left\{ \kappa_{ij}\right\} _{i,j=1,\dots,N}$, and it is constant.
The class of non-adaptive networks is extremely useful for modeling
many processes and phenomena in nature and technology \cite{PIK01,BOC18,Yanchuk2021a}, see also Secs.~\ref{SEC:SCHOELL},~\ref{SEC:HOEVEL},~\ref{SEC:MOELTER} and~\ref{SEC:LORENZSPREEN}.

When the structure depends on the function, we obtain an adaptive
system 
\begin{align}
\dot{X}(t) & =f(X,Y),\label{eq:sy-1-1}\\
\dot{Y}(t) & =g(X,Y),\label{eq:sy-2-1}
\end{align}
with a mutual structure-function interaction \cite{Cabral2022}.

An example of an adaptive dynamical network is 
\begin{align}
\dot{x_{i}}=f_{i}(x_{i},t)+\sum_{j=1}^{N}\kappa_{ij}g(x_{i},x_{j}),\label{eq:es-3-1}\\
\dot{\kappa}_{ij}=h(x_{i},x_{j},\kappa_{ij}),\label{eq:sy-3}
\end{align}
where the rule (\ref{eq:sy-3}) is responsible for the adaptation
and the temporal changes of the structure $Y$. The rule (\ref{eq:sy-3})
is the case when the connection weight between node $i$ and node
$j$ depends only on the function of these nodes $x_{i}(t)$ and $x_{j}(t)$.
Of course, this is not the only possible adaptation rule. Particular
realisations of the adaptation rule (\ref{eq:sy-3}) are neuronal
systems with plasticity. Specifically, when the plasticity is long-term,
i.e., the structural changes act on a slower timescale than the functional
dynamics (neuronal spiking) \cite{Abbott2000,Dan2004,MLHBT07,PYT13,LUE16},
this leads to systems with multiple timescales. As a representative
system, the paradigmatic adaptive network of phase oscillators
\begin{align}
\dot{\phi}_{i}=\omega_{i}-\sum_{j=1}^{N}\kappa_{ij}\sin(\phi_{i}-\phi_{j}+\alpha),\label{eq:sy-phase-1}\\
\dot{\kappa}_{ij}=-\varepsilon\left(\kappa_{ij}+\sin(\phi_{i}-\phi_{j}+\beta)\right),\label{eq:sy-phase-2}
\end{align}
appears to be very useful to study various phenomena in adaptive networks,
such as synchronization, frequency clustering, recurrent synchronization,
adaptivity-induced resistance to noise, and others \cite{Aoki2009,PYT13,KAS17,BER21b,Berner2021a,BER21f,Thiele2021}. Equations \eqref{eq:sy-phase-1}--\eqref{eq:sy-phase-2} are a special case of the more general Eqs. \eqref{eq:APO_phi}-- \eqref{eq:APO_kappa} in Sec.~\ref{SEC:SCHOELL}, see also the examples discussed there.
All of these phenomena are also revealed in more realistic and complex
models such as Hodgkin-Huxley neurons with spike-timing-dependent
plasticity \cite{ROE19a,PYT13}. Thus, paradigmatic models
of the type (\ref{eq:sy-phase-1})--(\ref{eq:sy-phase-2}) have demonstrated
their effectiveness in studying and predicting novel phenomena characteristic
for large classes of adaptive networks. 

The main challenges in studying the above classes of adaptive dynamical
networks are as follows: 
\begin{itemize}
\item High dimensionality. If the number of nodes in the network is $N$,
the number of possible connections is $N^{2}$. Thus, the dimensionality
of the model increases dramatically compared to dynamical networks
with a fixed structure. 
\item If the adaptation is slow, i.e., $\varepsilon\ll1$ in Eq.~(\ref{eq:sy-phase-2}),
the system becomes multi-scale with the slow manifold of dimension
$N^{2}$. This additional multiscale structure provides opportunities
for analysis \cite{K15}, but for large networks, it goes far
beyond the standard results employing geometric singular perturbation
theory. 
\end{itemize}
Despite recent advances in the study of dynamical adaptive networks,
many challenging problems remain unsolved. These problems include
mean-field theory, application to climate network modeling, understanding
the role of adaptivity in machine learning, developing dimensionality
reduction techniques, particularly methods for dealing with extremely
high-dimensional slow manifolds. Besides large networks, small networks
with adaptivity appear to have a highly nontrivial bifurcation structure
compared to their non-adaptive counterparts. Studying and finding
typical bifurcation scenarios in such systems (à la Eckhaus instability
or Busse-baloons in PDEs) is another open and challenging problem. 
\subsection{Adaptation, slow feedback and noise -- by Igor Franovi\'c}\label{SEC:FRANOVIC}
Adaptation is often qualitatively described as a slow evolution of network connectivity patterns due to a feedback from the nodal dynamics, drawing comparison to synaptic plasticity in neuronal systems \cite{CD08}, see also previous section. Nevertheless, one should bear in mind that adaptation may also directly impact the features of nodal dynamics, with examples ranging from frequency adaptation in clapping audiences or flashing fireflies \cite{TAY10} to scenarios where the limited availability of metabolic resources modulates neuronal excitability \cite{FL20,BB21} or contributes to maintaining neuronal systems near criticality \cite{RIVB14}. A detailed discussion concerning the two latter effects in relation to spike-frequency adaptation and short-term synaptic plasticity can be found in Sec.~\ref{SEC:OLMI}. While these two types of adaptation, affecting the coupling or nodal dynamics, may appear independently, it is also not uncommon that they act in concert guiding the system's self-organization \cite{VSRO16,LHG07}. So far, most of the systematic insights on the role of adaptation have been gained regarding its impact on synchronization, including how it gives rise to different states of (partial) synchrony \cite{MLHBT07,BER19,BER19a,BER21b,KMB21,TCS22}, or the way it modifies the order of synchronization transition \cite{TAY10} and the associated nucleation process \cite{FIA22}. Another active branch of research concerns adaptation as a general control mechanism, establishing its role in inducing critical transitions \cite{BB21,RIVB14} and triggering of alternating or cyclic activity patterns \cite{FEYB22,FYEBW20,BKNPF18}. Moreover, unfolding studies employing reservoir computing for design of controllers for nonlinear, and in particular chaotic systems, also hold a great promise, see Sec.~\ref{SEC:GAUTHIER}.

\paragraph{Interaction of adaptation and noise.}

An important, but still insufficiently understood problem concerns the interaction between adaptation
and noise, an issue naturally arising in applications to neuroscience. In spite of an apparently desynchronizing effect of noise, it has been shown that adaptation and noise may give rise to a self-organized network activity that promotes growth of overall synaptic strength \cite{PYT13},
thereby canceling the potentially desynchronizing stochastic effects. While this may seem counterintuitive, one should recall that classical synaptic plasticity rules, such as spike-timing dependent plasticity \cite{CD08}, support synaptic potentiation if coupled neurons are approximately (but not identically) synchronized and maintain their relative order of firing \cite{KHGKP12}. However, such self-organized resilience of synchronization to noise is so far
evinced for coupled oscillators rather than coupled excitable or mixed excitable-oscillatory populations. Addressing the two latter cases would be highly relevant for applications in neuroscience where local dynamics typically involves excitability and diversity \cite{PM06,LCT10,KF19}.

Apart from the mean effect on the overall coupling strength, an additional subtlety from the interaction of adaptation and noise concerns stochastic fluctuations, so far addressed mostly at the microscopic level. For motifs of coupled stochastic excitable units, such an interaction may induce switching dynamics, i.e. slow stochastic fluctuations between coexisting metastable states. The switching is naturally reflected both at the level of nodal dynamics and the effective motif coupling configuration, given by the coupling strengths \cite{BYWF18}. In particular, for the example of a system of two identical excitable units, the noise can induce two different oscillatory modes with a different prevailing order of firing between the units. In presence of slow adaptation, such metastable states engage in an alternating dynamics, accompanied by an alternation of coupling configurations characterized by a strong coupling in one direction and a strongly depressed one in the opposite direction. Translated to the language of neuroscience, the latter effect corresponds to a switching between two functional neuronal motifs with directed couplings on the same structural motif \cite{SK04}.

Concerning stochastic fluctuations at the level of a single excitable system, it has been shown that a slowly adapting feedback, acting as a low pass filter to affect the unit’s excitability \cite{FYEBW20}, may in an interaction with noise induce a novel form of behavior called stochastic bursting, an alternating activity involving episodes of relative silence interspersed with irregular spiking. Such stochastic bursting occurs in the parameter region that in the limit of an infinite scale separation between the units' dynamics and adaptation, supports bistability between noise-induced and noise-perturbed spiking. Apart from inducing a novel type of behavior, adaptation may also provide for a control mechanism of coherence resonance \cite{FYEBW20} or may make the noise-induced suppression of spiking frequency within inverse stochastic resonance more efficient \cite{BF20,BKNPF18}.

\paragraph{Impact of adaptation rate.}

An often overlooked feature of adaptation when elaborating its impact on emergent dynamics is the adaptation rate. Classically, adaptation rate is considered to be sufficiently slow such that the overall dynamics may be treated within the framework of singular perturbation theory \cite{K15}, separating between the fast local dynamics of units and the slow evolution of adaptation variables. However, the impact of adaptation rate has not been investigated systematically, mostly due to a lack of an appropriate analytical method. In certain examples, it has numerically been shown that intermediate adaptation rates can substantially deviate the system's behavior from the predictions of singular perturbation theory \cite{BYWF18}, and finding appropriate means to address this issue remains an open problem.

\paragraph{Mathematical approaches to adaptation.}

In a broader perspective, developing mathematical approaches to study adaptive networks is challenging because it requires reconciling different aspects of system behavior, such as criticality, feedback, multiple timescale dynamics, diversity and noise. So far, an extension of master stability function approach \cite{PEC98} has proven effective in reducing the synchronization problem by separating for dynamical and topological features, allowing for a classification of system states with respect to synchronization properties. For coupled phase or neural oscillators, such an approach has revealed
that adaptation may induce a desynchronization transition \cite{BER21b}, and support different multi-frequency hierarchical cluster states and chimera-like states of partial synchronization. Nevertheless, the general problem of the impact of adaptation on system's multistability remains
open. In certain cases, like the Kuramoto phase oscillators with an asymmetric spike-timing dependent plasticity-like plasticity rule, adaptation has been shown to induce multistability between the synchronized, desynchronized and
multiple partially synchronous states \cite{MLHBT07}. Also, for adaptively coupled identical phase oscillators, multicluster states have been shown to exhibit a high degree of multistability \cite{BER19,BER19a}. Apart from understanding the impact on synchronization problem, an important 
issue concerns the role of adaptation in inducing cyclic activity patterns by controlling critical transitions of the adaptation-free system. Treating such problems, like the onset of collective activity bursts in heterogeneous systems adaptively coupled to a pool of resources \cite{FEYB22}, requires combining different reduction approaches \cite{OA08,OA09,BGLM20} and multiple timescale methods. Nevertheless, developing rigorous mathematical approaches where mean-field methods apply to layer dynamics while adaptation is treated by a reduced system, is a vibrant field of investigation. In parallel, a hybrid approach for treating the interaction of adaptation and noise by combining the Fokker-Planck formalism with multiple timescale methods has recently been derived \cite{FYEBW20}. Further generalization of adaptation concept to cases where adaptation rate itself varies in time may additionally require including methods from nonequilibrium thermodynamics and information theory. This naturally applies to sensory adaptation \cite{LSNST12,CM22}, where information transmission is optimized under different constraints, including metabolic costs, dynamic range, and intrinsic stochasticity \cite{TB16}. From the perspective of nonequilibrium thermodynamics, sensory adaptation is a dissipative process ruled by an energy-speed-accuracy tradeoff \cite{LSNST12}, where one may exploit the relation between adaptation and irreversibility \cite{CM22}, quantified by the entropy production. 
\subsection{Adaptivity: a shared notion? -- by Nuria Brede and Nicola Botta}\label{SEC:BOTTA}

This article discusses notions of \emph{adaptivity} from the
perspective of different disciplines, ranging from non-linear dynamics to psychology, neuroscience and computer science.
Yet, while most authors would agree that adaptivity is a
\emph{property}, their answers to the question ``A property of {\bf what}?''\ presented in the various contributions seem to differ.
This is not accidental, but simply a consequence of the exploratory
nature of the paper, and it poses a challenge for future work: Can we
find a framework that is sufficiently generic to formulate and compare
the notions of \emph{adaptivity} in different research areas,
understand their differences and similarities, identify shared
concepts and computational methods and facilitate the communication
between disciplines?

We argue in this section that \emph{Dependent Type Theory} would be an
ideal candidate for (formulating) such a framework.  What do we mean
by this? The reader who is unfamiliar with Dependent Type Theory
should for the moment think of it as mathematical logic fused with a
programming language (we will explain more in paragraph b below).
Ionescu et al.~\cite{10.1007/978-3-030-03418-4} argue that type
theory fits most of the requirements for frameworks for
\emph{modelling} and \emph{programming} put forward by Broy et
al.~\cite{10.1007/978-3-319-47169-3}. In a research programme
originally initiated by Ionescu, type theory has been applied to
understand notions of \emph{vulnerability}, \emph{viability},
\emph{reachability}, \emph{avoidability} (discrete dynamical systems),
\emph{optimality} (control theory), \emph{climate sensitivity},
\emph{commitment}, and \emph{responsibility} (climate
policy)~\cite{ionescu+al2009, 2017_Botta_Jansson_Ionescu,
brede2021dsl, Botta2023MatterMost}.  The largest study of the above is Ionescu et al.~\cite{ionescu+al2009} where various notions
of \emph{vulnerability}, stemming from domains such as climate change,
food security or natural hazard studies are compared.

\paragraph{Notions of adaptivity.}

A key idea commonly put forward is that adaptivity is a ``feature of
natural and artificial complex systems''. Thus, from this perspective,
adaptivity is a \emph{property of a system}.
However, in their 1992 seminal paper ``Reinforcement learning is
direct adaptive optimal control'' \cite{SuttonBartoWilliams1992}
Sutton, Barto and Williams argue that what is \emph{adaptive} is a
\emph{method} for controlling a system, rather than the system
itself. This suggests to see adaptivity as a \emph{property of optimal
control methods}.

It is worth noticing that optimal control methods do not need to be
adaptive. At least since 1957 \cite{Bellman1957}, we know that many
deterministic and stochastic sequential decision problems can be
solved for optimal \emph{policies} via dynamic programming. And
dynamic programming can indeed be applied to also solve
non-deterministic, fuzzy and, more generally, \emph{monadic}
sequential decision problems \cite{botta+al2014a}, as long as the
uncertainty monads and the \emph{measures of uncertainty} (for
example, for stochastic uncertainty, the expected value measure)
satisfy certain \emph{compatibility} conditions
\cite{brede_botta_2021}.
But when the transition function (or the reward function) of a
sequential decision problem is not given, optimal policies have to be
learned by interacting, step by step, with an \emph{environment}: for
example, via Q-learning \cite{Watkins1992}. This is \emph{learning to
act optimally} rather than \emph{optimal planning}.

Even if we share the intuition that adaptivity is a property of a
\emph{system}, or of a \emph{method} for controlling a system that
interacts \emph{sequentially} with an \emph{environment}, it remains
to clarify whether the notions of adaptivity in different domains
arise as instances of the same abstract notion or whether they are
genuinely different, potentially even incompatible.
Such a clarification requires \emph{specifying} and comparing
different notions of adaptivity in a common framework.
As mentioned above, in prior work, we have employed type theory for
this purpose.
 
\paragraph{Logic and type theory.}\label{dtt} Most scientists are well
trained in applying elementary mathematics and first order logic to
formulate properties in specific \emph{domains}. E.g., in mathematics
you might define what it means for a function to be \emph{injective},
or in dynamical systems theory what it means for a function to be the
\emph{flow} of a dynamical system. So, for mathematically trained
people logic is a well-suited language to make precise and develop a
shared understanding of concepts.  Indeed, this purpose is at the
heart of modern mathematical logic, at least going back to Leibniz'
vision of a universal language that would not suffer from the
ambiguities of natural language.

Dependent Type Theory \cite{nordstrom+petersson1990}, takes the
advantages of mathematical logic one step further. It is a theory that
may be seen both as a higher order logic and as a pure functional
programming language with a static type system. It was developed as a
foundational theory for constructive mathematics by the Swedish
mathematician and philosopher Per Martin-Löf
\cite{martin-lf_intuitionistic_1984}.  Dependent type theory has solid
implementations~\cite{ABCEKLM06, the_coq_development_team_zenodo,
norell2007, idrisbook, botta1} and impeccable
mathematical credentials \cite{voevodsky2011univalent,
gonthier2008formal, gonthier2013machine, buzzard2020formalising} (see
also \cite{the_mathematical_library_of_the_future,
how_close_are_computers_to_automating_mathematical_reasoning} for
popular science accounts, including the voices of mathematicians who
have turned to computer-aided formalization).

Due to its double role as logic and programming language, 
Dependent Type Theory is well-suited as a framework for both
formulating and machine checking mathematical specifications. 
Because types can represent propositions and well-typed programs
correspond to proofs~\cite{botta2}, Dependent Type Theory is
also the key for writing programs that are correct ``by
construction'', bridging the gap between mathematical model and
implementation.
This is crucial for safety-critical applications
\cite{leroy_formal_2009, swamy_secure_2011, licata_2011,
brady_resource-safe_2012,chlipala2022certified} but also in research
areas in which testing model implementations is nearly impossible or
too expensive \cite{ionescujansson:LIPIcs:2013:3899}.

\paragraph{Monadic dynamical systems.}  Ionescu et al.'s vulnerability
study\cite{ionescu+al2009} led to the introduction of \emph{monadic
dynamical systems}, combining ideas from generic
programming\cite{DBLP:books/daglib/0096998,DBLP:conf/lics/1989} and
category theory\cite{maclane} with dynamical systems theory. Monadic
dynamical systems are sufficiently general to capture various
different definitions of vulnerability as instances of a common
abstract schema.  The framework for vulnerability was later extended
by Botta et al.\cite{botta+al2014a,
2017_Botta_Jansson_Ionescu,brede_botta_2021} to a framework for
specifying and solving \emph{sequential decision problems} within
Dependent Type Theory. We think that this framework could also be
applied and suitably extended to study different notions of
adaptivity.
\subsection{Partial synchronization patterns in adaptive networks -- by Eckehard Schöll}\label{SEC:SCHOELL}

This subsection explores the applications of network models as outlined in Sec.~\ref{SEC:YANCHUK} in different domains. From a complex networks perspective, the interplay between dynamics and network topology is in the center of interest. Collective dynamics in networks of nonlinear oscillators is often characterized by synchronization phenomena~\cite{PIK01,BOC18}, as already studied by Christiaan Huygens in 1656. Among these, partial synchronization patterns have become a major focus of research recently~\cite{SCHOELL21}. Examples are provided by cluster or group synchronization (where within each cluster all elements are completely synchronized, but between the clusters there is a phase lag, or even a difference in frequency), and many other forms. A particularly intriguing example of partial synchronization patterns, which has recently gained much attention, are \textit{chimera states}, i.e., symmetry-breaking states of partially coherent and partially incoherent behavior, for recent reviews see \cite{SCH20,SAW20,ZAK20}. Chimera states in dynamical networks consist of spatially separated, coexisting domains of synchronized (spatially coherent) and desynchronized (spatially incoherent) dynamics. They are a manifestation of spontaneous symmetry-breaking in systems of identical oscillators, and occur in a variety of physical, chemical, biological, neuronal, ecological, technological, or socio-economic systems. Other examples of partial synchronization include solitary states~\cite{MAI14a,JAR15,SEM15b}, where one single or a few elements behave differently compared with the behavior of the background group, i.e., the neighboring elements, or hierarchical multifrequency clusters~\cite{KAS17}.

In adaptive networks the coupling weights are not fixed, but are continuously adapted by feedback of the dynamics, and both the local dynamics and the coupling weights evolve in time as co-evolutionary processes, compare with discussions in Secs.~\ref{SEC:YANCHUK} or~\ref{SEC:FRANOVIC}.
Adaptive networks have been reported for chemical~\cite{JAI01}, epidemic~\cite{GRO06a} (see also Secs.~\ref{SEC:HOEVEL} and~\ref{SEC:MOELTER}), biological, and social systems~\cite{GRO08a} (see also Secs.~\ref{SEC:LORENZSPREEN}). A paradigmatic example of adaptively coupled phase oscillators has recently attracted much attention~\cite{GUT11,ZHA15a,KAS17,ASL18a,KAS18,KAS18a,BER19,BER19a,BER20,BER21b,FEK20,BER21b,BER21c} and it appears to be useful for predicting and describing phenomena in more realistic and detailed models~\cite{POP15,LUE16,CHA17a,ROE19a}. It describes $N$ adaptively coupled phase oscillators~\cite{KAS17,BER19} (as a general case of Eqs.\,\eqref{eq:sy-phase-1}--\eqref{eq:sy-phase-2} in Sec.~\ref{SEC:YANCHUK})
\begin{align}
\dot{\phi}_i &= \omega_i + \sum_{j=1}^N a_{ij}\kappa_{ij} f(\phi_i-\phi_j), \label{eq:APO_phi}\\
\dot{\kappa}_{ij} & = -\epsilon\left(\kappa_{ij} + g(\phi_i - \phi_j)\right), \label{eq:APO_kappa}
\end{align}
where $\phi_{i}\in[0,2\pi)$ represents the phase of the $i$th oscillator ($i=1,\dots,N$), $\omega_i$ is its natural frequency, and $\kappa_{ij}$ is the coupling weight of the connection from node $j$ to $i$. Further, $f$ and $g$ are $2\pi$-periodic functions where $f$ is the coupling function and $g$ is the adaptation rule, and $\epsilon \ll 1$ is the adaptation time constant. The connectivity between the oscillators is described by the entries $a_{ij}\in\{0,1\}$ of the adjacency matrix $A$. In particular, for the Kuramoto phase oscillator~\cite{KUR84}, the coupling function is $f(\phi)=-\sin \phi$
, and synaptic neuronal plasticity may be described by $g(\phi)=-\cos (\phi+\beta)$ where the parameter $\beta$ describes different adaptivity rules.

One purpose of this section is to provide a new perspective by demonstrating that a wide range of models ranging from neuronal networks with synaptic plasticity via power grids to physiological networks modeling tumor disease and sepsis can be viewed as adaptive oscillator networks, and partial synchronization patterns can be described on equal footing. This modeling approach allows one to transfer methods and results from one system to the other.

A common class of network models describing power grids is given by $N$ coupled phase oscillators with inertia~\cite{FIL08a}, also known as swing equation. It has been widely used in works on synchronization of complex networks and as a paradigm for the dynamics of modern power grids~\cite{DOE12,ROH12,MOT13a,ROD16,TUM18,TUM19,TAH19,HEL20,KUE19,TOT20,ZHA20c}:
\begin{align}\label{eq:KwI_2order}
M\ddot{\phi}_i +\gamma\dot{\phi}_i & = P_i + \sum_{j=1}^N a_{ij}h(\phi_i-\phi_j),
\end{align}
where $M$ is the inertia coefficient, $\gamma$ is the damping constant, $P_i$ is the power of the $i$th oscillator (related to the natural frequency $\omega_i = {P_i}/{\gamma}$), $h$ is the coupling function, and $a_{ij}$ is the adjacency matrix as defined in Eq.~\eqref{eq:APO_phi}. Another view on the role of adaptivity for power grid systems can be also found in Sec.~\ref{SEC:ANVARI}.

It has been shown~\cite{BER21a} that the class of phase oscillator models with inertia is a natural subclass of systems with adaptive coupling weights where the weights denote the power flows between the corresponding nodes. We first write Eq.~\eqref{eq:KwI_2order} in the form
\begin{align}
\dot{\phi}_i &= \omega_i + \psi_i, \label{eq:KwI_1order_phi}\\
\dot{\psi}_i & = -\frac{\gamma}{M}\left({\psi}_i - \frac{1}{\gamma}\sum_{j=1}^N a_{ij}h(\phi_i - \phi_j)\right), \label{eq:KwI_1order_psi}
\end{align}
where $\psi_i$ is the deviation of the instantaneous phase velocity from the natural frequency $\omega_i$. We observe that this is a system of $N$ phase oscillators~\eqref{eq:KwI_1order_phi} augmented by the adaptation~\eqref{eq:KwI_1order_psi} of the frequency deviation $\psi_i$. Similar systems with a direct frequency adaptation have been studied in~\cite{ACE98,ACE05,TAY10,SKA13a}.  
Note that the coupling between the phase oscillators is realized in the frequency adaptation which is different from the classical Kuramoto system~\cite{KUR84}. 
In order to introduce coupling weights into system~\eqref{eq:KwI_1order_phi}--\eqref{eq:KwI_1order_psi}, we express the frequency deviation $\psi_i$ as the sum $\psi_i = \sum_{j=1}^N a_{ij}\chi_{ij}$ of the dynamical power flows $\chi_{ij}$ from the nodes $j$ that are coupled with node $i$. The power flows are governed by the equation $\dot{\chi}_{ij}  = -\epsilon\left({\chi}_{ij} + g(\phi_i - \phi_j)\right)$, where $g(\phi_i - \phi_j)\equiv-h(\phi_i - \phi_j)/\gamma$ are their stationary values~\cite{SCH18i} and $\epsilon =\gamma/M$. It is straightforward to check that $\psi_i$, defined in such a way, satisfies the dynamical equation~\eqref{eq:KwI_1order_psi}.

As a result, the swing equation~\eqref{eq:KwI_1order_phi}--\eqref{eq:KwI_1order_psi} can be written as the following system of adaptively coupled phase oscillators
\begin{align}
\dot{\phi}_i &= \omega_i+\sum_{j=1}^N a_{ij}\chi_{ij}, \label{eq:KwI_pseudo_phi}\\
\dot{\chi}_{ij} & = -\epsilon\left({\chi}_{ij} + g(\phi_i - \phi_j)\right). \label{eq:KwI_pseudo_kappa}
\end{align}
The obtained system corresponds to~\eqref{eq:APO_phi}--\eqref{eq:APO_kappa} with coupling weights $\chi_{ij}$ and coupling function $f(\phi_i-\phi_j) \equiv 1$. The coupling weights form a pseudo coupling matrix $\chi$ describing the power flow between the nodes. Note that the base network topology $a_{ij}$ of the phase oscillator system with inertia Eq.~\eqref{eq:KwI_2order} is unaffected by the transformation.

In adaptive phase oscillator networks there exists a diversity of multifrequency cluster states~\cite{KAS17,BER19a,BER21c}, including chimera states~\cite{KAS17} and solitary states~\cite{BER20c}. In a multifrequency cluster state, all oscillators split into $M$ groups (called clusters) each of which is characterized by a common cluster frequency $\Omega_\mu$. In particular, the temporal behavior of the $i$th oscillator of the $\mu$th cluster ($\mu=1,\dots,M$) is given by $\phi_i^\mu (t)= \Omega_\mu t + \rho^{\mu}_i + s^{\mu}_i(t)$ where $\rho^{\mu}_i\in[0,2\pi)$ and $s^{\mu}_i(t)$ are bounded functions describing different types of phase clusters characterized by the phase relation within each cluster~\cite{BER19}. 

\begin{figure}
	\includegraphics{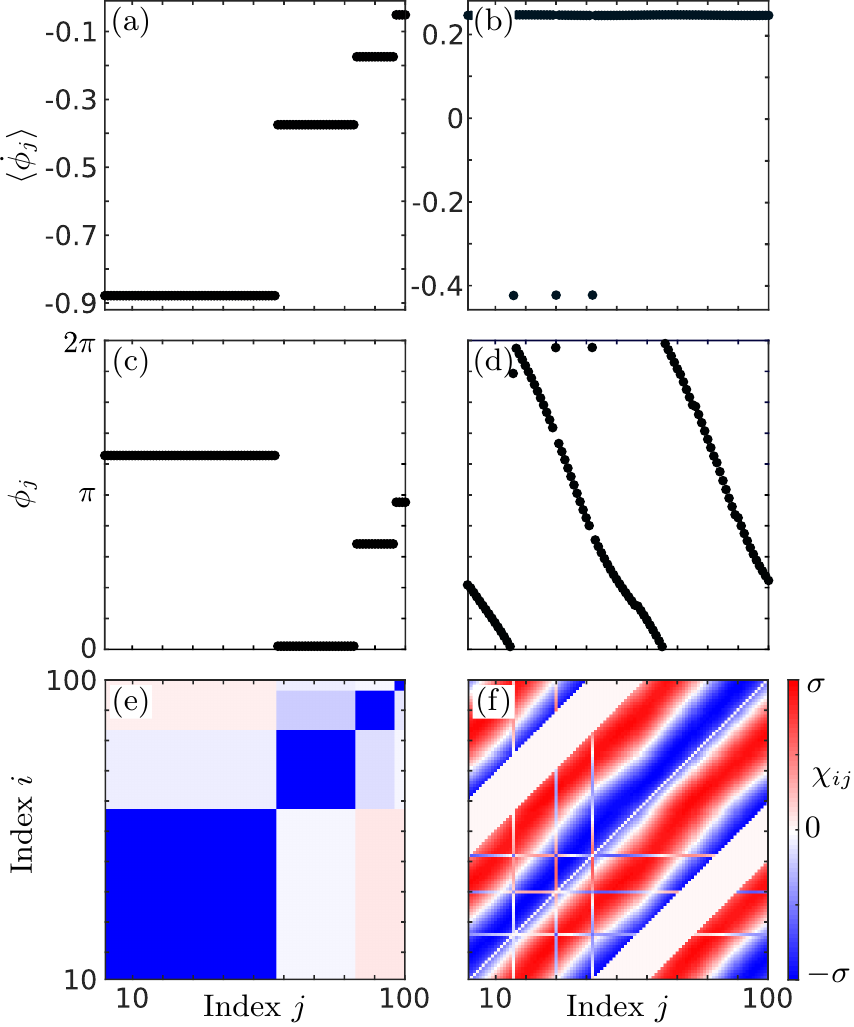}
	\caption{\label{fig:MC_KwI_MultiCluster}Hierarchical multicluster states in networks of coupled phase oscillators with inertia. The panels (a,b), (c,d) and (e,f) show the temporally averaged phase velocities $\langle\dot{\phi}_j\rangle$, phase snapshots $\phi_j(t)$ and the pseudo coupling matrices $\chi_{ij}(t)$, respectively, at $t=10000$. In (e) the oscillator indices are sorted in increasing order of their mean phase velocity. The states were found by numerical integration of~\eqref{eq:KwI_2order} with identical oscillators $P_i=0$, $h(\phi)=-\sigma\gamma\sin(\phi+\alpha)$, and uniform random initial conditions $\phi_i(0)\in(0,2\pi)$, $\psi_i(0)\in(-0.5,0.5)$. The parameter $\alpha$ is a phase-lag of the interaction~\cite{SAK86}. Parameters: (a,c,e) globally coupled network, $M=1$, $\gamma=0.05$, $\sigma=0.016$, $\alpha=0.46\pi$; (b,d,f) nonlocally coupled ring network with coupling radius $P=40$, $M=1$, $\gamma=0.3$, $\sigma=0.033$, $\alpha=0.8\pi$; $N=100$. After~\cite{BER21a}.}
\end{figure}

As an example, in Fig.~\ref{fig:MC_KwI_MultiCluster}(a,c), we present a $4$-cluster state of in-phase synchronous clusters on a globally coupled network. Hierarchical multicluster states are built out of single cluster states whose frequency scales approximately with the number $N_\mu$ of elements in the cluster. The coupling matrix displayed in Fig.~\ref{fig:MC_KwI_MultiCluster}(e) shows the characteristic block diagonal shape known for adaptive networks. In particular, the oscillators within each cluster are more strongly connected than the oscillators between different clusters.

A second example, which uses a splay state with $\phi_j = 2\pi k j/N$ and wavenumber $k\in \mathbb{N}$ as the building block for multiclusters, is shown in \ref{fig:MC_KwI_MultiCluster}(b,d,f). Splay states are characterized by the vanishing local order parameter $R_j=|\sum_{k=1}^N a_{jk}\exp(\mathrm{i}\phi_k)|=0$. Figure~\ref{fig:MC_KwI_MultiCluster}(b,d,f)presents a hierarchical mixed-type multicluster on a nonlocally coupled ring of phase oscillators. It consists of one large splay cluster with wavenumber $k=2$ and a small in-phase cluster consisting of three solitary states. 

In summary, the findings for partial synchronization of adaptively coupled phase oscillators can be transferred to networks of phase oscillators with inertia. This holds not only for simple homogeneous systems, but also for heterogeneous real-world networks, like the German ultra-high voltage power grid~\cite{BER21a}.

In recent years, studies on both types of models, oscillators with inertia and adaptively coupled oscillators, have revealed a plethora of common dynamical scenarios including solitary states~\cite{JAR18,TAH19,HEL20,BER20c}, multifrequency clusters~\cite{BEL16a,BER19,BER19a,TUM19}, chimera states~\cite{OLM15a,KAS17,KAS18}, hysteretic behavior and non-smooth synchronization transitions~\cite{OLM14a,ZHA15a,BAR16a,TUM18,FIA22}. 
Power grids, as well as neuronal networks with synaptic plasticity, and other adaptive networks describe real-world systems of tremendous importance for our daily life, which exhibit partial synchronization patterns that may be important for the understanding of the onset of instability. Neural systems and power grid networks are also discussed in the Secs.~\ref{SEC:NEURAL} and~\ref{SEC:SOCIO-ECONOMIC}, respectively. A particularly intriguing example and a future perspective is the functional modeling of physiological 2-layer networks of the immune system and the parenchyma coupled adaptively by cytokines \cite{SAW21b,BER22}. This can be used for the modeling of tumor disease and sepsis with the immune layer as reference point, where the healthy state is characterized by complete frequency synchronization and the pathological state is a multifrequency cluster state. 

%
%

\section{Perception and neural adaptivity}
\label{SEC:NEURAL}

In this section, the focus is on adaptive mechanisms in physiological systems. Here, basic regulatory principles are highlighted, fundamental concepts for a physical culture theory are developed, mechanisms and modeling of perception are described, and concrete medical applications on neural networks are presented.

\subsection{Design principles for adaptation in physiological systems -- by Omer Karin}\label{SEC:KARIN}

Here we will explore motifs for adaptation in physiological regulatory networks. The physiological properties of biological systems arise from the myriad of interactions of their underlying components. As an example, the production rate of proteins from a gene depends on the abundance of other proteins, known as transcription factors, whose production depends on the abundance of other transcription factors. Similarly, the secretion of a hormone to the bloodstream depends on the concentrations of other blood factors, which are themselves affected by the levels of other hormones. These complex networks of interactions are known as regulatory networks.

To study regulatory networks, it is useful to notice that evolution tends to come up with similar solutions to related problems. It is often the case that, under similar contexts, the regulatory networks of distinct systems share mathematical similarities - these are so called regulatory motifs or design principles \cite{1,2,3}. By identifying such \textit{design principles}, one can extract a deeper understanding of the functional significance of the regulatory interactions. We may therefore ask what are the design principles that support adaptation - the ability of the system to adjust itself to function properly, despite uncertainty in internal parameters or the external environment.

Consider the problem of maintaining homeostasis of a blood factor such as glucose (denoted $x$). Blood glucose needs to be maintained within a narrow range (around 5\,mM) with deviations being detrimental or even life-threatening. Our bodies have a natural mechanism to lower blood glucose - we have specialized cells called $\beta$-cells which can sense blood glucose and secrete the hormone insulin, which causes remote cells (fat cells, skeletal muscle cells, and liver cells) to reduce glucose levels. This mechanism can maintain glucose around some steady-state, which would depend sensitively on many parameters, including the abundances $\beta$-cells, plasma volume, and the responses of cells to insulin. These can (and do) vary greatly between individuals, yet we know that most individuals can maintain blood glucose within a narrow range\cite{4}.

\begin{figure}[th]
\includegraphics[width=1.0\linewidth]{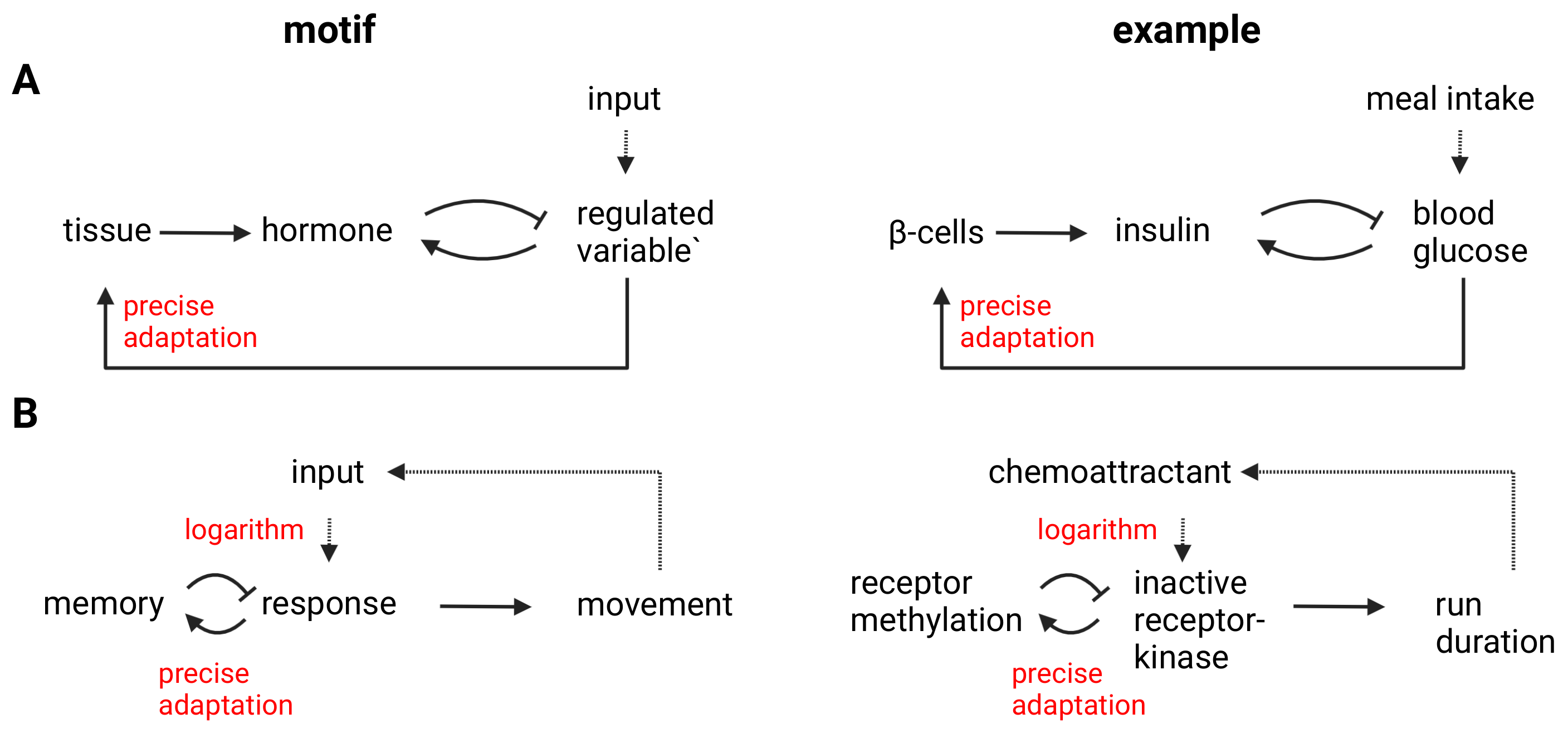}
\caption{\label{fig:Karin} Motifs for adaptation in physiological systems. (A) In hormone circuits, a hormone-regulated variable governs the growth rate of the tissue responsible for its secretion, enabling precise adaptation. This adaptation mechanism ensures that the dynamics of the regulated variable remain robust in the face of physiological variations. (B) Organisms employ a combination of logarithmic sensing, precise adaptation, coupled to movement regulation, to achieve robust sampling of an input field. This motif is observed in chemotaxis and potentially in the mammalian dopamine system.}
\end{figure}

A related problem occurs in bacterial chemotaxis. The bacteria \textit{E. coli} navigates with a strategy resembling a random walk, where it moves and reorients with some set rate $\phi$ (typically once every few seconds). This is known as the tumbling rate. Navigation is achieved by adjusting $\phi$  according to sensed ligand molecules known as attractors and repellants. A step increase in an attractant molecule transiently decreases $\phi$, leading to net drift towards areas with higher attractant concentration. However, at fixed attractant concentration $u$, over a wide sensed range, $\phi$ is constant and independent of $u$ \cite{5,6}. How is $\phi$ maintained constant, despite variations in the input activity of the circuit?

It has long been suggested that both problems are closely related to the engineering problem of disturbance rejection \cite{7,8,9}. This problem  is exemplified by how a cruise-control system of a car maintains a fixed speed on varying slopes, or how a thermostat maintains a fixed temperature in uncertain operating conditions. The solution requires integral feedback: the controller feedback increases with the error (it integrates the error), so at steady-state the error is zero.

How is integral feedback implemented in biological circuits? In hormone circuits there appears to be a simple answer (Fig.\,\ref{fig:Karin}A). Let $x$ be the regulated variable and $y$ to be its regulating hormone, with $Z$ the mass of the tissue that secretes the hormone. In the blood glucose system, $x$ is blood glucose, $y$ is blood insulin, and $Z$ is $\beta$-cell mass. The following motif is observed across hormone systems: there is a slow negative feedback where the main regulator of the growth dynamics of $Z$ is $x$, that is, $x$ adjusts the death-, growth-, and replication-rates of the cells of $Z$. Thus:

\begin{equation}
\label{Eq.Omer1}
	\dot{Z}=f(x)Z,
\end{equation}

where $f(x)$ is the $x$-dependent growth rate. The system will settle at the steady-state where $f(x)=0$ (denoted $x_0$) regardless of variation in the other physiological parameters, including plasma volume, secretion rate, and the responses of remote cells.

The ubiquity of the motif suggests that it is uniquely advantageous. Why is it so prevalent? Beyond integral feedback, another intriguing phenomena occurs. Consider for example the following simple model for the glucose system: 
\begin{align}
\begin{split}
	\dot{u}&=u-sxy,\\
	\dot{y}&=pZ-\gamma y,
\end{split}
\end{align}
where $s$ is the sensitivity to the response of the hormone, and $p$ is the product of the per-cell secretion and (inverse) plasma volume. $u$ is the time-dependent input, incorporating e.g. meal intake. Eq.\,\eqref{Eq.Omer1} not only sets the steady-state of $x$ to $x=x_0$, it makes the entire dynamics in response to any input $u$ invariant of $s$,$p$ \cite{10}. These scale invariant dynamics are evident in clinical data from distinct hormonal systems \cite{10,11,12,13}. Thus, in hormone systems, negative feedback from the regulated variable to its controlling tissue allows the system to adapt its dynamics to variability in key system parameters, which are uncertain and may be highly variable. 

Scale invariance also occurs in bacterial chemotaxis; in this case, the dynamics of the tumbling rate $\phi(t)$ are modulated by the attractant input $u(t)$ in a manner which depends only on relative, rather than absolute, changes in $u(t)$, a phenomena known as \textit{fold-change detection}\cite{14}. Fold-change detection is documented in the navigation systems of other simple organisms, including in worms and slime molds \cite{15,16}.

What about more complex organisms? In vertebrates, including mice and humans, movement is controlled by the transmission of dopamine in the mid-brain. Dopamine is secreted in response to surprise (or \textit{prediction error}) about rewards, such as food or drink; better outcomes than expected cause dopaminergic neurons to fire above their baseline rate, while worse outcomes transiently inhibit dopaminergic firing \cite{SchultzEtAl1997}. The responses are also scale-invariant \cite{18}. Finally, when the animal moves, dopamine changes in a way that is consistent with a response to the temporal derivative of a spatial input field \cite{19}.

Upon closer examination, the dopamine system shares key similarities with the chemotaxis system, where in the case of dopamine the input field corresponds to expectations about rewards \cite{20}. This input field decays spatially from actual locations where rewards are provided, similar to the decay of a chemical attractant from its source. Dopamine also invigorates movements in a manner analogous to the effect of attractants on bacterial movement.

We therefore identified another regulatory motif: fold-change detection of an input field, which modulates movement statistics (Fig.\,\ref{fig:Karin}B). What is the function of this motif? From the perspective of sensing, scale invariance allows us to remove uncertainty and retain sensitivity over a wide dynamic input range. An additional distinct advantage is apparent when we consider the coupling between sensing and movement. The fold-change detection circuit calculates the temporal logarithmic derivative of the input $u(t)$. In a spatial setting, we can consider a spatial input field $U(x)$; the movement dynamics of the organism over long time- and length- scales are captured by the stochastic dynamics:

\begin{equation}
	\dd x = \beta v^2 \phi \nabla \log U \enspace \dd t + v^2 \sqrt{2\phi}\enspace \dd W ,
\end{equation}
where $v$ is the typical movement speed, and $\beta$ depends on circuit parameters. The steady-state distribution of the organism location is $P(x)=U(x)^\beta$, which only depends on circuit parameters (rather than movement parameters); the motif thus provides a robust mechanism for sampling a power of the input field. This is again consistent with experimental observations on both chemotaxis and the dopamine system \cite{20}. Thus, in these systems, a motif that appears to support adaptation of sensing in the background of uncertain input levels, in fact provides a mechanism for robust sampling of uncertain environments.

The examples considered here suggest that adaptation motifs that allow for scale-invariant dynamics are prevalent; and that specific adaptation regulatory motifs, which recur in similar contexts, have distinct functional significance. Identifying these motifs, and comparing their behavior in different contexts, is due to improve our understanding of how adaptation is achieved by complex regulatory networks.

\subsection{Adaptation and neuronal coding -- by Christoph Miehl}\label{SEC:MIEHL}

``To live is to adapt to the world around us'' \cite{Whitmire2016}. The environment of an organism can change on vastly different timescales, ranging from, e.g.,~a change in lighting to climate change. Organisms, and hence their brains, have developed strategies to adapt to these modifications in the environment across timescales, from adaptation to sudden changes in sensory stimuli to long timescales of evolutionary processes. In the following, some key adaptive mechanisms in the brain on short timescales are highlighted.

\begin{figure}
    \centering
    \includegraphics[width=0.5\textwidth]{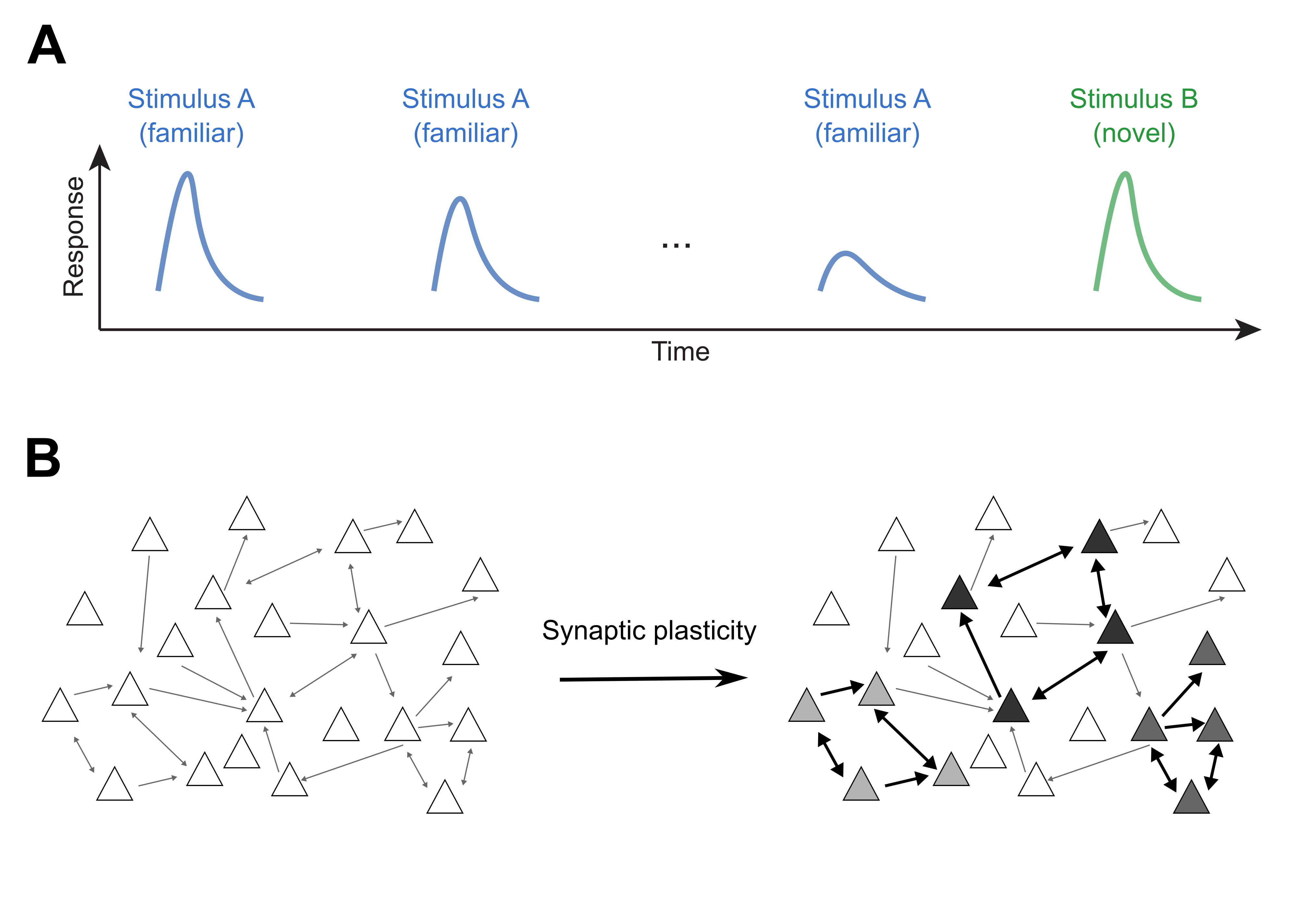}
    \caption{\textbf{A} Oddball paradigm. Presenting a stimulus repeatedly (stimulus A) leads to decrease of the neuronal response while the deviant stimulus (stimulus B) leads to high neuronal response. Panel adapted from \cite{Wu2022}. \textbf{B} Synaptic plasticity leading to strongly recurrently connected structures (assemblies). Panel adapted from \cite{Miehl2022b}.}
    \label{fig:fig_idea}
\end{figure}

In principle, single neurons can adapt to changes in the environment based on two strategies, either by modifying their intrinsic or extrinsic properties. Intrinsic changes include, e.g~increases or decreases in the excitability of a neuron \cite{Debanne2019}. Extrinsic changes are related to updates in the strength of the synaptic connections onto the neuron. An extrinsic mechanism that has been linked to adaptation on short timescales (tens to hundreds of milliseconds) is short-term synaptic plasticity. Input spikes that occur within short timescales can cause a transient decrease (short-term depression) or an increase (short-term facilitation) of the synaptic efficacy \cite{Zucker2002} (see Sec.\,\ref{SEC:OLMI}). The mechanism leading to a permanent increase or decrease in synaptic strength is long-term synaptic plasticity. In experiments, long-term changes in the synaptic strength can be induced via a `pairing protocol', a prominent example being spike-timing-dependent plasticity \cite{Feldman2012}. Repeatedly triggering a spike in the postsynaptic neuron following a spike in the presynaptic neuron within approx.~10ms leads to long-term potentiation, while presynaptic spikes following postsynaptic spikes within approx.~10 - 100\,ms leads to long-term depression \cite{Markram1997,Bi1998}. Both, short- and long-term plasticity have not only been identified at synapses between excitatory neurons but also at inhibitory-to-excitatory synapses (for more information see \cite{Motanis2018,Wu2022}).

A prominent experimental paradigm to test adaptation on short timescales is the `oddball paradigm' \cite{Weber2019a}. In this paradigm, one (usually visual or auditory) stimulus is presented many times, the standard (or familiar, predictable) stimulus. The second stimulus is only presented rarely, the deviant (or novel, unpredictable) stimulus. On the whole-brain level, electroencephalogram measurements reveal that presenting the deviant stimulus leads to a strong negative deflection in the EEG signal compared to the signal following from standard stimulus presentation, termed `mismatch negativity' \cite{Naatanen1982, Ross2020}. Similarly, measurements of either single neurons or neuronal populations in sensory cortices reveal elevated neuronal responses for deviant compared to the standard stimuli \cite{Ulanovsky2003, Natan2015, Homann2022} (Fig.~\ref{fig:fig_idea}A). Computational models have proven to be useful for understanding the mechanisms underlying short-term adaptation in the brain (see also Secs.\,\ref{SEC:HAJIZADEH} and \ref{SEC:OLMI}). Multiple studies suggest that short-term plasticity is a critical mechanism underlying adaptation to familiar stimuli \cite{Mill2011a,Mill2012,Hershenhoren2014}, and short-term plasticity at inhibitory synapses is important for controlling temporal context-dependent neuronal responses \cite{Park2020, Seay2020}. In a complementary approach, it has been suggested that long-term plasticity at inhibitory-to-excitatory synapses underlies the difference in responses to familiar and novel stimuli \cite{Schulz2021}. In this work, increase of inhibitory-to-excitatory synapses via long-term plasticity leads to a decrease in excitatory responses to familiar stimuli, while novel stimuli still lead to elevated responses. 

Many functional implications have been suggested for the role of reduced neuronal activity for familiar stimuli compared to elevated activity for novel stimuli, ranging from efficient coding and redundancy reduction, fast detection of unexpected events, to Bayesian inference \cite{Weber2019a, Whitmire2016}. Another highly considered implication is predictive coding. In this framework, it is thought that the goal of the brain is to minimize the difference between its internal prediction about the world and the sensory input \cite{Rao1999}. High responses to novel stimuli can be thought of as the prediction error. However, how exactly these computations are implemented in the brain, and how they are related to short- and long-term plasticity mechanisms is largely unresolved.

Neuronal circuits also need to be robust against perturbations. In experimental studies, disrupting the sensory inputs in the developing brain by performing deprivation experiments (e.g.,~closing the eye of an animal) leads to homeostatic adjustments of the respective neuronal circuits \cite{Turrigiano2004}. A related question is how tightly neuron intrinsic properties, like conductance densities, need to be regulated to maintain proper circuit function \cite{Marder2006}. For example, computational models and machine learning tools reveal that similar circuit dynamics can be found even for vastly different ion channel conductance densities and that this degeneracy allows to dynamically compensate perturbations on very fast timescales \cite{Onasch2020, Goncalves2020, Deistler2022}. Neuromodulators (like serotonin, dopamine, etc.) are the chemicals that control the neuron's intrinsic properties \cite{Marder2014}. Further computational studies have started investigating the combined effects of intrinsic and extrinsic neuron properties on neuronal activity and robust formation of switches between activity states, as found, e.g.,~in the sleep-wake cycle \cite{Jacquerie2021}.

Furthermore, learning and memory formation can be viewed as adaptive processes. Interestingly, it is suggested that learning in neuronal circuits relies on the same mechanisms as described above, short- and specifically long-term synaptic plasticity. While short-term plasticity might underlie working memory \cite{Mongillo2008}, long-term plasticity has been hypothesized as the basis for long-term memory storage \cite{Abbott2000}. One prominent idea is that groups of strongly interconnected neurons, so-called `assemblies', are the basic unit of representation in the brain and long-term plasticity has proven key for learning these connectivity structures in computational models \cite{Yuste2015, Miehl2022b} (Fig.~\ref{fig:fig_idea}B). Neuronal circuits face the problem of `stability-flexibility tradeoff', meaning that on the one hand synaptic connectivity should remain stable to allow for long-term memory storage and be robust against perturbations, while on the other hand circuits should remain flexible allowing re-learning, or learning of new representations \cite{Fusi2017}. Computational studies modeling neuronal networks have suggested different solutions, like reverberate neuronal activity \cite{Litwin-Kumar2014}, inhibitory-to-excitatory plasticity \cite{Miehl2022a} or a combination of multiple synaptic plasticity and homeostatic mechanisms \cite{Zenke2015}.

Despite recent promising developments, experimental and computational studies have only scratched the surface of understanding the role of intrinsic, short-, and long-term plasticity mechanisms in sensory adaptation. This endeavor is specifically important because deficits of information processing in neuropsychiatric diseases have been linked to disruptions in excitatory and inhibitory local circuits \cite{Hamm2017, Batista-Brito2018} and mismatch negativity has been suggested as a biomarker for psychotic disorders \cite{Light2013}. Therefore, uncovering the role of different cellular dynamics can have positive therapeutical impacts (see Sec.\,\ref{SEC:TASS}).

\subsection{A next generation neural mass approach to spike frequency adaptation and short term plasticity -- by Simona Olmi}\label{SEC:OLMI}

Neural mass models are mean field models developed to mimic the dynamics of homogenous populations of neurons. These models range from purely heuristic ones (as the well-known Wilson-Cowan model \cite{wilson1973}), to more refined versions obtained by considering the eigenfunction expansion of the Fokker-Planck equation for the distribution of the membrane potentials  \cite{mattia2002,schaffer2013}. However, quite recently, a {\it next generation neural mass model} has been derived in an exact manner for heterogeneous populations of spiking neurons \cite{luke2013, laing2014, montbrio2015}. This exact derivation is possible for networks of Quadratic Integrate and Fire (QIF) neurons, representing the normal form of Hodgkin’s class I excitable membranes \cite{ermentrout1986}, thanks to the analytic techniques developed for coupled phase oscillators \cite{ott2008}. Specifically, next generation neural mass models describe the dynamics of networks of spiking neurons in terms of macroscopic variables like the population firing rate  and the mean membrane potential, and they have already found various applications in many neuroscientific contexts \cite{byrne2017, schmidt2018, byrne2020, ceni2020,bi2020,taher2020,segneri2020, gast2020mean, gerster2021,gast2021}. Resuming the terminology introduced in Sec.\,\ref{SEC:MIEHL}, here we investigate the dynamics emergent in next generation neural mass models when   populations of neurons adapt to changes in the environment by modifying their intrinsic or extrinsic properties. In particular we present an overview of the emergence of collective dynamics (e.g. synchronous, bursting neural dynamics) in next generation neural mass models that arise from spike-frequency adaptation or post-synaptic plasticity. 

Spike-frequency adaptation is a widespread neurobiological phenomenon, exhibited by almost any type of neuron that generates action potentials. It occurs in vertebrates as well as in invertebrates, in peripheral as well as in central neurons, and may play an important role in neural information processing. As it will be clarified in the following, all biophysical mechanisms that can cause spike-frequency adaptation include a form of slow negative feedback to the excitability of the cell, therefore spike-frequency adaptation represents an intrinsic mechanism to adaptation. More in detail, experimental work suggests that it is a result of different balancing currents triggered at a single cell after it generates a spike \cite{fuhrmann2002, benda2003}. Three main types of ionic adaptation currents that influence spike generation are known: voltage-gated potassium currents, which are caused by voltage-dependent, high-threshold potassium channels \cite{brown1980}; the interplay of calcium currents and intracellular calcium dynamics with calcium-gated potassium channels \cite{madison1984}, and the slow recovery from inactivation of the fast sodium channel \cite{fleidervish1996}. As a result of these cellular mechanisms, many neurons show a reduction in the firing frequency of their spike response following an initial increase when stimulated with a square pulse or step. 

Short-term plasticity \cite{stevens1995, markram1996, abbott1997, Zucker2002,Abbott2004} refers to a phenomenon in which synaptic efficacy changes over time in a way that reflects the history of presynaptic activity, thus resulting to be an extrinsic mechanism of adaptation (see Sec.\,\ref{SEC:MIEHL}). Two types of short-term plasticity, with opposite effects on synaptic efficacy, have been observed in experiments: short-term depression and short-term facilitation. On one hand synaptic depression is caused by the depletion of neurotransmitters consumed during the synaptic signaling process at the axon terminal of a pre-synaptic neuron and it has been linked to various mechanisms such as receptor desensitization \cite{jones1996, wong2003}, receptor density reduction \cite{turrigiano2008, pozo2010}, or resource depletion at glial cells involved in synaptic transmission \cite{VSRO16, huang2017}. On the other hand synaptic facilitation is caused by the influx of calcium into the axon terminal after spike generation, which increases the release probability of neurotransmitters. Short-term plasticity has been found in various cortical regions and exhibits great diversity in properties \cite{markram1998, dittman2000, wang2006}. 

In the context of spike-frequency adaptation, first efforts in the direction of applying a neural mass model were made in a network of coupled linear integrate and fire neurons, employing the Fokker-Planck formalism and an adiabatic approximation given long spike-frequency adaptation timescales \cite{gigante2007}. Analyzing this mean-field description, Gigante et al. were able to identify different types of collective bursting. Recently, it has been shown that an excitatory next generation neural mass equipped with different short-term mechanisms of global adaptation can give rise to bursting behaviors \cite{gast2020mean}. Moreover, in \cite{ferrara2023}, the authors have studied the effect of this adaptation mechanism on the macroscopic dynamics of excitatory and inhibitory next generation neural mass models, by including in the original neural mass model proposed in \cite{montbrio2015}, an additional collective afterhyperpolarization current, which temporarily hyperpolarizes the cell upon spike emission. In a single population spike-frequency adaptation favours the emergence of population bursts in excitatory networks, while it hinders tonic population spiking for inhibitory ones. When considering two neural masses, symmetrically coupled in absence of adaptation, it is possible to observe the emergence of macroscopic solutions with broken symmetry: namely, chimera-like solutions in the inhibitory case and anti-phase population spikes in the excitatory one. Here the addition of spike-frequency adaptation leads to new collective dynamical regimes exhibiting cross-frequency coupling among the fast synaptic time scale and the slow adaptation one, ranging from anti-phase slow-fast nested oscillations to symmetric and asymmetric bursting phenomena.

In the context of short-term plasticity, a fundamental implementation has been first done by Mongillo et al. in \cite{Mongillo2008}, to explain the mechanisms underlying \emph{working memory}. Working memory is the ability to temporarily store and manipulate stimuli representations that are no longer available to the senses. In particular, in the model suggested by Mongillo and coauthors, synaptic facilitation allows the system to maintain an item stored for a certain period in working memory, without the need of an enhanced spiking activity. Furthermore, synaptic depression is responsible for the emergence of population bursts, which correspond to a sub-population of neurons firing almost synchronously within a short time window \cite{tsodyks2000, luccioli2014}. In this context, the bursting activity allows for item retrieval. The working memory mechanism is investigated in \cite{Mongillo2008} by means of a recurrent network of spiking neurons, while a simplified heuristic firing rate model is employed to gain some insight into the population dynamics.  A next generation neural mass model encompassing short-term synaptic facilitation and depression has been recently developed to revise the synaptic theory of working memory with a specific focus on the emergence of neural oscillations and their relevance for working memory operations \cite{taher2020}. In particular Taher and coauthors in  \cite{taher2020} consider multiple coupled excitatory populations, each coding for one item, and a single inhibitory population connected to all the excitatory neurons. This architecture is justified by recent experimental results indicating that GABAergic (i.e. inhibitory) interneurons in mouse frontal cortex are not arranged in sub-populations and that they densely innervate all pyramidal (i.e. excitatory) cells \cite{fino2011}. The role of inhibition is to avoid abnormal synchronization and to allow for a competition of different items once stored in the excitatory population activity. Furthermore, in order to mimic synaptic-based working memory only the excitatory-excitatory synapses are assumed to be plastic displaying short-term depression and facilitation (at the contrary with what shown in Sec.\,\ref{SEC:MIEHL} where examples of short-term plasticity in inhibitory-to-excitatory synapses are also considered). As a result, memory operations are joined to sustained or transient oscillations emerging in different frequency bands, in accordance with experimental results for primate and humans performing working memory tasks \cite{tallon1998, howard2003, vanvugt2010, roux2012}. Due to the possibility of reproducing working memory operations associated with population bursts delivered at different frequencies,  the neural mass model with short-term plasticity presented in \cite{taher2020} can represent a first building block for the development of an unified control mechanism for working memory, relying on the frequencies of deliverance of the self-emerging trains of population bursts. However, a development towards realistic neural architectures would require to design a multi-layer network topology to reproduce the interactions among superficial and deep cortical layers \cite{miller2018}.

Spike-frequency adaptation and post-synaptic plasticity can be modeled respectively as an additive and a multiplicative term in the evolution equation of the mean membrane potential in the exact neural mass model. The novelty of this neural mass model, besides not being heuristic, but derived in an exact manner from the microscopic underlying dynamics, is that it reproduces the evolution of the population firing rate as well as of the mean membrane potential. This allows us to get insight not only on the synchronized spiking activity, but also on the sub-threshold dynamics and to extract information correlated to local field potentials and electroencephalographic signals, that are usually measured to characterize the activity of the brain at a mesoscopic/macroscopic scale. Even though these adaptation mechanisms can express tremendously different timescales, ranging from a few hundred milliseconds (e.g., spike-frequency adaptation \cite{fuhrmann2002}) to days (e.g., postsynaptic receptor density reduction \cite{pozo2010}) the mean-field descriptions remain applicable. However, note that a macroscopic model of synaptic plasticity cannot express a vesicle depletion at the presynaptic site \cite{gast2021}, as introduced for single cell models in \cite{tsodyks1998}. Finally, thanks to the fact that adding spike-frequency adaptation leads to new collective dynamical regimes exhibiting cross-frequency coupling among the fast synaptic time scale and the slow adaptation one, the adaptive mechanisms in the framework of exact neural mass models could be useful to develop new models of self-organizing biological neural circuits that produce rhythmic outputs even in the absence of rhythmic input. An example could be the Central Pattern Generators, which are responsible for the generation of rhythmic movements, since these models are often based on two interacting oscillatory populations with adaptation, as reported for the spinal cord \cite{linden2021} and the respiratory system \cite{rubin2011}.

\subsection{Therapeutic reshaping of plastic networks -- by Peter A. Tass}\label{SEC:TASS}

Regular deep brain stimulation is the gold standard for treating medically refractory Parkinson’s patients \cite{Benabid1994,Pinter1999,Rodriguez-Oroz2005,Benabid2009,Lozano2019}. In patients with advanced Parkinson’s disease, it was shown that regular deep brain stimulation plus medication was superior to medication alone \cite{Deuschl2006}. Notwithstanding its therapeutic efficacy \cite{Weaver2009,Follett2010}, side effects are an issue \cite{Beric2001,Oh2002,Binder2003,Lyons2004}. In fact, regular deep brain stimulation may cause characteristic side effects denoted as deep brain stimulation-induced movement disorders \cite{Uc2007,Baizabal-Carvallo2016}. Treatment efficacy is another limitation. Regular deep brain stimulation administered to the standard targets, subthalamic nucleus or globus pallidus internus, is not effective for the therapy of gait and other so-called axial symptoms, e.g., balance and posture impairment, and hardly improves or even worsens speech as well as affective and cognitive symptoms \cite{Voon2007,Eisenstein2014,Merola2017,Healy2022}. 

Abnormal neuronal synchrony is a hallmark of Parkinson’s disease \cite{Hammond2007}. Based on computational modelling it was suggested to specifically counteract abnormal neuronal synchrony by desynchronizing stimulation with phase-dependent stimulus delivery \cite{Tass2007a} or by administering compound stimuli which cause a desynchronization irrespective of the initial dynamic condition \cite{Tass2001,Tass2003a}. By design, coordinated reset stimulation employs comparably weak, phase resetting stimuli and does not require sophisticated calibration procedures \cite{Tass2003a}. Accordingly, it was selected for pre-clinical studies (animal experiments) and clinical studies. Initially, coordinated reset stimuli were suggested to be delivered in a demand-controlled manner in a closed-loop setting, e.g., by delivering coordinated reset stimuli whenever a neuronal population gets resynchronized or by adapting the amplitude of the coordinated reset stimuli to the amount of synchrony \cite{Tass2003a}. At that time, no implantable pulse generators for coordinated reset stimulation were available for clinical tests \cite{Lozano2019,Adamchic2014}. Engineering-based concepts led to the development of closed-loop brain stimulation devices that recorded muscular or neuronal activity to suppress unwanted neuronal activity whenever detected \cite{Bouthour2019,Hoang2019}. Routine clinical applications of closed-loop deep brain stimulation still require a number of issues to be resolved \cite{Beudel2016}. 

In contrast, based on principles of adaptive dynamical systems, a qualitatively different stimulation approach was computationally developed \cite{Tass2006}. Adaptivity is a fundamental feature of the nervous system and, in fact, the entire body to cope with complex physiological processes subjected to environmental changes,  see Secs.\,\ref{SEC:KARIN}, \ref{SEC:MIEHL}, \ref{SEC:OLMI}, \ref{SEC:BADER} and \ref{SEC:HAJIZADEH}. By the same token, adaptive as well as maladaptive, i.e., less favorable responses to pathological changes, are key to disease mechanisms. For instance, in Parkinson’s disease a lack of dopamine initiates a cascade of functional and structural changes \cite{Madadi2022}. To specifically counteract disease-related adaptive changes, synaptic plasticity synaptic plasticity \cite{Zenke2015,Miehl2022b} (see also Secs.\,\ref{SEC:MIEHL} and \ref{SEC:OLMI}), specifically spike-timing-dependent plasticity \cite{Markram1997,Abbott2000,CD08,Feldman2012}, was incorporated in neuronal network models used to design therapeutic stimulation, giving rise to a radically new stimulation and treatment concept \cite{Tass2006}. 

\begin{figure}[hbt]
  \includegraphics[width=.4\textwidth]{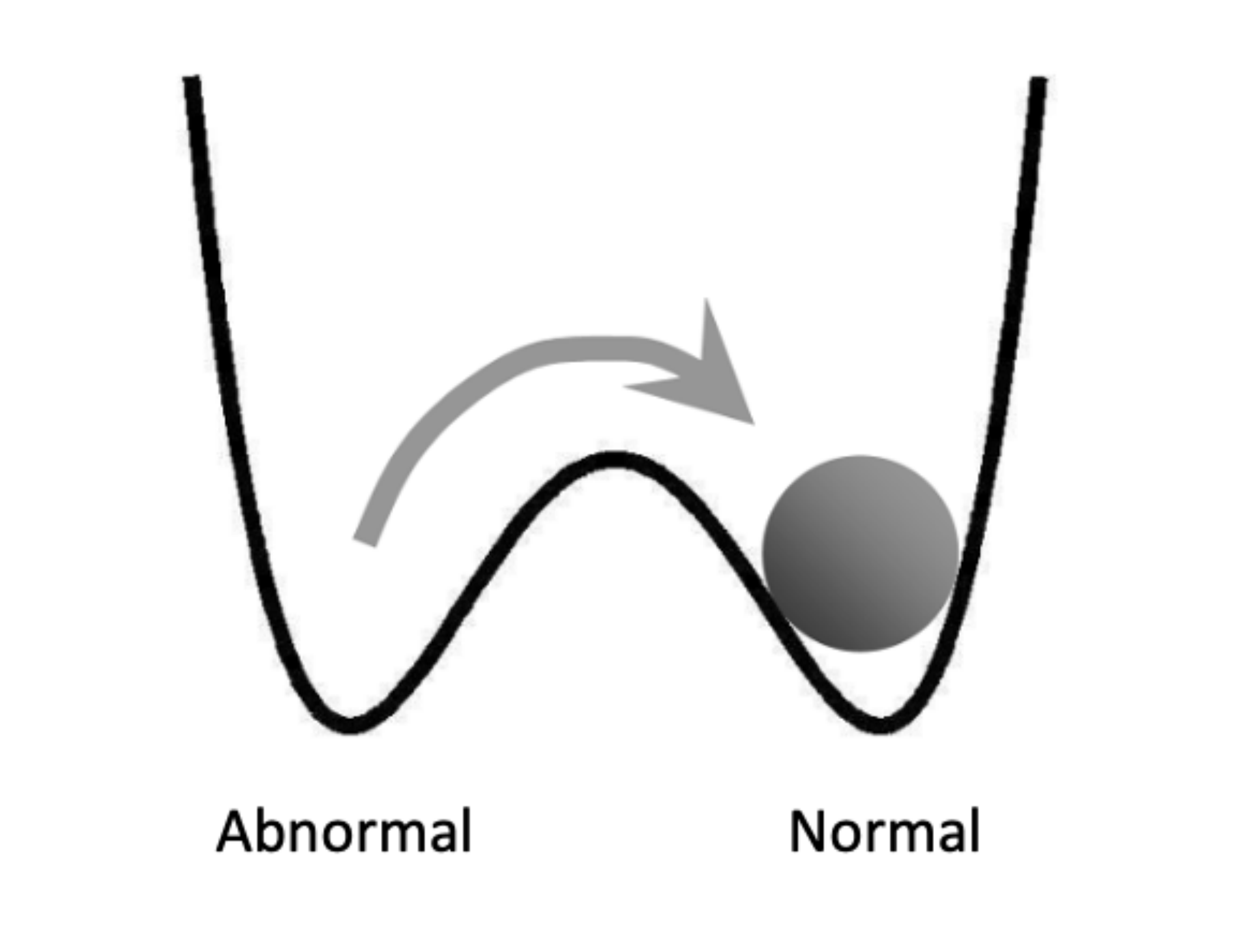}
  \caption{\label{fig_tass}Schematic illustrating how desynchronizing stimulation induces long-lasting therapeutic effects by leveraging plasticity. Spike-timing-dependent plasticity is a fundamental plasticity mechanism of the nervous system which adapts the synaptic strengths based on the relative timings of post- and presynaptic spikes \cite{Markram1997,Abbott2000,CD08}. Neural networks with spike-timing-dependent plasticity typically display bi- or multi-stability of stable states with stronger synchrony and synaptic connectivity and stable desynchronized states with weaker synaptic connectivity \cite{Tass2006,Maistrenko2007,Tass2007,Hauptmann2009,Manos2018,Kromer2020}, as illustrated by a simple double-well potential here. These states serve as models for pathological and physiological conditions. Coordinated reset stimulation may shift the network into the basin of attraction of a stable desynchronized state, in this way causing a long-lasting desynchronization \cite{Tass2006}.}
\end{figure}

It was observed that Coordinated Reset stimulation can shift a network from an unfavorable, synchronized attractor to a more favorable, desynchronized attractor (Fig.\,\ref{fig_tass}) \cite{Tass2006}. From then on, coordinated reset stimulation and further variants were computationally developed and optimized to robustly cause an ``unlearning'' of pathological synchrony and synaptic connectivity, in this way causing long-lasting therapeutic effects \cite{Tass2006,Tass2007,Zeitler2015,Manos2018,Tyulmankov2018,Kromer2020,Kromer2020,Khaledi-Nasab2020,Kromer2022,Madadi2023}. A series of computational studies revealed \textit{novel stimulus response characteristics} of neural networks with spike-timing-dependent plasticity: 
\begin{itemize}
	\item \textit{Rebound of synchrony after cessation of stimulation}: Directly after cessation of coordinated reset stimulation, synchrony may reemerge and then spontaneously fade while further approaching the desynchronized attractor \cite{Tass2007}.  
	\item \textit{Cumulative effects}: Effects of coordinated reset stimulation may accumulate over time \cite{Hauptmann2009}, and stimulation pauses may even improve the outcome \cite{POP15}. 
	\item \textit{Acute vs. long-term effects}: Acute stimulation effects (observed during stimulation) and long-term effects (emerging when the system relaxes into a stable state after cessation of stimulation) may differ substantially \cite{Manos2018,Kromer2020}. One can even decouple neurons, i.e., reduce their synaptic weights, without desynchronization during stimulation \cite{Kromer2020}. In fact, acute effects do not necessarily serve as predictive markers for long-term outcome \cite{Manos2018,Kromer2020}.  
	\item \textit{Transition to non-invasive stimulation}: Long-term effects are favorable because they enable to reduce stimulation time and, hence, potentially reduce side effects. However, a profound advantage of this type of stimulation is that it does not require implants to permanently deliver stimulation. Rather, as predicted theoretically \cite{Popovych2012,Tass2012}, non-invasive stimulation can be delivered occasionally or regularly for a few hours. Non-invasive therapies are typically less risky and more appropriate for larger patient populations. 
	\item \textit{Functional restoration}: Not only stimulation-induced unlearning of abnormal synaptic connectivity and neuronal synchronization \cite{Tass2006}, but also reshaping network connectivity by differentially up- or downregulating different synaptic connections \cite{Kromer2022} may contribute to a restoration of function. 
	\item \textit{Different plasticity mechanisms}: In Parkinson’s disease pathophysiology, both spike-timing-dependent plasticity and structural plasticity~\cite{BUT09,BUT13c} are important \cite{Madadi2022} and may induce different stimulation responses \cite{Manos2021,Chauhan2022}.
\end{itemize}
These computationally derived predictions and results enabled to design appropriate protocols for pre-clinical and clinical studies. 

\textit{Invasive coordinated reset studies}: Coordinated reset deep brain stimulation was successfully tested in Parkinsonian monkeys \cite{Tass2012,Wang2016,Wang2022,Chelangat-Bore2022}. For instance, a few hours of coordinated reset deep brain stimulation led to therapeutic effects lasting for one month \cite{Tass2012}. In addition, cumulative and long-lasting desynchronizing and therapeutic effects were observed in Parkinson’s patients treated with coordinated reset deep brain stimulation \cite{Adamchic2014}. 

\textit{Non-invasive coordinated reset studies}: Vibrotactile coordinated reset fingertip stimulation was developed to provide patients with a non-surgical and non-pharmacological treatment option \cite{Tass2017}. To this end, instead of administering electrical bursts through depth electrodes, weak, non-painful vibratory bursts were non-invasively delivered in a coordinated reset mode to patients’ fingertips \cite{Tass2017}. A first in human study \cite{Syrkin-Nikolau2018} as well as pilot studies \cite{Pfeifer2021} showed that vibrotactile Coordinated Reset stimulation is safe and tolerable and revealed a statistically and clinically significant reduction of Parkinson’s disease symptoms off medication together with a significant reduction of high beta (21-30 Hz) power in the sensorimotor cortex. Remarkably, also axial symptoms, difficult to treat with regular deep brain stimulation, responded well to vibrotactile coordinated reset in these studies \cite{Syrkin-Nikolau2018,Pfeifer2021}. For illustration, see patient videos in \cite{Pfeifer2021}. Of note, Parkinson’s disease patients improved during a months-long vibrotactile coordinated reset treatment when evaluated after medication withdrawal, indicating a substantial improvement of the patients’ conditions \cite{Pfeifer2021}. These findings indicate that vibrotactile coordinated reset treatment might even have an impact on metabolic and degenerative processes \cite{Pfeifer2021,Tass2022}, e.g., by slowing or even counteracting degeneration-related processes, e.g., vicious circles giving rise to oxidant stress and mitochondrial impairment, causing a bioenergetic crisis and the death of dopamine neurons in the substantia nigra \cite{Esposito2007,Haelterman2014,McGregor2019}.  

In summary, instead of simply suppressing unwanted neuronal activity, based on principles of adaptive dynamical systems, appropriately designed stimulation techniques intend to induce sustained therapeutic effects by moving affected neural systems to more favorable attractors (Fig.\,\ref{fig_tass}).

\subsection{Music and adaptivity – A Physical Culture Theory -- by Rolf Bader}\label{SEC:BADER}

Understanding music is an interdisciplinary task \cite{Bader2019b}. Musical instruments are built such that we can listen to them, actively play them, use them in social contexts, or use them in terms of individual demands and tasks. Therefore, scientific disciplines like the physics of musical instruments, music psychology and neuromusicology, music sociology, or political science must interact to arrive at a holistic understanding of music. Furthermore, the role of music in culture, technology, economy, ethnicity, or its interactions with natural resources like wood or alternative material for musical instrument building needs to be considered.

Thereby, music is a constant adaptation process. Listeners adapt to new musical pieces. Musicians adapt to audiences, new musical instruments available, or new ideas of compositional techniques. Instrument builders adapt to contemporary sound and performance demands, new materials, or new technologies. Society adapts to new musical pieces, genres, or ways of music presentations like mass media or streaming platforms. Such adaptations are processes, including changing strategies, emotional reactions, or the development of new abilities. The participants of such adaptations might welcome and deal with or might try to reject and oppose new developments.

In contemporary research, each scientific discipline uses its own methods for understanding and predicting music \cite{Bader2019b}. Music Psychology often uses statistics or Bayesian methods. Musical Acoustics involves mainly analytical equations and discretization methods like Finite-Element or Finite-Difference methods. Music ethnology is still dominated by heuristic and historical methodology, while computational or analytical ethnomusicology also include mathematical modeling, e.g., of tonal systems. In all fields, machine learning methods have become more and more important like connectionist models nearly always used for composition (see e.g., Briot et al.\cite{Briot2020} for an overview), or self-organizing Kohonen maps \cite{Kohonen1995}, often used for the analytical purpose \cite{Bla2020,Bader2019a,Leman1997}.

The methodologies used, therefore, strongly depend on the subfields, but some also intertwine, e.g., in the field of psycho-acoustics, relating physics to perception, using algorithms calculating loudness, brightness, pitch, spatial audio, or the like. Still, to arrive at a common, robust, suitable algorithm able to model music in a global, holistic way in the future, also including extra-musical players like ecology, economy, or politics, a common ground is needed, not debatable among the very diverse disciplines involved. For example, a Physical Culture Theory suggests music as an adaptive system to consist of impulses, physical energy bursts sent out, returning with a specific damping, thereby causing new impulses \cite{Bader2021}. In its most general form, the Impulse Pattern Formulation can be written as a system parameter $g$ representing an impulse sent out by one subsystem. This impulse is reflected at $n$ other subsystems with damping parameters $\alpha$ and $\beta_k$ for each reflection point $k$ like \cite{Bader2013,Linke2019}:

\begin{align}\label{eq:Bader}
g_+ = g - \ln\Bigl(\tfrac{1}{\alpha}\bigl(g-\sum_{k=1}^{n} \beta_k e^{g-g_{k-}}\bigr)\Bigr).
\end{align}

The system parameter $g$ is updated at each iteration step to $g_+$, taking the most recent $g$ and the previous $g_{k-}$ into consideration. The logarithm reflects the exponential damping found in most systems. Adaptation is present for $g_+ = g$. E.g. with musical instruments, $g$ can be taken as a periodicity of a musical tone. During the initial transient phase, $g_+ \mathrel{\mathtt{!=}} g$, and the system struggles, leading to a complex initial transient sound. After the initial phase, a stable periodicity is reached, and a musical pitch is heard. E.g., with a guitar, two subsystems are present, the string and the guitar body, both with their own eigenfrequencies. Still, when playing a note, the string's vibration takes over the guitar body's vibration, i.e., the body adapts to the string's pitch. Thereby, the Impulse Pattern Formulation is able to model the guitar tone very precisely, which is especially reflected in the length and complexity of the initial sound phase \cite{Bader2013}.

Such an Impulse Pattern Formulation algorithm is scale-free and therefore able to model and predict very small networks as well as overall or general behavior fast and precise in musical acoustics \cite{Linke2019,Linke2021} or music perception and action \cite{Linke2021a}. Such a self-organizing system is found as a basis for all musical instrument families. Moreover, it is the basis of brain dynamics \cite{Bader2022} and all interactions in society or politics.

For such a system to work for aesthetic and artistic matters, consciousness and conscious content, like experiencing sound, vision, emotion, or any kind of cognition, need to be incorporated. The Physical Culture Theory assumes conscious content to be spatio-temporal electric fields in the brain, complex enough to arrive at experiences of all kinds. Such a spatio-temporal field, again is nothing but a complex impulse pattern. Brain dynamics is thereby no longer taken as an interplay of bottom-up and top-down processes but as a complex, self-organizing system. Localization of brain regions processing certain tasks, like audition, vision, or thinking, is still evident in this picture, as auditory input enters the brain through the ear, cochlear, and auditory pathway to end in the auditory cortex (as e.g. in the auditory oddball paradigm, see Fig.\,\ref{fig:fig_idea} and Secs.\,\ref{SEC:MIEHL} and \ref{SEC:HAJIZADEH}). Still, already within this brain network, circular neural processing is often present, nearly directly connecting the cortex to the cochlear in the inner ear and back up to the cortex. Therefore, adaptation of the brain to external input is an active process involving the whole brain, although the input of sensory information can clearly be located.

In such global musical networks, stable, bi-stable, bifurcating, complex, or chaotic scenarios occur \cite{Bader2013}. In terms of musical instrument sounds, a stable musical pitch is only established after a complex initial transient sound phase. Each new tone of a melody needs to undergo such changes. This also holds for brain activity \cite{Kozma2016}. In ensemble playing, the interaction of musicians reacting to co-musicians’ performances is also undergoing such complex changes. Therefore, the whole system is a constant interplay of surprise and adaptation to changing scenarios. Although such adaptation might work, leading to a steady state, it also might fail to arrive at more extended times of chaos, noise, or bifurcating sounds. Adaptation and disruption are, therefore, two essential and ever-repeating sides of music on all levels, with sound, musical pieces, musical genre formation, or music history.

\subsection{Adaptation in auditory cortex explained by modulations of synaptic coupling -- by Aida Hajizadeh}\label{SEC:HAJIZADEH}

Most sounds like speech and music evolve and unfold in time and, yet, the brain perceives them as one whole continuous entity (see also Sec.\,\ref{SEC:BADER}). For this, the brain needs to exhibit a memory mechanism whereby incoming stimuli are represented and integrated with the trace of the stimuli extending to the immediate past. This ability is termed temporal integration. Whilst source localization and spectral analysis are suggested to be the task of subcortical areas, temporal integration of sounds is proposed to occur in the auditory cortex \citep{Nelken2004}. In attempts to understand how auditory cortex performs temporal binding, it was shown by intracranial and extracranial measurements that neural responses in auditory cortex are context sensitive \citep{Butler1968,Brosch2000}. That is, the neural response to a stimulus is modified when the same stimulus is presented in the context of different stimuli where this sensitivity is a function of both temporal occurrence and spectral content of the preceding stimuli \citep[see, for example,][]{Naatanen1978,Ulanovsky2004,Zacharias2012}. The simplest form of context sensitivity in the auditory cortex occurs when the same stimulus is presented repetitively with a constant stimulus onset interval. The result is a gradual reduction of the magnitude of the neural responses and is termed adaptation. Adaptation is stimulus specific and a function of the interval between the stimulus onset interval \citep{Gonzalez2014}. 

The stimulus-specificity of adaptation was shown in oddball paradigms, where the repetitive presentation of a frequent standard stimulus is interrupted by an infrequent deviant stimulus (see also Fig.\ref{fig:fig_idea} in Sec.\,\ref{SEC:MIEHL}). The magnitude of the neural responses to the standards is smaller than the magnitude of the responses to the deviants \citep{Butler1968,Ulanovsky2004}. This is known as stimulus-specific adaptation and the mismatch responses in invasive and noninvasive measurements, respectively \citep{Naatanen1978,Ulanovsky2004}. Despite decades of research on adaptation and its relevance for stimulus-specific adaptation and mismatch responses, understanding how adaptation takes place in auditory cortex remains challenging. Already single neurons, due to their intrinsic properties, show adaptation, which is termed spike frequency adaptation (see also Sec.\,\ref{SEC:OLMI}) \citep{Gonzalez2014}. Adaptation is observed in the auditory nerve fibers of the cochlea as well as in the inferior colliculus and thalamus which act as relay stations between the cochlea and the auditory cortex. There are reasons why adaptation in auditory cortex is neither only the result of single neurons adapting to the stimulus statistics nor just inherited from the subcortical regions \citep[for reviews, see, for example,][]{Gonzalez2014,Whitmire2016}. The time scales at which single neurons in different stations along the auditory pathway exhibit adaptation is different from those occurring in the auditory cortex \citep{Gonzalez2014,Malmierca2015}. Unlike in the nonlemniscal pathway, adaptation does not occur in those subdivisions of the inferior colliculus and thalamus in the lemniscal pathway which target the primary auditory cortex (i.e., core area) \citep{Ulanovsky2004,Gonzalez2014}. Along the auditory pathway, adaptation manifests itself in more complex ways with its time scales in the auditory cortex adapting to the time scales of the stimulation \citep{Ulanovsky2004,Gonzalez2014}.

Neurons in the brain form 
networks and do not appear in isolation. 
The contact points between neurons are 
synapses whose dynamics are highly plastic. One prevailing view on the underlying mechanisms of adaptation in auditory cortex is that it
is due to modulations of synaptic coupling between 
neurons. However, what accounts for modulations 
of synaptic coupling is an ongoing debate 
\citep{Friston2005,May2021}. Short-term 
synaptic depression has been hypothesized to be one plausible 
physiological mechanism \citep[see, for example,][]{May2010,Wang2013,Kudela2018,Seay2020} (see also Secs.\,\ref{SEC:MIEHL} and \ref{SEC:OLMI}). 
This type of synaptic plasticity, which occurs 
due to the repetitive stimulation of the 
pre-synaptic neurons, is mainly based on 
vesicle depletion and desensitization of release 
site and calcium channels on the synapses of the 
pre-synaptic neurons \citep{Zucker2002}. Short-term 
synaptic depression occurs at time scales that are 
similar to the time scales of context sensitive 
responses, and it has a high functional relevance for 
temporal filtering \citep{Fortune2001}, 
gain control \citep{abbott1997}, and, although 
counterintuitively, efficient information transfer 
between neurons \citep{Salmasi2019}.

In our research, we implemented dynamics of short-term 
synaptic depression in a computational model whose 
network structure is based on the anatomy of the 
mammalian auditory cortex \cite{May2013,Hajizadeh2019,Hajizadeh2022}. 
The auditory cortex of mammals is characterized by the hierarchical 
core-belt-parabelt structure, where each of these three 
areas is subdivided into tonotopically organized 
fields \citep{Kaas2000,Hackett2015}. The model 
comprises mean-field excitatory and mean-field inhibitory 
cell populations, which are characterized by nonlinear firing 
rates. The interconnection between cell populations 
are modulated by short-term synaptic depression 
according to the spectrotemporal pattern of the stimulation.
The linearized form of the state equations together 
with the slow-fast approximation of the equation for 
short-term synaptic depression allows for the analysis 
of the model dynamics in terms of damped harmonic oscillators, i.e.,
normal modes \citep{Hajizadeh2019,Hajizadeh2022}. We could show that the 
properties of the normal modes (i.e., frequency, phase, 
initial amplitude, spatial wave pattern, and decay rate) 
are functions of the macro- (gross anatomy) and 
micro-structure (synaptic weight values) of the auditory cortex network 
as well as of the spectrotemporal pattern of the 
stimulation. In this approach, the auditory cortex is viewed as a spatially 
extended structure and the activity elicited by an 
external stimulus propagates in time and space. The dynamics 
of short-term synaptic depression, which locally trace 
the stimulus history at the synapses, determine the 
oscillations that are spread over the entire auditory cortex. In 
this view, local and global population activity that 
are revealed by intracranial and extracranial recordings, respectively, 
emerge from the constructive and destructive 
interference patterns of superimposed normal modes.
This contrasts with the traditional view where, for 
example, electromagnetic activity in the brain measured 
by means of magnetoencephalography reflects the summed activity of discrete
local generators distributed over the auditory cortex. In the 
normal-mode view, adaptation in the auditory cortex can be described 
as modulations of the properties of these normal modes 
due to the modulations of synaptic coupling, where the 
reduction of response magnitude is just a by-product 
\citep{Hajizadeh2022}.


%
%

\section{Adaptivity and artificial learning}
\label{SEC:ARTIFICIAL-LEARNING}

In this section, different authors reflect on the meaning of adaptivity in the context of artificial learning. Among other topics, fundamental open problems in machine learning are discussed, as well as some perspectives on how machine learning can be used to solve physics problems, and to create new control strategies for nonlinear (chaotic) systems are given. Towards the end of this section, the role of artificial learning to understand and control complex many-body systems and cooperative behavior is discussed.

\subsection{Adaptivity is the key to success of neural networks -- by Sebastian Goldt}\label{SEC:GOLDT}%
\label{sec:neural networks}

Deep neural networks have powered a series of breakthroughs in machine
learning over the last ten years. Since their early success in computer
vision~\cite{krizhevsky2012imagenet, lecun2015deep, simonyan2015very,
  he2016deep, dosovitskiy2021image}, they have set new standards in natural
language processing~\cite{hinton2012deep, sutskever2014sequence,
  vaswani2017attention, devlin2019bert} and the playing of complex games like
Go~\cite{silver2016, silver2017mastering} or
Poker~\cite{bowling2015heads, brown2018superhuman, brown2019superhuman}. Deep
learning also increasingly impacts the natural
sciences~\cite{carleo2019machine}; for example, deep neural networks recently helped predicting the
3D-structure of nearly every human protein~\cite{tunyasuvunakool2021highly} in a
breakthrough for structural biology. 
Further applications of machine learning to solve physics problems are also given in Sec.~\ref{SEC:SEIF}.

While neural networks used in machine learning are inspired by biological neural circuits such as the ones described in Secs.~\ref{SEC:MIEHL} and \ref{SEC:OLMI}, 
the neurons in machine learning are much simpler than biological neurons. Yet it turns out that a different form of adaptivity is behind the success of deep learning.
We illustrate this point using the classic machine learning task of
recognising whether a given image shows a cat or a dog. Given an image $x$,
represented by an array of pixel values, the classical approach was to compute a
vector $\tilde x$ of \emph{features}~\cite{oliva2001modeling,
  dalal2005histograms, bay2006surf} that represents the image, which is then fed
into a classifier. Features could be the location of edges in an image, or the
correlations between patches of the same image. These features were designed
\emph{a priori} and required extensive domain knowledge.

The key idea of deep learning is instead to learn the relevant features directly
from data.  So rather than computing a feature vector using a predefined set of
transformations, we try to learn a function $f_\theta(x)$ that maps the raw
images $x$ directly to a ``label'' $y=\pm 1$ indicating whether the image shows
a cat or a dog. A neural network is a particular functional form for
$f_\theta(x)$, usually consisting of a series of alternating linear
transformations and point-wise non-linear functions.\footnote{There are
  different architectures which are appropriate for certain types of data or
  certain tasks; for an overview, cf.~\cite{goodfellow2016deep}.} The
adjustable parameters~$\theta$, called weights, determine what the
transformations compute exactly. They are found by maximizing the prediction
accuracy of the network on a given set of
images
, which is called ``training'' the network~\cite{mackay2003}. In practice,
simple first-order optimization methods such as stochastic gradient descent work
best~\cite{hardt2016train, wilson2017marginal}. Training a neural network is
thus a general-purpose procedure to obtain features that are well-adapted to the
input data and the task at hand.

From a theoretical point of view, the success of this approach is surprising for
several reasons. For example, fitting a function in a high-dimensional space,
such as the space of natural images, suffers from the curse of dimensionality:
the number of samples required to estimate such a function accurately scales
exponentially in the input dimension~\cite{luxburg2004distance}. A lot of
current research activity, for example in statistical
physics~\cite{carleo2019machine, bahri2020statistical,
  zdeborova2020understanding, geiger2021landscape}, is currently working to
reconcile the success of neural networks with the curse of dimensionality.

\begin{figure}[t]
  \centering
  \includegraphics[width=\linewidth]{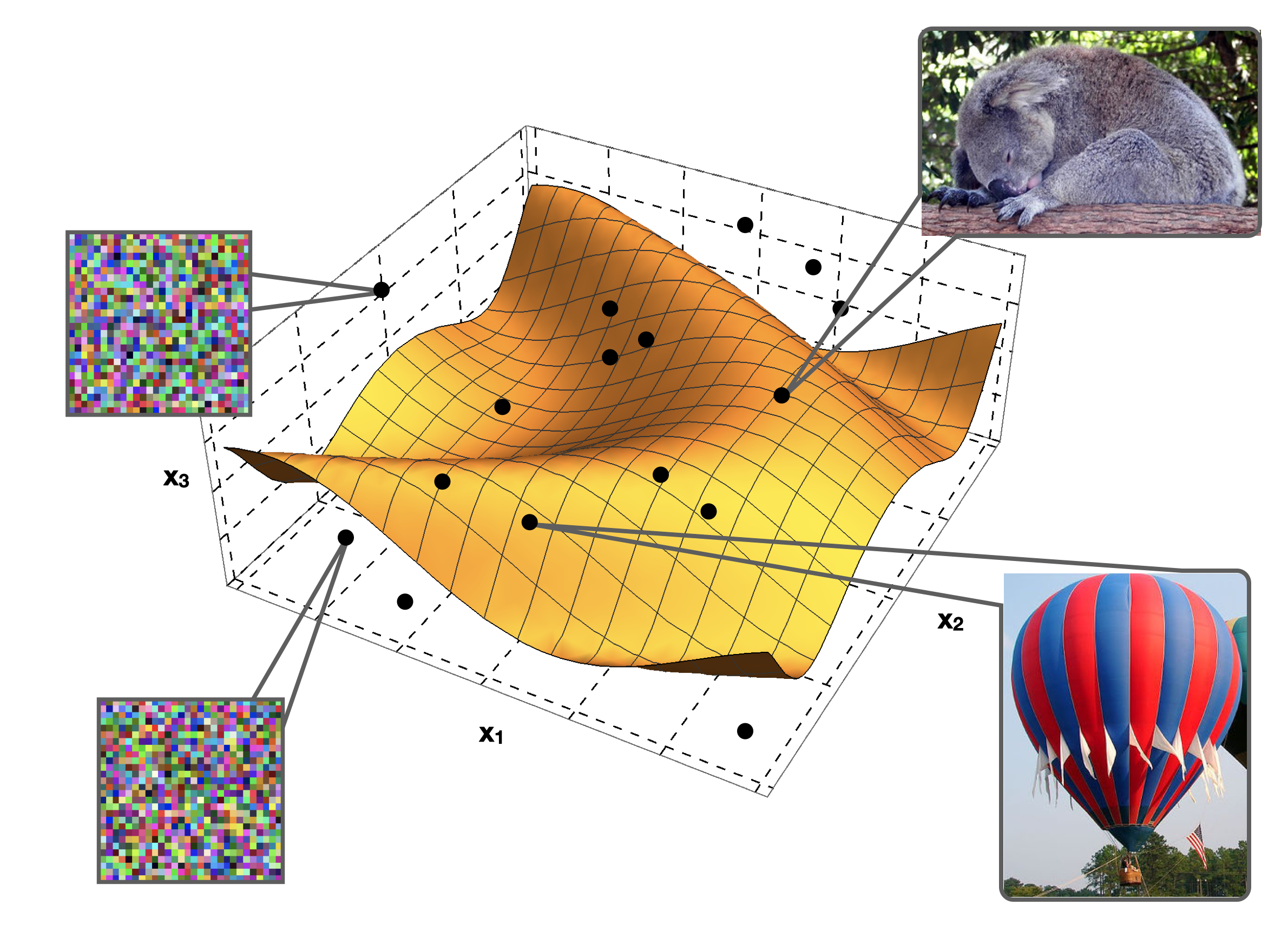}
    \caption{\textbf{The manifold structure of realistic images.} 
Each black dot indicates a point in a high-
dimensional space which could be an input for a neural networks. In the eye of a human observer, most inputs in this space resemble random noise, like the ``images'' shown on the left. Neural networks exploit the fact that realistic images tend to concentrate on a lower-dimensional manifold in input space, sketched here as a two-dimensional curved surface. Figure adapted
from\textcite{goldt2020modelling}, images taken from ImageNet~\cite{deng2009imagenet} data set.}\label{fig:manifold}
\end{figure}

One key to this puzzle is that images are not as high-dimensional as they
seem. Most of the points in the high-dimensional input space do not represent
images (at least not to a human observer) and instead look like random
noise. The points that do represent real images tend to concentrate on a
lower-dimensional \emph{manifold} in input space, sketched as a two-dimensional
curved surface in Fig.~\ref{fig:manifold}. While the manifold is not easily defined,
it is tangible: its dimension has been estimated
numerically
~\cite{grassberger1983measuring, Costa2004, Levina2004a, spigler2019asymptotic,
  pope2021intrinsic}, and found to be 10-100 times smaller than the image
dimension.

It is difficult to analyse the impact of the low intrinsic dimension of images
on neural networks theoretically, because we lack the mathematical tools to
reason about real-world data. A series of works therefore introduced models of
data with low intrinsic dimension, such as object
manifolds~\cite{chung2018classification}, the hidden
manifold~\cite{goldt2020modelling, goldt2022gaussian}, or the spiked covariate
model~\cite{ghorbani2020neural, richards21asymptotics}. Each of these models
offers a controlled environment in which the adaptivity of neural networks can
be studied, using tools from statistics or statistical physics. One result of
these studies is that neural networks can indeed adapt to lower-dimensional
manifolds in their data better than classical methods of machine learning, like
kernel methods~\cite{bach2017breaking, ghorbani2019limitations,
  chizat2020implicit, ghorbani2020neural, geiger2020disentangling,
  daniely2020learning, refinetti2021classifying}.

These results set the blueprint for a research program that aims to understand
the interplay of neural networks and the data on which they operate. What are
the (potentially) low-dimensional structures in other data modalities such as
human language, or amino acid sequences, that neural networks can exploit?

\subsection{Machine learning applications in physics -- by Alireza Seif}\label{SEC:SEIF}
	Machine learning tools have found extensive use in the study of physical problems~\cite{carleo2019machine}. While it is not possible to provide an exhaustive list of these applications in this perspective, we highlight a few examples related to statistical physics, namely learning and sampling from equilibrium distributions~\cite{torlai2016learning}, classifying phases of matter~\cite{carrasquilla2017machine,van2017learning}, estimating free energy differences~\cite{wirnsberger2020targeted}, identifying the direction of time's arrow~\cite{seif2021machine}, and estimating entropy production~\cite{kim2020entropy}. For a comprehensive review of machine learning in physical sciences, readers may refer to Ref.~\cite{carleo2019machine}. However, the relationship between physics and machine learning is not one-sided. Tools from theoretical physics have illuminated how machine learning tools function ~\cite{bahri2020statistical} (see also Sec.~\ref{SEC:GOLDT}).  In the following, we examine these two directions through the lens of adaptivity.

	First, we examine how machine learning can be applied to solve physics problems, with a focus on the role of adaptivity. In particular, we consider supervised learning tasks where input-output pairs are provided, and the objective is to train a neural network to accurately predict the target output value given an input. As discussed in Sec.~\ref{SEC:GOLDT}, adaptivity plays a crucial role in training the networks. In the optimization process, the network's weights are adjusted to minimize the difference between the predicted and target output values, so that the network can make accurate predictions. However, as we discuss in this section, the network's prediction can be further enhanced by adapting to the history of previous inputs. This additional degree of adaptivity is particularly useful when working with sequential data.  Recurrent neural networks allow for this type of adaptive inference by using an internal state that depends on the input at the previous step. Given a sequence of input tokens $\mathbf{x}_t\in\mathbb{R}^{n_v}$ and the hidden state $\mathbf{h}_t \in \mathbb{R}^{n_h}$ at timestep $t$, this dependency can be captured as~\cite{goodfellow2016deep}
	\begin{equation}\label{eq:rnnint}
		\mathbf{h}_{t} = f(\mathbf{x}_t,\mathbf{h}_{t-1};\boldsymbol{\theta}),
	\end{equation}
	where $f$ represents a neural network parameterized by $\mathbf{\theta}$. In the most basic form, the output of the network $\mathbf{y}_t$ can be calculated by applying another parameterized function to  $\mathbf{h}_{t}$. While in principle these networks can capture long-term dependencies in a sequence, it has been shown that training them can be challenging due to vanishing or exploding gradients~\cite{bengio1994learning}. More complicated constructions of recurrent neural networks, such as long short-term memory networks solve this problem using a self-loop that allows the gradient to flow for longer~\cite{hochreiter1997long}. Modern machine translation tools build on these networks to map sequences in one language to sequences in another (seq2seq)~\cite{sutskever2014sequence}. 
	
	Among many applications of these models in physics, we briefly discuss inferring force fields from the trajectory of particles~\cite{argun2020enhanced} and chaotic time-series forecasting~\cite{pathak2018model}. Ref.~\cite{argun2020enhanced} considers the problem of inferring the force field in overdamped Brownian motion. Specifically, the input $\mathbf{x}_t$ represents the position of the Brownian particle, and the output is the parameter(s) that describe the functional form of the potential. For example, in the case of a harmonic potential $U(x)=\frac{1}{2}kx^2$, the output of the network at the final step represents the inferred value of $k$. The recurrent neural network is shown to outperform conventional methods with limited data and can remarkably infer non-conservative time-dependent force fields, which conventional methods cannot handle. Ref.~\cite{pathak2018model} focuses on forecasting the dynamics of chaotic systems following the Kuramoto-Sivashinsky equation~\cite{kuramoto1978diffusion,sivashinsky1977nonlinear,sivashinsky1980flame}. The input is a discretized scalar field in space at step $t$, and the desired output is the value of the field at step $t+1$. The authors use the framework of reservoir computing~\cite{lukovsevivcius2009reservoir} (a recurrent neural network with an untrainable input-to-internal-state mapping) to forecast the dynamics far beyond the Lyapunov time. In addition, see Sec.~\ref{SEC:GAUTHIER} for a discussion on using reservoir computing to control chaotic dynamical systems. In both of these examples, the network's internal state is adjusted based on the input history (see Eq.~\eqref{eq:rnnint}), allowing it to capture temporal dependencies in the input data sequence. 
	
	The two examples discussed earlier demonstrate applications of recurrent neural networks in solving physics problems. However, it is also important to examine the reverse direction, where physics problems can be used to better understand recurrent neural networks. Ref.~\cite{seif2022impact} provides a case study of this approach, where a simple model for seq2seq tasks is used to investigate the impact of data distribution in learning using a physical problem. Specifically, it considers the stochastic switching-Ornstein-Uhlenbeck process, which is a latent variable model that describes the trajectories of a Brownian particle in a harmonic potential with a time-dependent center that stochastically alternates between two values. The non-Markovianity of the input sequence is controlled by varying the distribution of waiting times between these alternations. The goal is to infer the current location of the center from the particle's past trajectory. The authors use several machine learning models for this task and demonstrate that increasing the memory of the learning model always improves the accuracy of the predictions, whereas increasing the non-Markovianity of the input sequences can either improve or degrade performance. They also identify an intriguing relationship between the performance of a learning model and distinct phases in the stationary state of the stochastic switching-Ornstein-Uhlenbeck process. In this case, as the memory of the learning model is increased, the network becomes more adaptable to longer-term dependencies in the input sequence, which in turn leads to improved performance.
	
	The two-sided relationship between physics and machine learning is still in its early stages of development, leaving plenty of opportunities for further exploration. On the one hand, artificial intelligence can aid in discovering and explaining scientific phenomena, with emerging techniques such as natural language processing models potentially facilitating communication between users and algorithms~\cite{krenn2022scientific}. On the other hand, statistical physics has already been used to provide theoretical insights into the behavior of deep learning~\cite{bahri2020statistical}, and the theory of adaptive systems could prove particularly valuable in understanding the role of data structure and the dynamics of learning in recurrent neural networks.
	
\subsection{Controlling Dynamical Systems  -- by Daniel Gauthier}\label{SEC:GAUTHIER}

\begin{figure}[th]
\includegraphics[width=0.8\linewidth]{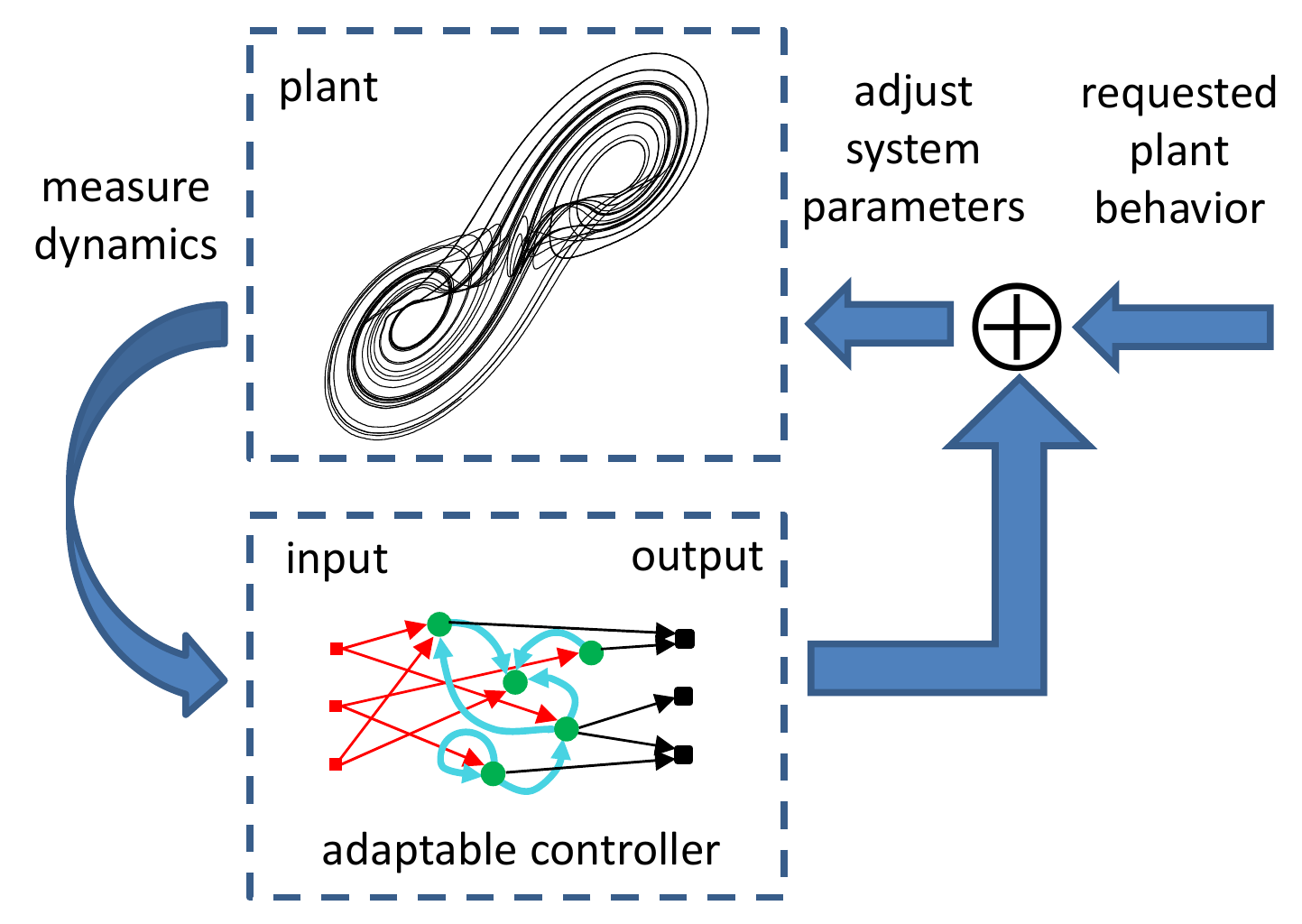}
\caption{\label{fig:Gauthier} A complex dynamical system controlled using closed-loop feedback. The controller is designed using an adaptive machine learning approach.}
\end{figure}

In this Section, we consider controlling complex dynamical systems using closed-loop feedback based on a machine learning approach known as reservoir computing.  Here, the concept of adaptivity appears in at least two guises: the dynamical system being controlled, often call the \textit{plant}, and the the controller.  For a plant to be controlled to a desired behavior, we need to have access to some signals generated by transducers attached to the plant that can be used to infer its dynamical state, and have access to one or more parameters that adjust the state of the plant as illustrated in Fig.~\ref{fig:Gauthier}.

The controller need to process plant signals and perform inference to estimate the state, compare this to the requested plant behavior, and generate control perturbations that are applied to the adjustable system parameters. For complex dynamical systems, especially those that display chaos, the control perturbations are a nonlinear function of the plant's state and requested behavior and thus fall in the category of a nonlinear controller.  Traditionally, nonlinear controllers require an accurate model of the plant, which often entails substantial effort from expert control engineers and mathematical model builders.

One highly successful alternative that was developed decades ago for controlling chaotic systems is to take advantage of unstable sets that are the backbone of the chaotic system in phase space, such as unstable periodic orbits.\cite{Ott1990,Romeiras1992}  A chaotic system naturally visits these unstable sets and control perturbations are designed using a linear algorithm that is valid in a local neighborhood of these sets.  Controlling other behaviors, however, requires a fully nonlinear controller.

One approach for realizing a fully nonlinear controller is to use machine learning to learn a model of the plant,\cite{Sarangapani2006} referred to as nonlinear system identification in the control engineering literature. Artificial deep neural networks in a feed-forward geometry {are} known to be universal approximators of functions (see Sec.~\ref{SEC:GOLDT}) and hence should be able to learn how to map measurements and requested state to control perturbations.  Here, a multi-layer network of artificial neurons with nonlinear input-output functions is trained by adjusting the network link weights using a supervised learning.  While there has been good success using this approach, the amount of data needed to train the network can be substantial making it difficult for the controller to adapt to changes in the plant.

Reservoir computing is a fast and low-data machine learning approach especially well suited for learning models of dynamical systems\cite{Jaeger2004} because it is also a dynamical system and it holds great promise for controlling dynamical systems.  
As seen in the lower dashed box of Fig.~\ref{fig:Gauthier}, the reservoir computer consists of an input layer (red squares), a pool of neurons (green dots, the ``reservoir''), and an output layer (black squares).  The neuron dynamics are described by a differential equation that is driven by a nonlinear function of the signals from the input layer and the output of other neurons in the the reservoir, and has a simple exponential time constant.
Thus, it has short-term memory that can be matched to the plant dynamics.  The link weights on the input layer and internal `reservoir' of neurons are not trained; they are assigned randomly at the outset and only the weights of the output layer are trained.  This dramatically reduces the size of the training data as well as the training computation time. Furthermore, the neural network can perform multiple tasks by combining a single reservoir with different trained output layers.
One approach for controlling dynamical systems with a reservoir computer is to train it to learn the inverse of a dynamical system in the presence of control.\cite{Waegeman2012} That is, we train it to learn the perturbations required to guide the system to the desired state sometime in the future.  This approach works well for systems such as a robotic arm that display constrained low-dimensional behavior, but a parallel deep architecture appears to be required for controlling complex systems that display chaos.\cite{Canaday2021}  The training data required for reservoir-computing inverse control appears to be on the order of 10,000 data points and modest computation time, suggesting that it can be used for real-time adaption of the controller as the underlying plant changes its dynamics because of non-stationarity or a damage event.

An open question is whether the data requirements can be reduced further so that a small microprocessor typically found on internet-of-things devices can be used to retrain the controller.  Our recent work\cite{Gauthier2021} that reformulates the reservoir computer as delay lines of the measured plant signals followed by a nonlinear output layer may be promising for this application because it reduces the amount of training data by a factor of ten or more.  But it is not yet clear whether this new approach gives up some adaptivity.  We are working on extensions of this work to balance the desire for fast training with wide adaptivity.


\subsection{Modeling complex adaptive human-environment systems with multi-agent reinforcement learning dynamics -- by Wolfram Barfuss}\label{SEC:BARFUSS}

Rapid and large-scale collective action is required to enter sustainable development pathways in coupled human--environment systems safely away from dangerous tipping elements\cite{SteffenEtAl2018} (also see Sec. \ref{SEC:KURTHS}).
The question, however, of how \textit{collective} or \textit{cooperative} behavior — in which agents seek ways to improve their welfare jointly — emerges is unresolved\cite{Levin2020}.
Evolutionary game theory has produced a sound equation-based analytical understanding of the mechanisms for the evolution of cooperation\cite{Nowak2006a}. Yet, this was primarily done with highly simplified models, lacking environmental context and cognitive processes\cite{McNamara2013}. These elements are the center of artificial intelligence and cognitive neuroscience research\cite{BotvinickEtAl2020, SuttonBarto2018}, which only recently emphasized the need for developing cooperative intelligence\cite{DafoeEtAl2020, DafoeEtAl2021}. Moreover, analyzing systems composed of multiple intelligent agents typically requires expensive computer simulations, which are not straightforward to understand\cite{Holland2006, SchulzeEtAl2017, GottsEtAl2019, Hernandez-LealEtAl2019}.
Thus, little is known about how cooperative behavior emerges from and influences a collective of individually intelligent agents in complex environments.

There is a unique opportunity for \textit{adaptivity} in non-linear dynamical systems to help solve this challenge. Based on the link between evolutionary game theory and reinforcement learning\cite{TuylsNowe2005, TuylsParsons2007}, we can model a collective of reinforcement learning agents \textit{as} a dynamical system. Doing so provides improved, qualitative insights into the emerging collective learning dynamics\cite{BloembergenEtAl2015}, enabling equation-based analytical tractability of agent-based simulations.

Here, reinforcement learning is the central adaptive mechanism (cf. ``Reinforcement learning is direct adaptive optimal control''\cite{SuttonBartoWilliams1992}, also see Sec. \ref{SEC:BOTTA}). Reinforcement learning is a trial-and-error method of mapping observations to actions in order to maximize a numerical reward signal. The challenge is that those actions can change the environment's state, and rewards may be delayed. Reinforcement learning is not only an artificial learning algorithm\cite{SuttonBarto2018}, is has also wide empirical support from  neuroscience\cite{SchultzEtAl1997,DayanNiv2008}, psychology \cite{BushMosteller1951} and economics\cite{Cross1973,ErevRoth1998,SchultzEtAl2017,Burton-ChellewWest2021}. It is, therefore, ideally suited to model coupled human-nature systems.

In their seminal work, Börgers and Sarin showed how one of the most basic reinforcement learning update schemes, cross-learning\cite{Cross1973}, can converge to the deterministic replicator dynamics of evolutionary games theory\cite{BorgersSarin1997}. The relationship between the two fields is as follows: one population with a frequency over phenotypes in the evolutionary setting corresponds to one agent with a frequency over actions in the learning setting\cite{TuylsNowe2005}.
Since then, this analogy has been extended to other reinforcement learning variants, such as stateless Q-learning\cite{TuylsEtAl2003, SatoCrutchfield2003}, regret-minimization \cite{KlosEtAl2010}, and fictitious play\cite{Hopkins2002}.
Of particular relevance to modeling coupled human-nature systems is the dynamic formulation of the general and widely used class of temporal-difference learning\cite{BarfussEtAl2019}, which is able to learn in changing state-full environments.

Typically, the learning dynamics are derived by performing a mathematical separation of timescales of the interacting process with the other agents and the environment and the process of adapting the agents' policy to gain more reward over time\cite{Barfuss2020}. Under the complete separation of timescales, the dynamics become deterministic.
One can understand such learning dynamics as an idealized model of the multi-agent learning process, in which agents learn as if they have a perfect model of the current environment, including the other agents' current behavior\cite{Barfuss2021}.

This learning-dynamic approach offers a formal yet practical, lightweight, and deterministically reproducible way to uncover the principles of collective cooperation emerging from intelligent agents in changing environments. We briefly highlight three examples.
For instance, it was found that, in contrast to non-changing static environments, no social reciprocity is required for cooperation to emerge in changing environments\cite{BarfussEtAl2020}. The individual attitude of how much the agents care for the future alone can adjust the setting from a tragedy of the commons to a comedy, where agents predominantly learn to cooperate. However, for this mechanism to work, the severity of an environmental collapse must be sufficiently severe.
Another work showed how the agents' irreducible uncertainty about the actual environmental state can induce a tipping point towards mutually high-rewarding cooperation. However, this is only valid when all agents are equally uncertain about the environment\cite{BarfussMann2022}.
The last example highlights how the same temporal-difference learning dynamics can be used to model agents that not only learn to react to their physical but also to their social environment, which is likewise a pathway to mutually high-rewarding cooperation\cite{BarfussMeylahn2022}.

Such learning dynamic studies focus on understanding the underlying principles of collective cooperation from intelligent agents in complex environments. Therefore, these models are reduced as much as possible to capture only the most essential features. However, evidence from deep multi-agent reinforcement learning studies shows that sustainable and cooperative behavior can likewise emerge from intelligent agents in high-dimensional environments\cite{StrnadEtAl2019, PerolatEtAl2017, ZhengEtAl2020}.

The advantage of the learning dynamics approach is that it opens up all the tools of dynamical systems theory to the study of collective learning. For instance, the learning dynamics have been found to exhibit multiple dynamic regimes, such as the convergence to fixed points, limit cycles, and chaos\cite{SatoCrutchfield2003, BarfussEtAl2019}, critical transitions with a slowing down of the learning processes at the tipping point\cite{BarfussMann2022}, and the separation of the learning dynamics into fast and slow eigendirections\cite{BarfussMann2022}.


Future work in many directions is required to build this approach of  \textit{adaptivity} in non-linear dynamical systems into a new way of modeling human-environment interactions and socio-economic systems (see Sec.~\ref{SEC:SOCIO-ECONOMIC}).
First, the presented learning dynamics need to become applicable to the system with many agents, using various types of mean-field approaches\cite{YangEtAl2018, HuEtAl2019, PerrinEtAl2021a}.
Second, the learning dynamics need to consider the effect of intrinsic noise, which can substantially alter their collective behavior \cite{Galla2009, BarfussMeylahn2022} (see also Sec.~\ref{SEC:FRANOVIC}).
Third, the learning dynamics needs to be advanced to be able to model more refined notions of cognition, such as representation learning, learning and using intrinsic world models, and intrinsic motivations (see also Sec.~\ref{SEC:NEURAL}).
A social-ecological resilience paradigm of multi-agent environment interactions, in turn, can benefit such endeavors\cite{DongesBarfuss2017, BarfussEtAl2018}.

\subsection{Biomimetic intelligence for active matter -- by Giovanni Volpe}\label{SEC:VOLPE}
\begin{figure*}[ht!] 
\centering
\includegraphics{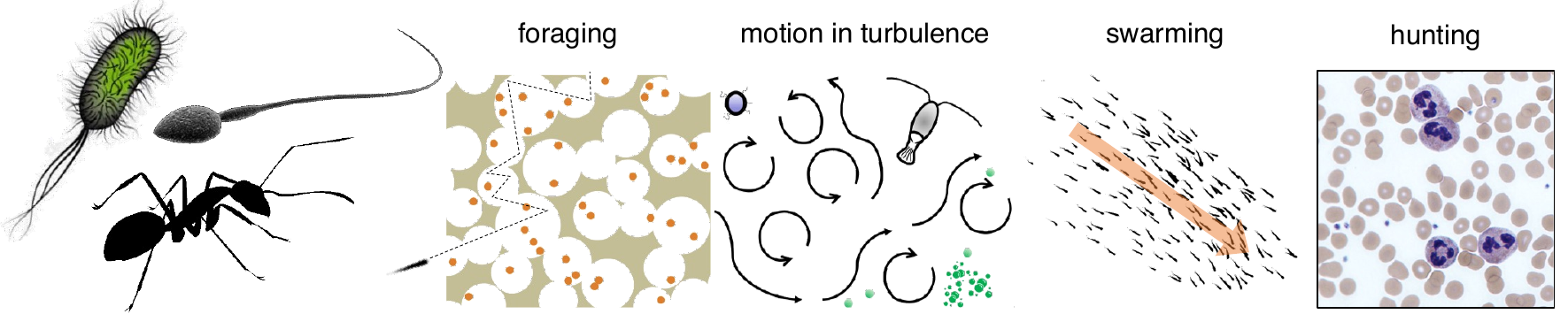}
\caption{
{\bf Active matter with embodied intelligence.}
Bacteria, sperm cells and ants are biological examples of active particles with embodied intelligence. They feature intelligent behaviors that permit them to survive and thrive in their ecosystem thanks to the integration of sensors, actuators, and information processing. Their behaviors also adapt to complex environments (e.g., foraging for food) and their dynamic interactions lead to collective emerging behaviors (e.g., swarming and hunting). The challenge is now to draw inspiration from Nature to create microscopic artificial active particles with embodied intelligence that mimic these adaptive and dynamic emerging behaviors. Adapted from Ref.~\citenum{cichos2020machine}.
}
\label{fig:volpe}
\end{figure*} 

Over billions of years of evolution, motile microorganisms have developed complex strategies to survive and thrive in their environment by integrating three components (Fig.~\ref{fig:volpe}): \emph{sensors}, \emph{actuators}, and \emph{information processing}. Their biochemical networks and sensory systems are optimized to excel at specific tasks, such as to climb chemical gradients \cite{berg2004coli}, to cope with ocean turbulence \cite{sengupta2017phytoplankton}, and to efficiently forage for food \cite{viswanathan2011physics,benichou2011intermittent}. They have also acquired complex strategies to interact with their environment and with other microorganisms, leading to the emergence of macroscopic collective patterns \cite{vicsek2012collective} (also see Sec. \ref{SEC:BARFUSS}). These patterns are driven by energy conversion from the smallest to the largest scales, and permit microorganisms to break free of some of their physical limits. For example, dense systems of bacteria develop ``active turbulence'' at length scales where only laminar flows are expected from the underlying physical laws \cite{yeomans2017nature,urzay2017multi}. 

There are both scientific and technological reasons that are driving the quest towards biomimetic artificial active matter. Scientifically, biomimetic systems capable of harnessing energy and information flows are ideal model systems to investigate and test physics far from equilibrium, which is one of the greatest challenges for physics in the 21st Century. Technologically, biomimetic active particles hold a tremendous potential as autonomous agents for healthcare, sustainability and security applications: for example, enabling the targeted localization, pick-up and delivery of microscopic objects in bioremediation, catalysis, chemical sensing and drug delivery \cite{bechinger2016active}.

In the last two decades, the active-matter research field has tried to replicate the evolutionary success of microorganisms in artificial systems~\cite{bechinger2016active}. Researchers have replicated the actuators by developing artificial active particles that extract energy from their environment to perform mechanical work \cite{ebbens2010pursuit,moran2017phoretic}. Albeit to a lesser extent, they have also been able to replicate the sensors by making these active particles adjust their motion properties (e.g., their speed) to chemical, thermal or illumination stimuli \cite{villa2019fuel,palagi2018bioinspired}. However, these artificial particles are still largely incapable of autonomous information processing, which is dramatically limiting the potential of artificial microscopic active matter to provide scientific insight and technological applications \cite{cichos2020machine}.
Thus, the active-matter research field is now confronted with several open challenges to create truly autonomous active particles.


\paragraph{Make active particles capable of autonomous information processing.}
Currently available active particles lack the complexity necessary for autonomous information processing. In fact, active particles are still rather simple in shape and behavior \cite{bechinger2016active}. They are often Janus microspheres or microrods with different materials on their two sides, which can self-propel and sterically interact with each other. This physical simplicity is a consequence of the relative simplicity of the employed design and fabrication processes, which in turn limits the range of behaviors achievable by the active particles. Despite this simplicity, the study of active particles has already led to major breakthroughs, such as to understand how plankton copes with turbulence \cite{sengupta2017phytoplankton,gustavsson2016preferential,reddy2016learning} and to program self-assembling robotic swarms \cite{rubenstein2014programmable,bayindir2016review}.

Motile microorganisms exhibit more powerful and flexible strategies to survive and thrive in their environment. Even the simplest motile bacteria have evolved intelligent behaviors by following powerful adaptive strategies encoded in their shape, biophysical properties, and signal-processing networks: not only can they extract energy from their environment to move and interact with other bacteria, but they can also extract information by sensing their environment and adjust their behavior accordingly \cite{berg2004coli}.
The challenge is now to make active particles capable of autonomous information processing, like living motile microorganisms.
This can be addressed by pushing the boundaries of  design and microfabrication techniques to build microscopic active particles with embodied intelligence (\emph{microbots}) \cite{andren2021microscopic}. 
These microbots will be able to sense their environment, to differentiate stimuli, and to adapt their behavior towards determinate goals.
 
\paragraph{Optimize the behavioral strategies adopted by individual active particles.}
The behavioral strategies that can be adopted by active particles are still very limited. There have been several studies on the behaviors of active particles in response to the properties of their environment \cite{bechinger2016active,galajda2007wall,volpe2011microswimmers,simmchen2016topographical}. For example, the presence of periodic arrays of static obstacles alters the preferential swimming direction of self-propelling active particles, a fact that permits one to sort microswimmers on the basis of their swimming style \cite{volpe2011microswimmers}. However, these behaviors are still rather simple and rely on in-built properties of the active particles that cannot be changed at will or adapted to different environmental conditions. This is a consequence of their limited capability of gaining information about their environment and reacting accordingly.

More complex behaviors have been achieved using microorganisms instead of active particles. For example, the presence of obstacles in the environment has permitted to alter the pathway toward the formation of multicellular colonies of bacteria \cite{ramos2021environment}. Also, genetically modified bacteria whose speed is controllable by light have been arranged into complex and re-configurable density patterns using a digital light projector \cite{frangipane2018dynamic,arlt2018painting}.
The optimal behaviors in complex environments are often not obvious. 
For example, let us consider the foraging problem \cite{viswanathan2011physics,benichou2011intermittent}, where an active particle performs a blind search to catch some sparse targets. When the environment does not present spatial features, the number of caught targets is maximum for a L\'{e}vy-search strategy \cite{viswanathan2011physics,benichou2011intermittent}  (even though this is still an active research field \cite{levernier2020inverse}). Surprisingly, in a porous medium, the optimal strategy mixes L\'{e}vy runs and Brownian diffusion \cite{volpe2017topography}. 

The challenge is now to discover, understand and engineer intelligent behavioral strategies that can be autonomously adopted by active particles with embodied intelligence.
This can be addressed by designing and engineering the behavior of microbots to enable them to autonomously perform directed tasks in complex environments, such as efficient navigation, target localization, environment monitoring, and conditional execution of actions.

\paragraph{Optimize the interactions between active particles.}
Currently, active particles cannot communicate with each other beyond interacting through some simple physical interactions. Natural systems, such as swarms of midges, schools of fish and flocks of birds, have evolved powerful sensing capabilities to gain information about their environments and to communicate \cite{charlesworth2019intrinsically,strandburg2013visual}.
The underlying behavioral rules are often hard to identify \cite{vicsek2012collective,cichos2020machine,loos2020irreversibility,loos2023long}. Active-matter studies provide the testing grounds for new non-equilibrium descriptions, which are by necessity often computational \cite{argun2021simulation}. They are either based on hypothesized mechanistic models for local interactions \cite{vicsek2012collective}, upon coarse-grained hydrodynamic approximations \cite{marchetti2013hydrodynamics} or on basic fluctuation theorems \cite{falasco2016exact}. The question is often how local energy input and physical interactions determine the macroscopic spatio-temporal patterns. Answers may be sought, e.g., by computational techniques \cite{frenkel2001understanding,rosenbluth2003genesis,wolfram1984cellular,lauga2009hydrodynamics}.

Differently from computational studies, most active-matter experiments rely on very simple steric and short-range physical interactions. Even these simple interactions can lead to interesting complex behaviors and self-organization whose onset is often observed in artificial systems where increased energy input above a threshold density drives a phase transition to an aggregated state. An example of such behaviors is the formation of ``living crystals'', which are metastable clusters of active particles \cite{palacci2013living,buttinoni2013dynamical}.

Much more interesting behaviors are observed when the interactions between the active particles can be tuned at will. This can be achieved by externally imposing interaction rules on the active particles. For example, external feedback control loops have been used to create information-based individual dynamical behavior \cite{lavergne2019group} of, or interactions\cite{khadka2018active} between active particles, which explicitly depends on the information about the position of other particles.
Such complex forms of interaction can also be achieved using macroscopic robots. In fact, the field of robotics can serve as a major source of inspiration for the development of active matter at the microscale \cite{doncieux2015evolutionary,bayindir2016review,jones2019onboard}. For example, some robots (5 cm in diameter) have been programmed to respond to sensorial inputs with a delay and shown that, by controlling the delay, we can control the aggregation vs. dispersion of the robots \cite{mijalkov2016engineering,volpe2016effective,leyman2018tuning}.
 
The challenge is now to identify and engineer optimal interaction rules that can be embodied in active particles interacting with other particles and with their environment.
 This can be addressed by programming microbots with embodied interaction strategies beyond the simple steric and short-range interactions employed by current active particles. This will permit researchers to realize microscopic swarms of artificial active particles capable of collective intelligent behaviors and to engineer microscopic ecosystems where multiple species of microbots and particles interact.

%
%
\section{Adaptivity in socio-economic systems}
\label{SEC:SOCIO-ECONOMIC}
In this section, we provide a perspective on adaptivity from socio-economic systems including topic such as the conception of modern power grids, adaptive social interactions and the role of adaptive mechanisms in epidemiological and climatic models.
\subsection{Adaptive networks and their importance for epidemiology -- by Philipp H\"ovel}\label{SEC:HOEVEL}
\begin{figure}[h!]
 \includegraphics[width=\linewidth]{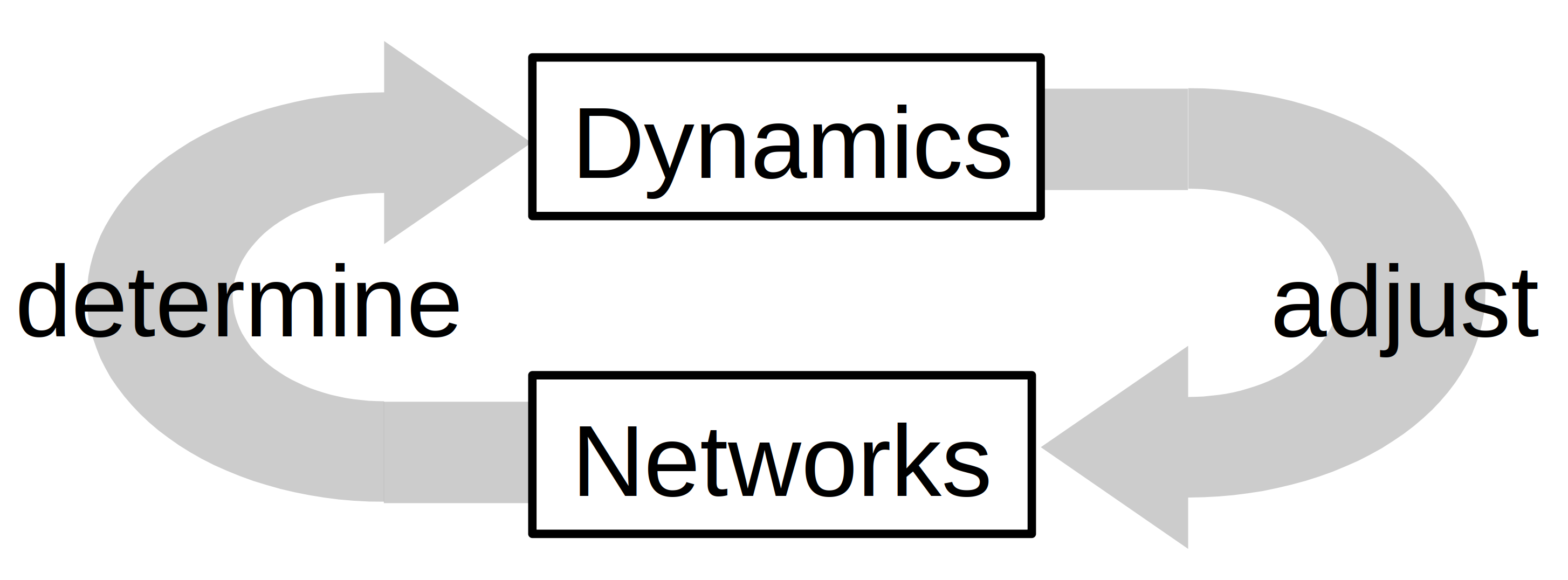}
 \caption{Schematics of the interplay between dynamics and networks.}
 \label{fig-PH:schematics}
\end{figure}

Network epidemiology is a prime example of adaptive networks at work. Many infectious diseases spread via direct contacts. These contacts can be captured by social, transportation, and other logistic networks. They provide a mathematical framework to formalize the interaction of individuals (humans or animals) and hence, potential paths of disease transmission. Locally, e.g., within a population or between groups individuals, the dynamics of pathogens are often described by compartment models such as the widely used susceptible-infected-recovered model originally introduced by Kermack and McKendrick \cite{KER27}. Adaptivity must be considered if the state of the networked system, say, the number of infected, triggers an adjustment of the network structure with the aim to mitigate an outbreak and to contain the disease. This closes the mutually influencing feedback loop of the dynamics on and of networks as depicted in Fig.~\ref{fig-PH:schematics}: (i) The network structure governs the spreading of the disease (\textit{dynamics}). (ii) In turn, the current state of the system leads to changes in the structure of interactions (\textit{networks}).

The dynamics-induced changes to the network are often akin to control schemes that involve minimizing a goal function to reach a target state \cite{LEH14}. Similarly, non-pharmaceutical containment protocols, which demand a reduction of social contacts or restriction of movement, can be based on, for instance, the number of new infections. Prominent cases, where such applications of adaptive networks have been successfully implemented, include the H1N1 pandemics in 2009 \cite{BAL09b,CHU15}, the Ebola epidemic in 2014 \cite{POL14}, and -- of course -- the on-going SARS-CoV-2 pandemic \cite{KYR20,MAI20,PRE20,HUM21}. In these examples, one prominent path of transmission was the global airline transportation network, which has been accounted for in many studies \cite{BRO13,IAN17,GOL19a}.

Extensive numerical simulations are able to explore possible interventions and quantify their impact. Key findings might be that international travel bans yield a limited delay of spreading as demonstrated for Ebola in 2014 \cite{POL14} or the feasibility of \textit{Zero-COVID} or \textit{low-incidence} strategies \cite{COU20,CZY22}. They are also able to provide insight into less than optimal adherence to containment measures \cite{HUM21}. In any case, these studies are valuable tools for policy makers to reach evidence-based and data-driven decisions and to inform the public about their potential impact.

The concept of adaptive networks for the study of epidemic spreading of infections diseases has a long history and dates back beyond the most recent examples of public health emergency of international concern. Rewiring of susceptibles to avoid contact with infected has been studied, for instance, by Gross \textit{et al.} \cite{GRO06b,GRO08a,GRO08b}. They employed a low-dimensional compartment model, which allowed an exhaustive bifurcation analysis, and identified dynamical patterns such as first-order transitions and hysteresis. In short, as long as a node remained healthy, its network neighborhood evolved gradually. However, the moment an infection occurs, the degree of a node drops rapidly, and the node finds itself isolated until recovery. Note that due to the small-worldness of many social networks \cite{WAT98}, there could be situations, where rewiring would potentially deteriorate the situation because it might create new shortcuts through the network that could -- unintentionally -- bring nodes closer to other, distant regions of infection.

Besides travel restrictions, surveillance and monitoring of incidence numbers are key ingredients for a rapid identification of an outbreak. For that purpose, the introduction of \textit{sentinel nodes} on temporal networks has proven to be insightful and demonstrated in the case of animal diseases \cite{BAJ12,SCH21}. These nodes should be monitored because of their position in the network that allows early detection and reliable identification of the origin of the outbreak for many different initial conditions. Therefore, they provide helpful and detailed clues where the network could be best adjusted. Similarly, screening a fraction of incoming patients has been shown to be effective as a potential control measure nosocomial infectious diseases and the spread via hospital-referral networks \cite{BEL16,BEL17}. The impact of a rapid response has been exemplified during the early stages of the COVID-19 pandemic, where -- in mainland China -- containment policies effectively depleted the susceptible population and resulted in a subexponential growth of infection cases \cite{MAI20}. Upon successful containment, restriction can be relaxed again, and the network returns to its original state.

Adaptive networks are a special case of time-varying or temporal networks, where every edge has a time stamp and is active for a certain amount of time \cite{MAS17a,HOL19a}. In epidemiology, in particular, the sequence of contacts is crucially important. Only time-respecting paths contribute to the transmission of a pathogen and the spreading of a disease. Any interaction with contacts/neighbors in the social network before their infection carries no risk of transmission. Luckily, concepts like network controllability \cite{LIU11} can be easily extended for temporal and multiplex networks \cite{POS14,POS16}. From a mathematical point of view, the temporal nature of networks -- including changes of their structure due to adaptation -- can be implemented by time-dependent adjacency matrices, which give rise to modelling frameworks for the spreading of epidemics such as the individual-based and pair-based models \cite{VAL15,FRA16a,KOH19,HUM21a}.

To sum up, adaptive networks play a central role not only for realistic investigations of spreading dynamics but can help to study and design interventions for disease containment, mitigation, and eradication. With further increase of data availability (often in real time), models of network epidemiology become more and more realistic and informative in their predictive power. Future challenges include the integration of purely epidemiological models and a mathematical framework for the dynamics of social behavior and opinion formation. This would lead the way for a holistic description of disease spreading on adaptive networks.
\subsection{Coevolutionary network dynamics in social and epidemic systems-- by Jan M\"olter}\label{SEC:MOELTER}
In the context of dynamical systems on networks, one manifestation of adaptivity is in so-called adaptive or coevolutionary networks~\citep{GRO08a,gross2009adaptive}.

A network is a collection of entities together with a relation between these entities that are generally represented as nodes and links, respectively. In a dynamical setting, every node is a dynamical system that not only depends on its internal dynamics but also on the dynamics of its neighborhood in the network, i.e., the set of nodes it is linked to. Constituting for an adaptive network is the idea that the topology of the network and therefore the interactions between the individual nodes of the network are not static but rather also dynamic, coupled to the dynamics of the nodes. As such, we have a closed feedback loop in which the topology of the network influences the dynamics of the nodes and the state of the nodes influences the dynamics of the topology~\citep{gross2008adaptive}. Combining the so-called dynamics on the network with the dynamics of the network in that way is what makes the system \emph{fully adaptive} (see Fig.\,\ref{fig-PH:schematics} in Sec.\,\ref{SEC:HOEVEL}).

To make this more concrete, let us consider the paradigmatic example of the adaptive voter model~\citep{holme2006nonequilibrium,demirel2014momentclosure,zschaler2012adaptive}, which is an extension of classical models of opinion or consensus formation~\citep{clifford1973model,holley1975ergodic}. In this model, one considers a population in which every individual subscribes to one of two contradictory opinions and in which the social relationships are encoded in some social network. As for the dynamics, one assumes that in each time step, individuals either adapt their opinion to the opinion of individuals in their neighborhood or that they break off their relationship with individuals of opposing opinions and rather connect with others of the same opinion. While the former corresponds to the dynamics on the network, the latter corresponds to the dynamics of the network. Depending on the relative strength of these two processes, in expectation, the population eventually reaches either a dynamic equilibrium characterized by a non-vanishing prevalence of pairs of connected individuals with opposing opinions or a static equilibrium where the underlying social network fragments so that in every component only one opinion prevails~\citep{vazquez2008generic,demirel2014momentclosure,durrett2012graph}.

Another paradigmatic example besides the adaptive voter model is that of an adaptive susceptible-infected-susceptible (SIS) epidemic~\citep{GRO06a}. One considers again a population on an underlying social network that encodes the relationship between individuals. Every individual is then exposed to an SIS epidemic, meaning that individuals start off as susceptible, become infected at some rate when individuals in their neighborhood are infected, and upon recovery at another rate are susceptible again~\citep{kiss2017mathematics}. In addition to these epidemic transitions, one allows, similar to the adaptive voter model, that susceptible individuals can break off the relationship with infected individuals and instead connect to a susceptible individual~\citep{GRO06a}. Now, for the SIS epidemic and in expectation, it is well-known that at a critical infection rate the system exhibits a supercritical transcritical bifurcation and beyond which the system eventually reaches an endemic dynamic equilibrium as opposed to the epidemic dying out. In contrast, due to the adaptivity, this bifurcation can turn from supercritical to subcritical, the consequence being that a region of bistability emerges and the transition to an endemic equilibrium is not continuous anymore~\citep{GRO06a,kuehn2021universal}.

While these examples both illustrate the idea behind adaptive or coevolutionary networks in the sense that dynamics on the network and of the network depend on each other, they also highlight the fact that adaptivity can induce fundamental changes in the phenomenology. This suggests that, when developing models of the natural world, it can be paramount to take adaptive dynamics into account.

Recognizing the importance of adaptive networks, a lot of research has been done focussed on different aspects of the phenomenology that comes with adaptivity or extending existing models by introducing adaptivity. Hence, in the following, we are going to highlight some works from the last decade -- without any claim to comprehensiveness.

In relation to the adaptive voter model we have introduced before, it has been reported that if one considers directed as opposed to directed networks in an adaptive voter model, fragmentation might occur far below the critical value due to the formation of self-stabilizing structure~\citep{zschaler2012early,boehme2011analytical}. Moreover, there has also been work extending the model to allow for a continuum of opinions (see also Sec.\,\ref{SEC:LORENZSPREEN}), which in many cases is a more realistic assumption, demonstrating the emergence of communities with diverse opinions rather than leading to fragmentation \citep{kozma2008consensus,yu2017opinion}.

Further investigations in the adaptive SIS epidemic and adaptive epidemics, in general, have led to studies about the bifurcation behavior~\cite{zhang2019complex} and the epidemic threshold itself~\citep{ogura2016epidemic} as well as the dynamics near this threshold with an emphasis on early-warnings signs~\citep{horstmeyer2018network}. In the context of a pandemic (see also Sec.\,\ref{SEC:HOEVEL}), adaptive epidemics have also been studied to assess the relationship between containment strategies of quarantining and social distancing~\citep{horstmeyer2022balancing}. Besides rewiring as a mechanism for adaptivity~\citep{GRO06a,shaw2008fluctuating}, others have considered network growth due to birth and death processes\citep{demirel2017dynamics}, the latter in response to the epidemic upon being infected, and activation and deactivation of links following an adaptive strategy~\citep{clauss2022selfadapting}.

Apart from the adaptive voter model and adaptive epidemics, another frequently studied model system is that of coupled phase oscillators~\citep{KUR84,strogatz2000from} with adaptive coupling strengths (see also Secs.\,\ref{SEC:YANCHUK} and \ref{SEC:SCHOELL}). The main feature one is interested in these systems is the emergence of fully or partly synchronous states. Importantly, it has been shown that certain adaptivity rules promote the explosive transitions into synchrony~\citep{avalosgaytan2018emergent}. Moreover, others have reported that adaptivity can be used to control cluster synchronisation~\citep{LEH14} or that slow adaptation leads to the emergence of frequency clusters~\citep{BER19a,BER19}.

In recent years, there has been an increasing interest in generalizing the notion of networks to higher-order networks, i.e., simplicial complexes or more generally hypergraphs. Instead of only dyadic relations, these structures can also capture higher-order interactions. Consequently, evolutionary games~\citep{schlager2022stability} as well as consensus formation in the form an adaptive voter model have been investigated on simplicial complexes as well as hypergraphs~\citep{horstmeyer2020adaptive,papanikolaou2022consensus}. Due to their much more complex topology, these structures promise a much richer phenomenology while at the same time being considerably more complicated to handle, so that it will be interesting to see what the coming years will bring.
\subsection{How social dynamics and networks adapt to growing connectivity -- by Philipp Lorenz-Spreen}\label{SEC:LORENZSPREEN}

Online communication can be understood as an adaptive, nonlinear system, all the more so because it increasingly involves many-to-many interactions and is thus a highly coupled system. In my research on self-organized online discourse, I interpret adaptivity as the process of changing social systems through external influences, such as technological developments. Information technology has made various aspects of our lives more dynamic, both in spatial and temporal dimensions. Connections with others can be made across spatial and sociodemographic constraints, and messages can be recorded and spread across the globe in seconds. 

However, these increased dynamics and the resulting adaptations do not happen without values: As old boundaries are overcome, new ones emerge, if only because of finite amounts of available attention resulting from very simple limits on human processing capabilities, but also because of the implementation and commercial incentive structure of the technology. Here I will present two mechanisms we have recently proposed for how social systems adapt to these changes and how online platforms shape this process along commercial interests, since there is no apparent, neutral status quo in which social systems would evolve. To this end, I want to focus on two key questions that an individual decision maker faces online and their downstream consequences for macroscopic dynamics and the shape of public discourse.

First, connectivity is increasing through online platforms, and new connections can and are easily made. Since the famous six degrees of separation \cite{milgram1967small} on the U.S. social network, networks seem much better connected; Facebook reports 3.5 degrees of separation on its friendship graph\footnote{see, https://research.fb.com/blog/2016/02/three-and-a-half-degrees-of-separation/}. Nevertheless, there are consistent reports of segregated, homophilic network structures on nearly all online platforms, as well as related trends of increasing polarization (see \cite{lorenz2022systematic} for a recent overview). The mechanism that might resolve this apparent paradox may lie behind the question of whether we change our opinions according to our friends or whether we change our friends according to our opinions. In classical models of opinion dynamics, the network structure of is fixed and the core assumption is a constructive process of opinion change in social interaction \cite{deffuant2000mixing}. In the long run, this process would predict convergence to a global consensus opinion with increasing connectivity; only under the assumption of disconnected networks or limited trust are disconnected opinions conceivable, let alone an outward or distancing movement of these clusters possible. We have recently proposed an alternative mechanism that describes the dynamics of an agent's opinion $o_i(t)$ \cite{baumann2020modeling}: 

\begin{equation}\label{eq:model}
  \dot{o}_i = -o_i + K \sum_{j=1}^N A_{ij}(t) \tanh(\alpha o_j),
\end{equation}
which describes a process of mutual reinforcement of opinions within groups of shared stance (i.e., if $sgn(o_i) = sgn(o_j)$). The additive term $\tanh(\alpha o_j)$ moves both opinions in the same direction if they ave the same sign and moves them towards the neutral state $0$ if they have different sign. Who is interacting with whom is governed by the time dependent adjacency matrix $A_{ij}(t)$, which only has a non-zero entry if an interaction happens between $i$ and $j$ at time $t$. Its structure dynamically adapts to changing opinions, hence co-evolving and following a probability distribution ruled by homophily \cite{mcpherson2001birds}: 

\begin{equation}\label{eq:homophily}
 p_{ij} = \frac{|o_i(t)-o_j(t)| ^{-\beta}}{\sum_j|o_i(t)-o_j(t)| ^{-\beta}},
\end{equation}
which is a term that might be partly driven by algorithmic recommendations suggesting like-minded others as interaction partners on social media. This combination helps to explain the potential emergence of growing polarization dynamics even under increasing connectivity (i.e., if the average path length of $A_{ij}(t)$ decreases, at least for controversial topics (i.e., high $\alpha$). For more details, please see \cite{baumann2020modeling} and an extension into multi-dimensional opinion spaces see \cite{baumann2021emergence}.

Second, the increasing availability of information poses a challenge to the allocation of attention. So how does public discussion adapt to the increasing speed of available information? To describe this process, we quantified and modeled the dynamics of public interest for individual topics in various domains \cite{lorenz2019accelerating}. The main result can be described as an acceleration of the dynamics of public interest in a topic and a narrowing of the amount of time spent on each topic, while the overall amount of attention spent on a topic remained stable over the years. For a mechanistic understanding of these dynamics, we modeled them as an adaptation of a Lotka-Volterra process for species competing for a common resource, with finite memory:

\begin{equation}
    \dot{a_i} = r_{p}a_i\left( {1 - r_c{\int}_{ - \infty }^t {{\mathrm{e}}^{ - \alpha (t - t\prime )}} a_i(t\prime )dt\prime - c\mathop {\sum}\limits_{j = 1,j \ne i}^N {a_j}} \right),
\end{equation}
where $a_i(t)$ describes the dynamics of the collective attention or public interest to a topic $i$. It depends on a growth term $r_{p}a_i$ with an exponential growth rate $r_{p}$, if it is undisturbed. However, two terms are slowing and eventually reversing the growth process. That is $r_c{\int}_{ - \infty }^t {{\mathrm{e}}^{ - \alpha (t - t\prime )}} a_i(t\prime )dt\prime$, which grows proportionally with the attention to the topic itself by exhausting the available resources at rate $r_c$, and $c\mathop {\sum}\limits_{j = 1,j \ne i}^N {a_j}$, which describes the constantly ongoing competition with all other topics $j$ for that common resource. This we believe captures the essence of the idea of competitive attention economy originally formulated by \cite{simon1971designing} and describes well the empirical observations. It also captures the economic incentive structures to produce information faster in this competitive situation to have an advantage for gaining public interest.

In summary, I believe that these mechanisms may capture two adaptive mechanisms of social systems in response to increasing interconnectedness and information availability that are driven by fundamental limits of human cognition, namely the ability to maintain a certain number of social contacts as well as to process a finite amount of information in parallel, as well as economic incentives to capture those. Future research in this area should aim to put those assumptions of mechanism of social dynamics on an empirical, probably experimental, footing to understand the causal drivers of how social systems adapt to changes in our world, e.g., technological and political changes.

\subsection{Energy transition and moving towards the CO2-neutral power grid -- by Mehrnaz Anvari}\label{SEC:ANVARI}

The important role played by electricity in the daily life and activities causes a serious dependency of modern society on the reliable functionality of the power grid. Moreover, because of the interconnection of the power grid to other societal networks and systems, such as transportation \cite{bialek2020does}, telecommunication \cite{buldyrev2010catastrophic}, and health systems \cite{klinger2014power}, it is of great importance that the power grid adjusts itself to changing conditions or, indeed, mitigates any internal and external perturbations and fluctuations, as generally discussed in Sec.\,\ref{SEC:YANCHUK} for dynamic networks. Any failure in the power grid can quickly spread not only within the grid itself, but can set off a chain of failures, as a domino effect, in other social networks and systems. During energy transition and moving towards a CO2-neutral power grid, fossil fuels sources should be replaced by renewable energy sources, such as wind, sunlight, water and geothermal heat. The need for rapid CO2 reduction is comprehensively discussed in Sec.\,\ref{SEC:KURTHS}. Among renewable energy sources, wind- and solar power are sources inherently time-varying. This means that a constant generator power in Eq.\,\eqref{eq:KwI_2order} in Sec.\ref{SEC:SCHOELL} will be replaced by  irregular, hardly predictable wind and solar power may constitute serious threats for power grid stability. Furthermore, the pattern of electricity consumption is changing due to the exploitation of green energy sources in other sectors such as transportation \cite{DW} and heating \cite{ferroukhi2020renewable}. Thereby, for being able to plan and operate future-compliant electricity grids with a continuously increasing share of renewable energy sources, it is vital to recognize the new origins of fluctuations in both supply and demand side, along with their statistical and stochastic characteristics to be able to adapt the power grid or to mitigate these fluctuations and thus maintain the energy balance in the grid.

The identification of these characteristics, along with the empirical data enable us to develop valid data-driven models to describe the underlying system dynamics. Lastly, the combination of data-driven models and the complex network science empowers us to indicate the impact of new sources of both supply and demand on the current power grid and, thereby, to determine how the power grid structure and control systems should be adapted in the future to keep the energy balance and, consequently, the stability in the system. 

In the following, we will review briefly some recent works related to the data analysis and data-driven models as well as their combination with the complex network science leading to a deep understanding of power grid dynamics. 

\paragraph{Data analysis} Wind and solar power are highly dependent on weather conditions and, therefore, can ramp up or down in just a few seconds. In a power grid with a high integration of variable energy sources, these extreme short-time fluctuations not only influence the energy availability, but also the stability of the power grid. The analysis of the data of wind and solar power recorded in different regions around the world demonstrates multiple universal types of variability and nonlinearity in the short-time scales \cite{anvari2016short,apt2007spectrum,curtright2008character,luo2006frequency}. Importantly, considering the aggregated variable energy sources of even country-wide installation of wind and solar fields shows that the data is still non-Gaussian and includes intermittent fluctuations \cite{anvari2016short}. Indeed, this is the direct consequence of the long-range correlations of the wind velocity and cloud size distribution that are 
approximately $600\,\si{km}$ and $1200\,\si{km}$, respectively\cite{baile2010spatial,wood2011distribution}. The footprint of these short-time intermittent fluctuations have been recently monitored in the power grid frequency variations \cite{haehne2018footprint}. 

The analysis of the highly resolved electricity consumption data of households that consume $29\%$ of all electricity in European Union \cite{EEA2021} shows that these data are highly intermittent. The intermittent fluctuations of electricity consumption can not be captured from the data with a resolution of one hour or even 15 minutes \cite{monacchi2014greend,wright2007nature,marszal2016household}. The variability of energy sources, along with the uncertainty of the electricity consumption can make it more difficult to balance supply and demand. Thereby, as the share of  feed-in is increasing, a deeper understanding of the variable energy source dynamics as well as advanced approach of balancing demand and supply by load shifting is required \cite{lopez2015demand,logenthiran2012demand}. 
\paragraph{Data-driven models} Identifying the stochastic behaviour of the short-time variable energy sources and electricity consumption fluctuations allows us to construct a dynamic equation that governs these stochastic processes. The dynamic equation should include two main terms as following:
\begin{equation}
    \dot X(t)=F(X,t)+G(X,t),
    \label{model}
\end{equation}
where $F(X,t)$ is the deterministic term showing the trend of a stochastic process $X(t)$ (which is here a variable source of energy or electricity consumption) versus time, and $G(X,t)$ is the stochastic term modelling the extreme fluctuations and, indeed, non-Gaussianity in the considered process. Equation~(\ref{model}) is known as a stochastic differential equation which is a non-parametric model. With the term `non-parametric', we mean that all of the functions and parameters in the model can be found directly from empirical time series. Recently, the {\it jump-diffusion} process \cite{anvari2017suppressing,anvari2016disentangling} and the {\it superstatistics} method \cite{anvari2022} have been introduced to model short-term variable energy sources and electricity consumption fluctuations, respectively. Moreover, in \cite{anvari2022} a data-driven load profile that is consistent with high-resolution electricity
consumption data is obtained. This data-driven load profile outperforms the standard load profile used by energy supplier \cite{SLP} and it does not require microscopic parameters for consumer behaviour, consumer appliances, house infrastructures or other features that other models depend on \cite{parti1980total}.

The data-driven models allow us not only to generate time series with identical statistics to empirical ones, but also by adjusting the parameters in the stochastic models, to consider the response of the power grid and control systems to different circumstances.

\paragraph{Complex network science} From a structural view point, the power grid is a complex network consisting of many units and agents that interact in a nonlinear way. Due to economic factors, power grids often run near their operational limits. The nature of renewable energies will add more and more fluctuations to this complex system, causing concerns about the reliability and stability of the power supply \cite{anvari2020introduction,auer2017stability,li2006wind}. Therefore, the probability of having grid instabilities will increase, which may result in more frequent occurrences of extreme events like cascading failures resulting in large blackouts. Any strategy under discussion, like upgrading the existing power grid, the formation of virtual power plants combining different power sources, introducing new storage capacities and intelligent ‘smart grid’ concepts, etc will further increase the complexity of the existing systems and have to be based on the detailed knowledge of the dynamics of variable energy sources and consumersvariable sources of energy. The data-driven models and the generation of data sets imitating the characteristics of the real data sets, empower us to consider accurately the interplay of the network structure and features with supply and demand fluctuations and, therefore, resulting in deep insight into how the future structure and control systems should be designed to mitigate the intermittent fluctuations and allows us to increase the share of variable sources of energy in the power grid without any restriction.
\subsection{Adaptability of the earth system: past success and present challenges -- by J\"{u}rgen Kurths}\label{SEC:KURTHS}

The Earth system is a highly complex system with various interactions, including positive and negative feedbacks. Its representation is sometimes even called a horrendogram. But it is also an open system that corresponds with its closer and farther surrounding. All these properties are crucial for the ability to \textit{adapt} in response to external as well as intrinsic changes and perturbations. 

There are outstanding examples of {adaptive} behavior in the history of the Earth system: About 66.000.000 years ago, a rather large asteroid struck Earth and formed the Chicxulub impactor crater with a diameter of about 180 km in the peninsular Yucatan in Mexico \cite{A1980}. This external shock induced titanic changes on the surface and in the atmosphere as megatsunamis, giant wildfires and a rapid strong decrease of the temperature. More important, it is now well accepted that it was the main cause of the \textit{Cretaceous–Paleogene extinction event}, a mass extinction of 75\% of plant and animal species on Earth, including all non-avian dinosaurs. However, it is important to emphasize that the Earth system was not destroyed due to this giant event, but it adapted and reached another stable regime after some time whose global climate was rather similar to the former one \cite{B16}. Another example of a shock-like but intrinsic event was the Toba supervolcanic eruption about 74.000 years back in Sumatra \cite{N1978}. It changed the climate situation drastically and, in particular, induced a strong temperature decrease $3-5^\circ\text{C}$. But the Earth system again adapted and reached via rather large fluctuations a stable climate regime whose global temperature was however clearly below the former one \cite{W09}. 
\begin{figure*}
\centering
	\includegraphics{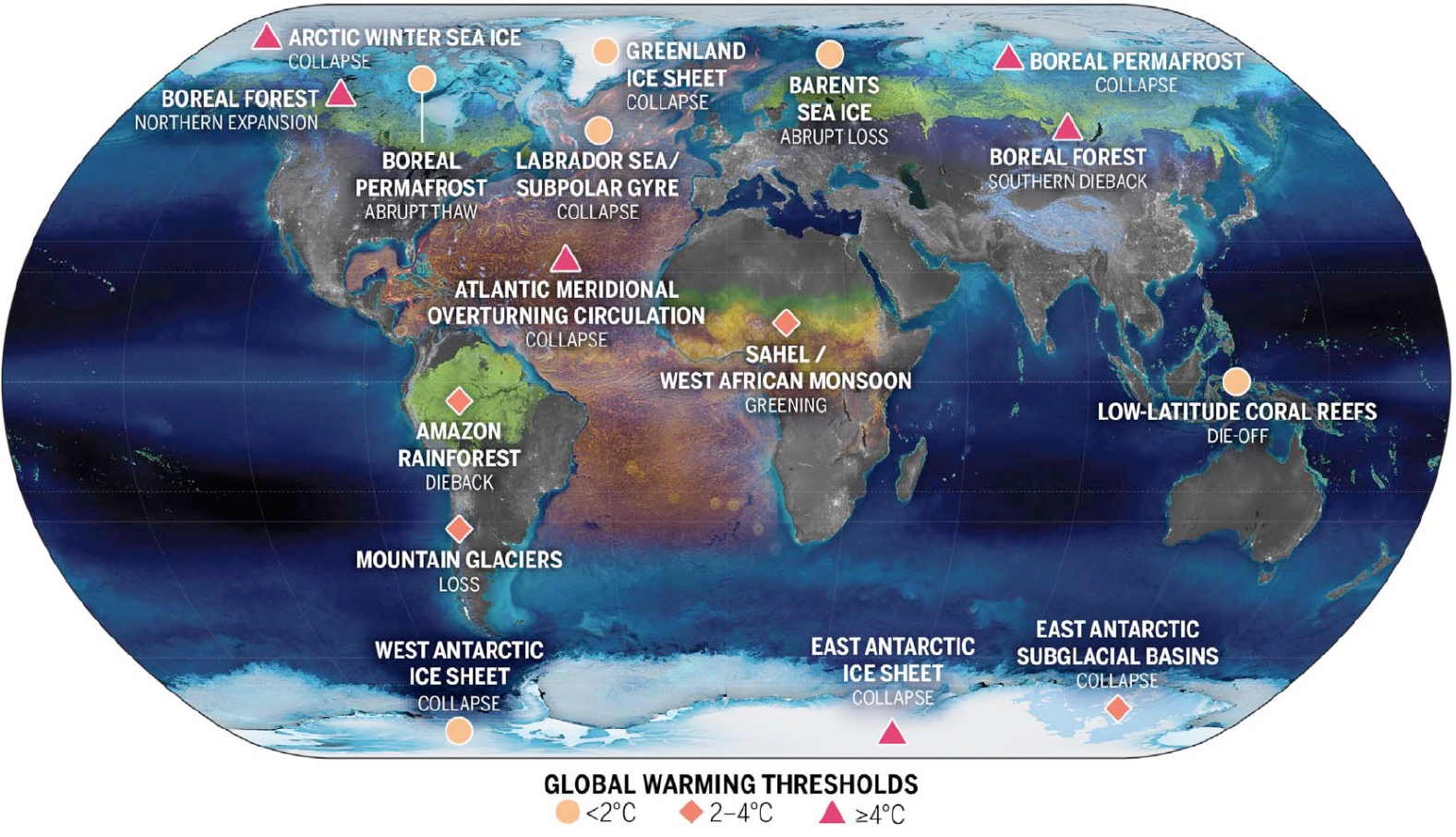}
	\caption{\label{fig:pole}The location of climate tipping elements in the cryosphere (blue), biosphere (green), and ocean/atmosphere (orange), and global warming levels at which their tipping points will likely be triggered. Pins are colored according to our central global warming threshold estimate being below $2^\circ\text{C}$, i.e., within the Paris Agreement range (light orange, circles); between 2 and 4$^\circ\text{C}$, i.e., accessible with current policies (orange, diamonds); and $4^\circ\text{C}$ and above (red, triangles) (figure from \cite{M2022}).}
\end{figure*} 

There are also recurrent-like strong influences on the Earth system over broad scales in time. On the one end, we have as long-term factors the Milankovic cycles, which are due to complex variations in eccentricity, axial tilt, and precession of the Earth' motion in the solar system leading to main components of 41.000 years, 95.000 years and others. These orbital forcing components have a strong influence on long-range climate dynamics, as the occurrence of glacials and interglacials.  On the other end, recurrent patterns as El Nino Southern Oscillations (ENSO) in the range of $3 - 7$ years have a powerful impact on the onset and intensity of monsoons and the formation of extreme climate events. However, the Earth system has been able to adapt to all these recurrent events and acts in stable regimes which can be even become different, e.g., switching from glacial to interglacial.

But one component of the Earth system has substantially increased its impact in the more recent past, the humans. The huge amount of greenhouse gas emission, as CO$_2$ and methane, is the most striking expression of this tremendous \textit{anthropogenic activity}. There is clear evidence and broad acceptance that this has already caused a distinct global warming and various other strong changes in the Earth system \cite{Intergouvernemental_panel2021}. Due to several reasons, the kind of adaptation of the Earth system in response to these emissions is hard to evaluate. One crucial uncertainty is the future development of these emissions despite the immense efforts for their serious reduction, e.g., via the UN Climate Change Conferences of the Parties (COP). 

Therefore, typical scenarios of future Earth system's adaption in dependence on different emission amounts are estimated based on combined models and measured data. But there are challenging problems in modeling of the corresponding processes and data acquisition. A very promising approach to treat these tasks is based on the study of \textit{tipping elements} because the Earth system comprises a number of such large-scale subsystems, which are vulnerable and can undergo large and possibly irreversible changes in response to anthropogenic perturbations beyond a critical threshold \cite{L2008, M2022}. The whole system of tipping elements including their interactions can be well described as a complex network in order to understand the spreading of tipping, i.e., will the tipping of one element exert only local effects or will it induce a cascading-like dynamics \cite{W2021}? This is a typical multistable system where phenomena as partial synchronization are typical (see also Sec.\,\ref{SEC:SCHOELL}). Additionally, intrinsic and external noise may strongly influence the dynamics of the Earth system (see also Sec.\,\ref{SEC:FRANOVIC}). We know the main elements of this network because they have been identified and described, such as dieback of Amazon forest or melting of poles (see Fig.\,\ref{fig:pole}). However, the kind of interactions as well as the intrinsic dynamics at each tipping area are only very partly known. 

To treat the first problem, connections between the Amazon forest area and other tipping points have been recently uncovered quantitatively by analyzing near-surface air temperature fields \cite{L2023}. This way, teleconnections between the Amazon forest area and the Tibetan plateau as well the West Antarctic ice sheet have been identified. In other studies based on conceptual models for selected tipping elements with complex structure-function interrelations as treated in Sec.\,\ref{SEC:YANCHUK} of this perspective, it has been shown that the polar ice sheets could be typically the initiators of tipping cascades, while the Atlantic Meridional overturning circulation acts as a mediator \cite{W2021}. However, these studies are in the beginning and there are several crucial problems to solve till getting a reliable predictability of tipping dynamics and, hence, on evaluating in detail the adaptability of the Earth system in particular to anthropogenic influences. A promising way to retrieve these interactions will be application of modern machine learning methods (see Sec.\,\ref{SEC:ARTIFICIAL-LEARNING}).

However, it is evident that the greenhouse gas emissions have to be strongly reduced. In Sec.\,\ref{SEC:ANVARI} of this perspective paper, problems and approaches for reaching this ambitious goal are discussed.

To summarize: the Earth system is an adaptive one as is obvious from the past. We have now clear evidence that the huge anthropogenic influences create a new kind of perturbation which have the power to induce a novel pathway of adaptation. This will end for sure in some stable regime, but it is very questionable whether we can live there.

\section{\label{sec:discussion}Concluding remarks}
\label{SEC:CONCLUSION}
The notion of \textit{adaptivity} is used in a variety of contexts, from nonlinear dynamics over socioeconomic systems to cognitive science and musicology.
%
%
This article presents various viewpoints on adaptive systems and the notion of adaptivity itself from different research disciplines aiming to open the dialogue between communities. 

The article shows that the terminology and definition of `adaptivity' may vary among the communities. While `adaptability' refers generally to the ability of a system to amend its properties according to dynamic (external or intrinsic) changes, the specific details of adaptive mechanisms depend on the context and the community. For example, how and which part of a system can amend (adaptation rules), or what strategies enable the perception (or sensing) of such changes. In addition, the mathematical framework for describing adaptive mechanisms and adaptive systems also varies across communities.

On the other hand, various commonalities become apparent throughout this article. For example, a common starting point in many contexts are descriptions based on \textit{networks}, where the notion of adaptivity is well established. Adaptive networks are applied in numerous fields, such as power grids, neural systems, and machine learning. Another commonality across disciplines is the link between adaptivity and \textit{feedback} mechanisms, which are ubiquitous in both natural systems and engineering.

We believe that the similarities and differences provide opportunities for further cross-fertilization between the research communities centered around the concept of adaptivity as a common mechanism. For example, adaptive networks can serve as a powerful modeling paradigm for realistic dynamical systems, possibly applicable to even more systems, e.g., in the context of cognitive sciences, musicology, or active matter. Furthermore, a great opportunity lies in utilizing the mechanisms that have emerged in nature as inspiration and guiding principles to engineer artificial (intelligent, cooperative) systems and to develop control strategies. In this spirit, for example, the cooperative behavior of animals may guide the way to engineer robots capable to perform collective motion reminiscent of swarms of insects or schools of fish. Or, the development of new machine learning algorithms may potentially profit from a deeper understanding of the brain provided by the field of neuroscience. Indeed, it has long been recognized that ``The adaptiveness of the human organism, the facility with which it acquires new representations and strategies and becomes adept in dealing with highly specialized environments, makes it an elusive and fascinating target of our scientific inquiries and the very prototype of the artificial.''\cite{SIM69}. 

This article follows the workshop on ``Adaptivity in nonlinear dynamical systems'', which brought together specialist from various disciplines to share their view on the abstract concept of \textit{adaptivity}. During the presentations and the coffee breaks, there was a lively exchange of ideas that highlighted the great interest in this topic. We hope that this perspective article will be a first step in promoting knowledge transfer between disciplines.

In order to conclude this perspective article, we collect the current open research questions for each section to stimulate future research on adaptivity in the different fields represented in this collection of perspectives and beyond. 
\begin{itemize}
    \item How does a mathematical theory of adaptive systems which includes cutting-edge applications such as, e.g., adaptive networks look like?
    \item How can knowledge about adaptive mechanisms be used to better understand and influence processes in neuronal, physiological and socio-economic systems? 
    \item Can the knowledge about neural plasticity of the human brain be used to inspire the development of new artificial learning algorithms?
    \item What are the capabilities of modeling real-world dynamical system by using adaptivity?
\end{itemize}

\begin{acknowledgments}
JS, RB, and SL thank the Joachim Herz Foundation for funding an interdisciplinary and international workshop on ``Adaptivity in nonlinear dynamical systems'', which was held from 20th to 23rd September 2022 and provided a platform for discussions that resulted in this article. We further thank the Potsdam Institute for Climate Impact Research for supporting and hosting the workshop.
JS acknowledges funding support by the Deutsche Forschungsgemeinschaft (DFG, German Research Foundation) – through the project 429685422.
RB acknowledges funding support by the Deutsche Forschungsgemeinschaft (DFG, German Research Foundation) – through the project 440145547.
SL acknowledges funding support by the Deutsche Forschungsgemeinschaft (DFG, German Research Foundation) – through the project 498288081.
IF acknowledges funding from the Institute of Physics Belgrade through a grant by Ministry of Education, Science and Technological Development of Republic of Serbia.
PH acknowledges further support by the Deutsche Forschungsgemeinschaft (DFG, German Research Foundation) under project ID 434434223 - SFB 1461.
JM acknowledges funding support by the Deutsche Forschungsgemeinschaft (DFG, German Research Foundation) – through the project 458548755.
PLS acknowledges financial support from the Volkswagen Foundation (grant `Reclaiming individual autonomy and democratic discourse online: How to rebalance human and algorithmic decision-making').
PT acknowledges funding support by John A. Blume Foundation and the Vaughn and Nancy Bryson fund. 
GV would like to acknowledge funding from the H2020 European Research Council (ERC) Starting Grant ComplexSwimmers (Grant No. 677511), the Horizon Europe ERC Consolidator Grant MAPEI (Grant No. 101001267), and the Knut and Alice Wallenberg Foundation (Grant No. 2019.0079).

\end{acknowledgments}

\section*{Data Availability Statement}
The data that support the findings of this study are available within the article.

\bibliography{ref_files_resub/refs_ANVARI,ref_files_resub/refs_BADER,ref_files_resub/refs_BARFUSS,ref_files_resub/refs_BOTTA_BREDE,ref_files_resub/refs_FRANOVIC,ref_files_resub/refs_GAUTHIER,ref_files_resub/refs_GOLDT,ref_files_resub/refs_HAJIZADEH,ref_files_resub/refs_HOEVEL,ref_files_resub/refs_KARIN,ref_files_resub/refs_KURTHS,ref_files_resub/refs_LORENZSPREEN,ref_files_resub/refs_MIEHL,ref_files_resub/refs_MOELTER,ref_files_resub/refs_OLMI,ref_files_resub/refs_SCHOELL,ref_files_resub/refs_SEIF,ref_files_resub/refs_TASS,ref_files_resub/refs_VOLPE,ref_files_resub/refs_YANCHUK,ref_files_resub/refs_INTRO}

\begin{thebibliography}{584}%
\makeatletter
\providecommand \@ifxundefined [1]{%
 \@ifx{#1\undefined}
}%
\providecommand \@ifnum [1]{%
 \ifnum #1\expandafter \@firstoftwo
 \else \expandafter \@secondoftwo
 \fi
}%
\providecommand \@ifx [1]{%
 \ifx #1\expandafter \@firstoftwo
 \else \expandafter \@secondoftwo
 \fi
}%
\providecommand \natexlab [1]{#1}%
\providecommand \enquote  [1]{``#1''}%
\providecommand \bibnamefont  [1]{#1}%
\providecommand \bibfnamefont [1]{#1}%
\providecommand \citenamefont [1]{#1}%
\providecommand \href@noop [0]{\@secondoftwo}%
\providecommand \href [0]{\begingroup \@sanitize@url \@href}%
\providecommand \@href[1]{\@@startlink{#1}\@@href}%
\providecommand \@@href[1]{\endgroup#1\@@endlink}%
\providecommand \@sanitize@url [0]{\catcode `\\12\catcode `\$12\catcode
  `\&12\catcode `\#12\catcode `\^12\catcode `\_12\catcode `\%12\relax}%
\providecommand \@@startlink[1]{}%
\providecommand \@@endlink[0]{}%
\providecommand \url  [0]{\begingroup\@sanitize@url \@url }%
\providecommand \@url [1]{\endgroup\@href {#1}{\urlprefix }}%
\providecommand \urlprefix  [0]{URL }%
\providecommand \Eprint [0]{\href }%
\providecommand \doibase [0]{https://doi.org/}%
\providecommand \selectlanguage [0]{\@gobble}%
\providecommand \bibinfo  [0]{\@secondoftwo}%
\providecommand \bibfield  [0]{\@secondoftwo}%
\providecommand \translation [1]{[#1]}%
\providecommand \BibitemOpen [0]{}%
\providecommand \bibitemStop [0]{}%
\providecommand \bibitemNoStop [0]{.\EOS\space}%
\providecommand \EOS [0]{\spacefactor3000\relax}%
\providecommand \BibitemShut  [1]{\csname bibitem#1\endcsname}%
\let\auto@bib@innerbib\@empty
\bibitem [{\citenamefont {Simon}(1969)}]{SIM69}%
  \BibitemOpen
  \bibfield  {author} {\bibinfo {author} {\bibfnamefont {H.~A.}\ \bibnamefont
  {Simon}},\ }\href@noop {} {\emph {\bibinfo {title} {The sciences of the
  artificial}}},\ Karl Taylor Compton lectures\ (\bibinfo  {publisher} {M.I.T.
  Press},\ \bibinfo {address} {Cambridge},\ \bibinfo {year} {1969})\BibitemShut
  {NoStop}%
\bibitem [{\citenamefont {Yakubovich}(1968{\natexlab{a}})}]{YAK68}%
  \BibitemOpen
  \bibfield  {author} {\bibinfo {author} {\bibfnamefont {V.}~\bibnamefont
  {Yakubovich}},\ }\bibfield  {title} {\enquote {\bibinfo {title} {Theory of
  adaptive systems.}}\ }\href@noop {} {\bibfield  {journal} {\bibinfo
  {journal} {Soviet Physics - Doklady}\ }\textbf {\bibinfo {volume} {13}},\
  \bibinfo {pages} {852--855} (\bibinfo {year}
  {1968}{\natexlab{a}})}\BibitemShut {NoStop}%
\bibitem [{\citenamefont {Yakubovich}(1968{\natexlab{b}})}]{YAK68a}%
  \BibitemOpen
  \bibfield  {author} {\bibinfo {author} {\bibfnamefont {V.}~\bibnamefont
  {Yakubovich}},\ }\bibfield  {title} {\enquote {\bibinfo {title} {Adaptive
  systems with multistep goal conditions},}\ }\href@noop {} {\bibfield
  {journal} {\bibinfo  {journal} {Soviet Physics - Doklady}\ }\textbf {\bibinfo
  {volume} {13}},\ \bibinfo {pages} {1096--1099} (\bibinfo {year}
  {1968}{\natexlab{b}})}\BibitemShut {NoStop}%
\bibitem [{\citenamefont {Fomin}, \citenamefont {Fradkov},\ and\ \citenamefont
  {Yakubovich}(1981)}]{fomin1981adaptive}%
  \BibitemOpen
  \bibfield  {author} {\bibinfo {author} {\bibfnamefont {V.}~\bibnamefont
  {Fomin}}, \bibinfo {author} {\bibfnamefont {A.}~\bibnamefont {Fradkov}},\
  and\ \bibinfo {author} {\bibfnamefont {V.}~\bibnamefont {Yakubovich}},\
  }\bibfield  {title} {\enquote {\bibinfo {title} {Adaptive control of
  dynamical systems},}\ }\href@noop {} {\ \bibinfo {series} {Nauka, Moscow}
  (\bibinfo {year} {1981})}\BibitemShut {NoStop}%
\bibitem [{\citenamefont {Annaswamy}\ and\ \citenamefont
  {Fradkov}(2021)}]{ANN21}%
  \BibitemOpen
  \bibfield  {author} {\bibinfo {author} {\bibfnamefont {A.~M.}\ \bibnamefont
  {Annaswamy}}\ and\ \bibinfo {author} {\bibfnamefont {A.~L.}\ \bibnamefont
  {Fradkov}},\ }\bibfield  {title} {\enquote {\bibinfo {title} {A historical
  perspective of adaptive control and learning},}\ }\href
  {https://doi.org/https://doi.org/10.1016/j .arcontrol.2021.10.014} {\bibfield
   {journal} {\bibinfo  {journal} {Annual Reviews in Control}\ }\textbf
  {\bibinfo {volume} {52}},\ \bibinfo {pages} {18--41} (\bibinfo {year}
  {2021})}\BibitemShut {NoStop}%
\bibitem [{\citenamefont {Fradkov}\ and\ \citenamefont
  {Shepeljavyi}(2022)}]{FRAD22}%
  \BibitemOpen
  \bibfield  {author} {\bibinfo {author} {\bibfnamefont {A.~L.}\ \bibnamefont
  {Fradkov}}\ and\ \bibinfo {author} {\bibfnamefont {A.~I.}\ \bibnamefont
  {Shepeljavyi}},\ }\bibfield  {title} {\enquote {\bibinfo {title} {The history
  of cybernetics and artificial intelligence: a view from {S}aint
  {P}etersburg},}\ }\href
  {https://doi.org/10.35470/2226-4116-2022-11-3-253-263} {\bibfield  {journal}
  {\bibinfo  {journal} {Cybernetics and Physics}\ ,\ \bibinfo {pages}
  {253--263}} (\bibinfo {year} {2022})}\BibitemShut {NoStop}%
\bibitem [{\citenamefont {Tsypkin}(1971)}]{tsoepkin1971adaptation}%
  \BibitemOpen
  \bibfield  {author} {\bibinfo {author} {\bibfnamefont {Y.~Z.}\ \bibnamefont
  {Tsypkin}},\ }\href {https://books.google.de/books?id=krZywwEACAAJ} {\emph
  {\bibinfo {title} {Adaptation and Learning in Automatic Systems}}},\
  Mathematics in science and engineering\ (\bibinfo  {publisher} {Academic
  Press},\ \bibinfo {year} {1971})\BibitemShut {NoStop}%
\bibitem [{\citenamefont {Fradkov}(2007)}]{Fradkov2007}%
  \BibitemOpen
  \bibfield  {author} {\bibinfo {author} {\bibfnamefont {A.}~\bibnamefont
  {Fradkov}},\ }\href {https://doi.org/10.1007/978-3-540-46277-4} {\emph
  {\bibinfo {title} {Cybernetical Physics: From Control of Chaos to Quantum
  Control}}}\ (\bibinfo  {publisher} {Springer Berlin Heidelberg},\ \bibinfo
  {year} {2007})\BibitemShut {NoStop}%
\bibitem [{\citenamefont {Lehnert}\ \emph {et~al.}(2014)\citenamefont
  {Lehnert}, \citenamefont {H{\"o}vel}, \citenamefont {Selivanov},
  \citenamefont {Fradkov},\ and\ \citenamefont {Sch{\"o}ll}}]{LEH14}%
  \BibitemOpen
  \bibfield  {author} {\bibinfo {author} {\bibfnamefont {J.}~\bibnamefont
  {Lehnert}}, \bibinfo {author} {\bibfnamefont {P.}~\bibnamefont {H{\"o}vel}},
  \bibinfo {author} {\bibfnamefont {A.~A.}\ \bibnamefont {Selivanov}}, \bibinfo
  {author} {\bibfnamefont {A.~L.}\ \bibnamefont {Fradkov}},\ and\ \bibinfo
  {author} {\bibfnamefont {E.}~\bibnamefont {Sch{\"o}ll}},\ }\bibfield  {title}
  {\enquote {\bibinfo {title} {Controlling cluster synchronization by adapting
  the topology},}\ }\href {https://doi.org/10.1103/physreve.90.042914}
  {\bibfield  {journal} {\bibinfo  {journal} {Phys. Rev. E}\ }\textbf {\bibinfo
  {volume} {90}},\ \bibinfo {pages} {042914} (\bibinfo {year}
  {2014})}\BibitemShut {NoStop}%
\bibitem [{\citenamefont {Pikovsky}, \citenamefont {Rosenblum},\ and\
  \citenamefont {Kurths}(2001)}]{PIK01}%
  \BibitemOpen
  \bibfield  {author} {\bibinfo {author} {\bibfnamefont {A.}~\bibnamefont
  {Pikovsky}}, \bibinfo {author} {\bibfnamefont {M.}~\bibnamefont
  {Rosenblum}},\ and\ \bibinfo {author} {\bibfnamefont {J.}~\bibnamefont
  {Kurths}},\ }\href@noop {} {\emph {\bibinfo {title} {Synchronization: a
  universal concept in nonlinear sciences}}},\ \bibinfo {edition} {1st}\ ed.\
  (\bibinfo  {publisher} {Cambridge University Press},\ \bibinfo {address}
  {Cambridge},\ \bibinfo {year} {2001})\BibitemShut {NoStop}%
\bibitem [{\citenamefont {Boccaletti}\ \emph {et~al.}(2018)\citenamefont
  {Boccaletti}, \citenamefont {Pisarchik}, \citenamefont {del Genio},\ and\
  \citenamefont {Amann}}]{BOC18}%
  \BibitemOpen
  \bibfield  {author} {\bibinfo {author} {\bibfnamefont {S.}~\bibnamefont
  {Boccaletti}}, \bibinfo {author} {\bibfnamefont {A.~N.}\ \bibnamefont
  {Pisarchik}}, \bibinfo {author} {\bibfnamefont {C.~I.}\ \bibnamefont {del
  Genio}},\ and\ \bibinfo {author} {\bibfnamefont {A.}~\bibnamefont {Amann}},\
  }\href@noop {} {\emph {\bibinfo {title} {Synchronization: From Coupled
  Systems to Complex Networks}}}\ (\bibinfo  {publisher} {Cambridge University
  Press},\ \bibinfo {address} {Cambridge},\ \bibinfo {year} {2018})\BibitemShut
  {NoStop}%
\bibitem [{\citenamefont {Yanchuk}\ \emph {et~al.}(2021)\citenamefont
  {Yanchuk}, \citenamefont {Roque}, \citenamefont {Macau},\ and\ \citenamefont
  {Kurths}}]{Yanchuk2021a}%
  \BibitemOpen
  \bibfield  {author} {\bibinfo {author} {\bibfnamefont {S.}~\bibnamefont
  {Yanchuk}}, \bibinfo {author} {\bibfnamefont {A.~C.}\ \bibnamefont {Roque}},
  \bibinfo {author} {\bibfnamefont {E.~E.~N.}\ \bibnamefont {Macau}},\ and\
  \bibinfo {author} {\bibfnamefont {J.}~\bibnamefont {Kurths}},\ }\bibfield
  {title} {\enquote {\bibinfo {title} {{Dynamical phenomena in complex
  networks: fundamentals and applications}},}\ }\href
  {https://doi.org/10.1140/epjs/s11734-021-00282-y} {\bibfield  {journal}
  {\bibinfo  {journal} {The European Physical Journal Special Topics}\ }\textbf
  {\bibinfo {volume} {230}},\ \bibinfo {pages} {2711--2716} (\bibinfo {year}
  {2021})}\BibitemShut {NoStop}%
\bibitem [{\citenamefont {Cabral}\ \emph {et~al.}(2022)\citenamefont {Cabral},
  \citenamefont {Jirsa}, \citenamefont {Popovych}, \citenamefont {Torcini},\
  and\ \citenamefont {Yanchuk}}]{Cabral2022}%
  \BibitemOpen
  \bibfield  {author} {\bibinfo {author} {\bibfnamefont {J.}~\bibnamefont
  {Cabral}}, \bibinfo {author} {\bibfnamefont {V.}~\bibnamefont {Jirsa}},
  \bibinfo {author} {\bibfnamefont {O.}~\bibnamefont {Popovych}}, \bibinfo
  {author} {\bibfnamefont {A.}~\bibnamefont {Torcini}},\ and\ \bibinfo {author}
  {\bibfnamefont {S.}~\bibnamefont {Yanchuk}},\ }\bibfield  {title} {\enquote
  {\bibinfo {title} {{Editorial: From Structure to Function in Neuronal
  Networks: Effects of Adaptation, Time-Delays, and Noise}},}\ }\href@noop {}
  {\bibfield  {journal} {\bibinfo  {journal} {Frontiers in Systems
  Neuroscience}\ } (\bibinfo {year} {2022})}\BibitemShut {NoStop}%
\bibitem [{\citenamefont {Abbott}\ and\ \citenamefont
  {Nelson}(2000)}]{Abbott2000}%
  \BibitemOpen
  \bibfield  {author} {\bibinfo {author} {\bibfnamefont {L.~F.}\ \bibnamefont
  {Abbott}}\ and\ \bibinfo {author} {\bibfnamefont {S.~B.}\ \bibnamefont
  {Nelson}},\ }\bibfield  {title} {\enquote {\bibinfo {title} {{Synaptic
  plasticity: taming the beast}},}\ }\href
  {https://doi.org/https://doi.org/10.1038/81453} {\bibfield  {journal}
  {\bibinfo  {journal} {Nature Neuroscience}\ }\textbf {\bibinfo {volume}
  {3}},\ \bibinfo {pages} {1178--1183} (\bibinfo {year} {2000})}\BibitemShut
  {NoStop}%
\bibitem [{\citenamefont {Dan}\ and\ \citenamefont {Poo}(2004)}]{Dan2004}%
  \BibitemOpen
  \bibfield  {author} {\bibinfo {author} {\bibfnamefont {Y.}~\bibnamefont
  {Dan}}\ and\ \bibinfo {author} {\bibfnamefont {M.-m.}\ \bibnamefont {Poo}},\
  }\bibfield  {title} {\enquote {\bibinfo {title} {{Spike Timing-Dependent
  Plasticity of Neural Circuits}},}\ }\href
  {https://doi.org/10.1016/j.neuron.2004.09.007} {\bibfield  {journal}
  {\bibinfo  {journal} {Neuron}\ }\textbf {\bibinfo {volume} {44}},\ \bibinfo
  {pages} {23--30} (\bibinfo {year} {2004})}\BibitemShut {NoStop}%
\bibitem [{\citenamefont {Maistrenko}\ \emph
  {et~al.}(2007{\natexlab{a}})\citenamefont {Maistrenko}, \citenamefont
  {Lysyansky}, \citenamefont {Hauptmann}, \citenamefont {Burylko},\ and\
  \citenamefont {Tass}}]{MLHBT07}%
  \BibitemOpen
  \bibfield  {author} {\bibinfo {author} {\bibfnamefont {Y.~L.}\ \bibnamefont
  {Maistrenko}}, \bibinfo {author} {\bibfnamefont {B.}~\bibnamefont
  {Lysyansky}}, \bibinfo {author} {\bibfnamefont {C.}~\bibnamefont
  {Hauptmann}}, \bibinfo {author} {\bibfnamefont {O.}~\bibnamefont {Burylko}},\
  and\ \bibinfo {author} {\bibfnamefont {P.~A.}\ \bibnamefont {Tass}},\
  }\bibfield  {title} {\enquote {\bibinfo {title} {Multistability in the
  Kuramoto model with synaptic plasticity},}\ }\href
  {https://doi.org/10.1103/PhysRevE.75.066207} {\bibfield  {journal} {\bibinfo
  {journal} {Phys. Rev. E}\ }\textbf {\bibinfo {volume} {75}},\ \bibinfo
  {pages} {066207} (\bibinfo {year} {2007}{\natexlab{a}})}\BibitemShut
  {NoStop}%
\bibitem [{\citenamefont {Popovych}, \citenamefont {Yanchuk},\ and\
  \citenamefont {Tass}(2013)}]{PYT13}%
  \BibitemOpen
  \bibfield  {author} {\bibinfo {author} {\bibfnamefont {O.~V.}\ \bibnamefont
  {Popovych}}, \bibinfo {author} {\bibfnamefont {S.}~\bibnamefont {Yanchuk}},\
  and\ \bibinfo {author} {\bibfnamefont {P.~A.}\ \bibnamefont {Tass}},\
  }\bibfield  {title} {\enquote {\bibinfo {title} {Self-organized noise
  resistance of oscillatory neural networks with spike timing-dependent
  plasticity},}\ }\href {https://doi.org/10.1038/srep02926} {\bibfield
  {journal} {\bibinfo  {journal} {Sci. Rep.}\ }\textbf {\bibinfo {volume}
  {3}},\ \bibinfo {pages} {2926} (\bibinfo {year} {2013})}\BibitemShut
  {NoStop}%
\bibitem [{\citenamefont {L{\"u}cken}\ \emph {et~al.}(2016)\citenamefont
  {L{\"u}cken}, \citenamefont {Popovych}, \citenamefont {Tass},\ and\
  \citenamefont {Yanchuk}}]{LUE16}%
  \BibitemOpen
  \bibfield  {author} {\bibinfo {author} {\bibfnamefont {L.}~\bibnamefont
  {L{\"u}cken}}, \bibinfo {author} {\bibfnamefont {O.~V.}\ \bibnamefont
  {Popovych}}, \bibinfo {author} {\bibfnamefont {P.~A.}\ \bibnamefont {Tass}},\
  and\ \bibinfo {author} {\bibfnamefont {S.}~\bibnamefont {Yanchuk}},\
  }\bibfield  {title} {\enquote {\bibinfo {title} {{N}oise-enhanced coupling
  between two oscillators with long-term plasticity},}\ }\href
  {https://doi.org/10.1103/physreve.93.032210} {\bibfield  {journal} {\bibinfo
  {journal} {Phys. Rev. E}\ }\textbf {\bibinfo {volume} {93}},\ \bibinfo
  {pages} {032210} (\bibinfo {year} {2016})}\BibitemShut {NoStop}%
\bibitem [{\citenamefont {Aoki}\ and\ \citenamefont {Aoyagi}(2009)}]{Aoki2009}%
  \BibitemOpen
  \bibfield  {author} {\bibinfo {author} {\bibfnamefont {T.}~\bibnamefont
  {Aoki}}\ and\ \bibinfo {author} {\bibfnamefont {T.}~\bibnamefont {Aoyagi}},\
  }\bibfield  {title} {\enquote {\bibinfo {title} {{Co-evolution of Phases and
  Connection Strengths in a Network of Phase Oscillators}},}\ }\href
  {https://doi.org/10.1103/PhysRevLett.102.034101} {\bibfield  {journal}
  {\bibinfo  {journal} {Physical Review Letters}\ }\textbf {\bibinfo {volume}
  {102}},\ \bibinfo {pages} {034101} (\bibinfo {year} {2009})}\BibitemShut
  {NoStop}%
\bibitem [{\citenamefont {Kasatkin}\ \emph {et~al.}(2017)\citenamefont
  {Kasatkin}, \citenamefont {Yanchuk}, \citenamefont {Sch{\"o}ll},\ and\
  \citenamefont {Nekorkin}}]{KAS17}%
  \BibitemOpen
  \bibfield  {author} {\bibinfo {author} {\bibfnamefont {D.~V.}\ \bibnamefont
  {Kasatkin}}, \bibinfo {author} {\bibfnamefont {S.}~\bibnamefont {Yanchuk}},
  \bibinfo {author} {\bibfnamefont {E.}~\bibnamefont {Sch{\"o}ll}},\ and\
  \bibinfo {author} {\bibfnamefont {V.~I.}\ \bibnamefont {Nekorkin}},\
  }\bibfield  {title} {\enquote {\bibinfo {title} {{S}elf-organized emergence
  of multi-layer structure and chimera states in dynamical networks with
  adaptive couplings},}\ }\href {https://doi.org/10.1103/physreve.96.062211}
  {\bibfield  {journal} {\bibinfo  {journal} {Phys. Rev. E}\ }\textbf {\bibinfo
  {volume} {96}},\ \bibinfo {pages} {062211} (\bibinfo {year}
  {2017})}\BibitemShut {NoStop}%
\bibitem [{\citenamefont {Berner}\ \emph
  {et~al.}(2021{\natexlab{a}})\citenamefont {Berner}, \citenamefont {Vock},
  \citenamefont {Sch{\"o}ll},\ and\ \citenamefont {Yanchuk}}]{BER21b}%
  \BibitemOpen
  \bibfield  {author} {\bibinfo {author} {\bibfnamefont {R.}~\bibnamefont
  {Berner}}, \bibinfo {author} {\bibfnamefont {S.}~\bibnamefont {Vock}},
  \bibinfo {author} {\bibfnamefont {E.}~\bibnamefont {Sch{\"o}ll}},\ and\
  \bibinfo {author} {\bibfnamefont {S.}~\bibnamefont {Yanchuk}},\ }\bibfield
  {title} {\enquote {\bibinfo {title} {Desynchronization transitions in
  adaptive networks},}\ }\href {https://doi.org/10.1103/physrevlett.126.028301}
  {\bibfield  {journal} {\bibinfo  {journal} {Phys. Rev. Lett.}\ }\textbf
  {\bibinfo {volume} {126}},\ \bibinfo {pages} {028301} (\bibinfo {year}
  {2021}{\natexlab{a}})}\BibitemShut {NoStop}%
\bibitem [{\citenamefont {Berner}\ \emph
  {et~al.}(2021{\natexlab{b}})\citenamefont {Berner}, \citenamefont {Mehrmann},
  \citenamefont {Sch{\"{o}}ll},\ and\ \citenamefont {Yanchuk}}]{Berner2021a}%
  \BibitemOpen
  \bibfield  {author} {\bibinfo {author} {\bibfnamefont {R.}~\bibnamefont
  {Berner}}, \bibinfo {author} {\bibfnamefont {V.}~\bibnamefont {Mehrmann}},
  \bibinfo {author} {\bibfnamefont {E.}~\bibnamefont {Sch{\"{o}}ll}},\ and\
  \bibinfo {author} {\bibfnamefont {S.}~\bibnamefont {Yanchuk}},\ }\bibfield
  {title} {\enquote {\bibinfo {title} {{The Multiplex Decomposition: An
  Analytic Framework for Multilayer Dynamical Networks}},}\ }\href
  {https://doi.org/10.1137/21M1406180} {\bibfield  {journal} {\bibinfo
  {journal} {SIAM Journal on Applied Dynamical Systems}\ }\textbf {\bibinfo
  {volume} {20}},\ \bibinfo {pages} {1752--1772} (\bibinfo {year}
  {2021}{\natexlab{b}})}\BibitemShut {NoStop}%
\bibitem [{\citenamefont {Berner}\ and\ \citenamefont
  {Yanchuk}(2021)}]{BER21f}%
  \BibitemOpen
  \bibfield  {author} {\bibinfo {author} {\bibfnamefont {R.}~\bibnamefont
  {Berner}}\ and\ \bibinfo {author} {\bibfnamefont {S.}~\bibnamefont
  {Yanchuk}},\ }\bibfield  {title} {\enquote {\bibinfo {title} {Synchronization
  in networks with heterogeneous adaptation rules and applications to
  distance-dependent synaptic plasticity},}\ }\href
  {https://doi.org/10.3389/fams.2021.714978} {\bibfield  {journal} {\bibinfo
  {journal} {Front. Appl. Math. Stat.}\ }\textbf {\bibinfo {volume} {7}},\
  \bibinfo {pages} {714978} (\bibinfo {year} {2021})}\BibitemShut {NoStop}%
\bibitem [{\citenamefont {Thiele}\ \emph {et~al.}(2023)\citenamefont {Thiele},
  \citenamefont {Berner}, \citenamefont {Tass}, \citenamefont {Sch{\"{o}}ll},\
  and\ \citenamefont {Yanchuk}}]{Thiele2021}%
  \BibitemOpen
  \bibfield  {author} {\bibinfo {author} {\bibfnamefont {M.}~\bibnamefont
  {Thiele}}, \bibinfo {author} {\bibfnamefont {R.}~\bibnamefont {Berner}},
  \bibinfo {author} {\bibfnamefont {P.~A.}\ \bibnamefont {Tass}}, \bibinfo
  {author} {\bibfnamefont {E.}~\bibnamefont {Sch{\"{o}}ll}},\ and\ \bibinfo
  {author} {\bibfnamefont {S.}~\bibnamefont {Yanchuk}},\ }\bibfield  {title}
  {\enquote {\bibinfo {title} {{Asymmetric Adaptivity induces Recurrent
  Synchronization in Complex Networks}},}\ }\href@noop {} {\bibfield  {journal}
  {\bibinfo  {journal} {Chaos}\ }\textbf {\bibinfo {volume} {33}},\ \bibinfo
  {pages} {023123} (\bibinfo {year} {2023})}\BibitemShut {NoStop}%
\bibitem [{\citenamefont {R{\"o}hr}\ \emph {et~al.}(2019)\citenamefont
  {R{\"o}hr}, \citenamefont {Berner}, \citenamefont {Lameu}, \citenamefont
  {Popovych},\ and\ \citenamefont {Yanchuk}}]{ROE19a}%
  \BibitemOpen
  \bibfield  {author} {\bibinfo {author} {\bibfnamefont {V.}~\bibnamefont
  {R{\"o}hr}}, \bibinfo {author} {\bibfnamefont {R.}~\bibnamefont {Berner}},
  \bibinfo {author} {\bibfnamefont {E.~L.}\ \bibnamefont {Lameu}}, \bibinfo
  {author} {\bibfnamefont {O.~V.}\ \bibnamefont {Popovych}},\ and\ \bibinfo
  {author} {\bibfnamefont {S.}~\bibnamefont {Yanchuk}},\ }\bibfield  {title}
  {\enquote {\bibinfo {title} {Frequency cluster formation and slow
  oscillations in neural populations with plasticity},}\ }\href
  {https://doi.org/10.1371/journal.pone.0225094} {\bibfield  {journal}
  {\bibinfo  {journal} {PLoS ONE}\ }\textbf {\bibinfo {volume} {14}},\ \bibinfo
  {pages} {e0225094} (\bibinfo {year} {2019})}\BibitemShut {NoStop}%
\bibitem [{\citenamefont {Kuehn}(2015)}]{K15}%
  \BibitemOpen
  \bibfield  {author} {\bibinfo {author} {\bibfnamefont {C.}~\bibnamefont
  {Kuehn}},\ }\href@noop {} {\emph {\bibinfo {title} {Multiple Time Scale
  Dynamics}}}\ (\bibinfo  {publisher} {Springer International Publishing},\
  \bibinfo {address} {Switzerland},\ \bibinfo {year} {2015})\BibitemShut
  {NoStop}%
\bibitem [{\citenamefont {Caporale}\ and\ \citenamefont {Dan}(2008)}]{CD08}%
  \BibitemOpen
  \bibfield  {author} {\bibinfo {author} {\bibfnamefont {N.}~\bibnamefont
  {Caporale}}\ and\ \bibinfo {author} {\bibfnamefont {Y.}~\bibnamefont {Dan}},\
  }\bibfield  {title} {\enquote {\bibinfo {title} {Spike timing--dependent
  plasticity: a hebbian learning rule},}\ }\href
  {https://doi.org/10.1146/annurev.neuro.31.060407.125639} {\bibfield
  {journal} {\bibinfo  {journal} {Annu. Rev. Neurosci.}\ }\textbf {\bibinfo
  {volume} {31}},\ \bibinfo {pages} {25} (\bibinfo {year} {2008})}\BibitemShut
  {NoStop}%
\bibitem [{\citenamefont {Taylor}, \citenamefont {Ott},\ and\ \citenamefont
  {Restrepo}(2010)}]{TAY10}%
  \BibitemOpen
  \bibfield  {author} {\bibinfo {author} {\bibfnamefont {D.}~\bibnamefont
  {Taylor}}, \bibinfo {author} {\bibfnamefont {E.}~\bibnamefont {Ott}},\ and\
  \bibinfo {author} {\bibfnamefont {J.~G.}\ \bibnamefont {Restrepo}},\
  }\bibfield  {title} {\enquote {\bibinfo {title} {Spontaneous synchronization
  of coupled oscillator systems with frequency adaptation},}\ }\href
  {https://doi.org/10.1103/physreve.81.046214} {\bibfield  {journal} {\bibinfo
  {journal} {Phys. Rev. E}\ }\textbf {\bibinfo {volume} {81}},\ \bibinfo
  {pages} {046214} (\bibinfo {year} {2010})}\BibitemShut {NoStop}%
\bibitem [{\citenamefont {Fardet}\ and\ \citenamefont {Levina}(2020)}]{FL20}%
  \BibitemOpen
  \bibfield  {author} {\bibinfo {author} {\bibfnamefont {T.}~\bibnamefont
  {Fardet}}\ and\ \bibinfo {author} {\bibfnamefont {A.}~\bibnamefont
  {Levina}},\ }\bibfield  {title} {\enquote {\bibinfo {title} {Simple models
  including energy and spike constraints reproduce complex activity patterns
  and metabolic disruptions},}\ }\href {https://doi.org/10.1371/journal.
  pcbi.1008503} {\bibfield  {journal} {\bibinfo  {journal} {PLoS Comput.
  Biol.}\ }\textbf {\bibinfo {volume} {16}},\ \bibinfo {pages} {e1008503}
  (\bibinfo {year} {2020})}\BibitemShut {NoStop}%
\bibitem [{\citenamefont {Bonvento}\ and\ \citenamefont
  {Bola\~{n}os}(2021)}]{BB21}%
  \BibitemOpen
  \bibfield  {author} {\bibinfo {author} {\bibfnamefont {G.}~\bibnamefont
  {Bonvento}}\ and\ \bibinfo {author} {\bibfnamefont {J.~P.}\ \bibnamefont
  {Bola\~{n}os}},\ }\bibfield  {title} {\enquote {\bibinfo {title}
  {Astrocyte-neuron metabolic cooperation shapes brain activity},}\ }\href
  {https://doi.org/10.1016/j. cmet.2021.07.006} {\bibfield  {journal} {\bibinfo
   {journal} {Cell Metab.}\ }\textbf {\bibinfo {volume} {33}},\ \bibinfo
  {pages} {1546} (\bibinfo {year} {2021})}\BibitemShut {NoStop}%
\bibitem [{\citenamefont {Roberts}\ \emph {et~al.}(2014)\citenamefont
  {Roberts}, \citenamefont {Iyer}, \citenamefont {Vanhatalo},\ and\
  \citenamefont {Breakspear}}]{RIVB14}%
  \BibitemOpen
  \bibfield  {author} {\bibinfo {author} {\bibfnamefont {J.~A.}\ \bibnamefont
  {Roberts}}, \bibinfo {author} {\bibfnamefont {K.~K.}\ \bibnamefont {Iyer}},
  \bibinfo {author} {\bibfnamefont {S.}~\bibnamefont {Vanhatalo}},\ and\
  \bibinfo {author} {\bibfnamefont {M.}~\bibnamefont {Breakspear}},\ }\bibfield
   {title} {\enquote {\bibinfo {title} {Critical role for resource constraints
  in neural models},}\ }\href {https://doi.org/10.3389/fnsys.2014.00154}
  {\bibfield  {journal} {\bibinfo  {journal} {Front. Syst. Neurosci.}\ }\textbf
  {\bibinfo {volume} {8}},\ \bibinfo {pages} {154} (\bibinfo {year}
  {2014})}\BibitemShut {NoStop}%
\bibitem [{\citenamefont {Virkar}\ \emph {et~al.}(2016)\citenamefont {Virkar},
  \citenamefont {Shew}, \citenamefont {Restrepo},\ and\ \citenamefont
  {Ott}}]{VSRO16}%
  \BibitemOpen
  \bibfield  {author} {\bibinfo {author} {\bibfnamefont {Y.~S.}\ \bibnamefont
  {Virkar}}, \bibinfo {author} {\bibfnamefont {W.~L.}\ \bibnamefont {Shew}},
  \bibinfo {author} {\bibfnamefont {J.~G.}\ \bibnamefont {Restrepo}},\ and\
  \bibinfo {author} {\bibfnamefont {E.}~\bibnamefont {Ott}},\ }\bibfield
  {title} {\enquote {\bibinfo {title} {Feedback control stabilization of
  critical dynamics via resource transport on multilayer networks: How glia
  enable learning dynamics in the brain},}\ }\href
  {https://doi.org/10.1103/PhysRevE.94.042310} {\bibfield  {journal} {\bibinfo
  {journal} {Phys. Rev. E}\ }\textbf {\bibinfo {volume} {94}},\ \bibinfo
  {pages} {042310} (\bibinfo {year} {2016})}\BibitemShut {NoStop}%
\bibitem [{\citenamefont {Levina}, \citenamefont {Herrmann},\ and\
  \citenamefont {Geisel}(2007)}]{LHG07}%
  \BibitemOpen
  \bibfield  {author} {\bibinfo {author} {\bibfnamefont {A.}~\bibnamefont
  {Levina}}, \bibinfo {author} {\bibfnamefont {J.~M.}\ \bibnamefont
  {Herrmann}},\ and\ \bibinfo {author} {\bibfnamefont {T.}~\bibnamefont
  {Geisel}},\ }\bibfield  {title} {\enquote {\bibinfo {title} {Dynamical
  synapses causing self-organized criticality in neural networks},}\ }\href
  {https://doi.org/10.1038/nphys758} {\bibfield  {journal} {\bibinfo  {journal}
  {Nat. Phys.}\ }\textbf {\bibinfo {volume} {3}},\ \bibinfo {pages} {857}
  (\bibinfo {year} {2007})}\BibitemShut {NoStop}%
\bibitem [{\citenamefont {Berner}, \citenamefont {Sch{\"o}ll},\ and\
  \citenamefont {Yanchuk}(2019)}]{BER19}%
  \BibitemOpen
  \bibfield  {author} {\bibinfo {author} {\bibfnamefont {R.}~\bibnamefont
  {Berner}}, \bibinfo {author} {\bibfnamefont {E.}~\bibnamefont {Sch{\"o}ll}},\
  and\ \bibinfo {author} {\bibfnamefont {S.}~\bibnamefont {Yanchuk}},\
  }\bibfield  {title} {\enquote {\bibinfo {title} {Multiclusters in networks of
  adaptively coupled phase oscillators},}\ }\href
  {https://doi.org/10.1137/18m1210150} {\bibfield  {journal} {\bibinfo
  {journal} {SIAM J. Appl. Dyn. Syst.}\ }\textbf {\bibinfo {volume} {18}},\
  \bibinfo {pages} {2227--2266} (\bibinfo {year} {2019})}\BibitemShut {NoStop}%
\bibitem [{\citenamefont {Berner}\ \emph {et~al.}(2019)\citenamefont {Berner},
  \citenamefont {Fialkowski}, \citenamefont {Kasatkin}, \citenamefont
  {Nekorkin}, \citenamefont {Yanchuk},\ and\ \citenamefont
  {Sch{\"o}ll}}]{BER19a}%
  \BibitemOpen
  \bibfield  {author} {\bibinfo {author} {\bibfnamefont {R.}~\bibnamefont
  {Berner}}, \bibinfo {author} {\bibfnamefont {J.}~\bibnamefont {Fialkowski}},
  \bibinfo {author} {\bibfnamefont {D.~V.}\ \bibnamefont {Kasatkin}}, \bibinfo
  {author} {\bibfnamefont {V.~I.}\ \bibnamefont {Nekorkin}}, \bibinfo {author}
  {\bibfnamefont {S.}~\bibnamefont {Yanchuk}},\ and\ \bibinfo {author}
  {\bibfnamefont {E.}~\bibnamefont {Sch{\"o}ll}},\ }\bibfield  {title}
  {\enquote {\bibinfo {title} {Hierarchical frequency clusters in adaptive
  networks of phase oscillators},}\ }\href {https://doi.org/10.1063/1.5097835}
  {\bibfield  {journal} {\bibinfo  {journal} {Chaos}\ }\textbf {\bibinfo
  {volume} {29}},\ \bibinfo {pages} {103134} (\bibinfo {year}
  {2019})}\BibitemShut {NoStop}%
\bibitem [{\citenamefont {Kroma-Wiley}, \citenamefont {Mucha},\ and\
  \citenamefont {Bassett}(2021)}]{KMB21}%
  \BibitemOpen
  \bibfield  {author} {\bibinfo {author} {\bibfnamefont {K.~A.}\ \bibnamefont
  {Kroma-Wiley}}, \bibinfo {author} {\bibfnamefont {P.~J.}\ \bibnamefont
  {Mucha}},\ and\ \bibinfo {author} {\bibfnamefont {D.~S.}\ \bibnamefont
  {Bassett}},\ }\bibfield  {title} {\enquote {\bibinfo {title} {Synchronization
  of coupled Kuramoto oscillators under resource constraints},}\ }\href
  {https://doi.org/10.1103/PhysRevE.104.014211} {\bibfield  {journal} {\bibinfo
   {journal} {Phys. Rev. E}\ }\textbf {\bibinfo {volume} {104}},\ \bibinfo
  {pages} {014211} (\bibinfo {year} {2021})}\BibitemShut {NoStop}%
\bibitem [{\citenamefont {Thamizharasan}\ \emph {et~al.}(2021)\citenamefont
  {Thamizharasan}, \citenamefont {Chandrasekar}, \citenamefont {Senthilvelan},
  \citenamefont {Berner}, \citenamefont {Sch\"{o}ll},\ and\ \citenamefont
  {Senthilkumar}}]{TCS22}%
  \BibitemOpen
  \bibfield  {author} {\bibinfo {author} {\bibfnamefont {S.}~\bibnamefont
  {Thamizharasan}}, \bibinfo {author} {\bibfnamefont {V.~K.}\ \bibnamefont
  {Chandrasekar}}, \bibinfo {author} {\bibfnamefont {M.}~\bibnamefont
  {Senthilvelan}}, \bibinfo {author} {\bibfnamefont {R.}~\bibnamefont
  {Berner}}, \bibinfo {author} {\bibfnamefont {E.}~\bibnamefont {Sch\"{o}ll}},\
  and\ \bibinfo {author} {\bibfnamefont {D.~V.}\ \bibnamefont {Senthilkumar}},\
  }\bibfield  {title} {\enquote {\bibinfo {title} {Exotic states induced by
  coevolving connection weights and phases in complex networks},}\ }\href
  {https://doi.org/10.1103/PhysRevE.105.034312} {\bibfield  {journal} {\bibinfo
   {journal} {Phys. Rev. E}\ }\textbf {\bibinfo {volume} {105}},\ \bibinfo
  {pages} {034312} (\bibinfo {year} {2021})}\BibitemShut {NoStop}%
\bibitem [{\citenamefont {Fialkowski}\ \emph {et~al.}(2022)\citenamefont
  {Fialkowski}, \citenamefont {Yanchuk}, \citenamefont {Sokolov}, \citenamefont
  {Sch{\"o}ll}, \citenamefont {Gottwald},\ and\ \citenamefont
  {Berner}}]{FIA22}%
  \BibitemOpen
  \bibfield  {author} {\bibinfo {author} {\bibfnamefont {J.}~\bibnamefont
  {Fialkowski}}, \bibinfo {author} {\bibfnamefont {S.}~\bibnamefont {Yanchuk}},
  \bibinfo {author} {\bibfnamefont {I.~M.}\ \bibnamefont {Sokolov}}, \bibinfo
  {author} {\bibfnamefont {E.}~\bibnamefont {Sch{\"o}ll}}, \bibinfo {author}
  {\bibfnamefont {G.~A.}\ \bibnamefont {Gottwald}},\ and\ \bibinfo {author}
  {\bibfnamefont {R.}~\bibnamefont {Berner}},\ }\bibfield  {title} {\enquote
  {\bibinfo {title} {Heterogeneous nucleation in finite size adaptive dynamical
  networks},}\ }\href {https://doi.org/10.1103/physrevlett.130.067402}
  {\bibfield  {journal} {\bibinfo  {journal} {Phys. Rev. Lett.}\ }\textbf
  {\bibinfo {volume} {130}},\ \bibinfo {pages} {067402} (\bibinfo {year}
  {2022})}\BibitemShut {NoStop}%
\bibitem [{\citenamefont {Franovi\'c}\ \emph {et~al.}(2022)\citenamefont
  {Franovi\'c}, \citenamefont {Eydam}, \citenamefont {Yanchuk},\ and\
  \citenamefont {Berner}}]{FEYB22}%
  \BibitemOpen
  \bibfield  {author} {\bibinfo {author} {\bibfnamefont {I.}~\bibnamefont
  {Franovi\'c}}, \bibinfo {author} {\bibfnamefont {S.~R.}\ \bibnamefont
  {Eydam}}, \bibinfo {author} {\bibfnamefont {S.}~\bibnamefont {Yanchuk}},\
  and\ \bibinfo {author} {\bibfnamefont {R.}~\bibnamefont {Berner}},\
  }\bibfield  {title} {\enquote {\bibinfo {title} {Collective activity bursting
  in a population of excitable units adaptively coupled to a pool of
  resources},}\ }\href {https://doi.org/10.3389/fnetp.2022.841829} {\bibfield
  {journal} {\bibinfo  {journal} {Front. Netw. Physiol.}\ }\textbf {\bibinfo
  {volume} {2}},\ \bibinfo {pages} {841829} (\bibinfo {year}
  {2022})}\BibitemShut {NoStop}%
\bibitem [{\citenamefont {Franovi\'c}\ \emph {et~al.}(2020)\citenamefont
  {Franovi\'c}, \citenamefont {Yanchuk}, \citenamefont {Eydam}, \citenamefont
  {Ba\v{c}i\'c},\ and\ \citenamefont {Wolfrum}}]{FYEBW20}%
  \BibitemOpen
  \bibfield  {author} {\bibinfo {author} {\bibfnamefont {I.}~\bibnamefont
  {Franovi\'c}}, \bibinfo {author} {\bibfnamefont {S.}~\bibnamefont {Yanchuk}},
  \bibinfo {author} {\bibfnamefont {S.~R.}\ \bibnamefont {Eydam}}, \bibinfo
  {author} {\bibfnamefont {I.}~\bibnamefont {Ba\v{c}i\'c}},\ and\ \bibinfo
  {author} {\bibfnamefont {M.}~\bibnamefont {Wolfrum}},\ }\bibfield  {title}
  {\enquote {\bibinfo {title} {Dynamics of a stochastic excitable system with
  slowly adapting feedback},}\ }\href {https://doi.org/10.1063/1.5145176}
  {\bibfield  {journal} {\bibinfo  {journal} {Chaos}\ }\textbf {\bibinfo
  {volume} {30}},\ \bibinfo {pages} {083109} (\bibinfo {year}
  {2020})}\BibitemShut {NoStop}%
\bibitem [{\citenamefont {Ba\v{c}i\'c}\ \emph
  {et~al.}(2018{\natexlab{a}})\citenamefont {Ba\v{c}i\'c}, \citenamefont
  {Klinshov}, \citenamefont {Nekorkin}, \citenamefont {Perc},\ and\
  \citenamefont {Franovi\'c}}]{BKNPF18}%
  \BibitemOpen
  \bibfield  {author} {\bibinfo {author} {\bibfnamefont {I.}~\bibnamefont
  {Ba\v{c}i\'c}}, \bibinfo {author} {\bibfnamefont {V.}~\bibnamefont
  {Klinshov}}, \bibinfo {author} {\bibfnamefont {V.~I.}\ \bibnamefont
  {Nekorkin}}, \bibinfo {author} {\bibfnamefont {M.}~\bibnamefont {Perc}},\
  and\ \bibinfo {author} {\bibfnamefont {I.}~\bibnamefont {Franovi\'c}},\
  }\bibfield  {title} {\enquote {\bibinfo {title} {Inverse stochastic resonance
  in a system of excitable active rotators with adaptive coupling},}\ }\href
  {https://doi.org/10.1209/0295-5075/124/40004} {\bibfield  {journal} {\bibinfo
   {journal} {EPL}\ }\textbf {\bibinfo {volume} {124}},\ \bibinfo {pages}
  {40004} (\bibinfo {year} {2018}{\natexlab{a}})}\BibitemShut {NoStop}%
\bibitem [{\citenamefont {Knoblauch}\ \emph {et~al.}(2012)\citenamefont
  {Knoblauch}, \citenamefont {Hauser}, \citenamefont {Gewaltig}, \citenamefont
  {K\"{o}rner},\ and\ \citenamefont {Palm}}]{KHGKP12}%
  \BibitemOpen
  \bibfield  {author} {\bibinfo {author} {\bibfnamefont {A.}~\bibnamefont
  {Knoblauch}}, \bibinfo {author} {\bibfnamefont {F.}~\bibnamefont {Hauser}},
  \bibinfo {author} {\bibfnamefont {M.-O.}\ \bibnamefont {Gewaltig}}, \bibinfo
  {author} {\bibfnamefont {E.}~\bibnamefont {K\"{o}rner}},\ and\ \bibinfo
  {author} {\bibfnamefont {G.}~\bibnamefont {Palm}},\ }\bibfield  {title}
  {\enquote {\bibinfo {title} {Does spike-timing-dependent synaptic plasticity
  couple or decouple neurons firing in synchrony?}}\ }\href
  {https://doi.org/10.3389/fncom.2012.00055} {\bibfield  {journal} {\bibinfo
  {journal} {Front. Comput. Neurosci.}\ }\textbf {\bibinfo {volume} {6}},\
  \bibinfo {pages} {55} (\bibinfo {year} {2012})}\BibitemShut {NoStop}%
\bibitem [{\citenamefont {Pazo}\ and\ \citenamefont
  {Montbri\'{o}}(2006)}]{PM06}%
  \BibitemOpen
  \bibfield  {author} {\bibinfo {author} {\bibfnamefont {D.}~\bibnamefont
  {Pazo}}\ and\ \bibinfo {author} {\bibfnamefont {E.}~\bibnamefont
  {Montbri\'{o}}},\ }\bibfield  {title} {\enquote {\bibinfo {title} {Universal
  behavior in populations composed of excitable and self-oscillatory
  elements},}\ }\href {https://doi.org/10.1103/PhysRevE.73.055202} {\bibfield
  {journal} {\bibinfo  {journal} {Phys. Rev. E}\ }\textbf {\bibinfo {volume}
  {73}},\ \bibinfo {pages} {055202(R)} (\bibinfo {year} {2006})}\BibitemShut
  {NoStop}%
\bibitem [{\citenamefont {Lafuerza}, \citenamefont {Colet},\ and\ \citenamefont
  {Toral}(2010)}]{LCT10}%
  \BibitemOpen
  \bibfield  {author} {\bibinfo {author} {\bibfnamefont {L.~F.}\ \bibnamefont
  {Lafuerza}}, \bibinfo {author} {\bibfnamefont {P.}~\bibnamefont {Colet}},\
  and\ \bibinfo {author} {\bibfnamefont {R.}~\bibnamefont {Toral}},\ }\bibfield
   {title} {\enquote {\bibinfo {title} {Nonuniversal results iinduced by
  diversity distribution in coupled excitable systems},}\ }\href
  {https://doi.org/10.1103/PhysRevLett.105.084101} {\bibfield  {journal}
  {\bibinfo  {journal} {Phys. Rev. Lett.}\ }\textbf {\bibinfo {volume} {105}},\
  \bibinfo {pages} {084101} (\bibinfo {year} {2010})}\BibitemShut {NoStop}%
\bibitem [{\citenamefont {Klinshov}\ and\ \citenamefont
  {Franovi\'c}(2019)}]{KF19}%
  \BibitemOpen
  \bibfield  {author} {\bibinfo {author} {\bibfnamefont {V.}~\bibnamefont
  {Klinshov}}\ and\ \bibinfo {author} {\bibfnamefont {I.}~\bibnamefont
  {Franovi\'c}},\ }\bibfield  {title} {\enquote {\bibinfo {title} {Two
  scenarios for the onset and suppression of collective oscillations in
  heterogeneous populations of active rotators},}\ }\href
  {https://doi.org/10.1103/PhysRevE.100.062211} {\bibfield  {journal} {\bibinfo
   {journal} {Phys. Rev. E}\ }\textbf {\bibinfo {volume} {100}},\ \bibinfo
  {pages} {062211} (\bibinfo {year} {2019})}\BibitemShut {NoStop}%
\bibitem [{\citenamefont {Ba\v{c}i\'c}\ \emph
  {et~al.}(2018{\natexlab{b}})\citenamefont {Ba\v{c}i\'c}, \citenamefont
  {Yanchuk}, \citenamefont {Wolfrum},\ and\ \citenamefont
  {Franovi\'c}}]{BYWF18}%
  \BibitemOpen
  \bibfield  {author} {\bibinfo {author} {\bibfnamefont {I.}~\bibnamefont
  {Ba\v{c}i\'c}}, \bibinfo {author} {\bibfnamefont {S.}~\bibnamefont
  {Yanchuk}}, \bibinfo {author} {\bibfnamefont {M.}~\bibnamefont {Wolfrum}},\
  and\ \bibinfo {author} {\bibfnamefont {I.}~\bibnamefont {Franovi\'c}},\
  }\bibfield  {title} {\enquote {\bibinfo {title} {Noise-induced switching in
  two adaptively coupled excitable systems},}\ }\href
  {https://doi.org/10.1140/epjst/e2018-800084-6} {\bibfield  {journal}
  {\bibinfo  {journal} {Eur. Phys. J. - Spec. Top.}\ }\textbf {\bibinfo
  {volume} {227}},\ \bibinfo {pages} {1077} (\bibinfo {year}
  {2018}{\natexlab{b}})}\BibitemShut {NoStop}%
\bibitem [{\citenamefont {Sporns}\ and\ \citenamefont
  {K\"{o}tter}(2004)}]{SK04}%
  \BibitemOpen
  \bibfield  {author} {\bibinfo {author} {\bibfnamefont {O.}~\bibnamefont
  {Sporns}}\ and\ \bibinfo {author} {\bibfnamefont {R.}~\bibnamefont
  {K\"{o}tter}},\ }\bibfield  {title} {\enquote {\bibinfo {title} {Motifs in
  brain networks},}\ }\href {https://doi.org/10.1371/journal. pbio.0020369}
  {\bibfield  {journal} {\bibinfo  {journal} {PLoS Biol.}\ }\textbf {\bibinfo
  {volume} {2}},\ \bibinfo {pages} {e369} (\bibinfo {year} {2004})}\BibitemShut
  {NoStop}%
\bibitem [{\citenamefont {Ba\v{c}i\'c}\ and\ \citenamefont
  {Franovi\'c}(2020)}]{BF20}%
  \BibitemOpen
  \bibfield  {author} {\bibinfo {author} {\bibfnamefont {I.}~\bibnamefont
  {Ba\v{c}i\'c}}\ and\ \bibinfo {author} {\bibfnamefont {I.}~\bibnamefont
  {Franovi\'c}},\ }\bibfield  {title} {\enquote {\bibinfo {title} {Two
  paradigmatic scenarios for inverse stochastic resonance},}\ }\href
  {https://doi.org/10.1063/1.5139628} {\bibfield  {journal} {\bibinfo
  {journal} {Chaos}\ }\textbf {\bibinfo {volume} {30}},\ \bibinfo {pages}
  {033123} (\bibinfo {year} {2020})}\BibitemShut {NoStop}%
\bibitem [{\citenamefont {Pecora}\ and\ \citenamefont {Carroll}(1998)}]{PEC98}%
  \BibitemOpen
  \bibfield  {author} {\bibinfo {author} {\bibfnamefont {L.~M.}\ \bibnamefont
  {Pecora}}\ and\ \bibinfo {author} {\bibfnamefont {T.~L.}\ \bibnamefont
  {Carroll}},\ }\bibfield  {title} {\enquote {\bibinfo {title} {{M}aster
  {S}tability {F}unctions for {S}ynchronized {C}oupled {S}ystems},}\ }\href
  {https://doi.org/10.1103/physrevlett.80.2109} {\bibfield  {journal} {\bibinfo
   {journal} {Phys. Rev. Lett.}\ }\textbf {\bibinfo {volume} {80}},\ \bibinfo
  {pages} {2109--2112} (\bibinfo {year} {1998})}\BibitemShut {NoStop}%
\bibitem [{\citenamefont {Ott}\ and\ \citenamefont
  {Antonsen}(2008{\natexlab{a}})}]{OA08}%
  \BibitemOpen
  \bibfield  {author} {\bibinfo {author} {\bibfnamefont {E.}~\bibnamefont
  {Ott}}\ and\ \bibinfo {author} {\bibfnamefont {T.~M.}\ \bibnamefont
  {Antonsen}},\ }\bibfield  {title} {\enquote {\bibinfo {title} {Low
  dimensional behavior of large systems of globally coupled oscillators},}\
  }\href {https://doi.org/10.1063/1.2930766} {\bibfield  {journal} {\bibinfo
  {journal} {Chaos}\ }\textbf {\bibinfo {volume} {18}},\ \bibinfo {pages}
  {037113} (\bibinfo {year} {2008}{\natexlab{a}})}\BibitemShut {NoStop}%
\bibitem [{\citenamefont {Ott}\ and\ \citenamefont {Antonsen}(2009)}]{OA09}%
  \BibitemOpen
  \bibfield  {author} {\bibinfo {author} {\bibfnamefont {E.}~\bibnamefont
  {Ott}}\ and\ \bibinfo {author} {\bibfnamefont {T.~M.}\ \bibnamefont
  {Antonsen}},\ }\bibfield  {title} {\enquote {\bibinfo {title} {Long time
  evolution of phase oscillator systems},}\ }\href
  {https://doi.org/10.1063/1.3136851} {\bibfield  {journal} {\bibinfo
  {journal} {Chaos}\ }\textbf {\bibinfo {volume} {19}},\ \bibinfo {pages}
  {023117} (\bibinfo {year} {2009})}\BibitemShut {NoStop}%
\bibitem [{\citenamefont {Bick}\ \emph {et~al.}(2020)\citenamefont {Bick},
  \citenamefont {Goodfellow}, \citenamefont {Laing},\ and\ \citenamefont
  {Martens}}]{BGLM20}%
  \BibitemOpen
  \bibfield  {author} {\bibinfo {author} {\bibfnamefont {C.}~\bibnamefont
  {Bick}}, \bibinfo {author} {\bibfnamefont {M.}~\bibnamefont {Goodfellow}},
  \bibinfo {author} {\bibfnamefont {C.~R.}\ \bibnamefont {Laing}},\ and\
  \bibinfo {author} {\bibfnamefont {E.~A.}\ \bibnamefont {Martens}},\
  }\bibfield  {title} {\enquote {\bibinfo {title} {Understanding the dynamics
  of biological and neural oscillator networks through exact mean-field
  reductions: a review},}\ }\href {https://doi.org/10.1186/s13408-020-00086-9}
  {\bibfield  {journal} {\bibinfo  {journal} {J. Math. Neurosci.}\ }\textbf
  {\bibinfo {volume} {10}},\ \bibinfo {pages} {9} (\bibinfo {year}
  {2020})}\BibitemShut {NoStop}%
\bibitem [{\citenamefont {Lan}\ \emph {et~al.}(2012)\citenamefont {Lan},
  \citenamefont {Sartori}, \citenamefont {Neumann}, \citenamefont {Sourjik},\
  and\ \citenamefont {Tu}}]{LSNST12}%
  \BibitemOpen
  \bibfield  {author} {\bibinfo {author} {\bibfnamefont {G.}~\bibnamefont
  {Lan}}, \bibinfo {author} {\bibfnamefont {P.}~\bibnamefont {Sartori}},
  \bibinfo {author} {\bibfnamefont {S.}~\bibnamefont {Neumann}}, \bibinfo
  {author} {\bibfnamefont {V.}~\bibnamefont {Sourjik}},\ and\ \bibinfo {author}
  {\bibfnamefont {Y.}~\bibnamefont {Tu}},\ }\bibfield  {title} {\enquote
  {\bibinfo {title} {The energy--speed--accuracy trade-off in sensory
  adaptation},}\ }\href {https://doi.org/10.1038/nphys2276} {\bibfield
  {journal} {\bibinfo  {journal} {Nat. Phys.}\ }\textbf {\bibinfo {volume}
  {8}},\ \bibinfo {pages} {422} (\bibinfo {year} {2012})}\BibitemShut {NoStop}%
\bibitem [{\citenamefont {Conti}\ and\ \citenamefont {Mora}(2022)}]{CM22}%
  \BibitemOpen
  \bibfield  {author} {\bibinfo {author} {\bibfnamefont {D.}~\bibnamefont
  {Conti}}\ and\ \bibinfo {author} {\bibfnamefont {T.}~\bibnamefont {Mora}},\
  }\bibfield  {title} {\enquote {\bibinfo {title} {Nonequilibrium dynamics of
  adaptation in sensory systems},}\ }\href
  {https://doi.org/10.1103/PhysRevE.106.054404} {\bibfield  {journal} {\bibinfo
   {journal} {Phys. Rev. E}\ }\textbf {\bibinfo {volume} {106}},\ \bibinfo
  {pages} {054404} (\bibinfo {year} {2022})}\BibitemShut {NoStop}%
\bibitem [{\citenamefont {Tka\v{c}ik}\ and\ \citenamefont
  {Bialek}(2016)}]{TB16}%
  \BibitemOpen
  \bibfield  {author} {\bibinfo {author} {\bibfnamefont {G.}~\bibnamefont
  {Tka\v{c}ik}}\ and\ \bibinfo {author} {\bibfnamefont {W.}~\bibnamefont
  {Bialek}},\ }\bibfield  {title} {\enquote {\bibinfo {title} {Information
  processing in living systems},}\ }\href
  {https://doi.org/10.1146/annurev-conmatphys-031214-014803} {\bibfield
  {journal} {\bibinfo  {journal} {Annu. Rev. Condens. Matter Phys.}\ }\textbf
  {\bibinfo {volume} {7}},\ \bibinfo {pages} {89} (\bibinfo {year}
  {2016})}\BibitemShut {NoStop}%
\bibitem [{\citenamefont {Ionescu}, \citenamefont {Jansson},\ and\
  \citenamefont {Botta}(2018)}]{10.1007/978-3-030-03418-4}%
  \BibitemOpen
  \bibfield  {author} {\bibinfo {author} {\bibfnamefont {C.}~\bibnamefont
  {Ionescu}}, \bibinfo {author} {\bibfnamefont {P.}~\bibnamefont {Jansson}},\
  and\ \bibinfo {author} {\bibfnamefont {N.}~\bibnamefont {Botta}},\ }\bibfield
   {title} {\enquote {\bibinfo {title} {Type theory as a framework for
  modelling and programming},}\ }in\ \href@noop {} {\emph {\bibinfo {booktitle}
  {Leveraging Applications of Formal Methods, Verification and Validation.
  Modeling}}},\ \bibinfo {editor} {edited by\ \bibinfo {editor} {\bibfnamefont
  {T.}~\bibnamefont {Margaria}}\ and\ \bibinfo {editor} {\bibfnamefont
  {B.}~\bibnamefont {Steffen}}}\ (\bibinfo  {publisher} {Springer International
  Publishing},\ \bibinfo {address} {Cham},\ \bibinfo {year} {2018})\ pp.\
  \bibinfo {pages} {119--133}\BibitemShut {NoStop}%
\bibitem [{\citenamefont {Broy}, \citenamefont {Havelund},\ and\ \citenamefont
  {Kumar}(2016)}]{10.1007/978-3-319-47169-3}%
  \BibitemOpen
  \bibfield  {author} {\bibinfo {author} {\bibfnamefont {M.}~\bibnamefont
  {Broy}}, \bibinfo {author} {\bibfnamefont {K.}~\bibnamefont {Havelund}},\
  and\ \bibinfo {author} {\bibfnamefont {R.}~\bibnamefont {Kumar}},\ }\bibfield
   {title} {\enquote {\bibinfo {title} {Towards a unified view of modeling and
  programming},}\ }in\ \href@noop {} {\emph {\bibinfo {booktitle} {Leveraging
  Applications of Formal Methods, Verification and Validation: Discussion,
  Dissemination, Applications}}},\ \bibinfo {editor} {edited by\ \bibinfo
  {editor} {\bibfnamefont {T.}~\bibnamefont {Margaria}}\ and\ \bibinfo {editor}
  {\bibfnamefont {B.}~\bibnamefont {Steffen}}}\ (\bibinfo  {publisher}
  {Springer International Publishing},\ \bibinfo {address} {Cham},\ \bibinfo
  {year} {2016})\ pp.\ \bibinfo {pages} {238--257}\BibitemShut {NoStop}%
\bibitem [{\citenamefont {Ionescu}\ \emph {et~al.}(2009)\citenamefont
  {Ionescu}, \citenamefont {Klein}, \citenamefont {Hinkel}, \citenamefont
  {{Kavi Kumar}},\ and\ \citenamefont {Klein}}]{ionescu+al2009}%
  \BibitemOpen
  \bibfield  {author} {\bibinfo {author} {\bibfnamefont {C.}~\bibnamefont
  {Ionescu}}, \bibinfo {author} {\bibfnamefont {R.~J.~T.}\ \bibnamefont
  {Klein}}, \bibinfo {author} {\bibfnamefont {J.}~\bibnamefont {Hinkel}},
  \bibinfo {author} {\bibfnamefont {K.~S.}\ \bibnamefont {{Kavi Kumar}}},\ and\
  \bibinfo {author} {\bibfnamefont {R.}~\bibnamefont {Klein}},\ }\bibfield
  {title} {\enquote {\bibinfo {title} {{Towards a formal framework of
  vulnerability to climate change}},}\ }\href@noop {} {\bibfield  {journal}
  {\bibinfo  {journal} {Environmental Modelling and Assessment}\ }\textbf
  {\bibinfo {volume} {14}},\ \bibinfo {pages} {1--16} (\bibinfo {year}
  {2009})}\BibitemShut {NoStop}%
\bibitem [{\citenamefont {Botta}, \citenamefont {Jansson},\ and\ \citenamefont
  {Ionescu}(2017)}]{2017_Botta_Jansson_Ionescu}%
  \BibitemOpen
  \bibfield  {author} {\bibinfo {author} {\bibfnamefont {N.}~\bibnamefont
  {Botta}}, \bibinfo {author} {\bibfnamefont {P.}~\bibnamefont {Jansson}},\
  and\ \bibinfo {author} {\bibfnamefont {C.}~\bibnamefont {Ionescu}},\
  }\bibfield  {title} {\enquote {\bibinfo {title} {Contributions to a
  computational theory of policy advice and avoidability},}\ }\href
  {https://doi.org/10.1017/S0956796817000156} {\bibfield  {journal} {\bibinfo
  {journal} {J. Funct. Program.}\ }\textbf {\bibinfo {volume} {27}},\ \bibinfo
  {pages} {e23} (\bibinfo {year} {2017})}\BibitemShut {NoStop}%
\bibitem [{\citenamefont {Brede}(2021)}]{brede2021dsl}%
  \BibitemOpen
  \bibfield  {author} {\bibinfo {author} {\bibfnamefont {N.}~\bibnamefont
  {Brede}},\ }\href@noop {} {\enquote {\bibinfo {title} {Toward a {DSL} for
  {S}equential {D}ecision {P}roblems with tipping point uncertainties},}\
  }\bibinfo {howpublished} {https://doi.org/10.5281/zenodo.6783894} (\bibinfo
  {year} {2021})\BibitemShut {NoStop}%
\bibitem [{\citenamefont {Botta}\ \emph {et~al.}(2023)\citenamefont {Botta},
  \citenamefont {Brede}, \citenamefont {Crucifix}, \citenamefont {Ionescu},
  \citenamefont {Jansson}, \citenamefont {Li}, \citenamefont {Mart{\'i}nez},\
  and\ \citenamefont {Richter}}]{Botta2023MatterMost}%
  \BibitemOpen
  \bibfield  {author} {\bibinfo {author} {\bibfnamefont {N.}~\bibnamefont
  {Botta}}, \bibinfo {author} {\bibfnamefont {N.}~\bibnamefont {Brede}},
  \bibinfo {author} {\bibfnamefont {M.}~\bibnamefont {Crucifix}}, \bibinfo
  {author} {\bibfnamefont {C.}~\bibnamefont {Ionescu}}, \bibinfo {author}
  {\bibfnamefont {P.}~\bibnamefont {Jansson}}, \bibinfo {author} {\bibfnamefont
  {Z.}~\bibnamefont {Li}}, \bibinfo {author} {\bibfnamefont {M.}~\bibnamefont
  {Mart{\'i}nez}},\ and\ \bibinfo {author} {\bibfnamefont {T.}~\bibnamefont
  {Richter}},\ }\bibfield  {title} {\enquote {\bibinfo {title} {Responsibility
  under uncertainty: Which climate decisions matter most?}}\ }\href
  {https://doi.org/10.1007/s10666-022-09867-w} {\bibfield  {journal} {\bibinfo
  {journal} {Environmental Modeling {\&} Assessment}\ } (\bibinfo {year}
  {2023}),\ 10.1007/s10666-022-09867-w}\BibitemShut {NoStop}%
\bibitem [{\citenamefont {Sutton}, \citenamefont {Barto},\ and\ \citenamefont
  {Williams}(1992)}]{SuttonBartoWilliams1992}%
  \BibitemOpen
  \bibfield  {author} {\bibinfo {author} {\bibfnamefont {R.}~\bibnamefont
  {Sutton}}, \bibinfo {author} {\bibfnamefont {A.}~\bibnamefont {Barto}},\ and\
  \bibinfo {author} {\bibfnamefont {R.}~\bibnamefont {Williams}},\ }\bibfield
  {title} {\enquote {\bibinfo {title} {Reinforcement learning is direct
  adaptive optimal control},}\ }\href {https://doi.org/10.1109/37.126844}
  {\bibfield  {journal} {\bibinfo  {journal} {IEEE Control Systems Magazine}\
  }\textbf {\bibinfo {volume} {12}},\ \bibinfo {pages} {19--22} (\bibinfo
  {year} {1992})}\BibitemShut {NoStop}%
\bibitem [{\citenamefont {Bellman}(1957)}]{Bellman1957}%
  \BibitemOpen
  \bibfield  {author} {\bibinfo {author} {\bibfnamefont {R.}~\bibnamefont
  {Bellman}},\ }\href@noop {} {\emph {\bibinfo {title} {Dynamic Programming}}}\
  (\bibinfo  {publisher} {Princeton University Press},\ \bibinfo {year}
  {1957})\BibitemShut {NoStop}%
\bibitem [{\citenamefont {Botta}\ \emph {et~al.}(2017)\citenamefont {Botta},
  \citenamefont {Jansson}, \citenamefont {Ionescu}, \citenamefont
  {Christiansen},\ and\ \citenamefont {Brady}}]{botta+al2014a}%
  \BibitemOpen
  \bibfield  {author} {\bibinfo {author} {\bibfnamefont {N.}~\bibnamefont
  {Botta}}, \bibinfo {author} {\bibfnamefont {P.}~\bibnamefont {Jansson}},
  \bibinfo {author} {\bibfnamefont {C.}~\bibnamefont {Ionescu}}, \bibinfo
  {author} {\bibfnamefont {D.~R.}\ \bibnamefont {Christiansen}},\ and\ \bibinfo
  {author} {\bibfnamefont {E.}~\bibnamefont {Brady}},\ }\bibfield  {title}
  {\enquote {\bibinfo {title} {Sequential decision problems, dependent types
  and generic solutions},}\ }\href {https://doi.org/10.23638/LMCS-13(1:7)2017}
  {\bibfield  {journal} {\bibinfo  {journal} {Logical Methods in Computer
  Science}\ }\textbf {\bibinfo {volume} {13}} (\bibinfo {year} {2017}),\
  10.23638/LMCS-13(1:7)2017}\BibitemShut {NoStop}%
\bibitem [{\citenamefont {Brede}\ and\ \citenamefont
  {Botta}(2021)}]{brede_botta_2021}%
  \BibitemOpen
  \bibfield  {author} {\bibinfo {author} {\bibfnamefont {N.}~\bibnamefont
  {Brede}}\ and\ \bibinfo {author} {\bibfnamefont {N.}~\bibnamefont {Botta}},\
  }\bibfield  {title} {\enquote {\bibinfo {title} {On the correctness of
  monadic backward induction},}\ }\href
  {https://doi.org/10.1017/S0956796821000228} {\bibfield  {journal} {\bibinfo
  {journal} {Journal of Functional Programming}\ }\textbf {\bibinfo {volume}
  {31}},\ \bibinfo {pages} {e26} (\bibinfo {year} {2021})}\BibitemShut
  {NoStop}%
\bibitem [{\citenamefont {Watkins}\ and\ \citenamefont
  {Dayan}(1992)}]{Watkins1992}%
  \BibitemOpen
  \bibfield  {author} {\bibinfo {author} {\bibfnamefont {C.~J. C.~H.}\
  \bibnamefont {Watkins}}\ and\ \bibinfo {author} {\bibfnamefont
  {P.}~\bibnamefont {Dayan}},\ }\bibfield  {title} {\enquote {\bibinfo {title}
  {Q-learning},}\ }\href {https://doi.org/10.1007/BF00992698} {\bibfield
  {journal} {\bibinfo  {journal} {Machine Learning}\ }\textbf {\bibinfo
  {volume} {8}},\ \bibinfo {pages} {279--292} (\bibinfo {year}
  {1992})}\BibitemShut {NoStop}%
\bibitem [{\citenamefont {Nordstr\"om}, \citenamefont {Petersson},\ and\
  \citenamefont {Smith}(1990)}]{nordstrom+petersson1990}%
  \BibitemOpen
  \bibfield  {author} {\bibinfo {author} {\bibfnamefont {B.}~\bibnamefont
  {Nordstr\"om}}, \bibinfo {author} {\bibfnamefont {K.}~\bibnamefont
  {Petersson}},\ and\ \bibinfo {author} {\bibfnamefont {J.}~\bibnamefont
  {Smith}},\ }\href@noop {} {\emph {\bibinfo {title} {{Programming in
  Martin-L\"of's Type Theory}}}}\ (\bibinfo  {publisher} {Oxford University
  Press},\ \bibinfo {year} {1990})\BibitemShut {NoStop}%
\bibitem [{\citenamefont {Martin-Löf}(1984)}]{martin-lf_intuitionistic_1984}%
  \BibitemOpen
  \bibfield  {author} {\bibinfo {author} {\bibfnamefont {P.}~\bibnamefont
  {Martin-Löf}},\ }\href@noop {} {\emph {\bibinfo {title} {Intuitionistic type
  theory}}}\ (\bibinfo  {publisher} {Bibliopolis},\ \bibinfo {year}
  {1984})\BibitemShut {NoStop}%
\bibitem [{\citenamefont {Allen}\ \emph {et~al.}(2006)\citenamefont {Allen},
  \citenamefont {Bickford}, \citenamefont {Constable}, \citenamefont {Eaton},
  \citenamefont {Kreitz}, \citenamefont {Lorigo},\ and\ \citenamefont
  {Moran}}]{ABCEKLM06}%
  \BibitemOpen
  \bibfield  {author} {\bibinfo {author} {\bibfnamefont {S.}~\bibnamefont
  {Allen}}, \bibinfo {author} {\bibfnamefont {M.}~\bibnamefont {Bickford}},
  \bibinfo {author} {\bibfnamefont {R.}~\bibnamefont {Constable}}, \bibinfo
  {author} {\bibfnamefont {R.}~\bibnamefont {Eaton}}, \bibinfo {author}
  {\bibfnamefont {C.}~\bibnamefont {Kreitz}}, \bibinfo {author} {\bibfnamefont
  {L.}~\bibnamefont {Lorigo}},\ and\ \bibinfo {author} {\bibfnamefont
  {E.}~\bibnamefont {Moran}},\ }\bibfield  {title} {\enquote {\bibinfo {title}
  {Innovations in computational type theory using {NuPRL}},}\ }\href@noop {}
  {\bibfield  {journal} {\bibinfo  {journal} {Journal of Applied Logic}\
  }\textbf {\bibinfo {volume} {4}},\ \bibinfo {pages} {428 -- 469} (\bibinfo
  {year} {2006})}\BibitemShut {NoStop}%
\bibitem [{\citenamefont {{The Coq Development
  Team}}(sion)}]{the_coq_development_team_zenodo}%
  \BibitemOpen
  \bibfield  {author} {\bibinfo {author} {\bibnamefont {{The Coq Development
  Team}}},\ }\href {https://doi.org/10.5281/zenodo.1003420} {\enquote {\bibinfo
  {title} {The coq proof assistant},}\ }\bibinfo {howpublished} {Zenodo:
  https://doi.org/10.5281/zenodo.1003420} (\bibinfo {year} {latest
  version})\BibitemShut {NoStop}%
\bibitem [{\citenamefont {Norell}(2007)}]{norell2007}%
  \BibitemOpen
  \bibfield  {author} {\bibinfo {author} {\bibfnamefont {U.}~\bibnamefont
  {Norell}},\ }\emph {\bibinfo {title} {Towards a practical programming
  language based on dependent type theory}},\ \href
  {http://citeseerx.ist.psu.edu/viewdoc/download?doi=10.1.1.65.7934&amp;rep=rep1&amp;type=pdf}
  {Ph.D. thesis},\ \bibinfo  {school} {Chalmers University of Technology}
  (\bibinfo {year} {2007})\BibitemShut {NoStop}%
\bibitem [{\citenamefont {Brady}(2017)}]{idrisbook}%
  \BibitemOpen
  \bibfield  {author} {\bibinfo {author} {\bibfnamefont {E.}~\bibnamefont
  {Brady}},\ }\href@noop {} {\emph {\bibinfo {title} {Type-{D}riven
  {D}evelopment in {I}dris}}}\ (\bibinfo  {publisher} {Manning Publications
  Co.},\ \bibinfo {year} {2017})\BibitemShut {NoStop}%
\bibitem [{\citenamefont {{de Moura}}\ \emph {et~al.}(2015)\citenamefont {{de
  Moura}}, \citenamefont {Kong}, \citenamefont {Avigad}, \citenamefont {van
  Doorn},\ and\ \citenamefont {von Raumer}}]{botta1}%
  \BibitemOpen
  \bibfield  {author} {\bibinfo {author} {\bibfnamefont {L.}~\bibnamefont {{de
  Moura}}}, \bibinfo {author} {\bibfnamefont {S.}~\bibnamefont {Kong}},
  \bibinfo {author} {\bibfnamefont {J.}~\bibnamefont {Avigad}}, \bibinfo
  {author} {\bibfnamefont {F.}~\bibnamefont {van Doorn}},\ and\ \bibinfo
  {author} {\bibfnamefont {J.}~\bibnamefont {von Raumer}},\ }\bibfield  {title}
  {\enquote {\bibinfo {title} {The {L}ean {T}heorem {P}rover ({S}ystem
  {D}escription)},}\ }in\ \href@noop {} {\emph {\bibinfo {booktitle}
  {{A}utomated {D}eduction - {CADE}-25}}}\ (\bibinfo  {publisher} {Springer
  International Publishing},\ \bibinfo {address} {Cham},\ \bibinfo {year}
  {2015})\ pp.\ \bibinfo {pages} {378--388}\BibitemShut {NoStop}%
\bibitem [{\citenamefont {Voevodsky}(2011)}]{voevodsky2011univalent}%
  \BibitemOpen
  \bibfield  {author} {\bibinfo {author} {\bibfnamefont {V.}~\bibnamefont
  {Voevodsky}},\ }\bibfield  {title} {\enquote {\bibinfo {title} {Univalent
  foundations of mathematics},}\ }in\ \href@noop {} {\emph {\bibinfo
  {booktitle} {Logic, Language, Information and Computation: 18th International
  Workshop, WoLLIC 2011, Philadelphia, PA, USA. Proceedings 18}}}\ (\bibinfo
  {organization} {Springer},\ \bibinfo {year} {2011})\ pp.\ \bibinfo {pages}
  {4--4}\BibitemShut {NoStop}%
\bibitem [{\citenamefont {Gonthier}\ \emph {et~al.}(2008)\citenamefont
  {Gonthier} \emph {et~al.}}]{gonthier2008formal}%
  \BibitemOpen
  \bibfield  {author} {\bibinfo {author} {\bibfnamefont {G.}~\bibnamefont
  {Gonthier}} \emph {et~al.},\ }\bibfield  {title} {\enquote {\bibinfo {title}
  {Formal proof--the four-color theorem},}\ }\href@noop {} {\bibfield
  {journal} {\bibinfo  {journal} {Notices of the AMS}\ }\textbf {\bibinfo
  {volume} {55}},\ \bibinfo {pages} {1382--1393} (\bibinfo {year}
  {2008})}\BibitemShut {NoStop}%
\bibitem [{\citenamefont {Gonthier}\ \emph {et~al.}(2013)\citenamefont
  {Gonthier}, \citenamefont {Asperti}, \citenamefont {Avigad}, \citenamefont
  {Bertot}, \citenamefont {Cohen}, \citenamefont {Garillot}, \citenamefont
  {Le~Roux}, \citenamefont {Mahboubi}, \citenamefont {O’Connor},
  \citenamefont {Ould~Biha} \emph {et~al.}}]{gonthier2013machine}%
  \BibitemOpen
  \bibfield  {author} {\bibinfo {author} {\bibfnamefont {G.}~\bibnamefont
  {Gonthier}}, \bibinfo {author} {\bibfnamefont {A.}~\bibnamefont {Asperti}},
  \bibinfo {author} {\bibfnamefont {J.}~\bibnamefont {Avigad}}, \bibinfo
  {author} {\bibfnamefont {Y.}~\bibnamefont {Bertot}}, \bibinfo {author}
  {\bibfnamefont {C.}~\bibnamefont {Cohen}}, \bibinfo {author} {\bibfnamefont
  {F.}~\bibnamefont {Garillot}}, \bibinfo {author} {\bibfnamefont
  {S.}~\bibnamefont {Le~Roux}}, \bibinfo {author} {\bibfnamefont
  {A.}~\bibnamefont {Mahboubi}}, \bibinfo {author} {\bibfnamefont
  {R.}~\bibnamefont {O’Connor}}, \bibinfo {author} {\bibfnamefont
  {S.}~\bibnamefont {Ould~Biha}}, \emph {et~al.},\ }\bibfield  {title}
  {\enquote {\bibinfo {title} {A machine-checked proof of the odd order
  theorem},}\ }in\ \href@noop {} {\emph {\bibinfo {booktitle} {Interactive
  Theorem Proving: 4th International Conference, ITP 2013, Rennes, France, July
  22-26, 2013. Proceedings 4}}}\ (\bibinfo {organization} {Springer},\ \bibinfo
  {year} {2013})\ pp.\ \bibinfo {pages} {163--179}\BibitemShut {NoStop}%
\bibitem [{\citenamefont {Buzzard}, \citenamefont {Commelin},\ and\
  \citenamefont {Massot}(2020)}]{buzzard2020formalising}%
  \BibitemOpen
  \bibfield  {author} {\bibinfo {author} {\bibfnamefont {K.}~\bibnamefont
  {Buzzard}}, \bibinfo {author} {\bibfnamefont {J.}~\bibnamefont {Commelin}},\
  and\ \bibinfo {author} {\bibfnamefont {P.}~\bibnamefont {Massot}},\
  }\bibfield  {title} {\enquote {\bibinfo {title} {Formalising perfectoid
  spaces},}\ }in\ \href@noop {} {\emph {\bibinfo {booktitle} {Proceedings of
  the 9th ACM SIGPLAN International Conference on Certified Programs and
  Proofs}}}\ (\bibinfo {year} {2020})\ pp.\ \bibinfo {pages}
  {299--312}\BibitemShut {NoStop}%
\bibitem [{\citenamefont {{Kevin
  Hartnett}}(2020)}]{the_mathematical_library_of_the_future}%
  \BibitemOpen
  \bibfield  {author} {\bibinfo {author} {\bibnamefont {{Kevin Hartnett}}},\
  }\href@noop {} {\enquote {\bibinfo {title} {Building the mathematical library
  of the future},}\ }\bibinfo {howpublished}
  {https://www.quantamagazine.org/building-the-mathematical-library-of-the-future-20201001}
  (\bibinfo {year} {2020})\BibitemShut {NoStop}%
\bibitem [{\citenamefont {{Stephen
  Ornes}}(2020)}]{how_close_are_computers_to_automating_mathematical_reasoning}%
  \BibitemOpen
  \bibfield  {author} {\bibinfo {author} {\bibnamefont {{Stephen Ornes}}},\
  }\href@noop {} {\enquote {\bibinfo {title} {How close are computers to
  automating mathematical reasoning},}\ }\bibinfo {howpublished}
  {https://www.quantamagazine.org/how-close-are-computers-to-automating-mathematical-reasoning-20200827/}
  (\bibinfo {year} {2020})\BibitemShut {NoStop}%
\bibitem [{\citenamefont {Wadler}(2015)}]{botta2}%
  \BibitemOpen
  \bibfield  {author} {\bibinfo {author} {\bibfnamefont {P.}~\bibnamefont
  {Wadler}},\ }\bibfield  {title} {\enquote {\bibinfo {title} {Propositions as
  types},}\ }\href {https://doi.org/10.1145/2699407} {\bibfield  {journal}
  {\bibinfo  {journal} {Commun. ACM}\ }\textbf {\bibinfo {volume} {58}},\
  \bibinfo {pages} {75–84} (\bibinfo {year} {2015})}\BibitemShut {NoStop}%
\bibitem [{\citenamefont {Leroy}(2009)}]{leroy_formal_2009}%
  \BibitemOpen
  \bibfield  {author} {\bibinfo {author} {\bibfnamefont {X.}~\bibnamefont
  {Leroy}},\ }\bibfield  {title} {\enquote {\bibinfo {title} {Formal
  verification of a realistic compiler},}\ }\href
  {https://doi.org/10.1145/1538788.1538814} {\bibfield  {journal} {\bibinfo
  {journal} {Communications of the ACM}\ }\textbf {\bibinfo {volume} {52}},\
  \bibinfo {pages} {107--115} (\bibinfo {year} {2009})}\BibitemShut {NoStop}%
\bibitem [{\citenamefont {Swamy}\ \emph {et~al.}(2011)\citenamefont {Swamy},
  \citenamefont {Chen}, \citenamefont {Fournet}, \citenamefont {Strub},
  \citenamefont {Bhargavan},\ and\ \citenamefont {Yang}}]{swamy_secure_2011}%
  \BibitemOpen
  \bibfield  {author} {\bibinfo {author} {\bibfnamefont {N.}~\bibnamefont
  {Swamy}}, \bibinfo {author} {\bibfnamefont {J.}~\bibnamefont {Chen}},
  \bibinfo {author} {\bibfnamefont {C.}~\bibnamefont {Fournet}}, \bibinfo
  {author} {\bibfnamefont {P.-Y.}\ \bibnamefont {Strub}}, \bibinfo {author}
  {\bibfnamefont {K.}~\bibnamefont {Bhargavan}},\ and\ \bibinfo {author}
  {\bibfnamefont {J.}~\bibnamefont {Yang}},\ }\bibfield  {title} {\enquote
  {\bibinfo {title} {Secure distributed programming with value-dependent
  types},}\ }in\ \href {https://doi.org/10.1145/2034773.2034811} {\emph
  {\bibinfo {booktitle} {Proc. of ICFP 2011}}}\ (\bibinfo {year} {2011})\ pp.\
  \bibinfo {pages} {266--278}\BibitemShut {NoStop}%
\bibitem [{\citenamefont {Morgenstern}\ and\ \citenamefont
  {Licata}(2010)}]{licata_2011}%
  \BibitemOpen
  \bibfield  {author} {\bibinfo {author} {\bibfnamefont {J.}~\bibnamefont
  {Morgenstern}}\ and\ \bibinfo {author} {\bibfnamefont {D.}~\bibnamefont
  {Licata}},\ }\bibfield  {title} {\enquote {\bibinfo {title} {Security-typed
  programming within dependently-typed programming},}\ }in\ \href
  {https://doi.org/10.1145/1863543.1863569} {\emph {\bibinfo {booktitle}
  {International Conference on Functional Programming}}}\ (\bibinfo
  {publisher} {ACM},\ \bibinfo {year} {2010})\BibitemShut {NoStop}%
\bibitem [{\citenamefont {Brady}\ and\ \citenamefont
  {Hammond}(2012)}]{brady_resource-safe_2012}%
  \BibitemOpen
  \bibfield  {author} {\bibinfo {author} {\bibfnamefont {E.}~\bibnamefont
  {Brady}}\ and\ \bibinfo {author} {\bibfnamefont {K.}~\bibnamefont
  {Hammond}},\ }\bibfield  {title} {\enquote {\bibinfo {title} {Resource-safe
  systems programming with embedded domain specific languages},}\ }in\ \href
  {https://doi.org/10.1007/978-3-642-27694-1_18} {\emph {\bibinfo {booktitle}
  {Practical Aspects of Declarative Languages}}}\ (\bibinfo  {publisher}
  {Springer},\ \bibinfo {year} {2012})\ pp.\ \bibinfo {pages}
  {242--257}\BibitemShut {NoStop}%
\bibitem [{\citenamefont {Chlipala}(2022)}]{chlipala2022certified}%
  \BibitemOpen
  \bibfield  {author} {\bibinfo {author} {\bibfnamefont {A.}~\bibnamefont
  {Chlipala}},\ }\href@noop {} {\emph {\bibinfo {title} {Certified programming
  with dependent types: a pragmatic introduction to the Coq proof assistant}}}\
  (\bibinfo  {publisher} {MIT Press},\ \bibinfo {year} {2022})\BibitemShut
  {NoStop}%
\bibitem [{\citenamefont {Ionescu}\ and\ \citenamefont
  {Jansson}(2013)}]{ionescujansson:LIPIcs:2013:3899}%
  \BibitemOpen
  \bibfield  {author} {\bibinfo {author} {\bibfnamefont {C.}~\bibnamefont
  {Ionescu}}\ and\ \bibinfo {author} {\bibfnamefont {P.}~\bibnamefont
  {Jansson}},\ }\bibfield  {title} {\enquote {\bibinfo {title} {Testing versus
  proving in climate impact research},}\ }in\ \href
  {https://doi.org/10.4230/LIPIcs.TYPES.2011.41} {\emph {\bibinfo {booktitle}
  {Proc. TYPES 2011}}},\ \bibinfo {series} {Leibniz International Proceedings
  in Informatics (LIPIcs)}, Vol.~\bibinfo {volume} {19}\ (\bibinfo  {publisher}
  {Schloss Dagstuhl--Leibniz-Zentrum fuer Informatik},\ \bibinfo {address}
  {Dagstuhl, Germany},\ \bibinfo {year} {2013})\ pp.\ \bibinfo {pages}
  {41--54}\BibitemShut {NoStop}%
\bibitem [{\citenamefont {Bird}\ and\ \citenamefont
  {de~Moor}(1997)}]{DBLP:books/daglib/0096998}%
  \BibitemOpen
  \bibfield  {author} {\bibinfo {author} {\bibfnamefont {R.~S.}\ \bibnamefont
  {Bird}}\ and\ \bibinfo {author} {\bibfnamefont {O.}~\bibnamefont {de~Moor}},\
  }\href@noop {} {\emph {\bibinfo {title} {Algebra of programming}}},\ Prentice
  Hall International series in computer science\ (\bibinfo  {publisher}
  {Prentice Hall},\ \bibinfo {year} {1997})\BibitemShut {NoStop}%
\bibitem [{DBL(1989)}]{DBLP:conf/lics/1989}%
  \BibitemOpen
  \href {https://ieeexplore.ieee.org/xpl/conhome/249/proceeding} {\emph
  {\bibinfo {title} {Proceedings of the Fourth Annual Symposium on Logic in
  Computer Science {(LICS} '89), Pacific Grove, California, USA, June 5-8,
  1989}}}\ (\bibinfo  {publisher} {{IEEE} Computer Society},\ \bibinfo {year}
  {1989})\BibitemShut {NoStop}%
\bibitem [{\citenamefont {MacLane}(1978)}]{maclane}%
  \BibitemOpen
  \bibfield  {author} {\bibinfo {author} {\bibfnamefont {S.}~\bibnamefont
  {MacLane}},\ }\href@noop {} {\emph {\bibinfo {title} {Categories for the
  {W}orking {M}athematician}}},\ \bibinfo {edition} {2nd}\ ed.,\ Graduate Texts
  in Mathematics\ (\bibinfo  {publisher} {Springer},\ \bibinfo {year}
  {1978})\BibitemShut {NoStop}%
\bibitem [{\citenamefont {Sch\"{o}ll}(2021)}]{SCHOELL21}%
  \BibitemOpen
  \bibfield  {author} {\bibinfo {author} {\bibfnamefont {E.}~\bibnamefont
  {Sch\"{o}ll}},\ }\bibfield  {title} {\enquote {\bibinfo {title} {Partial
  synchronization patterns in brain networks},}\ }\href
  {https://doi.org/10.1209/0295-5075/ac3b97} {\bibfield  {journal} {\bibinfo
  {journal} {Europhysics Letters}\ }\textbf {\bibinfo {volume} {136}},\
  \bibinfo {pages} {18001} (\bibinfo {year} {2021})}\BibitemShut {NoStop}%
\bibitem [{\citenamefont {Sch{\"o}ll}(2020)}]{SCH20}%
  \BibitemOpen
  \bibfield  {author} {\bibinfo {author} {\bibfnamefont {E.}~\bibnamefont
  {Sch{\"o}ll}},\ }\bibfield  {title} {\enquote {\bibinfo {title} {Chimeras in
  physics and biology: Synchronization and desynchronization of rhythms},}\
  }\href {https://doi.org/10.26164/leopoldina_10_00275} {\bibfield  {journal}
  {\bibinfo  {journal} {Nova Acta Leopoldina}\ }\textbf {\bibinfo {volume}
  {425}},\ \bibinfo {pages} {67--95} (\bibinfo {year} {2020})},\ \bibinfo
  {note} {invited contribution}\BibitemShut {NoStop}%
\bibitem [{\citenamefont {Sawicki}(2019)}]{SAW20}%
  \BibitemOpen
  \bibfield  {author} {\bibinfo {author} {\bibfnamefont {J.}~\bibnamefont
  {Sawicki}},\ }\href {https://doi.org/10.1007/978-3-030-34076-6_5} {\emph
  {\bibinfo {title} {Delay controlled partial synchronization in complex
  networks}}},\ Springer Theses\ (\bibinfo  {publisher} {Springer},\ \bibinfo
  {address} {Heidelberg},\ \bibinfo {year} {2019})\BibitemShut {NoStop}%
\bibitem [{\citenamefont {Zakharova}(2020)}]{ZAK20}%
  \BibitemOpen
  \bibfield  {author} {\bibinfo {author} {\bibfnamefont {A.}~\bibnamefont
  {Zakharova}},\ }\href {https://doi.org/10.1007/978-3-030-21714-3} {\emph
  {\bibinfo {title} {Chimera Patterns in Networks: Interplay between Dynamics,
  Structure, Noise, and Delay}}},\ Understanding Complex Systems\ (\bibinfo
  {publisher} {Springer},\ \bibinfo {address} {Cham},\ \bibinfo {year}
  {2020})\BibitemShut {NoStop}%
\bibitem [{\citenamefont {Maistrenko}, \citenamefont {Penkovsky},\ and\
  \citenamefont {Rosenblum}(2014)}]{MAI14a}%
  \BibitemOpen
  \bibfield  {author} {\bibinfo {author} {\bibfnamefont {Y.}~\bibnamefont
  {Maistrenko}}, \bibinfo {author} {\bibfnamefont {B.}~\bibnamefont
  {Penkovsky}},\ and\ \bibinfo {author} {\bibfnamefont {M.}~\bibnamefont
  {Rosenblum}},\ }\bibfield  {title} {\enquote {\bibinfo {title} {Solitary
  state at the edge of synchrony in ensembles with attractive and repulsive
  interactions},}\ }\href {https://doi.org/10.1103/physreve.89.060901}
  {\bibfield  {journal} {\bibinfo  {journal} {Phys. Rev. E}\ }\textbf {\bibinfo
  {volume} {89}},\ \bibinfo {pages} {060901} (\bibinfo {year}
  {2014})}\BibitemShut {NoStop}%
\bibitem [{\citenamefont {Jaros}, \citenamefont {Maistrenko},\ and\
  \citenamefont {Kapitaniak}(2015)}]{JAR15}%
  \BibitemOpen
  \bibfield  {author} {\bibinfo {author} {\bibfnamefont {P.}~\bibnamefont
  {Jaros}}, \bibinfo {author} {\bibfnamefont {Y.}~\bibnamefont {Maistrenko}},\
  and\ \bibinfo {author} {\bibfnamefont {T.}~\bibnamefont {Kapitaniak}},\
  }\bibfield  {title} {\enquote {\bibinfo {title} {{C}himera states on the
  route from coherence to rotating waves},}\ }\href
  {https://doi.org/10.1103/physreve.91.022907} {\bibfield  {journal} {\bibinfo
  {journal} {Phys. Rev. E}\ }\textbf {\bibinfo {volume} {91}},\ \bibinfo
  {pages} {022907} (\bibinfo {year} {2015})}\BibitemShut {NoStop}%
\bibitem [{\citenamefont {Semenov}\ \emph {et~al.}(2016)\citenamefont
  {Semenov}, \citenamefont {Zakharova}, \citenamefont {Maistrenko},\ and\
  \citenamefont {Sch{\"o}ll}}]{SEM15b}%
  \BibitemOpen
  \bibfield  {author} {\bibinfo {author} {\bibfnamefont {V.}~\bibnamefont
  {Semenov}}, \bibinfo {author} {\bibfnamefont {A.}~\bibnamefont {Zakharova}},
  \bibinfo {author} {\bibfnamefont {Y.}~\bibnamefont {Maistrenko}},\ and\
  \bibinfo {author} {\bibfnamefont {E.}~\bibnamefont {Sch{\"o}ll}},\ }\bibfield
   {title} {\enquote {\bibinfo {title} {Delayed-feedback chimera states: Forced
  multiclusters and stochastic resonance},}\ }\href
  {https://doi.org/10.1209/0295-5075/115/10005} {\bibfield  {journal} {\bibinfo
   {journal} {Europhys. Lett.}\ }\textbf {\bibinfo {volume} {115}},\ \bibinfo
  {pages} {10005} (\bibinfo {year} {2016})}\BibitemShut {NoStop}%
\bibitem [{\citenamefont {Jain}\ and\ \citenamefont {Krishna}(2001)}]{JAI01}%
  \BibitemOpen
  \bibfield  {author} {\bibinfo {author} {\bibfnamefont {S.}~\bibnamefont
  {Jain}}\ and\ \bibinfo {author} {\bibfnamefont {S.}~\bibnamefont {Krishna}},\
  }\bibfield  {title} {\enquote {\bibinfo {title} {A model for the emergence of
  cooperation, interdependence, and structure in evolving networks},}\ }\href
  {https://doi.org/10.1073/pnas.98.2.543} {\bibfield  {journal} {\bibinfo
  {journal} {Proc. Natl. Acad. Sci.}\ }\textbf {\bibinfo {volume} {98}},\
  \bibinfo {pages} {543--547} (\bibinfo {year} {2001})}\BibitemShut {NoStop}%
\bibitem [{\citenamefont {Gross}, \citenamefont {Dommar~D'Lima},\ and\
  \citenamefont {Blasius}(2006)}]{GRO06a}%
  \BibitemOpen
  \bibfield  {author} {\bibinfo {author} {\bibfnamefont {T.}~\bibnamefont
  {Gross}}, \bibinfo {author} {\bibfnamefont {C.~J.}\ \bibnamefont
  {Dommar~D'Lima}},\ and\ \bibinfo {author} {\bibfnamefont {B.}~\bibnamefont
  {Blasius}},\ }\bibfield  {title} {\enquote {\bibinfo {title} {Epidemic
  dynamics on an adaptive network},}\ }\href
  {https://doi.org/10.1103/PhysRevLett.96.208701} {\bibfield  {journal}
  {\bibinfo  {journal} {Phys. Rev. Lett.}\ }\textbf {\bibinfo {volume} {96}}
  (\bibinfo {year} {2006}),\ 10.1103/PhysRevLett.96.208701}\BibitemShut
  {NoStop}%
\bibitem [{\citenamefont {Gross}\ and\ \citenamefont
  {Blasius}(2008{\natexlab{a}})}]{GRO08a}%
  \BibitemOpen
  \bibfield  {author} {\bibinfo {author} {\bibfnamefont {T.}~\bibnamefont
  {Gross}}\ and\ \bibinfo {author} {\bibfnamefont {B.}~\bibnamefont
  {Blasius}},\ }\bibfield  {title} {\enquote {\bibinfo {title} {Adaptive
  coevolutionary networks: a review},}\ }\href
  {https://doi.org/10.1098/rsif.2007.1229} {\bibfield  {journal} {\bibinfo
  {journal} {J. R. Soc. Interface}\ }\textbf {\bibinfo {volume} {5}},\ \bibinfo
  {pages} {259--271} (\bibinfo {year} {2008}{\natexlab{a}})}\BibitemShut
  {NoStop}%
\bibitem [{\citenamefont {Guti{\'e}rrez}\ \emph {et~al.}(2011)\citenamefont
  {Guti{\'e}rrez}, \citenamefont {Amann}, \citenamefont {Assenza},
  \citenamefont {G\'omez-Garde\~nes}, \citenamefont {Latora},\ and\
  \citenamefont {Boccaletti}}]{GUT11}%
  \BibitemOpen
  \bibfield  {author} {\bibinfo {author} {\bibfnamefont {R.}~\bibnamefont
  {Guti{\'e}rrez}}, \bibinfo {author} {\bibfnamefont {A.}~\bibnamefont
  {Amann}}, \bibinfo {author} {\bibfnamefont {S.}~\bibnamefont {Assenza}},
  \bibinfo {author} {\bibfnamefont {J.}~\bibnamefont {G\'omez-Garde\~nes}},
  \bibinfo {author} {\bibfnamefont {V.}~\bibnamefont {Latora}},\ and\ \bibinfo
  {author} {\bibfnamefont {S.}~\bibnamefont {Boccaletti}},\ }\bibfield  {title}
  {\enquote {\bibinfo {title} {Emerging meso- and macroscales from
  synchronization of adaptive networks},}\ }\href@noop {} {\bibfield  {journal}
  {\bibinfo  {journal} {Phys. Rev. Lett.}\ }\textbf {\bibinfo {volume} {107}},\
  \bibinfo {pages} {234103} (\bibinfo {year} {2011})}\BibitemShut {NoStop}%
\bibitem [{\citenamefont {Zhang}\ \emph {et~al.}(2015)\citenamefont {Zhang},
  \citenamefont {Boccaletti}, \citenamefont {Guan},\ and\ \citenamefont
  {Liu}}]{ZHA15a}%
  \BibitemOpen
  \bibfield  {author} {\bibinfo {author} {\bibfnamefont {X.}~\bibnamefont
  {Zhang}}, \bibinfo {author} {\bibfnamefont {S.}~\bibnamefont {Boccaletti}},
  \bibinfo {author} {\bibfnamefont {S.}~\bibnamefont {Guan}},\ and\ \bibinfo
  {author} {\bibfnamefont {Z.}~\bibnamefont {Liu}},\ }\bibfield  {title}
  {\enquote {\bibinfo {title} {Explosive synchronization in adaptive and
  multilayer networks},}\ }\href
  {https://doi.org/10.1103/physrevlett.114.038701} {\bibfield  {journal}
  {\bibinfo  {journal} {Phys. Rev. Lett.}\ }\textbf {\bibinfo {volume} {114}},\
  \bibinfo {pages} {038701} (\bibinfo {year} {2015})}\BibitemShut {NoStop}%
\bibitem [{\citenamefont {Madadi~Asl}, \citenamefont {Valizadeh},\ and\
  \citenamefont {Tass}(2018)}]{ASL18a}%
  \BibitemOpen
  \bibfield  {author} {\bibinfo {author} {\bibfnamefont {M.}~\bibnamefont
  {Madadi~Asl}}, \bibinfo {author} {\bibfnamefont {A.}~\bibnamefont
  {Valizadeh}},\ and\ \bibinfo {author} {\bibfnamefont {P.~A.}\ \bibnamefont
  {Tass}},\ }\bibfield  {title} {\enquote {\bibinfo {title} {Dendritic and
  axonal propagation delays may shape neuronal networks with plastic
  synapses},}\ }\href {https://doi.org/10.3389/fphys.2018.01849} {\bibfield
  {journal} {\bibinfo  {journal} {Front. Physiol.}\ }\textbf {\bibinfo {volume}
  {9}},\ \bibinfo {pages} {1849} (\bibinfo {year} {2018})}\BibitemShut
  {NoStop}%
\bibitem [{\citenamefont {Kasatkin}\ and\ \citenamefont
  {Nekorkin}(2018{\natexlab{a}})}]{KAS18}%
  \BibitemOpen
  \bibfield  {author} {\bibinfo {author} {\bibfnamefont {D.~V.}\ \bibnamefont
  {Kasatkin}}\ and\ \bibinfo {author} {\bibfnamefont {V.~I.}\ \bibnamefont
  {Nekorkin}},\ }\bibfield  {title} {\enquote {\bibinfo {title}
  {Synchronization of chimera states in a multiplex system of phase oscillators
  with adaptive couplings},}\ }\href {https://doi.org/10.1063/1.5031681}
  {\bibfield  {journal} {\bibinfo  {journal} {Chaos}\ }\textbf {\bibinfo
  {volume} {28}},\ \bibinfo {pages} {093115} (\bibinfo {year}
  {2018}{\natexlab{a}})}\BibitemShut {NoStop}%
\bibitem [{\citenamefont {Kasatkin}\ and\ \citenamefont
  {Nekorkin}(2018{\natexlab{b}})}]{KAS18a}%
  \BibitemOpen
  \bibfield  {author} {\bibinfo {author} {\bibfnamefont {D.~V.}\ \bibnamefont
  {Kasatkin}}\ and\ \bibinfo {author} {\bibfnamefont {V.~I.}\ \bibnamefont
  {Nekorkin}},\ }\bibfield  {title} {\enquote {\bibinfo {title} {The effect of
  topology on organization of synchronous behavior in dynamical networks with
  adaptive couplings},}\ }\href@noop {} {\bibfield  {journal} {\bibinfo
  {journal} {Eur. Phys. J. Spec. Top.}\ }\textbf {\bibinfo {volume} {227}},\
  \bibinfo {pages} {1051} (\bibinfo {year} {2018}{\natexlab{b}})}\BibitemShut
  {NoStop}%
\bibitem [{\citenamefont {Berner}, \citenamefont {Sawicki},\ and\ \citenamefont
  {Sch{\"o}ll}(2020)}]{BER20}%
  \BibitemOpen
  \bibfield  {author} {\bibinfo {author} {\bibfnamefont {R.}~\bibnamefont
  {Berner}}, \bibinfo {author} {\bibfnamefont {J.}~\bibnamefont {Sawicki}},\
  and\ \bibinfo {author} {\bibfnamefont {E.}~\bibnamefont {Sch{\"o}ll}},\
  }\bibfield  {title} {\enquote {\bibinfo {title} {Birth and stabilization of
  phase clusters by multiplexing of adaptive networks},}\ }\href
  {https://doi.org/10.1103/physrevlett.124.088301} {\bibfield  {journal}
  {\bibinfo  {journal} {Phys. Rev. Lett.}\ }\textbf {\bibinfo {volume} {124}},\
  \bibinfo {pages} {088301} (\bibinfo {year} {2020})}\BibitemShut {NoStop}%
\bibitem [{\citenamefont {Feketa}, \citenamefont {Schaum},\ and\ \citenamefont
  {Meurer}(2020)}]{FEK20}%
  \BibitemOpen
  \bibfield  {author} {\bibinfo {author} {\bibfnamefont {P.}~\bibnamefont
  {Feketa}}, \bibinfo {author} {\bibfnamefont {A.}~\bibnamefont {Schaum}},\
  and\ \bibinfo {author} {\bibfnamefont {T.}~\bibnamefont {Meurer}},\
  }\bibfield  {title} {\enquote {\bibinfo {title} {Synchronization and
  multi-cluster capabilities of oscillatory networks with adaptive coupling},}\
  }\href {https://doi.org/10.1109/tac.2020.3012528} {\bibfield  {journal}
  {\bibinfo  {journal} {IEEE Trans. Autom. Control}\ }\textbf {\bibinfo
  {volume} {66}},\ \bibinfo {pages} {3084--3096} (\bibinfo {year}
  {2020})}\BibitemShut {NoStop}%
\bibitem [{\citenamefont {Berner}(2021)}]{BER21c}%
  \BibitemOpen
  \bibfield  {author} {\bibinfo {author} {\bibfnamefont {R.}~\bibnamefont
  {Berner}},\ }\href {https://doi.org/10.1007/978-3-030-74938-5} {\emph
  {\bibinfo {title} {{Patterns of Synchrony in Complex Networks of Adaptively
  Coupled Oscillators}}}},\ Springer Theses\ (\bibinfo  {publisher}
  {Springer},\ \bibinfo {address} {Cham},\ \bibinfo {year} {2021})\BibitemShut
  {NoStop}%
\bibitem [{\citenamefont {Popovych}, \citenamefont {Xenakis},\ and\
  \citenamefont {Tass}(2015)}]{POP15}%
  \BibitemOpen
  \bibfield  {author} {\bibinfo {author} {\bibfnamefont {O.~V.}\ \bibnamefont
  {Popovych}}, \bibinfo {author} {\bibfnamefont {M.~N.}\ \bibnamefont
  {Xenakis}},\ and\ \bibinfo {author} {\bibfnamefont {P.~A.}\ \bibnamefont
  {Tass}},\ }\bibfield  {title} {\enquote {\bibinfo {title} {The spacing
  principle for unlearning abnormal neuronal synchrony},}\ }\href
  {https://doi.org/10.1371/journal.pone.0117205} {\bibfield  {journal}
  {\bibinfo  {journal} {PLoS ONE}\ }\textbf {\bibinfo {volume} {10}},\ \bibinfo
  {pages} {e0117205} (\bibinfo {year} {2015})}\BibitemShut {NoStop}%
\bibitem [{\citenamefont {Chakravartula}\ \emph {et~al.}(2017)\citenamefont
  {Chakravartula}, \citenamefont {Indic}, \citenamefont {Sundaram},\ and\
  \citenamefont {Killingback}}]{CHA17a}%
  \BibitemOpen
  \bibfield  {author} {\bibinfo {author} {\bibfnamefont {S.}~\bibnamefont
  {Chakravartula}}, \bibinfo {author} {\bibfnamefont {P.}~\bibnamefont
  {Indic}}, \bibinfo {author} {\bibfnamefont {B.}~\bibnamefont {Sundaram}},\
  and\ \bibinfo {author} {\bibfnamefont {T.}~\bibnamefont {Killingback}},\
  }\bibfield  {title} {\enquote {\bibinfo {title} {Emergence of local
  synchronization in neuronal networks with adaptive couplings},}\ }\href
  {https://doi.org/10.1371/journal.pone.0178975} {\bibfield  {journal}
  {\bibinfo  {journal} {PLoS ONE}\ }\textbf {\bibinfo {volume} {12}},\ \bibinfo
  {pages} {e0178975} (\bibinfo {year} {2017})}\BibitemShut {NoStop}%
\bibitem [{\citenamefont {Kuramoto}(1984)}]{KUR84}%
  \BibitemOpen
  \bibfield  {author} {\bibinfo {author} {\bibfnamefont {Y.}~\bibnamefont
  {Kuramoto}},\ }\href@noop {} {\emph {\bibinfo {title} {Chemical Oscillations,
  Waves and Turbulence}}}\ (\bibinfo  {publisher} {Springer-Verlag},\ \bibinfo
  {address} {Berlin},\ \bibinfo {year} {1984})\BibitemShut {NoStop}%
\bibitem [{\citenamefont {Filatrella}, \citenamefont {Nielsen},\ and\
  \citenamefont {Pedersen}(2008)}]{FIL08a}%
  \BibitemOpen
  \bibfield  {author} {\bibinfo {author} {\bibfnamefont {G.}~\bibnamefont
  {Filatrella}}, \bibinfo {author} {\bibfnamefont {A.~H.}\ \bibnamefont
  {Nielsen}},\ and\ \bibinfo {author} {\bibfnamefont {N.~F.}\ \bibnamefont
  {Pedersen}},\ }\bibfield  {title} {\enquote {\bibinfo {title} {Analysis of a
  power grid using a {K}uramoto-like model},}\ }\href@noop {} {\bibfield
  {journal} {\bibinfo  {journal} {Eur. Phys. J. B}\ }\textbf {\bibinfo {volume}
  {61}},\ \bibinfo {pages} {485--491} (\bibinfo {year} {2008})}\BibitemShut
  {NoStop}%
\bibitem [{\citenamefont {D{\"o}rfler}\ and\ \citenamefont
  {Bullo}(2012)}]{DOE12}%
  \BibitemOpen
  \bibfield  {author} {\bibinfo {author} {\bibfnamefont {F.}~\bibnamefont
  {D{\"o}rfler}}\ and\ \bibinfo {author} {\bibfnamefont {F.}~\bibnamefont
  {Bullo}},\ }\bibfield  {title} {\enquote {\bibinfo {title} {Synchronization
  and transient stability in power networks and nonuniform Kuramoto
  oscillators},}\ }\href {https://doi.org/10.1137/110851584} {\bibfield
  {journal} {\bibinfo  {journal} {SIAM J. Control Optim.}\ }\textbf {\bibinfo
  {volume} {50}},\ \bibinfo {pages} {1616--1642} (\bibinfo {year}
  {2012})}\BibitemShut {NoStop}%
\bibitem [{\citenamefont {Rohden}\ \emph {et~al.}(2012)\citenamefont {Rohden},
  \citenamefont {Sorge}, \citenamefont {Timme},\ and\ \citenamefont
  {Witthaut}}]{ROH12}%
  \BibitemOpen
  \bibfield  {author} {\bibinfo {author} {\bibfnamefont {M.}~\bibnamefont
  {Rohden}}, \bibinfo {author} {\bibfnamefont {A.}~\bibnamefont {Sorge}},
  \bibinfo {author} {\bibfnamefont {M.}~\bibnamefont {Timme}},\ and\ \bibinfo
  {author} {\bibfnamefont {D.}~\bibnamefont {Witthaut}},\ }\bibfield  {title}
  {\enquote {\bibinfo {title} {Self-organized synchronization in decentralized
  power grids},}\ }\href {https://doi.org/10.1103/physrevlett.109.064101}
  {\bibfield  {journal} {\bibinfo  {journal} {Phys. Rev. Lett.}\ }\textbf
  {\bibinfo {volume} {109}},\ \bibinfo {pages} {064101} (\bibinfo {year}
  {2012})}\BibitemShut {NoStop}%
\bibitem [{\citenamefont {Motter}\ \emph {et~al.}(2013)\citenamefont {Motter},
  \citenamefont {Myers}, \citenamefont {Anghel},\ and\ \citenamefont
  {Nishikawa}}]{MOT13a}%
  \BibitemOpen
  \bibfield  {author} {\bibinfo {author} {\bibfnamefont {A.~E.}\ \bibnamefont
  {Motter}}, \bibinfo {author} {\bibfnamefont {S.~A.}\ \bibnamefont {Myers}},
  \bibinfo {author} {\bibfnamefont {M.}~\bibnamefont {Anghel}},\ and\ \bibinfo
  {author} {\bibfnamefont {T.}~\bibnamefont {Nishikawa}},\ }\bibfield  {title}
  {\enquote {\bibinfo {title} {Spontaneous synchrony in power-grid networks},}\
  }\href {https://doi.org/doi:10.1038/nphys2535} {\bibfield  {journal}
  {\bibinfo  {journal} {Nat. Phys.}\ }\textbf {\bibinfo {volume} {9}},\
  \bibinfo {pages} {191--197} (\bibinfo {year} {2013})}\BibitemShut {NoStop}%
\bibitem [{\citenamefont {Rodrigues}\ \emph {et~al.}(2016)\citenamefont
  {Rodrigues}, \citenamefont {Peron}, \citenamefont {Ji},\ and\ \citenamefont
  {Kurths}}]{ROD16}%
  \BibitemOpen
  \bibfield  {author} {\bibinfo {author} {\bibfnamefont {F.~A.}\ \bibnamefont
  {Rodrigues}}, \bibinfo {author} {\bibfnamefont {T.~K. D.~M.}\ \bibnamefont
  {Peron}}, \bibinfo {author} {\bibfnamefont {P.}~\bibnamefont {Ji}},\ and\
  \bibinfo {author} {\bibfnamefont {J.}~\bibnamefont {Kurths}},\ }\bibfield
  {title} {\enquote {\bibinfo {title} {The {Kuramoto} model in complex
  networks},}\ }\href {https://doi.org/10.1016/j.physrep.2015.10.008}
  {\bibfield  {journal} {\bibinfo  {journal} {Phys. Rep.}\ }\textbf {\bibinfo
  {volume} {610}},\ \bibinfo {pages} {1--98} (\bibinfo {year}
  {2016})}\BibitemShut {NoStop}%
\bibitem [{\citenamefont {Tumash}, \citenamefont {Olmi},\ and\ \citenamefont
  {Sch{\"o}ll}(2018)}]{TUM18}%
  \BibitemOpen
  \bibfield  {author} {\bibinfo {author} {\bibfnamefont {L.}~\bibnamefont
  {Tumash}}, \bibinfo {author} {\bibfnamefont {S.}~\bibnamefont {Olmi}},\ and\
  \bibinfo {author} {\bibfnamefont {E.}~\bibnamefont {Sch{\"o}ll}},\ }\bibfield
   {title} {\enquote {\bibinfo {title} {{E}ffect of disorder and noise in
  shaping the dynamics of power grids},}\ }\href@noop {} {\bibfield  {journal}
  {\bibinfo  {journal} {Europhys. Lett.}\ }\textbf {\bibinfo {volume} {123}},\
  \bibinfo {pages} {20001} (\bibinfo {year} {2018})}\BibitemShut {NoStop}%
\bibitem [{\citenamefont {Tumash}, \citenamefont {Olmi},\ and\ \citenamefont
  {Sch{\"o}ll}(2019)}]{TUM19}%
  \BibitemOpen
  \bibfield  {author} {\bibinfo {author} {\bibfnamefont {L.}~\bibnamefont
  {Tumash}}, \bibinfo {author} {\bibfnamefont {S.}~\bibnamefont {Olmi}},\ and\
  \bibinfo {author} {\bibfnamefont {E.}~\bibnamefont {Sch{\"o}ll}},\ }\bibfield
   {title} {\enquote {\bibinfo {title} {{S}tability and control of power grids
  with diluted network topology},}\ }\href {https://doi.org/10.1063/1.5111686}
  {\bibfield  {journal} {\bibinfo  {journal} {Chaos}\ }\textbf {\bibinfo
  {volume} {29}},\ \bibinfo {pages} {123105} (\bibinfo {year}
  {2019})}\BibitemShut {NoStop}%
\bibitem [{\citenamefont {Taher}, \citenamefont {Olmi},\ and\ \citenamefont
  {Sch{\"o}ll}(2019)}]{TAH19}%
  \BibitemOpen
  \bibfield  {author} {\bibinfo {author} {\bibfnamefont {H.}~\bibnamefont
  {Taher}}, \bibinfo {author} {\bibfnamefont {S.}~\bibnamefont {Olmi}},\ and\
  \bibinfo {author} {\bibfnamefont {E.}~\bibnamefont {Sch{\"o}ll}},\ }\bibfield
   {title} {\enquote {\bibinfo {title} {Enhancing power grid synchronization
  and stability through time delayed feedback control},}\ }\href
  {https://doi.org/10.1103/physreve.100.062306} {\bibfield  {journal} {\bibinfo
   {journal} {Phys. Rev. E}\ }\textbf {\bibinfo {volume} {100}},\ \bibinfo
  {pages} {062306} (\bibinfo {year} {2019})}\BibitemShut {NoStop}%
\bibitem [{\citenamefont {Hellmann}\ \emph {et~al.}(2020)\citenamefont
  {Hellmann}, \citenamefont {Schultz}, \citenamefont {Jaros}, \citenamefont
  {Levchenko}, \citenamefont {Kapitaniak}, \citenamefont {Kurths},\ and\
  \citenamefont {Maistrenko}}]{HEL20}%
  \BibitemOpen
  \bibfield  {author} {\bibinfo {author} {\bibfnamefont {F.}~\bibnamefont
  {Hellmann}}, \bibinfo {author} {\bibfnamefont {P.}~\bibnamefont {Schultz}},
  \bibinfo {author} {\bibfnamefont {P.}~\bibnamefont {Jaros}}, \bibinfo
  {author} {\bibfnamefont {R.}~\bibnamefont {Levchenko}}, \bibinfo {author}
  {\bibfnamefont {T.}~\bibnamefont {Kapitaniak}}, \bibinfo {author}
  {\bibfnamefont {J.}~\bibnamefont {Kurths}},\ and\ \bibinfo {author}
  {\bibfnamefont {Y.}~\bibnamefont {Maistrenko}},\ }\bibfield  {title}
  {\enquote {\bibinfo {title} {Network-induced multistability through lossy
  coupling and exotic solitary states},}\ }\href
  {https://doi.org/10.1038/s41467-020-14417-7} {\bibfield  {journal} {\bibinfo
  {journal} {Nat. Commun.}\ }\textbf {\bibinfo {volume} {11}},\ \bibinfo
  {pages} {592} (\bibinfo {year} {2020})}\BibitemShut {NoStop}%
\bibitem [{\citenamefont {Kuehn}\ and\ \citenamefont {Throm}(2019)}]{KUE19}%
  \BibitemOpen
  \bibfield  {author} {\bibinfo {author} {\bibfnamefont {C.}~\bibnamefont
  {Kuehn}}\ and\ \bibinfo {author} {\bibfnamefont {S.}~\bibnamefont {Throm}},\
  }\bibfield  {title} {\enquote {\bibinfo {title} {Power network dynamics on
  graphons},}\ }\href {https://doi.org/10.1137/18m1200002} {\bibfield
  {journal} {\bibinfo  {journal} {SIAM J. Appl. Math.}\ }\textbf {\bibinfo
  {volume} {79}},\ \bibinfo {pages} {1271--1292} (\bibinfo {year}
  {2019})}\BibitemShut {NoStop}%
\bibitem [{\citenamefont {Totz}, \citenamefont {Olmi},\ and\ \citenamefont
  {Sch{\"o}ll}(2020)}]{TOT20}%
  \BibitemOpen
  \bibfield  {author} {\bibinfo {author} {\bibfnamefont {C.~H.}\ \bibnamefont
  {Totz}}, \bibinfo {author} {\bibfnamefont {S.}~\bibnamefont {Olmi}},\ and\
  \bibinfo {author} {\bibfnamefont {E.}~\bibnamefont {Sch{\"o}ll}},\ }\bibfield
   {title} {\enquote {\bibinfo {title} {Control of synchronization in two-layer
  power grids},}\ }\href@noop {} {\bibfield  {journal} {\bibinfo  {journal}
  {Phys. Rev. E}\ }\textbf {\bibinfo {volume} {102}},\ \bibinfo {pages}
  {022311} (\bibinfo {year} {2020})}\BibitemShut {NoStop}%
\bibitem [{\citenamefont {Zhang}, \citenamefont {Ma},\ and\ \citenamefont
  {Timme}(2020)}]{ZHA20c}%
  \BibitemOpen
  \bibfield  {author} {\bibinfo {author} {\bibfnamefont {X.}~\bibnamefont
  {Zhang}}, \bibinfo {author} {\bibfnamefont {C.}~\bibnamefont {Ma}},\ and\
  \bibinfo {author} {\bibfnamefont {M.}~\bibnamefont {Timme}},\ }\bibfield
  {title} {\enquote {\bibinfo {title} {Vulnerability in dynamically driven
  oscillatory networks and power grids},}\ }\href
  {https://doi.org/10.1063/1.5122963} {\bibfield  {journal} {\bibinfo
  {journal} {Chaos}\ }\textbf {\bibinfo {volume} {30}},\ \bibinfo {pages}
  {063111} (\bibinfo {year} {2020})}\BibitemShut {NoStop}%
\bibitem [{\citenamefont {Berner}, \citenamefont {Yanchuk},\ and\ \citenamefont
  {Sch{\"o}ll}(2021)}]{BER21a}%
  \BibitemOpen
  \bibfield  {author} {\bibinfo {author} {\bibfnamefont {R.}~\bibnamefont
  {Berner}}, \bibinfo {author} {\bibfnamefont {S.}~\bibnamefont {Yanchuk}},\
  and\ \bibinfo {author} {\bibfnamefont {E.}~\bibnamefont {Sch{\"o}ll}},\
  }\bibfield  {title} {\enquote {\bibinfo {title} {What adaptive neuronal
  networks teach us about power grids},}\ }\href
  {https://doi.org/10.1103/physreve.103.042315} {\bibfield  {journal} {\bibinfo
   {journal} {Phys. Rev. E}\ }\textbf {\bibinfo {volume} {103}},\ \bibinfo
  {pages} {042315} (\bibinfo {year} {2021})}\BibitemShut {NoStop}%
\bibitem [{\citenamefont {Acebr{\'o}n}\ and\ \citenamefont
  {Spigler}(1998)}]{ACE98}%
  \BibitemOpen
  \bibfield  {author} {\bibinfo {author} {\bibfnamefont {J.~A.}\ \bibnamefont
  {Acebr{\'o}n}}\ and\ \bibinfo {author} {\bibfnamefont {R.}~\bibnamefont
  {Spigler}},\ }\bibfield  {title} {\enquote {\bibinfo {title} {Adaptive
  frequency model for phase-frequency synchronization in large populations of
  globally coupled nonlinear oscillators},}\ }\href
  {https://doi.org/10.1103/physrevlett.81.2229} {\bibfield  {journal} {\bibinfo
   {journal} {Phys. Rev. Lett.}\ }\textbf {\bibinfo {volume} {81}},\ \bibinfo
  {pages} {2229} (\bibinfo {year} {1998})}\BibitemShut {NoStop}%
\bibitem [{\citenamefont {Acebr{\'o}n}\ \emph {et~al.}(2005)\citenamefont
  {Acebr{\'o}n}, \citenamefont {Bonilla}, \citenamefont {P\'{e}rez~Vicente},
  \citenamefont {Ritort},\ and\ \citenamefont {Spigler}}]{ACE05}%
  \BibitemOpen
  \bibfield  {author} {\bibinfo {author} {\bibfnamefont {J.~A.}\ \bibnamefont
  {Acebr{\'o}n}}, \bibinfo {author} {\bibfnamefont {L.~L.}\ \bibnamefont
  {Bonilla}}, \bibinfo {author} {\bibfnamefont {C.~J.}\ \bibnamefont
  {P\'{e}rez~Vicente}}, \bibinfo {author} {\bibfnamefont {F.}~\bibnamefont
  {Ritort}},\ and\ \bibinfo {author} {\bibfnamefont {R.}~\bibnamefont
  {Spigler}},\ }\bibfield  {title} {\enquote {\bibinfo {title} {The {K}uramoto
  model: A simple paradigm for synchronization phenomena},}\ }\href
  {https://doi.org/10.1103/revmodphys.77.137} {\bibfield  {journal} {\bibinfo
  {journal} {Rev. Mod. Phys.}\ }\textbf {\bibinfo {volume} {77}},\ \bibinfo
  {pages} {137--185} (\bibinfo {year} {2005})}\BibitemShut {NoStop}%
\bibitem [{\citenamefont {Skardal}, \citenamefont {Taylor},\ and\ \citenamefont
  {Restrepo}(2013)}]{SKA13a}%
  \BibitemOpen
  \bibfield  {author} {\bibinfo {author} {\bibfnamefont {P.~S.}\ \bibnamefont
  {Skardal}}, \bibinfo {author} {\bibfnamefont {D.}~\bibnamefont {Taylor}},\
  and\ \bibinfo {author} {\bibfnamefont {J.~G.}\ \bibnamefont {Restrepo}},\
  }\bibfield  {title} {\enquote {\bibinfo {title} {Complex macroscopic behavior
  in systems of phase oscillators with adaptive coupling},}\ }\href
  {https://doi.org/10.1016/j.physd.2013.01.012} {\bibfield  {journal} {\bibinfo
   {journal} {Physica D}\ }\textbf {\bibinfo {volume} {267}},\ \bibinfo {pages}
  {27--35} (\bibinfo {year} {2013})}\BibitemShut {NoStop}%
\bibitem [{\citenamefont {Sch\"afer}\ \emph {et~al.}(2018)\citenamefont
  {Sch\"afer}, \citenamefont {Witthaut}, \citenamefont {Timme},\ and\
  \citenamefont {Latora}}]{SCH18i}%
  \BibitemOpen
  \bibfield  {author} {\bibinfo {author} {\bibfnamefont {B.}~\bibnamefont
  {Sch\"afer}}, \bibinfo {author} {\bibfnamefont {D.}~\bibnamefont {Witthaut}},
  \bibinfo {author} {\bibfnamefont {M.}~\bibnamefont {Timme}},\ and\ \bibinfo
  {author} {\bibfnamefont {V.}~\bibnamefont {Latora}},\ }\bibfield  {title}
  {\enquote {\bibinfo {title} {Dynamically induced cascading failures in power
  grids},}\ }\href {https://doi.org/10.1038/s41467-018-04287-5} {\bibfield
  {journal} {\bibinfo  {journal} {Nat. Commun.}\ }\textbf {\bibinfo {volume}
  {9}},\ \bibinfo {pages} {1975} (\bibinfo {year} {2018})}\BibitemShut
  {NoStop}%
\bibitem [{\citenamefont {Berner}\ \emph {et~al.}(2020)\citenamefont {Berner},
  \citenamefont {Polanska}, \citenamefont {Sch{\"o}ll},\ and\ \citenamefont
  {Yanchuk}}]{BER20c}%
  \BibitemOpen
  \bibfield  {author} {\bibinfo {author} {\bibfnamefont {R.}~\bibnamefont
  {Berner}}, \bibinfo {author} {\bibfnamefont {A.}~\bibnamefont {Polanska}},
  \bibinfo {author} {\bibfnamefont {E.}~\bibnamefont {Sch{\"o}ll}},\ and\
  \bibinfo {author} {\bibfnamefont {S.}~\bibnamefont {Yanchuk}},\ }\bibfield
  {title} {\enquote {\bibinfo {title} {Solitary states in adaptive nonlocal
  oscillator networks},}\ }\href
  {https://doi.org/https://doi.org/10.1140/epjst/e2020-900253-0} {\bibfield
  {journal} {\bibinfo  {journal} {Eur. Phys. J. Spec. Top.}\ }\textbf {\bibinfo
  {volume} {229}},\ \bibinfo {pages} {2183--2203} (\bibinfo {year}
  {2020})}\BibitemShut {NoStop}%
\bibitem [{\citenamefont {Sakaguchi}\ and\ \citenamefont
  {Kuramoto}(1986)}]{SAK86}%
  \BibitemOpen
  \bibfield  {author} {\bibinfo {author} {\bibfnamefont {H.}~\bibnamefont
  {Sakaguchi}}\ and\ \bibinfo {author} {\bibfnamefont {Y.}~\bibnamefont
  {Kuramoto}},\ }\bibfield  {title} {\enquote {\bibinfo {title} {A soluble
  active rotater model showing phase transitions via mutual entertainment},}\
  }\href@noop {} {\bibfield  {journal} {\bibinfo  {journal} {Prog. Theor.
  Phys}\ }\textbf {\bibinfo {volume} {76}},\ \bibinfo {pages} {576--581}
  (\bibinfo {year} {1986})}\BibitemShut {NoStop}%
\bibitem [{\citenamefont {Jaros}\ \emph {et~al.}(2018)\citenamefont {Jaros},
  \citenamefont {Brezetsky}, \citenamefont {Levchenko}, \citenamefont
  {Dudkowski}, \citenamefont {Kapitaniak},\ and\ \citenamefont
  {Maistrenko}}]{JAR18}%
  \BibitemOpen
  \bibfield  {author} {\bibinfo {author} {\bibfnamefont {P.}~\bibnamefont
  {Jaros}}, \bibinfo {author} {\bibfnamefont {S.}~\bibnamefont {Brezetsky}},
  \bibinfo {author} {\bibfnamefont {R.}~\bibnamefont {Levchenko}}, \bibinfo
  {author} {\bibfnamefont {D.}~\bibnamefont {Dudkowski}}, \bibinfo {author}
  {\bibfnamefont {T.}~\bibnamefont {Kapitaniak}},\ and\ \bibinfo {author}
  {\bibfnamefont {Y.}~\bibnamefont {Maistrenko}},\ }\bibfield  {title}
  {\enquote {\bibinfo {title} {Solitary states for coupled oscillators with
  inertia},}\ }\href@noop {} {\bibfield  {journal} {\bibinfo  {journal}
  {Chaos}\ }\textbf {\bibinfo {volume} {28}},\ \bibinfo {pages} {011103}
  (\bibinfo {year} {2018})}\BibitemShut {NoStop}%
\bibitem [{\citenamefont {Belykh}, \citenamefont {Brister},\ and\ \citenamefont
  {Belykh}(2016)}]{BEL16a}%
  \BibitemOpen
  \bibfield  {author} {\bibinfo {author} {\bibfnamefont {I.~V.}\ \bibnamefont
  {Belykh}}, \bibinfo {author} {\bibfnamefont {B.~N.}\ \bibnamefont
  {Brister}},\ and\ \bibinfo {author} {\bibfnamefont {V.~N.}\ \bibnamefont
  {Belykh}},\ }\bibfield  {title} {\enquote {\bibinfo {title} {Bistability of
  patterns of synchrony in {Kuramoto} oscillators with inertia},}\ }\href
  {https://doi.org/10.1063/1.4961435} {\bibfield  {journal} {\bibinfo
  {journal} {Chaos}\ }\textbf {\bibinfo {volume} {26}},\ \bibinfo {pages}
  {094822} (\bibinfo {year} {2016})}\BibitemShut {NoStop}%
\bibitem [{\citenamefont {Olmi}(2015)}]{OLM15a}%
  \BibitemOpen
  \bibfield  {author} {\bibinfo {author} {\bibfnamefont {S.}~\bibnamefont
  {Olmi}},\ }\bibfield  {title} {\enquote {\bibinfo {title} {Chimera states in
  coupled {Kuramoto} oscillators with inertia},}\ }\href
  {https://doi.org/10.1063/1.4938734} {\bibfield  {journal} {\bibinfo
  {journal} {Chaos}\ }\textbf {\bibinfo {volume} {25}},\ \bibinfo {pages}
  {123125} (\bibinfo {year} {2015})}\BibitemShut {NoStop}%
\bibitem [{\citenamefont {Olmi}\ \emph {et~al.}(2014)\citenamefont {Olmi},
  \citenamefont {Navas}, \citenamefont {Boccaletti},\ and\ \citenamefont
  {Torcini}}]{OLM14a}%
  \BibitemOpen
  \bibfield  {author} {\bibinfo {author} {\bibfnamefont {S.}~\bibnamefont
  {Olmi}}, \bibinfo {author} {\bibfnamefont {A.}~\bibnamefont {Navas}},
  \bibinfo {author} {\bibfnamefont {S.}~\bibnamefont {Boccaletti}},\ and\
  \bibinfo {author} {\bibfnamefont {A.}~\bibnamefont {Torcini}},\ }\bibfield
  {title} {\enquote {\bibinfo {title} {Hysteretic transitions in the {Kuramoto}
  model with inertia},}\ }\href {https://doi.org/10.1103/physreve.90.042905}
  {\bibfield  {journal} {\bibinfo  {journal} {Phys. Rev. E}\ }\textbf {\bibinfo
  {volume} {90}},\ \bibinfo {pages} {042905} (\bibinfo {year}
  {2014})}\BibitemShut {NoStop}%
\bibitem [{\citenamefont {Barr{\'e}}\ and\ \citenamefont
  {M{\'e}tivier}(2016)}]{BAR16a}%
  \BibitemOpen
  \bibfield  {author} {\bibinfo {author} {\bibfnamefont {J.}~\bibnamefont
  {Barr{\'e}}}\ and\ \bibinfo {author} {\bibfnamefont {D.}~\bibnamefont
  {M{\'e}tivier}},\ }\bibfield  {title} {\enquote {\bibinfo {title}
  {Bifurcations and singularities for coupled oscillators with inertia and
  frustration},}\ }\href {https://doi.org/10.1103/physrevlett.117.214102}
  {\bibfield  {journal} {\bibinfo  {journal} {Phys. Rev. Lett.}\ }\textbf
  {\bibinfo {volume} {117}},\ \bibinfo {pages} {214102} (\bibinfo {year}
  {2016})}\BibitemShut {NoStop}%
\bibitem [{\citenamefont {Sawicki}\ \emph {et~al.}(2022)\citenamefont
  {Sawicki}, \citenamefont {Berner}, \citenamefont {L{\"o}ser},\ and\
  \citenamefont {Sch{\"o}ll}}]{SAW21b}%
  \BibitemOpen
  \bibfield  {author} {\bibinfo {author} {\bibfnamefont {J.}~\bibnamefont
  {Sawicki}}, \bibinfo {author} {\bibfnamefont {R.}~\bibnamefont {Berner}},
  \bibinfo {author} {\bibfnamefont {T.}~\bibnamefont {L{\"o}ser}},\ and\
  \bibinfo {author} {\bibfnamefont {E.}~\bibnamefont {Sch{\"o}ll}},\ }\bibfield
   {title} {\enquote {\bibinfo {title} {Modelling tumor disease and sepsis by
  networks of adaptively coupled phase oscillators},}\ }\href
  {https://doi.org/10.3389/fnetp.2021.730385} {\bibfield  {journal} {\bibinfo
  {journal} {Front. Netw. Physiol.}\ }\textbf {\bibinfo {volume} {1}},\
  \bibinfo {pages} {730385} (\bibinfo {year} {2022})}\BibitemShut {NoStop}%
\bibitem [{\citenamefont {Berner}\ \emph {et~al.}(2022)\citenamefont {Berner},
  \citenamefont {Sawicki}, \citenamefont {Thiele}, \citenamefont {L{\"o}ser},\
  and\ \citenamefont {Sch{\"o}ll}}]{BER22}%
  \BibitemOpen
  \bibfield  {author} {\bibinfo {author} {\bibfnamefont {R.}~\bibnamefont
  {Berner}}, \bibinfo {author} {\bibfnamefont {J.}~\bibnamefont {Sawicki}},
  \bibinfo {author} {\bibfnamefont {M.}~\bibnamefont {Thiele}}, \bibinfo
  {author} {\bibfnamefont {T.}~\bibnamefont {L{\"o}ser}},\ and\ \bibinfo
  {author} {\bibfnamefont {E.}~\bibnamefont {Sch{\"o}ll}},\ }\bibfield  {title}
  {\enquote {\bibinfo {title} {Critical parameters in dynamic network modeling
  of sepsis},}\ }\href {https://doi.org/10.3389/fnetp.2022.904480} {\bibfield
  {journal} {\bibinfo  {journal} {Front. Netw. Physiol.}\ }\textbf {\bibinfo
  {volume} {2}},\ \bibinfo {pages} {904480} (\bibinfo {year}
  {2022})}\BibitemShut {NoStop}%
\bibitem [{\citenamefont {Tyson}, \citenamefont {Chen},\ and\ \citenamefont
  {Novak}(2003)}]{1}%
  \BibitemOpen
  \bibfield  {author} {\bibinfo {author} {\bibfnamefont {J.~J.}\ \bibnamefont
  {Tyson}}, \bibinfo {author} {\bibfnamefont {K.~C.}\ \bibnamefont {Chen}},\
  and\ \bibinfo {author} {\bibfnamefont {B.}~\bibnamefont {Novak}},\ }\bibfield
   {title} {\enquote {\bibinfo {title} {Sniffers, buzzers, toggles and
  blinkers: dynamics of regulatory and signaling pathways in the cell.}}\
  }\href {https://doi.org/10.1016/s0955-0674(03)00017-6} {\bibfield  {journal}
  {\bibinfo  {journal} {Curr Opin Cell Biol}\ }\textbf {\bibinfo {volume}
  {15}},\ \bibinfo {pages} {221--231} (\bibinfo {year} {2003})}\BibitemShut
  {NoStop}%
\bibitem [{\citenamefont {Alon}(2019)}]{2}%
  \BibitemOpen
  \bibfield  {author} {\bibinfo {author} {\bibfnamefont {U.}~\bibnamefont
  {Alon}},\ }\href@noop {} {\emph {\bibinfo {title} {An introduction to systems
  biology an introduction to systems biology}}},\ \bibinfo {edition} {2nd}\
  ed.,\ Chapman \& Hall/CRC Computational Biology Series\ (\bibinfo
  {publisher} {Taylor \& Francis},\ \bibinfo {address} {Philadelphia, PA},\
  \bibinfo {year} {2019})\BibitemShut {NoStop}%
\bibitem [{\citenamefont {Ferrell}(2021)}]{3}%
  \BibitemOpen
  \bibfield  {author} {\bibinfo {author} {\bibfnamefont {J.}~\bibnamefont
  {Ferrell}},\ }\href@noop {} {\emph {\bibinfo {title} {Systems biology of cell
  signaling}}}\ (\bibinfo  {publisher} {CRC Press},\ \bibinfo {address} {Boca
  Raton, FL},\ \bibinfo {year} {2021})\BibitemShut {NoStop}%
\bibitem [{\citenamefont {Kahn}\ \emph {et~al.}(1993)\citenamefont {Kahn},
  \citenamefont {Prigeon}, \citenamefont {McCulloch}, \citenamefont {Boyko},
  \citenamefont {Bergman}, \citenamefont {Schwartz}, \citenamefont {Neifing},
  \citenamefont {Ward}, \citenamefont {Beard},\ and\ \citenamefont
  {Palmer}}]{4}%
  \BibitemOpen
  \bibfield  {author} {\bibinfo {author} {\bibfnamefont {S.~E.}\ \bibnamefont
  {Kahn}}, \bibinfo {author} {\bibfnamefont {R.~L.}\ \bibnamefont {Prigeon}},
  \bibinfo {author} {\bibfnamefont {D.~K.}\ \bibnamefont {McCulloch}}, \bibinfo
  {author} {\bibfnamefont {E.~J.}\ \bibnamefont {Boyko}}, \bibinfo {author}
  {\bibfnamefont {R.~N.}\ \bibnamefont {Bergman}}, \bibinfo {author}
  {\bibfnamefont {M.~W.}\ \bibnamefont {Schwartz}}, \bibinfo {author}
  {\bibfnamefont {J.~L.}\ \bibnamefont {Neifing}}, \bibinfo {author}
  {\bibfnamefont {W.~K.}\ \bibnamefont {Ward}}, \bibinfo {author}
  {\bibfnamefont {J.~C.}\ \bibnamefont {Beard}},\ and\ \bibinfo {author}
  {\bibfnamefont {J.~P.}\ \bibnamefont {Palmer}},\ }\bibfield  {title}
  {\enquote {\bibinfo {title} {Quantification of the relationship between
  insulin sensitivity and beta-cell function in human subjects. evidence for a
  hyperbolic function.}}\ }\href {https://doi.org/10.2337/diab.42.11.1663}
  {\bibfield  {journal} {\bibinfo  {journal} {Diabetes}\ }\textbf {\bibinfo
  {volume} {42}},\ \bibinfo {pages} {1663--1672} (\bibinfo {year}
  {1993})}\BibitemShut {NoStop}%
\bibitem [{\citenamefont {Macnab}\ and\ \citenamefont {Koshland}(1972)}]{5}%
  \BibitemOpen
  \bibfield  {author} {\bibinfo {author} {\bibfnamefont {R.~M.}\ \bibnamefont
  {Macnab}}\ and\ \bibinfo {author} {\bibfnamefont {D.~E.~J.}\ \bibnamefont
  {Koshland}},\ }\bibfield  {title} {\enquote {\bibinfo {title} {The
  gradient-sensing mechanism in bacterial chemotaxis.}}\ }\href
  {https://doi.org/10.1073/pnas.69.9.2509} {\bibfield  {journal} {\bibinfo
  {journal} {Proc Natl Acad Sci U S A}\ }\textbf {\bibinfo {volume} {69}},\
  \bibinfo {pages} {2509--2512} (\bibinfo {year} {1972})}\BibitemShut {NoStop}%
\bibitem [{\citenamefont {Berg}\ and\ \citenamefont {Tedesco}(1975)}]{6}%
  \BibitemOpen
  \bibfield  {author} {\bibinfo {author} {\bibfnamefont {H.~C.}\ \bibnamefont
  {Berg}}\ and\ \bibinfo {author} {\bibfnamefont {P.~M.}\ \bibnamefont
  {Tedesco}},\ }\bibfield  {title} {\enquote {\bibinfo {title} {Transient
  response to chemotactic stimuli in escherichia coli.}}\ }\href
  {https://doi.org/10.1073/pnas.72.8.3235} {\bibfield  {journal} {\bibinfo
  {journal} {Proc Natl Acad Sci U S A}\ }\textbf {\bibinfo {volume} {72}},\
  \bibinfo {pages} {3235--3239} (\bibinfo {year} {1975})}\BibitemShut {NoStop}%
\bibitem [{\citenamefont {Yi}\ \emph {et~al.}(2000)\citenamefont {Yi},
  \citenamefont {Huang}, \citenamefont {Simon},\ and\ \citenamefont
  {Doyle}}]{7}%
  \BibitemOpen
  \bibfield  {author} {\bibinfo {author} {\bibfnamefont {T.~M.}\ \bibnamefont
  {Yi}}, \bibinfo {author} {\bibfnamefont {Y.}~\bibnamefont {Huang}}, \bibinfo
  {author} {\bibfnamefont {M.~I.}\ \bibnamefont {Simon}},\ and\ \bibinfo
  {author} {\bibfnamefont {J.}~\bibnamefont {Doyle}},\ }\bibfield  {title}
  {\enquote {\bibinfo {title} {Robust perfect adaptation in bacterial
  chemotaxis through integral feedback control.}}\ }\href
  {https://doi.org/10.1073/pnas.97.9.4649} {\bibfield  {journal} {\bibinfo
  {journal} {Proc Natl Acad Sci U S A}\ }\textbf {\bibinfo {volume} {97}},\
  \bibinfo {pages} {4649--4653} (\bibinfo {year} {2000})}\BibitemShut {NoStop}%
\bibitem [{\citenamefont {El-Samad}, \citenamefont {Goff},\ and\ \citenamefont
  {Khammash}(2002)}]{8}%
  \BibitemOpen
  \bibfield  {author} {\bibinfo {author} {\bibfnamefont {H.}~\bibnamefont
  {El-Samad}}, \bibinfo {author} {\bibfnamefont {J.~P.}\ \bibnamefont {Goff}},\
  and\ \bibinfo {author} {\bibfnamefont {M.}~\bibnamefont {Khammash}},\
  }\bibfield  {title} {\enquote {\bibinfo {title} {Calcium homeostasis and
  parturient hypocalcemia: an integral feedback perspective.}}\ }\href
  {https://doi.org/10.1006/jtbi.2001.2422} {\bibfield  {journal} {\bibinfo
  {journal} {J Theor Biol}\ }\textbf {\bibinfo {volume} {214}},\ \bibinfo
  {pages} {17--29} (\bibinfo {year} {2002})}\BibitemShut {NoStop}%
\bibitem [{\citenamefont {Briat}, \citenamefont {Gupta},\ and\ \citenamefont
  {Khammash}(2016)}]{9}%
  \BibitemOpen
  \bibfield  {author} {\bibinfo {author} {\bibfnamefont {C.}~\bibnamefont
  {Briat}}, \bibinfo {author} {\bibfnamefont {A.}~\bibnamefont {Gupta}},\ and\
  \bibinfo {author} {\bibfnamefont {M.}~\bibnamefont {Khammash}},\ }\bibfield
  {title} {\enquote {\bibinfo {title} {Antithetic integral feedback ensures
  robust perfect adaptation in noisy biomolecular networks.}}\ }\href
  {https://doi.org/10.1016/j.cels.2016.01.004} {\bibfield  {journal} {\bibinfo
  {journal} {Cell Syst}\ }\textbf {\bibinfo {volume} {2}},\ \bibinfo {pages}
  {15--26} (\bibinfo {year} {2016})}\BibitemShut {NoStop}%
\bibitem [{\citenamefont {Karin}\ \emph {et~al.}(2016)\citenamefont {Karin},
  \citenamefont {Swisa}, \citenamefont {Glaser}, \citenamefont {Dor},\ and\
  \citenamefont {Alon}}]{10}%
  \BibitemOpen
  \bibfield  {author} {\bibinfo {author} {\bibfnamefont {O.}~\bibnamefont
  {Karin}}, \bibinfo {author} {\bibfnamefont {A.}~\bibnamefont {Swisa}},
  \bibinfo {author} {\bibfnamefont {B.}~\bibnamefont {Glaser}}, \bibinfo
  {author} {\bibfnamefont {Y.}~\bibnamefont {Dor}},\ and\ \bibinfo {author}
  {\bibfnamefont {U.}~\bibnamefont {Alon}},\ }\bibfield  {title} {\enquote
  {\bibinfo {title} {Dynamical compensation in physiological circuits.}}\
  }\href {https://doi.org/10.15252/msb.20167216} {\bibfield  {journal}
  {\bibinfo  {journal} {Mol Syst Biol}\ }\textbf {\bibinfo {volume} {12}},\
  \bibinfo {pages} {886} (\bibinfo {year} {2016})}\BibitemShut {NoStop}%
\bibitem [{\citenamefont {Karin}\ \emph {et~al.}(2020)\citenamefont {Karin},
  \citenamefont {Raz}, \citenamefont {Tendler}, \citenamefont {Bar},
  \citenamefont {Korem~Kohanim}, \citenamefont {Milo},\ and\ \citenamefont
  {Alon}}]{11}%
  \BibitemOpen
  \bibfield  {author} {\bibinfo {author} {\bibfnamefont {O.}~\bibnamefont
  {Karin}}, \bibinfo {author} {\bibfnamefont {M.}~\bibnamefont {Raz}}, \bibinfo
  {author} {\bibfnamefont {A.}~\bibnamefont {Tendler}}, \bibinfo {author}
  {\bibfnamefont {A.}~\bibnamefont {Bar}}, \bibinfo {author} {\bibfnamefont
  {Y.}~\bibnamefont {Korem~Kohanim}}, \bibinfo {author} {\bibfnamefont
  {T.}~\bibnamefont {Milo}},\ and\ \bibinfo {author} {\bibfnamefont
  {U.}~\bibnamefont {Alon}},\ }\bibfield  {title} {\enquote {\bibinfo {title}
  {A new model for the hpa axis explains dysregulation of stress hormones on
  the timescale of weeks.}}\ }\href {https://doi.org/10.15252/msb.20209510}
  {\bibfield  {journal} {\bibinfo  {journal} {Mol Syst Biol}\ }\textbf
  {\bibinfo {volume} {16}},\ \bibinfo {pages} {e9510} (\bibinfo {year}
  {2020})}\BibitemShut {NoStop}%
\bibitem [{\citenamefont {Korem~Kohanim}\ \emph {et~al.}(2022)\citenamefont
  {Korem~Kohanim}, \citenamefont {Milo}, \citenamefont {Raz}, \citenamefont
  {Karin}, \citenamefont {Bar}, \citenamefont {Mayo}, \citenamefont
  {Mendelson~Cohen}, \citenamefont {Toledano},\ and\ \citenamefont
  {Alon}}]{12}%
  \BibitemOpen
  \bibfield  {author} {\bibinfo {author} {\bibfnamefont {Y.}~\bibnamefont
  {Korem~Kohanim}}, \bibinfo {author} {\bibfnamefont {T.}~\bibnamefont {Milo}},
  \bibinfo {author} {\bibfnamefont {M.}~\bibnamefont {Raz}}, \bibinfo {author}
  {\bibfnamefont {O.}~\bibnamefont {Karin}}, \bibinfo {author} {\bibfnamefont
  {A.}~\bibnamefont {Bar}}, \bibinfo {author} {\bibfnamefont {A.}~\bibnamefont
  {Mayo}}, \bibinfo {author} {\bibfnamefont {N.}~\bibnamefont
  {Mendelson~Cohen}}, \bibinfo {author} {\bibfnamefont {Y.}~\bibnamefont
  {Toledano}},\ and\ \bibinfo {author} {\bibfnamefont {U.}~\bibnamefont
  {Alon}},\ }\bibfield  {title} {\enquote {\bibinfo {title} {Dynamics of
  thyroid diseases and thyroid-axis gland masses.}}\ }\href
  {https://doi.org/10.15252/msb.202210919} {\bibfield  {journal} {\bibinfo
  {journal} {Mol Syst Biol}\ }\textbf {\bibinfo {volume} {18}},\ \bibinfo
  {pages} {e10919} (\bibinfo {year} {2022})}\BibitemShut {NoStop}%
\bibitem [{\citenamefont {Tendler}\ \emph {et~al.}(2021)\citenamefont
  {Tendler}, \citenamefont {Bar}, \citenamefont {Mendelsohn-Cohen},
  \citenamefont {Karin}, \citenamefont {Kohanim}, \citenamefont {Maimon},
  \citenamefont {Milo}, \citenamefont {Raz}, \citenamefont {Mayo},
  \citenamefont {Tanay},\ and\ \citenamefont {Alon}}]{13}%
  \BibitemOpen
  \bibfield  {author} {\bibinfo {author} {\bibfnamefont {A.}~\bibnamefont
  {Tendler}}, \bibinfo {author} {\bibfnamefont {A.}~\bibnamefont {Bar}},
  \bibinfo {author} {\bibfnamefont {N.}~\bibnamefont {Mendelsohn-Cohen}},
  \bibinfo {author} {\bibfnamefont {O.}~\bibnamefont {Karin}}, \bibinfo
  {author} {\bibfnamefont {Y.~K.}\ \bibnamefont {Kohanim}}, \bibinfo {author}
  {\bibfnamefont {L.}~\bibnamefont {Maimon}}, \bibinfo {author} {\bibfnamefont
  {T.}~\bibnamefont {Milo}}, \bibinfo {author} {\bibfnamefont {M.}~\bibnamefont
  {Raz}}, \bibinfo {author} {\bibfnamefont {A.}~\bibnamefont {Mayo}}, \bibinfo
  {author} {\bibfnamefont {A.}~\bibnamefont {Tanay}},\ and\ \bibinfo {author}
  {\bibfnamefont {U.}~\bibnamefont {Alon}},\ }\bibfield  {title} {\enquote
  {\bibinfo {title} {Hormone seasonality in medical records suggests circannual
  endocrine circuits},}\ }\href {https://doi.org/10.1073/pnas.2003926118}
  {\bibfield  {journal} {\bibinfo  {journal} {Proceedings of the National
  Academy of Sciences}\ }\textbf {\bibinfo {volume} {118}} (\bibinfo {year}
  {2021}),\ 10.1073/pnas.2003926118}\BibitemShut {NoStop}%
\bibitem [{\citenamefont {Lazova}\ \emph {et~al.}(2011)\citenamefont {Lazova},
  \citenamefont {Ahmed}, \citenamefont {Bellomo}, \citenamefont {Stocker},\
  and\ \citenamefont {Shimizu}}]{14}%
  \BibitemOpen
  \bibfield  {author} {\bibinfo {author} {\bibfnamefont {M.~D.}\ \bibnamefont
  {Lazova}}, \bibinfo {author} {\bibfnamefont {T.}~\bibnamefont {Ahmed}},
  \bibinfo {author} {\bibfnamefont {D.}~\bibnamefont {Bellomo}}, \bibinfo
  {author} {\bibfnamefont {R.}~\bibnamefont {Stocker}},\ and\ \bibinfo {author}
  {\bibfnamefont {T.~S.}\ \bibnamefont {Shimizu}},\ }\bibfield  {title}
  {\enquote {\bibinfo {title} {Response rescaling in bacterial chemotaxis},}\
  }\href {https://doi.org/10.1073/pnas.1108608108} {\bibfield  {journal}
  {\bibinfo  {journal} {Proceedings of the National Academy of Sciences}\
  }\textbf {\bibinfo {volume} {108}},\ \bibinfo {pages} {13870--13875}
  (\bibinfo {year} {2011})}\BibitemShut {NoStop}%
\bibitem [{\citenamefont {Larsch}\ \emph {et~al.}(2015)\citenamefont {Larsch},
  \citenamefont {Flavell}, \citenamefont {Liu}, \citenamefont {Gordus},
  \citenamefont {Albrecht},\ and\ \citenamefont {Bargmann}}]{15}%
  \BibitemOpen
  \bibfield  {author} {\bibinfo {author} {\bibfnamefont {J.}~\bibnamefont
  {Larsch}}, \bibinfo {author} {\bibfnamefont {S.~W.}\ \bibnamefont {Flavell}},
  \bibinfo {author} {\bibfnamefont {Q.}~\bibnamefont {Liu}}, \bibinfo {author}
  {\bibfnamefont {A.}~\bibnamefont {Gordus}}, \bibinfo {author} {\bibfnamefont
  {D.~R.}\ \bibnamefont {Albrecht}},\ and\ \bibinfo {author} {\bibfnamefont
  {C.~I.}\ \bibnamefont {Bargmann}},\ }\bibfield  {title} {\enquote {\bibinfo
  {title} {A circuit for gradient climbing in c. elegans chemotaxis.}}\ }\href
  {https://doi.org/10.1016/j.celrep.2015.08.032} {\bibfield  {journal}
  {\bibinfo  {journal} {Cell Rep}\ }\textbf {\bibinfo {volume} {12}},\ \bibinfo
  {pages} {1748--1760} (\bibinfo {year} {2015})}\BibitemShut {NoStop}%
\bibitem [{\citenamefont {Kamino}\ \emph {et~al.}(2017)\citenamefont {Kamino},
  \citenamefont {Kondo}, \citenamefont {Nakajima}, \citenamefont
  {Honda-Kitahara}, \citenamefont {Kaneko},\ and\ \citenamefont {Sawai}}]{16}%
  \BibitemOpen
  \bibfield  {author} {\bibinfo {author} {\bibfnamefont {K.}~\bibnamefont
  {Kamino}}, \bibinfo {author} {\bibfnamefont {Y.}~\bibnamefont {Kondo}},
  \bibinfo {author} {\bibfnamefont {A.}~\bibnamefont {Nakajima}}, \bibinfo
  {author} {\bibfnamefont {M.}~\bibnamefont {Honda-Kitahara}}, \bibinfo
  {author} {\bibfnamefont {K.}~\bibnamefont {Kaneko}},\ and\ \bibinfo {author}
  {\bibfnamefont {S.}~\bibnamefont {Sawai}},\ }\bibfield  {title} {\enquote
  {\bibinfo {title} {Fold-change detection and scale invariance of
  cell{\textendash}cell signaling in social amoeba},}\ }\href
  {https://doi.org/10.1073/pnas.1702181114} {\bibfield  {journal} {\bibinfo
  {journal} {Proceedings of the National Academy of Sciences}\ }\textbf
  {\bibinfo {volume} {114}} (\bibinfo {year} {2017}),\
  10.1073/pnas.1702181114}\BibitemShut {NoStop}%
\bibitem [{\citenamefont {Schultz}, \citenamefont {Dayan},\ and\ \citenamefont
  {Montague}(1997)}]{SchultzEtAl1997}%
  \BibitemOpen
  \bibfield  {author} {\bibinfo {author} {\bibfnamefont {W.}~\bibnamefont
  {Schultz}}, \bibinfo {author} {\bibfnamefont {P.}~\bibnamefont {Dayan}},\
  and\ \bibinfo {author} {\bibfnamefont {P.~R.}\ \bibnamefont {Montague}},\
  }\bibfield  {title} {\enquote {\bibinfo {title} {A {{Neural Substrate}} of
  {{Prediction}} and {{Reward}}},}\ }\href
  {https://doi.org/10.1126/science.275.5306.1593} {\bibfield  {journal}
  {\bibinfo  {journal} {Science}\ }\textbf {\bibinfo {volume} {275}},\ \bibinfo
  {pages} {1593--1599} (\bibinfo {year} {1997})}\BibitemShut {NoStop}%
\bibitem [{\citenamefont {Tobler}, \citenamefont {Fiorillo},\ and\
  \citenamefont {Schultz}(2005)}]{18}%
  \BibitemOpen
  \bibfield  {author} {\bibinfo {author} {\bibfnamefont {P.~N.}\ \bibnamefont
  {Tobler}}, \bibinfo {author} {\bibfnamefont {C.~D.}\ \bibnamefont
  {Fiorillo}},\ and\ \bibinfo {author} {\bibfnamefont {W.}~\bibnamefont
  {Schultz}},\ }\bibfield  {title} {\enquote {\bibinfo {title} {Adaptive coding
  of reward value by dopamine neurons.}}\ }\href
  {https://doi.org/10.1126/science.1105370} {\bibfield  {journal} {\bibinfo
  {journal} {Science}\ }\textbf {\bibinfo {volume} {307}},\ \bibinfo {pages}
  {1642--1645} (\bibinfo {year} {2005})}\BibitemShut {NoStop}%
\bibitem [{\citenamefont {Kim}\ \emph {et~al.}(2020{\natexlab{a}})\citenamefont
  {Kim}, \citenamefont {Malik}, \citenamefont {Mikhael}, \citenamefont {Bech},
  \citenamefont {Tsutsui-Kimura}, \citenamefont {Sun}, \citenamefont {Zhang},
  \citenamefont {Li}, \citenamefont {Watabe-Uchida}, \citenamefont {Gershman},\
  and\ \citenamefont {Uchida}}]{19}%
  \BibitemOpen
  \bibfield  {author} {\bibinfo {author} {\bibfnamefont {H.~R.}\ \bibnamefont
  {Kim}}, \bibinfo {author} {\bibfnamefont {A.~N.}\ \bibnamefont {Malik}},
  \bibinfo {author} {\bibfnamefont {J.~G.}\ \bibnamefont {Mikhael}}, \bibinfo
  {author} {\bibfnamefont {P.}~\bibnamefont {Bech}}, \bibinfo {author}
  {\bibfnamefont {I.}~\bibnamefont {Tsutsui-Kimura}}, \bibinfo {author}
  {\bibfnamefont {F.}~\bibnamefont {Sun}}, \bibinfo {author} {\bibfnamefont
  {Y.}~\bibnamefont {Zhang}}, \bibinfo {author} {\bibfnamefont
  {Y.}~\bibnamefont {Li}}, \bibinfo {author} {\bibfnamefont {M.}~\bibnamefont
  {Watabe-Uchida}}, \bibinfo {author} {\bibfnamefont {S.~J.}\ \bibnamefont
  {Gershman}},\ and\ \bibinfo {author} {\bibfnamefont {N.}~\bibnamefont
  {Uchida}},\ }\bibfield  {title} {\enquote {\bibinfo {title} {A unified
  framework for dopamine signals across timescales.}}\ }\href
  {https://doi.org/10.1016/j.cell.2020.11.013} {\bibfield  {journal} {\bibinfo
  {journal} {Cell}\ }\textbf {\bibinfo {volume} {183}},\ \bibinfo {pages}
  {1600--1616} (\bibinfo {year} {2020}{\natexlab{a}})}\BibitemShut {NoStop}%
\bibitem [{\citenamefont {Karin}\ and\ \citenamefont {Alon}(2022)}]{20}%
  \BibitemOpen
  \bibfield  {author} {\bibinfo {author} {\bibfnamefont {O.}~\bibnamefont
  {Karin}}\ and\ \bibinfo {author} {\bibfnamefont {U.}~\bibnamefont {Alon}},\
  }\bibfield  {title} {\enquote {\bibinfo {title} {The dopamine circuit as a
  reward-taxis navigation system},}\ }\href {https://doi.org/10.1371/journal.
  pcbi.1010340} {\bibfield  {journal} {\bibinfo  {journal} {PLOS Computational
  Biology}\ }\textbf {\bibinfo {volume} {18}},\ \bibinfo {pages} {1--24}
  (\bibinfo {year} {2022})}\BibitemShut {NoStop}%
\bibitem [{\citenamefont {Whitmire}\ and\ \citenamefont
  {Stanley}(2016)}]{Whitmire2016}%
  \BibitemOpen
  \bibfield  {author} {\bibinfo {author} {\bibfnamefont {C.~J.}\ \bibnamefont
  {Whitmire}}\ and\ \bibinfo {author} {\bibfnamefont {G.~B.}\ \bibnamefont
  {Stanley}},\ }\bibfield  {title} {\enquote {\bibinfo {title} {Rapid sensory
  adaptation redux: a circuit perspective},}\ }\href@noop {} {\bibfield
  {journal} {\bibinfo  {journal} {Neuron}\ }\textbf {\bibinfo {volume} {92}},\
  \bibinfo {pages} {298--315} (\bibinfo {year} {2016})}\BibitemShut {NoStop}%
\bibitem [{\citenamefont {Wu}, \citenamefont {Miehl},\ and\ \citenamefont
  {Gjorgjieva}(2022)}]{Wu2022}%
  \BibitemOpen
  \bibfield  {author} {\bibinfo {author} {\bibfnamefont {Y.~K.}\ \bibnamefont
  {Wu}}, \bibinfo {author} {\bibfnamefont {C.}~\bibnamefont {Miehl}},\ and\
  \bibinfo {author} {\bibfnamefont {J.}~\bibnamefont {Gjorgjieva}},\ }\bibfield
   {title} {\enquote {\bibinfo {title} {{Regulation of circuit organization and
  function through inhibitory synaptic plasticity}},}\ }\href@noop {}
  {\bibfield  {journal} {\bibinfo  {journal} {Trends in Neurosciences}\
  }\textbf {\bibinfo {volume} {45}},\ \bibinfo {pages} {884--898} (\bibinfo
  {year} {2022})}\BibitemShut {NoStop}%
\bibitem [{\citenamefont {Miehl}\ \emph {et~al.}(2022)\citenamefont {Miehl},
  \citenamefont {Onasch}, \citenamefont {Festa},\ and\ \citenamefont
  {Gjorgjieva}}]{Miehl2022b}%
  \BibitemOpen
  \bibfield  {author} {\bibinfo {author} {\bibfnamefont {C.}~\bibnamefont
  {Miehl}}, \bibinfo {author} {\bibfnamefont {S.}~\bibnamefont {Onasch}},
  \bibinfo {author} {\bibfnamefont {D.}~\bibnamefont {Festa}},\ and\ \bibinfo
  {author} {\bibfnamefont {J.}~\bibnamefont {Gjorgjieva}},\ }\bibfield  {title}
  {\enquote {\bibinfo {title} {{Formation and computational implications of
  assemblies in neural circuits}},}\ }\href
  {https://doi.org/https://doi.org/10.1113/JP282750} {\bibfield  {journal}
  {\bibinfo  {journal} {Journal of Physiology}\ } (\bibinfo {year} {2022}),\
  https://doi.org/10.1113/JP282750}\BibitemShut {NoStop}%
\bibitem [{\citenamefont {Debanne}, \citenamefont {Inglebert},\ and\
  \citenamefont {Russier}(2019)}]{Debanne2019}%
  \BibitemOpen
  \bibfield  {author} {\bibinfo {author} {\bibfnamefont {D.}~\bibnamefont
  {Debanne}}, \bibinfo {author} {\bibfnamefont {Y.}~\bibnamefont {Inglebert}},\
  and\ \bibinfo {author} {\bibfnamefont {M.}~\bibnamefont {Russier}},\
  }\bibfield  {title} {\enquote {\bibinfo {title} {{Plasticity of intrinsic
  neuronal excitability}},}\ }\href
  {https://doi.org/https://doi.org/10.1016/j.conb.2018.09.001} {\bibfield
  {journal} {\bibinfo  {journal} {Current Opinion in Neurobiology}\ }\textbf
  {\bibinfo {volume} {54}},\ \bibinfo {pages} {73--82} (\bibinfo {year}
  {2019})}\BibitemShut {NoStop}%
\bibitem [{\citenamefont {Zucker}\ and\ \citenamefont
  {Regehr}(2002)}]{Zucker2002}%
  \BibitemOpen
  \bibfield  {author} {\bibinfo {author} {\bibfnamefont {R.~S.}\ \bibnamefont
  {Zucker}}\ and\ \bibinfo {author} {\bibfnamefont {W.~G.}\ \bibnamefont
  {Regehr}},\ }\bibfield  {title} {\enquote {\bibinfo {title} {Short-term
  synaptic plasticity},}\ }\href@noop {} {\bibfield  {journal} {\bibinfo
  {journal} {Annual Review of Physiology}\ }\textbf {\bibinfo {volume} {64}},\
  \bibinfo {pages} {355--405} (\bibinfo {year} {2002})}\BibitemShut {NoStop}%
\bibitem [{\citenamefont {Feldman}(2012)}]{Feldman2012}%
  \BibitemOpen
  \bibfield  {author} {\bibinfo {author} {\bibfnamefont {D.~E.}\ \bibnamefont
  {Feldman}},\ }\bibfield  {title} {\enquote {\bibinfo {title} {{The
  Spike-Timing Dependence of Plasticity}},}\ }\href
  {https://doi.org/10.1016/j.neuron.2012.08.001} {\bibfield  {journal}
  {\bibinfo  {journal} {Neuron}\ }\textbf {\bibinfo {volume} {75}},\ \bibinfo
  {pages} {556--571} (\bibinfo {year} {2012})}\BibitemShut {NoStop}%
\bibitem [{\citenamefont {Markram}\ \emph {et~al.}(1997)\citenamefont
  {Markram}, \citenamefont {L{\"{u}}bke}, \citenamefont {Frotscher},\ and\
  \citenamefont {Sakmann}}]{Markram1997}%
  \BibitemOpen
  \bibfield  {author} {\bibinfo {author} {\bibfnamefont {H.}~\bibnamefont
  {Markram}}, \bibinfo {author} {\bibfnamefont {J.}~\bibnamefont
  {L{\"{u}}bke}}, \bibinfo {author} {\bibfnamefont {M.}~\bibnamefont
  {Frotscher}},\ and\ \bibinfo {author} {\bibfnamefont {B.}~\bibnamefont
  {Sakmann}},\ }\bibfield  {title} {\enquote {\bibinfo {title} {{Regulation of
  Synaptic Efficacy by Coincidence of Postsynaptic APs and EPSPs}},}\ }\href
  {https://doi.org/10.1126/science.275.5297.213} {\bibfield  {journal}
  {\bibinfo  {journal} {Science}\ }\textbf {\bibinfo {volume} {275}},\ \bibinfo
  {pages} {213--215} (\bibinfo {year} {1997})}\BibitemShut {NoStop}%
\bibitem [{\citenamefont {Bi}\ and\ \citenamefont {Poo}(1998)}]{Bi1998}%
  \BibitemOpen
  \bibfield  {author} {\bibinfo {author} {\bibfnamefont {G.-g.}\ \bibnamefont
  {Bi}}\ and\ \bibinfo {author} {\bibfnamefont {M.-M.}\ \bibnamefont {Poo}},\
  }\bibfield  {title} {\enquote {\bibinfo {title} {{Synaptic modifications in
  cultured hippocampal neurons: dependence on spike timing, synaptic strength,
  and postsynaptic cell type}},}\ }\href@noop {} {\bibfield  {journal}
  {\bibinfo  {journal} {Journal of Neuroscience}\ }\textbf {\bibinfo {volume}
  {18}},\ \bibinfo {pages} {10464--72} (\bibinfo {year} {1998})}\BibitemShut
  {NoStop}%
\bibitem [{\citenamefont {Motanis}, \citenamefont {Seay},\ and\ \citenamefont
  {Buonomano}(2018)}]{Motanis2018}%
  \BibitemOpen
  \bibfield  {author} {\bibinfo {author} {\bibfnamefont {H.}~\bibnamefont
  {Motanis}}, \bibinfo {author} {\bibfnamefont {M.~J.}\ \bibnamefont {Seay}},\
  and\ \bibinfo {author} {\bibfnamefont {D.~V.}\ \bibnamefont {Buonomano}},\
  }\bibfield  {title} {\enquote {\bibinfo {title} {{Short-Term Synaptic
  Plasticity as a Mechanism for Sensory Timing}},}\ }\href
  {https://doi.org/10.1016/j.tins.2018.08.001} {\bibfield  {journal} {\bibinfo
  {journal} {Trends in Neurosciences}\ }\textbf {\bibinfo {volume} {41}},\
  \bibinfo {pages} {701--711} (\bibinfo {year} {2018})}\BibitemShut {NoStop}%
\bibitem [{\citenamefont {Weber}, \citenamefont {Krishnamurthy},\ and\
  \citenamefont {Fairhall}(2019)}]{Weber2019a}%
  \BibitemOpen
  \bibfield  {author} {\bibinfo {author} {\bibfnamefont {A.~I.}\ \bibnamefont
  {Weber}}, \bibinfo {author} {\bibfnamefont {K.}~\bibnamefont
  {Krishnamurthy}},\ and\ \bibinfo {author} {\bibfnamefont {A.~L.}\
  \bibnamefont {Fairhall}},\ }\bibfield  {title} {\enquote {\bibinfo {title}
  {{Coding Principles in Adaptation}},}\ }\href
  {https://doi.org/https://doi.org/10.1146/annurev-vision-091718-014818}
  {\bibfield  {journal} {\bibinfo  {journal} {Annual Review of Vision Science}\
  }\textbf {\bibinfo {volume} {5}},\ \bibinfo {pages} {427--449} (\bibinfo
  {year} {2019})}\BibitemShut {NoStop}%
\bibitem [{\citenamefont {N{\"{a}}{\"{a}}t{\"{a}}nen}, \citenamefont
  {Simpson},\ and\ \citenamefont {Loveless}(1982)}]{Naatanen1982}%
  \BibitemOpen
  \bibfield  {author} {\bibinfo {author} {\bibfnamefont {R.}~\bibnamefont
  {N{\"{a}}{\"{a}}t{\"{a}}nen}}, \bibinfo {author} {\bibfnamefont
  {M.}~\bibnamefont {Simpson}},\ and\ \bibinfo {author} {\bibfnamefont {N.~E.}\
  \bibnamefont {Loveless}},\ }\bibfield  {title} {\enquote {\bibinfo {title}
  {{Stimulus deviance and evoked potentials}},}\ }\href
  {https://doi.org/https://doi.org/10.1016/0301-0511(82)90017-5} {\bibfield
  {journal} {\bibinfo  {journal} {Biological Psychology}\ }\textbf {\bibinfo
  {volume} {14}},\ \bibinfo {pages} {53--98} (\bibinfo {year}
  {1982})}\BibitemShut {NoStop}%
\bibitem [{\citenamefont {Ross}\ and\ \citenamefont {Hamm}(2020)}]{Ross2020}%
  \BibitemOpen
  \bibfield  {author} {\bibinfo {author} {\bibfnamefont {J.~M.}\ \bibnamefont
  {Ross}}\ and\ \bibinfo {author} {\bibfnamefont {J.~P.}\ \bibnamefont
  {Hamm}},\ }\bibfield  {title} {\enquote {\bibinfo {title} {{Cortical
  Microcircuit Mechanisms of Mismatch Negativity and Its Underlying
  Subcomponents}},}\ }\href
  {https://doi.org/https://doi.org/10.3389/fncir.2020.00013} {\bibfield
  {journal} {\bibinfo  {journal} {Frontiers in Neural Circuits}\ }\textbf
  {\bibinfo {volume} {14}} (\bibinfo {year} {2020}),\
  https://doi.org/10.3389/fncir.2020.00013}\BibitemShut {NoStop}%
\bibitem [{\citenamefont {Ulanovsky}, \citenamefont {Las},\ and\ \citenamefont
  {Nelken}(2003)}]{Ulanovsky2003}%
  \BibitemOpen
  \bibfield  {author} {\bibinfo {author} {\bibfnamefont {N.}~\bibnamefont
  {Ulanovsky}}, \bibinfo {author} {\bibfnamefont {L.}~\bibnamefont {Las}},\
  and\ \bibinfo {author} {\bibfnamefont {I.}~\bibnamefont {Nelken}},\
  }\bibfield  {title} {\enquote {\bibinfo {title} {{Processing of
  low-probability sounds by cortical neurons}},}\ }\href
  {https://doi.org/https://doi.org/10.1038/nn1032} {\bibfield  {journal}
  {\bibinfo  {journal} {Nature Neuroscience}\ }\textbf {\bibinfo {volume}
  {6}},\ \bibinfo {pages} {391--398} (\bibinfo {year} {2003})}\BibitemShut
  {NoStop}%
\bibitem [{\citenamefont {Natan}\ \emph {et~al.}(2015)\citenamefont {Natan},
  \citenamefont {Briguglio}, \citenamefont {Mwilambwe-Tshilobo}, \citenamefont
  {Jones}, \citenamefont {Aizenberg}, \citenamefont {Goldberg},\ and\
  \citenamefont {Geffen}}]{Natan2015}%
  \BibitemOpen
  \bibfield  {author} {\bibinfo {author} {\bibfnamefont {R.~G.}\ \bibnamefont
  {Natan}}, \bibinfo {author} {\bibfnamefont {J.~J.}\ \bibnamefont
  {Briguglio}}, \bibinfo {author} {\bibfnamefont {L.}~\bibnamefont
  {Mwilambwe-Tshilobo}}, \bibinfo {author} {\bibfnamefont {S.~I.}\ \bibnamefont
  {Jones}}, \bibinfo {author} {\bibfnamefont {M.}~\bibnamefont {Aizenberg}},
  \bibinfo {author} {\bibfnamefont {E.~M.}\ \bibnamefont {Goldberg}},\ and\
  \bibinfo {author} {\bibfnamefont {M.~N.}\ \bibnamefont {Geffen}},\ }\bibfield
   {title} {\enquote {\bibinfo {title} {{Complementary control of sensory
  adaptation by two types of cortical interneurons}},}\ }\href
  {https://doi.org/https://doi.org/10.7554/eLife.09868} {\bibfield  {journal}
  {\bibinfo  {journal} {eLife}\ }\textbf {\bibinfo {volume} {4}},\ \bibinfo
  {pages} {e09868} (\bibinfo {year} {2015})}\BibitemShut {NoStop}%
\bibitem [{\citenamefont {Homann}\ \emph {et~al.}(2022)\citenamefont {Homann},
  \citenamefont {Ann}, \citenamefont {Chen}, \citenamefont {Tank},\ and\
  \citenamefont {Berry}}]{Homann2022}%
  \BibitemOpen
  \bibfield  {author} {\bibinfo {author} {\bibfnamefont {J.}~\bibnamefont
  {Homann}}, \bibinfo {author} {\bibfnamefont {S.}~\bibnamefont {Ann}},
  \bibinfo {author} {\bibfnamefont {K.~S.}\ \bibnamefont {Chen}}, \bibinfo
  {author} {\bibfnamefont {D.~W.}\ \bibnamefont {Tank}},\ and\ \bibinfo
  {author} {\bibfnamefont {M.~J.}\ \bibnamefont {Berry}},\ }\bibfield  {title}
  {\enquote {\bibinfo {title} {{Novel stimuli evoke excess activity in the
  mouse primary visual cortex}},}\ }\href
  {https://doi.org/10.1073/pnas.2108882119/-/DCSupplemental.Published}
  {\bibfield  {journal} {\bibinfo  {journal} {Proceedings of the National
  Academy of Sciences of the United States of America}\ }\textbf {\bibinfo
  {volume} {119}},\ \bibinfo {pages} {e2108882119} (\bibinfo {year}
  {2022})}\BibitemShut {NoStop}%
\bibitem [{\citenamefont {Mill}\ \emph {et~al.}(2011)\citenamefont {Mill},
  \citenamefont {Coath}, \citenamefont {Wennekers},\ and\ \citenamefont
  {Denham}}]{Mill2011a}%
  \BibitemOpen
  \bibfield  {author} {\bibinfo {author} {\bibfnamefont {R.}~\bibnamefont
  {Mill}}, \bibinfo {author} {\bibfnamefont {M.}~\bibnamefont {Coath}},
  \bibinfo {author} {\bibfnamefont {T.}~\bibnamefont {Wennekers}},\ and\
  \bibinfo {author} {\bibfnamefont {S.~L.}\ \bibnamefont {Denham}},\ }\bibfield
   {title} {\enquote {\bibinfo {title} {{Abstract stimulus-specific adaptation
  models}},}\ }\href {https://doi.org/https://doi.org/10.1162/NECO_a_00077}
  {\bibfield  {journal} {\bibinfo  {journal} {Neural Computation}\ }\textbf
  {\bibinfo {volume} {23}},\ \bibinfo {pages} {435--476} (\bibinfo {year}
  {2011})}\BibitemShut {NoStop}%
\bibitem [{\citenamefont {Mill}\ \emph {et~al.}(2012)\citenamefont {Mill},
  \citenamefont {Coath}, \citenamefont {Wennekers},\ and\ \citenamefont
  {Denham}}]{Mill2012}%
  \BibitemOpen
  \bibfield  {author} {\bibinfo {author} {\bibfnamefont {R.}~\bibnamefont
  {Mill}}, \bibinfo {author} {\bibfnamefont {M.}~\bibnamefont {Coath}},
  \bibinfo {author} {\bibfnamefont {T.}~\bibnamefont {Wennekers}},\ and\
  \bibinfo {author} {\bibfnamefont {S.~L.}\ \bibnamefont {Denham}},\ }\bibfield
   {title} {\enquote {\bibinfo {title} {{Characterising stimulus-specific
  adaptation using a multi-layer field model}},}\ }\href
  {https://doi.org/http://dx.doi.org/10.1016/j.brainres.2011.08.063} {\bibfield
   {journal} {\bibinfo  {journal} {Brain Research}\ }\textbf {\bibinfo {volume}
  {1434}},\ \bibinfo {pages} {178--188} (\bibinfo {year} {2012})}\BibitemShut
  {NoStop}%
\bibitem [{\citenamefont {Hershenhoren}\ \emph {et~al.}(2014)\citenamefont
  {Hershenhoren}, \citenamefont {Taaseh}, \citenamefont {Antunes},\ and\
  \citenamefont {Nelken}}]{Hershenhoren2014}%
  \BibitemOpen
  \bibfield  {author} {\bibinfo {author} {\bibfnamefont {I.}~\bibnamefont
  {Hershenhoren}}, \bibinfo {author} {\bibfnamefont {N.}~\bibnamefont
  {Taaseh}}, \bibinfo {author} {\bibfnamefont {F.~M.}\ \bibnamefont
  {Antunes}},\ and\ \bibinfo {author} {\bibfnamefont {I.}~\bibnamefont
  {Nelken}},\ }\bibfield  {title} {\enquote {\bibinfo {title} {{Intracellular
  correlates of stimulus-specific adaptation}},}\ }\href
  {https://doi.org/https://doi.org/10.1523/JNEUROSCI.2166-13.2014} {\bibfield
  {journal} {\bibinfo  {journal} {Journal of Neuroscience}\ }\textbf {\bibinfo
  {volume} {34}},\ \bibinfo {pages} {3303--3319} (\bibinfo {year}
  {2014})}\BibitemShut {NoStop}%
\bibitem [{\citenamefont {Park}\ and\ \citenamefont {Geffen}(2020)}]{Park2020}%
  \BibitemOpen
  \bibfield  {author} {\bibinfo {author} {\bibfnamefont {Y.}~\bibnamefont
  {Park}}\ and\ \bibinfo {author} {\bibfnamefont {M.~N.}\ \bibnamefont
  {Geffen}},\ }\bibfield  {title} {\enquote {\bibinfo {title} {{A circuit model
  of auditory cortex}},}\ }\href
  {https://doi.org/http://dx.doi.org/10.1371/journal.pcbi.1008016} {\bibfield
  {journal} {\bibinfo  {journal} {PLoS Computational Biology}\ }\textbf
  {\bibinfo {volume} {16}},\ \bibinfo {pages} {e1008016} (\bibinfo {year}
  {2020})}\BibitemShut {NoStop}%
\bibitem [{\citenamefont {Seay}\ \emph {et~al.}(2020)\citenamefont {Seay},
  \citenamefont {Natan}, \citenamefont {Geffen},\ and\ \citenamefont
  {Buonomano}}]{Seay2020}%
  \BibitemOpen
  \bibfield  {author} {\bibinfo {author} {\bibfnamefont {M.~J.}\ \bibnamefont
  {Seay}}, \bibinfo {author} {\bibfnamefont {R.~G.}\ \bibnamefont {Natan}},
  \bibinfo {author} {\bibfnamefont {M.~N.}\ \bibnamefont {Geffen}},\ and\
  \bibinfo {author} {\bibfnamefont {D.~V.}\ \bibnamefont {Buonomano}},\
  }\bibfield  {title} {\enquote {\bibinfo {title} {Differential short-term
  plasticity of {PV} and {SST} neurons accounts for adaptation and facilitation
  of cortical neurons to auditory tones},}\ }\href@noop {} {\bibfield
  {journal} {\bibinfo  {journal} {Journal of Neuroscience}\ }\textbf {\bibinfo
  {volume} {40}},\ \bibinfo {pages} {9224--9235} (\bibinfo {year}
  {2020})}\BibitemShut {NoStop}%
\bibitem [{\citenamefont {Schulz}\ \emph {et~al.}(2021)\citenamefont {Schulz},
  \citenamefont {Miehl}, \citenamefont {{Berry II}},\ and\ \citenamefont
  {Gjorgjieva}}]{Schulz2021}%
  \BibitemOpen
  \bibfield  {author} {\bibinfo {author} {\bibfnamefont {A.}~\bibnamefont
  {Schulz}}, \bibinfo {author} {\bibfnamefont {C.}~\bibnamefont {Miehl}},
  \bibinfo {author} {\bibfnamefont {M.~J.}\ \bibnamefont {{Berry II}}},\ and\
  \bibinfo {author} {\bibfnamefont {J.}~\bibnamefont {Gjorgjieva}},\ }\bibfield
   {title} {\enquote {\bibinfo {title} {{The generation of cortical novelty
  responses through inhibitory plasticity}},}\ }\href
  {https://doi.org/https://doi.org/10.7554/eLife.65309} {\bibfield  {journal}
  {\bibinfo  {journal} {eLife}\ }\textbf {\bibinfo {volume} {10}},\ \bibinfo
  {pages} {e65309} (\bibinfo {year} {2021})}\BibitemShut {NoStop}%
\bibitem [{\citenamefont {Rao}\ and\ \citenamefont {Ballard}(1999)}]{Rao1999}%
  \BibitemOpen
  \bibfield  {author} {\bibinfo {author} {\bibfnamefont {R.~P.~N.}\
  \bibnamefont {Rao}}\ and\ \bibinfo {author} {\bibfnamefont {D.~H.}\
  \bibnamefont {Ballard}},\ }\bibfield  {title} {\enquote {\bibinfo {title}
  {{Predictive coding in the visual cortex: a functional interpretation of some
  extra-classical receptive-field effects}},}\ }\href
  {https://doi.org/https://doi.org/10.1038/4580} {\bibfield  {journal}
  {\bibinfo  {journal} {Nature Neuroscience}\ }\textbf {\bibinfo {volume}
  {2}},\ \bibinfo {pages} {79--87} (\bibinfo {year} {1999})}\BibitemShut
  {NoStop}%
\bibitem [{\citenamefont {Turrigiano}\ and\ \citenamefont
  {Nelson}(2004)}]{Turrigiano2004}%
  \BibitemOpen
  \bibfield  {author} {\bibinfo {author} {\bibfnamefont {G.~G.}\ \bibnamefont
  {Turrigiano}}\ and\ \bibinfo {author} {\bibfnamefont {S.~B.}\ \bibnamefont
  {Nelson}},\ }\bibfield  {title} {\enquote {\bibinfo {title} {{Homeostatic
  plasticity in the developing nervous system}},}\ }\href
  {https://doi.org/https://doi.org/10.1038/nrn1327} {\bibfield  {journal}
  {\bibinfo  {journal} {Nature Reviews Neuroscience}\ }\textbf {\bibinfo
  {volume} {5}},\ \bibinfo {pages} {97--107} (\bibinfo {year}
  {2004})}\BibitemShut {NoStop}%
\bibitem [{\citenamefont {Marder}\ and\ \citenamefont
  {Goaillard}(2006)}]{Marder2006}%
  \BibitemOpen
  \bibfield  {author} {\bibinfo {author} {\bibfnamefont {E.}~\bibnamefont
  {Marder}}\ and\ \bibinfo {author} {\bibfnamefont {J.-M.}\ \bibnamefont
  {Goaillard}},\ }\bibfield  {title} {\enquote {\bibinfo {title} {{Variability,
  compensation and homeostasis in neuron and network function}},}\ }\href
  {https://doi.org/10.1038/nrn1949} {\bibfield  {journal} {\bibinfo  {journal}
  {Nature reviews}\ }\textbf {\bibinfo {volume} {7}},\ \bibinfo {pages}
  {563--574} (\bibinfo {year} {2006})}\BibitemShut {NoStop}%
\bibitem [{\citenamefont {Onasch}\ and\ \citenamefont
  {Gjorgjieva}(2020)}]{Onasch2020}%
  \BibitemOpen
  \bibfield  {author} {\bibinfo {author} {\bibfnamefont {S.}~\bibnamefont
  {Onasch}}\ and\ \bibinfo {author} {\bibfnamefont {J.}~\bibnamefont
  {Gjorgjieva}},\ }\bibfield  {title} {\enquote {\bibinfo {title} {{Circuit
  stability to perturbations reveals hidden variability in the balance of
  intrinsic and synaptic conductances}},}\ }\href
  {https://doi.org/10.1523/JNEUROSCI.0985-19.2020} {\bibfield  {journal}
  {\bibinfo  {journal} {Journal of Neuroscience}\ }\textbf {\bibinfo {volume}
  {40}},\ \bibinfo {pages} {3186--3202} (\bibinfo {year} {2020})}\BibitemShut
  {NoStop}%
\bibitem [{\citenamefont {Gon{\c{c}}alves}\ \emph {et~al.}(2020)\citenamefont
  {Gon{\c{c}}alves}, \citenamefont {Lueckmann}, \citenamefont {Deistler},
  \citenamefont {Nonnenmacher}, \citenamefont {{\"{O}}cal}, \citenamefont
  {Bassetto}, \citenamefont {Chintaluri}, \citenamefont {Podlaski},
  \citenamefont {Haddad}, \citenamefont {Vogels}, \citenamefont {Greenberg},\
  and\ \citenamefont {Macke}}]{Goncalves2020}%
  \BibitemOpen
  \bibfield  {author} {\bibinfo {author} {\bibfnamefont {P.~J.}\ \bibnamefont
  {Gon{\c{c}}alves}}, \bibinfo {author} {\bibfnamefont {J.~M.}\ \bibnamefont
  {Lueckmann}}, \bibinfo {author} {\bibfnamefont {M.}~\bibnamefont {Deistler}},
  \bibinfo {author} {\bibfnamefont {M.}~\bibnamefont {Nonnenmacher}}, \bibinfo
  {author} {\bibfnamefont {K.}~\bibnamefont {{\"{O}}cal}}, \bibinfo {author}
  {\bibfnamefont {G.}~\bibnamefont {Bassetto}}, \bibinfo {author}
  {\bibfnamefont {C.}~\bibnamefont {Chintaluri}}, \bibinfo {author}
  {\bibfnamefont {W.~F.}\ \bibnamefont {Podlaski}}, \bibinfo {author}
  {\bibfnamefont {S.~A.}\ \bibnamefont {Haddad}}, \bibinfo {author}
  {\bibfnamefont {T.~P.}\ \bibnamefont {Vogels}}, \bibinfo {author}
  {\bibfnamefont {D.~S.}\ \bibnamefont {Greenberg}},\ and\ \bibinfo {author}
  {\bibfnamefont {J.~H.}\ \bibnamefont {Macke}},\ }\bibfield  {title} {\enquote
  {\bibinfo {title} {{Training deep neural density estimators to identify
  mechanistic models of neural dynamics}},}\ }\href
  {https://doi.org/10.7554/ELIFE.56261} {\bibfield  {journal} {\bibinfo
  {journal} {eLife}\ }\textbf {\bibinfo {volume} {9}},\ \bibinfo {pages}
  {1--46} (\bibinfo {year} {2020})}\BibitemShut {NoStop}%
\bibitem [{\citenamefont {Deistler}, \citenamefont {Macke},\ and\ \citenamefont
  {Gon{\c{c}}alves}(2022)}]{Deistler2022}%
  \BibitemOpen
  \bibfield  {author} {\bibinfo {author} {\bibfnamefont {M.}~\bibnamefont
  {Deistler}}, \bibinfo {author} {\bibfnamefont {J.~H.}\ \bibnamefont
  {Macke}},\ and\ \bibinfo {author} {\bibfnamefont {P.~J.}\ \bibnamefont
  {Gon{\c{c}}alves}},\ }\bibfield  {title} {\enquote {\bibinfo {title} {{Energy
  efficient network activity from disparate circuit parameters}},}\ }\href
  {https://doi.org/https://doi.org/10.1073/pnas.2207632119} {\bibfield
  {journal} {\bibinfo  {journal} {Proceedings of the National Academy of
  Sciences}\ }\textbf {\bibinfo {volume} {119}},\ \bibinfo {pages}
  {e2207632119} (\bibinfo {year} {2022})}\BibitemShut {NoStop}%
\bibitem [{\citenamefont {Marder}, \citenamefont {O'Leary},\ and\ \citenamefont
  {Shruti}(2014)}]{Marder2014}%
  \BibitemOpen
  \bibfield  {author} {\bibinfo {author} {\bibfnamefont {E.}~\bibnamefont
  {Marder}}, \bibinfo {author} {\bibfnamefont {T.}~\bibnamefont {O'Leary}},\
  and\ \bibinfo {author} {\bibfnamefont {S.}~\bibnamefont {Shruti}},\
  }\bibfield  {title} {\enquote {\bibinfo {title} {{Neuromodulation of circuits
  with variable parameters: Single neurons and small circuits reveal principles
  of state-dependent and robust neuromodulation}},}\ }\href
  {https://doi.org/10.1146/annurev-neuro-071013-013958} {\bibfield  {journal}
  {\bibinfo  {journal} {Annual Review of Neuroscience}\ }\textbf {\bibinfo
  {volume} {37}},\ \bibinfo {pages} {329--346} (\bibinfo {year}
  {2014})}\BibitemShut {NoStop}%
\bibitem [{\citenamefont {Jacquerie}\ and\ \citenamefont
  {Drion}(2021)}]{Jacquerie2021}%
  \BibitemOpen
  \bibfield  {author} {\bibinfo {author} {\bibfnamefont {K.}~\bibnamefont
  {Jacquerie}}\ and\ \bibinfo {author} {\bibfnamefont {G.}~\bibnamefont
  {Drion}},\ }\bibfield  {title} {\enquote {\bibinfo {title} {{Robust switches
  in thalamic network activity require a timescale separation between sodium
  and T-type calcium channel activations}},}\ }\href
  {https://doi.org/10.1371/journal.pcbi.1008997} {\bibfield  {journal}
  {\bibinfo  {journal} {PLoS Computational Biology}\ }\textbf {\bibinfo
  {volume} {17}},\ \bibinfo {pages} {e1008997} (\bibinfo {year}
  {2021})}\BibitemShut {NoStop}%
\bibitem [{\citenamefont {Mongillo}, \citenamefont {Barak},\ and\ \citenamefont
  {Tsodyks}(2008)}]{Mongillo2008}%
  \BibitemOpen
  \bibfield  {author} {\bibinfo {author} {\bibfnamefont {G.}~\bibnamefont
  {Mongillo}}, \bibinfo {author} {\bibfnamefont {O.}~\bibnamefont {Barak}},\
  and\ \bibinfo {author} {\bibfnamefont {M.}~\bibnamefont {Tsodyks}},\
  }\bibfield  {title} {\enquote {\bibinfo {title} {{Synaptic Theory of Working
  Memory}},}\ }\href {https://doi.org/10.1126/science.1150769} {\bibfield
  {journal} {\bibinfo  {journal} {Science}\ }\textbf {\bibinfo {volume}
  {319}},\ \bibinfo {pages} {1543--1546} (\bibinfo {year} {2008})}\BibitemShut
  {NoStop}%
\bibitem [{\citenamefont {Yuste}(2015)}]{Yuste2015}%
  \BibitemOpen
  \bibfield  {author} {\bibinfo {author} {\bibfnamefont {R.}~\bibnamefont
  {Yuste}},\ }\bibfield  {title} {\enquote {\bibinfo {title} {{From the neuron
  doctrine to neural networks}},}\ }\href {https://doi.org/10.1038/nrn3962}
  {\bibfield  {journal} {\bibinfo  {journal} {Nature Reviews Neuroscience}\
  }\textbf {\bibinfo {volume} {16}},\ \bibinfo {pages} {487--497} (\bibinfo
  {year} {2015})}\BibitemShut {NoStop}%
\bibitem [{\citenamefont {Fusi}(2017)}]{Fusi2017}%
  \BibitemOpen
  \bibfield  {author} {\bibinfo {author} {\bibfnamefont {S.}~\bibnamefont
  {Fusi}},\ }\bibfield  {title} {\enquote {\bibinfo {title} {{Computational
  models of long term plasticity and memory}},}\ }\href
  {https://doi.org/https://doi.org/10.48550/arXiv.1706.04946} {\bibfield
  {journal} {\bibinfo  {journal} {arXiv}\ } (\bibinfo {year} {2017}),\
  https://doi.org/10.48550/arXiv.1706.04946},\ \Eprint
  {https://arxiv.org/abs/1706.04946} {arXiv:1706.04946} \BibitemShut {NoStop}%
\bibitem [{\citenamefont {Litwin-Kumar}\ and\ \citenamefont
  {Doiron}(2014)}]{Litwin-Kumar2014}%
  \BibitemOpen
  \bibfield  {author} {\bibinfo {author} {\bibfnamefont {A.}~\bibnamefont
  {Litwin-Kumar}}\ and\ \bibinfo {author} {\bibfnamefont {B.}~\bibnamefont
  {Doiron}},\ }\bibfield  {title} {\enquote {\bibinfo {title} {{Formation and
  maintenance of neuronal assemblies through synaptic plasticity}},}\ }\href
  {https://doi.org/http://dx.doi.org/10.1038/ncomms6319} {\bibfield  {journal}
  {\bibinfo  {journal} {Nature Communications}\ }\textbf {\bibinfo {volume}
  {5}} (\bibinfo {year} {2014}),\
  http://dx.doi.org/10.1038/ncomms6319}\BibitemShut {NoStop}%
\bibitem [{\citenamefont {Miehl}\ and\ \citenamefont
  {Gjorgjieva}(2022)}]{Miehl2022a}%
  \BibitemOpen
  \bibfield  {author} {\bibinfo {author} {\bibfnamefont {C.}~\bibnamefont
  {Miehl}}\ and\ \bibinfo {author} {\bibfnamefont {J.}~\bibnamefont
  {Gjorgjieva}},\ }\bibfield  {title} {\enquote {\bibinfo {title} {{Stability
  and learning in excitatory synapses by nonlinear inhibitory plasticity}},}\
  }\href@noop {} {\bibfield  {journal} {\bibinfo  {journal} {PLoS Computational
  Biology}\ }\textbf {\bibinfo {volume} {18}},\ \bibinfo {pages} {e1010682}
  (\bibinfo {year} {2022})}\BibitemShut {NoStop}%
\bibitem [{\citenamefont {Zenke}, \citenamefont {Agnes},\ and\ \citenamefont
  {Gerstner}(2015)}]{Zenke2015}%
  \BibitemOpen
  \bibfield  {author} {\bibinfo {author} {\bibfnamefont {F.}~\bibnamefont
  {Zenke}}, \bibinfo {author} {\bibfnamefont {E.~J.}\ \bibnamefont {Agnes}},\
  and\ \bibinfo {author} {\bibfnamefont {W.}~\bibnamefont {Gerstner}},\
  }\bibfield  {title} {\enquote {\bibinfo {title} {{Diverse synaptic plasticity
  mechanisms orchestrated to form and retrieve memories in spiking neural
  networks}},}\ }\href {https://doi.org/10.1038/ncomms7922} {\bibfield
  {journal} {\bibinfo  {journal} {Nature Communications}\ }\textbf {\bibinfo
  {volume} {6}} (\bibinfo {year} {2015}),\ 10.1038/ncomms7922}\BibitemShut
  {NoStop}%
\bibitem [{\citenamefont {Hamm}\ \emph {et~al.}(2017)\citenamefont {Hamm},
  \citenamefont {Peterka}, \citenamefont {Gogos},\ and\ \citenamefont
  {Yuste}}]{Hamm2017}%
  \BibitemOpen
  \bibfield  {author} {\bibinfo {author} {\bibfnamefont {J.~P.}\ \bibnamefont
  {Hamm}}, \bibinfo {author} {\bibfnamefont {D.~S.}\ \bibnamefont {Peterka}},
  \bibinfo {author} {\bibfnamefont {J.~A.}\ \bibnamefont {Gogos}},\ and\
  \bibinfo {author} {\bibfnamefont {R.}~\bibnamefont {Yuste}},\ }\bibfield
  {title} {\enquote {\bibinfo {title} {{Altered Cortical Ensembles in Mouse
  Models of Schizophrenia}},}\ }\href
  {https://doi.org/http://dx.doi.org/10.1016/j.neuron.2017.03.019} {\bibfield
  {journal} {\bibinfo  {journal} {Neuron}\ }\textbf {\bibinfo {volume} {94}},\
  \bibinfo {pages} {153--167} (\bibinfo {year} {2017})}\BibitemShut {NoStop}%
\bibitem [{\citenamefont {Batista-Brito}\ \emph {et~al.}(2018)\citenamefont
  {Batista-Brito}, \citenamefont {Zagha}, \citenamefont {Ratliff},\ and\
  \citenamefont {Vinck}}]{Batista-Brito2018}%
  \BibitemOpen
  \bibfield  {author} {\bibinfo {author} {\bibfnamefont {R.}~\bibnamefont
  {Batista-Brito}}, \bibinfo {author} {\bibfnamefont {E.}~\bibnamefont
  {Zagha}}, \bibinfo {author} {\bibfnamefont {J.~M.}\ \bibnamefont {Ratliff}},\
  and\ \bibinfo {author} {\bibfnamefont {M.}~\bibnamefont {Vinck}},\ }\bibfield
   {title} {\enquote {\bibinfo {title} {{Modulation of cortical circuits by
  top-down processing and arousal state in health and disease}},}\ }\href
  {https://doi.org/10.1016/j.conb.2018.06.008} {\bibfield  {journal} {\bibinfo
  {journal} {Current Opinion in Neurobiology}\ }\textbf {\bibinfo {volume}
  {52}},\ \bibinfo {pages} {172--181} (\bibinfo {year} {2018})}\BibitemShut
  {NoStop}%
\bibitem [{\citenamefont {Light}\ and\ \citenamefont
  {N{\"{a}}{\"{a}}t{\"{a}}nen}(2013)}]{Light2013}%
  \BibitemOpen
  \bibfield  {author} {\bibinfo {author} {\bibfnamefont {G.~A.}\ \bibnamefont
  {Light}}\ and\ \bibinfo {author} {\bibfnamefont {R.}~\bibnamefont
  {N{\"{a}}{\"{a}}t{\"{a}}nen}},\ }\bibfield  {title} {\enquote {\bibinfo
  {title} {{Mismatch negativity is a breakthrough biomarker for understanding
  and treating psychotic disorders}},}\ }\href
  {https://doi.org/10.1073/pnas.1313287110} {\bibfield  {journal} {\bibinfo
  {journal} {Proceedings of the National Academy of Sciences of the United
  States of America}\ }\textbf {\bibinfo {volume} {110}},\ \bibinfo {pages}
  {15175--15176} (\bibinfo {year} {2013})}\BibitemShut {NoStop}%
\bibitem [{\citenamefont {Wilson}\ and\ \citenamefont
  {Cowan}(1973)}]{wilson1973}%
  \BibitemOpen
  \bibfield  {author} {\bibinfo {author} {\bibfnamefont {H.~R.}\ \bibnamefont
  {Wilson}}\ and\ \bibinfo {author} {\bibfnamefont {J.~D.}\ \bibnamefont
  {Cowan}},\ }\bibfield  {title} {\enquote {\bibinfo {title} {A mathematical
  theory of the functional dynamics of cortical and thalamic nervous tissue},}\
  }\href@noop {} {\bibfield  {journal} {\bibinfo  {journal} {Kybernetik}\
  }\textbf {\bibinfo {volume} {13}},\ \bibinfo {pages} {55--80} (\bibinfo
  {year} {1973})}\BibitemShut {NoStop}%
\bibitem [{\citenamefont {Mattia}\ and\ \citenamefont
  {Del~Giudice}(2002)}]{mattia2002}%
  \BibitemOpen
  \bibfield  {author} {\bibinfo {author} {\bibfnamefont {M.}~\bibnamefont
  {Mattia}}\ and\ \bibinfo {author} {\bibfnamefont {P.}~\bibnamefont
  {Del~Giudice}},\ }\bibfield  {title} {\enquote {\bibinfo {title} {Population
  dynamics of interacting spiking neurons},}\ }\href@noop {} {\bibfield
  {journal} {\bibinfo  {journal} {Physical Review E}\ }\textbf {\bibinfo
  {volume} {66}},\ \bibinfo {pages} {051917} (\bibinfo {year}
  {2002})}\BibitemShut {NoStop}%
\bibitem [{\citenamefont {Schaffer}, \citenamefont {Ostojic},\ and\
  \citenamefont {Abbott}(2013)}]{schaffer2013}%
  \BibitemOpen
  \bibfield  {author} {\bibinfo {author} {\bibfnamefont {E.~S.}\ \bibnamefont
  {Schaffer}}, \bibinfo {author} {\bibfnamefont {S.}~\bibnamefont {Ostojic}},\
  and\ \bibinfo {author} {\bibfnamefont {L.~F.}\ \bibnamefont {Abbott}},\
  }\bibfield  {title} {\enquote {\bibinfo {title} {A complex-valued firing-rate
  model that approximates the dynamics of spiking networks},}\ }\href@noop {}
  {\bibfield  {journal} {\bibinfo  {journal} {PLoS computational biology}\
  }\textbf {\bibinfo {volume} {9}},\ \bibinfo {pages} {e1003301} (\bibinfo
  {year} {2013})}\BibitemShut {NoStop}%
\bibitem [{\citenamefont {Luke}, \citenamefont {Barreto},\ and\ \citenamefont
  {So}(2013)}]{luke2013}%
  \BibitemOpen
  \bibfield  {author} {\bibinfo {author} {\bibfnamefont {T.~B.}\ \bibnamefont
  {Luke}}, \bibinfo {author} {\bibfnamefont {E.}~\bibnamefont {Barreto}},\ and\
  \bibinfo {author} {\bibfnamefont {P.}~\bibnamefont {So}},\ }\bibfield
  {title} {\enquote {\bibinfo {title} {Complete classification of the
  macroscopic behavior of a heterogeneous network of theta neurons},}\
  }\href@noop {} {\bibfield  {journal} {\bibinfo  {journal} {Neural
  computation}\ }\textbf {\bibinfo {volume} {25}},\ \bibinfo {pages}
  {3207--3234} (\bibinfo {year} {2013})}\BibitemShut {NoStop}%
\bibitem [{\citenamefont {Laing}(2014)}]{laing2014}%
  \BibitemOpen
  \bibfield  {author} {\bibinfo {author} {\bibfnamefont {C.~R.}\ \bibnamefont
  {Laing}},\ }\bibfield  {title} {\enquote {\bibinfo {title} {Derivation of a
  neural field model from a network of theta neurons},}\ }\href@noop {}
  {\bibfield  {journal} {\bibinfo  {journal} {Physical Review E}\ }\textbf
  {\bibinfo {volume} {90}},\ \bibinfo {pages} {010901} (\bibinfo {year}
  {2014})}\BibitemShut {NoStop}%
\bibitem [{\citenamefont {Montbri{\'o}}, \citenamefont {Paz{\'o}},\ and\
  \citenamefont {Roxin}(2015)}]{montbrio2015}%
  \BibitemOpen
  \bibfield  {author} {\bibinfo {author} {\bibfnamefont {E.}~\bibnamefont
  {Montbri{\'o}}}, \bibinfo {author} {\bibfnamefont {D.}~\bibnamefont
  {Paz{\'o}}},\ and\ \bibinfo {author} {\bibfnamefont {A.}~\bibnamefont
  {Roxin}},\ }\bibfield  {title} {\enquote {\bibinfo {title} {Macroscopic
  description for networks of spiking neurons},}\ }\href@noop {} {\bibfield
  {journal} {\bibinfo  {journal} {Physical Review X}\ }\textbf {\bibinfo
  {volume} {5}},\ \bibinfo {pages} {021028} (\bibinfo {year}
  {2015})}\BibitemShut {NoStop}%
\bibitem [{\citenamefont {Ermentrout}\ and\ \citenamefont
  {Kopell}(1986)}]{ermentrout1986}%
  \BibitemOpen
  \bibfield  {author} {\bibinfo {author} {\bibfnamefont {G.~B.}\ \bibnamefont
  {Ermentrout}}\ and\ \bibinfo {author} {\bibfnamefont {N.}~\bibnamefont
  {Kopell}},\ }\bibfield  {title} {\enquote {\bibinfo {title} {Parabolic
  bursting in an excitable system coupled with a slow oscillation},}\
  }\href@noop {} {\bibfield  {journal} {\bibinfo  {journal} {SIAM journal on
  applied mathematics}\ }\textbf {\bibinfo {volume} {46}},\ \bibinfo {pages}
  {233--253} (\bibinfo {year} {1986})}\BibitemShut {NoStop}%
\bibitem [{\citenamefont {Ott}\ and\ \citenamefont
  {Antonsen}(2008{\natexlab{b}})}]{ott2008}%
  \BibitemOpen
  \bibfield  {author} {\bibinfo {author} {\bibfnamefont {E.}~\bibnamefont
  {Ott}}\ and\ \bibinfo {author} {\bibfnamefont {T.~M.}\ \bibnamefont
  {Antonsen}},\ }\bibfield  {title} {\enquote {\bibinfo {title} {Low
  dimensional behavior of large systems of globally coupled oscillators},}\
  }\href@noop {} {\bibfield  {journal} {\bibinfo  {journal} {Chaos: An
  Interdisciplinary Journal of Nonlinear Science}\ }\textbf {\bibinfo {volume}
  {18}},\ \bibinfo {pages} {037113} (\bibinfo {year}
  {2008}{\natexlab{b}})}\BibitemShut {NoStop}%
\bibitem [{\citenamefont {Byrne}, \citenamefont {Brookes},\ and\ \citenamefont
  {Coombes}(2017)}]{byrne2017}%
  \BibitemOpen
  \bibfield  {author} {\bibinfo {author} {\bibfnamefont {A.}~\bibnamefont
  {Byrne}}, \bibinfo {author} {\bibfnamefont {M.~J.}\ \bibnamefont {Brookes}},\
  and\ \bibinfo {author} {\bibfnamefont {S.}~\bibnamefont {Coombes}},\
  }\bibfield  {title} {\enquote {\bibinfo {title} {A mean field model for
  movement induced changes in the beta rhythm},}\ }\href@noop {} {\bibfield
  {journal} {\bibinfo  {journal} {Journal of computational neuroscience}\
  }\textbf {\bibinfo {volume} {43}},\ \bibinfo {pages} {143--158} (\bibinfo
  {year} {2017})}\BibitemShut {NoStop}%
\bibitem [{\citenamefont {Schmidt}\ \emph {et~al.}(2018)\citenamefont
  {Schmidt}, \citenamefont {Avitabile}, \citenamefont {Montbri{\'o}},\ and\
  \citenamefont {Roxin}}]{schmidt2018}%
  \BibitemOpen
  \bibfield  {author} {\bibinfo {author} {\bibfnamefont {H.}~\bibnamefont
  {Schmidt}}, \bibinfo {author} {\bibfnamefont {D.}~\bibnamefont {Avitabile}},
  \bibinfo {author} {\bibfnamefont {E.}~\bibnamefont {Montbri{\'o}}},\ and\
  \bibinfo {author} {\bibfnamefont {A.}~\bibnamefont {Roxin}},\ }\bibfield
  {title} {\enquote {\bibinfo {title} {Network mechanisms underlying the role
  of oscillations in cognitive tasks},}\ }\href@noop {} {\bibfield  {journal}
  {\bibinfo  {journal} {PLoS computational biology}\ }\textbf {\bibinfo
  {volume} {14}},\ \bibinfo {pages} {e1006430} (\bibinfo {year}
  {2018})}\BibitemShut {NoStop}%
\bibitem [{\citenamefont {Byrne}\ \emph {et~al.}(2020)\citenamefont {Byrne},
  \citenamefont {O'Dea}, \citenamefont {Forrester}, \citenamefont {Ross},\ and\
  \citenamefont {Coombes}}]{byrne2020}%
  \BibitemOpen
  \bibfield  {author} {\bibinfo {author} {\bibfnamefont {{\'A}.}~\bibnamefont
  {Byrne}}, \bibinfo {author} {\bibfnamefont {R.~D.}\ \bibnamefont {O'Dea}},
  \bibinfo {author} {\bibfnamefont {M.}~\bibnamefont {Forrester}}, \bibinfo
  {author} {\bibfnamefont {J.}~\bibnamefont {Ross}},\ and\ \bibinfo {author}
  {\bibfnamefont {S.}~\bibnamefont {Coombes}},\ }\bibfield  {title} {\enquote
  {\bibinfo {title} {Next-generation neural mass and field modeling},}\
  }\href@noop {} {\bibfield  {journal} {\bibinfo  {journal} {Journal of
  neurophysiology}\ }\textbf {\bibinfo {volume} {123}},\ \bibinfo {pages}
  {726--742} (\bibinfo {year} {2020})}\BibitemShut {NoStop}%
\bibitem [{\citenamefont {Ceni}\ \emph {et~al.}(2020)\citenamefont {Ceni},
  \citenamefont {Olmi}, \citenamefont {Torcini},\ and\ \citenamefont
  {Angulo-Garcia}}]{ceni2020}%
  \BibitemOpen
  \bibfield  {author} {\bibinfo {author} {\bibfnamefont {A.}~\bibnamefont
  {Ceni}}, \bibinfo {author} {\bibfnamefont {S.}~\bibnamefont {Olmi}}, \bibinfo
  {author} {\bibfnamefont {A.}~\bibnamefont {Torcini}},\ and\ \bibinfo {author}
  {\bibfnamefont {D.}~\bibnamefont {Angulo-Garcia}},\ }\bibfield  {title}
  {\enquote {\bibinfo {title} {Cross frequency coupling in next generation
  inhibitory neural mass models},}\ }\href@noop {} {\bibfield  {journal}
  {\bibinfo  {journal} {Chaos: An Interdisciplinary Journal of Nonlinear
  Science}\ }\textbf {\bibinfo {volume} {30}},\ \bibinfo {pages} {053121}
  (\bibinfo {year} {2020})}\BibitemShut {NoStop}%
\bibitem [{\citenamefont {Bi}\ \emph {et~al.}(2020)\citenamefont {Bi},
  \citenamefont {Segneri}, \citenamefont {Di~Volo},\ and\ \citenamefont
  {Torcini}}]{bi2020}%
  \BibitemOpen
  \bibfield  {author} {\bibinfo {author} {\bibfnamefont {H.}~\bibnamefont
  {Bi}}, \bibinfo {author} {\bibfnamefont {M.}~\bibnamefont {Segneri}},
  \bibinfo {author} {\bibfnamefont {M.}~\bibnamefont {Di~Volo}},\ and\ \bibinfo
  {author} {\bibfnamefont {A.}~\bibnamefont {Torcini}},\ }\bibfield  {title}
  {\enquote {\bibinfo {title} {Coexistence of fast and slow gamma oscillations
  in one population of inhibitory spiking neurons},}\ }\href@noop {} {\bibfield
   {journal} {\bibinfo  {journal} {Physical Review Research}\ }\textbf
  {\bibinfo {volume} {2}},\ \bibinfo {pages} {013042} (\bibinfo {year}
  {2020})}\BibitemShut {NoStop}%
\bibitem [{\citenamefont {Taher}, \citenamefont {Torcini},\ and\ \citenamefont
  {Olmi}(2020)}]{taher2020}%
  \BibitemOpen
  \bibfield  {author} {\bibinfo {author} {\bibfnamefont {H.}~\bibnamefont
  {Taher}}, \bibinfo {author} {\bibfnamefont {A.}~\bibnamefont {Torcini}},\
  and\ \bibinfo {author} {\bibfnamefont {S.}~\bibnamefont {Olmi}},\ }\bibfield
  {title} {\enquote {\bibinfo {title} {Exact neural mass model for
  synaptic-based working memory},}\ }\href@noop {} {\bibfield  {journal}
  {\bibinfo  {journal} {PLoS Computational Biology}\ }\textbf {\bibinfo
  {volume} {16}},\ \bibinfo {pages} {e1008533} (\bibinfo {year}
  {2020})}\BibitemShut {NoStop}%
\bibitem [{\citenamefont {Segneri}\ \emph {et~al.}(2020)\citenamefont
  {Segneri}, \citenamefont {Bi}, \citenamefont {Olmi},\ and\ \citenamefont
  {Torcini}}]{segneri2020}%
  \BibitemOpen
  \bibfield  {author} {\bibinfo {author} {\bibfnamefont {M.}~\bibnamefont
  {Segneri}}, \bibinfo {author} {\bibfnamefont {H.}~\bibnamefont {Bi}},
  \bibinfo {author} {\bibfnamefont {S.}~\bibnamefont {Olmi}},\ and\ \bibinfo
  {author} {\bibfnamefont {A.}~\bibnamefont {Torcini}},\ }\bibfield  {title}
  {\enquote {\bibinfo {title} {Theta-nested gamma oscillations in next
  generation neural mass models},}\ }\href@noop {} {\bibfield  {journal}
  {\bibinfo  {journal} {Frontiers in computational neuroscience}\ }\textbf
  {\bibinfo {volume} {14}},\ \bibinfo {pages} {47} (\bibinfo {year}
  {2020})}\BibitemShut {NoStop}%
\bibitem [{\citenamefont {Gast}, \citenamefont {Schmidt},\ and\ \citenamefont
  {Kn{\"o}sche}(2020)}]{gast2020mean}%
  \BibitemOpen
  \bibfield  {author} {\bibinfo {author} {\bibfnamefont {R.}~\bibnamefont
  {Gast}}, \bibinfo {author} {\bibfnamefont {H.}~\bibnamefont {Schmidt}},\ and\
  \bibinfo {author} {\bibfnamefont {T.~R.}\ \bibnamefont {Kn{\"o}sche}},\
  }\bibfield  {title} {\enquote {\bibinfo {title} {A mean-field description of
  bursting dynamics in spiking neural networks with short-term adaptation},}\
  }\href@noop {} {\bibfield  {journal} {\bibinfo  {journal} {Neural
  Computation}\ }\textbf {\bibinfo {volume} {32}},\ \bibinfo {pages}
  {1615--1634} (\bibinfo {year} {2020})}\BibitemShut {NoStop}%
\bibitem [{\citenamefont {Gerster}\ \emph {et~al.}(2021)\citenamefont
  {Gerster}, \citenamefont {Taher}, \citenamefont {{\v{S}}koch}, \citenamefont
  {Hlinka}, \citenamefont {Guye}, \citenamefont {Bartolomei}, \citenamefont
  {Jirsa}, \citenamefont {Zakharova},\ and\ \citenamefont
  {Olmi}}]{gerster2021}%
  \BibitemOpen
  \bibfield  {author} {\bibinfo {author} {\bibfnamefont {M.}~\bibnamefont
  {Gerster}}, \bibinfo {author} {\bibfnamefont {H.}~\bibnamefont {Taher}},
  \bibinfo {author} {\bibfnamefont {A.}~\bibnamefont {{\v{S}}koch}}, \bibinfo
  {author} {\bibfnamefont {J.}~\bibnamefont {Hlinka}}, \bibinfo {author}
  {\bibfnamefont {M.}~\bibnamefont {Guye}}, \bibinfo {author} {\bibfnamefont
  {F.}~\bibnamefont {Bartolomei}}, \bibinfo {author} {\bibfnamefont
  {V.}~\bibnamefont {Jirsa}}, \bibinfo {author} {\bibfnamefont
  {A.}~\bibnamefont {Zakharova}},\ and\ \bibinfo {author} {\bibfnamefont
  {S.}~\bibnamefont {Olmi}},\ }\bibfield  {title} {\enquote {\bibinfo {title}
  {Patient-specific network connectivity combined with a next generation neural
  mass model to test clinical hypothesis of seizure propagation},}\ }\href@noop
  {} {\bibfield  {journal} {\bibinfo  {journal} {Frontiers in Systems
  Neuroscience}\ ,\ \bibinfo {pages} {79}} (\bibinfo {year}
  {2021})}\BibitemShut {NoStop}%
\bibitem [{\citenamefont {Gast}, \citenamefont {Kn{\"o}sche},\ and\
  \citenamefont {Schmidt}(2021)}]{gast2021}%
  \BibitemOpen
  \bibfield  {author} {\bibinfo {author} {\bibfnamefont {R.}~\bibnamefont
  {Gast}}, \bibinfo {author} {\bibfnamefont {T.~R.}\ \bibnamefont
  {Kn{\"o}sche}},\ and\ \bibinfo {author} {\bibfnamefont {H.}~\bibnamefont
  {Schmidt}},\ }\bibfield  {title} {\enquote {\bibinfo {title} {Mean-field
  approximations of networks of spiking neurons with short-term synaptic
  plasticity},}\ }\href@noop {} {\bibfield  {journal} {\bibinfo  {journal}
  {Physical Review E}\ }\textbf {\bibinfo {volume} {104}},\ \bibinfo {pages}
  {044310} (\bibinfo {year} {2021})}\BibitemShut {NoStop}%
\bibitem [{\citenamefont {Fuhrmann}, \citenamefont {Markram},\ and\
  \citenamefont {Tsodyks}(2002)}]{fuhrmann2002}%
  \BibitemOpen
  \bibfield  {author} {\bibinfo {author} {\bibfnamefont {G.}~\bibnamefont
  {Fuhrmann}}, \bibinfo {author} {\bibfnamefont {H.}~\bibnamefont {Markram}},\
  and\ \bibinfo {author} {\bibfnamefont {M.}~\bibnamefont {Tsodyks}},\
  }\bibfield  {title} {\enquote {\bibinfo {title} {Spike frequency adaptation
  and neocortical rhythms},}\ }\href@noop {} {\bibfield  {journal} {\bibinfo
  {journal} {Journal of neurophysiology}\ }\textbf {\bibinfo {volume} {88}},\
  \bibinfo {pages} {761--770} (\bibinfo {year} {2002})}\BibitemShut {NoStop}%
\bibitem [{\citenamefont {Benda}\ and\ \citenamefont {Herz}(2003)}]{benda2003}%
  \BibitemOpen
  \bibfield  {author} {\bibinfo {author} {\bibfnamefont {J.}~\bibnamefont
  {Benda}}\ and\ \bibinfo {author} {\bibfnamefont {A.~V.}\ \bibnamefont
  {Herz}},\ }\bibfield  {title} {\enquote {\bibinfo {title} {A universal model
  for spike-frequency adaptation},}\ }\href@noop {} {\bibfield  {journal}
  {\bibinfo  {journal} {Neural computation}\ }\textbf {\bibinfo {volume}
  {15}},\ \bibinfo {pages} {2523--2564} (\bibinfo {year} {2003})}\BibitemShut
  {NoStop}%
\bibitem [{\citenamefont {Brown}\ and\ \citenamefont
  {Adams}(1980)}]{brown1980}%
  \BibitemOpen
  \bibfield  {author} {\bibinfo {author} {\bibfnamefont {D.}~\bibnamefont
  {Brown}}\ and\ \bibinfo {author} {\bibfnamefont {P.}~\bibnamefont {Adams}},\
  }\bibfield  {title} {\enquote {\bibinfo {title} {Muscarinic suppression of a
  novel voltage-sensitive k+ current in a vertebrate neurone},}\ }\href@noop {}
  {\bibfield  {journal} {\bibinfo  {journal} {Nature}\ }\textbf {\bibinfo
  {volume} {283}},\ \bibinfo {pages} {673--676} (\bibinfo {year}
  {1980})}\BibitemShut {NoStop}%
\bibitem [{\citenamefont {Madison}\ and\ \citenamefont
  {Nicoll}(1984)}]{madison1984}%
  \BibitemOpen
  \bibfield  {author} {\bibinfo {author} {\bibfnamefont {D.}~\bibnamefont
  {Madison}}\ and\ \bibinfo {author} {\bibfnamefont {R.}~\bibnamefont
  {Nicoll}},\ }\bibfield  {title} {\enquote {\bibinfo {title} {Control of the
  repetitive discharge of rat ca 1 pyramidal neurones in vitro.}}\ }\href@noop
  {} {\bibfield  {journal} {\bibinfo  {journal} {The Journal of physiology}\
  }\textbf {\bibinfo {volume} {354}},\ \bibinfo {pages} {319--331} (\bibinfo
  {year} {1984})}\BibitemShut {NoStop}%
\bibitem [{\citenamefont {Fleidervish}, \citenamefont {Friedman},\ and\
  \citenamefont {Gutnick}(1996)}]{fleidervish1996}%
  \BibitemOpen
  \bibfield  {author} {\bibinfo {author} {\bibfnamefont {I.~A.}\ \bibnamefont
  {Fleidervish}}, \bibinfo {author} {\bibfnamefont {A.}~\bibnamefont
  {Friedman}},\ and\ \bibinfo {author} {\bibfnamefont {M.~J.}\ \bibnamefont
  {Gutnick}},\ }\bibfield  {title} {\enquote {\bibinfo {title} {Slow
  inactivation of na+ current and slow cumulative spike adaptation in mouse and
  guinea-pig neocortical neurones in slices.}}\ }\href@noop {} {\bibfield
  {journal} {\bibinfo  {journal} {The Journal of physiology}\ }\textbf
  {\bibinfo {volume} {493}},\ \bibinfo {pages} {83--97} (\bibinfo {year}
  {1996})}\BibitemShut {NoStop}%
\bibitem [{\citenamefont {Stevens}\ and\ \citenamefont
  {Wang}(1995)}]{stevens1995}%
  \BibitemOpen
  \bibfield  {author} {\bibinfo {author} {\bibfnamefont {C.~F.}\ \bibnamefont
  {Stevens}}\ and\ \bibinfo {author} {\bibfnamefont {Y.}~\bibnamefont {Wang}},\
  }\bibfield  {title} {\enquote {\bibinfo {title} {Facilitation and depression
  at single central synapses},}\ }\href@noop {} {\bibfield  {journal} {\bibinfo
   {journal} {Neuron}\ }\textbf {\bibinfo {volume} {14}},\ \bibinfo {pages}
  {795--802} (\bibinfo {year} {1995})}\BibitemShut {NoStop}%
\bibitem [{\citenamefont {Markram}\ and\ \citenamefont
  {Tsodyks}(1996)}]{markram1996}%
  \BibitemOpen
  \bibfield  {author} {\bibinfo {author} {\bibfnamefont {H.}~\bibnamefont
  {Markram}}\ and\ \bibinfo {author} {\bibfnamefont {M.}~\bibnamefont
  {Tsodyks}},\ }\bibfield  {title} {\enquote {\bibinfo {title} {Redistribution
  of synaptic efficacy between neocortical pyramidal neurons},}\ }\href@noop {}
  {\bibfield  {journal} {\bibinfo  {journal} {Nature}\ }\textbf {\bibinfo
  {volume} {382}},\ \bibinfo {pages} {807--810} (\bibinfo {year}
  {1996})}\BibitemShut {NoStop}%
\bibitem [{\citenamefont {Abbott}\ \emph {et~al.}(1997)\citenamefont {Abbott},
  \citenamefont {Varela}, \citenamefont {Sen},\ and\ \citenamefont
  {Nelson}}]{abbott1997}%
  \BibitemOpen
  \bibfield  {author} {\bibinfo {author} {\bibfnamefont {L.~F.}\ \bibnamefont
  {Abbott}}, \bibinfo {author} {\bibfnamefont {J.}~\bibnamefont {Varela}},
  \bibinfo {author} {\bibfnamefont {K.}~\bibnamefont {Sen}},\ and\ \bibinfo
  {author} {\bibfnamefont {S.}~\bibnamefont {Nelson}},\ }\bibfield  {title}
  {\enquote {\bibinfo {title} {Synaptic depression and cortical gain
  control},}\ }\href@noop {} {\bibfield  {journal} {\bibinfo  {journal}
  {Science}\ }\textbf {\bibinfo {volume} {275}},\ \bibinfo {pages} {221--224}
  (\bibinfo {year} {1997})}\BibitemShut {NoStop}%
\bibitem [{\citenamefont {Abbott}\ and\ \citenamefont
  {Regehr}(2004)}]{Abbott2004}%
  \BibitemOpen
  \bibfield  {author} {\bibinfo {author} {\bibfnamefont {L.}~\bibnamefont
  {Abbott}}\ and\ \bibinfo {author} {\bibfnamefont {W.~G.}\ \bibnamefont
  {Regehr}},\ }\bibfield  {title} {\enquote {\bibinfo {title} {Synaptic
  computation},}\ }\href@noop {} {\bibfield  {journal} {\bibinfo  {journal}
  {Nature}\ }\textbf {\bibinfo {volume} {431}},\ \bibinfo {pages} {796--803}
  (\bibinfo {year} {2004})}\BibitemShut {NoStop}%
\bibitem [{\citenamefont {Jones}\ and\ \citenamefont
  {Westbrook}(1996)}]{jones1996}%
  \BibitemOpen
  \bibfield  {author} {\bibinfo {author} {\bibfnamefont {M.~V.}\ \bibnamefont
  {Jones}}\ and\ \bibinfo {author} {\bibfnamefont {G.~L.}\ \bibnamefont
  {Westbrook}},\ }\bibfield  {title} {\enquote {\bibinfo {title} {The impact of
  receptor desensitization on fast synaptic transmission},}\ }\href@noop {}
  {\bibfield  {journal} {\bibinfo  {journal} {Trends in neurosciences}\
  }\textbf {\bibinfo {volume} {19}},\ \bibinfo {pages} {96--101} (\bibinfo
  {year} {1996})}\BibitemShut {NoStop}%
\bibitem [{\citenamefont {Wong}\ \emph {et~al.}(2003)\citenamefont {Wong},
  \citenamefont {Graham}, \citenamefont {Billups},\ and\ \citenamefont
  {Forsythe}}]{wong2003}%
  \BibitemOpen
  \bibfield  {author} {\bibinfo {author} {\bibfnamefont {A.~Y.}\ \bibnamefont
  {Wong}}, \bibinfo {author} {\bibfnamefont {B.~P.}\ \bibnamefont {Graham}},
  \bibinfo {author} {\bibfnamefont {B.}~\bibnamefont {Billups}},\ and\ \bibinfo
  {author} {\bibfnamefont {I.~D.}\ \bibnamefont {Forsythe}},\ }\bibfield
  {title} {\enquote {\bibinfo {title} {Distinguishing between presynaptic and
  postsynaptic mechanisms of short-term depression during action potential
  trains},}\ }\href@noop {} {\bibfield  {journal} {\bibinfo  {journal} {Journal
  of Neuroscience}\ }\textbf {\bibinfo {volume} {23}},\ \bibinfo {pages}
  {4868--4877} (\bibinfo {year} {2003})}\BibitemShut {NoStop}%
\bibitem [{\citenamefont {Turrigiano}(2008)}]{turrigiano2008}%
  \BibitemOpen
  \bibfield  {author} {\bibinfo {author} {\bibfnamefont {G.~G.}\ \bibnamefont
  {Turrigiano}},\ }\bibfield  {title} {\enquote {\bibinfo {title} {The
  self-tuning neuron: synaptic scaling of excitatory synapses},}\ }\href@noop
  {} {\bibfield  {journal} {\bibinfo  {journal} {Cell}\ }\textbf {\bibinfo
  {volume} {135}},\ \bibinfo {pages} {422--435} (\bibinfo {year}
  {2008})}\BibitemShut {NoStop}%
\bibitem [{\citenamefont {Pozo}\ and\ \citenamefont {Goda}(2010)}]{pozo2010}%
  \BibitemOpen
  \bibfield  {author} {\bibinfo {author} {\bibfnamefont {K.}~\bibnamefont
  {Pozo}}\ and\ \bibinfo {author} {\bibfnamefont {Y.}~\bibnamefont {Goda}},\
  }\bibfield  {title} {\enquote {\bibinfo {title} {Unraveling mechanisms of
  homeostatic synaptic plasticity},}\ }\href@noop {} {\bibfield  {journal}
  {\bibinfo  {journal} {Neuron}\ }\textbf {\bibinfo {volume} {66}},\ \bibinfo
  {pages} {337--351} (\bibinfo {year} {2010})}\BibitemShut {NoStop}%
\bibitem [{\citenamefont {Huang}\ \emph {et~al.}(2017)\citenamefont {Huang},
  \citenamefont {Chang}, \citenamefont {Chen}, \citenamefont {Lai},\ and\
  \citenamefont {Chan}}]{huang2017}%
  \BibitemOpen
  \bibfield  {author} {\bibinfo {author} {\bibfnamefont {Y.-T.}\ \bibnamefont
  {Huang}}, \bibinfo {author} {\bibfnamefont {Y.-L.}\ \bibnamefont {Chang}},
  \bibinfo {author} {\bibfnamefont {C.-C.}\ \bibnamefont {Chen}}, \bibinfo
  {author} {\bibfnamefont {P.-Y.}\ \bibnamefont {Lai}},\ and\ \bibinfo {author}
  {\bibfnamefont {C.}~\bibnamefont {Chan}},\ }\bibfield  {title} {\enquote
  {\bibinfo {title} {Positive feedback and synchronized bursts in neuronal
  cultures},}\ }\href@noop {} {\bibfield  {journal} {\bibinfo  {journal} {PloS
  one}\ }\textbf {\bibinfo {volume} {12}},\ \bibinfo {pages} {e0187276}
  (\bibinfo {year} {2017})}\BibitemShut {NoStop}%
\bibitem [{\citenamefont {Markram}, \citenamefont {Wang},\ and\ \citenamefont
  {Tsodyks}(1998)}]{markram1998}%
  \BibitemOpen
  \bibfield  {author} {\bibinfo {author} {\bibfnamefont {H.}~\bibnamefont
  {Markram}}, \bibinfo {author} {\bibfnamefont {Y.}~\bibnamefont {Wang}},\ and\
  \bibinfo {author} {\bibfnamefont {M.}~\bibnamefont {Tsodyks}},\ }\bibfield
  {title} {\enquote {\bibinfo {title} {Differential signaling via the same axon
  of neocortical pyramidal neurons},}\ }\href@noop {} {\bibfield  {journal}
  {\bibinfo  {journal} {Proceedings of the National Academy of Sciences}\
  }\textbf {\bibinfo {volume} {95}},\ \bibinfo {pages} {5323--5328} (\bibinfo
  {year} {1998})}\BibitemShut {NoStop}%
\bibitem [{\citenamefont {Dittman}, \citenamefont {Kreitzer},\ and\
  \citenamefont {Regehr}(2000)}]{dittman2000}%
  \BibitemOpen
  \bibfield  {author} {\bibinfo {author} {\bibfnamefont {J.~S.}\ \bibnamefont
  {Dittman}}, \bibinfo {author} {\bibfnamefont {A.~C.}\ \bibnamefont
  {Kreitzer}},\ and\ \bibinfo {author} {\bibfnamefont {W.~G.}\ \bibnamefont
  {Regehr}},\ }\bibfield  {title} {\enquote {\bibinfo {title} {Interplay
  between facilitation, depression, and residual calcium at three presynaptic
  terminals},}\ }\href@noop {} {\bibfield  {journal} {\bibinfo  {journal}
  {Journal of Neuroscience}\ }\textbf {\bibinfo {volume} {20}},\ \bibinfo
  {pages} {1374--1385} (\bibinfo {year} {2000})}\BibitemShut {NoStop}%
\bibitem [{\citenamefont {Wang}\ \emph {et~al.}(2006)\citenamefont {Wang},
  \citenamefont {Markram}, \citenamefont {Goodman}, \citenamefont {Berger},
  \citenamefont {Ma},\ and\ \citenamefont {Goldman-Rakic}}]{wang2006}%
  \BibitemOpen
  \bibfield  {author} {\bibinfo {author} {\bibfnamefont {Y.}~\bibnamefont
  {Wang}}, \bibinfo {author} {\bibfnamefont {H.}~\bibnamefont {Markram}},
  \bibinfo {author} {\bibfnamefont {P.~H.}\ \bibnamefont {Goodman}}, \bibinfo
  {author} {\bibfnamefont {T.~K.}\ \bibnamefont {Berger}}, \bibinfo {author}
  {\bibfnamefont {J.}~\bibnamefont {Ma}},\ and\ \bibinfo {author}
  {\bibfnamefont {P.~S.}\ \bibnamefont {Goldman-Rakic}},\ }\bibfield  {title}
  {\enquote {\bibinfo {title} {Heterogeneity in the pyramidal network of the
  medial prefrontal cortex},}\ }\href@noop {} {\bibfield  {journal} {\bibinfo
  {journal} {Nature neuroscience}\ }\textbf {\bibinfo {volume} {9}},\ \bibinfo
  {pages} {534--542} (\bibinfo {year} {2006})}\BibitemShut {NoStop}%
\bibitem [{\citenamefont {Gigante}, \citenamefont {Mattia},\ and\ \citenamefont
  {Del~Giudice}(2007)}]{gigante2007}%
  \BibitemOpen
  \bibfield  {author} {\bibinfo {author} {\bibfnamefont {G.}~\bibnamefont
  {Gigante}}, \bibinfo {author} {\bibfnamefont {M.}~\bibnamefont {Mattia}},\
  and\ \bibinfo {author} {\bibfnamefont {P.}~\bibnamefont {Del~Giudice}},\
  }\bibfield  {title} {\enquote {\bibinfo {title} {Diverse population-bursting
  modes of adapting spiking neurons},}\ }\href@noop {} {\bibfield  {journal}
  {\bibinfo  {journal} {Physical Review Letters}\ }\textbf {\bibinfo {volume}
  {98}},\ \bibinfo {pages} {148101} (\bibinfo {year} {2007})}\BibitemShut
  {NoStop}%
\bibitem [{\citenamefont {Ferrara}\ \emph {et~al.}(2023)\citenamefont
  {Ferrara}, \citenamefont {Angulo-Garcia}, \citenamefont {Torcini},\ and\
  \citenamefont {Olmi}}]{ferrara2023}%
  \BibitemOpen
  \bibfield  {author} {\bibinfo {author} {\bibfnamefont {A.}~\bibnamefont
  {Ferrara}}, \bibinfo {author} {\bibfnamefont {D.}~\bibnamefont
  {Angulo-Garcia}}, \bibinfo {author} {\bibfnamefont {A.}~\bibnamefont
  {Torcini}},\ and\ \bibinfo {author} {\bibfnamefont {S.}~\bibnamefont
  {Olmi}},\ }\bibfield  {title} {\enquote {\bibinfo {title} {Population spiking
  and bursting in next-generation neural masses with spike-frequency
  adaptation},}\ }\href@noop {} {\bibfield  {journal} {\bibinfo  {journal}
  {Physical Review E}\ }\textbf {\bibinfo {volume} {107}},\ \bibinfo {pages}
  {024311} (\bibinfo {year} {2023})}\BibitemShut {NoStop}%
\bibitem [{\citenamefont {Tsodyks}\ \emph {et~al.}(2000)\citenamefont
  {Tsodyks}, \citenamefont {Uziel}, \citenamefont {Markram} \emph
  {et~al.}}]{tsodyks2000}%
  \BibitemOpen
  \bibfield  {author} {\bibinfo {author} {\bibfnamefont {M.}~\bibnamefont
  {Tsodyks}}, \bibinfo {author} {\bibfnamefont {A.}~\bibnamefont {Uziel}},
  \bibinfo {author} {\bibfnamefont {H.}~\bibnamefont {Markram}}, \emph
  {et~al.},\ }\bibfield  {title} {\enquote {\bibinfo {title} {Synchrony
  generation in recurrent networks with frequency-dependent synapses},}\
  }\href@noop {} {\bibfield  {journal} {\bibinfo  {journal} {Journal of
  Neuroscience}\ }\textbf {\bibinfo {volume} {20}},\ \bibinfo {pages}
  {RC50--RC50} (\bibinfo {year} {2000})}\BibitemShut {NoStop}%
\bibitem [{\citenamefont {Luccioli}\ \emph {et~al.}(2014)\citenamefont
  {Luccioli}, \citenamefont {Ben-Jacob}, \citenamefont {Barzilai},
  \citenamefont {Bonifazi},\ and\ \citenamefont {Torcini}}]{luccioli2014}%
  \BibitemOpen
  \bibfield  {author} {\bibinfo {author} {\bibfnamefont {S.}~\bibnamefont
  {Luccioli}}, \bibinfo {author} {\bibfnamefont {E.}~\bibnamefont {Ben-Jacob}},
  \bibinfo {author} {\bibfnamefont {A.}~\bibnamefont {Barzilai}}, \bibinfo
  {author} {\bibfnamefont {P.}~\bibnamefont {Bonifazi}},\ and\ \bibinfo
  {author} {\bibfnamefont {A.}~\bibnamefont {Torcini}},\ }\bibfield  {title}
  {\enquote {\bibinfo {title} {Clique of functional hubs orchestrates
  population bursts in developmentally regulated neural networks},}\
  }\href@noop {} {\bibfield  {journal} {\bibinfo  {journal} {PLoS Computational
  Biology}\ }\textbf {\bibinfo {volume} {10}},\ \bibinfo {pages} {e1003823}
  (\bibinfo {year} {2014})}\BibitemShut {NoStop}%
\bibitem [{\citenamefont {Fino}\ and\ \citenamefont {Yuste}(2011)}]{fino2011}%
  \BibitemOpen
  \bibfield  {author} {\bibinfo {author} {\bibfnamefont {E.}~\bibnamefont
  {Fino}}\ and\ \bibinfo {author} {\bibfnamefont {R.}~\bibnamefont {Yuste}},\
  }\bibfield  {title} {\enquote {\bibinfo {title} {Dense inhibitory
  connectivity in neocortex},}\ }\href@noop {} {\bibfield  {journal} {\bibinfo
  {journal} {Neuron}\ }\textbf {\bibinfo {volume} {69}},\ \bibinfo {pages}
  {1188--1203} (\bibinfo {year} {2011})}\BibitemShut {NoStop}%
\bibitem [{\citenamefont {Tallon-Baudry}\ \emph {et~al.}(1998)\citenamefont
  {Tallon-Baudry}, \citenamefont {Bertrand}, \citenamefont {Peronnet},\ and\
  \citenamefont {Pernier}}]{tallon1998}%
  \BibitemOpen
  \bibfield  {author} {\bibinfo {author} {\bibfnamefont {C.}~\bibnamefont
  {Tallon-Baudry}}, \bibinfo {author} {\bibfnamefont {O.}~\bibnamefont
  {Bertrand}}, \bibinfo {author} {\bibfnamefont {F.}~\bibnamefont {Peronnet}},\
  and\ \bibinfo {author} {\bibfnamefont {J.}~\bibnamefont {Pernier}},\
  }\bibfield  {title} {\enquote {\bibinfo {title} {Induced $\gamma$-band
  activity during the delay of a visual short-term memory task in humans},}\
  }\href@noop {} {\bibfield  {journal} {\bibinfo  {journal} {Journal of
  Neuroscience}\ }\textbf {\bibinfo {volume} {18}},\ \bibinfo {pages}
  {4244--4254} (\bibinfo {year} {1998})}\BibitemShut {NoStop}%
\bibitem [{\citenamefont {Howard}\ \emph {et~al.}(2003)\citenamefont {Howard},
  \citenamefont {Rizzuto}, \citenamefont {Caplan}, \citenamefont {Madsen},
  \citenamefont {Lisman}, \citenamefont {Aschenbrenner-Scheibe}, \citenamefont
  {Schulze-Bonhage},\ and\ \citenamefont {Kahana}}]{howard2003}%
  \BibitemOpen
  \bibfield  {author} {\bibinfo {author} {\bibfnamefont {M.~W.}\ \bibnamefont
  {Howard}}, \bibinfo {author} {\bibfnamefont {D.~S.}\ \bibnamefont {Rizzuto}},
  \bibinfo {author} {\bibfnamefont {J.~B.}\ \bibnamefont {Caplan}}, \bibinfo
  {author} {\bibfnamefont {J.~R.}\ \bibnamefont {Madsen}}, \bibinfo {author}
  {\bibfnamefont {J.}~\bibnamefont {Lisman}}, \bibinfo {author} {\bibfnamefont
  {R.}~\bibnamefont {Aschenbrenner-Scheibe}}, \bibinfo {author} {\bibfnamefont
  {A.}~\bibnamefont {Schulze-Bonhage}},\ and\ \bibinfo {author} {\bibfnamefont
  {M.~J.}\ \bibnamefont {Kahana}},\ }\bibfield  {title} {\enquote {\bibinfo
  {title} {Gamma oscillations correlate with working memory load in humans},}\
  }\href@noop {} {\bibfield  {journal} {\bibinfo  {journal} {Cerebral cortex}\
  }\textbf {\bibinfo {volume} {13}},\ \bibinfo {pages} {1369--1374} (\bibinfo
  {year} {2003})}\BibitemShut {NoStop}%
\bibitem [{\citenamefont {Van~Vugt}\ \emph {et~al.}(2010)\citenamefont
  {Van~Vugt}, \citenamefont {Schulze-Bonhage}, \citenamefont {Litt},
  \citenamefont {Brandt},\ and\ \citenamefont {Kahana}}]{vanvugt2010}%
  \BibitemOpen
  \bibfield  {author} {\bibinfo {author} {\bibfnamefont {M.~K.}\ \bibnamefont
  {Van~Vugt}}, \bibinfo {author} {\bibfnamefont {A.}~\bibnamefont
  {Schulze-Bonhage}}, \bibinfo {author} {\bibfnamefont {B.}~\bibnamefont
  {Litt}}, \bibinfo {author} {\bibfnamefont {A.}~\bibnamefont {Brandt}},\ and\
  \bibinfo {author} {\bibfnamefont {M.~J.}\ \bibnamefont {Kahana}},\ }\bibfield
   {title} {\enquote {\bibinfo {title} {Hippocampal gamma oscillations increase
  with memory load},}\ }\href@noop {} {\bibfield  {journal} {\bibinfo
  {journal} {Journal of Neuroscience}\ }\textbf {\bibinfo {volume} {30}},\
  \bibinfo {pages} {2694--2699} (\bibinfo {year} {2010})}\BibitemShut {NoStop}%
\bibitem [{\citenamefont {Roux}\ \emph {et~al.}(2012)\citenamefont {Roux},
  \citenamefont {Wibral}, \citenamefont {Mohr}, \citenamefont {Singer},\ and\
  \citenamefont {Uhlhaas}}]{roux2012}%
  \BibitemOpen
  \bibfield  {author} {\bibinfo {author} {\bibfnamefont {F.}~\bibnamefont
  {Roux}}, \bibinfo {author} {\bibfnamefont {M.}~\bibnamefont {Wibral}},
  \bibinfo {author} {\bibfnamefont {H.~M.}\ \bibnamefont {Mohr}}, \bibinfo
  {author} {\bibfnamefont {W.}~\bibnamefont {Singer}},\ and\ \bibinfo {author}
  {\bibfnamefont {P.~J.}\ \bibnamefont {Uhlhaas}},\ }\bibfield  {title}
  {\enquote {\bibinfo {title} {Gamma-band activity in human prefrontal cortex
  codes for the number of relevant items maintained in working memory},}\
  }\href@noop {} {\bibfield  {journal} {\bibinfo  {journal} {Journal of
  Neuroscience}\ }\textbf {\bibinfo {volume} {32}},\ \bibinfo {pages}
  {12411--12420} (\bibinfo {year} {2012})}\BibitemShut {NoStop}%
\bibitem [{\citenamefont {Miller}, \citenamefont {Lundqvist},\ and\
  \citenamefont {Bastos}(2018)}]{miller2018}%
  \BibitemOpen
  \bibfield  {author} {\bibinfo {author} {\bibfnamefont {E.~K.}\ \bibnamefont
  {Miller}}, \bibinfo {author} {\bibfnamefont {M.}~\bibnamefont {Lundqvist}},\
  and\ \bibinfo {author} {\bibfnamefont {A.~M.}\ \bibnamefont {Bastos}},\
  }\bibfield  {title} {\enquote {\bibinfo {title} {Working memory 2.0},}\
  }\href@noop {} {\bibfield  {journal} {\bibinfo  {journal} {Neuron}\ }\textbf
  {\bibinfo {volume} {100}},\ \bibinfo {pages} {463--475} (\bibinfo {year}
  {2018})}\BibitemShut {NoStop}%
\bibitem [{\citenamefont {Tsodyks}, \citenamefont {Pawelzik},\ and\
  \citenamefont {Markram}(1998)}]{tsodyks1998}%
  \BibitemOpen
  \bibfield  {author} {\bibinfo {author} {\bibfnamefont {M.}~\bibnamefont
  {Tsodyks}}, \bibinfo {author} {\bibfnamefont {K.}~\bibnamefont {Pawelzik}},\
  and\ \bibinfo {author} {\bibfnamefont {H.}~\bibnamefont {Markram}},\
  }\bibfield  {title} {\enquote {\bibinfo {title} {Neural networks with dynamic
  synapses},}\ }\href@noop {} {\bibfield  {journal} {\bibinfo  {journal}
  {Neural computation}\ }\textbf {\bibinfo {volume} {10}},\ \bibinfo {pages}
  {821--835} (\bibinfo {year} {1998})}\BibitemShut {NoStop}%
\bibitem [{\citenamefont {Lind{\'e}n}\ and\ \citenamefont
  {Berg}(2021)}]{linden2021}%
  \BibitemOpen
  \bibfield  {author} {\bibinfo {author} {\bibfnamefont {H.}~\bibnamefont
  {Lind{\'e}n}}\ and\ \bibinfo {author} {\bibfnamefont {R.~W.}\ \bibnamefont
  {Berg}},\ }\bibfield  {title} {\enquote {\bibinfo {title} {Why firing rate
  distributions are important for understanding spinal central pattern
  generators},}\ }\href@noop {} {\bibfield  {journal} {\bibinfo  {journal}
  {Frontiers in Human Neuroscience}\ ,\ \bibinfo {pages} {504}} (\bibinfo
  {year} {2021})}\BibitemShut {NoStop}%
\bibitem [{\citenamefont {Rubin}\ \emph {et~al.}(2011)\citenamefont {Rubin},
  \citenamefont {Bacak}, \citenamefont {Molkov}, \citenamefont {Shevtsova},
  \citenamefont {Smith},\ and\ \citenamefont {Rybak}}]{rubin2011}%
  \BibitemOpen
  \bibfield  {author} {\bibinfo {author} {\bibfnamefont {J.~E.}\ \bibnamefont
  {Rubin}}, \bibinfo {author} {\bibfnamefont {B.~J.}\ \bibnamefont {Bacak}},
  \bibinfo {author} {\bibfnamefont {Y.~I.}\ \bibnamefont {Molkov}}, \bibinfo
  {author} {\bibfnamefont {N.~A.}\ \bibnamefont {Shevtsova}}, \bibinfo {author}
  {\bibfnamefont {J.~C.}\ \bibnamefont {Smith}},\ and\ \bibinfo {author}
  {\bibfnamefont {I.~A.}\ \bibnamefont {Rybak}},\ }\bibfield  {title} {\enquote
  {\bibinfo {title} {Interacting oscillations in neural control of breathing:
  modeling and qualitative analysis},}\ }\href@noop {} {\bibfield  {journal}
  {\bibinfo  {journal} {Journal of computational neuroscience}\ }\textbf
  {\bibinfo {volume} {30}},\ \bibinfo {pages} {607--632} (\bibinfo {year}
  {2011})}\BibitemShut {NoStop}%
\bibitem [{\citenamefont {Benabid}\ \emph {et~al.}(1994)\citenamefont
  {Benabid}, \citenamefont {Pollak}, \citenamefont {Gross}, \citenamefont
  {Hoffmann}, \citenamefont {Benazzouz}, \citenamefont {Gao}, \citenamefont
  {Laurent}, \citenamefont {Gentil},\ and\ \citenamefont
  {Perret}}]{Benabid1994}%
  \BibitemOpen
  \bibfield  {author} {\bibinfo {author} {\bibfnamefont {A.}~\bibnamefont
  {Benabid}}, \bibinfo {author} {\bibfnamefont {P.}~\bibnamefont {Pollak}},
  \bibinfo {author} {\bibfnamefont {C.}~\bibnamefont {Gross}}, \bibinfo
  {author} {\bibfnamefont {D.}~\bibnamefont {Hoffmann}}, \bibinfo {author}
  {\bibfnamefont {A.}~\bibnamefont {Benazzouz}}, \bibinfo {author}
  {\bibfnamefont {D.}~\bibnamefont {Gao}}, \bibinfo {author} {\bibfnamefont
  {A.}~\bibnamefont {Laurent}}, \bibinfo {author} {\bibfnamefont
  {M.}~\bibnamefont {Gentil}},\ and\ \bibinfo {author} {\bibfnamefont
  {J.}~\bibnamefont {Perret}},\ }\bibfield  {title} {\enquote {\bibinfo {title}
  {Acute and long-term effects of subthalamic nucleus stimulation in
  parkinson's disease},}\ }\href@noop {} {\bibfield  {journal} {\bibinfo
  {journal} {Stereotactic and functional neurosurgery}\ }\textbf {\bibinfo
  {volume} {62}},\ \bibinfo {pages} {76--84} (\bibinfo {year}
  {1994})}\BibitemShut {NoStop}%
\bibitem [{\citenamefont {Pinter}\ \emph {et~al.}(1999)\citenamefont {Pinter},
  \citenamefont {Alesch}, \citenamefont {Murg}, \citenamefont {Seiwald},
  \citenamefont {Helscher},\ and\ \citenamefont {Binder}}]{Pinter1999}%
  \BibitemOpen
  \bibfield  {author} {\bibinfo {author} {\bibfnamefont {M.}~\bibnamefont
  {Pinter}}, \bibinfo {author} {\bibfnamefont {F.}~\bibnamefont {Alesch}},
  \bibinfo {author} {\bibfnamefont {M.}~\bibnamefont {Murg}}, \bibinfo {author}
  {\bibfnamefont {M.}~\bibnamefont {Seiwald}}, \bibinfo {author} {\bibfnamefont
  {R.}~\bibnamefont {Helscher}},\ and\ \bibinfo {author} {\bibfnamefont
  {H.}~\bibnamefont {Binder}},\ }\bibfield  {title} {\enquote {\bibinfo {title}
  {Deep brain stimulation of the subthalamic nucleus for control of
  extrapyramidal features in advanced idiopathic parkinson's disease: one year
  follow-up},}\ }\href@noop {} {\bibfield  {journal} {\bibinfo  {journal}
  {Journal of neural transmission}\ }\textbf {\bibinfo {volume} {106}},\
  \bibinfo {pages} {693--709} (\bibinfo {year} {1999})}\BibitemShut {NoStop}%
\bibitem [{\citenamefont {Rodriguez-Oroz}\ \emph {et~al.}(2005)\citenamefont
  {Rodriguez-Oroz}, \citenamefont {Obeso}, \citenamefont {Lang}, \citenamefont
  {Houeto}, \citenamefont {Pollak}, \citenamefont {Rehncrona}, \citenamefont
  {Kulisevsky}, \citenamefont {Albanese}, \citenamefont {Volkmann},
  \citenamefont {Hariz}, \citenamefont {Quinn}, \citenamefont {Speelman},
  \citenamefont {Guridi}, \citenamefont {Zamarbide}, \citenamefont {Gironell},
  \citenamefont {Molet}, \citenamefont {Pascual-Sedano}, \citenamefont
  {Pidoux}, \citenamefont {Bonnet}, \citenamefont {Agid}, \citenamefont {Xie},
  \citenamefont {Benabid}, \citenamefont {Lozano}, \citenamefont {Saint-Cyr},
  \citenamefont {Romito}, \citenamefont {Contarino}, \citenamefont {Scerrati},
  \citenamefont {Fraix},\ and\ \citenamefont
  {Van~Blercom}}]{Rodriguez-Oroz2005}%
  \BibitemOpen
  \bibfield  {author} {\bibinfo {author} {\bibfnamefont {M.~C.}\ \bibnamefont
  {Rodriguez-Oroz}}, \bibinfo {author} {\bibfnamefont {J.~A.}\ \bibnamefont
  {Obeso}}, \bibinfo {author} {\bibfnamefont {A.~E.}\ \bibnamefont {Lang}},
  \bibinfo {author} {\bibfnamefont {J.-L.}\ \bibnamefont {Houeto}}, \bibinfo
  {author} {\bibfnamefont {P.}~\bibnamefont {Pollak}}, \bibinfo {author}
  {\bibfnamefont {S.}~\bibnamefont {Rehncrona}}, \bibinfo {author}
  {\bibfnamefont {J.}~\bibnamefont {Kulisevsky}}, \bibinfo {author}
  {\bibfnamefont {A.}~\bibnamefont {Albanese}}, \bibinfo {author}
  {\bibfnamefont {J.}~\bibnamefont {Volkmann}}, \bibinfo {author}
  {\bibfnamefont {M.~I.}\ \bibnamefont {Hariz}}, \bibinfo {author}
  {\bibfnamefont {N.~P.}\ \bibnamefont {Quinn}}, \bibinfo {author}
  {\bibfnamefont {J.~D.}\ \bibnamefont {Speelman}}, \bibinfo {author}
  {\bibfnamefont {J.}~\bibnamefont {Guridi}}, \bibinfo {author} {\bibfnamefont
  {I.}~\bibnamefont {Zamarbide}}, \bibinfo {author} {\bibfnamefont
  {A.}~\bibnamefont {Gironell}}, \bibinfo {author} {\bibfnamefont
  {J.}~\bibnamefont {Molet}}, \bibinfo {author} {\bibfnamefont
  {B.}~\bibnamefont {Pascual-Sedano}}, \bibinfo {author} {\bibfnamefont
  {B.}~\bibnamefont {Pidoux}}, \bibinfo {author} {\bibfnamefont {A.~M.}\
  \bibnamefont {Bonnet}}, \bibinfo {author} {\bibfnamefont {Y.}~\bibnamefont
  {Agid}}, \bibinfo {author} {\bibfnamefont {J.}~\bibnamefont {Xie}}, \bibinfo
  {author} {\bibfnamefont {A.-L.}\ \bibnamefont {Benabid}}, \bibinfo {author}
  {\bibfnamefont {A.~M.}\ \bibnamefont {Lozano}}, \bibinfo {author}
  {\bibfnamefont {J.}~\bibnamefont {Saint-Cyr}}, \bibinfo {author}
  {\bibfnamefont {L.}~\bibnamefont {Romito}}, \bibinfo {author} {\bibfnamefont
  {M.~F.}\ \bibnamefont {Contarino}}, \bibinfo {author} {\bibfnamefont
  {M.}~\bibnamefont {Scerrati}}, \bibinfo {author} {\bibfnamefont
  {V.}~\bibnamefont {Fraix}},\ and\ \bibinfo {author} {\bibfnamefont
  {N.}~\bibnamefont {Van~Blercom}},\ }\bibfield  {title} {\enquote {\bibinfo
  {title} {Bilateral deep brain stimulation in parkinson's disease: a
  multicentre study with 4 years follow-up.}}\ }\href
  {https://doi.org/10.1093/brain/awh571} {\bibfield  {journal} {\bibinfo
  {journal} {Brain}\ }\textbf {\bibinfo {volume} {128}},\ \bibinfo {pages}
  {2240--2249} (\bibinfo {year} {2005})}\BibitemShut {NoStop}%
\bibitem [{\citenamefont {Benabid}\ \emph {et~al.}(2009)\citenamefont
  {Benabid}, \citenamefont {Chabardes}, \citenamefont {Mitrofanis},\ and\
  \citenamefont {Pollak}}]{Benabid2009}%
  \BibitemOpen
  \bibfield  {author} {\bibinfo {author} {\bibfnamefont {A.~L.}\ \bibnamefont
  {Benabid}}, \bibinfo {author} {\bibfnamefont {S.}~\bibnamefont {Chabardes}},
  \bibinfo {author} {\bibfnamefont {J.}~\bibnamefont {Mitrofanis}},\ and\
  \bibinfo {author} {\bibfnamefont {P.}~\bibnamefont {Pollak}},\ }\bibfield
  {title} {\enquote {\bibinfo {title} {Deep brain stimulation of the
  subthalamic nucleus for the treatment of parkinson's disease.}}\ }\href
  {https://doi.org/10.1016/S1474-4422(08)70291-6} {\bibfield  {journal}
  {\bibinfo  {journal} {Lancet Neurol}\ }\textbf {\bibinfo {volume} {8}},\
  \bibinfo {pages} {67--81} (\bibinfo {year} {2009})}\BibitemShut {NoStop}%
\bibitem [{\citenamefont {Lozano}\ \emph {et~al.}(2019)\citenamefont {Lozano},
  \citenamefont {Lipsman}, \citenamefont {Bergman}, \citenamefont {Brown},
  \citenamefont {Chabardes}, \citenamefont {Chang}, \citenamefont {Matthews},
  \citenamefont {McIntyre}, \citenamefont {Schlaepfer}, \citenamefont
  {Schulder}, \citenamefont {Temel}, \citenamefont {Volkmann},\ and\
  \citenamefont {Krauss}}]{Lozano2019}%
  \BibitemOpen
  \bibfield  {author} {\bibinfo {author} {\bibfnamefont {A.~M.}\ \bibnamefont
  {Lozano}}, \bibinfo {author} {\bibfnamefont {N.}~\bibnamefont {Lipsman}},
  \bibinfo {author} {\bibfnamefont {H.}~\bibnamefont {Bergman}}, \bibinfo
  {author} {\bibfnamefont {P.}~\bibnamefont {Brown}}, \bibinfo {author}
  {\bibfnamefont {S.}~\bibnamefont {Chabardes}}, \bibinfo {author}
  {\bibfnamefont {J.~W.}\ \bibnamefont {Chang}}, \bibinfo {author}
  {\bibfnamefont {K.}~\bibnamefont {Matthews}}, \bibinfo {author}
  {\bibfnamefont {C.~C.}\ \bibnamefont {McIntyre}}, \bibinfo {author}
  {\bibfnamefont {T.~E.}\ \bibnamefont {Schlaepfer}}, \bibinfo {author}
  {\bibfnamefont {M.}~\bibnamefont {Schulder}}, \bibinfo {author}
  {\bibfnamefont {Y.}~\bibnamefont {Temel}}, \bibinfo {author} {\bibfnamefont
  {J.}~\bibnamefont {Volkmann}},\ and\ \bibinfo {author} {\bibfnamefont
  {J.~K.}\ \bibnamefont {Krauss}},\ }\bibfield  {title} {\enquote {\bibinfo
  {title} {Deep brain stimulation: current challenges and future directions},}\
  }\href {https://doi.org/10.1038/s41582-018-0128-2} {\bibfield  {journal}
  {\bibinfo  {journal} {Nature Reviews Neurology}\ }\textbf {\bibinfo {volume}
  {15}},\ \bibinfo {pages} {148--160} (\bibinfo {year} {2019})}\BibitemShut
  {NoStop}%
\bibitem [{\citenamefont {Deuschl}\ \emph {et~al.}(2006)\citenamefont
  {Deuschl}, \citenamefont {Schade-Brittinger}, \citenamefont {Krack},
  \citenamefont {Volkmann}, \citenamefont {Sch{\"a}fer}, \citenamefont
  {B{\"o}tzel}, \citenamefont {Daniels}, \citenamefont {Deutschl{\"a}nder},
  \citenamefont {Dillmann}, \citenamefont {Eisner}, \citenamefont {Gruber},
  \citenamefont {Hamel}, \citenamefont {Herzog}, \citenamefont {Hilker},
  \citenamefont {Klebe}, \citenamefont {Kloss}, \citenamefont {Koy},
  \citenamefont {Krause}, \citenamefont {Kupsch}, \citenamefont {Lorenz},
  \citenamefont {Lorenzl}, \citenamefont {Mehdorn}, \citenamefont {Moringlane},
  \citenamefont {Oertel}, \citenamefont {Pinsker}, \citenamefont {Reichmann},
  \citenamefont {Reuss}, \citenamefont {Schneider}, \citenamefont {Schnitzler},
  \citenamefont {Steude}, \citenamefont {Sturm}, \citenamefont {Timmermann},
  \citenamefont {Tronnier}, \citenamefont {Trottenberg}, \citenamefont
  {Wojtecki}, \citenamefont {Wolf}, \citenamefont {Poewe},\ and\ \citenamefont
  {Voges}}]{Deuschl2006}%
  \BibitemOpen
  \bibfield  {author} {\bibinfo {author} {\bibfnamefont {G.}~\bibnamefont
  {Deuschl}}, \bibinfo {author} {\bibfnamefont {C.}~\bibnamefont
  {Schade-Brittinger}}, \bibinfo {author} {\bibfnamefont {P.}~\bibnamefont
  {Krack}}, \bibinfo {author} {\bibfnamefont {J.}~\bibnamefont {Volkmann}},
  \bibinfo {author} {\bibfnamefont {H.}~\bibnamefont {Sch{\"a}fer}}, \bibinfo
  {author} {\bibfnamefont {K.}~\bibnamefont {B{\"o}tzel}}, \bibinfo {author}
  {\bibfnamefont {C.}~\bibnamefont {Daniels}}, \bibinfo {author} {\bibfnamefont
  {A.}~\bibnamefont {Deutschl{\"a}nder}}, \bibinfo {author} {\bibfnamefont
  {U.}~\bibnamefont {Dillmann}}, \bibinfo {author} {\bibfnamefont
  {W.}~\bibnamefont {Eisner}}, \bibinfo {author} {\bibfnamefont
  {D.}~\bibnamefont {Gruber}}, \bibinfo {author} {\bibfnamefont
  {W.}~\bibnamefont {Hamel}}, \bibinfo {author} {\bibfnamefont
  {J.}~\bibnamefont {Herzog}}, \bibinfo {author} {\bibfnamefont
  {R.}~\bibnamefont {Hilker}}, \bibinfo {author} {\bibfnamefont
  {S.}~\bibnamefont {Klebe}}, \bibinfo {author} {\bibfnamefont
  {M.}~\bibnamefont {Kloss}}, \bibinfo {author} {\bibfnamefont
  {J.}~\bibnamefont {Koy}}, \bibinfo {author} {\bibfnamefont {M.}~\bibnamefont
  {Krause}}, \bibinfo {author} {\bibfnamefont {A.}~\bibnamefont {Kupsch}},
  \bibinfo {author} {\bibfnamefont {D.}~\bibnamefont {Lorenz}}, \bibinfo
  {author} {\bibfnamefont {S.}~\bibnamefont {Lorenzl}}, \bibinfo {author}
  {\bibfnamefont {H.~M.}\ \bibnamefont {Mehdorn}}, \bibinfo {author}
  {\bibfnamefont {J.~R.}\ \bibnamefont {Moringlane}}, \bibinfo {author}
  {\bibfnamefont {W.}~\bibnamefont {Oertel}}, \bibinfo {author} {\bibfnamefont
  {M.~O.}\ \bibnamefont {Pinsker}}, \bibinfo {author} {\bibfnamefont
  {H.}~\bibnamefont {Reichmann}}, \bibinfo {author} {\bibfnamefont
  {A.}~\bibnamefont {Reuss}}, \bibinfo {author} {\bibfnamefont {G.-H.}\
  \bibnamefont {Schneider}}, \bibinfo {author} {\bibfnamefont {A.}~\bibnamefont
  {Schnitzler}}, \bibinfo {author} {\bibfnamefont {U.}~\bibnamefont {Steude}},
  \bibinfo {author} {\bibfnamefont {V.}~\bibnamefont {Sturm}}, \bibinfo
  {author} {\bibfnamefont {L.}~\bibnamefont {Timmermann}}, \bibinfo {author}
  {\bibfnamefont {V.}~\bibnamefont {Tronnier}}, \bibinfo {author}
  {\bibfnamefont {T.}~\bibnamefont {Trottenberg}}, \bibinfo {author}
  {\bibfnamefont {L.}~\bibnamefont {Wojtecki}}, \bibinfo {author}
  {\bibfnamefont {E.}~\bibnamefont {Wolf}}, \bibinfo {author} {\bibfnamefont
  {W.}~\bibnamefont {Poewe}},\ and\ \bibinfo {author} {\bibfnamefont
  {J.}~\bibnamefont {Voges}},\ }\bibfield  {title} {\enquote {\bibinfo {title}
  {A randomized trial of deep-brain stimulation for parkinson's disease.}}\
  }\href {https://doi.org/10.1056/NEJMoa060281} {\bibfield  {journal} {\bibinfo
   {journal} {N Engl J Med}\ }\textbf {\bibinfo {volume} {355}},\ \bibinfo
  {pages} {896--908} (\bibinfo {year} {2006})}\BibitemShut {NoStop}%
\bibitem [{\citenamefont {Weaver}\ \emph {et~al.}(2009)\citenamefont {Weaver},
  \citenamefont {Follett}, \citenamefont {Stern}, \citenamefont {Hur},
  \citenamefont {Harris}, \citenamefont {Marks}, \citenamefont {Rothlind},
  \citenamefont {Sagher}, \citenamefont {Reda}, \citenamefont {Moy} \emph
  {et~al.}}]{Weaver2009}%
  \BibitemOpen
  \bibfield  {author} {\bibinfo {author} {\bibfnamefont {F.~M.}\ \bibnamefont
  {Weaver}}, \bibinfo {author} {\bibfnamefont {K.}~\bibnamefont {Follett}},
  \bibinfo {author} {\bibfnamefont {M.}~\bibnamefont {Stern}}, \bibinfo
  {author} {\bibfnamefont {K.}~\bibnamefont {Hur}}, \bibinfo {author}
  {\bibfnamefont {C.}~\bibnamefont {Harris}}, \bibinfo {author} {\bibfnamefont
  {W.~J.}\ \bibnamefont {Marks}}, \bibinfo {author} {\bibfnamefont
  {J.}~\bibnamefont {Rothlind}}, \bibinfo {author} {\bibfnamefont
  {O.}~\bibnamefont {Sagher}}, \bibinfo {author} {\bibfnamefont
  {D.}~\bibnamefont {Reda}}, \bibinfo {author} {\bibfnamefont {C.~S.}\
  \bibnamefont {Moy}}, \emph {et~al.},\ }\bibfield  {title} {\enquote {\bibinfo
  {title} {Bilateral deep brain stimulation vs best medical therapy for
  patients with advanced parkinson disease: a randomized controlled trial},}\
  }\href@noop {} {\bibfield  {journal} {\bibinfo  {journal} {Jama}\ }\textbf
  {\bibinfo {volume} {301}},\ \bibinfo {pages} {63--73} (\bibinfo {year}
  {2009})}\BibitemShut {NoStop}%
\bibitem [{\citenamefont {Follett}\ \emph {et~al.}(2010)\citenamefont
  {Follett}, \citenamefont {Weaver}, \citenamefont {Stern}, \citenamefont
  {Hur}, \citenamefont {Harris}, \citenamefont {Luo}, \citenamefont {Marks},
  \citenamefont {Rothlind}, \citenamefont {Sagher}, \citenamefont {Moy},
  \citenamefont {Pahwa}, \citenamefont {Burchiel}, \citenamefont {Hogarth},
  \citenamefont {Lai}, \citenamefont {Duda}, \citenamefont {Holloway},
  \citenamefont {Samii}, \citenamefont {Horn}, \citenamefont {Bronstein},
  \citenamefont {Stoner}, \citenamefont {Starr}, \citenamefont {Simpson},
  \citenamefont {Baltuch}, \citenamefont {De~Salles}, \citenamefont {Huang},\
  and\ \citenamefont {Reda}}]{Follett2010}%
  \BibitemOpen
  \bibfield  {author} {\bibinfo {author} {\bibfnamefont {K.~A.}\ \bibnamefont
  {Follett}}, \bibinfo {author} {\bibfnamefont {F.~M.}\ \bibnamefont {Weaver}},
  \bibinfo {author} {\bibfnamefont {M.}~\bibnamefont {Stern}}, \bibinfo
  {author} {\bibfnamefont {K.}~\bibnamefont {Hur}}, \bibinfo {author}
  {\bibfnamefont {C.~L.}\ \bibnamefont {Harris}}, \bibinfo {author}
  {\bibfnamefont {P.}~\bibnamefont {Luo}}, \bibinfo {author} {\bibfnamefont
  {W.~J.~J.}\ \bibnamefont {Marks}}, \bibinfo {author} {\bibfnamefont
  {J.}~\bibnamefont {Rothlind}}, \bibinfo {author} {\bibfnamefont
  {O.}~\bibnamefont {Sagher}}, \bibinfo {author} {\bibfnamefont
  {C.}~\bibnamefont {Moy}}, \bibinfo {author} {\bibfnamefont {R.}~\bibnamefont
  {Pahwa}}, \bibinfo {author} {\bibfnamefont {K.}~\bibnamefont {Burchiel}},
  \bibinfo {author} {\bibfnamefont {P.}~\bibnamefont {Hogarth}}, \bibinfo
  {author} {\bibfnamefont {E.~C.}\ \bibnamefont {Lai}}, \bibinfo {author}
  {\bibfnamefont {J.~E.}\ \bibnamefont {Duda}}, \bibinfo {author}
  {\bibfnamefont {K.}~\bibnamefont {Holloway}}, \bibinfo {author}
  {\bibfnamefont {A.}~\bibnamefont {Samii}}, \bibinfo {author} {\bibfnamefont
  {S.}~\bibnamefont {Horn}}, \bibinfo {author} {\bibfnamefont {J.~M.}\
  \bibnamefont {Bronstein}}, \bibinfo {author} {\bibfnamefont {G.}~\bibnamefont
  {Stoner}}, \bibinfo {author} {\bibfnamefont {P.~A.}\ \bibnamefont {Starr}},
  \bibinfo {author} {\bibfnamefont {R.}~\bibnamefont {Simpson}}, \bibinfo
  {author} {\bibfnamefont {G.}~\bibnamefont {Baltuch}}, \bibinfo {author}
  {\bibfnamefont {A.}~\bibnamefont {De~Salles}}, \bibinfo {author}
  {\bibfnamefont {G.~D.}\ \bibnamefont {Huang}},\ and\ \bibinfo {author}
  {\bibfnamefont {D.~J.}\ \bibnamefont {Reda}},\ }\bibfield  {title} {\enquote
  {\bibinfo {title} {Pallidal versus subthalamic deep-brain stimulation for
  parkinson's disease.}}\ }\href {https://doi.org/10.1056/NEJMoa0907083}
  {\bibfield  {journal} {\bibinfo  {journal} {N Engl J Med}\ }\textbf {\bibinfo
  {volume} {362}},\ \bibinfo {pages} {2077--2091} (\bibinfo {year}
  {2010})}\BibitemShut {NoStop}%
\bibitem [{\citenamefont {Beric}\ \emph {et~al.}(2001)\citenamefont {Beric},
  \citenamefont {Kelly}, \citenamefont {Rezai}, \citenamefont {Sterio},
  \citenamefont {Mogilner}, \citenamefont {Zonenshayn},\ and\ \citenamefont
  {Kopell}}]{Beric2001}%
  \BibitemOpen
  \bibfield  {author} {\bibinfo {author} {\bibfnamefont {A.}~\bibnamefont
  {Beric}}, \bibinfo {author} {\bibfnamefont {P.~J.}\ \bibnamefont {Kelly}},
  \bibinfo {author} {\bibfnamefont {A.}~\bibnamefont {Rezai}}, \bibinfo
  {author} {\bibfnamefont {D.}~\bibnamefont {Sterio}}, \bibinfo {author}
  {\bibfnamefont {A.}~\bibnamefont {Mogilner}}, \bibinfo {author}
  {\bibfnamefont {M.}~\bibnamefont {Zonenshayn}},\ and\ \bibinfo {author}
  {\bibfnamefont {B.}~\bibnamefont {Kopell}},\ }\bibfield  {title} {\enquote
  {\bibinfo {title} {Complications of deep brain stimulation surgery},}\
  }\href@noop {} {\bibfield  {journal} {\bibinfo  {journal} {Stereotactic and
  functional neurosurgery}\ }\textbf {\bibinfo {volume} {77}},\ \bibinfo
  {pages} {73--78} (\bibinfo {year} {2001})}\BibitemShut {NoStop}%
\bibitem [{\citenamefont {Oh}\ \emph {et~al.}(2002)\citenamefont {Oh},
  \citenamefont {Abosch}, \citenamefont {Kim}, \citenamefont {Lang},\ and\
  \citenamefont {Lozano}}]{Oh2002}%
  \BibitemOpen
  \bibfield  {author} {\bibinfo {author} {\bibfnamefont {M.~Y.}\ \bibnamefont
  {Oh}}, \bibinfo {author} {\bibfnamefont {A.}~\bibnamefont {Abosch}}, \bibinfo
  {author} {\bibfnamefont {S.~H.}\ \bibnamefont {Kim}}, \bibinfo {author}
  {\bibfnamefont {A.~E.}\ \bibnamefont {Lang}},\ and\ \bibinfo {author}
  {\bibfnamefont {A.~M.}\ \bibnamefont {Lozano}},\ }\bibfield  {title}
  {\enquote {\bibinfo {title} {Long-term hardware-related complications of deep
  brain stimulation},}\ }\href@noop {} {\bibfield  {journal} {\bibinfo
  {journal} {Neurosurgery}\ }\textbf {\bibinfo {volume} {50}},\ \bibinfo
  {pages} {1268--1276} (\bibinfo {year} {2002})}\BibitemShut {NoStop}%
\bibitem [{\citenamefont {Binder}, \citenamefont {Rau},\ and\ \citenamefont
  {Starr}(2003)}]{Binder2003}%
  \BibitemOpen
  \bibfield  {author} {\bibinfo {author} {\bibfnamefont {D.~K.}\ \bibnamefont
  {Binder}}, \bibinfo {author} {\bibfnamefont {G.}~\bibnamefont {Rau}},\ and\
  \bibinfo {author} {\bibfnamefont {P.~A.}\ \bibnamefont {Starr}},\ }\bibfield
  {title} {\enquote {\bibinfo {title} {Hemorrhagic complications of
  microelectrode-guided deep brain stimulation},}\ }\href@noop {} {\bibfield
  {journal} {\bibinfo  {journal} {Stereotactic and functional neurosurgery}\
  }\textbf {\bibinfo {volume} {80}},\ \bibinfo {pages} {28--31} (\bibinfo
  {year} {2003})}\BibitemShut {NoStop}%
\bibitem [{\citenamefont {Lyons}\ \emph {et~al.}(2004)\citenamefont {Lyons},
  \citenamefont {Wilkinson}, \citenamefont {Overman},\ and\ \citenamefont
  {Pahwa}}]{Lyons2004}%
  \BibitemOpen
  \bibfield  {author} {\bibinfo {author} {\bibfnamefont {K.~E.}\ \bibnamefont
  {Lyons}}, \bibinfo {author} {\bibfnamefont {S.~B.}\ \bibnamefont
  {Wilkinson}}, \bibinfo {author} {\bibfnamefont {J.}~\bibnamefont {Overman}},\
  and\ \bibinfo {author} {\bibfnamefont {R.}~\bibnamefont {Pahwa}},\ }\bibfield
   {title} {\enquote {\bibinfo {title} {Surgical and hardware complications of
  subthalamic stimulation: a series of 160 procedures},}\ }\href@noop {}
  {\bibfield  {journal} {\bibinfo  {journal} {Neurology}\ }\textbf {\bibinfo
  {volume} {63}},\ \bibinfo {pages} {612--616} (\bibinfo {year}
  {2004})}\BibitemShut {NoStop}%
\bibitem [{\citenamefont {Uc}\ and\ \citenamefont {Follett}(2007)}]{Uc2007}%
  \BibitemOpen
  \bibfield  {author} {\bibinfo {author} {\bibfnamefont {E.~Y.}\ \bibnamefont
  {Uc}}\ and\ \bibinfo {author} {\bibfnamefont {K.~A.}\ \bibnamefont
  {Follett}},\ }\bibfield  {title} {\enquote {\bibinfo {title} {Deep brain
  stimulation in movement disorders.}}\ }\href
  {https://doi.org/10.1055/s-2007-971175} {\bibfield  {journal} {\bibinfo
  {journal} {Semin Neurol}\ }\textbf {\bibinfo {volume} {27}},\ \bibinfo
  {pages} {170--182} (\bibinfo {year} {2007})}\BibitemShut {NoStop}%
\bibitem [{\citenamefont {Baizabal-Carvallo}\ and\ \citenamefont
  {Jankovic}(2016)}]{Baizabal-Carvallo2016}%
  \BibitemOpen
  \bibfield  {author} {\bibinfo {author} {\bibfnamefont {J.}~\bibnamefont
  {Baizabal-Carvallo}}\ and\ \bibinfo {author} {\bibfnamefont {J.}~\bibnamefont
  {Jankovic}},\ }\bibfield  {title} {\enquote {\bibinfo {title} {Movement
  disorders induced by deep brain stimulation.}}\ }\href
  {https://doi.org/10.1016/j.parkreldis.2016.01.014} {\bibfield  {journal}
  {\bibinfo  {journal} {Parkinsonism Relat Disord}\ }\textbf {\bibinfo {volume}
  {25}},\ \bibinfo {pages} {1--9} (\bibinfo {year} {2016})}\BibitemShut
  {NoStop}%
\bibitem [{\citenamefont {Voon}\ and\ \citenamefont {Fox}(2007)}]{Voon2007}%
  \BibitemOpen
  \bibfield  {author} {\bibinfo {author} {\bibfnamefont {V.}~\bibnamefont
  {Voon}}\ and\ \bibinfo {author} {\bibfnamefont {S.~H.}\ \bibnamefont {Fox}},\
  }\bibfield  {title} {\enquote {\bibinfo {title} {Medication-related impulse
  control and repetitive behaviors in parkinson disease},}\ }\href@noop {}
  {\bibfield  {journal} {\bibinfo  {journal} {Archives of neurology}\ }\textbf
  {\bibinfo {volume} {64}},\ \bibinfo {pages} {1089--1096} (\bibinfo {year}
  {2007})}\BibitemShut {NoStop}%
\bibitem [{\citenamefont {Eisenstein}\ \emph {et~al.}(2014)\citenamefont
  {Eisenstein}, \citenamefont {Dewispelaere}, \citenamefont {Campbell},
  \citenamefont {Lugar}, \citenamefont {Perlmutter}, \citenamefont {Black},\
  and\ \citenamefont {Hershey}}]{Eisenstein2014}%
  \BibitemOpen
  \bibfield  {author} {\bibinfo {author} {\bibfnamefont {S.~A.}\ \bibnamefont
  {Eisenstein}}, \bibinfo {author} {\bibfnamefont {W.~B.}\ \bibnamefont
  {Dewispelaere}}, \bibinfo {author} {\bibfnamefont {M.~C.}\ \bibnamefont
  {Campbell}}, \bibinfo {author} {\bibfnamefont {H.~M.}\ \bibnamefont {Lugar}},
  \bibinfo {author} {\bibfnamefont {J.~S.}\ \bibnamefont {Perlmutter}},
  \bibinfo {author} {\bibfnamefont {K.~J.}\ \bibnamefont {Black}},\ and\
  \bibinfo {author} {\bibfnamefont {T.}~\bibnamefont {Hershey}},\ }\bibfield
  {title} {\enquote {\bibinfo {title} {Acute changes in mood induced by
  subthalamic deep brain stimulation in parkinson disease are modulated by
  psychiatric diagnosis.}}\ }\href {https://doi.org/10.1016/j.brs.2014.06.002}
  {\bibfield  {journal} {\bibinfo  {journal} {Brain Stimul}\ }\textbf {\bibinfo
  {volume} {7}},\ \bibinfo {pages} {701--708} (\bibinfo {year}
  {2014})}\BibitemShut {NoStop}%
\bibitem [{\citenamefont {Merola}\ \emph {et~al.}(2017)\citenamefont {Merola},
  \citenamefont {Romagnolo}, \citenamefont {Rizzi}, \citenamefont {Rizzone},
  \citenamefont {Zibetti}, \citenamefont {Lanotte}, \citenamefont {Mandybur},
  \citenamefont {Duker}, \citenamefont {Espay},\ and\ \citenamefont
  {Lopiano}}]{Merola2017}%
  \BibitemOpen
  \bibfield  {author} {\bibinfo {author} {\bibfnamefont {A.}~\bibnamefont
  {Merola}}, \bibinfo {author} {\bibfnamefont {A.}~\bibnamefont {Romagnolo}},
  \bibinfo {author} {\bibfnamefont {L.}~\bibnamefont {Rizzi}}, \bibinfo
  {author} {\bibfnamefont {M.~G.}\ \bibnamefont {Rizzone}}, \bibinfo {author}
  {\bibfnamefont {M.}~\bibnamefont {Zibetti}}, \bibinfo {author} {\bibfnamefont
  {M.}~\bibnamefont {Lanotte}}, \bibinfo {author} {\bibfnamefont
  {G.}~\bibnamefont {Mandybur}}, \bibinfo {author} {\bibfnamefont {A.~P.}\
  \bibnamefont {Duker}}, \bibinfo {author} {\bibfnamefont {A.~J.}\ \bibnamefont
  {Espay}},\ and\ \bibinfo {author} {\bibfnamefont {L.}~\bibnamefont
  {Lopiano}},\ }\bibfield  {title} {\enquote {\bibinfo {title} {Impulse control
  behaviors and subthalamic deep brain stimulation in parkinson disease.}}\
  }\href {https://doi.org/10.1007/s00415-016-8314-x} {\bibfield  {journal}
  {\bibinfo  {journal} {J Neurol}\ }\textbf {\bibinfo {volume} {264}},\
  \bibinfo {pages} {40--48} (\bibinfo {year} {2017})}\BibitemShut {NoStop}%
\bibitem [{\citenamefont {Healy}\ \emph {et~al.}(2022)\citenamefont {Healy},
  \citenamefont {Shepherd}, \citenamefont {Mooney}, \citenamefont {Costa},
  \citenamefont {Osman-Farah},\ and\ \citenamefont {Macerollo}}]{Healy2022}%
  \BibitemOpen
  \bibfield  {author} {\bibinfo {author} {\bibfnamefont {S.}~\bibnamefont
  {Healy}}, \bibinfo {author} {\bibfnamefont {H.}~\bibnamefont {Shepherd}},
  \bibinfo {author} {\bibfnamefont {N.}~\bibnamefont {Mooney}}, \bibinfo
  {author} {\bibfnamefont {A.~D.}\ \bibnamefont {Costa}}, \bibinfo {author}
  {\bibfnamefont {J.}~\bibnamefont {Osman-Farah}},\ and\ \bibinfo {author}
  {\bibfnamefont {A.}~\bibnamefont {Macerollo}},\ }\bibfield  {title} {\enquote
  {\bibinfo {title} {The effect of deep brain stimulation on impulse control
  related disorders in parkinson{\textquotesingle}s disease {\textendash} a
  10-year retrospective study of 137 patients},}\ }\href
  {https://doi.org/10.1016/j.jns.2022.120339} {\bibfield  {journal} {\bibinfo
  {journal} {Journal of the Neurological Sciences}\ }\textbf {\bibinfo {volume}
  {440}},\ \bibinfo {pages} {120339} (\bibinfo {year} {2022})}\BibitemShut
  {NoStop}%
\bibitem [{\citenamefont {Hammond}, \citenamefont {Bergman},\ and\
  \citenamefont {Brown}(2007)}]{Hammond2007}%
  \BibitemOpen
  \bibfield  {author} {\bibinfo {author} {\bibfnamefont {C.}~\bibnamefont
  {Hammond}}, \bibinfo {author} {\bibfnamefont {H.}~\bibnamefont {Bergman}},\
  and\ \bibinfo {author} {\bibfnamefont {P.}~\bibnamefont {Brown}},\ }\bibfield
   {title} {\enquote {\bibinfo {title} {Pathological synchronization in
  parkinson's disease: networks, models and treatments.}}\ }\href
  {https://doi.org/10.1016/j.tins.2007.05.004} {\bibfield  {journal} {\bibinfo
  {journal} {Trends Neurosci}\ }\textbf {\bibinfo {volume} {30}},\ \bibinfo
  {pages} {357--364} (\bibinfo {year} {2007})}\BibitemShut {NoStop}%
\bibitem [{\citenamefont {Tass}(1999)}]{Tass2007a}%
  \BibitemOpen
  \bibfield  {author} {\bibinfo {author} {\bibfnamefont {P.~A.}\ \bibnamefont
  {Tass}},\ }\href@noop {} {\emph {\bibinfo {title} {Phase resetting in
  medicine and biology: stochastic modelling and data analysis}}}\ (\bibinfo
  {publisher} {Springer Verlag Berlin},\ \bibinfo {year} {1999})\BibitemShut
  {NoStop}%
\bibitem [{\citenamefont {Tass}(2001)}]{Tass2001}%
  \BibitemOpen
  \bibfield  {author} {\bibinfo {author} {\bibfnamefont {P.~A.}\ \bibnamefont
  {Tass}},\ }\bibfield  {title} {\enquote {\bibinfo {title} {Desynchronizing
  double-pulse phase resetting and application to deep brain stimulation.}}\
  }\href {https://doi.org/10.1007/s004220100268} {\bibfield  {journal}
  {\bibinfo  {journal} {Biol Cybern}\ }\textbf {\bibinfo {volume} {85}},\
  \bibinfo {pages} {343--354} (\bibinfo {year} {2001})}\BibitemShut {NoStop}%
\bibitem [{\citenamefont {Tass}(2003)}]{Tass2003a}%
  \BibitemOpen
  \bibfield  {author} {\bibinfo {author} {\bibfnamefont {P.~A.}\ \bibnamefont
  {Tass}},\ }\bibfield  {title} {\enquote {\bibinfo {title} {A model of
  desynchronizing deep brain stimulation with a demand-controlled coordinated
  reset of neural subpopulations.}}\ }\href
  {https://doi.org/10.1007/s00422-003-0425-7} {\bibfield  {journal} {\bibinfo
  {journal} {Biol Cybern}\ }\textbf {\bibinfo {volume} {89}},\ \bibinfo {pages}
  {81--88} (\bibinfo {year} {2003})}\BibitemShut {NoStop}%
\bibitem [{\citenamefont {Adamchic}\ \emph {et~al.}(2014)\citenamefont
  {Adamchic}, \citenamefont {Hauptmann}, \citenamefont {Barnikol},
  \citenamefont {Pawelczyk}, \citenamefont {Popovych}, \citenamefont
  {Barnikol}, \citenamefont {Silchenko}, \citenamefont {Volkmann},
  \citenamefont {Deuschl}, \citenamefont {Meissner}, \citenamefont {Maarouf},
  \citenamefont {Sturm}, \citenamefont {Freund},\ and\ \citenamefont
  {Tass}}]{Adamchic2014}%
  \BibitemOpen
  \bibfield  {author} {\bibinfo {author} {\bibfnamefont {I.}~\bibnamefont
  {Adamchic}}, \bibinfo {author} {\bibfnamefont {C.}~\bibnamefont {Hauptmann}},
  \bibinfo {author} {\bibfnamefont {U.~B.}\ \bibnamefont {Barnikol}}, \bibinfo
  {author} {\bibfnamefont {N.}~\bibnamefont {Pawelczyk}}, \bibinfo {author}
  {\bibfnamefont {O.}~\bibnamefont {Popovych}}, \bibinfo {author}
  {\bibfnamefont {T.~T.}\ \bibnamefont {Barnikol}}, \bibinfo {author}
  {\bibfnamefont {A.}~\bibnamefont {Silchenko}}, \bibinfo {author}
  {\bibfnamefont {J.}~\bibnamefont {Volkmann}}, \bibinfo {author}
  {\bibfnamefont {G.}~\bibnamefont {Deuschl}}, \bibinfo {author} {\bibfnamefont
  {W.~G.}\ \bibnamefont {Meissner}}, \bibinfo {author} {\bibfnamefont
  {M.}~\bibnamefont {Maarouf}}, \bibinfo {author} {\bibfnamefont
  {V.}~\bibnamefont {Sturm}}, \bibinfo {author} {\bibfnamefont {H.-J.}\
  \bibnamefont {Freund}},\ and\ \bibinfo {author} {\bibfnamefont {P.~A.}\
  \bibnamefont {Tass}},\ }\bibfield  {title} {\enquote {\bibinfo {title}
  {Coordinated reset neuromodulation for {Parkinson}'s disease:
  proof-of-concept study.}}\ }\href {https://doi.org/10.1002/mds.25923}
  {\bibfield  {journal} {\bibinfo  {journal} {Mov Disord}\ }\textbf {\bibinfo
  {volume} {29}},\ \bibinfo {pages} {1679--1684} (\bibinfo {year}
  {2014})}\BibitemShut {NoStop}%
\bibitem [{\citenamefont {Bouthour}\ \emph {et~al.}(2019)\citenamefont
  {Bouthour}, \citenamefont {M{\'e}gevand}, \citenamefont {Donoghue},
  \citenamefont {L{\"u}scher}, \citenamefont {Birbaumer},\ and\ \citenamefont
  {Krack}}]{Bouthour2019}%
  \BibitemOpen
  \bibfield  {author} {\bibinfo {author} {\bibfnamefont {W.}~\bibnamefont
  {Bouthour}}, \bibinfo {author} {\bibfnamefont {P.}~\bibnamefont
  {M{\'e}gevand}}, \bibinfo {author} {\bibfnamefont {J.}~\bibnamefont
  {Donoghue}}, \bibinfo {author} {\bibfnamefont {C.}~\bibnamefont
  {L{\"u}scher}}, \bibinfo {author} {\bibfnamefont {N.}~\bibnamefont
  {Birbaumer}},\ and\ \bibinfo {author} {\bibfnamefont {P.}~\bibnamefont
  {Krack}},\ }\bibfield  {title} {\enquote {\bibinfo {title} {Biomarkers for
  closed-loop deep brain stimulation in parkinson disease and beyond.}}\ }\href
  {https://doi.org/10.1038/s41582-019-0166-4} {\bibfield  {journal} {\bibinfo
  {journal} {Nat Rev Neurol}\ }\textbf {\bibinfo {volume} {15}},\ \bibinfo
  {pages} {343--352} (\bibinfo {year} {2019})}\BibitemShut {NoStop}%
\bibitem [{\citenamefont {Hoang}\ and\ \citenamefont
  {Turner}(2019)}]{Hoang2019}%
  \BibitemOpen
  \bibfield  {author} {\bibinfo {author} {\bibfnamefont {K.~B.}\ \bibnamefont
  {Hoang}}\ and\ \bibinfo {author} {\bibfnamefont {D.~A.}\ \bibnamefont
  {Turner}},\ }\bibfield  {title} {\enquote {\bibinfo {title} {The emerging
  role of biomarkers in adaptive modulation of clinical brain stimulation.}}\
  }\href {https://doi.org/10.1093/neuros/nyz096} {\bibfield  {journal}
  {\bibinfo  {journal} {Neurosurgery}\ }\textbf {\bibinfo {volume} {85}},\
  \bibinfo {pages} {E430--E439} (\bibinfo {year} {2019})}\BibitemShut {NoStop}%
\bibitem [{\citenamefont {Beudel}\ and\ \citenamefont
  {Brown}(2016)}]{Beudel2016}%
  \BibitemOpen
  \bibfield  {author} {\bibinfo {author} {\bibfnamefont {M.}~\bibnamefont
  {Beudel}}\ and\ \bibinfo {author} {\bibfnamefont {P.}~\bibnamefont {Brown}},\
  }\bibfield  {title} {\enquote {\bibinfo {title} {Adaptive deep brain
  stimulation in parkinson's disease.}}\ }\href
  {https://doi.org/10.1016/j.parkreldis.2015.09.028} {\bibfield  {journal}
  {\bibinfo  {journal} {Parkinsonism Relat Disord}\ }\textbf {\bibinfo {volume}
  {22 Suppl 1}},\ \bibinfo {pages} {S123--6} (\bibinfo {year}
  {2016})}\BibitemShut {NoStop}%
\bibitem [{\citenamefont {Tass}\ and\ \citenamefont
  {Majtanik}(2006)}]{Tass2006}%
  \BibitemOpen
  \bibfield  {author} {\bibinfo {author} {\bibfnamefont {P.~A.}\ \bibnamefont
  {Tass}}\ and\ \bibinfo {author} {\bibfnamefont {M.}~\bibnamefont
  {Majtanik}},\ }\bibfield  {title} {\enquote {\bibinfo {title} {Long-term
  anti-kindling effects of desynchronizing brain stimulation: a theoretical
  study.}}\ }\href {https://doi.org/10.1007/s00422-005-0028-6} {\bibfield
  {journal} {\bibinfo  {journal} {Biol Cybern}\ }\textbf {\bibinfo {volume}
  {94}},\ \bibinfo {pages} {58--66} (\bibinfo {year} {2006})}\BibitemShut
  {NoStop}%
\bibitem [{\citenamefont {Madadi~Asl}\ \emph {et~al.}(2022)\citenamefont
  {Madadi~Asl}, \citenamefont {Vahabie}, \citenamefont {Valizadeh},\ and\
  \citenamefont {Tass}}]{Madadi2022}%
  \BibitemOpen
  \bibfield  {author} {\bibinfo {author} {\bibfnamefont {M.}~\bibnamefont
  {Madadi~Asl}}, \bibinfo {author} {\bibfnamefont {A.-H.}\ \bibnamefont
  {Vahabie}}, \bibinfo {author} {\bibfnamefont {A.}~\bibnamefont {Valizadeh}},\
  and\ \bibinfo {author} {\bibfnamefont {P.~A.}\ \bibnamefont {Tass}},\
  }\bibfield  {title} {\enquote {\bibinfo {title} {Spike-timing-dependent
  plasticity mediated by dopamine and its role in parkinson's disease
  pathophysiology},}\ }\href {https://doi.org/10.3389/fnetp.2022.817524} {\
  \textbf {\bibinfo {volume} {2}},\ \bibinfo {pages} {817524} (\bibinfo {year}
  {2022})}\BibitemShut {NoStop}%
\bibitem [{\citenamefont {Tass}\ and\ \citenamefont
  {Hauptmann}(2007)}]{Tass2007}%
  \BibitemOpen
  \bibfield  {author} {\bibinfo {author} {\bibfnamefont {P.~A.}\ \bibnamefont
  {Tass}}\ and\ \bibinfo {author} {\bibfnamefont {C.}~\bibnamefont
  {Hauptmann}},\ }\bibfield  {title} {\enquote {\bibinfo {title} {Therapeutic
  modulation of synaptic connectivity with desynchronizing brain
  stimulation.}}\ }\href {https://doi.org/10.1016/j.ijpsycho.2006.07.013}
  {\bibfield  {journal} {\bibinfo  {journal} {Int J Psychophysiol}\ }\textbf
  {\bibinfo {volume} {64}},\ \bibinfo {pages} {53--61} (\bibinfo {year}
  {2007})}\BibitemShut {NoStop}%
\bibitem [{\citenamefont {Zeitler}\ and\ \citenamefont
  {Tass}(2015)}]{Zeitler2015}%
  \BibitemOpen
  \bibfield  {author} {\bibinfo {author} {\bibfnamefont {M.}~\bibnamefont
  {Zeitler}}\ and\ \bibinfo {author} {\bibfnamefont {P.~A.}\ \bibnamefont
  {Tass}},\ }\bibfield  {title} {\enquote {\bibinfo {title} {Augmented brain
  function by coordinated reset stimulation with slowly varying sequences.}}\
  }\href {https://doi.org/10.3389/fnsys.2015.00049} {\bibfield  {journal}
  {\bibinfo  {journal} {Front Syst Neurosci}\ }\textbf {\bibinfo {volume}
  {9}},\ \bibinfo {pages} {49} (\bibinfo {year} {2015})}\BibitemShut {NoStop}%
\bibitem [{\citenamefont {Manos}, \citenamefont {Zeitler},\ and\ \citenamefont
  {Tass}(2018)}]{Manos2018}%
  \BibitemOpen
  \bibfield  {author} {\bibinfo {author} {\bibfnamefont {T.}~\bibnamefont
  {Manos}}, \bibinfo {author} {\bibfnamefont {M.}~\bibnamefont {Zeitler}},\
  and\ \bibinfo {author} {\bibfnamefont {P.~A.}\ \bibnamefont {Tass}},\
  }\bibfield  {title} {\enquote {\bibinfo {title} {How stimulation frequency
  and intensity impact on the long-lasting effects of coordinated reset
  stimulation},}\ }\href {https://doi.org/10.1371/journal.pcbi.1006113}
  {\bibfield  {journal} {\bibinfo  {journal} {PLoS Comput Biol}\ }\textbf
  {\bibinfo {volume} {14}},\ \bibinfo {pages} {e1006113} (\bibinfo {year}
  {2018})}\BibitemShut {NoStop}%
\bibitem [{\citenamefont {Tyulmankov}, \citenamefont {Tass},\ and\
  \citenamefont {Bokil}(2018)}]{Tyulmankov2018}%
  \BibitemOpen
  \bibfield  {author} {\bibinfo {author} {\bibfnamefont {D.}~\bibnamefont
  {Tyulmankov}}, \bibinfo {author} {\bibfnamefont {P.~A.}\ \bibnamefont
  {Tass}},\ and\ \bibinfo {author} {\bibfnamefont {H.}~\bibnamefont {Bokil}},\
  }\bibfield  {title} {\enquote {\bibinfo {title} {Periodic flashing
  coordinated reset stimulation paradigm reduces sensitivity to on and off
  period durations},}\ }\href {https://doi.org/10.1371/journal.pone.0203782}
  {\bibfield  {journal} {\bibinfo  {journal} {PLoS ONE}\ }\textbf {\bibinfo
  {volume} {13}},\ \bibinfo {pages} {e0203782} (\bibinfo {year}
  {2018})}\BibitemShut {NoStop}%
\bibitem [{\citenamefont {Kromer}\ and\ \citenamefont
  {Tass}(2020)}]{Kromer2020}%
  \BibitemOpen
  \bibfield  {author} {\bibinfo {author} {\bibfnamefont {J.~A.}\ \bibnamefont
  {Kromer}}\ and\ \bibinfo {author} {\bibfnamefont {P.~A.}\ \bibnamefont
  {Tass}},\ }\bibfield  {title} {\enquote {\bibinfo {title} {Long-lasting
  desynchronization by decoupling stimulation},}\ }\href
  {https://doi.org/10.1103/PhysRevResearch.2.033101} {\bibfield  {journal}
  {\bibinfo  {journal} {Phys. Rev. Res.}\ }\textbf {\bibinfo {volume} {2}},\
  \bibinfo {pages} {033101} (\bibinfo {year} {2020})}\BibitemShut {NoStop}%
\bibitem [{\citenamefont {Khaledi-Nasab}, \citenamefont {Kromer},\ and\
  \citenamefont {Tass}(2021)}]{Khaledi-Nasab2020}%
  \BibitemOpen
  \bibfield  {author} {\bibinfo {author} {\bibfnamefont {A.}~\bibnamefont
  {Khaledi-Nasab}}, \bibinfo {author} {\bibfnamefont {J.~A.}\ \bibnamefont
  {Kromer}},\ and\ \bibinfo {author} {\bibfnamefont {P.~A.}\ \bibnamefont
  {Tass}},\ }\bibfield  {title} {\enquote {\bibinfo {title} {Long-lasting
  desynchronization of plastic neural networks by random reset stimulation.}}\
  }\href {https://doi.org/10.3389/fphys.2020.622620} {\bibfield  {journal}
  {\bibinfo  {journal} {Front Physiol}\ }\textbf {\bibinfo {volume} {11}},\
  \bibinfo {pages} {622620} (\bibinfo {year} {2021})}\BibitemShut {NoStop}%
\bibitem [{\citenamefont {Kromer}\ and\ \citenamefont
  {Tass}(2022)}]{Kromer2022}%
  \BibitemOpen
  \bibfield  {author} {\bibinfo {author} {\bibfnamefont {J.~A.}\ \bibnamefont
  {Kromer}}\ and\ \bibinfo {author} {\bibfnamefont {P.~A.}\ \bibnamefont
  {Tass}},\ }\bibfield  {title} {\enquote {\bibinfo {title} {Synaptic reshaping
  of plastic neuronal networks by periodic multichannel stimulation with
  single-pulse and burst stimuli},}\ }\href {https://doi.org/10.1371/journal.
  pcbi.1010568} {\bibfield  {journal} {\bibinfo  {journal} {PLoS Comput Biol}\
  }\textbf {\bibinfo {volume} {18}},\ \bibinfo {pages} {e1010568} (\bibinfo
  {year} {2022})}\BibitemShut {NoStop}%
\bibitem [{\citenamefont {Asl}, \citenamefont {Valizadeh},\ and\ \citenamefont
  {Tass}(2023)}]{Madadi2023}%
  \BibitemOpen
  \bibfield  {author} {\bibinfo {author} {\bibfnamefont {M.~M.}\ \bibnamefont
  {Asl}}, \bibinfo {author} {\bibfnamefont {A.}~\bibnamefont {Valizadeh}},\
  and\ \bibinfo {author} {\bibfnamefont {P.~A.}\ \bibnamefont {Tass}},\
  }\bibfield  {title} {\enquote {\bibinfo {title} {Decoupling of interacting
  neuronal populations by time-shifted stimulation through
  spike-timing-dependent plasticity},}\ }\href
  {https://doi.org/10.1371/journal.pcbi.1010853} {\bibfield  {journal}
  {\bibinfo  {journal} {PLoS Comput Biol}\ }\textbf {\bibinfo {volume} {19}},\
  \bibinfo {pages} {e1010853} (\bibinfo {year} {2023})}\BibitemShut {NoStop}%
\bibitem [{\citenamefont {Hauptmann}\ and\ \citenamefont
  {Tass}(2009)}]{Hauptmann2009}%
  \BibitemOpen
  \bibfield  {author} {\bibinfo {author} {\bibfnamefont {C.}~\bibnamefont
  {Hauptmann}}\ and\ \bibinfo {author} {\bibfnamefont {P.~A.}\ \bibnamefont
  {Tass}},\ }\bibfield  {title} {\enquote {\bibinfo {title} {Cumulative and
  after-effects of short and weak coordinated reset stimulation: a modeling
  study.}}\ }\href {https://doi.org/10.1088/1741-2560/6/1/016004} {\bibfield
  {journal} {\bibinfo  {journal} {J Neural Eng}\ }\textbf {\bibinfo {volume}
  {6}},\ \bibinfo {pages} {016004} (\bibinfo {year} {2009})}\BibitemShut
  {NoStop}%
\bibitem [{\citenamefont {Popovych}\ and\ \citenamefont
  {Tass}(2012)}]{Popovych2012}%
  \BibitemOpen
  \bibfield  {author} {\bibinfo {author} {\bibfnamefont {O.~V.}\ \bibnamefont
  {Popovych}}\ and\ \bibinfo {author} {\bibfnamefont {P.~A.}\ \bibnamefont
  {Tass}},\ }\bibfield  {title} {\enquote {\bibinfo {title} {Desynchronizing
  electrical and sensory coordinated reset neuromodulation.}}\ }\href
  {https://doi.org/10.3389/fnhum.2012.00058} {\bibfield  {journal} {\bibinfo
  {journal} {Front Hum Neurosci}\ }\textbf {\bibinfo {volume} {6}},\ \bibinfo
  {pages} {58} (\bibinfo {year} {2012})}\BibitemShut {NoStop}%
\bibitem [{\citenamefont {Tass}\ \emph {et~al.}(2012)\citenamefont {Tass},
  \citenamefont {Qin}, \citenamefont {Hauptmann}, \citenamefont {Dovero},
  \citenamefont {Bezard}, \citenamefont {Boraud},\ and\ \citenamefont
  {Meissner}}]{Tass2012}%
  \BibitemOpen
  \bibfield  {author} {\bibinfo {author} {\bibfnamefont {P.~A.}\ \bibnamefont
  {Tass}}, \bibinfo {author} {\bibfnamefont {L.}~\bibnamefont {Qin}}, \bibinfo
  {author} {\bibfnamefont {C.}~\bibnamefont {Hauptmann}}, \bibinfo {author}
  {\bibfnamefont {S.}~\bibnamefont {Dovero}}, \bibinfo {author} {\bibfnamefont
  {E.}~\bibnamefont {Bezard}}, \bibinfo {author} {\bibfnamefont
  {T.}~\bibnamefont {Boraud}},\ and\ \bibinfo {author} {\bibfnamefont {W.~G.}\
  \bibnamefont {Meissner}},\ }\bibfield  {title} {\enquote {\bibinfo {title}
  {Coordinated reset has sustained aftereffects in parkinsonian monkeys.}}\
  }\href {https://doi.org/10.1002/ana.23663} {\bibfield  {journal} {\bibinfo
  {journal} {Ann Neurol}\ }\textbf {\bibinfo {volume} {72}},\ \bibinfo {pages}
  {816--820} (\bibinfo {year} {2012})}\BibitemShut {NoStop}%
\bibitem [{\citenamefont {Butz}, \citenamefont {W{\"o}rg{\"o}tter},\ and\
  \citenamefont {van Ooyen}(2009)}]{BUT09}%
  \BibitemOpen
  \bibfield  {author} {\bibinfo {author} {\bibfnamefont {M.}~\bibnamefont
  {Butz}}, \bibinfo {author} {\bibfnamefont {F.}~\bibnamefont
  {W{\"o}rg{\"o}tter}},\ and\ \bibinfo {author} {\bibfnamefont
  {A.}~\bibnamefont {van Ooyen}},\ }\bibfield  {title} {\enquote {\bibinfo
  {title} {Activity-dependent structural plasticity},}\ }\href
  {https://doi.org/10.1016/j.brainresrev.2008.12.023} {\bibfield  {journal}
  {\bibinfo  {journal} {Brain Res. Rev.}\ }\textbf {\bibinfo {volume} {60}},\
  \bibinfo {pages} {287} (\bibinfo {year} {2009})}\BibitemShut {NoStop}%
\bibitem [{\citenamefont {Butz}\ and\ \citenamefont {van
  Ooyen}(2013)}]{BUT13c}%
  \BibitemOpen
  \bibfield  {author} {\bibinfo {author} {\bibfnamefont {M.}~\bibnamefont
  {Butz}}\ and\ \bibinfo {author} {\bibfnamefont {A.}~\bibnamefont {van
  Ooyen}},\ }\bibfield  {title} {\enquote {\bibinfo {title} {A simple rule for
  dendritic spine and axonal bouton formation can account for cortical
  reorganization after focal retinal lesions},}\ }\href
  {https://doi.org/10.1371/journal.pcbi.1003259} {\bibfield  {journal}
  {\bibinfo  {journal} {PLoS Comput. Biol.}\ }\textbf {\bibinfo {volume} {9}},\
  \bibinfo {pages} {e1003259} (\bibinfo {year} {2013})}\BibitemShut {NoStop}%
\bibitem [{\citenamefont {Manos}, \citenamefont {Diaz-Pier},\ and\
  \citenamefont {Tass}(2021)}]{Manos2021}%
  \BibitemOpen
  \bibfield  {author} {\bibinfo {author} {\bibfnamefont {T.}~\bibnamefont
  {Manos}}, \bibinfo {author} {\bibfnamefont {S.}~\bibnamefont {Diaz-Pier}},\
  and\ \bibinfo {author} {\bibfnamefont {P.~A.}\ \bibnamefont {Tass}},\
  }\bibfield  {title} {\enquote {\bibinfo {title} {Long-term desynchronization
  by coordinated reset stimulation in a neural network model with synaptic and
  structural plasticity.}}\ }\href {https://doi.org/10.3389/fphys.2021.716556}
  {\bibfield  {journal} {\bibinfo  {journal} {Front Physiol}\ }\textbf
  {\bibinfo {volume} {12}},\ \bibinfo {pages} {716556} (\bibinfo {year}
  {2021})}\BibitemShut {NoStop}%
\bibitem [{\citenamefont {Chauhan}\ \emph {et~al.}(2022)\citenamefont
  {Chauhan}, \citenamefont {Khaledi-Nasab}, \citenamefont {Neiman},\ and\
  \citenamefont {Tass}}]{Chauhan2022}%
  \BibitemOpen
  \bibfield  {author} {\bibinfo {author} {\bibfnamefont {K.}~\bibnamefont
  {Chauhan}}, \bibinfo {author} {\bibfnamefont {A.}~\bibnamefont
  {Khaledi-Nasab}}, \bibinfo {author} {\bibfnamefont {A.~B.}\ \bibnamefont
  {Neiman}},\ and\ \bibinfo {author} {\bibfnamefont {P.~A.}\ \bibnamefont
  {Tass}},\ }\bibfield  {title} {\enquote {\bibinfo {title} {Dynamics of phase
  oscillator networks with synaptic weight and structural plasticity},}\ }\href
  {https://doi.org/10.1038/s41598-022-19417-9} {\bibfield  {journal} {\bibinfo
  {journal} {Scientific Reports}\ }\textbf {\bibinfo {volume} {12}},\ \bibinfo
  {pages} {15003} (\bibinfo {year} {2022})}\BibitemShut {NoStop}%
\bibitem [{\citenamefont {Maistrenko}\ \emph
  {et~al.}(2007{\natexlab{b}})\citenamefont {Maistrenko}, \citenamefont
  {Lysyansky}, \citenamefont {Hauptmann}, \citenamefont {Burylko},\ and\
  \citenamefont {Tass}}]{Maistrenko2007}%
  \BibitemOpen
  \bibfield  {author} {\bibinfo {author} {\bibfnamefont {Y.~L.}\ \bibnamefont
  {Maistrenko}}, \bibinfo {author} {\bibfnamefont {B.}~\bibnamefont
  {Lysyansky}}, \bibinfo {author} {\bibfnamefont {C.}~\bibnamefont
  {Hauptmann}}, \bibinfo {author} {\bibfnamefont {O.}~\bibnamefont {Burylko}},\
  and\ \bibinfo {author} {\bibfnamefont {P.~A.}\ \bibnamefont {Tass}},\
  }\bibfield  {title} {\enquote {\bibinfo {title} {{Multistability in the
  Kuramoto model with synaptic plasticity}},}\ }\href
  {https://doi.org/10.1103/PhysRevE.75.066207} {\bibfield  {journal} {\bibinfo
  {journal} {Phys Rev E}\ }\textbf {\bibinfo {volume} {75}},\ \bibinfo {pages}
  {066207} (\bibinfo {year} {2007}{\natexlab{b}})}\BibitemShut {NoStop}%
\bibitem [{\citenamefont {Wang}\ \emph {et~al.}(2016)\citenamefont {Wang},
  \citenamefont {Nebeck}, \citenamefont {Muralidharan}, \citenamefont
  {Johnson}, \citenamefont {Vitek},\ and\ \citenamefont {Baker}}]{Wang2016}%
  \BibitemOpen
  \bibfield  {author} {\bibinfo {author} {\bibfnamefont {J.}~\bibnamefont
  {Wang}}, \bibinfo {author} {\bibfnamefont {S.}~\bibnamefont {Nebeck}},
  \bibinfo {author} {\bibfnamefont {A.}~\bibnamefont {Muralidharan}}, \bibinfo
  {author} {\bibfnamefont {M.~D.}\ \bibnamefont {Johnson}}, \bibinfo {author}
  {\bibfnamefont {J.~L.}\ \bibnamefont {Vitek}},\ and\ \bibinfo {author}
  {\bibfnamefont {K.~B.}\ \bibnamefont {Baker}},\ }\bibfield  {title} {\enquote
  {\bibinfo {title} {Coordinated reset deep brain stimulation of subthalamic
  nucleus produces long-lasting, dose-dependent motor improvements in the
  1-methyl-4-phenyl-1,2,3,6-tetrahydropyridine non-human primate model of
  parkinsonism.}}\ }\href {https://doi.org/10.1016/j.brs.2016.03.014}
  {\bibfield  {journal} {\bibinfo  {journal} {Brain Stimul}\ }\textbf {\bibinfo
  {volume} {9}},\ \bibinfo {pages} {609--617} (\bibinfo {year}
  {2016})}\BibitemShut {NoStop}%
\bibitem [{\citenamefont {Wang}\ \emph {et~al.}(2022)\citenamefont {Wang},
  \citenamefont {Fergus}, \citenamefont {Johnson}, \citenamefont {Nebeck},
  \citenamefont {Zhang}, \citenamefont {Kulkarni}, \citenamefont {Bokil},
  \citenamefont {Molnar},\ and\ \citenamefont {Vitek}}]{Wang2022}%
  \BibitemOpen
  \bibfield  {author} {\bibinfo {author} {\bibfnamefont {J.}~\bibnamefont
  {Wang}}, \bibinfo {author} {\bibfnamefont {S.~P.}\ \bibnamefont {Fergus}},
  \bibinfo {author} {\bibfnamefont {L.~A.}\ \bibnamefont {Johnson}}, \bibinfo
  {author} {\bibfnamefont {S.~D.}\ \bibnamefont {Nebeck}}, \bibinfo {author}
  {\bibfnamefont {J.}~\bibnamefont {Zhang}}, \bibinfo {author} {\bibfnamefont
  {S.}~\bibnamefont {Kulkarni}}, \bibinfo {author} {\bibfnamefont
  {H.}~\bibnamefont {Bokil}}, \bibinfo {author} {\bibfnamefont {G.~F.}\
  \bibnamefont {Molnar}},\ and\ \bibinfo {author} {\bibfnamefont {J.~L.}\
  \bibnamefont {Vitek}},\ }\bibfield  {title} {\enquote {\bibinfo {title}
  {Shuffling improves the acute and carryover effect of subthalamic coordinated
  reset deep brain stimulation.}}\ }\href
  {https://doi.org/10.3389/fneur.2022.716046} {\bibfield  {journal} {\bibinfo
  {journal} {Front Neurol}\ }\textbf {\bibinfo {volume} {13}},\ \bibinfo
  {pages} {716046} (\bibinfo {year} {2022})}\BibitemShut {NoStop}%
\bibitem [{\citenamefont {Chelangat~Bore}\ \emph {et~al.}(2022)\citenamefont
  {Chelangat~Bore}, \citenamefont {A~Campbell}, \citenamefont {Cho},
  \citenamefont {Pucci}, \citenamefont {Gopalakrishnan}, \citenamefont
  {G~Machado},\ and\ \citenamefont {B~Baker}}]{Chelangat-Bore2022}%
  \BibitemOpen
  \bibfield  {author} {\bibinfo {author} {\bibfnamefont {J.}~\bibnamefont
  {Chelangat~Bore}}, \bibinfo {author} {\bibfnamefont {B.}~\bibnamefont
  {A~Campbell}}, \bibinfo {author} {\bibfnamefont {H.}~\bibnamefont {Cho}},
  \bibinfo {author} {\bibfnamefont {F.}~\bibnamefont {Pucci}}, \bibinfo
  {author} {\bibfnamefont {R.}~\bibnamefont {Gopalakrishnan}}, \bibinfo
  {author} {\bibfnamefont {A.}~\bibnamefont {G~Machado}},\ and\ \bibinfo
  {author} {\bibfnamefont {K.}~\bibnamefont {B~Baker}},\ }\bibfield  {title}
  {\enquote {\bibinfo {title} {Long-lasting effects of subthalamic nucleus
  coordinated reset deep brain stimulation in the non-human primate model of
  parkinsonism: A case report.}}\ }\href
  {https://doi.org/10.1016/j.brs.2022.04.005} {\bibfield  {journal} {\bibinfo
  {journal} {Brain Stimul}\ }\textbf {\bibinfo {volume} {15}},\ \bibinfo
  {pages} {598--600} (\bibinfo {year} {2022})}\BibitemShut {NoStop}%
\bibitem [{\citenamefont {Tass}(2017)}]{Tass2017}%
  \BibitemOpen
  \bibfield  {author} {\bibinfo {author} {\bibfnamefont {P.~A.}\ \bibnamefont
  {Tass}},\ }\bibfield  {title} {\enquote {\bibinfo {title} {Vibrotactile
  coordinated reset stimulation for the treatment of neurological diseases:
  Concepts and device specifications.}}\ }\href
  {https://doi.org/10.7759/cureus.1535} {\bibfield  {journal} {\bibinfo
  {journal} {Cureus}\ }\textbf {\bibinfo {volume} {9}},\ \bibinfo {pages}
  {e1535} (\bibinfo {year} {2017})}\BibitemShut {NoStop}%
\bibitem [{\citenamefont {Syrkin-Nikolau}\ \emph {et~al.}(2018)\citenamefont
  {Syrkin-Nikolau}, \citenamefont {Neuville}, \citenamefont {O'Day},
  \citenamefont {Anidi}, \citenamefont {Miller~Koop}, \citenamefont {Martin},
  \citenamefont {Tass},\ and\ \citenamefont
  {Bronte-Stewart}}]{Syrkin-Nikolau2018}%
  \BibitemOpen
  \bibfield  {author} {\bibinfo {author} {\bibfnamefont {J.}~\bibnamefont
  {Syrkin-Nikolau}}, \bibinfo {author} {\bibfnamefont {R.}~\bibnamefont
  {Neuville}}, \bibinfo {author} {\bibfnamefont {J.}~\bibnamefont {O'Day}},
  \bibinfo {author} {\bibfnamefont {C.}~\bibnamefont {Anidi}}, \bibinfo
  {author} {\bibfnamefont {M.}~\bibnamefont {Miller~Koop}}, \bibinfo {author}
  {\bibfnamefont {T.}~\bibnamefont {Martin}}, \bibinfo {author} {\bibfnamefont
  {P.~A.}\ \bibnamefont {Tass}},\ and\ \bibinfo {author} {\bibfnamefont
  {H.}~\bibnamefont {Bronte-Stewart}},\ }\bibfield  {title} {\enquote {\bibinfo
  {title} {Coordinated reset vibrotactile stimulation shows prolonged
  improvement in {Parkinson}'s disease.}}\ }\href
  {https://doi.org/10.1002/mds.27223} {\bibfield  {journal} {\bibinfo
  {journal} {Mov Disord}\ }\textbf {\bibinfo {volume} {33}},\ \bibinfo {pages}
  {179--180} (\bibinfo {year} {2018})}\BibitemShut {NoStop}%
\bibitem [{\citenamefont {Pfeifer}\ \emph {et~al.}(2021)\citenamefont
  {Pfeifer}, \citenamefont {Kromer}, \citenamefont {Cook}, \citenamefont
  {Hornbeck}, \citenamefont {Lim}, \citenamefont {Mortimer}, \citenamefont
  {Fogarty}, \citenamefont {Han}, \citenamefont {Dhall}, \citenamefont
  {Halpern},\ and\ \citenamefont {Tass}}]{Pfeifer2021}%
  \BibitemOpen
  \bibfield  {author} {\bibinfo {author} {\bibfnamefont {K.~J.}\ \bibnamefont
  {Pfeifer}}, \bibinfo {author} {\bibfnamefont {J.~A.}\ \bibnamefont {Kromer}},
  \bibinfo {author} {\bibfnamefont {A.~J.}\ \bibnamefont {Cook}}, \bibinfo
  {author} {\bibfnamefont {T.}~\bibnamefont {Hornbeck}}, \bibinfo {author}
  {\bibfnamefont {E.~A.}\ \bibnamefont {Lim}}, \bibinfo {author} {\bibfnamefont
  {B.~J.~P.}\ \bibnamefont {Mortimer}}, \bibinfo {author} {\bibfnamefont
  {A.~S.}\ \bibnamefont {Fogarty}}, \bibinfo {author} {\bibfnamefont {S.~S.}\
  \bibnamefont {Han}}, \bibinfo {author} {\bibfnamefont {R.}~\bibnamefont
  {Dhall}}, \bibinfo {author} {\bibfnamefont {C.~H.}\ \bibnamefont {Halpern}},\
  and\ \bibinfo {author} {\bibfnamefont {P.~A.}\ \bibnamefont {Tass}},\
  }\bibfield  {title} {\enquote {\bibinfo {title} {Coordinated reset
  vibrotactile stimulation induces sustained cumulative benefits in
  {Parkinson}'s disease.}}\ }\href {https://doi.org/10.3389/fphys.2021.624317}
  {\bibfield  {journal} {\bibinfo  {journal} {Front Physiol}\ }\textbf
  {\bibinfo {volume} {12}},\ \bibinfo {pages} {624317} (\bibinfo {year}
  {2021})}\BibitemShut {NoStop}%
\bibitem [{\citenamefont {Tass}(2021)}]{Tass2022}%
  \BibitemOpen
  \bibfield  {author} {\bibinfo {author} {\bibfnamefont {P.~A.}\ \bibnamefont
  {Tass}},\ }\bibfield  {title} {\enquote {\bibinfo {title} {Vibrotactile
  coordinated reset stimulation for the treatment of {Parkinson}'s disease.}}\
  }\href {https://doi.org/10.4103/1673-5374.329001} {\bibfield  {journal}
  {\bibinfo  {journal} {Neural Regen Res}\ }\textbf {\bibinfo {volume} {17}},\
  \bibinfo {pages} {1495--1497} (\bibinfo {year} {2021})}\BibitemShut {NoStop}%
\bibitem [{\citenamefont {Esposito}, \citenamefont {Di~Matteo},\ and\
  \citenamefont {Di~Giovanni}(2007)}]{Esposito2007}%
  \BibitemOpen
  \bibfield  {author} {\bibinfo {author} {\bibfnamefont {E.}~\bibnamefont
  {Esposito}}, \bibinfo {author} {\bibfnamefont {V.}~\bibnamefont
  {Di~Matteo}},\ and\ \bibinfo {author} {\bibfnamefont {G.}~\bibnamefont
  {Di~Giovanni}},\ }\bibfield  {title} {\enquote {\bibinfo {title} {Death in
  the substantia nigra: a motor tragedy.}}\ }\href
  {https://doi.org/10.1586/14737175.7.6.677} {\bibfield  {journal} {\bibinfo
  {journal} {Expert Rev Neurother}\ }\textbf {\bibinfo {volume} {7}},\ \bibinfo
  {pages} {677--697} (\bibinfo {year} {2007})}\BibitemShut {NoStop}%
\bibitem [{\citenamefont {Haelterman}\ \emph {et~al.}(2014)\citenamefont
  {Haelterman}, \citenamefont {Yoon}, \citenamefont {Sandoval}, \citenamefont
  {Jaiswal}, \citenamefont {Shulman},\ and\ \citenamefont
  {Bellen}}]{Haelterman2014}%
  \BibitemOpen
  \bibfield  {author} {\bibinfo {author} {\bibfnamefont {N.~A.}\ \bibnamefont
  {Haelterman}}, \bibinfo {author} {\bibfnamefont {W.~H.}\ \bibnamefont
  {Yoon}}, \bibinfo {author} {\bibfnamefont {H.}~\bibnamefont {Sandoval}},
  \bibinfo {author} {\bibfnamefont {M.}~\bibnamefont {Jaiswal}}, \bibinfo
  {author} {\bibfnamefont {J.~M.}\ \bibnamefont {Shulman}},\ and\ \bibinfo
  {author} {\bibfnamefont {H.~J.}\ \bibnamefont {Bellen}},\ }\bibfield  {title}
  {\enquote {\bibinfo {title} {A mitocentric view of parkinson's disease.}}\
  }\href {https://doi.org/10.1146/annurev-neuro-071013-014317} {\bibfield
  {journal} {\bibinfo  {journal} {Annu Rev Neurosci}\ }\textbf {\bibinfo
  {volume} {37}},\ \bibinfo {pages} {137--159} (\bibinfo {year}
  {2014})}\BibitemShut {NoStop}%
\bibitem [{\citenamefont {McGregor}\ and\ \citenamefont
  {Nelson}(2019)}]{McGregor2019}%
  \BibitemOpen
  \bibfield  {author} {\bibinfo {author} {\bibfnamefont {M.~M.}\ \bibnamefont
  {McGregor}}\ and\ \bibinfo {author} {\bibfnamefont {A.~B.}\ \bibnamefont
  {Nelson}},\ }\bibfield  {title} {\enquote {\bibinfo {title} {Circuit
  mechanisms of parkinson's disease.}}\ }\href
  {https://doi.org/10.1016/j.neuron.2019.03.004} {\bibfield  {journal}
  {\bibinfo  {journal} {Neuron}\ }\textbf {\bibinfo {volume} {101}},\ \bibinfo
  {pages} {1042--1056} (\bibinfo {year} {2019})}\BibitemShut {NoStop}%
\bibitem [{\citenamefont {Bader}(2019{\natexlab{a}})}]{Bader2019b}%
  \BibitemOpen
  \bibinfo {editor} {\bibfnamefont {R.}~\bibnamefont {Bader}},\ ed.,\
  \href@noop {} {\emph {\bibinfo {title} {Springer handbook of systematic
  musicology}}},\ Springer Handbooks\ (\bibinfo  {publisher} {Springer},\
  \bibinfo {address} {Berlin, Germany},\ \bibinfo {year} {2019})\BibitemShut
  {NoStop}%
\bibitem [{\citenamefont {Briot}, \citenamefont {Hadjeres},\ and\ \citenamefont
  {Pachet}(2020)}]{Briot2020}%
  \BibitemOpen
  \bibfield  {author} {\bibinfo {author} {\bibfnamefont {J.-P.}\ \bibnamefont
  {Briot}}, \bibinfo {author} {\bibfnamefont {G.}~\bibnamefont {Hadjeres}},\
  and\ \bibinfo {author} {\bibfnamefont {F.-D.}\ \bibnamefont {Pachet}},\
  }\href {https://doi.org/10.1007/978-3-319-70163-9} {\emph {\bibinfo {title}
  {Deep Learning Techniques for Music Generation}}}\ (\bibinfo  {publisher}
  {Springer International Publishing},\ \bibinfo {year} {2020})\BibitemShut
  {NoStop}%
\bibitem [{\citenamefont {Kohonen}(1995)}]{Kohonen1995}%
  \BibitemOpen
  \bibfield  {author} {\bibinfo {author} {\bibfnamefont {T.}~\bibnamefont
  {Kohonen}},\ }\href {https://doi.org/10.1007/978-3-642-97610-0} {\emph
  {\bibinfo {title} {Self-Organizing Maps}}}\ (\bibinfo  {publisher} {Springer
  Berlin Heidelberg},\ \bibinfo {year} {1995})\BibitemShut {NoStop}%
\bibitem [{\citenamefont {Bla{\ss}}, \citenamefont {Fischer},\ and\
  \citenamefont {Plath}(2020)}]{Bla2020}%
  \BibitemOpen
  \bibfield  {author} {\bibinfo {author} {\bibfnamefont {M.}~\bibnamefont
  {Bla{\ss}}}, \bibinfo {author} {\bibfnamefont {J.~L.}\ \bibnamefont
  {Fischer}},\ and\ \bibinfo {author} {\bibfnamefont {N.}~\bibnamefont
  {Plath}},\ }\bibfield  {title} {\enquote {\bibinfo {title} {Computational
  phonogram archiving},}\ }\href {https://doi.org/10.1063/pt.3.4636} {\bibfield
   {journal} {\bibinfo  {journal} {Physics Today}\ }\textbf {\bibinfo {volume}
  {73}},\ \bibinfo {pages} {50--55} (\bibinfo {year} {2020})}\BibitemShut
  {NoStop}%
\bibitem [{\citenamefont {Bader}(2019{\natexlab{b}})}]{Bader2019a}%
  \BibitemOpen
  \bibinfo {editor} {\bibfnamefont {R.}~\bibnamefont {Bader}},\ ed.,\
  \href@noop {} {\emph {\bibinfo {title} {Computational Phonogram
  Archiving}}},\ \bibinfo {edition} {1st}\ ed.,\ Current Research in Systematic
  Musicology\ (\bibinfo  {publisher} {Springer Nature},\ \bibinfo {address}
  {Cham, Switzerland},\ \bibinfo {year} {2019})\BibitemShut {NoStop}%
\bibitem [{\citenamefont {Leman}\ and\ \citenamefont
  {Carreras}(1997)}]{Leman1997}%
  \BibitemOpen
  \bibfield  {author} {\bibinfo {author} {\bibfnamefont {M.}~\bibnamefont
  {Leman}}\ and\ \bibinfo {author} {\bibfnamefont {F.}~\bibnamefont
  {Carreras}},\ }\bibfield  {title} {\enquote {\bibinfo {title} {Schema and
  gestalt: Testing the hypothesis of psychoneural isomorphism by computer
  simulation},}\ }in\ \href {https://doi.org/10.1007/bfb0034112} {\emph
  {\bibinfo {booktitle} {Music, Gestalt, and Computing}}}\ (\bibinfo
  {publisher} {Springer Berlin Heidelberg},\ \bibinfo {year} {1997})\ pp.\
  \bibinfo {pages} {144--168}\BibitemShut {NoStop}%
\bibitem [{\citenamefont {Bader}(2021)}]{Bader2021}%
  \BibitemOpen
  \bibfield  {author} {\bibinfo {author} {\bibfnamefont {R.}~\bibnamefont
  {Bader}},\ }\href {https://doi.org/10.1007/978-3-030-67155-6} {\emph
  {\bibinfo {title} {How Music Works}}}\ (\bibinfo  {publisher} {Springer
  International Publishing},\ \bibinfo {year} {2021})\BibitemShut {NoStop}%
\bibitem [{\citenamefont {Bader}(2013)}]{Bader2013}%
  \BibitemOpen
  \bibfield  {author} {\bibinfo {author} {\bibfnamefont {R.}~\bibnamefont
  {Bader}},\ }\href {https://doi.org/10.1007/978-3-642-36098-5} {\emph
  {\bibinfo {title} {Nonlinearities and Synchronization in Musical Acoustics
  and Music Psychology}}}\ (\bibinfo  {publisher} {Springer Berlin
  Heidelberg},\ \bibinfo {year} {2013})\BibitemShut {NoStop}%
\bibitem [{\citenamefont {Linke}, \citenamefont {Bader},\ and\ \citenamefont
  {Mores}(2019)}]{Linke2019}%
  \BibitemOpen
  \bibfield  {author} {\bibinfo {author} {\bibfnamefont {S.}~\bibnamefont
  {Linke}}, \bibinfo {author} {\bibfnamefont {R.}~\bibnamefont {Bader}},\ and\
  \bibinfo {author} {\bibfnamefont {R.}~\bibnamefont {Mores}},\ }\bibfield
  {title} {\enquote {\bibinfo {title} {The impulse pattern formulation ({IPF})
  as a model of musical instruments{\textemdash}investigation of stability and
  limits},}\ }\href {https://doi.org/10.1063/1.5092511} {\bibfield  {journal}
  {\bibinfo  {journal} {Chaos: An Interdisciplinary Journal of Nonlinear
  Science}\ }\textbf {\bibinfo {volume} {29}},\ \bibinfo {pages} {103109}
  (\bibinfo {year} {2019})}\BibitemShut {NoStop}%
\bibitem [{\citenamefont {Linke}, \citenamefont {Bader},\ and\ \citenamefont
  {Mores}(2021{\natexlab{a}})}]{Linke2021}%
  \BibitemOpen
  \bibfield  {author} {\bibinfo {author} {\bibfnamefont {S.}~\bibnamefont
  {Linke}}, \bibinfo {author} {\bibfnamefont {R.}~\bibnamefont {Bader}},\ and\
  \bibinfo {author} {\bibfnamefont {R.}~\bibnamefont {Mores}},\ }\bibfield
  {title} {\enquote {\bibinfo {title} {Influence of the supporting table on
  initial transients of the fretted zither: An impulse pattern formulation
  model},}\ }in\ \href {https://doi.org/10.1121/2.0001494} {\emph {\bibinfo
  {booktitle} {Proceedings of Meetings on Acoustics}}}\ (\bibinfo  {publisher}
  {{ASA}},\ \bibinfo {year} {2021})\BibitemShut {NoStop}%
\bibitem [{\citenamefont {Linke}, \citenamefont {Bader},\ and\ \citenamefont
  {Mores}(2021{\natexlab{b}})}]{Linke2021a}%
  \BibitemOpen
  \bibfield  {author} {\bibinfo {author} {\bibfnamefont {S.}~\bibnamefont
  {Linke}}, \bibinfo {author} {\bibfnamefont {R.}~\bibnamefont {Bader}},\ and\
  \bibinfo {author} {\bibfnamefont {R.}~\bibnamefont {Mores}},\ }\href
  {https://doi.org/10.48550/ARXIV.2112.03218} {\enquote {\bibinfo {title}
  {Modeling synchronization in human musical rhythms using impulse pattern
  formulation (ipf)},}\ } (\bibinfo {year} {2021}{\natexlab{b}})\BibitemShut
  {NoStop}%
\bibitem [{\citenamefont {Bader}(2022)}]{Bader2022}%
  \BibitemOpen
  \bibfield  {author} {\bibinfo {author} {\bibfnamefont {R.}~\bibnamefont
  {Bader}},\ }\href {https://doi.org/10.48550/ARXIV.2212.11021} {\enquote
  {\bibinfo {title} {Impulse pattern formulation (ipf) brain model},}\ }
  (\bibinfo {year} {2022})\BibitemShut {NoStop}%
\bibitem [{\citenamefont {Kozma}\ and\ \citenamefont
  {Freeman}(2016)}]{Kozma2016}%
  \BibitemOpen
  \bibfield  {author} {\bibinfo {author} {\bibfnamefont {R.}~\bibnamefont
  {Kozma}}\ and\ \bibinfo {author} {\bibfnamefont {W.~J.}\ \bibnamefont
  {Freeman}},\ }\href {https://doi.org/10.1007/978-3-319-24406-8} {\emph
  {\bibinfo {title} {Cognitive Phase Transitions in the Cerebral Cortex -
  Enhancing the Neuron Doctrine by Modeling Neural Fields}}}\ (\bibinfo
  {publisher} {Springer International Publishing},\ \bibinfo {year}
  {2016})\BibitemShut {NoStop}%
\bibitem [{\citenamefont {Nelken}(2004)}]{Nelken2004}%
  \BibitemOpen
  \bibfield  {author} {\bibinfo {author} {\bibfnamefont {I.}~\bibnamefont
  {Nelken}},\ }\bibfield  {title} {\enquote {\bibinfo {title} {Processing of
  complex stimuli and natural scenes in the auditory cortex},}\ }\href@noop {}
  {\bibfield  {journal} {\bibinfo  {journal} {Current Opinion in Neurobiology}\
  }\textbf {\bibinfo {volume} {14}},\ \bibinfo {pages} {474--480} (\bibinfo
  {year} {2004})}\BibitemShut {NoStop}%
\bibitem [{\citenamefont {Butler}(1968)}]{Butler1968}%
  \BibitemOpen
  \bibfield  {author} {\bibinfo {author} {\bibfnamefont {R.~A.}\ \bibnamefont
  {Butler}},\ }\bibfield  {title} {\enquote {\bibinfo {title} {Effect of
  changes in stimulus frequency and intensity on habituation of the human
  vertex potential},}\ }\href@noop {} {\bibfield  {journal} {\bibinfo
  {journal} {The Journal of the Acoustical Society of America}\ }\textbf
  {\bibinfo {volume} {44}},\ \bibinfo {pages} {945--950} (\bibinfo {year}
  {1968})}\BibitemShut {NoStop}%
\bibitem [{\citenamefont {Brosch}\ and\ \citenamefont
  {Schreiner}(2000)}]{Brosch2000}%
  \BibitemOpen
  \bibfield  {author} {\bibinfo {author} {\bibfnamefont {M.}~\bibnamefont
  {Brosch}}\ and\ \bibinfo {author} {\bibfnamefont {C.~E.}\ \bibnamefont
  {Schreiner}},\ }\bibfield  {title} {\enquote {\bibinfo {title} {Sequence
  sensitivity of neurons in cat primary auditory cortex},}\ }\href@noop {}
  {\bibfield  {journal} {\bibinfo  {journal} {Cerebral Cortex}\ }\textbf
  {\bibinfo {volume} {10}},\ \bibinfo {pages} {1155--1167} (\bibinfo {year}
  {2000})}\BibitemShut {NoStop}%
\bibitem [{\citenamefont {N\"{a}\"{a}t\"{a}nen}, \citenamefont {Gaillard},\
  and\ \citenamefont {M\"{a}ntysalo}(1978)}]{Naatanen1978}%
  \BibitemOpen
  \bibfield  {author} {\bibinfo {author} {\bibfnamefont {R.}~\bibnamefont
  {N\"{a}\"{a}t\"{a}nen}}, \bibinfo {author} {\bibfnamefont {A.~W.}\
  \bibnamefont {Gaillard}},\ and\ \bibinfo {author} {\bibfnamefont
  {S.}~\bibnamefont {M\"{a}ntysalo}},\ }\bibfield  {title} {\enquote {\bibinfo
  {title} {Early selective-attention effect on evoked potential
  reinterpreted},}\ }\href@noop {} {\bibfield  {journal} {\bibinfo  {journal}
  {Acta Psychologica}\ }\textbf {\bibinfo {volume} {42}},\ \bibinfo {pages}
  {313--329} (\bibinfo {year} {1978})}\BibitemShut {NoStop}%
\bibitem [{\citenamefont {Ulanovsky}\ \emph {et~al.}(2004)\citenamefont
  {Ulanovsky}, \citenamefont {Las}, \citenamefont {Farkas},\ and\ \citenamefont
  {Nelken}}]{Ulanovsky2004}%
  \BibitemOpen
  \bibfield  {author} {\bibinfo {author} {\bibfnamefont {N.}~\bibnamefont
  {Ulanovsky}}, \bibinfo {author} {\bibfnamefont {L.}~\bibnamefont {Las}},
  \bibinfo {author} {\bibfnamefont {D.}~\bibnamefont {Farkas}},\ and\ \bibinfo
  {author} {\bibfnamefont {I.}~\bibnamefont {Nelken}},\ }\bibfield  {title}
  {\enquote {\bibinfo {title} {Multiple time scales of adaptation in auditory
  cortex neurons},}\ }\href@noop {} {\bibfield  {journal} {\bibinfo  {journal}
  {The Journal of Neuroscience}\ }\textbf {\bibinfo {volume} {24}},\ \bibinfo
  {pages} {10440--10453} (\bibinfo {year} {2004})}\BibitemShut {NoStop}%
\bibitem [{\citenamefont {Zacharias}, \citenamefont {K\"onig},\ and\
  \citenamefont {Heil}(2012)}]{Zacharias2012}%
  \BibitemOpen
  \bibfield  {author} {\bibinfo {author} {\bibfnamefont {N.}~\bibnamefont
  {Zacharias}}, \bibinfo {author} {\bibfnamefont {R.}~\bibnamefont {K\"onig}},\
  and\ \bibinfo {author} {\bibfnamefont {P.}~\bibnamefont {Heil}},\ }\bibfield
  {title} {\enquote {\bibinfo {title} {Stimulation-history effects on the
  {M100} revealed by its differential dependence on the stimulus onset
  interval},}\ }\href@noop {} {\bibfield  {journal} {\bibinfo  {journal}
  {Psychophysiology}\ }\textbf {\bibinfo {volume} {49}},\ \bibinfo {pages}
  {909--919} (\bibinfo {year} {2012})}\BibitemShut {NoStop}%
\bibitem [{\citenamefont {P\'erez-Gonz\'alez}\ and\ \citenamefont
  {Malmierca}(2014)}]{Gonzalez2014}%
  \BibitemOpen
  \bibfield  {author} {\bibinfo {author} {\bibfnamefont {D.}~\bibnamefont
  {P\'erez-Gonz\'alez}}\ and\ \bibinfo {author} {\bibfnamefont {M.~S.}\
  \bibnamefont {Malmierca}},\ }\bibfield  {title} {\enquote {\bibinfo {title}
  {Adaptation in the auditory system: an overview},}\ }\href@noop {} {\bibfield
   {journal} {\bibinfo  {journal} {Frontiers in Integrative Neuroscience}\
  }\textbf {\bibinfo {volume} {8}},\ \bibinfo {pages} {1--10} (\bibinfo {year}
  {2014})}\BibitemShut {NoStop}%
\bibitem [{\citenamefont {Malmierca}, \citenamefont {Anderson},\ and\
  \citenamefont {Antunes}(2015)}]{Malmierca2015}%
  \BibitemOpen
  \bibfield  {author} {\bibinfo {author} {\bibfnamefont {M.~S.}\ \bibnamefont
  {Malmierca}}, \bibinfo {author} {\bibfnamefont {L.~A.}\ \bibnamefont
  {Anderson}},\ and\ \bibinfo {author} {\bibfnamefont {F.~M.}\ \bibnamefont
  {Antunes}},\ }\bibfield  {title} {\enquote {\bibinfo {title} {The cortical
  modulation of stimulus-specific adaptation in the auditory midbrain and
  thalamus: a potential neuronal correlate for predictive coding},}\
  }\href@noop {} {\bibfield  {journal} {\bibinfo  {journal} {Frontiers in
  system neuroscience}\ }\textbf {\bibinfo {volume} {9}} (\bibinfo {year}
  {2015})}\BibitemShut {NoStop}%
\bibitem [{\citenamefont {Friston}(2005)}]{Friston2005}%
  \BibitemOpen
  \bibfield  {author} {\bibinfo {author} {\bibfnamefont {K.}~\bibnamefont
  {Friston}},\ }\bibfield  {title} {\enquote {\bibinfo {title} {A theory of
  cortical responses},}\ }\href@noop {} {\bibfield  {journal} {\bibinfo
  {journal} {Philosophical Transactions of the Royal Society}\ }\textbf
  {\bibinfo {volume} {360}},\ \bibinfo {pages} {815--836} (\bibinfo {year}
  {2005})}\BibitemShut {NoStop}%
\bibitem [{\citenamefont {May}(2021)}]{May2021}%
  \BibitemOpen
  \bibfield  {author} {\bibinfo {author} {\bibfnamefont {P.~J.~C.}\
  \bibnamefont {May}},\ }\bibfield  {title} {\enquote {\bibinfo {title} {The
  adaptation model offers a challenge for the predictive coding account of
  mismatch negativity},}\ }\href@noop {} {\bibfield  {journal} {\bibinfo
  {journal} {Frontiers in Human Neuroscience}\ }\textbf {\bibinfo {volume}
  {15}},\ \bibinfo {pages} {721574} (\bibinfo {year} {2021})}\BibitemShut
  {NoStop}%
\bibitem [{\citenamefont {May}\ and\ \citenamefont {Tiitinen}(2010)}]{May2010}%
  \BibitemOpen
  \bibfield  {author} {\bibinfo {author} {\bibfnamefont {P.~J.~C.}\
  \bibnamefont {May}}\ and\ \bibinfo {author} {\bibfnamefont {H.}~\bibnamefont
  {Tiitinen}},\ }\bibfield  {title} {\enquote {\bibinfo {title} {Mismatch
  negativity ({MMN}), the deviance-elicited auditory deflection, explained},}\
  }\href@noop {} {\bibfield  {journal} {\bibinfo  {journal} {Psychophysiology}\
  }\textbf {\bibinfo {volume} {47}},\ \bibinfo {pages} {66--122} (\bibinfo
  {year} {2010})}\BibitemShut {NoStop}%
\bibitem [{\citenamefont {Wang}\ and\ \citenamefont
  {Kn\"{o}sche}(2013)}]{Wang2013}%
  \BibitemOpen
  \bibfield  {author} {\bibinfo {author} {\bibfnamefont {P.}~\bibnamefont
  {Wang}}\ and\ \bibinfo {author} {\bibfnamefont {T.~R.}\ \bibnamefont
  {Kn\"{o}sche}},\ }\bibfield  {title} {\enquote {\bibinfo {title} {A realistic
  neural mass model of the cortex with laminar-specific connections and
  synaptic plasticity-evaluation with auditory habituation},}\ }\href@noop {}
  {\bibfield  {journal} {\bibinfo  {journal} {PLoS One}\ }\textbf {\bibinfo
  {volume} {8}},\ \bibinfo {pages} {e77876} (\bibinfo {year}
  {2013})}\BibitemShut {NoStop}%
\bibitem [{\citenamefont {Kudela}\ \emph {et~al.}(2018)\citenamefont {Kudela},
  \citenamefont {Boatman-Reich}, \citenamefont {Beeman},\ and\ \citenamefont
  {Anderson}}]{Kudela2018}%
  \BibitemOpen
  \bibfield  {author} {\bibinfo {author} {\bibfnamefont {P.}~\bibnamefont
  {Kudela}}, \bibinfo {author} {\bibfnamefont {D.}~\bibnamefont
  {Boatman-Reich}}, \bibinfo {author} {\bibfnamefont {D.}~\bibnamefont
  {Beeman}},\ and\ \bibinfo {author} {\bibfnamefont {W.~S.}\ \bibnamefont
  {Anderson}},\ }\bibfield  {title} {\enquote {\bibinfo {title} {Modeling
  neural adaptation in auditory cortex},}\ }\href@noop {} {\bibfield  {journal}
  {\bibinfo  {journal} {Frontiers in Neural Circuits}\ }\textbf {\bibinfo
  {volume} {12}} (\bibinfo {year} {2018})}\BibitemShut {NoStop}%
\bibitem [{\citenamefont {Fortune}\ and\ \citenamefont
  {Rose}(2001)}]{Fortune2001}%
  \BibitemOpen
  \bibfield  {author} {\bibinfo {author} {\bibfnamefont {E.~S.}\ \bibnamefont
  {Fortune}}\ and\ \bibinfo {author} {\bibfnamefont {G.~J.}\ \bibnamefont
  {Rose}},\ }\bibfield  {title} {\enquote {\bibinfo {title} {Short-term
  synaptic plasticity as a temporal filter},}\ }\href@noop {} {\bibfield
  {journal} {\bibinfo  {journal} {Trends in Neurosciences}\ }\textbf {\bibinfo
  {volume} {24}},\ \bibinfo {pages} {381--385} (\bibinfo {year}
  {2001})}\BibitemShut {NoStop}%
\bibitem [{\citenamefont {Salmasi}\ \emph {et~al.}(2019)\citenamefont
  {Salmasi}, \citenamefont {Loebel}, \citenamefont {Glasauer},\ and\
  \citenamefont {Stemmler}}]{Salmasi2019}%
  \BibitemOpen
  \bibfield  {author} {\bibinfo {author} {\bibfnamefont {M.}~\bibnamefont
  {Salmasi}}, \bibinfo {author} {\bibfnamefont {A.}~\bibnamefont {Loebel}},
  \bibinfo {author} {\bibfnamefont {S.}~\bibnamefont {Glasauer}},\ and\
  \bibinfo {author} {\bibfnamefont {M.}~\bibnamefont {Stemmler}},\ }\bibfield
  {title} {\enquote {\bibinfo {title} {Short-term synaptic depression can
  increase the rate of information transfer at a release site},}\ }\href@noop
  {} {\bibfield  {journal} {\bibinfo  {journal} {PLoS Computational Biology}\
  }\textbf {\bibinfo {volume} {15}},\ \bibinfo {pages} {1--21} (\bibinfo {year}
  {2019})}\BibitemShut {NoStop}%
\bibitem [{\citenamefont {May}\ and\ \citenamefont {Tiitinen}(2013)}]{May2013}%
  \BibitemOpen
  \bibfield  {author} {\bibinfo {author} {\bibfnamefont {P.~J.~C.}\
  \bibnamefont {May}}\ and\ \bibinfo {author} {\bibfnamefont {H.}~\bibnamefont
  {Tiitinen}},\ }\bibfield  {title} {\enquote {\bibinfo {title} {Temporal
  binding of sound emerges out of anatomical structure and synaptic dynamics of
  auditory cortex},}\ }\href@noop {} {\bibfield  {journal} {\bibinfo  {journal}
  {Frontiers in Computational Neuroscience}\ }\textbf {\bibinfo {volume} {7}}
  (\bibinfo {year} {2013})}\BibitemShut {NoStop}%
\bibitem [{\citenamefont {Hajizadeh}\ \emph {et~al.}(2019)\citenamefont
  {Hajizadeh}, \citenamefont {Matysiak}, \citenamefont {May},\ and\
  \citenamefont {K\"{o}nig}}]{Hajizadeh2019}%
  \BibitemOpen
  \bibfield  {author} {\bibinfo {author} {\bibfnamefont {A.}~\bibnamefont
  {Hajizadeh}}, \bibinfo {author} {\bibfnamefont {A.}~\bibnamefont {Matysiak}},
  \bibinfo {author} {\bibfnamefont {P.~J.~C.}\ \bibnamefont {May}},\ and\
  \bibinfo {author} {\bibfnamefont {R.}~\bibnamefont {K\"{o}nig}},\ }\bibfield
  {title} {\enquote {\bibinfo {title} {Explaining event-related fields by a
  mechanistic model encapsulating the anatomical structure of auditory
  cortex},}\ }\href@noop {} {\bibfield  {journal} {\bibinfo  {journal}
  {Biological Cybernetics}\ }\textbf {\bibinfo {volume} {113}},\ \bibinfo
  {pages} {321--345} (\bibinfo {year} {2019})}\BibitemShut {NoStop}%
\bibitem [{\citenamefont {Hajizadeh}\ \emph {et~al.}(2022)\citenamefont
  {Hajizadeh}, \citenamefont {Matysiak}, \citenamefont {Wolfrum}, \citenamefont
  {May},\ and\ \citenamefont {K\"{o}nig}}]{Hajizadeh2022}%
  \BibitemOpen
  \bibfield  {author} {\bibinfo {author} {\bibfnamefont {A.}~\bibnamefont
  {Hajizadeh}}, \bibinfo {author} {\bibfnamefont {A.}~\bibnamefont {Matysiak}},
  \bibinfo {author} {\bibfnamefont {M.}~\bibnamefont {Wolfrum}}, \bibinfo
  {author} {\bibfnamefont {P.~J.~C.}\ \bibnamefont {May}},\ and\ \bibinfo
  {author} {\bibfnamefont {R.}~\bibnamefont {K\"{o}nig}},\ }\bibfield  {title}
  {\enquote {\bibinfo {title} {Auditory cortex modelled as a dynamical network
  of oscillators: understanding event-related fields and their adaptation},}\
  }\href@noop {} {\bibfield  {journal} {\bibinfo  {journal} {Biological
  Cybernetics}\ }\textbf {\bibinfo {volume} {116}},\ \bibinfo {pages}
  {475--499} (\bibinfo {year} {2022})}\BibitemShut {NoStop}%
\bibitem [{\citenamefont {Kaas}\ and\ \citenamefont
  {Hackett}(2000)}]{Kaas2000}%
  \BibitemOpen
  \bibfield  {author} {\bibinfo {author} {\bibfnamefont {J.~H.}\ \bibnamefont
  {Kaas}}\ and\ \bibinfo {author} {\bibfnamefont {T.~A.}\ \bibnamefont
  {Hackett}},\ }\bibfield  {title} {\enquote {\bibinfo {title} {Subdivisions of
  auditory cortex and processing streams in primates},}\ }\href@noop {}
  {\bibfield  {journal} {\bibinfo  {journal} {Proceedings of the National
  Academy of Sciences of the United States of America}\ }\textbf {\bibinfo
  {volume} {97}},\ \bibinfo {pages} {11793--11799} (\bibinfo {year}
  {2000})}\BibitemShut {NoStop}%
\bibitem [{\citenamefont {Hackett}(2015)}]{Hackett2015}%
  \BibitemOpen
  \bibfield  {author} {\bibinfo {author} {\bibfnamefont {T.~A.}\ \bibnamefont
  {Hackett}},\ }\bibfield  {title} {\enquote {\bibinfo {title} {Anatomic
  organization of the auditory cortex},}\ }\href@noop {} {\bibfield  {journal}
  {\bibinfo  {journal} {Handbook of Clinical Neurology}\ }\textbf {\bibinfo
  {volume} {129}} (\bibinfo {year} {2015})}\BibitemShut {NoStop}%
\bibitem [{\citenamefont {Krizhevsky}, \citenamefont {Sutskever},\ and\
  \citenamefont {Hinton}(2012)}]{krizhevsky2012imagenet}%
  \BibitemOpen
  \bibfield  {author} {\bibinfo {author} {\bibfnamefont {A.}~\bibnamefont
  {Krizhevsky}}, \bibinfo {author} {\bibfnamefont {I.}~\bibnamefont
  {Sutskever}},\ and\ \bibinfo {author} {\bibfnamefont {G.}~\bibnamefont
  {Hinton}},\ }\bibfield  {title} {\enquote {\bibinfo {title} {Imagenet
  classification with deep convolutional neural networks},}\ }in\ \href@noop {}
  {\emph {\bibinfo {booktitle} {Advances in neural information processing
  systems}}}\ (\bibinfo {year} {2012})\ pp.\ \bibinfo {pages}
  {1097--1105}\BibitemShut {NoStop}%
\bibitem [{\citenamefont {LeCun}, \citenamefont {Bengio},\ and\ \citenamefont
  {Hinton}(2015)}]{lecun2015deep}%
  \BibitemOpen
  \bibfield  {author} {\bibinfo {author} {\bibfnamefont {Y.}~\bibnamefont
  {LeCun}}, \bibinfo {author} {\bibfnamefont {Y.}~\bibnamefont {Bengio}},\ and\
  \bibinfo {author} {\bibfnamefont {G.}~\bibnamefont {Hinton}},\ }\bibfield
  {title} {\enquote {\bibinfo {title} {{Deep learning}},}\ }\href@noop {}
  {\bibfield  {journal} {\bibinfo  {journal} {Nature}\ }\textbf {\bibinfo
  {volume} {521}},\ \bibinfo {pages} {436--444} (\bibinfo {year}
  {2015})}\BibitemShut {NoStop}%
\bibitem [{\citenamefont {Simonyan}\ and\ \citenamefont
  {Zisserman}(2015)}]{simonyan2015very}%
  \BibitemOpen
  \bibfield  {author} {\bibinfo {author} {\bibfnamefont {K.}~\bibnamefont
  {Simonyan}}\ and\ \bibinfo {author} {\bibfnamefont {A.}~\bibnamefont
  {Zisserman}},\ }\bibfield  {title} {\enquote {\bibinfo {title} {{Very Deep
  Convolutional Networks for Large-Scale Image Recognition}},}\ }in\ \href@noop
  {} {\emph {\bibinfo {booktitle} {International Conference on Learning
  Representations}}}\ (\bibinfo {year} {2015})\BibitemShut {NoStop}%
\bibitem [{\citenamefont {He}\ \emph {et~al.}(2016)\citenamefont {He},
  \citenamefont {Zhang}, \citenamefont {Ren},\ and\ \citenamefont
  {Sun}}]{he2016deep}%
  \BibitemOpen
  \bibfield  {author} {\bibinfo {author} {\bibfnamefont {K.}~\bibnamefont
  {He}}, \bibinfo {author} {\bibfnamefont {X.}~\bibnamefont {Zhang}}, \bibinfo
  {author} {\bibfnamefont {S.}~\bibnamefont {Ren}},\ and\ \bibinfo {author}
  {\bibfnamefont {J.}~\bibnamefont {Sun}},\ }\bibfield  {title} {\enquote
  {\bibinfo {title} {Deep residual learning for image recognition},}\ }in\
  \href@noop {} {\emph {\bibinfo {booktitle} {Proceedings of the IEEE
  conference on computer vision and pattern recognition}}}\ (\bibinfo {year}
  {2016})\ pp.\ \bibinfo {pages} {770--778}\BibitemShut {NoStop}%
\bibitem [{\citenamefont {Dosovitskiy}\ \emph {et~al.}(2021)\citenamefont
  {Dosovitskiy}, \citenamefont {Beyer}, \citenamefont {Kolesnikov},
  \citenamefont {Weissenborn}, \citenamefont {Zhai}, \citenamefont
  {Unterthiner}, \citenamefont {Dehghani}, \citenamefont {Minderer},
  \citenamefont {Heigold}, \citenamefont {Gelly}, \citenamefont {Uszkoreit},\
  and\ \citenamefont {Houlsby}}]{dosovitskiy2021image}%
  \BibitemOpen
  \bibfield  {author} {\bibinfo {author} {\bibfnamefont {A.}~\bibnamefont
  {Dosovitskiy}}, \bibinfo {author} {\bibfnamefont {L.}~\bibnamefont {Beyer}},
  \bibinfo {author} {\bibfnamefont {A.}~\bibnamefont {Kolesnikov}}, \bibinfo
  {author} {\bibfnamefont {D.}~\bibnamefont {Weissenborn}}, \bibinfo {author}
  {\bibfnamefont {X.}~\bibnamefont {Zhai}}, \bibinfo {author} {\bibfnamefont
  {T.}~\bibnamefont {Unterthiner}}, \bibinfo {author} {\bibfnamefont
  {M.}~\bibnamefont {Dehghani}}, \bibinfo {author} {\bibfnamefont
  {M.}~\bibnamefont {Minderer}}, \bibinfo {author} {\bibfnamefont
  {G.}~\bibnamefont {Heigold}}, \bibinfo {author} {\bibfnamefont
  {S.}~\bibnamefont {Gelly}}, \bibinfo {author} {\bibfnamefont
  {J.}~\bibnamefont {Uszkoreit}},\ and\ \bibinfo {author} {\bibfnamefont
  {N.}~\bibnamefont {Houlsby}},\ }\bibfield  {title} {\enquote {\bibinfo
  {title} {An image is worth 16x16 words: Transformers for image recognition at
  scale},}\ }in\ \href {https://openreview.net/forum?id=YicbFdNTTy} {\emph
  {\bibinfo {booktitle} {International Conference on Learning
  Representations}}}\ (\bibinfo {year} {2021})\BibitemShut {NoStop}%
\bibitem [{\citenamefont {Hinton}\ \emph {et~al.}(2012)\citenamefont {Hinton},
  \citenamefont {Deng}, \citenamefont {Yu}, \citenamefont {Dahl}, \citenamefont
  {Mohamed}, \citenamefont {Jaitly}, \citenamefont {Senior}, \citenamefont
  {Vanhoucke}, \citenamefont {Nguyen}, \citenamefont {Sainath} \emph
  {et~al.}}]{hinton2012deep}%
  \BibitemOpen
  \bibfield  {author} {\bibinfo {author} {\bibfnamefont {G.}~\bibnamefont
  {Hinton}}, \bibinfo {author} {\bibfnamefont {L.}~\bibnamefont {Deng}},
  \bibinfo {author} {\bibfnamefont {D.}~\bibnamefont {Yu}}, \bibinfo {author}
  {\bibfnamefont {G.~E.}\ \bibnamefont {Dahl}}, \bibinfo {author}
  {\bibfnamefont {A.-r.}\ \bibnamefont {Mohamed}}, \bibinfo {author}
  {\bibfnamefont {N.}~\bibnamefont {Jaitly}}, \bibinfo {author} {\bibfnamefont
  {A.}~\bibnamefont {Senior}}, \bibinfo {author} {\bibfnamefont
  {V.}~\bibnamefont {Vanhoucke}}, \bibinfo {author} {\bibfnamefont
  {P.}~\bibnamefont {Nguyen}}, \bibinfo {author} {\bibfnamefont {T.~N.}\
  \bibnamefont {Sainath}}, \emph {et~al.},\ }\bibfield  {title} {\enquote
  {\bibinfo {title} {Deep neural networks for acoustic modeling in speech
  recognition: The shared views of four research groups},}\ }\href@noop {}
  {\bibfield  {journal} {\bibinfo  {journal} {IEEE Signal processing magazine}\
  }\textbf {\bibinfo {volume} {29}},\ \bibinfo {pages} {82--97} (\bibinfo
  {year} {2012})}\BibitemShut {NoStop}%
\bibitem [{\citenamefont {Sutskever}, \citenamefont {Vinyals},\ and\
  \citenamefont {Le}(2014)}]{sutskever2014sequence}%
  \BibitemOpen
  \bibfield  {author} {\bibinfo {author} {\bibfnamefont {I.}~\bibnamefont
  {Sutskever}}, \bibinfo {author} {\bibfnamefont {O.}~\bibnamefont {Vinyals}},\
  and\ \bibinfo {author} {\bibfnamefont {Q.}~\bibnamefont {Le}},\ }\bibfield
  {title} {\enquote {\bibinfo {title} {Sequence to sequence learning with
  neural networks},}\ }in\ \href
  {http://papers.nips.cc/paper/5346-sequence-to-sequence-learning-with-neural-networks.pdf}
  {\emph {\bibinfo {booktitle} {Advances in Neural Information Processing
  Systems 27}}},\ \bibinfo {editor} {edited by\ \bibinfo {editor}
  {\bibfnamefont {Z.}~\bibnamefont {Ghahramani}}, \bibinfo {editor}
  {\bibfnamefont {M.}~\bibnamefont {Welling}}, \bibinfo {editor} {\bibfnamefont
  {C.}~\bibnamefont {Cortes}}, \bibinfo {editor} {\bibfnamefont {N.~D.}\
  \bibnamefont {Lawrence}},\ and\ \bibinfo {editor} {\bibfnamefont {K.~Q.}\
  \bibnamefont {Weinberger}}}\ (\bibinfo  {publisher} {Curran Associates,
  Inc.},\ \bibinfo {year} {2014})\ pp.\ \bibinfo {pages}
  {3104--3112}\BibitemShut {NoStop}%
\bibitem [{\citenamefont {Vaswani}\ \emph {et~al.}(2017)\citenamefont
  {Vaswani}, \citenamefont {Shazeer}, \citenamefont {Parmar}, \citenamefont
  {Uszkoreit}, \citenamefont {Jones}, \citenamefont {Gomez}, \citenamefont
  {Kaiser},\ and\ \citenamefont {Polosukhin}}]{vaswani2017attention}%
  \BibitemOpen
  \bibfield  {author} {\bibinfo {author} {\bibfnamefont {A.}~\bibnamefont
  {Vaswani}}, \bibinfo {author} {\bibfnamefont {N.}~\bibnamefont {Shazeer}},
  \bibinfo {author} {\bibfnamefont {N.}~\bibnamefont {Parmar}}, \bibinfo
  {author} {\bibfnamefont {J.}~\bibnamefont {Uszkoreit}}, \bibinfo {author}
  {\bibfnamefont {L.}~\bibnamefont {Jones}}, \bibinfo {author} {\bibfnamefont
  {A.~N.}\ \bibnamefont {Gomez}}, \bibinfo {author} {\bibfnamefont
  {{\L}.}~\bibnamefont {Kaiser}},\ and\ \bibinfo {author} {\bibfnamefont
  {I.}~\bibnamefont {Polosukhin}},\ }\bibfield  {title} {\enquote {\bibinfo
  {title} {Attention is all you need},}\ }in\ \href@noop {} {\emph {\bibinfo
  {booktitle} {Advances in neural information processing systems}}}\ (\bibinfo
  {year} {2017})\ pp.\ \bibinfo {pages} {5998--6008}\BibitemShut {NoStop}%
\bibitem [{\citenamefont {Devlin}\ \emph {et~al.}(2019)\citenamefont {Devlin},
  \citenamefont {Chang}, \citenamefont {Lee},\ and\ \citenamefont
  {Toutanova}}]{devlin2019bert}%
  \BibitemOpen
  \bibfield  {author} {\bibinfo {author} {\bibfnamefont {J.}~\bibnamefont
  {Devlin}}, \bibinfo {author} {\bibfnamefont {M.-W.}\ \bibnamefont {Chang}},
  \bibinfo {author} {\bibfnamefont {K.}~\bibnamefont {Lee}},\ and\ \bibinfo
  {author} {\bibfnamefont {K.}~\bibnamefont {Toutanova}},\ }\bibfield  {title}
  {\enquote {\bibinfo {title} {Bert: Pre-training of deep bidirectional
  transformers for language understanding},}\ }in\ \href
  {http://arxiv.org/abs/1810.04805} {\emph {\bibinfo {booktitle} {Proceedings
  of NAACL-HLT}}}\ (\bibinfo {year} {2019})\ pp.\ \bibinfo {pages}
  {4171--4186}\BibitemShut {NoStop}%
\bibitem [{\citenamefont {Silver}\ \emph {et~al.}(2016)\citenamefont {Silver},
  \citenamefont {Huang}, \citenamefont {Maddison}, \citenamefont {Guez},
  \citenamefont {Sifre}, \citenamefont {van~den Driessche}, \citenamefont
  {Schrittwieser}, \citenamefont {Antonoglou}, \citenamefont {Panneershelvam},
  \citenamefont {Lanctot}, \citenamefont {Dieleman}, \citenamefont {Grewe},
  \citenamefont {Nham}, \citenamefont {Kalchbrenner}, \citenamefont
  {Sutskever}, \citenamefont {Lillicrap}, \citenamefont {Leach}, \citenamefont
  {Kavukcuoglu}, \citenamefont {Graepel},\ and\ \citenamefont
  {Hassabis}}]{silver2016}%
  \BibitemOpen
  \bibfield  {author} {\bibinfo {author} {\bibfnamefont {D.}~\bibnamefont
  {Silver}}, \bibinfo {author} {\bibfnamefont {A.}~\bibnamefont {Huang}},
  \bibinfo {author} {\bibfnamefont {C.~J.}\ \bibnamefont {Maddison}}, \bibinfo
  {author} {\bibfnamefont {A.}~\bibnamefont {Guez}}, \bibinfo {author}
  {\bibfnamefont {L.}~\bibnamefont {Sifre}}, \bibinfo {author} {\bibfnamefont
  {G.}~\bibnamefont {van~den Driessche}}, \bibinfo {author} {\bibfnamefont
  {J.}~\bibnamefont {Schrittwieser}}, \bibinfo {author} {\bibfnamefont
  {I.}~\bibnamefont {Antonoglou}}, \bibinfo {author} {\bibfnamefont
  {V.}~\bibnamefont {Panneershelvam}}, \bibinfo {author} {\bibfnamefont
  {M.}~\bibnamefont {Lanctot}}, \bibinfo {author} {\bibfnamefont
  {S.}~\bibnamefont {Dieleman}}, \bibinfo {author} {\bibfnamefont
  {D.}~\bibnamefont {Grewe}}, \bibinfo {author} {\bibfnamefont
  {J.}~\bibnamefont {Nham}}, \bibinfo {author} {\bibfnamefont {N.}~\bibnamefont
  {Kalchbrenner}}, \bibinfo {author} {\bibfnamefont {I.}~\bibnamefont
  {Sutskever}}, \bibinfo {author} {\bibfnamefont {T.}~\bibnamefont
  {Lillicrap}}, \bibinfo {author} {\bibfnamefont {M.}~\bibnamefont {Leach}},
  \bibinfo {author} {\bibfnamefont {K.}~\bibnamefont {Kavukcuoglu}}, \bibinfo
  {author} {\bibfnamefont {T.}~\bibnamefont {Graepel}},\ and\ \bibinfo {author}
  {\bibfnamefont {D.}~\bibnamefont {Hassabis}},\ }\bibfield  {title} {\enquote
  {\bibinfo {title} {{Mastering the game of Go with deep neural networks and
  tree search}},}\ }\href@noop {} {\bibfield  {journal} {\bibinfo  {journal}
  {Nature}\ }\textbf {\bibinfo {volume} {529}},\ \bibinfo {pages} {484--489}
  (\bibinfo {year} {2016})}\BibitemShut {NoStop}%
\bibitem [{\citenamefont {Silver}\ \emph {et~al.}(2017)\citenamefont {Silver},
  \citenamefont {Schrittwieser}, \citenamefont {Simonyan}, \citenamefont
  {Antonoglou}, \citenamefont {Huang}, \citenamefont {Guez}, \citenamefont
  {Hubert}, \citenamefont {Baker}, \citenamefont {Lai}, \citenamefont {Bolton},
  \citenamefont {Chen}, \citenamefont {Lillicrap}, \citenamefont {Hui},
  \citenamefont {Sifre}, \citenamefont {van~den Driessche}, \citenamefont
  {Graepel},\ and\ \citenamefont {Hassabis}}]{silver2017mastering}%
  \BibitemOpen
  \bibfield  {author} {\bibinfo {author} {\bibfnamefont {D.}~\bibnamefont
  {Silver}}, \bibinfo {author} {\bibfnamefont {J.}~\bibnamefont
  {Schrittwieser}}, \bibinfo {author} {\bibfnamefont {K.}~\bibnamefont
  {Simonyan}}, \bibinfo {author} {\bibfnamefont {I.}~\bibnamefont
  {Antonoglou}}, \bibinfo {author} {\bibfnamefont {A.}~\bibnamefont {Huang}},
  \bibinfo {author} {\bibfnamefont {A.}~\bibnamefont {Guez}}, \bibinfo {author}
  {\bibfnamefont {T.}~\bibnamefont {Hubert}}, \bibinfo {author} {\bibfnamefont
  {L.}~\bibnamefont {Baker}}, \bibinfo {author} {\bibfnamefont
  {M.}~\bibnamefont {Lai}}, \bibinfo {author} {\bibfnamefont {A.}~\bibnamefont
  {Bolton}}, \bibinfo {author} {\bibfnamefont {Y.}~\bibnamefont {Chen}},
  \bibinfo {author} {\bibfnamefont {T.}~\bibnamefont {Lillicrap}}, \bibinfo
  {author} {\bibfnamefont {F.}~\bibnamefont {Hui}}, \bibinfo {author}
  {\bibfnamefont {L.}~\bibnamefont {Sifre}}, \bibinfo {author} {\bibfnamefont
  {G.}~\bibnamefont {van~den Driessche}}, \bibinfo {author} {\bibfnamefont
  {T.}~\bibnamefont {Graepel}},\ and\ \bibinfo {author} {\bibfnamefont
  {D.}~\bibnamefont {Hassabis}},\ }\bibfield  {title} {\enquote {\bibinfo
  {title} {{Mastering the game of Go without human knowledge}},}\ }\href
  {https://doi.org/10.1038/nature24270} {\bibfield  {journal} {\bibinfo
  {journal} {Nature}\ }\textbf {\bibinfo {volume} {550}},\ \bibinfo {pages}
  {354--359} (\bibinfo {year} {2017})}\BibitemShut {NoStop}%
\bibitem [{\citenamefont {Bowling}\ \emph {et~al.}(2015)\citenamefont
  {Bowling}, \citenamefont {Burch}, \citenamefont {Johanson},\ and\
  \citenamefont {Tammelin}}]{bowling2015heads}%
  \BibitemOpen
  \bibfield  {author} {\bibinfo {author} {\bibfnamefont {M.}~\bibnamefont
  {Bowling}}, \bibinfo {author} {\bibfnamefont {N.}~\bibnamefont {Burch}},
  \bibinfo {author} {\bibfnamefont {M.}~\bibnamefont {Johanson}},\ and\
  \bibinfo {author} {\bibfnamefont {O.}~\bibnamefont {Tammelin}},\ }\bibfield
  {title} {\enquote {\bibinfo {title} {Heads-up limit hold'em poker is
  solved},}\ }\href@noop {} {\bibfield  {journal} {\bibinfo  {journal}
  {Science}\ }\textbf {\bibinfo {volume} {347}},\ \bibinfo {pages} {145--149}
  (\bibinfo {year} {2015})}\BibitemShut {NoStop}%
\bibitem [{\citenamefont {Brown}\ and\ \citenamefont
  {Sandholm}(2018)}]{brown2018superhuman}%
  \BibitemOpen
  \bibfield  {author} {\bibinfo {author} {\bibfnamefont {N.}~\bibnamefont
  {Brown}}\ and\ \bibinfo {author} {\bibfnamefont {T.}~\bibnamefont
  {Sandholm}},\ }\bibfield  {title} {\enquote {\bibinfo {title} {Superhuman ai
  for heads-up no-limit poker: Libratus beats top professionals},}\ }\href@noop
  {} {\bibfield  {journal} {\bibinfo  {journal} {Science}\ }\textbf {\bibinfo
  {volume} {359}},\ \bibinfo {pages} {418--424} (\bibinfo {year}
  {2018})}\BibitemShut {NoStop}%
\bibitem [{\citenamefont {Brown}\ and\ \citenamefont
  {Sandholm}(2019)}]{brown2019superhuman}%
  \BibitemOpen
  \bibfield  {author} {\bibinfo {author} {\bibfnamefont {N.}~\bibnamefont
  {Brown}}\ and\ \bibinfo {author} {\bibfnamefont {T.}~\bibnamefont
  {Sandholm}},\ }\bibfield  {title} {\enquote {\bibinfo {title} {Superhuman ai
  for multiplayer poker},}\ }\href@noop {} {\bibfield  {journal} {\bibinfo
  {journal} {Science}\ }\textbf {\bibinfo {volume} {365}},\ \bibinfo {pages}
  {885--890} (\bibinfo {year} {2019})}\BibitemShut {NoStop}%
\bibitem [{\citenamefont {Carleo}\ \emph {et~al.}(2019)\citenamefont {Carleo},
  \citenamefont {Cirac}, \citenamefont {Cranmer}, \citenamefont {Daudet},
  \citenamefont {Schuld}, \citenamefont {Tishby}, \citenamefont
  {Vogt-Maranto},\ and\ \citenamefont {Zdeborov\'a}}]{carleo2019machine}%
  \BibitemOpen
  \bibfield  {author} {\bibinfo {author} {\bibfnamefont {G.}~\bibnamefont
  {Carleo}}, \bibinfo {author} {\bibfnamefont {I.}~\bibnamefont {Cirac}},
  \bibinfo {author} {\bibfnamefont {K.}~\bibnamefont {Cranmer}}, \bibinfo
  {author} {\bibfnamefont {L.}~\bibnamefont {Daudet}}, \bibinfo {author}
  {\bibfnamefont {M.}~\bibnamefont {Schuld}}, \bibinfo {author} {\bibfnamefont
  {N.}~\bibnamefont {Tishby}}, \bibinfo {author} {\bibfnamefont
  {L.}~\bibnamefont {Vogt-Maranto}},\ and\ \bibinfo {author} {\bibfnamefont
  {L.}~\bibnamefont {Zdeborov\'a}},\ }\bibfield  {title} {\enquote {\bibinfo
  {title} {Machine learning and the physical sciences},}\ }\href@noop {}
  {\bibfield  {journal} {\bibinfo  {journal} {Rev. Mod. Phys.}\ }\textbf
  {\bibinfo {volume} {91}},\ \bibinfo {pages} {045002} (\bibinfo {year}
  {2019})}\BibitemShut {NoStop}%
\bibitem [{\citenamefont {Tunyasuvunakool}\ \emph {et~al.}(2021)\citenamefont
  {Tunyasuvunakool}, \citenamefont {Adler}, \citenamefont {Wu}, \citenamefont
  {Green}, \citenamefont {Zielinski}, \citenamefont {{\v{Z}}{\'{i}}dek},
  \citenamefont {Bridgland}, \citenamefont {Cowie}, \citenamefont {Meyer},
  \citenamefont {Laydon}, \citenamefont {Velankar}, \citenamefont {Kleywegt},
  \citenamefont {Bateman}, \citenamefont {Evans}, \citenamefont {Pritzel},
  \citenamefont {Figurnov}, \citenamefont {Ronneberger}, \citenamefont {Bates},
  \citenamefont {Kohl}, \citenamefont {Potapenko}, \citenamefont {Ballard},
  \citenamefont {Romera-Paredes}, \citenamefont {Nikolov}, \citenamefont
  {Jain}, \citenamefont {Clancy}, \citenamefont {Reiman}, \citenamefont
  {Petersen}, \citenamefont {Senior}, \citenamefont {Kavukcuoglu},
  \citenamefont {Birney}, \citenamefont {Kohli}, \citenamefont {Jumper},\ and\
  \citenamefont {Hassabis}}]{tunyasuvunakool2021highly}%
  \BibitemOpen
  \bibfield  {author} {\bibinfo {author} {\bibfnamefont {K.}~\bibnamefont
  {Tunyasuvunakool}}, \bibinfo {author} {\bibfnamefont {J.}~\bibnamefont
  {Adler}}, \bibinfo {author} {\bibfnamefont {Z.}~\bibnamefont {Wu}}, \bibinfo
  {author} {\bibfnamefont {T.}~\bibnamefont {Green}}, \bibinfo {author}
  {\bibfnamefont {M.}~\bibnamefont {Zielinski}}, \bibinfo {author}
  {\bibfnamefont {A.}~\bibnamefont {{\v{Z}}{\'{i}}dek}}, \bibinfo {author}
  {\bibfnamefont {A.}~\bibnamefont {Bridgland}}, \bibinfo {author}
  {\bibfnamefont {A.}~\bibnamefont {Cowie}}, \bibinfo {author} {\bibfnamefont
  {C.}~\bibnamefont {Meyer}}, \bibinfo {author} {\bibfnamefont
  {A.}~\bibnamefont {Laydon}}, \bibinfo {author} {\bibfnamefont
  {S.}~\bibnamefont {Velankar}}, \bibinfo {author} {\bibfnamefont {G.~J.}\
  \bibnamefont {Kleywegt}}, \bibinfo {author} {\bibfnamefont {A.}~\bibnamefont
  {Bateman}}, \bibinfo {author} {\bibfnamefont {R.}~\bibnamefont {Evans}},
  \bibinfo {author} {\bibfnamefont {A.}~\bibnamefont {Pritzel}}, \bibinfo
  {author} {\bibfnamefont {M.}~\bibnamefont {Figurnov}}, \bibinfo {author}
  {\bibfnamefont {O.}~\bibnamefont {Ronneberger}}, \bibinfo {author}
  {\bibfnamefont {R.}~\bibnamefont {Bates}}, \bibinfo {author} {\bibfnamefont
  {S.~A.~A.}\ \bibnamefont {Kohl}}, \bibinfo {author} {\bibfnamefont
  {A.}~\bibnamefont {Potapenko}}, \bibinfo {author} {\bibfnamefont {A.~J.}\
  \bibnamefont {Ballard}}, \bibinfo {author} {\bibfnamefont {B.}~\bibnamefont
  {Romera-Paredes}}, \bibinfo {author} {\bibfnamefont {S.}~\bibnamefont
  {Nikolov}}, \bibinfo {author} {\bibfnamefont {R.}~\bibnamefont {Jain}},
  \bibinfo {author} {\bibfnamefont {E.}~\bibnamefont {Clancy}}, \bibinfo
  {author} {\bibfnamefont {D.}~\bibnamefont {Reiman}}, \bibinfo {author}
  {\bibfnamefont {S.}~\bibnamefont {Petersen}}, \bibinfo {author}
  {\bibfnamefont {A.~W.}\ \bibnamefont {Senior}}, \bibinfo {author}
  {\bibfnamefont {K.}~\bibnamefont {Kavukcuoglu}}, \bibinfo {author}
  {\bibfnamefont {E.}~\bibnamefont {Birney}}, \bibinfo {author} {\bibfnamefont
  {P.}~\bibnamefont {Kohli}}, \bibinfo {author} {\bibfnamefont
  {J.}~\bibnamefont {Jumper}},\ and\ \bibinfo {author} {\bibfnamefont
  {D.}~\bibnamefont {Hassabis}},\ }\bibfield  {title} {\enquote {\bibinfo
  {title} {{Highly accurate protein structure prediction for the human
  proteome}},}\ }\href {https://doi.org/10.1038/s41586-021-03828-1} {\bibfield
  {journal} {\bibinfo  {journal} {Nature}\ } (\bibinfo {year} {2021}),\
  10.1038/s41586-021-03828-1}\BibitemShut {NoStop}%
\bibitem [{\citenamefont {Oliva}\ and\ \citenamefont
  {Torralba}(2001)}]{oliva2001modeling}%
  \BibitemOpen
  \bibfield  {author} {\bibinfo {author} {\bibfnamefont {A.}~\bibnamefont
  {Oliva}}\ and\ \bibinfo {author} {\bibfnamefont {A.}~\bibnamefont
  {Torralba}},\ }\bibfield  {title} {\enquote {\bibinfo {title} {Modeling the
  shape of the scene: A holistic representation of the spatial envelope},}\
  }\href@noop {} {\bibfield  {journal} {\bibinfo  {journal} {International
  journal of computer vision}\ }\textbf {\bibinfo {volume} {42}},\ \bibinfo
  {pages} {145--175} (\bibinfo {year} {2001})}\BibitemShut {NoStop}%
\bibitem [{\citenamefont {Dalal}\ and\ \citenamefont
  {Triggs}(2005)}]{dalal2005histograms}%
  \BibitemOpen
  \bibfield  {author} {\bibinfo {author} {\bibfnamefont {N.}~\bibnamefont
  {Dalal}}\ and\ \bibinfo {author} {\bibfnamefont {B.}~\bibnamefont {Triggs}},\
  }\bibfield  {title} {\enquote {\bibinfo {title} {Histograms of oriented
  gradients for human detection},}\ }in\ \href@noop {} {\emph {\bibinfo
  {booktitle} {2005 IEEE computer society conference on computer vision and
  pattern recognition (CVPR'05)}}},\ Vol.~\bibinfo {volume} {1}\ (\bibinfo
  {organization} {Ieee},\ \bibinfo {year} {2005})\ pp.\ \bibinfo {pages}
  {886--893}\BibitemShut {NoStop}%
\bibitem [{\citenamefont {Bay}, \citenamefont {Tuytelaars},\ and\ \citenamefont
  {Van~Gool}(2006)}]{bay2006surf}%
  \BibitemOpen
  \bibfield  {author} {\bibinfo {author} {\bibfnamefont {H.}~\bibnamefont
  {Bay}}, \bibinfo {author} {\bibfnamefont {T.}~\bibnamefont {Tuytelaars}},\
  and\ \bibinfo {author} {\bibfnamefont {L.}~\bibnamefont {Van~Gool}},\
  }\bibfield  {title} {\enquote {\bibinfo {title} {Surf: Speeded up robust
  features},}\ }in\ \href@noop {} {\emph {\bibinfo {booktitle} {Computer Vision
  -- ECCV 2006}}},\ \bibinfo {editor} {edited by\ \bibinfo {editor}
  {\bibfnamefont {A.}~\bibnamefont {Leonardis}}, \bibinfo {editor}
  {\bibfnamefont {H.}~\bibnamefont {Bischof}},\ and\ \bibinfo {editor}
  {\bibfnamefont {A.}~\bibnamefont {Pinz}}}\ (\bibinfo  {publisher} {Springer
  Berlin Heidelberg},\ \bibinfo {address} {Berlin, Heidelberg},\ \bibinfo
  {year} {2006})\ pp.\ \bibinfo {pages} {404--417}\BibitemShut {NoStop}%
\bibitem [{Note1()}]{Note1}%
  \BibitemOpen
  \bibinfo {note} {There are different architectures which are appropriate for
  certain types of data or certain tasks; for an overview, cf.~\cite
  {goodfellow2016deep}.}\BibitemShut {Stop}%
\bibitem [{\citenamefont {MacKay}(2003)}]{mackay2003}%
  \BibitemOpen
  \bibfield  {author} {\bibinfo {author} {\bibfnamefont {D.~J.}\ \bibnamefont
  {MacKay}},\ }\href@noop {} {\emph {\bibinfo {title} {{Information Theory,
  Inference and Learning Algorithms}}}}\ (\bibinfo  {publisher} {Cambridge
  University Press},\ \bibinfo {year} {2003})\BibitemShut {NoStop}%
\bibitem [{\citenamefont {Hardt}, \citenamefont {Recht},\ and\ \citenamefont
  {Singer}(2016)}]{hardt2016train}%
  \BibitemOpen
  \bibfield  {author} {\bibinfo {author} {\bibfnamefont {M.}~\bibnamefont
  {Hardt}}, \bibinfo {author} {\bibfnamefont {B.}~\bibnamefont {Recht}},\ and\
  \bibinfo {author} {\bibfnamefont {Y.}~\bibnamefont {Singer}},\ }\bibfield
  {title} {\enquote {\bibinfo {title} {Train faster, generalize better:
  Stability of stochastic gradient descent},}\ }in\ \href
  {https://proceedings.mlr.press/v48/hardt16.html} {\emph {\bibinfo {booktitle}
  {Proceedings of The 33rd International Conference on Machine Learning}}},\
  \bibinfo {series} {Proceedings of Machine Learning Research}, Vol.~\bibinfo
  {volume} {48},\ \bibinfo {editor} {edited by\ \bibinfo {editor}
  {\bibfnamefont {M.~F.}\ \bibnamefont {Balcan}}\ and\ \bibinfo {editor}
  {\bibfnamefont {K.~Q.}\ \bibnamefont {Weinberger}}}\ (\bibinfo  {publisher}
  {PMLR},\ \bibinfo {address} {New York, New York, USA},\ \bibinfo {year}
  {2016})\ pp.\ \bibinfo {pages} {1225--1234}\BibitemShut {NoStop}%
\bibitem [{\citenamefont {Wilson}\ \emph {et~al.}(2017)\citenamefont {Wilson},
  \citenamefont {Roelofs}, \citenamefont {Stern}, \citenamefont {Srebro},\ and\
  \citenamefont {Recht}}]{wilson2017marginal}%
  \BibitemOpen
  \bibfield  {author} {\bibinfo {author} {\bibfnamefont {A.~C.}\ \bibnamefont
  {Wilson}}, \bibinfo {author} {\bibfnamefont {R.}~\bibnamefont {Roelofs}},
  \bibinfo {author} {\bibfnamefont {M.}~\bibnamefont {Stern}}, \bibinfo
  {author} {\bibfnamefont {N.}~\bibnamefont {Srebro}},\ and\ \bibinfo {author}
  {\bibfnamefont {B.}~\bibnamefont {Recht}},\ }\bibfield  {title} {\enquote
  {\bibinfo {title} {The marginal value of adaptive gradient methods in machine
  learning},}\ }in\ \href
  {https://proceedings.neurips.cc/paper/2017/file/81b3833e2504647f9d794f7d7b9bf341-Paper.pdf}
  {\emph {\bibinfo {booktitle} {Advances in Neural Information Processing
  Systems}}},\ Vol.~\bibinfo {volume} {30},\ \bibinfo {editor} {edited by\
  \bibinfo {editor} {\bibfnamefont {I.}~\bibnamefont {Guyon}}, \bibinfo
  {editor} {\bibfnamefont {U.~V.}\ \bibnamefont {Luxburg}}, \bibinfo {editor}
  {\bibfnamefont {S.}~\bibnamefont {Bengio}}, \bibinfo {editor} {\bibfnamefont
  {H.}~\bibnamefont {Wallach}}, \bibinfo {editor} {\bibfnamefont
  {R.}~\bibnamefont {Fergus}}, \bibinfo {editor} {\bibfnamefont
  {S.}~\bibnamefont {Vishwanathan}},\ and\ \bibinfo {editor} {\bibfnamefont
  {R.}~\bibnamefont {Garnett}}}\ (\bibinfo  {publisher} {Curran Associates,
  Inc.},\ \bibinfo {year} {2017})\BibitemShut {NoStop}%
\bibitem [{\citenamefont {von Luxburg}\ and\ \citenamefont
  {Bousquet}(2004)}]{luxburg2004distance}%
  \BibitemOpen
  \bibfield  {author} {\bibinfo {author} {\bibfnamefont {U.}~\bibnamefont {von
  Luxburg}}\ and\ \bibinfo {author} {\bibfnamefont {O.}~\bibnamefont
  {Bousquet}},\ }\bibfield  {title} {\enquote {\bibinfo {title} {Distance-based
  classification with lipschitz functions.}}\ }\href@noop {} {\bibfield
  {journal} {\bibinfo  {journal} {J. Mach. Learn. Res.}\ }\textbf {\bibinfo
  {volume} {5}},\ \bibinfo {pages} {669--695} (\bibinfo {year}
  {2004})}\BibitemShut {NoStop}%
\bibitem [{\citenamefont {Bahri}\ \emph {et~al.}(2020)\citenamefont {Bahri},
  \citenamefont {Kadmon}, \citenamefont {Pennington}, \citenamefont
  {Schoenholz}, \citenamefont {Sohl-Dickstein},\ and\ \citenamefont
  {Ganguli}}]{bahri2020statistical}%
  \BibitemOpen
  \bibfield  {author} {\bibinfo {author} {\bibfnamefont {Y.}~\bibnamefont
  {Bahri}}, \bibinfo {author} {\bibfnamefont {J.}~\bibnamefont {Kadmon}},
  \bibinfo {author} {\bibfnamefont {J.}~\bibnamefont {Pennington}}, \bibinfo
  {author} {\bibfnamefont {S.}~\bibnamefont {Schoenholz}}, \bibinfo {author}
  {\bibfnamefont {J.}~\bibnamefont {Sohl-Dickstein}},\ and\ \bibinfo {author}
  {\bibfnamefont {S.}~\bibnamefont {Ganguli}},\ }\bibfield  {title} {\enquote
  {\bibinfo {title} {{Statistical Mechanics of Deep Learning}},}\ }\href@noop
  {} {\bibfield  {journal} {\bibinfo  {journal} {Annual Review of Condensed
  Matter Physics}\ }\textbf {\bibinfo {volume} {11}},\ \bibinfo {pages}
  {501--528} (\bibinfo {year} {2020})}\BibitemShut {NoStop}%
\bibitem [{\citenamefont {Zdeborov{\'{a}}}(2020)}]{zdeborova2020understanding}%
  \BibitemOpen
  \bibfield  {author} {\bibinfo {author} {\bibfnamefont {L.}~\bibnamefont
  {Zdeborov{\'{a}}}},\ }\bibfield  {title} {\enquote {\bibinfo {title}
  {{Understanding deep learning is also a job for physicists}},}\ }\href@noop
  {} {\bibfield  {journal} {\bibinfo  {journal} {Nature Physics}\ } (\bibinfo
  {year} {2020})}\BibitemShut {NoStop}%
\bibitem [{\citenamefont {Geiger}, \citenamefont {Petrini},\ and\ \citenamefont
  {Wyart}(2021)}]{geiger2021landscape}%
  \BibitemOpen
  \bibfield  {author} {\bibinfo {author} {\bibfnamefont {M.}~\bibnamefont
  {Geiger}}, \bibinfo {author} {\bibfnamefont {L.}~\bibnamefont {Petrini}},\
  and\ \bibinfo {author} {\bibfnamefont {M.}~\bibnamefont {Wyart}},\ }\bibfield
   {title} {\enquote {\bibinfo {title} {Landscape and training regimes in deep
  learning},}\ }\href
  {https://doi.org/https://doi.org/10.1016/j.physrep.2021.04.001} {\bibfield
  {journal} {\bibinfo  {journal} {Physics Reports}\ }\textbf {\bibinfo {volume}
  {924}},\ \bibinfo {pages} {1--18} (\bibinfo {year} {2021})},\ \bibinfo {note}
  {landscape and training regimes in deep learning}\BibitemShut {NoStop}%
\bibitem [{\citenamefont {Grassberger}\ and\ \citenamefont
  {Procaccia}(1983)}]{grassberger1983measuring}%
  \BibitemOpen
  \bibfield  {author} {\bibinfo {author} {\bibfnamefont {P.}~\bibnamefont
  {Grassberger}}\ and\ \bibinfo {author} {\bibfnamefont {I.}~\bibnamefont
  {Procaccia}},\ }\bibfield  {title} {\enquote {\bibinfo {title} {{Measuring
  the strangeness of strange attractors}},}\ }\href@noop {} {\bibfield
  {journal} {\bibinfo  {journal} {Physica D: Nonlinear Phenomena}\ }\textbf
  {\bibinfo {volume} {9}},\ \bibinfo {pages} {189--208} (\bibinfo {year}
  {1983})}\BibitemShut {NoStop}%
\bibitem [{\citenamefont {{Costa}}\ and\ \citenamefont
  {{Hero}}(2004)}]{Costa2004}%
  \BibitemOpen
  \bibfield  {author} {\bibinfo {author} {\bibfnamefont {J.}~\bibnamefont
  {{Costa}}}\ and\ \bibinfo {author} {\bibfnamefont {A.}~\bibnamefont
  {{Hero}}},\ }\bibfield  {title} {\enquote {\bibinfo {title} {Learning
  intrinsic dimension and intrinsic entropy of high-dimensional datasets},}\
  }in\ \href@noop {} {\emph {\bibinfo {booktitle} {2004 12th European Signal
  Processing Conference}}}\ (\bibinfo {year} {2004})\ pp.\ \bibinfo {pages}
  {369--372}\BibitemShut {NoStop}%
\bibitem [{\citenamefont {Levina}\ and\ \citenamefont
  {Bickel}(2004)}]{Levina2004a}%
  \BibitemOpen
  \bibfield  {author} {\bibinfo {author} {\bibfnamefont {E.}~\bibnamefont
  {Levina}}\ and\ \bibinfo {author} {\bibfnamefont {P.}~\bibnamefont
  {Bickel}},\ }\bibfield  {title} {\enquote {\bibinfo {title} {{Maximum
  likelihood estimation of intrinsic dimension}},}\ }in\ \href@noop {} {\emph
  {\bibinfo {booktitle} {Advances in Neural Information Processing Systems
  17}}}\ (\bibinfo {year} {2004})\BibitemShut {NoStop}%
\bibitem [{\citenamefont {Spigler}, \citenamefont {Geiger},\ and\ \citenamefont
  {Wyart}(2020)}]{spigler2019asymptotic}%
  \BibitemOpen
  \bibfield  {author} {\bibinfo {author} {\bibfnamefont {S.}~\bibnamefont
  {Spigler}}, \bibinfo {author} {\bibfnamefont {M.}~\bibnamefont {Geiger}},\
  and\ \bibinfo {author} {\bibfnamefont {M.}~\bibnamefont {Wyart}},\ }\bibfield
   {title} {\enquote {\bibinfo {title} {Asymptotic learning curves of kernel
  methods: empirical data versus teacher--student paradigm},}\ }\href
  {https://doi.org/10.1088/1742-5468/abc61d} {\bibfield  {journal} {\bibinfo
  {journal} {Journal of Statistical Mechanics: Theory and Experiment}\ }\textbf
  {\bibinfo {volume} {2020}},\ \bibinfo {pages} {124001} (\bibinfo {year}
  {2020})}\BibitemShut {NoStop}%
\bibitem [{\citenamefont {Pope}\ \emph {et~al.}(2021)\citenamefont {Pope},
  \citenamefont {Zhu}, \citenamefont {Abdelkader}, \citenamefont {Goldblum},\
  and\ \citenamefont {Goldstein}}]{pope2021intrinsic}%
  \BibitemOpen
  \bibfield  {author} {\bibinfo {author} {\bibfnamefont {P.}~\bibnamefont
  {Pope}}, \bibinfo {author} {\bibfnamefont {C.}~\bibnamefont {Zhu}}, \bibinfo
  {author} {\bibfnamefont {A.}~\bibnamefont {Abdelkader}}, \bibinfo {author}
  {\bibfnamefont {M.}~\bibnamefont {Goldblum}},\ and\ \bibinfo {author}
  {\bibfnamefont {T.}~\bibnamefont {Goldstein}},\ }\bibfield  {title} {\enquote
  {\bibinfo {title} {The intrinsic dimension of images and its impact on
  learning},}\ }in\ \href {https://openreview.net/forum?id=XJk19XzGq2J} {\emph
  {\bibinfo {booktitle} {International Conference on Learning
  Representations}}}\ (\bibinfo {year} {2021})\BibitemShut {NoStop}%
\bibitem [{\citenamefont {Goldt}\ \emph {et~al.}(2020)\citenamefont {Goldt},
  \citenamefont {M{\'e}zard}, \citenamefont {Krzakala},\ and\ \citenamefont
  {Zdeborov{\'a}}}]{goldt2020modelling}%
  \BibitemOpen
  \bibfield  {author} {\bibinfo {author} {\bibfnamefont {S.}~\bibnamefont
  {Goldt}}, \bibinfo {author} {\bibfnamefont {M.}~\bibnamefont {M{\'e}zard}},
  \bibinfo {author} {\bibfnamefont {F.}~\bibnamefont {Krzakala}},\ and\
  \bibinfo {author} {\bibfnamefont {L.}~\bibnamefont {Zdeborov{\'a}}},\
  }\bibfield  {title} {\enquote {\bibinfo {title} {Modeling the influence of
  data structure on learning in neural networks: The hidden manifold model},}\
  }\href@noop {} {\bibfield  {journal} {\bibinfo  {journal} {Phys. Rev. X}\
  }\textbf {\bibinfo {volume} {10}},\ \bibinfo {pages} {041044} (\bibinfo
  {year} {2020})}\BibitemShut {NoStop}%
\bibitem [{\citenamefont {Deng}\ \emph {et~al.}(2009)\citenamefont {Deng},
  \citenamefont {Dong}, \citenamefont {Socher}, \citenamefont {Li},
  \citenamefont {Li},\ and\ \citenamefont {Fei-Fei}}]{deng2009imagenet}%
  \BibitemOpen
  \bibfield  {author} {\bibinfo {author} {\bibfnamefont {J.}~\bibnamefont
  {Deng}}, \bibinfo {author} {\bibfnamefont {W.}~\bibnamefont {Dong}}, \bibinfo
  {author} {\bibfnamefont {R.}~\bibnamefont {Socher}}, \bibinfo {author}
  {\bibfnamefont {L.-J.}\ \bibnamefont {Li}}, \bibinfo {author} {\bibfnamefont
  {K.}~\bibnamefont {Li}},\ and\ \bibinfo {author} {\bibfnamefont
  {L.}~\bibnamefont {Fei-Fei}},\ }\bibfield  {title} {\enquote {\bibinfo
  {title} {Imagenet: A large-scale hierarchical image database},}\ }in\
  \href@noop {} {\emph {\bibinfo {booktitle} {2009 IEEE conference on computer
  vision and pattern recognition}}}\ (\bibinfo {organization} {Ieee},\ \bibinfo
  {year} {2009})\ pp.\ \bibinfo {pages} {248--255}\BibitemShut {NoStop}%
\bibitem [{\citenamefont {Chung}, \citenamefont {Lee},\ and\ \citenamefont
  {Sompolinsky}(2018)}]{chung2018classification}%
  \BibitemOpen
  \bibfield  {author} {\bibinfo {author} {\bibfnamefont {S.}~\bibnamefont
  {Chung}}, \bibinfo {author} {\bibfnamefont {D.}~\bibnamefont {Lee}},\ and\
  \bibinfo {author} {\bibfnamefont {H.}~\bibnamefont {Sompolinsky}},\
  }\bibfield  {title} {\enquote {\bibinfo {title} {Classification and geometry
  of general perceptual manifolds},}\ }\href
  {https://doi.org/10.1103/PhysRevX.8.031003} {\bibfield  {journal} {\bibinfo
  {journal} {Phys. Rev. X}\ }\textbf {\bibinfo {volume} {8}},\ \bibinfo {pages}
  {031003} (\bibinfo {year} {2018})}\BibitemShut {NoStop}%
\bibitem [{\citenamefont {Goldt}\ \emph {et~al.}(2022)\citenamefont {Goldt},
  \citenamefont {Loureiro}, \citenamefont {Reeves}, \citenamefont {Krzakala},
  \citenamefont {M{\'e}zard},\ and\ \citenamefont
  {Zdeborov{\'a}}}]{goldt2022gaussian}%
  \BibitemOpen
  \bibfield  {author} {\bibinfo {author} {\bibfnamefont {S.}~\bibnamefont
  {Goldt}}, \bibinfo {author} {\bibfnamefont {B.}~\bibnamefont {Loureiro}},
  \bibinfo {author} {\bibfnamefont {G.}~\bibnamefont {Reeves}}, \bibinfo
  {author} {\bibfnamefont {F.}~\bibnamefont {Krzakala}}, \bibinfo {author}
  {\bibfnamefont {M.}~\bibnamefont {M{\'e}zard}},\ and\ \bibinfo {author}
  {\bibfnamefont {L.}~\bibnamefont {Zdeborov{\'a}}},\ }\bibfield  {title}
  {\enquote {\bibinfo {title} {The gaussian equivalence of generative models
  for learning with shallow neural networks},}\ }in\ \href
  {https://proceedings.mlr.press/v145/goldt22a.html} {\emph {\bibinfo
  {booktitle} {Proceedings of the 2nd Mathematical and Scientific Machine
  Learning Conference}}},\ \bibinfo {series} {Proceedings of Machine Learning
  Research}, Vol.\ \bibinfo {volume} {145},\ \bibinfo {editor} {edited by\
  \bibinfo {editor} {\bibfnamefont {J.}~\bibnamefont {Bruna}}, \bibinfo
  {editor} {\bibfnamefont {J.}~\bibnamefont {Hesthaven}},\ and\ \bibinfo
  {editor} {\bibfnamefont {L.}~\bibnamefont {Zdeborov{\'a}}}}\ (\bibinfo
  {publisher} {PMLR},\ \bibinfo {year} {2022})\ pp.\ \bibinfo {pages}
  {426--471}\BibitemShut {NoStop}%
\bibitem [{\citenamefont {Ghorbani}\ \emph {et~al.}(2020)\citenamefont
  {Ghorbani}, \citenamefont {Mei}, \citenamefont {Misiakiewicz},\ and\
  \citenamefont {Montanari}}]{ghorbani2020neural}%
  \BibitemOpen
  \bibfield  {author} {\bibinfo {author} {\bibfnamefont {B.}~\bibnamefont
  {Ghorbani}}, \bibinfo {author} {\bibfnamefont {S.}~\bibnamefont {Mei}},
  \bibinfo {author} {\bibfnamefont {T.}~\bibnamefont {Misiakiewicz}},\ and\
  \bibinfo {author} {\bibfnamefont {A.}~\bibnamefont {Montanari}},\ }\bibfield
  {title} {\enquote {\bibinfo {title} {When do neural networks outperform
  kernel methods?}}\ }in\ \href@noop {} {\emph {\bibinfo {booktitle} {Advances
  in Neural Information Processing Systems}}},\ Vol.~\bibinfo {volume} {33}\
  (\bibinfo {year} {2020})\BibitemShut {NoStop}%
\bibitem [{\citenamefont {Richards}, \citenamefont {Mourtada},\ and\
  \citenamefont {Rosasco}(2021)}]{richards21asymptotics}%
  \BibitemOpen
  \bibfield  {author} {\bibinfo {author} {\bibfnamefont {D.}~\bibnamefont
  {Richards}}, \bibinfo {author} {\bibfnamefont {J.}~\bibnamefont {Mourtada}},\
  and\ \bibinfo {author} {\bibfnamefont {L.}~\bibnamefont {Rosasco}},\
  }\bibfield  {title} {\enquote {\bibinfo {title} {Asymptotics of ridge(less)
  regression under general source condition},}\ }in\ \href
  {https://proceedings.mlr.press/v130/richards21b.html} {\emph {\bibinfo
  {booktitle} {Proceedings of The 24th International Conference on Artificial
  Intelligence and Statistics}}},\ \bibinfo {series} {Proceedings of Machine
  Learning Research}, Vol.\ \bibinfo {volume} {130},\ \bibinfo {editor} {edited
  by\ \bibinfo {editor} {\bibfnamefont {A.}~\bibnamefont {Banerjee}}\ and\
  \bibinfo {editor} {\bibfnamefont {K.}~\bibnamefont {Fukumizu}}}\ (\bibinfo
  {publisher} {PMLR},\ \bibinfo {year} {2021})\ pp.\ \bibinfo {pages}
  {3889--3897}\BibitemShut {NoStop}%
\bibitem [{\citenamefont {Bach}(2017)}]{bach2017breaking}%
  \BibitemOpen
  \bibfield  {author} {\bibinfo {author} {\bibfnamefont {F.}~\bibnamefont
  {Bach}},\ }\bibfield  {title} {\enquote {\bibinfo {title} {Breaking the curse
  of dimensionality with convex neural networks},}\ }\href@noop {} {\bibfield
  {journal} {\bibinfo  {journal} {The Journal of Machine Learning Research}\
  }\textbf {\bibinfo {volume} {18}},\ \bibinfo {pages} {629--681} (\bibinfo
  {year} {2017})}\BibitemShut {NoStop}%
\bibitem [{\citenamefont {Ghorbani}\ \emph {et~al.}(2019)\citenamefont
  {Ghorbani}, \citenamefont {Mei}, \citenamefont {Misiakiewicz},\ and\
  \citenamefont {Montanari}}]{ghorbani2019limitations}%
  \BibitemOpen
  \bibfield  {author} {\bibinfo {author} {\bibfnamefont {B.}~\bibnamefont
  {Ghorbani}}, \bibinfo {author} {\bibfnamefont {S.}~\bibnamefont {Mei}},
  \bibinfo {author} {\bibfnamefont {T.}~\bibnamefont {Misiakiewicz}},\ and\
  \bibinfo {author} {\bibfnamefont {A.}~\bibnamefont {Montanari}},\ }\bibfield
  {title} {\enquote {\bibinfo {title} {Limitations of lazy training of
  two-layers neural network},}\ }in\ \href@noop {} {\emph {\bibinfo {booktitle}
  {Advances in Neural Information Processing Systems}}},\ Vol.~\bibinfo
  {volume} {32}\ (\bibinfo {year} {2019})\ pp.\ \bibinfo {pages}
  {9111--9121}\BibitemShut {NoStop}%
\bibitem [{\citenamefont {Chizat}\ and\ \citenamefont
  {Bach}(2020)}]{chizat2020implicit}%
  \BibitemOpen
  \bibfield  {author} {\bibinfo {author} {\bibfnamefont {L.}~\bibnamefont
  {Chizat}}\ and\ \bibinfo {author} {\bibfnamefont {F.}~\bibnamefont {Bach}},\
  }\bibfield  {title} {\enquote {\bibinfo {title} {Implicit bias of gradient
  descent for wide two-layer neural networks trained with the logistic loss},}\
  }in\ \href@noop {} {\emph {\bibinfo {booktitle} {Conference on Learning
  Theory}}}\ (\bibinfo {organization} {PMLR},\ \bibinfo {year} {2020})\ pp.\
  \bibinfo {pages} {1305--1338}\BibitemShut {NoStop}%
\bibitem [{\citenamefont {Geiger}\ \emph {et~al.}(2020)\citenamefont {Geiger},
  \citenamefont {Spigler}, \citenamefont {Jacot},\ and\ \citenamefont
  {Wyart}}]{geiger2020disentangling}%
  \BibitemOpen
  \bibfield  {author} {\bibinfo {author} {\bibfnamefont {M.}~\bibnamefont
  {Geiger}}, \bibinfo {author} {\bibfnamefont {S.}~\bibnamefont {Spigler}},
  \bibinfo {author} {\bibfnamefont {A.}~\bibnamefont {Jacot}},\ and\ \bibinfo
  {author} {\bibfnamefont {M.}~\bibnamefont {Wyart}},\ }\bibfield  {title}
  {\enquote {\bibinfo {title} {Disentangling feature and lazy training in deep
  neural networks},}\ }\href@noop {} {\bibfield  {journal} {\bibinfo  {journal}
  {Journal of Statistical Mechanics: Theory and Experiment}\ }\textbf {\bibinfo
  {volume} {2020}},\ \bibinfo {pages} {113301} (\bibinfo {year}
  {2020})}\BibitemShut {NoStop}%
\bibitem [{\citenamefont {Daniely}\ and\ \citenamefont
  {Malach}(2020)}]{daniely2020learning}%
  \BibitemOpen
  \bibfield  {author} {\bibinfo {author} {\bibfnamefont {A.}~\bibnamefont
  {Daniely}}\ and\ \bibinfo {author} {\bibfnamefont {E.}~\bibnamefont
  {Malach}},\ }\bibfield  {title} {\enquote {\bibinfo {title} {Learning
  parities with neural networks},}\ }in\ \href@noop {} {\emph {\bibinfo
  {booktitle} {Advances in Neural Information Processing Systems}}},\
  Vol.~\bibinfo {volume} {33}\ (\bibinfo {year} {2020})\BibitemShut {NoStop}%
\bibitem [{\citenamefont {Refinetti}\ \emph {et~al.}(2021)\citenamefont
  {Refinetti}, \citenamefont {Goldt}, \citenamefont {Krzakala},\ and\
  \citenamefont {Zdeborova}}]{refinetti2021classifying}%
  \BibitemOpen
  \bibfield  {author} {\bibinfo {author} {\bibfnamefont {M.}~\bibnamefont
  {Refinetti}}, \bibinfo {author} {\bibfnamefont {S.}~\bibnamefont {Goldt}},
  \bibinfo {author} {\bibfnamefont {F.}~\bibnamefont {Krzakala}},\ and\
  \bibinfo {author} {\bibfnamefont {L.}~\bibnamefont {Zdeborova}},\ }\bibfield
  {title} {\enquote {\bibinfo {title} {Classifying high-dimensional gaussian
  mixtures: Where kernel methods fail and neural networks succeed},}\ }in\
  \href {http://proceedings.mlr.press/v139/refinetti21b.html} {\emph {\bibinfo
  {booktitle} {Proceedings of the 38th International Conference on Machine
  Learning}}},\ \bibinfo {series} {Proceedings of Machine Learning Research},
  Vol.\ \bibinfo {volume} {139},\ \bibinfo {editor} {edited by\ \bibinfo
  {editor} {\bibfnamefont {M.}~\bibnamefont {Meila}}\ and\ \bibinfo {editor}
  {\bibfnamefont {T.}~\bibnamefont {Zhang}}}\ (\bibinfo  {publisher} {PMLR},\
  \bibinfo {year} {2021})\ pp.\ \bibinfo {pages} {8936--8947}\BibitemShut
  {NoStop}%
\bibitem [{\citenamefont {Torlai}\ and\ \citenamefont
  {Melko}(2016)}]{torlai2016learning}%
  \BibitemOpen
  \bibfield  {author} {\bibinfo {author} {\bibfnamefont {G.}~\bibnamefont
  {Torlai}}\ and\ \bibinfo {author} {\bibfnamefont {R.~G.}\ \bibnamefont
  {Melko}},\ }\bibfield  {title} {\enquote {\bibinfo {title} {Learning
  thermodynamics with boltzmann machines},}\ }\href
  {https://doi.org/10.1103/PhysRevB.94.165134} {\bibfield  {journal} {\bibinfo
  {journal} {Phys. Rev. B}\ }\textbf {\bibinfo {volume} {94}},\ \bibinfo
  {pages} {165134} (\bibinfo {year} {2016})}\BibitemShut {NoStop}%
\bibitem [{\citenamefont {Carrasquilla}\ and\ \citenamefont
  {Melko}(2017)}]{carrasquilla2017machine}%
  \BibitemOpen
  \bibfield  {author} {\bibinfo {author} {\bibfnamefont {J.}~\bibnamefont
  {Carrasquilla}}\ and\ \bibinfo {author} {\bibfnamefont {R.~G.}\ \bibnamefont
  {Melko}},\ }\bibfield  {title} {\enquote {\bibinfo {title} {Machine learning
  phases of matter},}\ }\href@noop {} {\bibfield  {journal} {\bibinfo
  {journal} {Nature Physics}\ }\textbf {\bibinfo {volume} {13}},\ \bibinfo
  {pages} {431--434} (\bibinfo {year} {2017})}\BibitemShut {NoStop}%
\bibitem [{\citenamefont {Van~Nieuwenburg}, \citenamefont {Liu},\ and\
  \citenamefont {Huber}(2017)}]{van2017learning}%
  \BibitemOpen
  \bibfield  {author} {\bibinfo {author} {\bibfnamefont {E.~P.}\ \bibnamefont
  {Van~Nieuwenburg}}, \bibinfo {author} {\bibfnamefont {Y.-H.}\ \bibnamefont
  {Liu}},\ and\ \bibinfo {author} {\bibfnamefont {S.~D.}\ \bibnamefont
  {Huber}},\ }\bibfield  {title} {\enquote {\bibinfo {title} {Learning phase
  transitions by confusion},}\ }\href@noop {} {\bibfield  {journal} {\bibinfo
  {journal} {Nature Physics}\ }\textbf {\bibinfo {volume} {13}},\ \bibinfo
  {pages} {435--439} (\bibinfo {year} {2017})}\BibitemShut {NoStop}%
\bibitem [{\citenamefont {Wirnsberger}\ \emph {et~al.}(2020)\citenamefont
  {Wirnsberger}, \citenamefont {Ballard}, \citenamefont {Papamakarios},
  \citenamefont {Abercrombie}, \citenamefont {Racani{\`e}re}, \citenamefont
  {Pritzel}, \citenamefont {Jimenez~Rezende},\ and\ \citenamefont
  {Blundell}}]{wirnsberger2020targeted}%
  \BibitemOpen
  \bibfield  {author} {\bibinfo {author} {\bibfnamefont {P.}~\bibnamefont
  {Wirnsberger}}, \bibinfo {author} {\bibfnamefont {A.~J.}\ \bibnamefont
  {Ballard}}, \bibinfo {author} {\bibfnamefont {G.}~\bibnamefont
  {Papamakarios}}, \bibinfo {author} {\bibfnamefont {S.}~\bibnamefont
  {Abercrombie}}, \bibinfo {author} {\bibfnamefont {S.}~\bibnamefont
  {Racani{\`e}re}}, \bibinfo {author} {\bibfnamefont {A.}~\bibnamefont
  {Pritzel}}, \bibinfo {author} {\bibfnamefont {D.}~\bibnamefont
  {Jimenez~Rezende}},\ and\ \bibinfo {author} {\bibfnamefont {C.}~\bibnamefont
  {Blundell}},\ }\bibfield  {title} {\enquote {\bibinfo {title} {Targeted free
  energy estimation via learned mappings},}\ }\href@noop {} {\bibfield
  {journal} {\bibinfo  {journal} {The Journal of Chemical Physics}\ }\textbf
  {\bibinfo {volume} {153}},\ \bibinfo {pages} {144112} (\bibinfo {year}
  {2020})}\BibitemShut {NoStop}%
\bibitem [{\citenamefont {Seif}, \citenamefont {Hafezi},\ and\ \citenamefont
  {Jarzynski}(2021)}]{seif2021machine}%
  \BibitemOpen
  \bibfield  {author} {\bibinfo {author} {\bibfnamefont {A.}~\bibnamefont
  {Seif}}, \bibinfo {author} {\bibfnamefont {M.}~\bibnamefont {Hafezi}},\ and\
  \bibinfo {author} {\bibfnamefont {C.}~\bibnamefont {Jarzynski}},\ }\bibfield
  {title} {\enquote {\bibinfo {title} {Machine learning the thermodynamic arrow
  of time},}\ }\href@noop {} {\bibfield  {journal} {\bibinfo  {journal} {Nature
  Physics}\ }\textbf {\bibinfo {volume} {17}},\ \bibinfo {pages} {105--113}
  (\bibinfo {year} {2021})}\BibitemShut {NoStop}%
\bibitem [{\citenamefont {Kim}\ \emph {et~al.}(2020{\natexlab{b}})\citenamefont
  {Kim}, \citenamefont {Bae}, \citenamefont {Lee},\ and\ \citenamefont
  {Jeong}}]{kim2020entropy}%
  \BibitemOpen
  \bibfield  {author} {\bibinfo {author} {\bibfnamefont {D.-K.}\ \bibnamefont
  {Kim}}, \bibinfo {author} {\bibfnamefont {Y.}~\bibnamefont {Bae}}, \bibinfo
  {author} {\bibfnamefont {S.}~\bibnamefont {Lee}},\ and\ \bibinfo {author}
  {\bibfnamefont {H.}~\bibnamefont {Jeong}},\ }\bibfield  {title} {\enquote
  {\bibinfo {title} {Learning entropy production via neural networks},}\ }\href
  {https://doi.org/10.1103/PhysRevLett.125.140604} {\bibfield  {journal}
  {\bibinfo  {journal} {Phys. Rev. Lett.}\ }\textbf {\bibinfo {volume} {125}},\
  \bibinfo {pages} {140604} (\bibinfo {year} {2020}{\natexlab{b}})}\BibitemShut
  {NoStop}%
\bibitem [{\citenamefont {Goodfellow}, \citenamefont {Bengio},\ and\
  \citenamefont {Courville}(2016)}]{goodfellow2016deep}%
  \BibitemOpen
  \bibfield  {author} {\bibinfo {author} {\bibfnamefont {I.}~\bibnamefont
  {Goodfellow}}, \bibinfo {author} {\bibfnamefont {Y.}~\bibnamefont {Bengio}},\
  and\ \bibinfo {author} {\bibfnamefont {A.}~\bibnamefont {Courville}},\
  }\href@noop {} {\emph {\bibinfo {title} {Deep learning}}}\ (\bibinfo
  {publisher} {MIT Press},\ \bibinfo {year} {2016})\BibitemShut {NoStop}%
\bibitem [{\citenamefont {Bengio}, \citenamefont {Simard},\ and\ \citenamefont
  {Frasconi}(1994)}]{bengio1994learning}%
  \BibitemOpen
  \bibfield  {author} {\bibinfo {author} {\bibfnamefont {Y.}~\bibnamefont
  {Bengio}}, \bibinfo {author} {\bibfnamefont {P.}~\bibnamefont {Simard}},\
  and\ \bibinfo {author} {\bibfnamefont {P.}~\bibnamefont {Frasconi}},\
  }\bibfield  {title} {\enquote {\bibinfo {title} {Learning long-term
  dependencies with gradient descent is difficult},}\ }\href@noop {} {\bibfield
   {journal} {\bibinfo  {journal} {IEEE transactions on neural networks}\
  }\textbf {\bibinfo {volume} {5}},\ \bibinfo {pages} {157--166} (\bibinfo
  {year} {1994})}\BibitemShut {NoStop}%
\bibitem [{\citenamefont {Hochreiter}\ and\ \citenamefont
  {Schmidhuber}(1997)}]{hochreiter1997long}%
  \BibitemOpen
  \bibfield  {author} {\bibinfo {author} {\bibfnamefont {S.}~\bibnamefont
  {Hochreiter}}\ and\ \bibinfo {author} {\bibfnamefont {J.}~\bibnamefont
  {Schmidhuber}},\ }\bibfield  {title} {\enquote {\bibinfo {title} {Long
  short-term memory},}\ }\href@noop {} {\bibfield  {journal} {\bibinfo
  {journal} {Neural computation}\ }\textbf {\bibinfo {volume} {9}},\ \bibinfo
  {pages} {1735--1780} (\bibinfo {year} {1997})}\BibitemShut {NoStop}%
\bibitem [{\citenamefont {Argun}\ \emph {et~al.}(2020)\citenamefont {Argun},
  \citenamefont {Thalheim}, \citenamefont {Bo}, \citenamefont {Cichos},\ and\
  \citenamefont {Volpe}}]{argun2020enhanced}%
  \BibitemOpen
  \bibfield  {author} {\bibinfo {author} {\bibfnamefont {A.}~\bibnamefont
  {Argun}}, \bibinfo {author} {\bibfnamefont {T.}~\bibnamefont {Thalheim}},
  \bibinfo {author} {\bibfnamefont {S.}~\bibnamefont {Bo}}, \bibinfo {author}
  {\bibfnamefont {F.}~\bibnamefont {Cichos}},\ and\ \bibinfo {author}
  {\bibfnamefont {G.}~\bibnamefont {Volpe}},\ }\bibfield  {title} {\enquote
  {\bibinfo {title} {Enhanced force-field calibration via machine learning},}\
  }\href@noop {} {\bibfield  {journal} {\bibinfo  {journal} {Applied Physics
  Reviews}\ }\textbf {\bibinfo {volume} {7}},\ \bibinfo {pages} {041404}
  (\bibinfo {year} {2020})}\BibitemShut {NoStop}%
\bibitem [{\citenamefont {Pathak}\ \emph {et~al.}(2018)\citenamefont {Pathak},
  \citenamefont {Hunt}, \citenamefont {Girvan}, \citenamefont {Lu},\ and\
  \citenamefont {Ott}}]{pathak2018model}%
  \BibitemOpen
  \bibfield  {author} {\bibinfo {author} {\bibfnamefont {J.}~\bibnamefont
  {Pathak}}, \bibinfo {author} {\bibfnamefont {B.}~\bibnamefont {Hunt}},
  \bibinfo {author} {\bibfnamefont {M.}~\bibnamefont {Girvan}}, \bibinfo
  {author} {\bibfnamefont {Z.}~\bibnamefont {Lu}},\ and\ \bibinfo {author}
  {\bibfnamefont {E.}~\bibnamefont {Ott}},\ }\bibfield  {title} {\enquote
  {\bibinfo {title} {Model-free prediction of large spatiotemporally chaotic
  systems from data: A reservoir computing approach},}\ }\href@noop {}
  {\bibfield  {journal} {\bibinfo  {journal} {Physical review letters}\
  }\textbf {\bibinfo {volume} {120}},\ \bibinfo {pages} {024102} (\bibinfo
  {year} {2018})}\BibitemShut {NoStop}%
\bibitem [{\citenamefont {Kuramoto}(1978)}]{Kuramoto1978diffusion}%
  \BibitemOpen
  \bibfield  {author} {\bibinfo {author} {\bibfnamefont {Y.}~\bibnamefont
  {Kuramoto}},\ }\bibfield  {title} {\enquote {\bibinfo {title}
  {Diffusion-induced chaos in reaction systems},}\ }\href@noop {} {\bibfield
  {journal} {\bibinfo  {journal} {Progress of Theoretical Physics Supplement}\
  }\textbf {\bibinfo {volume} {64}},\ \bibinfo {pages} {346--367} (\bibinfo
  {year} {1978})}\BibitemShut {NoStop}%
\bibitem [{\citenamefont {Sivashinsky}(1977)}]{sivashinsky1977nonlinear}%
  \BibitemOpen
  \bibfield  {author} {\bibinfo {author} {\bibfnamefont {G.~I.}\ \bibnamefont
  {Sivashinsky}},\ }\bibfield  {title} {\enquote {\bibinfo {title} {Nonlinear
  analysis of hydrodynamic instability in laminar flames---i. derivation of
  basic equations},}\ }\href@noop {} {\bibfield  {journal} {\bibinfo  {journal}
  {Acta astronautica}\ }\textbf {\bibinfo {volume} {4}},\ \bibinfo {pages}
  {1177--1206} (\bibinfo {year} {1977})}\BibitemShut {NoStop}%
\bibitem [{\citenamefont {Sivashinsky}(1980)}]{sivashinsky1980flame}%
  \BibitemOpen
  \bibfield  {author} {\bibinfo {author} {\bibfnamefont {G.~I.}\ \bibnamefont
  {Sivashinsky}},\ }\bibfield  {title} {\enquote {\bibinfo {title} {On flame
  propagation under conditions of stoichiometry},}\ }\href@noop {} {\bibfield
  {journal} {\bibinfo  {journal} {SIAM Journal on Applied Mathematics}\
  }\textbf {\bibinfo {volume} {39}},\ \bibinfo {pages} {67--82} (\bibinfo
  {year} {1980})}\BibitemShut {NoStop}%
\bibitem [{\citenamefont {Luko{\v{s}}evi{\v{c}}ius}\ and\ \citenamefont
  {Jaeger}(2009)}]{lukovsevivcius2009reservoir}%
  \BibitemOpen
  \bibfield  {author} {\bibinfo {author} {\bibfnamefont {M.}~\bibnamefont
  {Luko{\v{s}}evi{\v{c}}ius}}\ and\ \bibinfo {author} {\bibfnamefont
  {H.}~\bibnamefont {Jaeger}},\ }\bibfield  {title} {\enquote {\bibinfo {title}
  {Reservoir computing approaches to recurrent neural network training},}\
  }\href@noop {} {\bibfield  {journal} {\bibinfo  {journal} {Computer Science
  Review}\ }\textbf {\bibinfo {volume} {3}},\ \bibinfo {pages} {127--149}
  (\bibinfo {year} {2009})}\BibitemShut {NoStop}%
\bibitem [{\citenamefont {Seif}\ \emph {et~al.}(2022)\citenamefont {Seif},
  \citenamefont {Loos}, \citenamefont {Tucci}, \citenamefont {Rold{\'a}n},\
  and\ \citenamefont {Goldt}}]{seif2022impact}%
  \BibitemOpen
  \bibfield  {author} {\bibinfo {author} {\bibfnamefont {A.}~\bibnamefont
  {Seif}}, \bibinfo {author} {\bibfnamefont {S.~A.}\ \bibnamefont {Loos}},
  \bibinfo {author} {\bibfnamefont {G.}~\bibnamefont {Tucci}}, \bibinfo
  {author} {\bibfnamefont {{\'E}.}~\bibnamefont {Rold{\'a}n}},\ and\ \bibinfo
  {author} {\bibfnamefont {S.}~\bibnamefont {Goldt}},\ }\bibfield  {title}
  {\enquote {\bibinfo {title} {The impact of memory on learning
  sequence-to-sequence tasks},}\ }\href@noop {} {\bibfield  {journal} {\bibinfo
   {journal} {arXiv preprint arXiv:2205.14683}\ } (\bibinfo {year}
  {2022})}\BibitemShut {NoStop}%
\bibitem [{\citenamefont {Krenn}\ \emph {et~al.}(2022)\citenamefont {Krenn},
  \citenamefont {Pollice}, \citenamefont {Guo}, \citenamefont {Aldeghi},
  \citenamefont {Cervera-Lierta}, \citenamefont {Friederich}, \citenamefont
  {dos Passos~Gomes}, \citenamefont {H{\"a}se}, \citenamefont {Jinich},
  \citenamefont {Nigam} \emph {et~al.}}]{krenn2022scientific}%
  \BibitemOpen
  \bibfield  {author} {\bibinfo {author} {\bibfnamefont {M.}~\bibnamefont
  {Krenn}}, \bibinfo {author} {\bibfnamefont {R.}~\bibnamefont {Pollice}},
  \bibinfo {author} {\bibfnamefont {S.~Y.}\ \bibnamefont {Guo}}, \bibinfo
  {author} {\bibfnamefont {M.}~\bibnamefont {Aldeghi}}, \bibinfo {author}
  {\bibfnamefont {A.}~\bibnamefont {Cervera-Lierta}}, \bibinfo {author}
  {\bibfnamefont {P.}~\bibnamefont {Friederich}}, \bibinfo {author}
  {\bibfnamefont {G.}~\bibnamefont {dos Passos~Gomes}}, \bibinfo {author}
  {\bibfnamefont {F.}~\bibnamefont {H{\"a}se}}, \bibinfo {author}
  {\bibfnamefont {A.}~\bibnamefont {Jinich}}, \bibinfo {author} {\bibfnamefont
  {A.}~\bibnamefont {Nigam}}, \emph {et~al.},\ }\bibfield  {title} {\enquote
  {\bibinfo {title} {On scientific understanding with artificial
  intelligence},}\ }\href@noop {} {\bibfield  {journal} {\bibinfo  {journal}
  {Nature Reviews Physics}\ ,\ \bibinfo {pages} {1--9}} (\bibinfo {year}
  {2022})}\BibitemShut {NoStop}%
\bibitem [{\citenamefont {Ott}, \citenamefont {Grebogi},\ and\ \citenamefont
  {Yorke}(1990)}]{Ott1990}%
  \BibitemOpen
  \bibfield  {author} {\bibinfo {author} {\bibfnamefont {E.}~\bibnamefont
  {Ott}}, \bibinfo {author} {\bibfnamefont {C.}~\bibnamefont {Grebogi}},\ and\
  \bibinfo {author} {\bibfnamefont {J.~A.}\ \bibnamefont {Yorke}},\ }\bibfield
  {title} {\enquote {\bibinfo {title} {Controlling chaos},}\ }\href@noop {}
  {\bibfield  {journal} {\bibinfo  {journal} {Phys. Rev. Lett.}\ }\textbf
  {\bibinfo {volume} {64}},\ \bibinfo {pages} {1196--1199} (\bibinfo {year}
  {1990})}\BibitemShut {NoStop}%
\bibitem [{\citenamefont {Romeiras}\ \emph {et~al.}(1992)\citenamefont
  {Romeiras}, \citenamefont {Grebogi}, \citenamefont {Ott},\ and\ \citenamefont
  {Dayawansa}}]{Romeiras1992}%
  \BibitemOpen
  \bibfield  {author} {\bibinfo {author} {\bibfnamefont {F.}~\bibnamefont
  {Romeiras}}, \bibinfo {author} {\bibfnamefont {C.}~\bibnamefont {Grebogi}},
  \bibinfo {author} {\bibfnamefont {E.}~\bibnamefont {Ott}},\ and\ \bibinfo
  {author} {\bibfnamefont {W.}~\bibnamefont {Dayawansa}},\ }\href@noop {}
  {\bibfield  {journal} {\bibinfo  {journal} {Physica D}\ }\textbf {\bibinfo
  {volume} {58}},\ \bibinfo {pages} {165--192} (\bibinfo {year}
  {1992})}\BibitemShut {NoStop}%
\bibitem [{\citenamefont {Sarangapani}(2006)}]{Sarangapani2006}%
  \BibitemOpen
  \bibfield  {author} {\bibinfo {author} {\bibfnamefont {J.}~\bibnamefont
  {Sarangapani}},\ }\href@noop {} {\emph {\bibinfo {title} {Neural network
  control of nonlinear discrete-time systems}}}\ (\bibinfo  {publisher} {Taylor
  \& Francis},\ \bibinfo {address} {Boca Raton},\ \bibinfo {year}
  {2006})\BibitemShut {NoStop}%
\bibitem [{\citenamefont {Jaeger}\ and\ \citenamefont
  {Haas}(2004)}]{Jaeger2004}%
  \BibitemOpen
  \bibfield  {author} {\bibinfo {author} {\bibfnamefont {H.}~\bibnamefont
  {Jaeger}}\ and\ \bibinfo {author} {\bibfnamefont {H.}~\bibnamefont {Haas}},\
  }\bibfield  {title} {\enquote {\bibinfo {title} {Harnessing nonlinearity:
  Predicting chaotic systems and saving energy in wireless communication},}\
  }\href@noop {} {\bibfield  {journal} {\bibinfo  {journal} {Science}\ }\textbf
  {\bibinfo {volume} {304}},\ \bibinfo {pages} {78--80} (\bibinfo {year}
  {2004})}\BibitemShut {NoStop}%
\bibitem [{\citenamefont {Waegeman}, \citenamefont {Wyffels},\ and\
  \citenamefont {B.}(2012)}]{Waegeman2012}%
  \BibitemOpen
  \bibfield  {author} {\bibinfo {author} {\bibfnamefont {T.}~\bibnamefont
  {Waegeman}}, \bibinfo {author} {\bibfnamefont {F.}~\bibnamefont {Wyffels}},\
  and\ \bibinfo {author} {\bibfnamefont {S.}~\bibnamefont {B.}},\ }\bibfield
  {title} {\enquote {\bibinfo {title} {Feedback control by online learning an
  inverse model},}\ }\href@noop {} {\bibfield  {journal} {\bibinfo  {journal}
  {IEEE Trans. Neural Netw. Learn. Syst.}\ }\textbf {\bibinfo {volume} {23}},\
  \bibinfo {pages} {1637--1648} (\bibinfo {year} {2012})}\BibitemShut {NoStop}%
\bibitem [{\citenamefont {Canaday}, \citenamefont {Pomerance},\ and\
  \citenamefont {Gauthier}(2021)}]{Canaday2021}%
  \BibitemOpen
  \bibfield  {author} {\bibinfo {author} {\bibfnamefont {D.}~\bibnamefont
  {Canaday}}, \bibinfo {author} {\bibfnamefont {A.}~\bibnamefont {Pomerance}},\
  and\ \bibinfo {author} {\bibfnamefont {D.~J.}\ \bibnamefont {Gauthier}},\
  }\bibfield  {title} {\enquote {\bibinfo {title} {Model-free control of
  dynamical systems with deep reservoir computing},}\ }\href@noop {} {\bibfield
   {journal} {\bibinfo  {journal} {J. Phys. Complex.}\ }\textbf {\bibinfo
  {volume} {2}},\ \bibinfo {pages} {035025} (\bibinfo {year}
  {2021})}\BibitemShut {NoStop}%
\bibitem [{\citenamefont {Gauthier}\ \emph {et~al.}(2021)\citenamefont
  {Gauthier}, \citenamefont {Bollt}, \citenamefont {Griffith},\ and\
  \citenamefont {Barbosa}}]{Gauthier2021}%
  \BibitemOpen
  \bibfield  {author} {\bibinfo {author} {\bibfnamefont {D.~J.}\ \bibnamefont
  {Gauthier}}, \bibinfo {author} {\bibfnamefont {E.}~\bibnamefont {Bollt}},
  \bibinfo {author} {\bibfnamefont {A.}~\bibnamefont {Griffith}},\ and\
  \bibinfo {author} {\bibfnamefont {W.~A.~S.}\ \bibnamefont {Barbosa}},\
  }\bibfield  {title} {\enquote {\bibinfo {title} {Next generation reservoir
  computing},}\ }\href@noop {} {\bibfield  {journal} {\bibinfo  {journal} {Nat.
  Commun.}\ }\textbf {\bibinfo {volume} {12}},\ \bibinfo {pages} {5564}
  (\bibinfo {year} {2021})}\BibitemShut {NoStop}%
\bibitem [{\citenamefont {Steffen}\ \emph {et~al.}(2018)\citenamefont
  {Steffen}, \citenamefont {Rockstr{\"o}m}, \citenamefont {Richardson},
  \citenamefont {Lenton}, \citenamefont {Folke}, \citenamefont {Liverman},
  \citenamefont {Summerhayes}, \citenamefont {Barnosky}, \citenamefont
  {Cornell}, \citenamefont {Crucifix}, \citenamefont {Donges}, \citenamefont
  {Fetzer}, \citenamefont {Lade}, \citenamefont {Scheffer}, \citenamefont
  {Winkelmann},\ and\ \citenamefont {Schellnhuber}}]{SteffenEtAl2018}%
  \BibitemOpen
  \bibfield  {author} {\bibinfo {author} {\bibfnamefont {W.}~\bibnamefont
  {Steffen}}, \bibinfo {author} {\bibfnamefont {J.}~\bibnamefont
  {Rockstr{\"o}m}}, \bibinfo {author} {\bibfnamefont {K.}~\bibnamefont
  {Richardson}}, \bibinfo {author} {\bibfnamefont {T.~M.}\ \bibnamefont
  {Lenton}}, \bibinfo {author} {\bibfnamefont {C.}~\bibnamefont {Folke}},
  \bibinfo {author} {\bibfnamefont {D.}~\bibnamefont {Liverman}}, \bibinfo
  {author} {\bibfnamefont {C.~P.}\ \bibnamefont {Summerhayes}}, \bibinfo
  {author} {\bibfnamefont {A.~D.}\ \bibnamefont {Barnosky}}, \bibinfo {author}
  {\bibfnamefont {S.~E.}\ \bibnamefont {Cornell}}, \bibinfo {author}
  {\bibfnamefont {M.}~\bibnamefont {Crucifix}}, \bibinfo {author}
  {\bibfnamefont {J.~F.}\ \bibnamefont {Donges}}, \bibinfo {author}
  {\bibfnamefont {I.}~\bibnamefont {Fetzer}}, \bibinfo {author} {\bibfnamefont
  {S.~J.}\ \bibnamefont {Lade}}, \bibinfo {author} {\bibfnamefont
  {M.}~\bibnamefont {Scheffer}}, \bibinfo {author} {\bibfnamefont
  {R.}~\bibnamefont {Winkelmann}},\ and\ \bibinfo {author} {\bibfnamefont
  {H.~J.}\ \bibnamefont {Schellnhuber}},\ }\bibfield  {title} {\enquote
  {\bibinfo {title} {Trajectories of the {{Earth System}} in the
  {{Anthropocene}}},}\ }\href {https://doi.org/10.1073/pnas.1810141115}
  {\bibfield  {journal} {\bibinfo  {journal} {Proceedings of the National
  Academy of Sciences}\ }\textbf {\bibinfo {volume} {115}},\ \bibinfo {pages}
  {8252--8259} (\bibinfo {year} {2018})}\BibitemShut {NoStop}%
\bibitem [{\citenamefont {Levin}(2020)}]{Levin2020}%
  \BibitemOpen
  \bibfield  {author} {\bibinfo {author} {\bibfnamefont {S.~A.}\ \bibnamefont
  {Levin}},\ }\bibfield  {title} {\enquote {\bibinfo {title} {Collective
  {{Cooperation}}: {{From Ecological Communities}} to {{Global Governance}} and
  {{Back}}},}\ }in\ \href {https://doi.org/10.1515/9780691195322-025} {\emph
  {\bibinfo {booktitle} {Collective {{Cooperation}}: {{From Ecological
  Communities}} to {{Global Governance}} and {{Back}}}}}\ (\bibinfo
  {publisher} {{Princeton University Press}},\ \bibinfo {year} {2020})\ pp.\
  \bibinfo {pages} {311--317}\BibitemShut {NoStop}%
\bibitem [{\citenamefont {Nowak}(2006)}]{Nowak2006a}%
  \BibitemOpen
  \bibfield  {author} {\bibinfo {author} {\bibfnamefont {M.~A.}\ \bibnamefont
  {Nowak}},\ }\href@noop {} {\emph {\bibinfo {title} {Evolutionary Dynamics:
  Exploring the Equations of Life}}}\ (\bibinfo  {publisher} {{Harvard
  university press}},\ \bibinfo {year} {2006})\BibitemShut {NoStop}%
\bibitem [{\citenamefont {McNamara}(2013)}]{McNamara2013}%
  \BibitemOpen
  \bibfield  {author} {\bibinfo {author} {\bibfnamefont {J.~M.}\ \bibnamefont
  {McNamara}},\ }\bibfield  {title} {\enquote {\bibinfo {title} {Towards a
  richer evolutionary game theory},}\ }\href
  {https://doi.org/10.1098/rsif.2013.0544} {\bibfield  {journal} {\bibinfo
  {journal} {Journal of The Royal Society Interface}\ }\textbf {\bibinfo
  {volume} {10}},\ \bibinfo {pages} {20130544} (\bibinfo {year}
  {2013})}\BibitemShut {NoStop}%
\bibitem [{\citenamefont {Botvinick}\ \emph {et~al.}(2020)\citenamefont
  {Botvinick}, \citenamefont {Wang}, \citenamefont {Dabney}, \citenamefont
  {Miller},\ and\ \citenamefont {{Kurth-Nelson}}}]{BotvinickEtAl2020}%
  \BibitemOpen
  \bibfield  {author} {\bibinfo {author} {\bibfnamefont {M.}~\bibnamefont
  {Botvinick}}, \bibinfo {author} {\bibfnamefont {J.~X.}\ \bibnamefont {Wang}},
  \bibinfo {author} {\bibfnamefont {W.}~\bibnamefont {Dabney}}, \bibinfo
  {author} {\bibfnamefont {K.~J.}\ \bibnamefont {Miller}},\ and\ \bibinfo
  {author} {\bibfnamefont {Z.}~\bibnamefont {{Kurth-Nelson}}},\ }\bibfield
  {title} {\enquote {\bibinfo {title} {Deep {{Reinforcement Learning}} and
  {{Its Neuroscientific Implications}}},}\ }\href
  {https://doi.org/10.1016/j.neuron.2020.06.014} {\bibfield  {journal}
  {\bibinfo  {journal} {Neuron}\ }\textbf {\bibinfo {volume} {107}},\ \bibinfo
  {pages} {603--616} (\bibinfo {year} {2020})}\BibitemShut {NoStop}%
\bibitem [{\citenamefont {Sutton}\ and\ \citenamefont
  {Barto}(2018)}]{SuttonBarto2018}%
  \BibitemOpen
  \bibfield  {author} {\bibinfo {author} {\bibfnamefont {R.~S.}\ \bibnamefont
  {Sutton}}\ and\ \bibinfo {author} {\bibfnamefont {A.~G.}\ \bibnamefont
  {Barto}},\ }\href@noop {} {\emph {\bibinfo {title} {Reinforcement Learning:
  An Introduction}}},\ \bibinfo {edition} {second edition}\ ed.,\ Adaptive
  Computation and Machine Learning Series\ (\bibinfo  {publisher} {{The MIT
  Press}},\ \bibinfo {address} {{Cambridge, Massachusetts}},\ \bibinfo {year}
  {2018})\BibitemShut {NoStop}%
\bibitem [{\citenamefont {Dafoe}\ \emph {et~al.}(2020)\citenamefont {Dafoe},
  \citenamefont {Hughes}, \citenamefont {Bachrach}, \citenamefont {Collins},
  \citenamefont {McKee}, \citenamefont {Leibo}, \citenamefont {Larson},\ and\
  \citenamefont {Graepel}}]{DafoeEtAl2020}%
  \BibitemOpen
  \bibfield  {author} {\bibinfo {author} {\bibfnamefont {A.}~\bibnamefont
  {Dafoe}}, \bibinfo {author} {\bibfnamefont {E.}~\bibnamefont {Hughes}},
  \bibinfo {author} {\bibfnamefont {Y.}~\bibnamefont {Bachrach}}, \bibinfo
  {author} {\bibfnamefont {T.}~\bibnamefont {Collins}}, \bibinfo {author}
  {\bibfnamefont {K.~R.}\ \bibnamefont {McKee}}, \bibinfo {author}
  {\bibfnamefont {J.~Z.}\ \bibnamefont {Leibo}}, \bibinfo {author}
  {\bibfnamefont {K.}~\bibnamefont {Larson}},\ and\ \bibinfo {author}
  {\bibfnamefont {T.}~\bibnamefont {Graepel}},\ }\href@noop {} {\enquote
  {\bibinfo {title} {Open {{Problems}} in {{Cooperative AI}}},}\ } (\bibinfo
  {year} {2020})\BibitemShut {NoStop}%
\bibitem [{\citenamefont {Dafoe}\ \emph {et~al.}(2021)\citenamefont {Dafoe},
  \citenamefont {Bachrach}, \citenamefont {Hadfield}, \citenamefont {Horvitz},
  \citenamefont {Larson},\ and\ \citenamefont {Graepel}}]{DafoeEtAl2021}%
  \BibitemOpen
  \bibfield  {author} {\bibinfo {author} {\bibfnamefont {A.}~\bibnamefont
  {Dafoe}}, \bibinfo {author} {\bibfnamefont {Y.}~\bibnamefont {Bachrach}},
  \bibinfo {author} {\bibfnamefont {G.}~\bibnamefont {Hadfield}}, \bibinfo
  {author} {\bibfnamefont {E.}~\bibnamefont {Horvitz}}, \bibinfo {author}
  {\bibfnamefont {K.}~\bibnamefont {Larson}},\ and\ \bibinfo {author}
  {\bibfnamefont {T.}~\bibnamefont {Graepel}},\ }\bibfield  {title} {\enquote
  {\bibinfo {title} {Cooperative {{AI}}: Machines must learn to find common
  ground},}\ }\href {https://doi.org/10.1038/d41586-021-01170-0} {\bibfield
  {journal} {\bibinfo  {journal} {Nature}\ }\textbf {\bibinfo {volume} {593}},\
  \bibinfo {pages} {33--36} (\bibinfo {year} {2021})}\BibitemShut {NoStop}%
\bibitem [{\citenamefont {Holland}(2006)}]{Holland2006}%
  \BibitemOpen
  \bibfield  {author} {\bibinfo {author} {\bibfnamefont {J.~H.}\ \bibnamefont
  {Holland}},\ }\bibfield  {title} {\enquote {\bibinfo {title} {Studying
  {{Complex Adaptive Systems}}},}\ }\href
  {https://doi.org/10.1007/s11424-006-0001-z} {\bibfield  {journal} {\bibinfo
  {journal} {Journal of Systems Science and Complexity}\ }\textbf {\bibinfo
  {volume} {19}},\ \bibinfo {pages} {1--8} (\bibinfo {year}
  {2006})}\BibitemShut {NoStop}%
\bibitem [{\citenamefont {Schulze}\ \emph {et~al.}(2017)\citenamefont
  {Schulze}, \citenamefont {M{\"u}ller}, \citenamefont {Groeneveld},\ and\
  \citenamefont {Grimm}}]{SchulzeEtAl2017}%
  \BibitemOpen
  \bibfield  {author} {\bibinfo {author} {\bibfnamefont {J.}~\bibnamefont
  {Schulze}}, \bibinfo {author} {\bibfnamefont {B.}~\bibnamefont {M{\"u}ller}},
  \bibinfo {author} {\bibfnamefont {J.}~\bibnamefont {Groeneveld}},\ and\
  \bibinfo {author} {\bibfnamefont {V.}~\bibnamefont {Grimm}},\ }\bibfield
  {title} {\enquote {\bibinfo {title} {Agent-{{Based Modelling}} of
  {{Social-Ecological Systems}}: {{Achievements}}, {{Challenges}}, and a {{Way
  Forward}}},}\ }\href@noop {} {\bibfield  {journal} {\bibinfo  {journal}
  {Journal of Artificial Societies and Social Simulation}\ }\textbf {\bibinfo
  {volume} {20}},\ \bibinfo {pages} {8} (\bibinfo {year} {2017})}\BibitemShut
  {NoStop}%
\bibitem [{\citenamefont {Gotts}\ \emph {et~al.}(2019)\citenamefont {Gotts},
  \citenamefont {{van Voorn}}, \citenamefont {Polhill}, \citenamefont
  {de~Jong}, \citenamefont {Edmonds}, \citenamefont {Hofstede},\ and\
  \citenamefont {Meyer}}]{GottsEtAl2019}%
  \BibitemOpen
  \bibfield  {author} {\bibinfo {author} {\bibfnamefont {N.~M.}\ \bibnamefont
  {Gotts}}, \bibinfo {author} {\bibfnamefont {G.~A.~K.}\ \bibnamefont {{van
  Voorn}}}, \bibinfo {author} {\bibfnamefont {J.~G.}\ \bibnamefont {Polhill}},
  \bibinfo {author} {\bibfnamefont {E.}~\bibnamefont {de~Jong}}, \bibinfo
  {author} {\bibfnamefont {B.}~\bibnamefont {Edmonds}}, \bibinfo {author}
  {\bibfnamefont {G.~J.}\ \bibnamefont {Hofstede}},\ and\ \bibinfo {author}
  {\bibfnamefont {R.}~\bibnamefont {Meyer}},\ }\bibfield  {title} {\enquote
  {\bibinfo {title} {Agent-based modelling of socio-ecological systems:
  {{Models}}, projects and ontologies},}\ }\href
  {https://doi.org/10.1016/j.ecocom.2018.07.007} {\bibfield  {journal}
  {\bibinfo  {journal} {Ecological Complexity}\ }\bibinfo {series} {Agent-Based
  Modelling to Study Resilience in Socio-Ecological Systems},\ \textbf
  {\bibinfo {volume} {40}},\ \bibinfo {pages} {100728} (\bibinfo {year}
  {2019})}\BibitemShut {NoStop}%
\bibitem [{\citenamefont {{Hernandez-Leal}}, \citenamefont {Kartal},\ and\
  \citenamefont {Taylor}(2019)}]{Hernandez-LealEtAl2019}%
  \BibitemOpen
  \bibfield  {author} {\bibinfo {author} {\bibfnamefont {P.}~\bibnamefont
  {{Hernandez-Leal}}}, \bibinfo {author} {\bibfnamefont {B.}~\bibnamefont
  {Kartal}},\ and\ \bibinfo {author} {\bibfnamefont {M.~E.}\ \bibnamefont
  {Taylor}},\ }\bibfield  {title} {\enquote {\bibinfo {title} {A survey and
  critique of multiagent deep reinforcement learning},}\ }\href
  {https://doi.org/10.1007/s10458-019-09421-1} {\bibfield  {journal} {\bibinfo
  {journal} {Autonomous Agents and Multi-Agent Systems}\ }\textbf {\bibinfo
  {volume} {33}},\ \bibinfo {pages} {750--797} (\bibinfo {year}
  {2019})}\BibitemShut {NoStop}%
\bibitem [{\citenamefont {Tuyls}\ and\ \citenamefont
  {Now{\'e}}(2005)}]{TuylsNowe2005}%
  \BibitemOpen
  \bibfield  {author} {\bibinfo {author} {\bibfnamefont {K.}~\bibnamefont
  {Tuyls}}\ and\ \bibinfo {author} {\bibfnamefont {A.}~\bibnamefont
  {Now{\'e}}},\ }\bibfield  {title} {\enquote {\bibinfo {title} {Evolutionary
  game theory and multi-agent reinforcement learning},}\ }\href
  {https://doi.org/10.1017/S026988890500041X} {\bibfield  {journal} {\bibinfo
  {journal} {The Knowledge Engineering Review}\ }\textbf {\bibinfo {volume}
  {20}},\ \bibinfo {pages} {63--90} (\bibinfo {year} {2005})}\BibitemShut
  {NoStop}%
\bibitem [{\citenamefont {Tuyls}\ and\ \citenamefont
  {Parsons}(2007)}]{TuylsParsons2007}%
  \BibitemOpen
  \bibfield  {author} {\bibinfo {author} {\bibfnamefont {K.}~\bibnamefont
  {Tuyls}}\ and\ \bibinfo {author} {\bibfnamefont {S.}~\bibnamefont
  {Parsons}},\ }\bibfield  {title} {\enquote {\bibinfo {title} {What
  evolutionary game theory tells us about multiagent learning},}\ }\href
  {https://doi.org/10.1016/j.artint.2007.01.004} {\bibfield  {journal}
  {\bibinfo  {journal} {Artificial Intelligence}\ }\bibinfo {series}
  {Foundations of {{Multi-Agent Learning}}},\ \textbf {\bibinfo {volume}
  {171}},\ \bibinfo {pages} {406--416} (\bibinfo {year} {2007})}\BibitemShut
  {NoStop}%
\bibitem [{\citenamefont {Bloembergen}\ \emph {et~al.}(2015)\citenamefont
  {Bloembergen}, \citenamefont {Tuyls}, \citenamefont {Hennes},\ and\
  \citenamefont {Kaisers}}]{BloembergenEtAl2015}%
  \BibitemOpen
  \bibfield  {author} {\bibinfo {author} {\bibfnamefont {D.}~\bibnamefont
  {Bloembergen}}, \bibinfo {author} {\bibfnamefont {K.}~\bibnamefont {Tuyls}},
  \bibinfo {author} {\bibfnamefont {D.}~\bibnamefont {Hennes}},\ and\ \bibinfo
  {author} {\bibfnamefont {M.}~\bibnamefont {Kaisers}},\ }\bibfield  {title}
  {\enquote {\bibinfo {title} {Evolutionary {{Dynamics}} of {{Multi-Agent
  Learning}}: {{A Survey}}},}\ }\href {https://doi.org/10.1613/jair.4818}
  {\bibfield  {journal} {\bibinfo  {journal} {Journal of Artificial
  Intelligence Research}\ }\textbf {\bibinfo {volume} {53}},\ \bibinfo {pages}
  {659--697} (\bibinfo {year} {2015})}\BibitemShut {NoStop}%
\bibitem [{\citenamefont {Dayan}\ and\ \citenamefont
  {Niv}(2008)}]{DayanNiv2008}%
  \BibitemOpen
  \bibfield  {author} {\bibinfo {author} {\bibfnamefont {P.}~\bibnamefont
  {Dayan}}\ and\ \bibinfo {author} {\bibfnamefont {Y.}~\bibnamefont {Niv}},\
  }\bibfield  {title} {\enquote {\bibinfo {title} {Reinforcement learning:
  {{The Good}}, {{The Bad}} and {{The Ugly}}},}\ }\href
  {https://doi.org/10.1016/j.conb.2008.08.003} {\bibfield  {journal} {\bibinfo
  {journal} {Current Opinion in Neurobiology}\ }\textbf {\bibinfo {volume}
  {18}},\ \bibinfo {pages} {185--196} (\bibinfo {year} {2008})}\BibitemShut
  {NoStop}%
\bibitem [{\citenamefont {Bush}\ and\ \citenamefont
  {Mosteller}(1951)}]{BushMosteller1951}%
  \BibitemOpen
  \bibfield  {author} {\bibinfo {author} {\bibfnamefont {R.~R.}\ \bibnamefont
  {Bush}}\ and\ \bibinfo {author} {\bibfnamefont {F.}~\bibnamefont
  {Mosteller}},\ }\bibfield  {title} {\enquote {\bibinfo {title} {A
  mathematical model for simple learning},}\ }\href
  {https://doi.org/10.1037/h0054388} {\bibfield  {journal} {\bibinfo  {journal}
  {Psychological Review}\ }\textbf {\bibinfo {volume} {58}},\ \bibinfo {pages}
  {313--323} (\bibinfo {year} {1951})}\BibitemShut {NoStop}%
\bibitem [{\citenamefont {Cross}(1973)}]{Cross1973}%
  \BibitemOpen
  \bibfield  {author} {\bibinfo {author} {\bibfnamefont {J.~G.}\ \bibnamefont
  {Cross}},\ }\bibfield  {title} {\enquote {\bibinfo {title} {A {{Stochastic
  Learning Model}} of {{Economic Behavior}}*},}\ }\href
  {https://doi.org/10.2307/1882186} {\bibfield  {journal} {\bibinfo  {journal}
  {The Quarterly Journal of Economics}\ }\textbf {\bibinfo {volume} {87}},\
  \bibinfo {pages} {239--266} (\bibinfo {year} {1973})}\BibitemShut {NoStop}%
\bibitem [{\citenamefont {Erev}\ and\ \citenamefont
  {Roth}(1998)}]{ErevRoth1998}%
  \BibitemOpen
  \bibfield  {author} {\bibinfo {author} {\bibfnamefont {I.}~\bibnamefont
  {Erev}}\ and\ \bibinfo {author} {\bibfnamefont {A.~E.}\ \bibnamefont
  {Roth}},\ }\bibfield  {title} {\enquote {\bibinfo {title} {Predicting {{How
  People Play Games}}: {{Reinforcement Learning}} in {{Experimental Games}}
  with {{Unique}}, {{Mixed Strategy Equilibria}}},}\ }\href@noop {} {\bibfield
  {journal} {\bibinfo  {journal} {The American Economic Review}\ }\textbf
  {\bibinfo {volume} {88}},\ \bibinfo {pages} {848--881} (\bibinfo {year}
  {1998})}\BibitemShut {NoStop}%
\bibitem [{\citenamefont {Schultz}, \citenamefont {Stauffer},\ and\
  \citenamefont {Lak}(2017)}]{SchultzEtAl2017}%
  \BibitemOpen
  \bibfield  {author} {\bibinfo {author} {\bibfnamefont {W.}~\bibnamefont
  {Schultz}}, \bibinfo {author} {\bibfnamefont {W.~R.}\ \bibnamefont
  {Stauffer}},\ and\ \bibinfo {author} {\bibfnamefont {A.}~\bibnamefont
  {Lak}},\ }\bibfield  {title} {\enquote {\bibinfo {title} {The phasic dopamine
  signal maturing: From reward via behavioural activation to formal economic
  utility},}\ }\href {https://doi.org/10.1016/j.conb.2017.03.013} {\bibfield
  {journal} {\bibinfo  {journal} {Current Opinion in Neurobiology}\ }\textbf
  {\bibinfo {volume} {43}},\ \bibinfo {pages} {139--148} (\bibinfo {year}
  {2017})}\BibitemShut {NoStop}%
\bibitem [{\citenamefont {{Burton-Chellew}}\ and\ \citenamefont
  {West}(2021)}]{Burton-ChellewWest2021}%
  \BibitemOpen
  \bibfield  {author} {\bibinfo {author} {\bibfnamefont {M.~N.}\ \bibnamefont
  {{Burton-Chellew}}}\ and\ \bibinfo {author} {\bibfnamefont {S.~A.}\
  \bibnamefont {West}},\ }\bibfield  {title} {\enquote {\bibinfo {title}
  {Payoff-based learning best explains the rate of decline in cooperation
  across 237 public-goods games},}\ }\href
  {https://doi.org/10.1038/s41562-021-01107-7} {\bibfield  {journal} {\bibinfo
  {journal} {Nature Human Behaviour}\ }\textbf {\bibinfo {volume} {5}},\
  \bibinfo {pages} {1330--1338} (\bibinfo {year} {2021})}\BibitemShut {NoStop}%
\bibitem [{\citenamefont {B{\"o}rgers}\ and\ \citenamefont
  {Sarin}(1997)}]{BorgersSarin1997}%
  \BibitemOpen
  \bibfield  {author} {\bibinfo {author} {\bibfnamefont {T.}~\bibnamefont
  {B{\"o}rgers}}\ and\ \bibinfo {author} {\bibfnamefont {R.}~\bibnamefont
  {Sarin}},\ }\bibfield  {title} {\enquote {\bibinfo {title} {Learning
  {{Through Reinforcement}} and {{Replicator Dynamics}}},}\ }\href
  {https://doi.org/10.1006/jeth.1997.2319} {\bibfield  {journal} {\bibinfo
  {journal} {Journal of Economic Theory}\ }\textbf {\bibinfo {volume} {77}},\
  \bibinfo {pages} {1--14} (\bibinfo {year} {1997})}\BibitemShut {NoStop}%
\bibitem [{\citenamefont {Tuyls}, \citenamefont {Verbeeck},\ and\ \citenamefont
  {Lenaerts}(2003)}]{TuylsEtAl2003}%
  \BibitemOpen
  \bibfield  {author} {\bibinfo {author} {\bibfnamefont {K.}~\bibnamefont
  {Tuyls}}, \bibinfo {author} {\bibfnamefont {K.}~\bibnamefont {Verbeeck}},\
  and\ \bibinfo {author} {\bibfnamefont {T.}~\bibnamefont {Lenaerts}},\
  }\bibfield  {title} {\enquote {\bibinfo {title} {A selection-mutation model
  for q-learning in multi-agent systems},}\ }in\ \href
  {https://doi.org/10.1145/860575.860687} {\emph {\bibinfo {booktitle}
  {Proceedings of the Second International Joint Conference on {{Autonomous}}
  Agents and Multiagent Systems - {{AAMAS}} '03}}}\ (\bibinfo  {publisher}
  {{ACM Press}},\ \bibinfo {address} {{Melbourne, Australia}},\ \bibinfo {year}
  {2003})\ p.\ \bibinfo {pages} {693}\BibitemShut {NoStop}%
\bibitem [{\citenamefont {Sato}\ and\ \citenamefont
  {Crutchfield}(2003)}]{SatoCrutchfield2003}%
  \BibitemOpen
  \bibfield  {author} {\bibinfo {author} {\bibfnamefont {Y.}~\bibnamefont
  {Sato}}\ and\ \bibinfo {author} {\bibfnamefont {J.~P.}\ \bibnamefont
  {Crutchfield}},\ }\bibfield  {title} {\enquote {\bibinfo {title} {Coupled
  replicator equations for the dynamics of learning in multiagent systems},}\
  }\href {https://doi.org/10.1103/PhysRevE.67.015206} {\bibfield  {journal}
  {\bibinfo  {journal} {Physical Review E}\ }\textbf {\bibinfo {volume} {67}},\
  \bibinfo {pages} {015206} (\bibinfo {year} {2003})}\BibitemShut {NoStop}%
\bibitem [{\citenamefont {Klos}, \citenamefont {{van Ahee}},\ and\
  \citenamefont {Tuyls}(2010)}]{KlosEtAl2010}%
  \BibitemOpen
  \bibfield  {author} {\bibinfo {author} {\bibfnamefont {T.}~\bibnamefont
  {Klos}}, \bibinfo {author} {\bibfnamefont {G.~J.}\ \bibnamefont {{van
  Ahee}}},\ and\ \bibinfo {author} {\bibfnamefont {K.}~\bibnamefont {Tuyls}},\
  }\bibfield  {title} {\enquote {\bibinfo {title} {Evolutionary {{Dynamics}} of
  {{Regret Minimization}}},}\ }in\ \href
  {https://doi.org/10.1007/978-3-642-15883-4_6} {\emph {\bibinfo {booktitle}
  {Machine {{Learning}} and {{Knowledge Discovery}} in {{Databases}}}}},\ Vol.\
  \bibinfo {volume} {6322},\ \bibinfo {editor} {edited by\ \bibinfo {editor}
  {\bibfnamefont {D.}~\bibnamefont {Hutchison}}, \bibinfo {editor}
  {\bibfnamefont {T.}~\bibnamefont {Kanade}}, \bibinfo {editor} {\bibfnamefont
  {J.}~\bibnamefont {Kittler}}, \bibinfo {editor} {\bibfnamefont {J.~M.}\
  \bibnamefont {Kleinberg}}, \bibinfo {editor} {\bibfnamefont {F.}~\bibnamefont
  {Mattern}}, \bibinfo {editor} {\bibfnamefont {J.~C.}\ \bibnamefont
  {Mitchell}}, \bibinfo {editor} {\bibfnamefont {M.}~\bibnamefont {Naor}},
  \bibinfo {editor} {\bibfnamefont {O.}~\bibnamefont {Nierstrasz}}, \bibinfo
  {editor} {\bibfnamefont {C.}~\bibnamefont {Pandu~Rangan}}, \bibinfo {editor}
  {\bibfnamefont {B.}~\bibnamefont {Steffen}}, \bibinfo {editor} {\bibfnamefont
  {M.}~\bibnamefont {Sudan}}, \bibinfo {editor} {\bibfnamefont
  {D.}~\bibnamefont {Terzopoulos}}, \bibinfo {editor} {\bibfnamefont
  {D.}~\bibnamefont {Tygar}}, \bibinfo {editor} {\bibfnamefont {M.~Y.}\
  \bibnamefont {Vardi}}, \bibinfo {editor} {\bibfnamefont {G.}~\bibnamefont
  {Weikum}}, \bibinfo {editor} {\bibfnamefont {J.~L.}\ \bibnamefont
  {Balc{\'a}zar}}, \bibinfo {editor} {\bibfnamefont {F.}~\bibnamefont
  {Bonchi}}, \bibinfo {editor} {\bibfnamefont {A.}~\bibnamefont {Gionis}},\
  and\ \bibinfo {editor} {\bibfnamefont {M.}~\bibnamefont {Sebag}}}\ (\bibinfo
  {publisher} {{Springer Berlin Heidelberg}},\ \bibinfo {address} {{Berlin,
  Heidelberg}},\ \bibinfo {year} {2010})\ pp.\ \bibinfo {pages}
  {82--96}\BibitemShut {NoStop}%
\bibitem [{\citenamefont {Hopkins}(2002)}]{Hopkins2002}%
  \BibitemOpen
  \bibfield  {author} {\bibinfo {author} {\bibfnamefont {E.}~\bibnamefont
  {Hopkins}},\ }\bibfield  {title} {\enquote {\bibinfo {title} {Two {{Competing
  Models}} of {{How People Learn}} in {{Games}}},}\ }\href
  {https://doi.org/10.1111/j.1468-0262.2002.00436.x} {\bibfield  {journal}
  {\bibinfo  {journal} {Econometrica}\ }\textbf {\bibinfo {volume} {70}},\
  \bibinfo {pages} {2141--2166} (\bibinfo {year} {2002})}\BibitemShut {NoStop}%
\bibitem [{\citenamefont {Barfuss}, \citenamefont {Donges},\ and\ \citenamefont
  {Kurths}(2019)}]{BarfussEtAl2019}%
  \BibitemOpen
  \bibfield  {author} {\bibinfo {author} {\bibfnamefont {W.}~\bibnamefont
  {Barfuss}}, \bibinfo {author} {\bibfnamefont {J.~F.}\ \bibnamefont
  {Donges}},\ and\ \bibinfo {author} {\bibfnamefont {J.}~\bibnamefont
  {Kurths}},\ }\bibfield  {title} {\enquote {\bibinfo {title} {Deterministic
  limit of temporal difference reinforcement learning for stochastic games},}\
  }\href {https://doi.org/10.1103/PhysRevE.99.043305} {\bibfield  {journal}
  {\bibinfo  {journal} {Physical Review E}\ }\textbf {\bibinfo {volume} {99}},\
  \bibinfo {pages} {043305} (\bibinfo {year} {2019})}\BibitemShut {NoStop}%
\bibitem [{\citenamefont {Barfuss}(2020)}]{Barfuss2020}%
  \BibitemOpen
  \bibfield  {author} {\bibinfo {author} {\bibfnamefont {W.}~\bibnamefont
  {Barfuss}},\ }\bibfield  {title} {\enquote {\bibinfo {title} {Reinforcement
  {{Learning Dynamics}} in the {{Infinite Memory Limit}}},}\ }in\ \href@noop {}
  {\emph {\bibinfo {booktitle} {Proceedings of the 19th {{International
  Conference}} on {{Autonomous Agents}} and {{MultiAgent Systems}}}}},\
  \bibinfo {series and number} {{{AAMAS}} '20}\ (\bibinfo  {publisher}
  {{International Foundation for Autonomous Agents and Multiagent Systems}},\
  \bibinfo {address} {{Richland, SC}},\ \bibinfo {year} {2020})\ pp.\ \bibinfo
  {pages} {1768--1770}\BibitemShut {NoStop}%
\bibitem [{\citenamefont {Barfuss}(2021)}]{Barfuss2021}%
  \BibitemOpen
  \bibfield  {author} {\bibinfo {author} {\bibfnamefont {W.}~\bibnamefont
  {Barfuss}},\ }\bibfield  {title} {\enquote {\bibinfo {title} {Dynamical
  systems as a level of cognitive analysis of multi-agent learning},}\ }\href
  {https://doi.org/10.1007/s00521-021-06117-0} {\bibfield  {journal} {\bibinfo
  {journal} {Neural Computing and Applications}\ } (\bibinfo {year} {2021}),\
  10.1007/s00521-021-06117-0}\BibitemShut {NoStop}%
\bibitem [{\citenamefont {Barfuss}\ \emph {et~al.}(2020)\citenamefont
  {Barfuss}, \citenamefont {Donges}, \citenamefont {Vasconcelos}, \citenamefont
  {Kurths},\ and\ \citenamefont {Levin}}]{BarfussEtAl2020}%
  \BibitemOpen
  \bibfield  {author} {\bibinfo {author} {\bibfnamefont {W.}~\bibnamefont
  {Barfuss}}, \bibinfo {author} {\bibfnamefont {J.~F.}\ \bibnamefont {Donges}},
  \bibinfo {author} {\bibfnamefont {V.~V.}\ \bibnamefont {Vasconcelos}},
  \bibinfo {author} {\bibfnamefont {J.}~\bibnamefont {Kurths}},\ and\ \bibinfo
  {author} {\bibfnamefont {S.~A.}\ \bibnamefont {Levin}},\ }\bibfield  {title}
  {\enquote {\bibinfo {title} {Caring for the future can turn tragedy into
  comedy for long-term collective action under risk of collapse},}\ }\href
  {https://doi.org/10.1073/pnas.1916545117} {\bibfield  {journal} {\bibinfo
  {journal} {Proceedings of the National Academy of Sciences}\ }\textbf
  {\bibinfo {volume} {117}},\ \bibinfo {pages} {12915--12922} (\bibinfo {year}
  {2020})}\BibitemShut {NoStop}%
\bibitem [{\citenamefont {Barfuss}\ and\ \citenamefont
  {Mann}(2022)}]{BarfussMann2022}%
  \BibitemOpen
  \bibfield  {author} {\bibinfo {author} {\bibfnamefont {W.}~\bibnamefont
  {Barfuss}}\ and\ \bibinfo {author} {\bibfnamefont {R.~P.}\ \bibnamefont
  {Mann}},\ }\bibfield  {title} {\enquote {\bibinfo {title} {Modeling the
  effects of environmental and perceptual uncertainty using deterministic
  reinforcement learning dynamics with partial observability},}\ }\href
  {https://doi.org/10.1103/PhysRevE.105.034409} {\bibfield  {journal} {\bibinfo
   {journal} {Physical Review E}\ }\textbf {\bibinfo {volume} {105}},\ \bibinfo
  {pages} {034409} (\bibinfo {year} {2022})}\BibitemShut {NoStop}%
\bibitem [{\citenamefont {Barfuss}\ and\ \citenamefont
  {Meylahn}(2023)}]{BarfussMeylahn2022}%
  \BibitemOpen
  \bibfield  {author} {\bibinfo {author} {\bibfnamefont {W.}~\bibnamefont
  {Barfuss}}\ and\ \bibinfo {author} {\bibfnamefont {J.~M.}\ \bibnamefont
  {Meylahn}},\ }\bibfield  {title} {\enquote {\bibinfo {title} {Intrinsic
  fluctuations of reinforcement learning promote cooperation},}\ }\href@noop {}
  {\bibfield  {journal} {\bibinfo  {journal} {Scientific Reports}\ }\textbf
  {\bibinfo {volume} {13}},\ \bibinfo {pages} {1309} (\bibinfo {year}
  {2023})}\BibitemShut {NoStop}%
\bibitem [{\citenamefont {Strnad}\ \emph {et~al.}(2019)\citenamefont {Strnad},
  \citenamefont {Barfuss}, \citenamefont {Donges},\ and\ \citenamefont
  {Heitzig}}]{StrnadEtAl2019}%
  \BibitemOpen
  \bibfield  {author} {\bibinfo {author} {\bibfnamefont {F.~M.}\ \bibnamefont
  {Strnad}}, \bibinfo {author} {\bibfnamefont {W.}~\bibnamefont {Barfuss}},
  \bibinfo {author} {\bibfnamefont {J.~F.}\ \bibnamefont {Donges}},\ and\
  \bibinfo {author} {\bibfnamefont {J.}~\bibnamefont {Heitzig}},\ }\bibfield
  {title} {\enquote {\bibinfo {title} {Deep reinforcement learning in
  {{World-Earth}} system models to discover sustainable management
  strategies},}\ }\href {https://doi.org/10.1063/1.5124673} {\bibfield
  {journal} {\bibinfo  {journal} {Chaos: An Interdisciplinary Journal of
  Nonlinear Science}\ }\textbf {\bibinfo {volume} {29}},\ \bibinfo {pages}
  {123122} (\bibinfo {year} {2019})}\BibitemShut {NoStop}%
\bibitem [{\citenamefont {Perolat}\ \emph {et~al.}(2017)\citenamefont
  {Perolat}, \citenamefont {Leibo}, \citenamefont {Zambaldi}, \citenamefont
  {Beattie}, \citenamefont {Tuyls},\ and\ \citenamefont
  {Graepel}}]{PerolatEtAl2017}%
  \BibitemOpen
  \bibfield  {author} {\bibinfo {author} {\bibfnamefont {J.}~\bibnamefont
  {Perolat}}, \bibinfo {author} {\bibfnamefont {J.~Z.}\ \bibnamefont {Leibo}},
  \bibinfo {author} {\bibfnamefont {V.}~\bibnamefont {Zambaldi}}, \bibinfo
  {author} {\bibfnamefont {C.}~\bibnamefont {Beattie}}, \bibinfo {author}
  {\bibfnamefont {K.}~\bibnamefont {Tuyls}},\ and\ \bibinfo {author}
  {\bibfnamefont {T.}~\bibnamefont {Graepel}},\ }\bibfield  {title} {\enquote
  {\bibinfo {title} {A multi-agent reinforcement learning model of common-pool
  resource appropriation},}\ }in\ \href@noop {} {\emph {\bibinfo {booktitle}
  {Proceedings of the 31st {{International Conference}} on {{Neural Information
  Processing Systems}}}}},\ \bibinfo {series and number} {{{NIPS}}'17}\
  (\bibinfo  {publisher} {{Curran Associates Inc.}},\ \bibinfo {address} {{Red
  Hook, NY, USA}},\ \bibinfo {year} {2017})\ pp.\ \bibinfo {pages}
  {3646--3655}\BibitemShut {NoStop}%
\bibitem [{\citenamefont {Zheng}\ \emph {et~al.}(2020)\citenamefont {Zheng},
  \citenamefont {Trott}, \citenamefont {Srinivasa}, \citenamefont {Naik},
  \citenamefont {Gruesbeck}, \citenamefont {Parkes},\ and\ \citenamefont
  {Socher}}]{ZhengEtAl2020}%
  \BibitemOpen
  \bibfield  {author} {\bibinfo {author} {\bibfnamefont {S.}~\bibnamefont
  {Zheng}}, \bibinfo {author} {\bibfnamefont {A.}~\bibnamefont {Trott}},
  \bibinfo {author} {\bibfnamefont {S.}~\bibnamefont {Srinivasa}}, \bibinfo
  {author} {\bibfnamefont {N.}~\bibnamefont {Naik}}, \bibinfo {author}
  {\bibfnamefont {M.}~\bibnamefont {Gruesbeck}}, \bibinfo {author}
  {\bibfnamefont {D.~C.}\ \bibnamefont {Parkes}},\ and\ \bibinfo {author}
  {\bibfnamefont {R.}~\bibnamefont {Socher}},\ }\bibfield  {title} {\enquote
  {\bibinfo {title} {The {{AI Economist}}: {{Improving Equality}} and
  {{Productivity}} with {{AI-Driven Tax Policies}}},}\ }\href@noop {}
  {\bibfield  {journal} {\bibinfo  {journal} {arXiv:2004.13332 [cs, econ,
  q-fin, stat]}\ } (\bibinfo {year} {2020})}\BibitemShut {NoStop}%
\bibitem [{\citenamefont {Yang}\ \emph {et~al.}(2018)\citenamefont {Yang},
  \citenamefont {Luo}, \citenamefont {Li}, \citenamefont {Zhou}, \citenamefont
  {Zhang},\ and\ \citenamefont {Wang}}]{YangEtAl2018}%
  \BibitemOpen
  \bibfield  {author} {\bibinfo {author} {\bibfnamefont {Y.}~\bibnamefont
  {Yang}}, \bibinfo {author} {\bibfnamefont {R.}~\bibnamefont {Luo}}, \bibinfo
  {author} {\bibfnamefont {M.}~\bibnamefont {Li}}, \bibinfo {author}
  {\bibfnamefont {M.}~\bibnamefont {Zhou}}, \bibinfo {author} {\bibfnamefont
  {W.}~\bibnamefont {Zhang}},\ and\ \bibinfo {author} {\bibfnamefont
  {J.}~\bibnamefont {Wang}},\ }\bibfield  {title} {\enquote {\bibinfo {title}
  {Mean {{Field Multi-Agent Reinforcement Learning}}},}\ }in\ \href@noop {}
  {\emph {\bibinfo {booktitle} {Proceedings of the 35th {{International
  Conference}} on {{Machine Learning}}}}}\ (\bibinfo  {publisher} {{PMLR}},\
  \bibinfo {year} {2018})\ pp.\ \bibinfo {pages} {5571--5580}\BibitemShut
  {NoStop}%
\bibitem [{\citenamefont {Hu}, \citenamefont {Leung},\ and\ \citenamefont
  {Leung}(2019)}]{HuEtAl2019}%
  \BibitemOpen
  \bibfield  {author} {\bibinfo {author} {\bibfnamefont {S.}~\bibnamefont
  {Hu}}, \bibinfo {author} {\bibfnamefont {C.-w.}\ \bibnamefont {Leung}},\ and\
  \bibinfo {author} {\bibfnamefont {H.-f.}\ \bibnamefont {Leung}},\ }\bibfield
  {title} {\enquote {\bibinfo {title} {Modelling the {{Dynamics}} of
  {{Multiagent Q-Learning}} in {{Repeated Symmetric Games}}: A {{Mean Field
  Theoretic Approach}}},}\ }\href@noop {} {\ ,\ \bibinfo {pages} {11} (\bibinfo
  {year} {2019})}\BibitemShut {NoStop}%
\bibitem [{\citenamefont {Perrin}\ \emph {et~al.}(2021)\citenamefont {Perrin},
  \citenamefont {Lauri{\`e}re}, \citenamefont {P{\'e}rolat}, \citenamefont
  {{\'E}lie}, \citenamefont {Geist},\ and\ \citenamefont
  {Pietquin}}]{PerrinEtAl2021a}%
  \BibitemOpen
  \bibfield  {author} {\bibinfo {author} {\bibfnamefont {S.}~\bibnamefont
  {Perrin}}, \bibinfo {author} {\bibfnamefont {M.}~\bibnamefont
  {Lauri{\`e}re}}, \bibinfo {author} {\bibfnamefont {J.}~\bibnamefont
  {P{\'e}rolat}}, \bibinfo {author} {\bibfnamefont {R.}~\bibnamefont
  {{\'E}lie}}, \bibinfo {author} {\bibfnamefont {M.}~\bibnamefont {Geist}},\
  and\ \bibinfo {author} {\bibfnamefont {O.}~\bibnamefont {Pietquin}},\
  }\href@noop {} {\enquote {\bibinfo {title} {Generalization in {{Mean Field
  Games}} by {{Learning Master Policies}}},}\ } (\bibinfo {year}
  {2021})\BibitemShut {NoStop}%
\bibitem [{\citenamefont {Galla}(2009)}]{Galla2009}%
  \BibitemOpen
  \bibfield  {author} {\bibinfo {author} {\bibfnamefont {T.}~\bibnamefont
  {Galla}},\ }\bibfield  {title} {\enquote {\bibinfo {title} {Intrinsic
  {{Noise}} in {{Game Dynamical Learning}}},}\ }\href
  {https://doi.org/10.1103/PhysRevLett.103.198702} {\bibfield  {journal}
  {\bibinfo  {journal} {Physical Review Letters}\ }\textbf {\bibinfo {volume}
  {103}},\ \bibinfo {pages} {198702} (\bibinfo {year} {2009})}\BibitemShut
  {NoStop}%
\bibitem [{\citenamefont {Donges}\ and\ \citenamefont
  {Barfuss}(2017)}]{DongesBarfuss2017}%
  \BibitemOpen
  \bibfield  {author} {\bibinfo {author} {\bibfnamefont {J.~F.}\ \bibnamefont
  {Donges}}\ and\ \bibinfo {author} {\bibfnamefont {W.}~\bibnamefont
  {Barfuss}},\ }\bibfield  {title} {\enquote {\bibinfo {title} {From {{Math}}
  to {{Metaphors}} and {{Back Again}}: {{Social-Ecological Resilience}} from a
  {{Multi-Agent-Environment Perspective}}},}\ }\href
  {https://doi.org/10.14512/gaia.26.S1.5} {\bibfield  {journal} {\bibinfo
  {journal} {GAIA - Ecological Perspectives for Science and Society}\ }\textbf
  {\bibinfo {volume} {26}},\ \bibinfo {pages} {182--190} (\bibinfo {year}
  {2017})}\BibitemShut {NoStop}%
\bibitem [{\citenamefont {Barfuss}\ \emph {et~al.}(2018)\citenamefont
  {Barfuss}, \citenamefont {Donges}, \citenamefont {Lade},\ and\ \citenamefont
  {Kurths}}]{BarfussEtAl2018}%
  \BibitemOpen
  \bibfield  {author} {\bibinfo {author} {\bibfnamefont {W.}~\bibnamefont
  {Barfuss}}, \bibinfo {author} {\bibfnamefont {J.~F.}\ \bibnamefont {Donges}},
  \bibinfo {author} {\bibfnamefont {S.~J.}\ \bibnamefont {Lade}},\ and\
  \bibinfo {author} {\bibfnamefont {J.}~\bibnamefont {Kurths}},\ }\bibfield
  {title} {\enquote {\bibinfo {title} {When optimization for governing
  human-environment tipping elements is neither sustainable nor safe},}\ }\href
  {https://doi.org/10.1038/s41467-018-04738-z} {\bibfield  {journal} {\bibinfo
  {journal} {Nature Communications}\ }\textbf {\bibinfo {volume} {9}},\
  \bibinfo {pages} {2354} (\bibinfo {year} {2018})}\BibitemShut {NoStop}%
\bibitem [{\citenamefont {Cichos}\ \emph {et~al.}(2020)\citenamefont {Cichos},
  \citenamefont {Gustavsson}, \citenamefont {Mehlig},\ and\ \citenamefont
  {Volpe}}]{cichos2020machine}%
  \BibitemOpen
  \bibfield  {author} {\bibinfo {author} {\bibfnamefont {F.}~\bibnamefont
  {Cichos}}, \bibinfo {author} {\bibfnamefont {K.}~\bibnamefont {Gustavsson}},
  \bibinfo {author} {\bibfnamefont {B.}~\bibnamefont {Mehlig}},\ and\ \bibinfo
  {author} {\bibfnamefont {G.}~\bibnamefont {Volpe}},\ }\bibfield  {title}
  {\enquote {\bibinfo {title} {Machine learning for active matter},}\
  }\href@noop {} {\bibfield  {journal} {\bibinfo  {journal} {Nature Machine
  Intelligence}\ }\textbf {\bibinfo {volume} {2}},\ \bibinfo {pages} {94--103}
  (\bibinfo {year} {2020})}\BibitemShut {NoStop}%
\bibitem [{\citenamefont {Berg}(2004)}]{berg2004coli}%
  \BibitemOpen
  \bibfield  {author} {\bibinfo {author} {\bibfnamefont {H.~C.}\ \bibnamefont
  {Berg}},\ }\href@noop {} {\emph {\bibinfo {title} {E. coli in Motion}}}\
  (\bibinfo  {publisher} {Springer},\ \bibinfo {year} {2004})\BibitemShut
  {NoStop}%
\bibitem [{\citenamefont {Sengupta}, \citenamefont {Carrara},\ and\
  \citenamefont {Stocker}(2017)}]{sengupta2017phytoplankton}%
  \BibitemOpen
  \bibfield  {author} {\bibinfo {author} {\bibfnamefont {A.}~\bibnamefont
  {Sengupta}}, \bibinfo {author} {\bibfnamefont {F.}~\bibnamefont {Carrara}},\
  and\ \bibinfo {author} {\bibfnamefont {R.}~\bibnamefont {Stocker}},\
  }\bibfield  {title} {\enquote {\bibinfo {title} {Phytoplankton can actively
  diversify their migration strategy in response to turbulent cues},}\
  }\href@noop {} {\bibfield  {journal} {\bibinfo  {journal} {Nature}\ }\textbf
  {\bibinfo {volume} {543}},\ \bibinfo {pages} {555--558} (\bibinfo {year}
  {2017})}\BibitemShut {NoStop}%
\bibitem [{\citenamefont {Viswanathan}\ \emph {et~al.}(2011)\citenamefont
  {Viswanathan}, \citenamefont {Da~Luz}, \citenamefont {Raposo},\ and\
  \citenamefont {Stanley}}]{viswanathan2011physics}%
  \BibitemOpen
  \bibfield  {author} {\bibinfo {author} {\bibfnamefont {G.~M.}\ \bibnamefont
  {Viswanathan}}, \bibinfo {author} {\bibfnamefont {M.~G.}\ \bibnamefont
  {Da~Luz}}, \bibinfo {author} {\bibfnamefont {E.~P.}\ \bibnamefont {Raposo}},\
  and\ \bibinfo {author} {\bibfnamefont {H.~E.}\ \bibnamefont {Stanley}},\
  }\href@noop {} {\emph {\bibinfo {title} {The physics of foraging: an
  introduction to random searches and biological encounters}}}\ (\bibinfo
  {publisher} {Cambridge University Press},\ \bibinfo {year}
  {2011})\BibitemShut {NoStop}%
\bibitem [{\citenamefont {B{\'e}nichou}\ \emph {et~al.}(2011)\citenamefont
  {B{\'e}nichou}, \citenamefont {Loverdo}, \citenamefont {Moreau},\ and\
  \citenamefont {Voituriez}}]{benichou2011intermittent}%
  \BibitemOpen
  \bibfield  {author} {\bibinfo {author} {\bibfnamefont {O.}~\bibnamefont
  {B{\'e}nichou}}, \bibinfo {author} {\bibfnamefont {C.}~\bibnamefont
  {Loverdo}}, \bibinfo {author} {\bibfnamefont {M.}~\bibnamefont {Moreau}},\
  and\ \bibinfo {author} {\bibfnamefont {R.}~\bibnamefont {Voituriez}},\
  }\bibfield  {title} {\enquote {\bibinfo {title} {Intermittent search
  strategies},}\ }\href@noop {} {\bibfield  {journal} {\bibinfo  {journal}
  {Reviews of Modern Physics}\ }\textbf {\bibinfo {volume} {83}},\ \bibinfo
  {pages} {81} (\bibinfo {year} {2011})}\BibitemShut {NoStop}%
\bibitem [{\citenamefont {Vicsek}\ and\ \citenamefont
  {Zafeiris}(2012)}]{vicsek2012collective}%
  \BibitemOpen
  \bibfield  {author} {\bibinfo {author} {\bibfnamefont {T.}~\bibnamefont
  {Vicsek}}\ and\ \bibinfo {author} {\bibfnamefont {A.}~\bibnamefont
  {Zafeiris}},\ }\bibfield  {title} {\enquote {\bibinfo {title} {Collective
  motion},}\ }\href@noop {} {\bibfield  {journal} {\bibinfo  {journal} {Physics
  reports}\ }\textbf {\bibinfo {volume} {517}},\ \bibinfo {pages} {71--140}
  (\bibinfo {year} {2012})}\BibitemShut {NoStop}%
\bibitem [{\citenamefont {Yeomans}(2017)}]{yeomans2017nature}%
  \BibitemOpen
  \bibfield  {author} {\bibinfo {author} {\bibfnamefont {J.~M.}\ \bibnamefont
  {Yeomans}},\ }\bibfield  {title} {\enquote {\bibinfo {title} {Nature’s
  engines: active matter},}\ }\href@noop {} {\bibfield  {journal} {\bibinfo
  {journal} {Europhysics News}\ }\textbf {\bibinfo {volume} {48}},\ \bibinfo
  {pages} {21--25} (\bibinfo {year} {2017})}\BibitemShut {NoStop}%
\bibitem [{\citenamefont {Urzay}, \citenamefont {Doostmohammadi},\ and\
  \citenamefont {Yeomans}(2017)}]{urzay2017multi}%
  \BibitemOpen
  \bibfield  {author} {\bibinfo {author} {\bibfnamefont {J.}~\bibnamefont
  {Urzay}}, \bibinfo {author} {\bibfnamefont {A.}~\bibnamefont
  {Doostmohammadi}},\ and\ \bibinfo {author} {\bibfnamefont {J.~M.}\
  \bibnamefont {Yeomans}},\ }\bibfield  {title} {\enquote {\bibinfo {title}
  {Multi-scale statistics of turbulence motorized by active matter},}\
  }\href@noop {} {\bibfield  {journal} {\bibinfo  {journal} {Journal of Fluid
  Mechanics}\ }\textbf {\bibinfo {volume} {822}},\ \bibinfo {pages} {762--773}
  (\bibinfo {year} {2017})}\BibitemShut {NoStop}%
\bibitem [{\citenamefont {Bechinger}\ \emph {et~al.}(2016)\citenamefont
  {Bechinger}, \citenamefont {Di~Leonardo}, \citenamefont {L{\"o}wen},
  \citenamefont {Reichhardt}, \citenamefont {Volpe},\ and\ \citenamefont
  {Volpe}}]{bechinger2016active}%
  \BibitemOpen
  \bibfield  {author} {\bibinfo {author} {\bibfnamefont {C.}~\bibnamefont
  {Bechinger}}, \bibinfo {author} {\bibfnamefont {R.}~\bibnamefont
  {Di~Leonardo}}, \bibinfo {author} {\bibfnamefont {H.}~\bibnamefont
  {L{\"o}wen}}, \bibinfo {author} {\bibfnamefont {C.}~\bibnamefont
  {Reichhardt}}, \bibinfo {author} {\bibfnamefont {G.}~\bibnamefont {Volpe}},\
  and\ \bibinfo {author} {\bibfnamefont {G.}~\bibnamefont {Volpe}},\ }\bibfield
   {title} {\enquote {\bibinfo {title} {Active particles in complex and crowded
  environments},}\ }\href@noop {} {\bibfield  {journal} {\bibinfo  {journal}
  {Reviews of Modern Physics}\ }\textbf {\bibinfo {volume} {88}},\ \bibinfo
  {pages} {045006} (\bibinfo {year} {2016})}\BibitemShut {NoStop}%
\bibitem [{\citenamefont {Ebbens}\ and\ \citenamefont
  {Howse}(2010)}]{ebbens2010pursuit}%
  \BibitemOpen
  \bibfield  {author} {\bibinfo {author} {\bibfnamefont {S.~J.}\ \bibnamefont
  {Ebbens}}\ and\ \bibinfo {author} {\bibfnamefont {J.~R.}\ \bibnamefont
  {Howse}},\ }\bibfield  {title} {\enquote {\bibinfo {title} {In pursuit of
  propulsion at the nanoscale},}\ }\href@noop {} {\bibfield  {journal}
  {\bibinfo  {journal} {Soft Matter}\ }\textbf {\bibinfo {volume} {6}},\
  \bibinfo {pages} {726--738} (\bibinfo {year} {2010})}\BibitemShut {NoStop}%
\bibitem [{\citenamefont {Moran}\ and\ \citenamefont
  {Posner}(2017)}]{moran2017phoretic}%
  \BibitemOpen
  \bibfield  {author} {\bibinfo {author} {\bibfnamefont {J.~L.}\ \bibnamefont
  {Moran}}\ and\ \bibinfo {author} {\bibfnamefont {J.~D.}\ \bibnamefont
  {Posner}},\ }\bibfield  {title} {\enquote {\bibinfo {title} {Phoretic
  self-propulsion},}\ }\href@noop {} {\bibfield  {journal} {\bibinfo  {journal}
  {Annual Review of Fluid Mechanics}\ }\textbf {\bibinfo {volume} {49}},\
  \bibinfo {pages} {511--540} (\bibinfo {year} {2017})}\BibitemShut {NoStop}%
\bibitem [{\citenamefont {Villa}\ and\ \citenamefont
  {Pumera}(2019)}]{villa2019fuel}%
  \BibitemOpen
  \bibfield  {author} {\bibinfo {author} {\bibfnamefont {K.}~\bibnamefont
  {Villa}}\ and\ \bibinfo {author} {\bibfnamefont {M.}~\bibnamefont {Pumera}},\
  }\bibfield  {title} {\enquote {\bibinfo {title} {Fuel-free light-driven
  micro/nanomachines: artificial active matter mimicking nature},}\ }\href@noop
  {} {\bibfield  {journal} {\bibinfo  {journal} {Chemical Society Reviews}\
  }\textbf {\bibinfo {volume} {48}},\ \bibinfo {pages} {4966--4978} (\bibinfo
  {year} {2019})}\BibitemShut {NoStop}%
\bibitem [{\citenamefont {Palagi}\ and\ \citenamefont
  {Fischer}(2018)}]{palagi2018bioinspired}%
  \BibitemOpen
  \bibfield  {author} {\bibinfo {author} {\bibfnamefont {S.}~\bibnamefont
  {Palagi}}\ and\ \bibinfo {author} {\bibfnamefont {P.}~\bibnamefont
  {Fischer}},\ }\bibfield  {title} {\enquote {\bibinfo {title} {Bioinspired
  microrobots},}\ }\href@noop {} {\bibfield  {journal} {\bibinfo  {journal}
  {Nature Reviews Materials}\ }\textbf {\bibinfo {volume} {3}},\ \bibinfo
  {pages} {113--124} (\bibinfo {year} {2018})}\BibitemShut {NoStop}%
\bibitem [{\citenamefont {Gustavsson}\ \emph {et~al.}(2016)\citenamefont
  {Gustavsson}, \citenamefont {Berglund}, \citenamefont {Jonsson},\ and\
  \citenamefont {Mehlig}}]{gustavsson2016preferential}%
  \BibitemOpen
  \bibfield  {author} {\bibinfo {author} {\bibfnamefont {K.}~\bibnamefont
  {Gustavsson}}, \bibinfo {author} {\bibfnamefont {F.}~\bibnamefont
  {Berglund}}, \bibinfo {author} {\bibfnamefont {P.}~\bibnamefont {Jonsson}},\
  and\ \bibinfo {author} {\bibfnamefont {B.}~\bibnamefont {Mehlig}},\
  }\bibfield  {title} {\enquote {\bibinfo {title} {Preferential sampling and
  small-scale clustering of gyrotactic microswimmers in turbulence},}\
  }\href@noop {} {\bibfield  {journal} {\bibinfo  {journal} {Physical review
  letters}\ }\textbf {\bibinfo {volume} {116}},\ \bibinfo {pages} {108104}
  (\bibinfo {year} {2016})}\BibitemShut {NoStop}%
\bibitem [{\citenamefont {Reddy}\ \emph {et~al.}(2016)\citenamefont {Reddy},
  \citenamefont {Celani}, \citenamefont {Sejnowski},\ and\ \citenamefont
  {Vergassola}}]{reddy2016learning}%
  \BibitemOpen
  \bibfield  {author} {\bibinfo {author} {\bibfnamefont {G.}~\bibnamefont
  {Reddy}}, \bibinfo {author} {\bibfnamefont {A.}~\bibnamefont {Celani}},
  \bibinfo {author} {\bibfnamefont {T.~J.}\ \bibnamefont {Sejnowski}},\ and\
  \bibinfo {author} {\bibfnamefont {M.}~\bibnamefont {Vergassola}},\ }\bibfield
   {title} {\enquote {\bibinfo {title} {Learning to soar in turbulent
  environments},}\ }\href@noop {} {\bibfield  {journal} {\bibinfo  {journal}
  {Proceedings of the National Academy of Sciences}\ }\textbf {\bibinfo
  {volume} {113}},\ \bibinfo {pages} {E4877--E4884} (\bibinfo {year}
  {2016})}\BibitemShut {NoStop}%
\bibitem [{\citenamefont {Rubenstein}, \citenamefont {Cornejo},\ and\
  \citenamefont {Nagpal}(2014)}]{rubenstein2014programmable}%
  \BibitemOpen
  \bibfield  {author} {\bibinfo {author} {\bibfnamefont {M.}~\bibnamefont
  {Rubenstein}}, \bibinfo {author} {\bibfnamefont {A.}~\bibnamefont
  {Cornejo}},\ and\ \bibinfo {author} {\bibfnamefont {R.}~\bibnamefont
  {Nagpal}},\ }\bibfield  {title} {\enquote {\bibinfo {title} {Programmable
  self-assembly in a thousand-robot swarm},}\ }\href@noop {} {\bibfield
  {journal} {\bibinfo  {journal} {Science}\ }\textbf {\bibinfo {volume}
  {345}},\ \bibinfo {pages} {795--799} (\bibinfo {year} {2014})}\BibitemShut
  {NoStop}%
\bibitem [{\citenamefont {Bay{\i}nd{\i}r}(2016)}]{bayindir2016review}%
  \BibitemOpen
  \bibfield  {author} {\bibinfo {author} {\bibfnamefont {L.}~\bibnamefont
  {Bay{\i}nd{\i}r}},\ }\bibfield  {title} {\enquote {\bibinfo {title} {A review
  of swarm robotics tasks},}\ }\href@noop {} {\bibfield  {journal} {\bibinfo
  {journal} {Neurocomputing}\ }\textbf {\bibinfo {volume} {172}},\ \bibinfo
  {pages} {292--321} (\bibinfo {year} {2016})}\BibitemShut {NoStop}%
\bibitem [{\citenamefont {Andr{\'e}n}\ \emph {et~al.}(2021)\citenamefont
  {Andr{\'e}n}, \citenamefont {Baranov}, \citenamefont {Jones}, \citenamefont
  {Volpe}, \citenamefont {Verre},\ and\ \citenamefont
  {K{\"a}ll}}]{andren2021microscopic}%
  \BibitemOpen
  \bibfield  {author} {\bibinfo {author} {\bibfnamefont {D.}~\bibnamefont
  {Andr{\'e}n}}, \bibinfo {author} {\bibfnamefont {D.~G.}\ \bibnamefont
  {Baranov}}, \bibinfo {author} {\bibfnamefont {S.}~\bibnamefont {Jones}},
  \bibinfo {author} {\bibfnamefont {G.}~\bibnamefont {Volpe}}, \bibinfo
  {author} {\bibfnamefont {R.}~\bibnamefont {Verre}},\ and\ \bibinfo {author}
  {\bibfnamefont {M.}~\bibnamefont {K{\"a}ll}},\ }\bibfield  {title} {\enquote
  {\bibinfo {title} {Microscopic metavehicles powered and steered by embedded
  optical metasurfaces},}\ }\href@noop {} {\bibfield  {journal} {\bibinfo
  {journal} {Nature Nanotechnology}\ }\textbf {\bibinfo {volume} {16}},\
  \bibinfo {pages} {970--974} (\bibinfo {year} {2021})}\BibitemShut {NoStop}%
\bibitem [{\citenamefont {Galajda}\ \emph {et~al.}(2007)\citenamefont
  {Galajda}, \citenamefont {Keymer}, \citenamefont {Chaikin},\ and\
  \citenamefont {Austin}}]{galajda2007wall}%
  \BibitemOpen
  \bibfield  {author} {\bibinfo {author} {\bibfnamefont {P.}~\bibnamefont
  {Galajda}}, \bibinfo {author} {\bibfnamefont {J.}~\bibnamefont {Keymer}},
  \bibinfo {author} {\bibfnamefont {P.}~\bibnamefont {Chaikin}},\ and\ \bibinfo
  {author} {\bibfnamefont {R.}~\bibnamefont {Austin}},\ }\bibfield  {title}
  {\enquote {\bibinfo {title} {A wall of funnels concentrates swimming
  bacteria},}\ }\href@noop {} {\bibfield  {journal} {\bibinfo  {journal}
  {Journal of bacteriology}\ }\textbf {\bibinfo {volume} {189}},\ \bibinfo
  {pages} {8704--8707} (\bibinfo {year} {2007})}\BibitemShut {NoStop}%
\bibitem [{\citenamefont {Volpe}\ \emph {et~al.}(2011)\citenamefont {Volpe},
  \citenamefont {Buttinoni}, \citenamefont {Vogt}, \citenamefont
  {K{\"u}mmerer},\ and\ \citenamefont {Bechinger}}]{volpe2011microswimmers}%
  \BibitemOpen
  \bibfield  {author} {\bibinfo {author} {\bibfnamefont {G.}~\bibnamefont
  {Volpe}}, \bibinfo {author} {\bibfnamefont {I.}~\bibnamefont {Buttinoni}},
  \bibinfo {author} {\bibfnamefont {D.}~\bibnamefont {Vogt}}, \bibinfo {author}
  {\bibfnamefont {H.-J.}\ \bibnamefont {K{\"u}mmerer}},\ and\ \bibinfo {author}
  {\bibfnamefont {C.}~\bibnamefont {Bechinger}},\ }\bibfield  {title} {\enquote
  {\bibinfo {title} {Microswimmers in patterned environments},}\ }\href@noop {}
  {\bibfield  {journal} {\bibinfo  {journal} {Soft Matter}\ }\textbf {\bibinfo
  {volume} {7}},\ \bibinfo {pages} {8810--8815} (\bibinfo {year}
  {2011})}\BibitemShut {NoStop}%
\bibitem [{\citenamefont {Simmchen}\ \emph {et~al.}(2016)\citenamefont
  {Simmchen}, \citenamefont {Katuri}, \citenamefont {Uspal}, \citenamefont
  {Popescu}, \citenamefont {Tasinkevych},\ and\ \citenamefont
  {S{\'a}nchez}}]{simmchen2016topographical}%
  \BibitemOpen
  \bibfield  {author} {\bibinfo {author} {\bibfnamefont {J.}~\bibnamefont
  {Simmchen}}, \bibinfo {author} {\bibfnamefont {J.}~\bibnamefont {Katuri}},
  \bibinfo {author} {\bibfnamefont {W.~E.}\ \bibnamefont {Uspal}}, \bibinfo
  {author} {\bibfnamefont {M.~N.}\ \bibnamefont {Popescu}}, \bibinfo {author}
  {\bibfnamefont {M.}~\bibnamefont {Tasinkevych}},\ and\ \bibinfo {author}
  {\bibfnamefont {S.}~\bibnamefont {S{\'a}nchez}},\ }\bibfield  {title}
  {\enquote {\bibinfo {title} {Topographical pathways guide chemical
  microswimmers},}\ }\href@noop {} {\bibfield  {journal} {\bibinfo  {journal}
  {Nature communications}\ }\textbf {\bibinfo {volume} {7}},\ \bibinfo {pages}
  {10598} (\bibinfo {year} {2016})}\BibitemShut {NoStop}%
\bibitem [{\citenamefont {Ramos}\ \emph {et~al.}(2021)\citenamefont {Ramos},
  \citenamefont {Rodr{\'\i}guez-S{\'a}nchez}, \citenamefont {Del~Angel},
  \citenamefont {Arzola}, \citenamefont {Ben{\'\i}tez}, \citenamefont
  {Escalante}, \citenamefont {Franci}, \citenamefont {Volpe},\ and\
  \citenamefont {Rivera-Yoshida}}]{ramos2021environment}%
  \BibitemOpen
  \bibfield  {author} {\bibinfo {author} {\bibfnamefont {C.~H.}\ \bibnamefont
  {Ramos}}, \bibinfo {author} {\bibfnamefont {E.}~\bibnamefont
  {Rodr{\'\i}guez-S{\'a}nchez}}, \bibinfo {author} {\bibfnamefont {J.~A.~A.}\
  \bibnamefont {Del~Angel}}, \bibinfo {author} {\bibfnamefont {A.~V.}\
  \bibnamefont {Arzola}}, \bibinfo {author} {\bibfnamefont {M.}~\bibnamefont
  {Ben{\'\i}tez}}, \bibinfo {author} {\bibfnamefont {A.~E.}\ \bibnamefont
  {Escalante}}, \bibinfo {author} {\bibfnamefont {A.}~\bibnamefont {Franci}},
  \bibinfo {author} {\bibfnamefont {G.}~\bibnamefont {Volpe}},\ and\ \bibinfo
  {author} {\bibfnamefont {N.}~\bibnamefont {Rivera-Yoshida}},\ }\bibfield
  {title} {\enquote {\bibinfo {title} {The environment topography alters the
  way to multicellularity in {\it myxococcus xanthus}},}\ }\href@noop {}
  {\bibfield  {journal} {\bibinfo  {journal} {Science Advances}\ }\textbf
  {\bibinfo {volume} {7}},\ \bibinfo {pages} {eabh2278} (\bibinfo {year}
  {2021})}\BibitemShut {NoStop}%
\bibitem [{\citenamefont {Frangipane}\ \emph {et~al.}(2018)\citenamefont
  {Frangipane}, \citenamefont {Dell'Arciprete}, \citenamefont {Petracchini},
  \citenamefont {Maggi}, \citenamefont {Saglimbeni}, \citenamefont {Bianchi},
  \citenamefont {Vizsnyiczai}, \citenamefont {Bernardini},\ and\ \citenamefont
  {Di~Leonardo}}]{frangipane2018dynamic}%
  \BibitemOpen
  \bibfield  {author} {\bibinfo {author} {\bibfnamefont {G.}~\bibnamefont
  {Frangipane}}, \bibinfo {author} {\bibfnamefont {D.}~\bibnamefont
  {Dell'Arciprete}}, \bibinfo {author} {\bibfnamefont {S.}~\bibnamefont
  {Petracchini}}, \bibinfo {author} {\bibfnamefont {C.}~\bibnamefont {Maggi}},
  \bibinfo {author} {\bibfnamefont {F.}~\bibnamefont {Saglimbeni}}, \bibinfo
  {author} {\bibfnamefont {S.}~\bibnamefont {Bianchi}}, \bibinfo {author}
  {\bibfnamefont {G.}~\bibnamefont {Vizsnyiczai}}, \bibinfo {author}
  {\bibfnamefont {M.~L.}\ \bibnamefont {Bernardini}},\ and\ \bibinfo {author}
  {\bibfnamefont {R.}~\bibnamefont {Di~Leonardo}},\ }\bibfield  {title}
  {\enquote {\bibinfo {title} {Dynamic density shaping of photokinetic {\it e.
  coli}},}\ }\href@noop {} {\bibfield  {journal} {\bibinfo  {journal} {Elife}\
  }\textbf {\bibinfo {volume} {7}},\ \bibinfo {pages} {e36608} (\bibinfo {year}
  {2018})}\BibitemShut {NoStop}%
\bibitem [{\citenamefont {Arlt}\ \emph {et~al.}(2018)\citenamefont {Arlt},
  \citenamefont {Martinez}, \citenamefont {Dawson}, \citenamefont {Pilizota},\
  and\ \citenamefont {Poon}}]{arlt2018painting}%
  \BibitemOpen
  \bibfield  {author} {\bibinfo {author} {\bibfnamefont {J.}~\bibnamefont
  {Arlt}}, \bibinfo {author} {\bibfnamefont {V.~A.}\ \bibnamefont {Martinez}},
  \bibinfo {author} {\bibfnamefont {A.}~\bibnamefont {Dawson}}, \bibinfo
  {author} {\bibfnamefont {T.}~\bibnamefont {Pilizota}},\ and\ \bibinfo
  {author} {\bibfnamefont {W.~C.}\ \bibnamefont {Poon}},\ }\bibfield  {title}
  {\enquote {\bibinfo {title} {Painting with light-powered bacteria},}\
  }\href@noop {} {\bibfield  {journal} {\bibinfo  {journal} {Nature
  Communications}\ }\textbf {\bibinfo {volume} {9}},\ \bibinfo {pages} {768}
  (\bibinfo {year} {2018})}\BibitemShut {NoStop}%
\bibitem [{\citenamefont {Levernier}\ \emph {et~al.}(2020)\citenamefont
  {Levernier}, \citenamefont {Textor}, \citenamefont {B{\'e}nichou},\ and\
  \citenamefont {Voituriez}}]{levernier2020inverse}%
  \BibitemOpen
  \bibfield  {author} {\bibinfo {author} {\bibfnamefont {N.}~\bibnamefont
  {Levernier}}, \bibinfo {author} {\bibfnamefont {J.}~\bibnamefont {Textor}},
  \bibinfo {author} {\bibfnamefont {O.}~\bibnamefont {B{\'e}nichou}},\ and\
  \bibinfo {author} {\bibfnamefont {R.}~\bibnamefont {Voituriez}},\ }\bibfield
  {title} {\enquote {\bibinfo {title} {Inverse square l{\'e}vy walks are not
  optimal search strategies for d $\geq$ 2},}\ }\href@noop {} {\bibfield
  {journal} {\bibinfo  {journal} {Physical Review Letters}\ }\textbf {\bibinfo
  {volume} {124}},\ \bibinfo {pages} {080601} (\bibinfo {year}
  {2020})}\BibitemShut {NoStop}%
\bibitem [{\citenamefont {Volpe}\ and\ \citenamefont
  {Volpe}(2017)}]{volpe2017topography}%
  \BibitemOpen
  \bibfield  {author} {\bibinfo {author} {\bibfnamefont {G.}~\bibnamefont
  {Volpe}}\ and\ \bibinfo {author} {\bibfnamefont {G.}~\bibnamefont {Volpe}},\
  }\bibfield  {title} {\enquote {\bibinfo {title} {The topography of the
  environment alters the optimal search strategy for active particles},}\
  }\href@noop {} {\bibfield  {journal} {\bibinfo  {journal} {Proceedings of the
  National Academy of Sciences}\ }\textbf {\bibinfo {volume} {114}},\ \bibinfo
  {pages} {11350--11355} (\bibinfo {year} {2017})}\BibitemShut {NoStop}%
\bibitem [{\citenamefont {Charlesworth}\ and\ \citenamefont
  {Turner}(2019)}]{charlesworth2019intrinsically}%
  \BibitemOpen
  \bibfield  {author} {\bibinfo {author} {\bibfnamefont {H.~J.}\ \bibnamefont
  {Charlesworth}}\ and\ \bibinfo {author} {\bibfnamefont {M.~S.}\ \bibnamefont
  {Turner}},\ }\bibfield  {title} {\enquote {\bibinfo {title} {Intrinsically
  motivated collective motion},}\ }\href@noop {} {\bibfield  {journal}
  {\bibinfo  {journal} {Proceedings of the National Academy of Sciences}\
  }\textbf {\bibinfo {volume} {116}},\ \bibinfo {pages} {15362--15367}
  (\bibinfo {year} {2019})}\BibitemShut {NoStop}%
\bibitem [{\citenamefont {Strandburg-Peshkin}\ \emph
  {et~al.}(2013)\citenamefont {Strandburg-Peshkin}, \citenamefont {Twomey},
  \citenamefont {Bode}, \citenamefont {Kao}, \citenamefont {Katz},
  \citenamefont {Ioannou}, \citenamefont {Rosenthal}, \citenamefont {Torney},
  \citenamefont {Wu}, \citenamefont {Levin} \emph
  {et~al.}}]{strandburg2013visual}%
  \BibitemOpen
  \bibfield  {author} {\bibinfo {author} {\bibfnamefont {A.}~\bibnamefont
  {Strandburg-Peshkin}}, \bibinfo {author} {\bibfnamefont {C.~R.}\ \bibnamefont
  {Twomey}}, \bibinfo {author} {\bibfnamefont {N.~W.}\ \bibnamefont {Bode}},
  \bibinfo {author} {\bibfnamefont {A.~B.}\ \bibnamefont {Kao}}, \bibinfo
  {author} {\bibfnamefont {Y.}~\bibnamefont {Katz}}, \bibinfo {author}
  {\bibfnamefont {C.~C.}\ \bibnamefont {Ioannou}}, \bibinfo {author}
  {\bibfnamefont {S.~B.}\ \bibnamefont {Rosenthal}}, \bibinfo {author}
  {\bibfnamefont {C.~J.}\ \bibnamefont {Torney}}, \bibinfo {author}
  {\bibfnamefont {H.~S.}\ \bibnamefont {Wu}}, \bibinfo {author} {\bibfnamefont
  {S.~A.}\ \bibnamefont {Levin}}, \emph {et~al.},\ }\bibfield  {title}
  {\enquote {\bibinfo {title} {Visual sensory networks and effective
  information transfer in animal groups},}\ }\href@noop {} {\bibfield
  {journal} {\bibinfo  {journal} {Current Biology}\ }\textbf {\bibinfo {volume}
  {23}},\ \bibinfo {pages} {R709--R711} (\bibinfo {year} {2013})}\BibitemShut
  {NoStop}%
\bibitem [{\citenamefont {Loos}\ and\ \citenamefont
  {Klapp}(2020)}]{loos2020irreversibility}%
  \BibitemOpen
  \bibfield  {author} {\bibinfo {author} {\bibfnamefont {S.~A.~M.}\
  \bibnamefont {Loos}}\ and\ \bibinfo {author} {\bibfnamefont {S.~H.~L.}\
  \bibnamefont {Klapp}},\ }\bibfield  {title} {\enquote {\bibinfo {title}
  {Irreversibility, heat and information flows induced by non-reciprocal
  interactions},}\ }\href@noop {} {\bibfield  {journal} {\bibinfo  {journal}
  {New Journal of Physics}\ }\textbf {\bibinfo {volume} {22}},\ \bibinfo
  {pages} {123051} (\bibinfo {year} {2020})}\BibitemShut {NoStop}%
\bibitem [{\citenamefont {Loos}, \citenamefont {Klapp},\ and\ \citenamefont
  {Martynec}(2023)}]{loos2023long}%
  \BibitemOpen
  \bibfield  {author} {\bibinfo {author} {\bibfnamefont {S.~A.~M.}\
  \bibnamefont {Loos}}, \bibinfo {author} {\bibfnamefont {S.~H.~L.}\
  \bibnamefont {Klapp}},\ and\ \bibinfo {author} {\bibfnamefont
  {T.}~\bibnamefont {Martynec}},\ }\bibfield  {title} {\enquote {\bibinfo
  {title} {Long-{R}ange {O}rder and {D}irectional {D}efect {P}ropagation in the
  {N}onreciprocal {XY} {M}odel with {V}ision {C}one {I}nteractions},}\
  }\href@noop {} {\bibfield  {journal} {\bibinfo  {journal} {Physical review
  letters}\ }\textbf {\bibinfo {volume} {130}},\ \bibinfo {pages} {198301}
  (\bibinfo {year} {2023})}\BibitemShut {NoStop}%
\bibitem [{\citenamefont {Argun}, \citenamefont {Callegari},\ and\
  \citenamefont {Volpe}(2021)}]{argun2021simulation}%
  \BibitemOpen
  \bibfield  {author} {\bibinfo {author} {\bibfnamefont {A.}~\bibnamefont
  {Argun}}, \bibinfo {author} {\bibfnamefont {A.}~\bibnamefont {Callegari}},\
  and\ \bibinfo {author} {\bibfnamefont {G.}~\bibnamefont {Volpe}},\
  }\href@noop {} {\emph {\bibinfo {title} {Simulation of Complex Systems}}}\
  (\bibinfo {year} {2021})\BibitemShut {NoStop}%
\bibitem [{\citenamefont {Marchetti}\ \emph {et~al.}(2013)\citenamefont
  {Marchetti}, \citenamefont {Joanny}, \citenamefont {Ramaswamy}, \citenamefont
  {Liverpool}, \citenamefont {Prost}, \citenamefont {Rao},\ and\ \citenamefont
  {Simha}}]{marchetti2013hydrodynamics}%
  \BibitemOpen
  \bibfield  {author} {\bibinfo {author} {\bibfnamefont {M.~C.}\ \bibnamefont
  {Marchetti}}, \bibinfo {author} {\bibfnamefont {J.-F.}\ \bibnamefont
  {Joanny}}, \bibinfo {author} {\bibfnamefont {S.}~\bibnamefont {Ramaswamy}},
  \bibinfo {author} {\bibfnamefont {T.~B.}\ \bibnamefont {Liverpool}}, \bibinfo
  {author} {\bibfnamefont {J.}~\bibnamefont {Prost}}, \bibinfo {author}
  {\bibfnamefont {M.}~\bibnamefont {Rao}},\ and\ \bibinfo {author}
  {\bibfnamefont {R.~A.}\ \bibnamefont {Simha}},\ }\bibfield  {title} {\enquote
  {\bibinfo {title} {Hydrodynamics of soft active matter},}\ }\href@noop {}
  {\bibfield  {journal} {\bibinfo  {journal} {Reviews of modern physics}\
  }\textbf {\bibinfo {volume} {85}},\ \bibinfo {pages} {1143} (\bibinfo {year}
  {2013})}\BibitemShut {NoStop}%
\bibitem [{\citenamefont {Falasco}\ \emph {et~al.}(2016)\citenamefont
  {Falasco}, \citenamefont {Pfaller}, \citenamefont {Bregulla}, \citenamefont
  {Cichos},\ and\ \citenamefont {Kroy}}]{falasco2016exact}%
  \BibitemOpen
  \bibfield  {author} {\bibinfo {author} {\bibfnamefont {G.}~\bibnamefont
  {Falasco}}, \bibinfo {author} {\bibfnamefont {R.}~\bibnamefont {Pfaller}},
  \bibinfo {author} {\bibfnamefont {A.~P.}\ \bibnamefont {Bregulla}}, \bibinfo
  {author} {\bibfnamefont {F.}~\bibnamefont {Cichos}},\ and\ \bibinfo {author}
  {\bibfnamefont {K.}~\bibnamefont {Kroy}},\ }\bibfield  {title} {\enquote
  {\bibinfo {title} {Exact symmetries in the velocity fluctuations of a hot
  brownian swimmer},}\ }\href@noop {} {\bibfield  {journal} {\bibinfo
  {journal} {Physical Review E}\ }\textbf {\bibinfo {volume} {94}},\ \bibinfo
  {pages} {030602} (\bibinfo {year} {2016})}\BibitemShut {NoStop}%
\bibitem [{\citenamefont {Frenkel}\ and\ \citenamefont
  {Smit}(2001)}]{frenkel2001understanding}%
  \BibitemOpen
  \bibfield  {author} {\bibinfo {author} {\bibfnamefont {D.}~\bibnamefont
  {Frenkel}}\ and\ \bibinfo {author} {\bibfnamefont {B.}~\bibnamefont {Smit}},\
  }\href@noop {} {\emph {\bibinfo {title} {Understanding molecular simulation:
  from algorithms to applications}}},\ Vol.~\bibinfo {volume} {1}\ (\bibinfo
  {publisher} {Elsevier},\ \bibinfo {year} {2001})\BibitemShut {NoStop}%
\bibitem [{\citenamefont {Rosenbluth}(2003)}]{rosenbluth2003genesis}%
  \BibitemOpen
  \bibfield  {author} {\bibinfo {author} {\bibfnamefont {M.~N.}\ \bibnamefont
  {Rosenbluth}},\ }\bibfield  {title} {\enquote {\bibinfo {title} {Genesis of
  the monte carlo algorithm for statistical mechanics},}\ }in\ \href@noop {}
  {\emph {\bibinfo {booktitle} {AIP Conference Proceedings}}},\ Vol.\ \bibinfo
  {volume} {690}\ (\bibinfo {organization} {American Institute of Physics},\
  \bibinfo {year} {2003})\ pp.\ \bibinfo {pages} {22--30}\BibitemShut {NoStop}%
\bibitem [{\citenamefont {Wolfram}(1984)}]{wolfram1984cellular}%
  \BibitemOpen
  \bibfield  {author} {\bibinfo {author} {\bibfnamefont {S.}~\bibnamefont
  {Wolfram}},\ }\bibfield  {title} {\enquote {\bibinfo {title} {Cellular
  automata as models of complexity},}\ }\href@noop {} {\bibfield  {journal}
  {\bibinfo  {journal} {Nature}\ }\textbf {\bibinfo {volume} {311}},\ \bibinfo
  {pages} {419--424} (\bibinfo {year} {1984})}\BibitemShut {NoStop}%
\bibitem [{\citenamefont {Lauga}\ and\ \citenamefont
  {Powers}(2009)}]{lauga2009hydrodynamics}%
  \BibitemOpen
  \bibfield  {author} {\bibinfo {author} {\bibfnamefont {E.}~\bibnamefont
  {Lauga}}\ and\ \bibinfo {author} {\bibfnamefont {T.~R.}\ \bibnamefont
  {Powers}},\ }\bibfield  {title} {\enquote {\bibinfo {title} {The
  hydrodynamics of swimming microorganisms},}\ }\href@noop {} {\bibfield
  {journal} {\bibinfo  {journal} {Reports on progress in physics}\ }\textbf
  {\bibinfo {volume} {72}},\ \bibinfo {pages} {096601} (\bibinfo {year}
  {2009})}\BibitemShut {NoStop}%
\bibitem [{\citenamefont {Palacci}\ \emph {et~al.}(2013)\citenamefont
  {Palacci}, \citenamefont {Sacanna}, \citenamefont {Steinberg}, \citenamefont
  {Pine},\ and\ \citenamefont {Chaikin}}]{palacci2013living}%
  \BibitemOpen
  \bibfield  {author} {\bibinfo {author} {\bibfnamefont {J.}~\bibnamefont
  {Palacci}}, \bibinfo {author} {\bibfnamefont {S.}~\bibnamefont {Sacanna}},
  \bibinfo {author} {\bibfnamefont {A.~P.}\ \bibnamefont {Steinberg}}, \bibinfo
  {author} {\bibfnamefont {D.~J.}\ \bibnamefont {Pine}},\ and\ \bibinfo
  {author} {\bibfnamefont {P.~M.}\ \bibnamefont {Chaikin}},\ }\bibfield
  {title} {\enquote {\bibinfo {title} {Living crystals of light-activated
  colloidal surfers},}\ }\href@noop {} {\bibfield  {journal} {\bibinfo
  {journal} {Science}\ }\textbf {\bibinfo {volume} {339}},\ \bibinfo {pages}
  {936--940} (\bibinfo {year} {2013})}\BibitemShut {NoStop}%
\bibitem [{\citenamefont {Buttinoni}\ \emph {et~al.}(2013)\citenamefont
  {Buttinoni}, \citenamefont {Bialk{\'e}}, \citenamefont {K{\"u}mmel},
  \citenamefont {L{\"o}wen}, \citenamefont {Bechinger},\ and\ \citenamefont
  {Speck}}]{buttinoni2013dynamical}%
  \BibitemOpen
  \bibfield  {author} {\bibinfo {author} {\bibfnamefont {I.}~\bibnamefont
  {Buttinoni}}, \bibinfo {author} {\bibfnamefont {J.}~\bibnamefont
  {Bialk{\'e}}}, \bibinfo {author} {\bibfnamefont {F.}~\bibnamefont
  {K{\"u}mmel}}, \bibinfo {author} {\bibfnamefont {H.}~\bibnamefont
  {L{\"o}wen}}, \bibinfo {author} {\bibfnamefont {C.}~\bibnamefont
  {Bechinger}},\ and\ \bibinfo {author} {\bibfnamefont {T.}~\bibnamefont
  {Speck}},\ }\bibfield  {title} {\enquote {\bibinfo {title} {Dynamical
  clustering and phase separation in suspensions of self-propelled colloidal
  particles},}\ }\href@noop {} {\bibfield  {journal} {\bibinfo  {journal}
  {Physical review letters}\ }\textbf {\bibinfo {volume} {110}},\ \bibinfo
  {pages} {238301} (\bibinfo {year} {2013})}\BibitemShut {NoStop}%
\bibitem [{\citenamefont {Lavergne}\ \emph {et~al.}(2019)\citenamefont
  {Lavergne}, \citenamefont {Wendehenne}, \citenamefont {B{\"a}uerle},\ and\
  \citenamefont {Bechinger}}]{lavergne2019group}%
  \BibitemOpen
  \bibfield  {author} {\bibinfo {author} {\bibfnamefont {F.~A.}\ \bibnamefont
  {Lavergne}}, \bibinfo {author} {\bibfnamefont {H.}~\bibnamefont
  {Wendehenne}}, \bibinfo {author} {\bibfnamefont {T.}~\bibnamefont
  {B{\"a}uerle}},\ and\ \bibinfo {author} {\bibfnamefont {C.}~\bibnamefont
  {Bechinger}},\ }\bibfield  {title} {\enquote {\bibinfo {title} {Group
  formation and cohesion of active particles with visual perception--dependent
  motility},}\ }\href@noop {} {\bibfield  {journal} {\bibinfo  {journal}
  {Science}\ }\textbf {\bibinfo {volume} {364}},\ \bibinfo {pages} {70--74}
  (\bibinfo {year} {2019})}\BibitemShut {NoStop}%
\bibitem [{\citenamefont {Khadka}\ \emph {et~al.}(2018)\citenamefont {Khadka},
  \citenamefont {Holubec}, \citenamefont {Yang},\ and\ \citenamefont
  {Cichos}}]{khadka2018active}%
  \BibitemOpen
  \bibfield  {author} {\bibinfo {author} {\bibfnamefont {U.}~\bibnamefont
  {Khadka}}, \bibinfo {author} {\bibfnamefont {V.}~\bibnamefont {Holubec}},
  \bibinfo {author} {\bibfnamefont {H.}~\bibnamefont {Yang}},\ and\ \bibinfo
  {author} {\bibfnamefont {F.}~\bibnamefont {Cichos}},\ }\bibfield  {title}
  {\enquote {\bibinfo {title} {Active particles bound by information flows},}\
  }\href@noop {} {\bibfield  {journal} {\bibinfo  {journal} {Nature
  communications}\ }\textbf {\bibinfo {volume} {9}},\ \bibinfo {pages} {1--9}
  (\bibinfo {year} {2018})}\BibitemShut {NoStop}%
\bibitem [{\citenamefont {Doncieux}\ \emph {et~al.}(2015)\citenamefont
  {Doncieux}, \citenamefont {Bredeche}, \citenamefont {Mouret},\ and\
  \citenamefont {Eiben}}]{doncieux2015evolutionary}%
  \BibitemOpen
  \bibfield  {author} {\bibinfo {author} {\bibfnamefont {S.}~\bibnamefont
  {Doncieux}}, \bibinfo {author} {\bibfnamefont {N.}~\bibnamefont {Bredeche}},
  \bibinfo {author} {\bibfnamefont {J.~B.}\ \bibnamefont {Mouret}},\ and\
  \bibinfo {author} {\bibfnamefont {A.~E.~G.}\ \bibnamefont {Eiben}},\
  }\bibfield  {title} {\enquote {\bibinfo {title} {Evolutionary robotics: What,
  why, and where to},}\ }\href@noop {} {\bibfield  {journal} {\bibinfo
  {journal} {Front. Robot. AI}\ }\textbf {\bibinfo {volume} {2}},\ \bibinfo
  {pages} {4} (\bibinfo {year} {2015})}\BibitemShut {NoStop}%
\bibitem [{\citenamefont {Jones}\ \emph {et~al.}(2019)\citenamefont {Jones},
  \citenamefont {Winfield}, \citenamefont {Hauert},\ and\ \citenamefont
  {M.Studley}}]{jones2019onboard}%
  \BibitemOpen
  \bibfield  {author} {\bibinfo {author} {\bibfnamefont {S.}~\bibnamefont
  {Jones}}, \bibinfo {author} {\bibfnamefont {A.~F.}\ \bibnamefont {Winfield}},
  \bibinfo {author} {\bibfnamefont {S.}~\bibnamefont {Hauert}},\ and\ \bibinfo
  {author} {\bibnamefont {M.Studley}},\ }\bibfield  {title} {\enquote {\bibinfo
  {title} {Onboard evolution of understandable swarm behaviors},}\ }\href@noop
  {} {\bibfield  {journal} {\bibinfo  {journal} {Adv. Intell. Systems}\ ,\
  \bibinfo {pages} {1900031}} (\bibinfo {year} {2019})}\BibitemShut {NoStop}%
\bibitem [{\citenamefont {Mijalkov}\ \emph {et~al.}(2016)\citenamefont
  {Mijalkov}, \citenamefont {McDaniel}, \citenamefont {Wehr},\ and\
  \citenamefont {Volpe}}]{mijalkov2016engineering}%
  \BibitemOpen
  \bibfield  {author} {\bibinfo {author} {\bibfnamefont {M.}~\bibnamefont
  {Mijalkov}}, \bibinfo {author} {\bibfnamefont {A.}~\bibnamefont {McDaniel}},
  \bibinfo {author} {\bibfnamefont {J.}~\bibnamefont {Wehr}},\ and\ \bibinfo
  {author} {\bibfnamefont {G.}~\bibnamefont {Volpe}},\ }\bibfield  {title}
  {\enquote {\bibinfo {title} {Engineering sensorial delay to control
  phototaxis and emergent collective behaviors},}\ }\href@noop {} {\bibfield
  {journal} {\bibinfo  {journal} {Physical Review X}\ }\textbf {\bibinfo
  {volume} {6}},\ \bibinfo {pages} {011008} (\bibinfo {year}
  {2016})}\BibitemShut {NoStop}%
\bibitem [{\citenamefont {Volpe}\ and\ \citenamefont
  {Wehr}(2016)}]{volpe2016effective}%
  \BibitemOpen
  \bibfield  {author} {\bibinfo {author} {\bibfnamefont {G.}~\bibnamefont
  {Volpe}}\ and\ \bibinfo {author} {\bibfnamefont {J.}~\bibnamefont {Wehr}},\
  }\bibfield  {title} {\enquote {\bibinfo {title} {Effective drifts in
  dynamical systems with multiplicative noise: a review of recent progress},}\
  }\href@noop {} {\bibfield  {journal} {\bibinfo  {journal} {Reports on
  Progress in Physics}\ }\textbf {\bibinfo {volume} {79}},\ \bibinfo {pages}
  {053901} (\bibinfo {year} {2016})}\BibitemShut {NoStop}%
\bibitem [{\citenamefont {Leyman}\ \emph {et~al.}(2018)\citenamefont {Leyman},
  \citenamefont {Ogemark}, \citenamefont {Wehr},\ and\ \citenamefont
  {Volpe}}]{leyman2018tuning}%
  \BibitemOpen
  \bibfield  {author} {\bibinfo {author} {\bibfnamefont {M.}~\bibnamefont
  {Leyman}}, \bibinfo {author} {\bibfnamefont {F.}~\bibnamefont {Ogemark}},
  \bibinfo {author} {\bibfnamefont {J.}~\bibnamefont {Wehr}},\ and\ \bibinfo
  {author} {\bibfnamefont {G.}~\bibnamefont {Volpe}},\ }\bibfield  {title}
  {\enquote {\bibinfo {title} {Tuning phototactic robots with sensorial
  delays},}\ }\href@noop {} {\bibfield  {journal} {\bibinfo  {journal}
  {Physical Review E}\ }\textbf {\bibinfo {volume} {98}},\ \bibinfo {pages}
  {052606} (\bibinfo {year} {2018})}\BibitemShut {NoStop}%
\bibitem [{\citenamefont {Kermack}\ and\ \citenamefont
  {McKendrick}(1927)}]{KER27}%
  \BibitemOpen
  \bibfield  {author} {\bibinfo {author} {\bibfnamefont {W.~O.}\ \bibnamefont
  {Kermack}}\ and\ \bibinfo {author} {\bibfnamefont {A.~G.}\ \bibnamefont
  {McKendrick}},\ }\bibfield  {title} {\enquote {\bibinfo {title} {A
  contribution to the mathematical theory of epidemics},}\ }\href@noop {}
  {\bibfield  {journal} {\bibinfo  {journal} {Proc. R. Soc. A}\ }\textbf
  {\bibinfo {volume} {115}},\ \bibinfo {pages} {700--721} (\bibinfo {year}
  {1927})}\BibitemShut {NoStop}%
\bibitem [{\citenamefont {Balcan}\ \emph {et~al.}(2009)\citenamefont {Balcan},
  \citenamefont {Hu}, \citenamefont {Goncalves}, \citenamefont {Bajardi},
  \citenamefont {Poletto}, \citenamefont {Ramasco}, \citenamefont {Paolotti},
  \citenamefont {Perra}, \citenamefont {Tizzoni}, \citenamefont {Van~den
  Broeck}, \citenamefont {Colizza},\ and\ \citenamefont {Vespignani}}]{BAL09b}%
  \BibitemOpen
  \bibfield  {author} {\bibinfo {author} {\bibfnamefont {D.}~\bibnamefont
  {Balcan}}, \bibinfo {author} {\bibfnamefont {H.}~\bibnamefont {Hu}}, \bibinfo
  {author} {\bibfnamefont {B.}~\bibnamefont {Goncalves}}, \bibinfo {author}
  {\bibfnamefont {P.}~\bibnamefont {Bajardi}}, \bibinfo {author} {\bibfnamefont
  {C.}~\bibnamefont {Poletto}}, \bibinfo {author} {\bibfnamefont {J.~J.}\
  \bibnamefont {Ramasco}}, \bibinfo {author} {\bibfnamefont {D.}~\bibnamefont
  {Paolotti}}, \bibinfo {author} {\bibfnamefont {N.}~\bibnamefont {Perra}},
  \bibinfo {author} {\bibfnamefont {M.}~\bibnamefont {Tizzoni}}, \bibinfo
  {author} {\bibfnamefont {W.}~\bibnamefont {Van~den Broeck}}, \bibinfo
  {author} {\bibfnamefont {V.}~\bibnamefont {Colizza}},\ and\ \bibinfo {author}
  {\bibfnamefont {A.}~\bibnamefont {Vespignani}},\ }\bibfield  {title}
  {\enquote {\bibinfo {title} {Seasonal transmission potential and activity
  peaks of the new influenza a(h1n1): a monte carlo likelihood analysis based
  on human mobility},}\ }\href {https://doi.org/10.1186/1741-7015-7-45}
  {\bibfield  {journal} {\bibinfo  {journal} {BMC Med.}\ }\textbf {\bibinfo
  {volume} {7}},\ \bibinfo {pages} {45} (\bibinfo {year} {2009})}\BibitemShut
  {NoStop}%
\bibitem [{\citenamefont {Chung}(2015)}]{CHU15}%
  \BibitemOpen
  \bibfield  {author} {\bibinfo {author} {\bibfnamefont {L.~H.}\ \bibnamefont
  {Chung}},\ }\bibfield  {title} {\enquote {\bibinfo {title} {Impact of
  pandemic control over airport economics: Reconciling public health with
  airport business through a streamlined approach in pandemic control},}\
  }\href@noop {} {\bibfield  {journal} {\bibinfo  {journal} {J. Air Transp.
  Manag.}\ }\textbf {\bibinfo {volume} {44}},\ \bibinfo {pages} {42--53}
  (\bibinfo {year} {2015})}\BibitemShut {NoStop}%
\bibitem [{\citenamefont {Poletto}\ \emph {et~al.}(2014)\citenamefont
  {Poletto}, \citenamefont {Gomes}, \citenamefont {y~Piontti}, \citenamefont
  {Rossi}, \citenamefont {Bioglio}, \citenamefont {Chao}, \citenamefont
  {Longini~Jr}, \citenamefont {Halloran}, \citenamefont {Colizza},\ and\
  \citenamefont {Vespignani}}]{POL14}%
  \BibitemOpen
  \bibfield  {author} {\bibinfo {author} {\bibfnamefont {C.}~\bibnamefont
  {Poletto}}, \bibinfo {author} {\bibfnamefont {M.}~\bibnamefont {Gomes}},
  \bibinfo {author} {\bibfnamefont {A.~P.}\ \bibnamefont {y~Piontti}}, \bibinfo
  {author} {\bibfnamefont {L.}~\bibnamefont {Rossi}}, \bibinfo {author}
  {\bibfnamefont {L.}~\bibnamefont {Bioglio}}, \bibinfo {author} {\bibfnamefont
  {D.~L.}\ \bibnamefont {Chao}}, \bibinfo {author} {\bibfnamefont
  {I.}~\bibnamefont {Longini~Jr}}, \bibinfo {author} {\bibfnamefont {M.~E.}\
  \bibnamefont {Halloran}}, \bibinfo {author} {\bibfnamefont {V.}~\bibnamefont
  {Colizza}},\ and\ \bibinfo {author} {\bibfnamefont {A.}~\bibnamefont
  {Vespignani}},\ }\bibfield  {title} {\enquote {\bibinfo {title} {Assessing
  the impact of travel restrictions on international spread of the 2014 {West
  African Ebola} epidemic},}\ }\href@noop {} {\bibfield  {journal} {\bibinfo
  {journal} {Eurosurveillance}\ }\textbf {\bibinfo {volume} {19}},\ \bibinfo
  {pages} {20936} (\bibinfo {year} {2014})}\BibitemShut {NoStop}%
\bibitem [{\citenamefont {Kyrychko}, \citenamefont {Blyuss},\ and\
  \citenamefont {Brovchenko}(2020)}]{KYR20}%
  \BibitemOpen
  \bibfield  {author} {\bibinfo {author} {\bibfnamefont {Y.~N.}\ \bibnamefont
  {Kyrychko}}, \bibinfo {author} {\bibfnamefont {K.~B.}\ \bibnamefont
  {Blyuss}},\ and\ \bibinfo {author} {\bibfnamefont {I.}~\bibnamefont
  {Brovchenko}},\ }\bibfield  {title} {\enquote {\bibinfo {title} {Mathematical
  modelling of the dynamics and containment of {COVID-19} in {U}kraine},}\
  }\href@noop {} {\bibfield  {journal} {\bibinfo  {journal} {Sci. Rep.}\
  }\textbf {\bibinfo {volume} {10}},\ \bibinfo {pages} {19662} (\bibinfo {year}
  {2020})}\BibitemShut {NoStop}%
\bibitem [{\citenamefont {Maier}\ and\ \citenamefont
  {Brockmann}(2020)}]{MAI20}%
  \BibitemOpen
  \bibfield  {author} {\bibinfo {author} {\bibfnamefont {B.~F.}\ \bibnamefont
  {Maier}}\ and\ \bibinfo {author} {\bibfnamefont {D.}~\bibnamefont
  {Brockmann}},\ }\bibfield  {title} {\enquote {\bibinfo {title} {Effective
  containment explains subexponential growth in recent confirmed {COVID-19}
  cases in china},}\ }\href@noop {} {\bibfield  {journal} {\bibinfo  {journal}
  {Science}\ }\textbf {\bibinfo {volume} {368}},\ \bibinfo {pages} {742--746}
  (\bibinfo {year} {2020})}\BibitemShut {NoStop}%
\bibitem [{\citenamefont {Prem}\ \emph {et~al.}(2020)\citenamefont {Prem},
  \citenamefont {Liu}, \citenamefont {Russell}, \citenamefont {Kucharski},
  \citenamefont {Eggo}, \citenamefont {Davies}, \citenamefont {Flasche},
  \citenamefont {Clifford}, \citenamefont {Pearson}, \citenamefont {Munday}
  \emph {et~al.}}]{PRE20}%
  \BibitemOpen
  \bibfield  {author} {\bibinfo {author} {\bibfnamefont {K.}~\bibnamefont
  {Prem}}, \bibinfo {author} {\bibfnamefont {Y.}~\bibnamefont {Liu}}, \bibinfo
  {author} {\bibfnamefont {T.~W.}\ \bibnamefont {Russell}}, \bibinfo {author}
  {\bibfnamefont {A.~J.}\ \bibnamefont {Kucharski}}, \bibinfo {author}
  {\bibfnamefont {R.~M.}\ \bibnamefont {Eggo}}, \bibinfo {author}
  {\bibfnamefont {N.}~\bibnamefont {Davies}}, \bibinfo {author} {\bibfnamefont
  {S.}~\bibnamefont {Flasche}}, \bibinfo {author} {\bibfnamefont
  {S.}~\bibnamefont {Clifford}}, \bibinfo {author} {\bibfnamefont {C.~A.}\
  \bibnamefont {Pearson}}, \bibinfo {author} {\bibfnamefont {J.~D.}\
  \bibnamefont {Munday}}, \emph {et~al.},\ }\bibfield  {title} {\enquote
  {\bibinfo {title} {The effect of control strategies to reduce social mixing
  on outcomes of the {COVID-19} epidemic in {Wuhan, China}: a modelling
  study},}\ }\href@noop {} {\bibfield  {journal} {\bibinfo  {journal} {The
  Lancet Public Health}\ }\textbf {\bibinfo {volume} {5}},\ \bibinfo {pages}
  {e261--e270} (\bibinfo {year} {2020})}\BibitemShut {NoStop}%
\bibitem [{\citenamefont {Humphries}\ \emph
  {et~al.}(2021{\natexlab{a}})\citenamefont {Humphries}, \citenamefont
  {Spillane}, \citenamefont {Mulchrone}, \citenamefont {Wieczorek},
  \citenamefont {O'Riordain},\ and\ \citenamefont {H{\"o}vel}}]{HUM21}%
  \BibitemOpen
  \bibfield  {author} {\bibinfo {author} {\bibfnamefont {R.}~\bibnamefont
  {Humphries}}, \bibinfo {author} {\bibfnamefont {M.}~\bibnamefont {Spillane}},
  \bibinfo {author} {\bibfnamefont {K.}~\bibnamefont {Mulchrone}}, \bibinfo
  {author} {\bibfnamefont {S.}~\bibnamefont {Wieczorek}}, \bibinfo {author}
  {\bibfnamefont {M.}~\bibnamefont {O'Riordain}},\ and\ \bibinfo {author}
  {\bibfnamefont {P.}~\bibnamefont {H{\"o}vel}},\ }\bibfield  {title} {\enquote
  {\bibinfo {title} {A metapopulation network model for the spreading of
  {SARS-CoV}-2: Case study for {I}reland},}\ }\href@noop {} {\bibfield
  {journal} {\bibinfo  {journal} {Infectious Disease Modelling}\ }\textbf
  {\bibinfo {volume} {6}},\ \bibinfo {pages} {420--437} (\bibinfo {year}
  {2021}{\natexlab{a}})}\BibitemShut {NoStop}%
\bibitem [{\citenamefont {Brockmann}\ and\ \citenamefont
  {Helbing}(2013)}]{BRO13}%
  \BibitemOpen
  \bibfield  {author} {\bibinfo {author} {\bibfnamefont {D.}~\bibnamefont
  {Brockmann}}\ and\ \bibinfo {author} {\bibfnamefont {D.}~\bibnamefont
  {Helbing}},\ }\bibfield  {title} {\enquote {\bibinfo {title} {The hidden
  geometry of complex, network-driven contagion phenomena},}\ }\href@noop {}
  {\bibfield  {journal} {\bibinfo  {journal} {Science}\ }\textbf {\bibinfo
  {volume} {342}},\ \bibinfo {pages} {1337--1342} (\bibinfo {year}
  {2013})}\BibitemShut {NoStop}%
\bibitem [{\citenamefont {Iannelli}\ \emph {et~al.}(2017)\citenamefont
  {Iannelli}, \citenamefont {Koher}, \citenamefont {Brockmann}, \citenamefont
  {H{\"o}vel},\ and\ \citenamefont {Sokolov}}]{IAN17}%
  \BibitemOpen
  \bibfield  {author} {\bibinfo {author} {\bibfnamefont {F.}~\bibnamefont
  {Iannelli}}, \bibinfo {author} {\bibfnamefont {A.}~\bibnamefont {Koher}},
  \bibinfo {author} {\bibfnamefont {D.}~\bibnamefont {Brockmann}}, \bibinfo
  {author} {\bibfnamefont {P.}~\bibnamefont {H{\"o}vel}},\ and\ \bibinfo
  {author} {\bibfnamefont {I.~M.}\ \bibnamefont {Sokolov}},\ }\bibfield
  {title} {\enquote {\bibinfo {title} {Effective distances for epidemics
  spreading on complex networks},}\ }\href
  {https://doi.org/10.1103/physreve.95.012313} {\bibfield  {journal} {\bibinfo
  {journal} {Phys. Rev. E}\ }\textbf {\bibinfo {volume} {95}},\ \bibinfo
  {pages} {012313} (\bibinfo {year} {2017})}\BibitemShut {NoStop}%
\bibitem [{\citenamefont {Gold}\ \emph {et~al.}(2019)\citenamefont {Gold},
  \citenamefont {Balal}, \citenamefont {Horak}, \citenamefont {Cheu},
  \citenamefont {Mehmetoglu},\ and\ \citenamefont {Gurbuz}}]{GOL19a}%
  \BibitemOpen
  \bibfield  {author} {\bibinfo {author} {\bibfnamefont {L.}~\bibnamefont
  {Gold}}, \bibinfo {author} {\bibfnamefont {E.}~\bibnamefont {Balal}},
  \bibinfo {author} {\bibfnamefont {T.}~\bibnamefont {Horak}}, \bibinfo
  {author} {\bibfnamefont {R.~L.}\ \bibnamefont {Cheu}}, \bibinfo {author}
  {\bibfnamefont {T.}~\bibnamefont {Mehmetoglu}},\ and\ \bibinfo {author}
  {\bibfnamefont {O.}~\bibnamefont {Gurbuz}},\ }\bibfield  {title} {\enquote
  {\bibinfo {title} {Health screening strategies for international air
  travelers during an epidemic or pandemic},}\ }\href@noop {} {\bibfield
  {journal} {\bibinfo  {journal} {J. Air Transp. Manag.}\ }\textbf {\bibinfo
  {volume} {75}},\ \bibinfo {pages} {27--38} (\bibinfo {year}
  {2019})}\BibitemShut {NoStop}%
\bibitem [{\citenamefont {Cousins}(2020)}]{COU20}%
  \BibitemOpen
  \bibfield  {author} {\bibinfo {author} {\bibfnamefont {S.}~\bibnamefont
  {Cousins}},\ }\bibfield  {title} {\enquote {\bibinfo {title} {New {Z}ealand
  eliminates {COVID-19}},}\ }\href@noop {} {\bibfield  {journal} {\bibinfo
  {journal} {The Lancet}\ }\textbf {\bibinfo {volume} {395}},\ \bibinfo {pages}
  {1474} (\bibinfo {year} {2020})}\BibitemShut {NoStop}%
\bibitem [{\citenamefont {Czypionka}\ \emph {et~al.}(2022)\citenamefont
  {Czypionka}, \citenamefont {Iftekhar}, \citenamefont {Prainsack},
  \citenamefont {Priesemann}, \citenamefont {Bauer}, \citenamefont {Valdez},
  \citenamefont {Cuschieri}, \citenamefont {Glaab}, \citenamefont {Grill},
  \citenamefont {Krutzinna} \emph {et~al.}}]{CZY22}%
  \BibitemOpen
  \bibfield  {author} {\bibinfo {author} {\bibfnamefont {T.}~\bibnamefont
  {Czypionka}}, \bibinfo {author} {\bibfnamefont {E.~N.}\ \bibnamefont
  {Iftekhar}}, \bibinfo {author} {\bibfnamefont {B.}~\bibnamefont {Prainsack}},
  \bibinfo {author} {\bibfnamefont {V.}~\bibnamefont {Priesemann}}, \bibinfo
  {author} {\bibfnamefont {S.}~\bibnamefont {Bauer}}, \bibinfo {author}
  {\bibfnamefont {A.~C.}\ \bibnamefont {Valdez}}, \bibinfo {author}
  {\bibfnamefont {S.}~\bibnamefont {Cuschieri}}, \bibinfo {author}
  {\bibfnamefont {E.}~\bibnamefont {Glaab}}, \bibinfo {author} {\bibfnamefont
  {E.}~\bibnamefont {Grill}}, \bibinfo {author} {\bibfnamefont
  {J.}~\bibnamefont {Krutzinna}}, \emph {et~al.},\ }\bibfield  {title}
  {\enquote {\bibinfo {title} {The benefits, costs and feasibility of a low
  incidence {COVID-19} strategy},}\ }\href@noop {} {\bibfield  {journal}
  {\bibinfo  {journal} {The Lancet Regional Health-Europe}\ }\textbf {\bibinfo
  {volume} {13}},\ \bibinfo {pages} {100294} (\bibinfo {year}
  {2022})}\BibitemShut {NoStop}%
\bibitem [{\citenamefont {Gross}, \citenamefont {D'Lima},\ and\ \citenamefont
  {Blasius}(2006)}]{GRO06b}%
  \BibitemOpen
  \bibfield  {author} {\bibinfo {author} {\bibfnamefont {T.}~\bibnamefont
  {Gross}}, \bibinfo {author} {\bibfnamefont {C.~J.~D.}\ \bibnamefont
  {D'Lima}},\ and\ \bibinfo {author} {\bibfnamefont {B.}~\bibnamefont
  {Blasius}},\ }\bibfield  {title} {\enquote {\bibinfo {title} {Epidemic
  dynamics on an adaptive network},}\ }\href
  {https://doi.org/10.1103/physrevlett.96.208701} {\bibfield  {journal}
  {\bibinfo  {journal} {Phys. Rev. Lett.}\ }\textbf {\bibinfo {volume} {96}},\
  \bibinfo {pages} {208701} (\bibinfo {year} {2006})}\BibitemShut {NoStop}%
\bibitem [{\citenamefont {Gross}\ and\ \citenamefont
  {Blasius}(2008{\natexlab{b}})}]{GRO08b}%
  \BibitemOpen
  \bibfield  {author} {\bibinfo {author} {\bibfnamefont {T.}~\bibnamefont
  {Gross}}\ and\ \bibinfo {author} {\bibfnamefont {B.}~\bibnamefont
  {Blasius}},\ }\bibfield  {title} {\enquote {\bibinfo {title} {Adaptive
  coevolutionary networks: a review},}\ }\href
  {https://doi.org/10.1098/rsif.2007.1229} {\bibfield  {journal} {\bibinfo
  {journal} {J. R. Soc. Interface}\ }\textbf {\bibinfo {volume} {5}} (\bibinfo
  {year} {2008}{\natexlab{b}}),\ 10.1098/rsif.2007.1229}\BibitemShut {NoStop}%
\bibitem [{\citenamefont {Watts}\ and\ \citenamefont {Strogatz}(1998)}]{WAT98}%
  \BibitemOpen
  \bibfield  {author} {\bibinfo {author} {\bibfnamefont {D.~J.}\ \bibnamefont
  {Watts}}\ and\ \bibinfo {author} {\bibfnamefont {S.~H.}\ \bibnamefont
  {Strogatz}},\ }\bibfield  {title} {\enquote {\bibinfo {title} {{{C}ollective
  dynamics of 'small-world' networks}},}\ }\href@noop {} {\bibfield  {journal}
  {\bibinfo  {journal} {Nature}\ }\textbf {\bibinfo {volume} {393}},\ \bibinfo
  {pages} {440--442} (\bibinfo {year} {1998})}\BibitemShut {NoStop}%
\bibitem [{\citenamefont {Bajardi}\ \emph {et~al.}(2012)\citenamefont
  {Bajardi}, \citenamefont {Barrat}, \citenamefont {Savini},\ and\
  \citenamefont {Colizza}}]{BAJ12}%
  \BibitemOpen
  \bibfield  {author} {\bibinfo {author} {\bibfnamefont {P.}~\bibnamefont
  {Bajardi}}, \bibinfo {author} {\bibfnamefont {A.}~\bibnamefont {Barrat}},
  \bibinfo {author} {\bibfnamefont {L.}~\bibnamefont {Savini}},\ and\ \bibinfo
  {author} {\bibfnamefont {V.}~\bibnamefont {Colizza}},\ }\bibfield  {title}
  {\enquote {\bibinfo {title} {Optimizing surveillance for livestock disease
  spreading through animal movements},}\ }\href@noop {} {\bibfield  {journal}
  {\bibinfo  {journal} {J. Royal Soc. Interface}\ }\textbf {\bibinfo {volume}
  {9}},\ \bibinfo {pages} {2814--2825} (\bibinfo {year} {2012})}\BibitemShut
  {NoStop}%
\bibitem [{\citenamefont {Sch{\"o}ll}(2021)}]{SCH21}%
  \BibitemOpen
  \bibfield  {author} {\bibinfo {author} {\bibfnamefont {E.}~\bibnamefont
  {Sch{\"o}ll}},\ }\bibfield  {title} {\enquote {\bibinfo {title} {Partial
  synchronization patterns in brain networks},}\ }\href
  {https://doi.org/10.1209/0295-5075/ac3b97} {\bibfield  {journal} {\bibinfo
  {journal} {Europhys. Lett.}\ }\textbf {\bibinfo {volume} {136}},\ \bibinfo
  {pages} {18001} (\bibinfo {year} {2021})}\BibitemShut {NoStop}%
\bibitem [{\citenamefont {Belik}, \citenamefont {Mikolajczyk},\ and\
  \citenamefont {H{\"o}vel}(2016)}]{BEL16}%
  \BibitemOpen
  \bibfield  {author} {\bibinfo {author} {\bibfnamefont {V.}~\bibnamefont
  {Belik}}, \bibinfo {author} {\bibfnamefont {R.}~\bibnamefont {Mikolajczyk}},\
  and\ \bibinfo {author} {\bibfnamefont {P.}~\bibnamefont {H{\"o}vel}},\
  }\bibfield  {title} {\enquote {\bibinfo {title} {Control of epidemics on
  hospital networks},}\ }in\ \href
  {https://doi.org/10.1007/978-3-319-28028-8_22} {\emph {\bibinfo {booktitle}
  {Control of Self-Organizing Nonlinear Systems}}},\ \bibinfo {editor} {edited
  by\ \bibinfo {editor} {\bibfnamefont {E.}~\bibnamefont {Sch{\"o}ll}},
  \bibinfo {editor} {\bibfnamefont {S.~H.~L.}\ \bibnamefont {Klapp}},\ and\
  \bibinfo {editor} {\bibfnamefont {P.}~\bibnamefont {H{\"o}vel}}}\ (\bibinfo
  {publisher} {Springer},\ \bibinfo {address} {Berlin, Heidelberg},\ \bibinfo
  {year} {2016})\ Chap.~\bibinfo {chapter} {22}, pp.\ \bibinfo {pages}
  {431--440}\BibitemShut {NoStop}%
\bibitem [{\citenamefont {Belik}\ \emph {et~al.}(2017)\citenamefont {Belik},
  \citenamefont {Karch}, \citenamefont {H{\"o}vel},\ and\ \citenamefont
  {Mikolajczyk}}]{BEL17}%
  \BibitemOpen
  \bibfield  {author} {\bibinfo {author} {\bibfnamefont {V.}~\bibnamefont
  {Belik}}, \bibinfo {author} {\bibfnamefont {A.}~\bibnamefont {Karch}},
  \bibinfo {author} {\bibfnamefont {P.}~\bibnamefont {H{\"o}vel}},\ and\
  \bibinfo {author} {\bibfnamefont {R.}~\bibnamefont {Mikolajczyk}},\
  }\bibfield  {title} {\enquote {\bibinfo {title} {Leveraging topological and
  temporal structure of hospital referral networks for epidemic control},}\
  }in\ \href {https://doi.org/https://doi.org/10.1007/978-981-10-5287-3_9}
  {\emph {\bibinfo {booktitle} {Temporal Network Epidemiology. Theoretical
  Biology.}}},\ \bibinfo {editor} {edited by\ \bibinfo {editor} {\bibfnamefont
  {N.}~\bibnamefont {Masuda}}\ and\ \bibinfo {editor} {\bibfnamefont
  {P.}~\bibnamefont {Holme}}}\ (\bibinfo  {publisher} {Springer},\ \bibinfo
  {address} {Singapore},\ \bibinfo {year} {2017})\ Chap.~\bibinfo {chapter}
  {9}, pp.\ \bibinfo {pages} {199--214}\BibitemShut {NoStop}%
\bibitem [{\citenamefont {Masuda}\ and\ \citenamefont {Holme}(2017)}]{MAS17a}%
  \BibitemOpen
  \bibfield  {author} {\bibinfo {author} {\bibfnamefont {N.}~\bibnamefont
  {Masuda}}\ and\ \bibinfo {author} {\bibfnamefont {P.}~\bibnamefont {Holme}},\
  }\href@noop {} {\emph {\bibinfo {title} {Temporal Network Epidemiology}}}\
  (\bibinfo  {publisher} {Springer},\ \bibinfo {address} {Singapore},\ \bibinfo
  {year} {2017})\BibitemShut {NoStop}%
\bibitem [{\citenamefont {Holme}\ and\ \citenamefont
  {Saram\"aki}(2019)}]{HOL19a}%
  \BibitemOpen
  \bibfield  {author} {\bibinfo {author} {\bibfnamefont {P.}~\bibnamefont
  {Holme}}\ and\ \bibinfo {author} {\bibfnamefont {J.}~\bibnamefont
  {Saram\"aki}},\ }\href@noop {} {\emph {\bibinfo {title} {Temporal network
  theory}}},\ Vol.~\bibinfo {volume} {2}\ (\bibinfo  {publisher} {Springer},\
  \bibinfo {year} {2019})\BibitemShut {NoStop}%
\bibitem [{\citenamefont {Liu}, \citenamefont {Slotine},\ and\ \citenamefont
  {Barab{\'a}si}(2011)}]{LIU11}%
  \BibitemOpen
  \bibfield  {author} {\bibinfo {author} {\bibfnamefont {Y.~Y.}\ \bibnamefont
  {Liu}}, \bibinfo {author} {\bibfnamefont {J.~J.}\ \bibnamefont {Slotine}},\
  and\ \bibinfo {author} {\bibfnamefont {A.~L.}\ \bibnamefont {Barab{\'a}si}},\
  }\bibfield  {title} {\enquote {\bibinfo {title} {Controllability of complex
  networks},}\ }\href {https://doi.org/10.1038/nature10011} {\bibfield
  {journal} {\bibinfo  {journal} {Nature}\ }\textbf {\bibinfo {volume} {473}},\
  \bibinfo {pages} {167--173} (\bibinfo {year} {2011})}\BibitemShut {NoStop}%
\bibitem [{\citenamefont {P{\'o}sfai}\ and\ \citenamefont
  {H{\"o}vel}(2014)}]{POS14}%
  \BibitemOpen
  \bibfield  {author} {\bibinfo {author} {\bibfnamefont {M.}~\bibnamefont
  {P{\'o}sfai}}\ and\ \bibinfo {author} {\bibfnamefont {P.}~\bibnamefont
  {H{\"o}vel}},\ }\bibfield  {title} {\enquote {\bibinfo {title} {Structural
  controllability of temporal networks},}\ }\href@noop {} {\bibfield  {journal}
  {\bibinfo  {journal} {New Journal of Physics}\ }\textbf {\bibinfo {volume}
  {16}},\ \bibinfo {pages} {123055} (\bibinfo {year} {2014})}\BibitemShut
  {NoStop}%
\bibitem [{\citenamefont {P{\'o}sfai}\ \emph {et~al.}(2016)\citenamefont
  {P{\'o}sfai}, \citenamefont {Gao}, \citenamefont {Cornelius}, \citenamefont
  {Barab{\'a}si},\ and\ \citenamefont {D'Souza}}]{POS16}%
  \BibitemOpen
  \bibfield  {author} {\bibinfo {author} {\bibfnamefont {M.}~\bibnamefont
  {P{\'o}sfai}}, \bibinfo {author} {\bibfnamefont {J.}~\bibnamefont {Gao}},
  \bibinfo {author} {\bibfnamefont {S.~P.}\ \bibnamefont {Cornelius}}, \bibinfo
  {author} {\bibfnamefont {A.-L.}\ \bibnamefont {Barab{\'a}si}},\ and\ \bibinfo
  {author} {\bibfnamefont {R.~M.}\ \bibnamefont {D'Souza}},\ }\bibfield
  {title} {\enquote {\bibinfo {title} {Controllability of multiplex,
  multi-time-scale networks},}\ }\href@noop {} {\bibfield  {journal} {\bibinfo
  {journal} {Physical Review E}\ }\textbf {\bibinfo {volume} {94}},\ \bibinfo
  {pages} {032316} (\bibinfo {year} {2016})}\BibitemShut {NoStop}%
\bibitem [{\citenamefont {Valdano}\ \emph {et~al.}(2015)\citenamefont
  {Valdano}, \citenamefont {Ferreri}, \citenamefont {Poletto},\ and\
  \citenamefont {Colizza}}]{VAL15}%
  \BibitemOpen
  \bibfield  {author} {\bibinfo {author} {\bibfnamefont {E.}~\bibnamefont
  {Valdano}}, \bibinfo {author} {\bibfnamefont {L.}~\bibnamefont {Ferreri}},
  \bibinfo {author} {\bibfnamefont {C.}~\bibnamefont {Poletto}},\ and\ \bibinfo
  {author} {\bibfnamefont {V.}~\bibnamefont {Colizza}},\ }\bibfield  {title}
  {\enquote {\bibinfo {title} {Analytical computation of the epidemic threshold
  on temporal networks},}\ }\href@noop {} {\bibfield  {journal} {\bibinfo
  {journal} {Physical Review X}\ }\textbf {\bibinfo {volume} {5}},\ \bibinfo
  {pages} {021005} (\bibinfo {year} {2015})}\BibitemShut {NoStop}%
\bibitem [{\citenamefont {Frasca}\ and\ \citenamefont
  {Sharkey}(2016)}]{FRA16a}%
  \BibitemOpen
  \bibfield  {author} {\bibinfo {author} {\bibfnamefont {M.}~\bibnamefont
  {Frasca}}\ and\ \bibinfo {author} {\bibfnamefont {K.~J.}\ \bibnamefont
  {Sharkey}},\ }\bibfield  {title} {\enquote {\bibinfo {title} {Discrete-time
  moment closure models for epidemic spreading in populations of interacting
  individuals},}\ }\href@noop {} {\bibfield  {journal} {\bibinfo  {journal} {J.
  Theor. Biol.}\ }\textbf {\bibinfo {volume} {399}},\ \bibinfo {pages} {13--21}
  (\bibinfo {year} {2016})}\BibitemShut {NoStop}%
\bibitem [{\citenamefont {Koher}\ \emph {et~al.}(2019)\citenamefont {Koher},
  \citenamefont {Lentz}, \citenamefont {Gleeson},\ and\ \citenamefont
  {H{\"o}vel}}]{KOH19}%
  \BibitemOpen
  \bibfield  {author} {\bibinfo {author} {\bibfnamefont {A.}~\bibnamefont
  {Koher}}, \bibinfo {author} {\bibfnamefont {H.~H.~K.}\ \bibnamefont {Lentz}},
  \bibinfo {author} {\bibfnamefont {J.~P.}\ \bibnamefont {Gleeson}},\ and\
  \bibinfo {author} {\bibfnamefont {P.}~\bibnamefont {H{\"o}vel}},\ }\bibfield
  {title} {\enquote {\bibinfo {title} {Contact-based model for epidemic
  spreading on temporal networks},}\ }\href
  {https://doi.org/10.1103/physrevx.9.031017} {\bibfield  {journal} {\bibinfo
  {journal} {Phys. Rev. X}\ }\textbf {\bibinfo {volume} {9}},\ \bibinfo {pages}
  {031017} (\bibinfo {year} {2019})}\BibitemShut {NoStop}%
\bibitem [{\citenamefont {Humphries}\ \emph
  {et~al.}(2021{\natexlab{b}})\citenamefont {Humphries}, \citenamefont
  {Mulchrone}, \citenamefont {Tratalos}, \citenamefont {More},\ and\
  \citenamefont {H{\"o}vel}}]{HUM21a}%
  \BibitemOpen
  \bibfield  {author} {\bibinfo {author} {\bibfnamefont {R.}~\bibnamefont
  {Humphries}}, \bibinfo {author} {\bibfnamefont {K.}~\bibnamefont
  {Mulchrone}}, \bibinfo {author} {\bibfnamefont {J.}~\bibnamefont {Tratalos}},
  \bibinfo {author} {\bibfnamefont {S.~J.}\ \bibnamefont {More}},\ and\
  \bibinfo {author} {\bibfnamefont {P.}~\bibnamefont {H{\"o}vel}},\ }\bibfield
  {title} {\enquote {\bibinfo {title} {A systematic framework of modelling
  epidemics on temporal networks},}\ }\href@noop {} {\bibfield  {journal}
  {\bibinfo  {journal} {Applied Network Science}\ }\textbf {\bibinfo {volume}
  {6}},\ \bibinfo {pages} {1--19} (\bibinfo {year}
  {2021}{\natexlab{b}})}\BibitemShut {NoStop}%
\bibitem [{\citenamefont {Gross}\ and\ \citenamefont
  {Sayama}(2009)}]{gross2009adaptive}%
  \BibitemOpen
  \bibinfo {editor} {\bibfnamefont {T.}~\bibnamefont {Gross}}\ and\ \bibinfo
  {editor} {\bibfnamefont {H.}~\bibnamefont {Sayama}},\ eds.,\ \href
  {https://doi.org/10.1007/978-3-642-01284-6} {\emph {\bibinfo {title}
  {Adaptive Networks}}},\ \bibinfo {edition} {1st}\ ed.,\ Understanding Complex
  Systems\ (\bibinfo  {publisher} {Springer},\ \bibinfo {year}
  {2009})\BibitemShut {NoStop}%
\bibitem [{\citenamefont {Gross}\ and\ \citenamefont
  {Blasius}(2008{\natexlab{c}})}]{gross2008adaptive}%
  \BibitemOpen
  \bibfield  {author} {\bibinfo {author} {\bibfnamefont {T.}~\bibnamefont
  {Gross}}\ and\ \bibinfo {author} {\bibfnamefont {B.}~\bibnamefont
  {Blasius}},\ }\bibfield  {title} {\enquote {\bibinfo {title} {Adaptive
  coevolutionary networks: a review},}\ }\href
  {https://doi.org/10.1098/rsif.2007.1229} {\bibfield  {journal} {\bibinfo
  {journal} {J. R. Soc. Interface}\ }\textbf {\bibinfo {volume} {5}},\ \bibinfo
  {pages} {259--271} (\bibinfo {year} {2008}{\natexlab{c}})}\BibitemShut
  {NoStop}%
\bibitem [{\citenamefont {Holme}\ and\ \citenamefont
  {Newman}(2006)}]{holme2006nonequilibrium}%
  \BibitemOpen
  \bibfield  {author} {\bibinfo {author} {\bibfnamefont {P.}~\bibnamefont
  {Holme}}\ and\ \bibinfo {author} {\bibfnamefont {M.~E.~J.}\ \bibnamefont
  {Newman}},\ }\bibfield  {title} {\enquote {\bibinfo {title} {Nonequilibrium
  phase transition in the coevolution of networks and opinions},}\ }\href
  {https://doi.org/10.1103/PhysRevE.74.056108} {\bibfield  {journal} {\bibinfo
  {journal} {Phys. Rev. E}\ }\textbf {\bibinfo {volume} {74}} (\bibinfo {year}
  {2006}),\ 10.1103/PhysRevE.74.056108}\BibitemShut {NoStop}%
\bibitem [{\citenamefont {Demirel}\ \emph {et~al.}(2014)\citenamefont
  {Demirel}, \citenamefont {Vazquez}, \citenamefont {B{\"o}hme},\ and\
  \citenamefont {Gross}}]{demirel2014momentclosure}%
  \BibitemOpen
  \bibfield  {author} {\bibinfo {author} {\bibfnamefont {G.}~\bibnamefont
  {Demirel}}, \bibinfo {author} {\bibfnamefont {F.}~\bibnamefont {Vazquez}},
  \bibinfo {author} {\bibfnamefont {G.~A.}\ \bibnamefont {B{\"o}hme}},\ and\
  \bibinfo {author} {\bibfnamefont {T.}~\bibnamefont {Gross}},\ }\bibfield
  {title} {\enquote {\bibinfo {title} {Moment-closure approximations for
  discrete adaptive networks},}\ }\href
  {https://doi.org/10.1016/j.physd.2013.07.003} {\bibfield  {journal} {\bibinfo
   {journal} {Physica D}\ }\textbf {\bibinfo {volume} {267}},\ \bibinfo {pages}
  {68--80} (\bibinfo {year} {2014})}\BibitemShut {NoStop}%
\bibitem [{\citenamefont {Zschaler}(2012)}]{zschaler2012adaptive}%
  \BibitemOpen
  \bibfield  {author} {\bibinfo {author} {\bibfnamefont {G.}~\bibnamefont
  {Zschaler}},\ }\bibfield  {title} {\enquote {\bibinfo {title}
  {Adaptive-network models of collective dynamics},}\ }\href
  {https://doi.org/10.1140/epjst/e2012-01648-5} {\bibfield  {journal} {\bibinfo
   {journal} {Eur. Phys. J. Spec. Top.}\ }\textbf {\bibinfo {volume} {211}},\
  \bibinfo {pages} {1--101} (\bibinfo {year} {2012})}\BibitemShut {NoStop}%
\bibitem [{\citenamefont {Clifford}\ and\ \citenamefont
  {Sudbury}(1973)}]{clifford1973model}%
  \BibitemOpen
  \bibfield  {author} {\bibinfo {author} {\bibfnamefont {P.}~\bibnamefont
  {Clifford}}\ and\ \bibinfo {author} {\bibfnamefont {A.}~\bibnamefont
  {Sudbury}},\ }\bibfield  {title} {\enquote {\bibinfo {title} {A model for
  spatial conflict},}\ }\href {https://doi.org/10.1093/biomet/60.3.581}
  {\bibfield  {journal} {\bibinfo  {journal} {Biometrika}\ }\textbf {\bibinfo
  {volume} {60}},\ \bibinfo {pages} {581--588} (\bibinfo {year}
  {1973})}\BibitemShut {NoStop}%
\bibitem [{\citenamefont {Holley}\ and\ \citenamefont
  {Liggett}(1975)}]{holley1975ergodic}%
  \BibitemOpen
  \bibfield  {author} {\bibinfo {author} {\bibfnamefont {R.~A.}\ \bibnamefont
  {Holley}}\ and\ \bibinfo {author} {\bibfnamefont {T.~M.}\ \bibnamefont
  {Liggett}},\ }\bibfield  {title} {\enquote {\bibinfo {title} {Ergodic
  theorems for weakly interacting infinite systems and the voter model},}\
  }\href {https://doi.org/10.1214/aop/1176996306} {\bibfield  {journal}
  {\bibinfo  {journal} {Ann. Probab.}\ }\textbf {\bibinfo {volume} {3}},\
  \bibinfo {pages} {643--663} (\bibinfo {year} {1975})}\BibitemShut {NoStop}%
\bibitem [{\citenamefont {Vazquez}, \citenamefont {Egu{\'\i}luz},\ and\
  \citenamefont {Miguel}(2008)}]{vazquez2008generic}%
  \BibitemOpen
  \bibfield  {author} {\bibinfo {author} {\bibfnamefont {F.}~\bibnamefont
  {Vazquez}}, \bibinfo {author} {\bibfnamefont {V.~M.}\ \bibnamefont
  {Egu{\'\i}luz}},\ and\ \bibinfo {author} {\bibfnamefont {M.~S.}\ \bibnamefont
  {Miguel}},\ }\bibfield  {title} {\enquote {\bibinfo {title} {Generic
  absorbing transition in coevolution dynamics},}\ }\href
  {https://doi.org/10.1103/PhysRevLett.100.108702} {\bibfield  {journal}
  {\bibinfo  {journal} {Phys. Rev. Lett.}\ }\textbf {\bibinfo {volume} {100}}
  (\bibinfo {year} {2008}),\ 10.1103/PhysRevLett.100.108702}\BibitemShut
  {NoStop}%
\bibitem [{\citenamefont {Durrett}\ \emph {et~al.}(2012)\citenamefont
  {Durrett}, \citenamefont {Gleeson}, \citenamefont {Lloyd}, \citenamefont
  {Mucha}, \citenamefont {Shi}, \citenamefont {Sivakoff}, \citenamefont
  {Socolar},\ and\ \citenamefont {Varghese}}]{durrett2012graph}%
  \BibitemOpen
  \bibfield  {author} {\bibinfo {author} {\bibfnamefont {R.}~\bibnamefont
  {Durrett}}, \bibinfo {author} {\bibfnamefont {J.~P.}\ \bibnamefont
  {Gleeson}}, \bibinfo {author} {\bibfnamefont {A.~L.}\ \bibnamefont {Lloyd}},
  \bibinfo {author} {\bibfnamefont {P.~J.}\ \bibnamefont {Mucha}}, \bibinfo
  {author} {\bibfnamefont {F.}~\bibnamefont {Shi}}, \bibinfo {author}
  {\bibfnamefont {D.}~\bibnamefont {Sivakoff}}, \bibinfo {author}
  {\bibfnamefont {J.~E.~S.}\ \bibnamefont {Socolar}},\ and\ \bibinfo {author}
  {\bibfnamefont {C.}~\bibnamefont {Varghese}},\ }\bibfield  {title} {\enquote
  {\bibinfo {title} {Graph fission in an evolving voter model},}\ }\href
  {https://doi.org/10.1073/pnas.1200709109} {\bibfield  {journal} {\bibinfo
  {journal} {Proc. Natl. Acad. Sci. USA}\ }\textbf {\bibinfo {volume} {109}},\
  \bibinfo {pages} {3682--3687} (\bibinfo {year} {2012})}\BibitemShut {NoStop}%
\bibitem [{\citenamefont {Kiss}, \citenamefont {Miller},\ and\ \citenamefont
  {Simon}(2017)}]{kiss2017mathematics}%
  \BibitemOpen
  \bibfield  {author} {\bibinfo {author} {\bibfnamefont {I.~Z.}\ \bibnamefont
  {Kiss}}, \bibinfo {author} {\bibfnamefont {J.}~\bibnamefont {Miller}},\ and\
  \bibinfo {author} {\bibfnamefont {P.~L.}\ \bibnamefont {Simon}},\ }\href
  {https://doi.org/10.1007/978-3-319-50806-1} {\emph {\bibinfo {title}
  {Mathematics of Epidemics on Networks}}},\ \bibinfo {edition} {1st}\ ed.,\
  Interdisciplinary Applied Mathematics\ (\bibinfo  {publisher} {Springer},\
  \bibinfo {year} {2017})\BibitemShut {NoStop}%
\bibitem [{\citenamefont {Kuehn}\ and\ \citenamefont
  {Bick}(2021)}]{kuehn2021universal}%
  \BibitemOpen
  \bibfield  {author} {\bibinfo {author} {\bibfnamefont {C.}~\bibnamefont
  {Kuehn}}\ and\ \bibinfo {author} {\bibfnamefont {C.}~\bibnamefont {Bick}},\
  }\bibfield  {title} {\enquote {\bibinfo {title} {A universal route to
  explosive phenomena},}\ }\href {https://doi.org/10.1126/sciadv.abe3824}
  {\bibfield  {journal} {\bibinfo  {journal} {Sci. Adv.}\ }\textbf {\bibinfo
  {volume} {7}} (\bibinfo {year} {2021}),\ 10.1126/sciadv.abe3824}\BibitemShut
  {NoStop}%
\bibitem [{\citenamefont {Zschaler}\ \emph {et~al.}(2012)\citenamefont
  {Zschaler}, \citenamefont {B{\"o}hme}, \citenamefont {Sei{\ss}inger},
  \citenamefont {Huepe},\ and\ \citenamefont {Gross}}]{zschaler2012early}%
  \BibitemOpen
  \bibfield  {author} {\bibinfo {author} {\bibfnamefont {G.}~\bibnamefont
  {Zschaler}}, \bibinfo {author} {\bibfnamefont {G.~A.}\ \bibnamefont
  {B{\"o}hme}}, \bibinfo {author} {\bibfnamefont {M.}~\bibnamefont
  {Sei{\ss}inger}}, \bibinfo {author} {\bibfnamefont {C.}~\bibnamefont
  {Huepe}},\ and\ \bibinfo {author} {\bibfnamefont {T.}~\bibnamefont {Gross}},\
  }\bibfield  {title} {\enquote {\bibinfo {title} {Early fragmentation in the
  adaptive voter model on directed networks},}\ }\href
  {https://doi.org/10.1103/PhysRevE.85.046107} {\bibfield  {journal} {\bibinfo
  {journal} {Phys. Rev. E}\ }\textbf {\bibinfo {volume} {85}} (\bibinfo {year}
  {2012}),\ 10.1103/PhysRevE.85.046107}\BibitemShut {NoStop}%
\bibitem [{\citenamefont {B{\"o}hme}\ and\ \citenamefont
  {Gross}(2011)}]{boehme2011analytical}%
  \BibitemOpen
  \bibfield  {author} {\bibinfo {author} {\bibfnamefont {G.~A.}\ \bibnamefont
  {B{\"o}hme}}\ and\ \bibinfo {author} {\bibfnamefont {T.}~\bibnamefont
  {Gross}},\ }\bibfield  {title} {\enquote {\bibinfo {title} {Analytical
  calculation of fragmentation transitions in adaptive networks},}\ }\href
  {https://doi.org/10.1103/PhysRevE.83.035101} {\bibfield  {journal} {\bibinfo
  {journal} {Phys. Rev. E}\ }\textbf {\bibinfo {volume} {83}},\ \bibinfo
  {pages} {035101} (\bibinfo {year} {2011})}\BibitemShut {NoStop}%
\bibitem [{\citenamefont {Kozma}\ and\ \citenamefont
  {Barrat}(2008)}]{kozma2008consensus}%
  \BibitemOpen
  \bibfield  {author} {\bibinfo {author} {\bibfnamefont {B.}~\bibnamefont
  {Kozma}}\ and\ \bibinfo {author} {\bibfnamefont {A.}~\bibnamefont {Barrat}},\
  }\bibfield  {title} {\enquote {\bibinfo {title} {Consensus formation on
  adaptive networks},}\ }\href {https://doi.org/10.1103/PhysRevE.77.016102}
  {\bibfield  {journal} {\bibinfo  {journal} {Phys. Rev. E}\ }\textbf {\bibinfo
  {volume} {77}},\ \bibinfo {pages} {016102} (\bibinfo {year}
  {2008})}\BibitemShut {NoStop}%
\bibitem [{\citenamefont {Yu}\ \emph {et~al.}(2017)\citenamefont {Yu},
  \citenamefont {Xiao}, \citenamefont {Li}, \citenamefont {Tay},\ and\
  \citenamefont {Teoh}}]{yu2017opinion}%
  \BibitemOpen
  \bibfield  {author} {\bibinfo {author} {\bibfnamefont {Y.}~\bibnamefont
  {Yu}}, \bibinfo {author} {\bibfnamefont {G.}~\bibnamefont {Xiao}}, \bibinfo
  {author} {\bibfnamefont {G.}~\bibnamefont {Li}}, \bibinfo {author}
  {\bibfnamefont {W.~P.}\ \bibnamefont {Tay}},\ and\ \bibinfo {author}
  {\bibfnamefont {H.~F.}\ \bibnamefont {Teoh}},\ }\bibfield  {title} {\enquote
  {\bibinfo {title} {Opinion diversity and community formation in adaptive
  networks},}\ }\href {https://doi.org/10.1063/1.4989668} {\bibfield  {journal}
  {\bibinfo  {journal} {Chaos}\ }\textbf {\bibinfo {volume} {27}} (\bibinfo
  {year} {2017}),\ 10.1063/1.4989668}\BibitemShut {NoStop}%
\bibitem [{\citenamefont {Zhang}\ \emph {et~al.}(2019)\citenamefont {Zhang},
  \citenamefont {Shan}, \citenamefont {Jin},\ and\ \citenamefont
  {Zhu}}]{zhang2019complex}%
  \BibitemOpen
  \bibfield  {author} {\bibinfo {author} {\bibfnamefont {X.}~\bibnamefont
  {Zhang}}, \bibinfo {author} {\bibfnamefont {C.}~\bibnamefont {Shan}},
  \bibinfo {author} {\bibfnamefont {Z.}~\bibnamefont {Jin}},\ and\ \bibinfo
  {author} {\bibfnamefont {H.}~\bibnamefont {Zhu}},\ }\bibfield  {title}
  {\enquote {\bibinfo {title} {Complex dynamics of epidemic models on adaptive
  networks},}\ }\href {https://doi.org/10.1016/j.jde.2018.07.054} {\bibfield
  {journal} {\bibinfo  {journal} {J. Differ. Equ.}\ }\textbf {\bibinfo {volume}
  {266}},\ \bibinfo {pages} {803--832} (\bibinfo {year} {2019})}\BibitemShut
  {NoStop}%
\bibitem [{\citenamefont {Ogura}\ and\ \citenamefont
  {Preciado}(2016)}]{ogura2016epidemic}%
  \BibitemOpen
  \bibfield  {author} {\bibinfo {author} {\bibfnamefont {M.}~\bibnamefont
  {Ogura}}\ and\ \bibinfo {author} {\bibfnamefont {V.~M.}\ \bibnamefont
  {Preciado}},\ }\bibfield  {title} {\enquote {\bibinfo {title} {Epidemic
  processes over adaptive state-dependent networks},}\ }\href
  {https://doi.org/10.1103/PhysRevE.93.062316} {\bibfield  {journal} {\bibinfo
  {journal} {Phys. Rev. E}\ }\textbf {\bibinfo {volume} {93}} (\bibinfo {year}
  {2016}),\ 10.1103/PhysRevE.93.062316}\BibitemShut {NoStop}%
\bibitem [{\citenamefont {Horstmeyer}, \citenamefont {Kuehn},\ and\
  \citenamefont {Thurner}(2018)}]{horstmeyer2018network}%
  \BibitemOpen
  \bibfield  {author} {\bibinfo {author} {\bibfnamefont {L.}~\bibnamefont
  {Horstmeyer}}, \bibinfo {author} {\bibfnamefont {C.}~\bibnamefont {Kuehn}},\
  and\ \bibinfo {author} {\bibfnamefont {S.}~\bibnamefont {Thurner}},\
  }\bibfield  {title} {\enquote {\bibinfo {title} {Network topology near
  criticality in adaptive epidemics},}\ }\href
  {https://doi.org/10.1103/PhysRevE.98.042313} {\bibfield  {journal} {\bibinfo
  {journal} {Phys. Rev. E}\ }\textbf {\bibinfo {volume} {98}} (\bibinfo {year}
  {2018}),\ 10.1103/PhysRevE.98.042313}\BibitemShut {NoStop}%
\bibitem [{\citenamefont {Horstmeyer}, \citenamefont {Kuehn},\ and\
  \citenamefont {Thurner}(2022)}]{horstmeyer2022balancing}%
  \BibitemOpen
  \bibfield  {author} {\bibinfo {author} {\bibfnamefont {L.}~\bibnamefont
  {Horstmeyer}}, \bibinfo {author} {\bibfnamefont {C.}~\bibnamefont {Kuehn}},\
  and\ \bibinfo {author} {\bibfnamefont {S.}~\bibnamefont {Thurner}},\
  }\bibfield  {title} {\enquote {\bibinfo {title} {Balancing quarantine and
  self-distancing measures in adaptive epidemic networks},}\ }\href
  {https://doi.org/10.1007/s11538-022-01033-3} {\bibfield  {journal} {\bibinfo
  {journal} {Bull. Math. Biol.}\ }\textbf {\bibinfo {volume} {84}} (\bibinfo
  {year} {2022}),\ 10.1007/s11538-022-01033-3}\BibitemShut {NoStop}%
\bibitem [{\citenamefont {Shaw}\ and\ \citenamefont
  {Schwartz}(2008)}]{shaw2008fluctuating}%
  \BibitemOpen
  \bibfield  {author} {\bibinfo {author} {\bibfnamefont {L.~B.}\ \bibnamefont
  {Shaw}}\ and\ \bibinfo {author} {\bibfnamefont {I.~B.}\ \bibnamefont
  {Schwartz}},\ }\bibfield  {title} {\enquote {\bibinfo {title} {Fluctuating
  epidemics on adaptive networks},}\ }\href
  {https://doi.org/10.1103/PhysRevE.77.066101} {\bibfield  {journal} {\bibinfo
  {journal} {Phys. Rev. E}\ }\textbf {\bibinfo {volume} {77}} (\bibinfo {year}
  {2008}),\ 10.1103/PhysRevE.77.066101}\BibitemShut {NoStop}%
\bibitem [{\citenamefont {Demirel}, \citenamefont {Barter},\ and\ \citenamefont
  {Gross}(2017)}]{demirel2017dynamics}%
  \BibitemOpen
  \bibfield  {author} {\bibinfo {author} {\bibfnamefont {G.}~\bibnamefont
  {Demirel}}, \bibinfo {author} {\bibfnamefont {E.}~\bibnamefont {Barter}},\
  and\ \bibinfo {author} {\bibfnamefont {T.}~\bibnamefont {Gross}},\ }\bibfield
   {title} {\enquote {\bibinfo {title} {Dynamics of epidemic diseases on a
  growing adaptive network},}\ }\href {https://doi.org/10.1038/srep42352}
  {\bibfield  {journal} {\bibinfo  {journal} {Sci. Rep.}\ }\textbf {\bibinfo
  {volume} {7}} (\bibinfo {year} {2017}),\ 10.1038/srep42352}\BibitemShut
  {NoStop}%
\bibitem [{\citenamefont {Clau{\ss}}\ and\ \citenamefont
  {Kuehn}(2022)}]{clauss2022selfadapting}%
  \BibitemOpen
  \bibfield  {author} {\bibinfo {author} {\bibfnamefont {K.}~\bibnamefont
  {Clau{\ss}}}\ and\ \bibinfo {author} {\bibfnamefont {C.}~\bibnamefont
  {Kuehn}},\ }\bibfield  {title} {\enquote {\bibinfo {title} {Self-adapting
  infectious dynamics on random networks},}\ }\href@noop {} {\  (\bibinfo
  {year} {2022})},\ \Eprint {https://arxiv.org/abs/arXiv: 2203.16949 [nlin.AO]}
  {arXiv: 2203.16949 [nlin.AO]} \BibitemShut {NoStop}%
\bibitem [{\citenamefont {Strogatz}(2000)}]{strogatz2000from}%
  \BibitemOpen
  \bibfield  {author} {\bibinfo {author} {\bibfnamefont {S.~H.}\ \bibnamefont
  {Strogatz}},\ }\bibfield  {title} {\enquote {\bibinfo {title} {From Kuramoto
  to crawford: exploring the onset of synchronization in populations of coupled
  oscillators},}\ }\href {https://doi.org/10.1016/S0167-2789(00)00094-4}
  {\bibfield  {journal} {\bibinfo  {journal} {Physica D}\ }\textbf {\bibinfo
  {volume} {143}},\ \bibinfo {pages} {1--20} (\bibinfo {year}
  {2000})}\BibitemShut {NoStop}%
\bibitem [{\citenamefont {Avalos-Gayt{\'a}n}\ \emph {et~al.}(2018)\citenamefont
  {Avalos-Gayt{\'a}n}, \citenamefont {Almendral}, \citenamefont {Leyva},
  \citenamefont {Battiston}, \citenamefont {Nicosia}, \citenamefont {Latora},\
  and\ \citenamefont {Boccaletti}}]{avalosgaytan2018emergent}%
  \BibitemOpen
  \bibfield  {author} {\bibinfo {author} {\bibfnamefont {V.}~\bibnamefont
  {Avalos-Gayt{\'a}n}}, \bibinfo {author} {\bibfnamefont {J.~A.}\ \bibnamefont
  {Almendral}}, \bibinfo {author} {\bibfnamefont {I.}~\bibnamefont {Leyva}},
  \bibinfo {author} {\bibfnamefont {F.}~\bibnamefont {Battiston}}, \bibinfo
  {author} {\bibfnamefont {V.}~\bibnamefont {Nicosia}}, \bibinfo {author}
  {\bibfnamefont {V.}~\bibnamefont {Latora}},\ and\ \bibinfo {author}
  {\bibfnamefont {S.}~\bibnamefont {Boccaletti}},\ }\bibfield  {title}
  {\enquote {\bibinfo {title} {Emergent explosive synchronization in adaptive
  complex networks},}\ }\href {https://doi.org/10.1103/PhysRevE.97.042301}
  {\bibfield  {journal} {\bibinfo  {journal} {Phys. Rev. E}\ }\textbf {\bibinfo
  {volume} {97}} (\bibinfo {year} {2018}),\
  10.1103/PhysRevE.97.042301}\BibitemShut {NoStop}%
\bibitem [{\citenamefont {Schlager}, \citenamefont {Clau{\ss}},\ and\
  \citenamefont {Kuehn}(2022)}]{schlager2022stability}%
  \BibitemOpen
  \bibfield  {author} {\bibinfo {author} {\bibfnamefont {D.}~\bibnamefont
  {Schlager}}, \bibinfo {author} {\bibfnamefont {K.}~\bibnamefont
  {Clau{\ss}}},\ and\ \bibinfo {author} {\bibfnamefont {C.}~\bibnamefont
  {Kuehn}},\ }\bibfield  {title} {\enquote {\bibinfo {title} {Stability
  analysis of multiplayer games on adaptive simplicial complexes},}\ }\href
  {https://doi.org/10.1063/5.0078863} {\bibfield  {journal} {\bibinfo
  {journal} {Chaos}\ }\textbf {\bibinfo {volume} {32}} (\bibinfo {year}
  {2022}),\ 10.1063/5.0078863}\BibitemShut {NoStop}%
\bibitem [{\citenamefont {Horstmeyer}\ and\ \citenamefont
  {Kuehn}(2020)}]{horstmeyer2020adaptive}%
  \BibitemOpen
  \bibfield  {author} {\bibinfo {author} {\bibfnamefont {L.}~\bibnamefont
  {Horstmeyer}}\ and\ \bibinfo {author} {\bibfnamefont {C.}~\bibnamefont
  {Kuehn}},\ }\bibfield  {title} {\enquote {\bibinfo {title} {Adaptive voter
  model on simplicial complexes},}\ }\href
  {https://doi.org/10.1103/PhysRevE.101.022305} {\bibfield  {journal} {\bibinfo
   {journal} {Phys. Rev. E}\ }\textbf {\bibinfo {volume} {101}} (\bibinfo
  {year} {2020}),\ 10.1103/PhysRevE.101.022305}\BibitemShut {NoStop}%
\bibitem [{\citenamefont {Papanikolaou}\ \emph {et~al.}(2022)\citenamefont
  {Papanikolaou}, \citenamefont {Vaccario}, \citenamefont {Hormann},
  \citenamefont {Lambiotte},\ and\ \citenamefont
  {Schweitzer}}]{papanikolaou2022consensus}%
  \BibitemOpen
  \bibfield  {author} {\bibinfo {author} {\bibfnamefont {N.}~\bibnamefont
  {Papanikolaou}}, \bibinfo {author} {\bibfnamefont {G.}~\bibnamefont
  {Vaccario}}, \bibinfo {author} {\bibfnamefont {E.}~\bibnamefont {Hormann}},
  \bibinfo {author} {\bibfnamefont {R.}~\bibnamefont {Lambiotte}},\ and\
  \bibinfo {author} {\bibfnamefont {F.}~\bibnamefont {Schweitzer}},\ }\bibfield
   {title} {\enquote {\bibinfo {title} {Consensus from group interactions: An
  adaptive voter model on hypergraphs},}\ }\href
  {https://doi.org/10.1103/PhysRevE.105.054307} {\bibfield  {journal} {\bibinfo
   {journal} {Phys. Rev. E}\ }\textbf {\bibinfo {volume} {105}} (\bibinfo
  {year} {2022}),\ 10.1103/PhysRevE.105.054307}\BibitemShut {NoStop}%
\bibitem [{\citenamefont {Milgram}(1967)}]{milgram1967small}%
  \BibitemOpen
  \bibfield  {author} {\bibinfo {author} {\bibfnamefont {S.}~\bibnamefont
  {Milgram}},\ }\bibfield  {title} {\enquote {\bibinfo {title} {The small world
  problem},}\ }\href@noop {} {\bibfield  {journal} {\bibinfo  {journal}
  {Psychology today}\ }\textbf {\bibinfo {volume} {2}},\ \bibinfo {pages}
  {60--67} (\bibinfo {year} {1967})}\BibitemShut {NoStop}%
\bibitem [{Note2()}]{Note2}%
  \BibitemOpen
  \bibinfo {note} {See,
  https://research.fb.com/blog/2016/02/three-and-a-half-degrees-of-separation/}\BibitemShut
  {NoStop}%
\bibitem [{\citenamefont {Lorenz-Spreen}\ \emph {et~al.}(2022)\citenamefont
  {Lorenz-Spreen}, \citenamefont {Oswald}, \citenamefont {Lewandowsky},\ and\
  \citenamefont {Hertwig}}]{lorenz2022systematic}%
  \BibitemOpen
  \bibfield  {author} {\bibinfo {author} {\bibfnamefont {P.}~\bibnamefont
  {Lorenz-Spreen}}, \bibinfo {author} {\bibfnamefont {L.}~\bibnamefont
  {Oswald}}, \bibinfo {author} {\bibfnamefont {S.}~\bibnamefont
  {Lewandowsky}},\ and\ \bibinfo {author} {\bibfnamefont {R.}~\bibnamefont
  {Hertwig}},\ }\bibfield  {title} {\enquote {\bibinfo {title} {A systematic
  review of worldwide causal and correlational evidence on digital media and
  democracy},}\ }\href@noop {} {\bibfield  {journal} {\bibinfo  {journal}
  {Nature human behaviour}\ ,\ \bibinfo {pages} {1--28}} (\bibinfo {year}
  {2022})}\BibitemShut {NoStop}%
\bibitem [{\citenamefont {Deffuant}\ \emph {et~al.}(2000)\citenamefont
  {Deffuant}, \citenamefont {Neau}, \citenamefont {Amblard},\ and\
  \citenamefont {Weisbuch}}]{deffuant2000mixing}%
  \BibitemOpen
  \bibfield  {author} {\bibinfo {author} {\bibfnamefont {G.}~\bibnamefont
  {Deffuant}}, \bibinfo {author} {\bibfnamefont {D.}~\bibnamefont {Neau}},
  \bibinfo {author} {\bibfnamefont {F.}~\bibnamefont {Amblard}},\ and\ \bibinfo
  {author} {\bibfnamefont {G.}~\bibnamefont {Weisbuch}},\ }\bibfield  {title}
  {\enquote {\bibinfo {title} {Mixing beliefs among interacting agents},}\
  }\href@noop {} {\bibfield  {journal} {\bibinfo  {journal} {Advances in
  Complex Systems}\ }\textbf {\bibinfo {volume} {3}},\ \bibinfo {pages}
  {87--98} (\bibinfo {year} {2000})}\BibitemShut {NoStop}%
\bibitem [{\citenamefont {Baumann}\ \emph {et~al.}(2020)\citenamefont
  {Baumann}, \citenamefont {Lorenz-Spreen}, \citenamefont {Sokolov},\ and\
  \citenamefont {Starnini}}]{baumann2020modeling}%
  \BibitemOpen
  \bibfield  {author} {\bibinfo {author} {\bibfnamefont {F.}~\bibnamefont
  {Baumann}}, \bibinfo {author} {\bibfnamefont {P.}~\bibnamefont
  {Lorenz-Spreen}}, \bibinfo {author} {\bibfnamefont {I.~M.}\ \bibnamefont
  {Sokolov}},\ and\ \bibinfo {author} {\bibfnamefont {M.}~\bibnamefont
  {Starnini}},\ }\bibfield  {title} {\enquote {\bibinfo {title} {Modeling echo
  chambers and polarization dynamics in social networks},}\ }\href@noop {}
  {\bibfield  {journal} {\bibinfo  {journal} {Physical Review Letters}\
  }\textbf {\bibinfo {volume} {124}},\ \bibinfo {pages} {048301} (\bibinfo
  {year} {2020})}\BibitemShut {NoStop}%
\bibitem [{\citenamefont {McPherson}, \citenamefont {Smith-Lovin},\ and\
  \citenamefont {Cook}(2001)}]{mcpherson2001birds}%
  \BibitemOpen
  \bibfield  {author} {\bibinfo {author} {\bibfnamefont {M.}~\bibnamefont
  {McPherson}}, \bibinfo {author} {\bibfnamefont {L.}~\bibnamefont
  {Smith-Lovin}},\ and\ \bibinfo {author} {\bibfnamefont {J.~M.}\ \bibnamefont
  {Cook}},\ }\bibfield  {title} {\enquote {\bibinfo {title} {Birds of a
  feather: Homophily in social networks},}\ }\href@noop {} {\bibfield
  {journal} {\bibinfo  {journal} {Annual review of sociology}\ ,\ \bibinfo
  {pages} {415--444}} (\bibinfo {year} {2001})}\BibitemShut {NoStop}%
\bibitem [{\citenamefont {Baumann}\ \emph {et~al.}(2021)\citenamefont
  {Baumann}, \citenamefont {Lorenz-Spreen}, \citenamefont {Sokolov},\ and\
  \citenamefont {Starnini}}]{baumann2021emergence}%
  \BibitemOpen
  \bibfield  {author} {\bibinfo {author} {\bibfnamefont {F.}~\bibnamefont
  {Baumann}}, \bibinfo {author} {\bibfnamefont {P.}~\bibnamefont
  {Lorenz-Spreen}}, \bibinfo {author} {\bibfnamefont {I.~M.}\ \bibnamefont
  {Sokolov}},\ and\ \bibinfo {author} {\bibfnamefont {M.}~\bibnamefont
  {Starnini}},\ }\bibfield  {title} {\enquote {\bibinfo {title} {Emergence of
  polarized ideological opinions in multidimensional topic spaces},}\
  }\href@noop {} {\bibfield  {journal} {\bibinfo  {journal} {Physical Review
  X}\ }\textbf {\bibinfo {volume} {11}},\ \bibinfo {pages} {011012} (\bibinfo
  {year} {2021})}\BibitemShut {NoStop}%
\bibitem [{\citenamefont {Lorenz-Spreen}\ \emph {et~al.}(2019)\citenamefont
  {Lorenz-Spreen}, \citenamefont {M{\o}nsted}, \citenamefont {H{\"o}vel},\ and\
  \citenamefont {Lehmann}}]{lorenz2019accelerating}%
  \BibitemOpen
  \bibfield  {author} {\bibinfo {author} {\bibfnamefont {P.}~\bibnamefont
  {Lorenz-Spreen}}, \bibinfo {author} {\bibfnamefont {B.~M.}\ \bibnamefont
  {M{\o}nsted}}, \bibinfo {author} {\bibfnamefont {P.}~\bibnamefont
  {H{\"o}vel}},\ and\ \bibinfo {author} {\bibfnamefont {S.}~\bibnamefont
  {Lehmann}},\ }\bibfield  {title} {\enquote {\bibinfo {title} {Accelerating
  dynamics of collective attention},}\ }\href@noop {} {\bibfield  {journal}
  {\bibinfo  {journal} {Nature communications}\ }\textbf {\bibinfo {volume}
  {10}},\ \bibinfo {pages} {1--9} (\bibinfo {year} {2019})}\BibitemShut
  {NoStop}%
\bibitem [{\citenamefont {Simon}\ \emph {et~al.}(1971)\citenamefont {Simon}
  \emph {et~al.}}]{simon1971designing}%
  \BibitemOpen
  \bibfield  {author} {\bibinfo {author} {\bibfnamefont {H.~A.}\ \bibnamefont
  {Simon}} \emph {et~al.},\ }\bibfield  {title} {\enquote {\bibinfo {title}
  {Designing organizations for an information-rich world},}\ }\href@noop {}
  {\bibfield  {journal} {\bibinfo  {journal} {Computers, communications, and
  the public interest}\ }\textbf {\bibinfo {volume} {72}},\ \bibinfo {pages}
  {37} (\bibinfo {year} {1971})}\BibitemShut {NoStop}%
\bibitem [{\citenamefont {Bialek}(2020)}]{bialek2020does}%
  \BibitemOpen
  \bibfield  {author} {\bibinfo {author} {\bibfnamefont {J.}~\bibnamefont
  {Bialek}},\ }\bibfield  {title} {\enquote {\bibinfo {title} {What does the gb
  power outage on 9 august 2019 tell us about the current state of decarbonised
  power systems?}}\ }\href@noop {} {\bibfield  {journal} {\bibinfo  {journal}
  {Energy Policy}\ }\textbf {\bibinfo {volume} {146}},\ \bibinfo {pages}
  {111821} (\bibinfo {year} {2020})}\BibitemShut {NoStop}%
\bibitem [{\citenamefont {Buldyrev}\ \emph {et~al.}(2010)\citenamefont
  {Buldyrev}, \citenamefont {Parshani}, \citenamefont {Paul}, \citenamefont
  {Stanley},\ and\ \citenamefont {Havlin}}]{buldyrev2010catastrophic}%
  \BibitemOpen
  \bibfield  {author} {\bibinfo {author} {\bibfnamefont {S.~V.}\ \bibnamefont
  {Buldyrev}}, \bibinfo {author} {\bibfnamefont {R.}~\bibnamefont {Parshani}},
  \bibinfo {author} {\bibfnamefont {G.}~\bibnamefont {Paul}}, \bibinfo {author}
  {\bibfnamefont {H.~E.}\ \bibnamefont {Stanley}},\ and\ \bibinfo {author}
  {\bibfnamefont {S.}~\bibnamefont {Havlin}},\ }\bibfield  {title} {\enquote
  {\bibinfo {title} {Catastrophic cascade of failures in interdependent
  networks},}\ }\href@noop {} {\bibfield  {journal} {\bibinfo  {journal}
  {Nature}\ }\textbf {\bibinfo {volume} {464}},\ \bibinfo {pages} {1025--1028}
  (\bibinfo {year} {2010})}\BibitemShut {NoStop}%
\bibitem [{\citenamefont {Klinger}, \citenamefont {Landeg},\ and\ \citenamefont
  {Murray}(2014)}]{klinger2014power}%
  \BibitemOpen
  \bibfield  {author} {\bibinfo {author} {\bibfnamefont {C.}~\bibnamefont
  {Klinger}}, \bibinfo {author} {\bibfnamefont {O.}~\bibnamefont {Landeg}},\
  and\ \bibinfo {author} {\bibfnamefont {V.}~\bibnamefont {Murray}},\
  }\bibfield  {title} {\enquote {\bibinfo {title} {Power outages, extreme
  events and health: a systematic review of the literature from 2011-2012.}}\
  }\href@noop {} {\bibfield  {journal} {\bibinfo  {journal} {PLoS currents}\
  }\textbf {\bibinfo {volume} {6}},\ \bibinfo {pages} {ecurrents--dis}
  (\bibinfo {year} {2014})}\BibitemShut {NoStop}%
\bibitem [{\citenamefont {DW}(2021)}]{DW}%
  \BibitemOpen
  \bibfield  {author} {\bibinfo {author} {\bibnamefont {DW}},\ }\href@noop {}
  {\enquote {\bibinfo {title} {High price and range anxiety stops germans from
  buying e-cars},}\ }\bibinfo {howpublished} {Available at
  \href{https://www.dw.com/en/germany-cars-e-cars-e-mobility-charging-stations-electric-cars-emissions/a-57921124}{https://www.dw.com/en/germany-cars-e-cars-e-mobility-charging-stations-electric-cars-emissions/a-57921124}}
  (\bibinfo {year} {2021})\BibitemShut {NoStop}%
\bibitem [{\citenamefont {Ferroukhi}\ \emph {et~al.}(2020)\citenamefont
  {Ferroukhi}, \citenamefont {Frankl}, \citenamefont {Adib} \emph
  {et~al.}}]{ferroukhi2020renewable}%
  \BibitemOpen
  \bibfield  {author} {\bibinfo {author} {\bibfnamefont {R.}~\bibnamefont
  {Ferroukhi}}, \bibinfo {author} {\bibfnamefont {P.}~\bibnamefont {Frankl}},
  \bibinfo {author} {\bibfnamefont {R.}~\bibnamefont {Adib}}, \emph {et~al.},\
  }\bibfield  {title} {\enquote {\bibinfo {title} {Renewable energy policies in
  a time of transition: Heating and cooling},}\ }\href@noop {} {\  (\bibinfo
  {year} {2020})}\BibitemShut {NoStop}%
\bibitem [{\citenamefont {Anvari}\ \emph
  {et~al.}(2016{\natexlab{a}})\citenamefont {Anvari}, \citenamefont {Lohmann},
  \citenamefont {W{\"a}chter}, \citenamefont {Milan}, \citenamefont {Lorenz},
  \citenamefont {Heinemann}, \citenamefont {Tabar},\ and\ \citenamefont
  {Peinke}}]{anvari2016short}%
  \BibitemOpen
  \bibfield  {author} {\bibinfo {author} {\bibfnamefont {M.}~\bibnamefont
  {Anvari}}, \bibinfo {author} {\bibfnamefont {G.}~\bibnamefont {Lohmann}},
  \bibinfo {author} {\bibfnamefont {M.}~\bibnamefont {W{\"a}chter}}, \bibinfo
  {author} {\bibfnamefont {P.}~\bibnamefont {Milan}}, \bibinfo {author}
  {\bibfnamefont {E.}~\bibnamefont {Lorenz}}, \bibinfo {author} {\bibfnamefont
  {D.}~\bibnamefont {Heinemann}}, \bibinfo {author} {\bibfnamefont {M.~R.~R.}\
  \bibnamefont {Tabar}},\ and\ \bibinfo {author} {\bibfnamefont
  {J.}~\bibnamefont {Peinke}},\ }\bibfield  {title} {\enquote {\bibinfo {title}
  {Short term fluctuations of wind and solar power systems},}\ }\href@noop {}
  {\bibfield  {journal} {\bibinfo  {journal} {New Journal of Physics}\ }\textbf
  {\bibinfo {volume} {18}},\ \bibinfo {pages} {063027} (\bibinfo {year}
  {2016}{\natexlab{a}})}\BibitemShut {NoStop}%
\bibitem [{\citenamefont {Apt}(2007)}]{apt2007spectrum}%
  \BibitemOpen
  \bibfield  {author} {\bibinfo {author} {\bibfnamefont {J.}~\bibnamefont
  {Apt}},\ }\bibfield  {title} {\enquote {\bibinfo {title} {The spectrum of
  power from wind turbines},}\ }\href@noop {} {\bibfield  {journal} {\bibinfo
  {journal} {Journal of Power Sources}\ }\textbf {\bibinfo {volume} {169}},\
  \bibinfo {pages} {369--374} (\bibinfo {year} {2007})}\BibitemShut {NoStop}%
\bibitem [{\citenamefont {Curtright}\ and\ \citenamefont
  {Apt}(2008)}]{curtright2008character}%
  \BibitemOpen
  \bibfield  {author} {\bibinfo {author} {\bibfnamefont {A.~E.}\ \bibnamefont
  {Curtright}}\ and\ \bibinfo {author} {\bibfnamefont {J.}~\bibnamefont
  {Apt}},\ }\bibfield  {title} {\enquote {\bibinfo {title} {The character of
  power output from utility-scale photovoltaic systems},}\ }\href@noop {}
  {\bibfield  {journal} {\bibinfo  {journal} {Progress in Photovoltaics:
  Research and Applications}\ }\textbf {\bibinfo {volume} {16}},\ \bibinfo
  {pages} {241--247} (\bibinfo {year} {2008})}\BibitemShut {NoStop}%
\bibitem [{\citenamefont {Luo}\ and\ \citenamefont
  {Ooi}(2006)}]{luo2006frequency}%
  \BibitemOpen
  \bibfield  {author} {\bibinfo {author} {\bibfnamefont {C.}~\bibnamefont
  {Luo}}\ and\ \bibinfo {author} {\bibfnamefont {B.-T.}\ \bibnamefont {Ooi}},\
  }\bibfield  {title} {\enquote {\bibinfo {title} {Frequency deviation of
  thermal power plants due to wind farms},}\ }\href@noop {} {\bibfield
  {journal} {\bibinfo  {journal} {IEEE Transactions on Energy Conversion}\
  }\textbf {\bibinfo {volume} {21}},\ \bibinfo {pages} {708--716} (\bibinfo
  {year} {2006})}\BibitemShut {NoStop}%
\bibitem [{\citenamefont {Baile}\ and\ \citenamefont
  {Muzy}(2010)}]{baile2010spatial}%
  \BibitemOpen
  \bibfield  {author} {\bibinfo {author} {\bibfnamefont {R.}~\bibnamefont
  {Baile}}\ and\ \bibinfo {author} {\bibfnamefont {J.-F.}\ \bibnamefont
  {Muzy}},\ }\bibfield  {title} {\enquote {\bibinfo {title} {Spatial
  intermittency of surface layer wind fluctuations at mesoscale range},}\
  }\href@noop {} {\bibfield  {journal} {\bibinfo  {journal} {Physical review
  letters}\ }\textbf {\bibinfo {volume} {105}},\ \bibinfo {pages} {254501}
  (\bibinfo {year} {2010})}\BibitemShut {NoStop}%
\bibitem [{\citenamefont {Wood}\ and\ \citenamefont
  {Field}(2011)}]{wood2011distribution}%
  \BibitemOpen
  \bibfield  {author} {\bibinfo {author} {\bibfnamefont {R.}~\bibnamefont
  {Wood}}\ and\ \bibinfo {author} {\bibfnamefont {P.~R.}\ \bibnamefont
  {Field}},\ }\bibfield  {title} {\enquote {\bibinfo {title} {The distribution
  of cloud horizontal sizes},}\ }\href@noop {} {\bibfield  {journal} {\bibinfo
  {journal} {Journal of Climate}\ }\textbf {\bibinfo {volume} {24}},\ \bibinfo
  {pages} {4800--4816} (\bibinfo {year} {2011})}\BibitemShut {NoStop}%
\bibitem [{\citenamefont {Haehne}\ \emph {et~al.}(2018)\citenamefont {Haehne},
  \citenamefont {Schottler}, \citenamefont {Waechter}, \citenamefont {Peinke},\
  and\ \citenamefont {Kamps}}]{haehne2018footprint}%
  \BibitemOpen
  \bibfield  {author} {\bibinfo {author} {\bibfnamefont {H.}~\bibnamefont
  {Haehne}}, \bibinfo {author} {\bibfnamefont {J.}~\bibnamefont {Schottler}},
  \bibinfo {author} {\bibfnamefont {M.}~\bibnamefont {Waechter}}, \bibinfo
  {author} {\bibfnamefont {J.}~\bibnamefont {Peinke}},\ and\ \bibinfo {author}
  {\bibfnamefont {O.}~\bibnamefont {Kamps}},\ }\bibfield  {title} {\enquote
  {\bibinfo {title} {The footprint of atmospheric turbulence in power grid
  frequency measurements},}\ }\href@noop {} {\bibfield  {journal} {\bibinfo
  {journal} {EPL (Europhysics Letters)}\ }\textbf {\bibinfo {volume} {121}},\
  \bibinfo {pages} {30001} (\bibinfo {year} {2018})}\BibitemShut {NoStop}%
\bibitem [{\citenamefont {Schilling}(2021)}]{EEA2021}%
  \BibitemOpen
  \bibfield  {author} {\bibinfo {author} {\bibfnamefont {S.}~\bibnamefont
  {Schilling}},\ }\href@noop {} {\enquote {\bibinfo {title} {Final energy
  consumption by sector and fuel. tech. rep., european environment agency.}}\
  }\bibinfo {howpublished} {Available at
  \url{https://www.eea.europa.eu/ims/primary-and-final-energy-consumption}}
  (\bibinfo {year} {2021})\BibitemShut {NoStop}%
\bibitem [{\citenamefont {Monacchi}\ \emph {et~al.}(2014)\citenamefont
  {Monacchi}, \citenamefont {Egarter}, \citenamefont {Elmenreich},
  \citenamefont {D'Alessandro},\ and\ \citenamefont
  {Tonello}}]{monacchi2014greend}%
  \BibitemOpen
  \bibfield  {author} {\bibinfo {author} {\bibfnamefont {A.}~\bibnamefont
  {Monacchi}}, \bibinfo {author} {\bibfnamefont {D.}~\bibnamefont {Egarter}},
  \bibinfo {author} {\bibfnamefont {W.}~\bibnamefont {Elmenreich}}, \bibinfo
  {author} {\bibfnamefont {S.}~\bibnamefont {D'Alessandro}},\ and\ \bibinfo
  {author} {\bibfnamefont {A.~M.}\ \bibnamefont {Tonello}},\ }\bibfield
  {title} {\enquote {\bibinfo {title} {Greend: An energy consumption dataset of
  households in italy and austria},}\ }in\ \href@noop {} {\emph {\bibinfo
  {booktitle} {2014 IEEE International Conference on Smart Grid Communications
  (SmartGridComm)}}}\ (\bibinfo {organization} {IEEE},\ \bibinfo {year}
  {2014})\ pp.\ \bibinfo {pages} {511--516}\BibitemShut {NoStop}%
\bibitem [{\citenamefont {Wright}\ and\ \citenamefont
  {Firth}(2007)}]{wright2007nature}%
  \BibitemOpen
  \bibfield  {author} {\bibinfo {author} {\bibfnamefont {A.}~\bibnamefont
  {Wright}}\ and\ \bibinfo {author} {\bibfnamefont {S.}~\bibnamefont {Firth}},\
  }\bibfield  {title} {\enquote {\bibinfo {title} {The nature of domestic
  electricity-loads and effects of time averaging on statistics and on-site
  generation calculations},}\ }\href@noop {} {\bibfield  {journal} {\bibinfo
  {journal} {Applied Energy}\ }\textbf {\bibinfo {volume} {84}},\ \bibinfo
  {pages} {389--403} (\bibinfo {year} {2007})}\BibitemShut {NoStop}%
\bibitem [{\citenamefont {Marszal-Pomianowska}, \citenamefont {Heiselberg},\
  and\ \citenamefont {Larsen}(2016)}]{marszal2016household}%
  \BibitemOpen
  \bibfield  {author} {\bibinfo {author} {\bibfnamefont {A.}~\bibnamefont
  {Marszal-Pomianowska}}, \bibinfo {author} {\bibfnamefont {P.}~\bibnamefont
  {Heiselberg}},\ and\ \bibinfo {author} {\bibfnamefont {O.~K.}\ \bibnamefont
  {Larsen}},\ }\bibfield  {title} {\enquote {\bibinfo {title} {Household
  electricity demand profiles--a high-resolution load model to facilitate
  modelling of energy flexible buildings},}\ }\href@noop {} {\bibfield
  {journal} {\bibinfo  {journal} {Energy}\ }\textbf {\bibinfo {volume} {103}},\
  \bibinfo {pages} {487--501} (\bibinfo {year} {2016})}\BibitemShut {NoStop}%
\bibitem [{\citenamefont {L{\'o}pez}\ \emph {et~al.}(2015)\citenamefont
  {L{\'o}pez}, \citenamefont {De~La~Torre}, \citenamefont {Mart{\'\i}n},\ and\
  \citenamefont {Aguado}}]{lopez2015demand}%
  \BibitemOpen
  \bibfield  {author} {\bibinfo {author} {\bibfnamefont {M.~A.}\ \bibnamefont
  {L{\'o}pez}}, \bibinfo {author} {\bibfnamefont {S.}~\bibnamefont
  {De~La~Torre}}, \bibinfo {author} {\bibfnamefont {S.}~\bibnamefont
  {Mart{\'\i}n}},\ and\ \bibinfo {author} {\bibfnamefont {J.~A.}\ \bibnamefont
  {Aguado}},\ }\bibfield  {title} {\enquote {\bibinfo {title} {Demand-side
  management in smart grid operation considering electric vehicles load
  shifting and vehicle-to-grid support},}\ }\href@noop {} {\bibfield  {journal}
  {\bibinfo  {journal} {International Journal of Electrical Power \& Energy
  Systems}\ }\textbf {\bibinfo {volume} {64}},\ \bibinfo {pages} {689--698}
  (\bibinfo {year} {2015})}\BibitemShut {NoStop}%
\bibitem [{\citenamefont {Logenthiran}, \citenamefont {Srinivasan},\ and\
  \citenamefont {Shun}(2012)}]{logenthiran2012demand}%
  \BibitemOpen
  \bibfield  {author} {\bibinfo {author} {\bibfnamefont {T.}~\bibnamefont
  {Logenthiran}}, \bibinfo {author} {\bibfnamefont {D.}~\bibnamefont
  {Srinivasan}},\ and\ \bibinfo {author} {\bibfnamefont {T.~Z.}\ \bibnamefont
  {Shun}},\ }\bibfield  {title} {\enquote {\bibinfo {title} {Demand side
  management in smart grid using heuristic optimization},}\ }\href@noop {}
  {\bibfield  {journal} {\bibinfo  {journal} {IEEE transactions on smart grid}\
  }\textbf {\bibinfo {volume} {3}},\ \bibinfo {pages} {1244--1252} (\bibinfo
  {year} {2012})}\BibitemShut {NoStop}%
\bibitem [{\citenamefont {Anvari}\ \emph {et~al.}(2017)\citenamefont {Anvari},
  \citenamefont {Werther}, \citenamefont {Lohmann}, \citenamefont
  {W{\"a}chter}, \citenamefont {Peinke},\ and\ \citenamefont
  {Beck}}]{anvari2017suppressing}%
  \BibitemOpen
  \bibfield  {author} {\bibinfo {author} {\bibfnamefont {M.}~\bibnamefont
  {Anvari}}, \bibinfo {author} {\bibfnamefont {B.}~\bibnamefont {Werther}},
  \bibinfo {author} {\bibfnamefont {G.}~\bibnamefont {Lohmann}}, \bibinfo
  {author} {\bibfnamefont {M.}~\bibnamefont {W{\"a}chter}}, \bibinfo {author}
  {\bibfnamefont {J.}~\bibnamefont {Peinke}},\ and\ \bibinfo {author}
  {\bibfnamefont {H.-P.}\ \bibnamefont {Beck}},\ }\bibfield  {title} {\enquote
  {\bibinfo {title} {Suppressing power output fluctuations of photovoltaic
  power plants},}\ }\href@noop {} {\bibfield  {journal} {\bibinfo  {journal}
  {Solar Energy}\ }\textbf {\bibinfo {volume} {157}},\ \bibinfo {pages}
  {735--743} (\bibinfo {year} {2017})}\BibitemShut {NoStop}%
\bibitem [{\citenamefont {Anvari}\ \emph
  {et~al.}(2016{\natexlab{b}})\citenamefont {Anvari}, \citenamefont {Tabar},
  \citenamefont {Peinke},\ and\ \citenamefont
  {Lehnertz}}]{anvari2016disentangling}%
  \BibitemOpen
  \bibfield  {author} {\bibinfo {author} {\bibfnamefont {M.}~\bibnamefont
  {Anvari}}, \bibinfo {author} {\bibfnamefont {M.}~\bibnamefont {Tabar}},
  \bibinfo {author} {\bibfnamefont {J.}~\bibnamefont {Peinke}},\ and\ \bibinfo
  {author} {\bibfnamefont {K.}~\bibnamefont {Lehnertz}},\ }\bibfield  {title}
  {\enquote {\bibinfo {title} {Disentangling the stochastic behavior of complex
  time series},}\ }\href@noop {} {\bibfield  {journal} {\bibinfo  {journal}
  {Scientific reports}\ }\textbf {\bibinfo {volume} {6}},\ \bibinfo {pages}
  {1--12} (\bibinfo {year} {2016}{\natexlab{b}})}\BibitemShut {NoStop}%
\bibitem [{\citenamefont {Anvari}, \citenamefont {Proedrou},\ and\
  \citenamefont {Schäfer}(2022)}]{anvari2022}%
  \BibitemOpen
  \bibfield  {author} {\bibinfo {author} {\bibfnamefont {M.}~\bibnamefont
  {Anvari}}, \bibinfo {author} {\bibfnamefont {E.}~\bibnamefont {Proedrou}},\
  and\ \bibinfo {author} {\bibfnamefont {B.~e.~a.}\ \bibnamefont {Schäfer}},\
  }\bibfield  {title} {\enquote {\bibinfo {title} {Data-driven load profiles
  and the dynamics of residential electricity consumption},}\ }\href@noop {}
  {\bibfield  {journal} {\bibinfo  {journal} {Nature Communication}\ }\textbf
  {\bibinfo {volume} {13}},\ \bibinfo {pages} {4593} (\bibinfo {year}
  {2022})}\BibitemShut {NoStop}%
\bibitem [{\citenamefont {Bitterer}\ and\ \citenamefont {{Prof. Dr. habil. B.
  Schieferdecker}}(2001)}]{SLP}%
  \BibitemOpen
  \bibfield  {author} {\bibinfo {author} {\bibfnamefont {R.}~\bibnamefont
  {Bitterer}}\ and\ \bibinfo {author} {\bibnamefont {{Prof. Dr. habil. B.
  Schieferdecker}}},\ }\href@noop {} {\enquote {\bibinfo {title}
  {{Repr{\"{a}}sentative VDEW-Lastprofile Aktionsplan Wettbewerb, M-32/99}},}\
  }\bibinfo {type} {Tech. Rep.}\ (\bibinfo  {institution} {VDEW},\ \bibinfo
  {address} {Stresemannallee 23 D-60596 Frankfurt /M},\ \bibinfo {year}
  {2001})\BibitemShut {NoStop}%
\bibitem [{\citenamefont {Parti}\ and\ \citenamefont
  {Parti}(1980)}]{parti1980total}%
  \BibitemOpen
  \bibfield  {author} {\bibinfo {author} {\bibfnamefont {M.}~\bibnamefont
  {Parti}}\ and\ \bibinfo {author} {\bibfnamefont {C.}~\bibnamefont {Parti}},\
  }\bibfield  {title} {\enquote {\bibinfo {title} {The total and
  appliance-specific conditional demand for electricity in the household
  sector},}\ }\href@noop {} {\bibfield  {journal} {\bibinfo  {journal} {The
  Bell journal of economics}\ ,\ \bibinfo {pages} {309--321}} (\bibinfo {year}
  {1980})}\BibitemShut {NoStop}%
\bibitem [{\citenamefont {Anvari}, \citenamefont {Hellmann},\ and\
  \citenamefont {Zhang}(2020)}]{anvari2020introduction}%
  \BibitemOpen
  \bibfield  {author} {\bibinfo {author} {\bibfnamefont {M.}~\bibnamefont
  {Anvari}}, \bibinfo {author} {\bibfnamefont {F.}~\bibnamefont {Hellmann}},\
  and\ \bibinfo {author} {\bibfnamefont {X.}~\bibnamefont {Zhang}},\ }\bibfield
   {title} {\enquote {\bibinfo {title} {Introduction to focus issue: Dynamics
  of modern power grids},}\ }\href@noop {} {\bibfield  {journal} {\bibinfo
  {journal} {Chaos: An Interdisciplinary Journal of Nonlinear Science}\
  }\textbf {\bibinfo {volume} {30}},\ \bibinfo {pages} {063140} (\bibinfo
  {year} {2020})}\BibitemShut {NoStop}%
\bibitem [{\citenamefont {Auer}\ \emph {et~al.}(2017)\citenamefont {Auer},
  \citenamefont {Hellmann}, \citenamefont {Krause},\ and\ \citenamefont
  {Kurths}}]{auer2017stability}%
  \BibitemOpen
  \bibfield  {author} {\bibinfo {author} {\bibfnamefont {S.}~\bibnamefont
  {Auer}}, \bibinfo {author} {\bibfnamefont {F.}~\bibnamefont {Hellmann}},
  \bibinfo {author} {\bibfnamefont {M.}~\bibnamefont {Krause}},\ and\ \bibinfo
  {author} {\bibfnamefont {J.}~\bibnamefont {Kurths}},\ }\bibfield  {title}
  {\enquote {\bibinfo {title} {Stability of synchrony against local
  intermittent fluctuations in tree-like power grids},}\ }\href@noop {}
  {\bibfield  {journal} {\bibinfo  {journal} {Chaos: An Interdisciplinary
  Journal of Nonlinear Science}\ }\textbf {\bibinfo {volume} {27}},\ \bibinfo
  {pages} {127003} (\bibinfo {year} {2017})}\BibitemShut {NoStop}%
\bibitem [{\citenamefont {Li}, \citenamefont {Jo{\'o}s},\ and\ \citenamefont
  {Abbey}(2006)}]{li2006wind}%
  \BibitemOpen
  \bibfield  {author} {\bibinfo {author} {\bibfnamefont {W.}~\bibnamefont
  {Li}}, \bibinfo {author} {\bibfnamefont {G.}~\bibnamefont {Jo{\'o}s}},\ and\
  \bibinfo {author} {\bibfnamefont {C.}~\bibnamefont {Abbey}},\ }\bibfield
  {title} {\enquote {\bibinfo {title} {Wind power impact on system frequency
  deviation and an ess based power filtering algorithm solution},}\ }in\
  \href@noop {} {\emph {\bibinfo {booktitle} {2006 IEEE PES Power Systems
  Conference and Exposition}}}\ (\bibinfo {organization} {IEEE},\ \bibinfo
  {year} {2006})\ pp.\ \bibinfo {pages} {2077--2084}\BibitemShut {NoStop}%
\bibitem [{\citenamefont {Alvarez}\ \emph {et~al.}(1980)\citenamefont
  {Alvarez}, \citenamefont {Alvarez}, \citenamefont {Asaro},\ and\
  \citenamefont {Michel}}]{A1980}%
  \BibitemOpen
  \bibfield  {author} {\bibinfo {author} {\bibfnamefont {L.~W.}\ \bibnamefont
  {Alvarez}}, \bibinfo {author} {\bibfnamefont {W.}~\bibnamefont {Alvarez}},
  \bibinfo {author} {\bibfnamefont {F.}~\bibnamefont {Asaro}},\ and\ \bibinfo
  {author} {\bibfnamefont {H.~V.}\ \bibnamefont {Michel}},\ }\bibfield  {title}
  {\enquote {\bibinfo {title} {Extraterrestrial cause for the
  cretaceous-tertiary extinction},}\ }\href
  {https://doi.org/10.1126/science.208.4448.1095} {\bibfield  {journal}
  {\bibinfo  {journal} {Science}\ }\textbf {\bibinfo {volume} {208}},\ \bibinfo
  {pages} {1095--1108} (\bibinfo {year} {1980})},\ \Eprint
  {https://arxiv.org/abs/https://www.science.org/doi/pdf/10.1126/science.208.4448.1095}
  {https://www.science.org/doi/pdf/10.1126/science.208.4448.1095} \BibitemShut
  {NoStop}%
\bibitem [{\citenamefont {Brugger}, \citenamefont {Feulner},\ and\
  \citenamefont {Petri}(2017)}]{B16}%
  \BibitemOpen
  \bibfield  {author} {\bibinfo {author} {\bibfnamefont {J.}~\bibnamefont
  {Brugger}}, \bibinfo {author} {\bibfnamefont {G.}~\bibnamefont {Feulner}},\
  and\ \bibinfo {author} {\bibfnamefont {S.}~\bibnamefont {Petri}},\ }\bibfield
   {title} {\enquote {\bibinfo {title} {Baby, it's cold outside: Climate model
  simulations of the effects of the asteroid impact at the end of the
  cretaceous},}\ }\href@noop {} {\bibfield  {journal} {\bibinfo  {journal}
  {Geophysical Research Letters}\ }\textbf {\bibinfo {volume} {44}},\ \bibinfo
  {pages} {419--427} (\bibinfo {year} {2017})}\BibitemShut {NoStop}%
\bibitem [{\citenamefont {Sparks}(1978)}]{N1978}%
  \BibitemOpen
  \bibfield  {author} {\bibinfo {author} {\bibfnamefont {R.}~\bibnamefont
  {Sparks}},\ }\bibfield  {title} {\enquote {\bibinfo {title} {Gas release
  rates from pyroclastic flows: a assessment of the role of fluidisation in
  their emplacement},}\ }\href@noop {} {\bibfield  {journal} {\bibinfo
  {journal} {Bulletin Volcanologique}\ }\textbf {\bibinfo {volume} {41}},\
  \bibinfo {pages} {1--13} (\bibinfo {year} {1978})}\BibitemShut {NoStop}%
\bibitem [{\citenamefont {Williams}\ \emph {et~al.}(2009)\citenamefont
  {Williams}, \citenamefont {Ambrose}, \citenamefont {van~der Kaars},
  \citenamefont {Ruehlemann}, \citenamefont {Chattopadhyaya}, \citenamefont
  {Pal},\ and\ \citenamefont {Chauhan}}]{W09}%
  \BibitemOpen
  \bibfield  {author} {\bibinfo {author} {\bibfnamefont {M.~A.}\ \bibnamefont
  {Williams}}, \bibinfo {author} {\bibfnamefont {S.~H.}\ \bibnamefont
  {Ambrose}}, \bibinfo {author} {\bibfnamefont {S.}~\bibnamefont {van~der
  Kaars}}, \bibinfo {author} {\bibfnamefont {C.}~\bibnamefont {Ruehlemann}},
  \bibinfo {author} {\bibfnamefont {U.}~\bibnamefont {Chattopadhyaya}},
  \bibinfo {author} {\bibfnamefont {J.}~\bibnamefont {Pal}},\ and\ \bibinfo
  {author} {\bibfnamefont {P.~R.}\ \bibnamefont {Chauhan}},\ }\bibfield
  {title} {\enquote {\bibinfo {title} {Environmental impact of the 73 ka toba
  super-eruption in south asia},}\ }\href@noop {} {\bibfield  {journal}
  {\bibinfo  {journal} {Palaeogeography, Palaeoclimatology, Palaeoecology}\
  }\textbf {\bibinfo {volume} {284}},\ \bibinfo {pages} {295--314} (\bibinfo
  {year} {2009})}\BibitemShut {NoStop}%
\bibitem [{\citenamefont {Armstrong~McKay}\ \emph {et~al.}(2022)\citenamefont
  {Armstrong~McKay}, \citenamefont {Staal}, \citenamefont {Abrams},
  \citenamefont {Winkelmann}, \citenamefont {Sakschewski}, \citenamefont
  {Loriani}, \citenamefont {Fetzer}, \citenamefont {Cornell}, \citenamefont
  {Rockstr{\"o}m},\ and\ \citenamefont {Lenton}}]{M2022}%
  \BibitemOpen
  \bibfield  {author} {\bibinfo {author} {\bibfnamefont {D.~I.}\ \bibnamefont
  {Armstrong~McKay}}, \bibinfo {author} {\bibfnamefont {A.}~\bibnamefont
  {Staal}}, \bibinfo {author} {\bibfnamefont {J.~F.}\ \bibnamefont {Abrams}},
  \bibinfo {author} {\bibfnamefont {R.}~\bibnamefont {Winkelmann}}, \bibinfo
  {author} {\bibfnamefont {B.}~\bibnamefont {Sakschewski}}, \bibinfo {author}
  {\bibfnamefont {S.}~\bibnamefont {Loriani}}, \bibinfo {author} {\bibfnamefont
  {I.}~\bibnamefont {Fetzer}}, \bibinfo {author} {\bibfnamefont {S.~E.}\
  \bibnamefont {Cornell}}, \bibinfo {author} {\bibfnamefont {J.}~\bibnamefont
  {Rockstr{\"o}m}},\ and\ \bibinfo {author} {\bibfnamefont {T.~M.}\
  \bibnamefont {Lenton}},\ }\bibfield  {title} {\enquote {\bibinfo {title}
  {Exceeding 1.5$\,^{\circ}$c global warming could trigger multiple climate
  tipping points.}}\ }\href {https://doi.org/10.1126/science.abn7950}
  {\bibfield  {journal} {\bibinfo  {journal} {Science}\ }\textbf {\bibinfo
  {volume} {377}},\ \bibinfo {pages} {eabn7950} (\bibinfo {year}
  {2022})}\BibitemShut {NoStop}%
\bibitem [{\citenamefont {{Masson-Delmotte, V. et
  al.}}(2021)}]{Intergouvernemental_panel2021}%
  \BibitemOpen
  \bibfield  {author} {\bibinfo {author} {\bibnamefont {{Masson-Delmotte, V. et
  al.}}},\ }\href@noop {} {\emph {\bibinfo {title} {Climate change 2021: : The
  Physical Science Basis. Contribution of Working Group I to the Sixth
  Assessment Report of the Intergovernmental Panel on Climate Change}}}\
  (\bibinfo {year} {2021})\BibitemShut {NoStop}%
\bibitem [{\citenamefont {Lenton}\ \emph {et~al.}(2008)\citenamefont {Lenton},
  \citenamefont {Held}, \citenamefont {Kriegler}, \citenamefont {Hall},
  \citenamefont {Lucht}, \citenamefont {Rahmstorf},\ and\ \citenamefont
  {Schellnhuber}}]{L2008}%
  \BibitemOpen
  \bibfield  {author} {\bibinfo {author} {\bibfnamefont {T.~M.}\ \bibnamefont
  {Lenton}}, \bibinfo {author} {\bibfnamefont {H.}~\bibnamefont {Held}},
  \bibinfo {author} {\bibfnamefont {E.}~\bibnamefont {Kriegler}}, \bibinfo
  {author} {\bibfnamefont {J.~W.}\ \bibnamefont {Hall}}, \bibinfo {author}
  {\bibfnamefont {W.}~\bibnamefont {Lucht}}, \bibinfo {author} {\bibfnamefont
  {S.}~\bibnamefont {Rahmstorf}},\ and\ \bibinfo {author} {\bibfnamefont
  {H.~J.}\ \bibnamefont {Schellnhuber}},\ }\bibfield  {title} {\enquote
  {\bibinfo {title} {Tipping elements in the earth's climate system},}\
  }\href@noop {} {\bibfield  {journal} {\bibinfo  {journal} {Proceedings of the
  national Academy of Sciences}\ }\textbf {\bibinfo {volume} {105}},\ \bibinfo
  {pages} {1786--1793} (\bibinfo {year} {2008})}\BibitemShut {NoStop}%
\bibitem [{\citenamefont {Wunderling}\ \emph {et~al.}(2021)\citenamefont
  {Wunderling}, \citenamefont {Donges}, \citenamefont {Kurths},\ and\
  \citenamefont {Winkelmann}}]{W2021}%
  \BibitemOpen
  \bibfield  {author} {\bibinfo {author} {\bibfnamefont {N.}~\bibnamefont
  {Wunderling}}, \bibinfo {author} {\bibfnamefont {J.~F.}\ \bibnamefont
  {Donges}}, \bibinfo {author} {\bibfnamefont {J.}~\bibnamefont {Kurths}},\
  and\ \bibinfo {author} {\bibfnamefont {R.}~\bibnamefont {Winkelmann}},\
  }\bibfield  {title} {\enquote {\bibinfo {title} {Interacting tipping elements
  increase risk of climate domino effects under global warming},}\ }\href
  {https://doi.org/10.5194/esd-12-601-2021} {\bibfield  {journal} {\bibinfo
  {journal} {Earth System Dynamics}\ }\textbf {\bibinfo {volume} {12}},\
  \bibinfo {pages} {601--619} (\bibinfo {year} {2021})}\BibitemShut {NoStop}%
\bibitem [{\citenamefont {Liu}\ \emph {et~al.}(2023)\citenamefont {Liu},
  \citenamefont {Chen}, \citenamefont {Yang}, \citenamefont {Meng},
  \citenamefont {Wang}, \citenamefont {Ludescher}, \citenamefont {Fan},
  \citenamefont {Yang}, \citenamefont {Chen}, \citenamefont {Kurths},
  \citenamefont {Chen}, \citenamefont {Havlin},\ and\ \citenamefont
  {Schellnhuber}}]{L2023}%
  \BibitemOpen
  \bibfield  {author} {\bibinfo {author} {\bibfnamefont {T.}~\bibnamefont
  {Liu}}, \bibinfo {author} {\bibfnamefont {D.}~\bibnamefont {Chen}}, \bibinfo
  {author} {\bibfnamefont {L.}~\bibnamefont {Yang}}, \bibinfo {author}
  {\bibfnamefont {J.}~\bibnamefont {Meng}}, \bibinfo {author} {\bibfnamefont
  {Z.}~\bibnamefont {Wang}}, \bibinfo {author} {\bibfnamefont {J.}~\bibnamefont
  {Ludescher}}, \bibinfo {author} {\bibfnamefont {J.}~\bibnamefont {Fan}},
  \bibinfo {author} {\bibfnamefont {S.}~\bibnamefont {Yang}}, \bibinfo {author}
  {\bibfnamefont {D.}~\bibnamefont {Chen}}, \bibinfo {author} {\bibfnamefont
  {J.}~\bibnamefont {Kurths}}, \bibinfo {author} {\bibfnamefont
  {X.}~\bibnamefont {Chen}}, \bibinfo {author} {\bibfnamefont {S.}~\bibnamefont
  {Havlin}},\ and\ \bibinfo {author} {\bibfnamefont {H.~J.}\ \bibnamefont
  {Schellnhuber}},\ }\bibfield  {title} {\enquote {\bibinfo {title}
  {Teleconnections among tipping elements in the earth system},}\ }\href
  {https://doi.org/10.1038/s41558-022-01558-4} {\bibfield  {journal} {\bibinfo
  {journal} {Nature Climate Change}\ } (\bibinfo {year} {2023}),\
  10.1038/s41558-022-01558-4}\BibitemShut {NoStop}%
\end{thebibliography}%
\end{document}